%% file: rop2.tex
\documentclass[12pt,final]{iopart}
\usepackage{iopams}  
\usepackage{har2nat}
\usepackage{hyperref}

\usepackage{graphicx}

\usepackage{epstopdf}

\usepackage[table]{xcolor}
\usepackage{array}
\usepackage{longtable}

  \newenvironment{Table_app}[3]{%
  \longtable{%
    |A{#1}{1.5}
    |>{\centering$\displaystyle}A{#2}{1}<{$}
    |>{\centering$\displaystyle}A{#3}{1}<{$}
    |}\hline\ignorespaces}{%
  \endlongtable\ignorespacesafterend}

\graphicspath{{.}}

\def\be{\begin{equation}}
\def\ee{\end{equation}}
\def\ba{\begin{eqnarray}}
\def\ea{\end{eqnarray}}
\def\bas{\begin{eqnarray*}}
\def\eas{\end{eqnarray*}}

\newcommand{\vlowk}{\ensuremath{V_{{\rm low\,}k}}}
\newcommand{\fmi}{\, \rm{fm}^{-1}}

\newcommand{\la}{\langle}
\newcommand{\ra}{\rangle}
\newcommand{\adag}{a^\dagger}

\begin{document}

\title[Coupled-cluster computations of atomic nuclei]{Coupled-cluster computations of atomic nuclei}

\author{G.~Hagen$^{1,2}$, T.~Papenbrock$^{2,1}$, M.~Hjorth-Jensen$^{3,4}$, and 
D.~J.~Dean$^1$}

\address{$^1$Physics Division, Oak Ridge National Laboratory, Oak
  Ridge, TN 37831 USA}

\address{$^2$Department of Physics and Astronomy, University of
  Tennessee, Knoxville, TN 37996, USA}

\address{$^3$National Superconducting Cyclotron Laboratory and
  Department of Physics and Astronomy, Michigan State University, East
  Lansing 48824 MI, USA}

\address{$^4$Department of Physics and Center of Mathematics for
  Applications, University of Oslo, N-0316 Oslo, Norway}

\eads{\mailto{hageng@ornl.gov}, \mailto{tpapenbr@utk.edu}, 
  \mailto{morten.hjorth-jensen@fys.uio.no}, \mailto{deandj@ornl.gov}}
\begin{abstract}
  In the past decade, coupled-cluster theory has seen a renaissance in
  nuclear physics, with computations of neutron-rich and medium-mass
  nuclei. The method is efficient for nuclei with product-state
  references, and it describes many aspects of weakly bound and
  unbound nuclei.  This report reviews the technical and conceptual
  developments of this method in nuclear physics, and the results of
  coupled-cluster calculations for nucleonic matter, and for exotic
  isotopes of helium, oxygen, calcium, and some of their neighbors.
\end{abstract}

\maketitle

\tableofcontents
\newpage

\section{Introduction}
\label{sec:intro}
\input{Intro2}

\section{Conceptual and technical details}
\label{sec:tech}
\input{TechDetails}

\section{Results for finite nuclei} 
\label{sec:res}
\input{Results}

\section{Other developments} 
\label{sec:misc}
\input{Misc}

\section{Summary}
\label{sec:sum}
\input{Summary}

\ack We thank G. Baardsen, J.~Gour, G.~R.~Jansen, {\O}.~Jensen,
S.~Kvaal, K.~Kowalski, H.~Nam, P.~Piecuch, D.~Pigg, B.~Velamur Asokan,
and M.~W{\l}och for their collaboration on the nuclear coupled-cluster
project.  This work was
supported in part by the U.  S. Department of Energy under Grants Nos.
DE-FG02-96ER40963 (University of Tennessee), DE-FC02-07ER41457
(SciDAC-2 Collaboration UNEDF), DE-SC0008499 (SciDAC-3 Collaboration
NUCLEI), DEAC05-00OR22725 (Oak Ridge National Laboratory), and the
Research Council of Norway.

\appendix
\renewcommand*{\thesection}{\Alph{section}}
\input{app1}

\bibliographystyle{apsrmp4-1} 

\addcontentsline{toc}{section}{References}
\bibliography{thomas}

\end{document}

%% file: Intro2.tex
First-principles computations play an important role in nuclear
physics. These calculations start from a given Hamiltonian and aim at
solving the nuclear $A$-body problem without any uncontrolled
approximations. This ambitious task has been carried out only for
selected (and small) regions of the nuclear chart. The recent review
by \citeasnoun{leidemann2013} summarizes the accomplishments and
challenges in few-nucleon systems. Here, virtually exact methods exist
that compute bound states of few-nucleon systems, and precision tests
of nuclear interactions are possible.  The Green's function Monte
Carlo (GFMC) computations~\cite{carlson1987,pudliner1997,pieper2001}
and no-core shell-model (NCSM)
computations~\cite{navratil2000,navratil2009,barrett2013} convincingly
demonstrated that $p$-shell nuclei can be computed from scratch.

First-principles calculations of relevant nuclear properties continues
to play an important role. As examples we mention the calculation of
the Hoyle state in $^{12}$C~\cite{epelbaum2011}, the understanding of
the origin of the anomalous long lifetime of $^{14}$C
\cite{maris2011}, photoabsorption on $^4$He \cite{gazit2006}, the
description of light-ion scattering~\cite{nollett2007, quaglioni2008},
and the computation of halo states~\cite{hagen2010a}. The calculation
of such finely tuned states probes interactions and challenges
computational methods. First-principles calculations also guide and
interpret experiments. Examples are the structure of
$^9$Li~\cite{wuosma2005}, the determination of the mass radius
$^{23}$O~\cite{kanungo2011}, and electromagnetic transitions in
neutron-rich carbon~\cite{voss2012}.  Finally, first-principles
calculations make predictions where no data is yet available. Examples
are the prediction of the charge radius of
$^{8}$He~\cite{caurier2006}, the spectrum of the neutron-deficient
nucleus $^{14}$F~\cite{maris2010}, and the structure of the exotic
nucleus $^{54}$Ca~\cite{hagen2012b}.

The results of first-principles computations are expected to have
errors of the order of less than a few percent for binding energies.
While this is impressive, it is nowhere close to the agreement with
data that more phenomenological approaches such as shell-model
calculations with interactions adjusted to many-body
data~\cite{caurier2005,brown2006} or nuclear density functional
methods~\cite{bender2003,goriely2009,niksic2011,erler2012}
achieve. First-principles calculations are relevant because they probe
our understanding of nuclear interactions, and because they are
expected to yield reliable predictions where data is lacking to build
a model. They are also the lowest rung on a ladder that reaches from
nuclear interactions with some roots in quantum chromodynamics to more
phenomenological (and computationally less expensive) approaches that
cover the entire nuclear chart~\cite{nam2012,bogner2013}.

The currently available first-principles methods complement each other
in their capabilities to compute specific nuclei and observables, and
they differ in their computational cost, and their flexibility to deal
with a variety of Hamiltonians. More than a decade ago, several {\it
ab initio} methods successfully benchmarked the $\alpha$ particle and
agreed on its binding energy and radius~\cite{kamada2001}. Since then,
many $p$-shell nuclei were computed, and more recently also some
medium-mass nuclei.  Examples of {\it ab initio} approaches are the
GFMC method~\cite{pieper2001}, the
NCSM~\cite{navratil2009,maris2009,barrett2013}, the self-consistent
Green's function method \cite{dickhoff2004,barbieri2009}, lattice
simulations~\cite{lee2009}, the in-medium
similarity-renormalization-group method
\cite{tsukiyama2011,hergert2013}, and coupled-cluster theory. Here, we
limit ourselves to a review of the coupled-cluster method and refer
the reader to the literature cited above for details about the other
approaches.

\citeasnoun{coester1958} and \citeasnoun{coester1960} invented the
coupled-cluster method (originally termed ``$\exp S$'') half a century
ago. \citeasnoun{cizek1966} and \citeasnoun{cizek1971} developed the
method further and applied it to problems in quantum chemistry. In
nuclear physics, the Bochum group computed nuclear matter, and the
structure of doubly magic nuclei $^4$He , $^{16}$O, and $^{40}$Ca. The
Bochum method was particularly suited to deal with the hard core of
local nucleon-nucleon ($NN$) interactions, and the relevance of
three-nucleon forces is stated prominently in the abstract of their
review~\cite{kuemmel1978}.

While the coupled-cluster method flourished in quantum chemistry (See,
e.g., the recent review by \citeasnoun{bartlett2007}), it saw only
sporadic applications in nuclear theory during the 1980s and 1990s,
see~\cite{bishop1991}. \citeasnoun{heisenberg1999} were the first to
employ high-precision $NN$ interactions and three-nucleon forces with
the coupled-cluster approach, and the method has seen a renaissance in
recent
years~\cite{dean2004,kowalski2004,hagen2010b,roth2012,kohno2012}. This
renaissance is due to several conceptual developments regarding the
development of soft interactions via renormalization group
transformations~\cite{bogner2003,bogner2010}, the description of
weakly bound and unbound nuclei~\cite{michel2009}, the inclusion of
three-nucleon forces, and -- last not least -- due to the dramatic
increase in available computational cycles. In the past decade, the
coupled-cluster method has studied medium-mass nuclei with
high-precision $NN$ interactions~\cite{hagen2008}. This method is most
efficient for doubly magic nuclei or nuclei with a closed subshell
structure. Thus, it is an ideal tool to address shell evolution in
semi-magic nuclei. The very recent coupled-cluster computations of
neutron-rich isotopes of oxygen~\cite{hagen2012a} and
calcium~\cite{hagen2012b} make several predictions for the spectra of
these important elements, some of which have been confirmed
experimentally~\cite{steppenbeck2013b}.

In this work, we review the developments of coupled-cluster theory in
nuclear physics that happened since the last reviews by
\citeasnoun{kuemmel1978} and \citeasnoun{bishop1991}. By necessity, we
also have to describe some of the recent advances in the fields of
nuclear interactions or the treatment of weakly bound and unbound
systems. On such topics, we do not present a review. Instead, we have
limited ourselves to citing some of the original work and often refer
to the pertinent reviews. We apologize to those readers who feel that
their work has been misinterpreted, overlooked, or omitted.

This review is organized as follows. Section~\ref{sec:tech} describes
the technical aspects of the coupled-cluster method including model
spaces, interactions, and the treatment of the center of mass. In
Section~\ref{sec:res} we describe the main results of coupled-cluster
computations of atomic nuclei. Other related developments are
presented in Sect.~\ref{sec:misc}. Finally, we present a summary and
discuss open problems in Section~\ref{sec:sum}. The appendices present
some technical details.

%% file: TechDetails.tex
In this Section, we present some of the technical aspects that arise
when applying coupled-cluster theory to atomic
nuclei. Subsection~\ref{subsec:techdetails_scheme} is dedicated to the
main ideas and formal developments of coupled-cluster theory
itself. Model spaces, including the Berggren basis for the description
of weakly bound and unbound nuclei, are presented in
Subsection~\ref{subsec:techdetails_berggren}. In
Subsection~\ref{subsec:techdetails_interact} we discuss the employed
interactions and approximation schemes for including three-nucleon
forces. The treatment of the center-of-mass problem within
coupled-cluster theory is presented in
Subsection~\ref{subsec:techdetails_com}.

\subsection{Nuclear coupled-cluster theory}
\label{subsec:techdetails_scheme}
\input{TechDetails_CCM}

\subsection{Model spaces}
\label{subsec:techdetails_berggren}
\input{TechDetails_Berggren}

\subsection{Interactions} 
\label{subsec:techdetails_interact}
\input{TechDetails_Interact}

\subsection{Center-of-mass problem}
\label{subsec:techdetails_com}
\input{TechDetails_CoM}

%% file: TechDetails_CCM.tex
In this Subsection, we review the essentials of coupled-cluster
theory. We follow the standard approach from quantum
chemistry~\cite{bartlett2007,crawford2007,shavittbartlett2009} that
\citeasnoun{dean2004} adapted for nuclear physics. This is not to say
that the implementation of coupled-cluster theory in nuclear theory
does not differ from quantum chemistry. In atomic nuclei, considerable
computational efficiency is gained from a spherical $j$-coupled
implementation of the method~\cite{hagen2008}, and the challenges of
three-nucleon forces, pairing, deformation, weak binding, and the
treatment of the center of mass are unique to this field, too. There
are also formulations of coupled-cluster theory that are particularly
suited for hard-core
potentials~\cite{kuemmel1978,bishop1992,heisenberg1999}. However,
$G$-matrices~\cite{hjorthjensen1995}, ``bare'' potentials from chiral
effective field theory~\cite{entem2003,epelbaum2009,machleidt2011},
low-momentum interactions~\cite{bogner2003} and
similarity-renormalization group transformations~\cite{bogner2010} are
sufficiently soft and can be used directly within the approach we
describe.

Coupled-cluster theory is formulated in second quantization. Let
$\adag_p$ and $a_p$ create and annihilate a fermion in
state $|p\rangle$, respectively. Here, $p$ denotes a set of quantum
numbers such as $p=(n, l, j, \tau_z)$ in the angular-momentum-coupled
$j$-scheme, or $p=(n,l,j,j_z,\tau_z)$ in the $m$-scheme. As usual, $n,
l, j, j_z$ and $\tau_z$ label the radial quantum number, the orbital
angular momentum, the total angular momentum, its $z$-projection, and
the projection of the isospin, respectively.

\subsubsection{Computation of the ground state}

Coupled-cluster theory is based on an $A$-body product state 
\be
\label{ref}
|\phi\rangle = \prod_{i=1}^A \adag_i |0\rangle \ , 
\ee 
that serves as a reference. The reference can result from  a
Hartree-Fock calculation, or from a naive filling of the orbitals of
the harmonic oscillator. Throughout this review we use the convention
that $i,j,k,\ldots$ refer to states occupied in the reference state
$|\phi\ra$, while $a, b, c,\ldots$ refer to the valence space. Labels
$p,q,r,s$ refer to any orbital. It is useful to normal order the
Hamiltonian with respect to the reference state~(\ref{ref}). In the
case of a two-body Hamiltonian
\[
H=\sum_{pq}\varepsilon_{pq} a^\dagger_p a_q + {1\over 4} \sum_{pqrs} \langle pq||rs\rangle
a^\dagger_p a^\dagger_q a_s \hat{a}_r \ . 
\]
the normal ordered Hamiltonian $H_N$ is defined by $H=H_N+E_0$ with 
\[
E_0=\sum_i\varepsilon_{ii} + {1\over 2} \sum_{ij} \langle ij||ij\rangle
\]
being the vacuum expectation value (or Hartree-Fock energy if the
Hartree-Fock basis is employed), and
\ba
\label{normal2}
H_N&=& \sum_{pq}f_{pq} \{a^\dagger_p a_q \} 
+ {1\over 4} \sum_{pqrs} \langle pq||rs\rangle
\{ a^\dagger_p a^\dagger_q a_s a_r \} \ . 
\ea
Here, the brackets denote normal ordering, and the Fock matrix is  
\be
\label{fockmat}
f_{pq}\equiv \varepsilon_{pq} + {1\over 2} \sum_{i} \langle ip||iq\rangle  \ .
\ee
For a three-body interaction, the corresponding expression is presented
below in Eq.~(\ref{normal3}). 
Note that $\langle\phi|H_N|\phi\rangle=0$ by construction.

The similarity-transformed Hamiltonian
\be
\label{hsim}
\overline{H}\equiv e^{-T}H_N e^T 
\ee
is at the heart of coupled-cluster theory. The
cluster operator
\be
\label{t_expan}
T=T_1 +T_2 +\ldots + T_A
\ee
is defined with respect to the reference. Here
\ba
\label{cluster}
T_1 &=& \sum_{ia} t_i^a \adag_a a_i \ , \nonumber\\
T_2 &=& {1\over 4}\sum_{ijab} t_{ij}^{ab} 
\adag_a \adag_b a_j a_i \ , 
\ea 
generate $1p$-$1h$ and $2p$-$2h$ excitations of the reference state,
respectively, and the cluster operator $T_n$ generates $np$-$nh$
excitations. 

Note that the similarity-transformed Hamiltonian~(\ref{hsim}) is not
Hermitian because $e^T$ is not unitary. Coupled-cluster theory can be
viewed from two perspectives -- based on a bi-variational principle or
as an eigenvalue problem of the similarity-transformed Hamiltonian. In
the bi-variational perspective one minimizes the energy
functional~\cite{arponen1983} 
\ba
\label{bi}
E(T,L) \equiv \la\phi| L e^{-T}H e^T|\phi\ra = \la\phi|L \overline{H}|\phi\ra
\ea
with respect to $T$ and $L$. Here, $L$ is a de-excitation operator
\be
\label{lam_expan}
L = l_0 + L_1 + L_2 +\ldots + L_A \ , 
\ee
with
\ba
\label{lambda}
L_1 &=& \sum_{ia} l^i_a \adag_i a_a \ , \nonumber\\
L_2 &=& {1\over 4}\sum_{ijab} 
l^{ij}_{ab} \adag_i \adag_j a_b a_a \ , 
\ea
and similar definitions for the $n$-body de-excitation operator $L_n$. 
The variation of
the functional~(\ref{bi}) can be viewed as an independent variation of
the bra state $\la\phi|L e^{-T}$ and the ket state
$e^T|\phi\ra$. Note that the functional~(\ref{bi}) is normalized 
\[
\la\phi|L e^{-T} e^T|\phi\ra = \la\phi|\phi\ra = 1  
\]
for $l_0=1$, because $L_n|\phi\ra = 0$ for $n=1,2,\ldots$.  

In practice, one truncates the expansions~(\ref{t_expan}) and
(\ref{lam_expan}). In the coupled cluster with singles and doubles
(CCSD) approximation on truncates $T_3=T_4=\ldots
=T_A=0=L_3=L_4=\ldots=L_A$. The variation of the
functional~(\ref{bi}) with respect to $L$ yields the CCSD equations
\ba
\label{ccsd}
\la\phi_{i}^{a}|\overline{H}|\phi\ra  & =&  0 \ , \nonumber\\
\la\phi_{ij}^{ab}|\overline{H}|\phi\ra &=& 0 \ .
\ea 
Here, $|\phi_i^a\ra\equiv \adag_a a_i|\phi\ra$, and
$|\phi_{ij}^{ab}\ra\equiv \adag_a \adag_b a_j a_i|\phi\ra$. The
Eqs.~(\ref{ccsd}) do not depend on $L$, and their solution
yields the cluster amplitudes $t_i^a$ and $t_{ij}^{ab}$. For these
cluster amplitudes, the reference $|\phi\ra$ becomes an eigenstate of
the similarity-transformed Hamiltonian in the space of $1p$-$1h$ and
$2p$-$2h$ excited states. In other words, in the CCSD approximation,
the similarity-transformed Hamiltonian generates no $1p$-$1h$ and no
$2p$-$2h$ excitations of the reference state.  Thus,
$\la\phi|L_n \overline{H}|\phi\ra=0$, and the energy
functional~(\ref{bi}) yields the energy
\be
\label{erg}
E = \la\phi|\overline{H}|\phi\ra \ .  
\ee 
The computational cost for solving the CCSD equations~(\ref{ccsd}) in
the $m$ scheme for a nucleus with mass number $A$ is $A^2 n^4$, with
$n$ being the number of single-particle valence states. Typically, $n\gg A$. 
This is much
more affordable than other {\it ab initio} methods such as GFMC or
NCSM but is much more expensive than mean-field methods.

The variation of the functional~(\ref{bi})
with respect to the cluster amplitudes yields
\ba
\la\phi|L\left[\overline{H},\adag_a a_i\right]|\phi\ra &=&0\ , \nonumber\\
\la\phi|L\left[\overline{H},\adag_a \adag_b a_j a_i\right]|\phi\ra &=&0\ \nonumber
\ea
and the solution of these equations determines $l_a^i$ and
$l_{ab}^{ij}$. These equations can be simplified to
\ba
\label{left}
\la\phi|L\overline{H} |\phi_i^a\rangle &=& \omega\la\phi|L|\phi_i^a\rangle  \ , \nonumber\\
\la\phi|L \overline{H} |\phi_{ij}^{ab}\rangle &=& \omega\la\phi|L|\phi_{ij}^{ab}\rangle  \ , 
\ea
Clearly $\la\phi|L$ is the left eigenstate of the
similarity-transformed Hamiltonian in the space of 1$p$-1$h$ and
2$p$-2$h$ excited states, and the excitation energy is $\omega$ 
\be
\label{leftev}
\langle\phi|L \overline{H}=\omega \la\phi|L \ .
\ee
Note that the solution of the Eq.~(\ref{left}) is only necessary when
one is interested in excited states or in computing expectation values
other than the energy (see below).

An alternative approach to the above  derivation would be to  insert the ket state
$e^T|\phi\ra$ into the time-independent Schr{\"o}dinger equation
$He^T|\phi\ra=E e^T |\phi\ra$, left-multiply with $e^{-T}$ and project
onto the bra states $\la\phi|$, and $\la\phi_i^a|$,
$\la\phi_{ij}^{ab}|$. This yields Eq.~(\ref{erg})
and the CCSD equations~(\ref{ccsd}), respectively. 

The computation of the similarity-transformed Hamiltonian~(\ref{hsim})
is based on the Baker-Campbell-Hausdorff expansion 
\ba 
\label{bch}
\overline{H} = H_N + \left[H_N,T\right] + {1\over
2!}\left[\left[H_N,T\right],T\right]+ {1\over
3!}\left[\left[\left[H_N,T\right],T\right],T\right] +\ldots .  
\ea 
Because the individual terms of the $T$ operator commute among
themselves, the commutators in Eq.~(\ref{bch}) ensure that each $T$
connects to the Hamiltonian $H_N$. For two-body (three-body)
Hamiltonians, the expansion~(\ref{bch}) thus terminates at fourfold
(sixfold) nested commutators, and the similarity transformation can be
evaluated exactly. In contrast, similarity-transformations of the form
$e^{T^\dagger}H_N e^T$ would remain Hermitian but would lead to the
evaluation of an infinite number of nested commutators, see
e.g. \cite{szalay1995}. When solving the CCSD equations~(\ref{ccsd}),
one only needs to compute those terms of $\overline{H}$ that are of
two-body nature. For instance, terms such as
$\left[\left[\left[H_N,T_1\right],T_1\right],T_1\right]$ contribute
but $\left[\left[\left[H_N,T_2\right],T_2\right],T_2\right]$ are of
three-body rank or higher, and vanish exactly in the space of
1$p$-1$h$ and 2$p$-2$h$ excitations. The actual computation of the
similarity-transformed Hamiltonian~(\ref{bch}) requires the user to
perform Wick contractions. This is best done with diagrammatic
methods, see, e.g.,~\cite{crawford2007,shavittbartlett2009}. Some
details are presented in Appendix~\ref{app1}. For an efficient
numerical implementation of the method, one computes and re-uses
matrix elements of the similarity-transformed Hamiltonian and
appropriately defined intermediates~\cite{kucharski1991}.

Size extensivity, i.e. the proper scaling of the energy with the size
of the system, is a key property of coupled-cluster
theory~\cite{bartlett1978,bartlett1981}. It implies that the particle
separation energy tends to a constant in the limit of infinite system
size, i.e. the total energy must be proportional to the number of
particles. This concept goes back to computations of nuclear matter,
for which \citeasnoun{goldstone1957} showed that only linked clusters
enter the computation of the energy in many-body
theory. Coupled-cluster theory is size-extensive because only linked
diagrams enter in the similarity transformed
Hamiltonian~(\ref{bch}). Size extensivity is a relevant concept also
for finite nuclei because nuclei along the valley of $\beta$ stability
exhibit a constant energy per particle.  A related concept is size
consistency, i.e. the energy for a system of two well separated,
non-interacting subsystems is the sum of the energies of the two
subsystems. Because of the short-ranged strong force, size consistency
obviously also is valid concept for nuclei. We refer the reader to the
very readable review by~\cite{nooijen2005} for more details on these
two topics.

The CCSD approximation with its inherent truncation to two-body terms
of the similarity-transformed Hamiltonian is computationally very
attractive. In this approximation, the solution of an $A$-body
Hamiltonian only requires two-body technology, and this explains the
efficiency of the coupled-cluster method. In nuclear structure,
three-body forces play an important role. Within the CCSD
approximation it is thus advantageous to treat their main
contributions as medium-dependent two-body forces, or as
normal-ordered two-body forces. Details are presented in
Subsection~\ref{subsec:techdetails_interact} when dealing with finite
nuclei. For nucleonic matter, the full inclusion of three-nucleon
forces is simpler, see Subsection~\ref{nucleonic}.

\subsubsection{Computation of excited states}

In this Subsection, the equation-of-motion (EOM)
approach~\cite{stanton1993} is used for the computation of ground- and
excited states in and around nuclei with closed sub-shells.  States in
neighbors of closed-shell nuclei can be viewed as generalized
excitations of the closed-shell reference. In this approach one
assumes that the CCSD equations~(\ref{ccsd}) have been solved, and
that all matrix elements of the similarity-transformed
Hamiltonian~(\ref{hsim}) are available.

For the right eigenstates of the similarity-transformed Hamiltonian,
one makes the following ansatz
\be
\label{right1}
\vert \psi_\mu \rangle = R_\mu |\phi\ra \ .
\ee 
Here, $R$ is an excitation operator and assumes a variety of forms. One uses 
\be
\label{eom}
R=r_0+ \sum_{ia} r_i^a a_a^\dagger a_i 
+ {1\over 4} \sum_{ijab}r_{ij}^{ab} a^\dagger_a a^\dagger_b a_j a_i +\cdots  
\ee
for excitations in the $A$-body nucleus, 
\be
\label{paeom}
R=\sum_{ia} r^a a_a^\dagger  
+ {1\over 2} \sum_{iab}r_{i}^{ab} a^\dagger_a a^\dagger_b a_i +\cdots  
\ee
for states of the nucleus with mass number $A+1$, 
\be
\label{preom}
R=\sum_{ia} r_i^a a_i 
+ {1\over 2} \sum_{ija}r_{ij}^{a} a^\dagger_a a_j a_i +\cdots 
\ee
for the computation of states in the nucleus $A-1$, and 
\be
\label{2paeom}
R={1\over 2} \sum_{ab} r^{ab} a^\dagger_a a^\dagger_b 
+ {1\over 6} \sum_{iabc}r_{i}^{abc} 
a^\dagger_a a^\dagger_b a^\dagger_c a_i +\cdots   \ .
\ee
for states in the nucleus $A+2$.   

The wave function ansatz ~(\ref{right1}) yields the following eigenvalue
problem for the similarity transformed Hamiltonian 
\be
\label{right2}
\overline{H}R_\mu |\phi\ra = E_\mu R_\mu |\phi\ra \ .  
\ee 
In order to
avoid disconnected terms one can subtract $ R_\mu\overline{H} |\phi\ra
= E_0 R_\mu |\phi\ra $ from Eq.~(\ref{right2}) and obtains 
\be
\label{right}
\left[ \overline{H},R_\mu \right] |\phi\ra = \omega_\mu R_\mu |\phi\ra \ .  
\ee 
Here $\omega_\mu= E_\mu - E_0$ is the $\mu$'th excited energy with
respect to the ground-state energy $E_0$ of the closed shell
nucleus $A$.

A few comments are in order. The right eigenvalue
problem~(\ref{right}) is analogous to the left eigenvalue
problem~(\ref{left}). It can be derived formally within linear
response theory from the time-dependent coupled-cluster
method~\cite{DalMon83}. Depending on the form of $R$, we deal with the
particle-attached and particle-removed EOM~\cite{gour2006} in
Eq.~(\ref{paeom}) and Eq.~(\ref{preom}), respectively, and the
two-particle-attached EOM in Eq.~(\ref{2paeom}) \cite{jansen2011}. The
expansions~(\ref{eom}) to (\ref{2paeom}) need to be truncated in
practice, and this truncation determines what kind of states can be
computed successfully. For example, a truncation of Eq.~(\ref{eom}) at
the 2$p$-2$h$ level is expected to be useful in the computation of
states that are 1$p$-1$h$ excitations of the ground state. As a rule
of thumb, the appropriate truncation level of the cluster operator
should exceed the dominant level of particle-hole excitation in the
targeted state by at least one unit. For example, low-lying
$J^\pi=0^+$ states in even-even nuclei that correspond to
alpha-particle excitations are out of reach at a truncation of the
excitation operator below the 4$p$-4$h$ level
\cite{brown1966,haxton1990}. Because $\overline{H}$ is not Hermitian,
the left eigenvector $\la\phi|L_\alpha$ and the right eigenvector
$R_\beta|\phi\ra$ corresponding to the eigenvalues $\omega_\alpha$ and
$\omega_\beta$ form a bi-orthonormal set, i.e. $\la\phi|L_\alpha
R_\beta|\phi\ra=\delta_{\alpha\beta}$. Thus, for the ground state we
have $r_0=1$, and $r^a_i=0=r_{ij}^{ab}$.

We also want to comment on the number-changing EOMs. To avoid problems
with the center of mass, it is useful to employ the intrinsic
Hamiltonian~(\ref{ham_in}), see
Subsection~\ref{subsec:techdetails_com} for details. The intrinsic
Hamiltonian depends on the mass number $A$, and a choice has to be
made. In the particle-removed EOM, one computes the
similarity-transformed intrinsic Hamiltonian with the mass number
$A-1$ when computing $\overline{H}$ for the reference with mass $A$ in
the first step. Thus, $\overline{H}$ will not fully capture the
intrinsic physics of the $A$-body problem. However, when solving the
eigenvalue problem of the particle-removed EOM, the resulting solution
approximately displays the factorization of the intrinsic and
center-of-mass wave function as discussed in
Subsection~\ref{subsec:techdetails_com}. Similar comments apply to the
computation of the excited states of the $A+1$ nucleus within particle
attached EOM. Here, the first step consists of computing the
similarity transformed intrinsic Hamiltonian for the $A$-body
reference, and $(A+1)$ is the mass number in the kinetic energy of the
center of mass.
 
\subsubsection{Observables other than the energy}

The computation of expectation values and transition matrix elements
can be based on the similarity-transformed one-body density matrix 
\[  \overline{\rho}^p_q \equiv e^{-T}\rho^p_q e^T \equiv
e^{-T}a^\dagger_p a_qe^T \] 
and the two-body matrix 
\[
\overline{\rho}^{pq}_{rs}\equiv e^{-T}\rho^{pq}_{rs} e^T \equiv
e^{-T}a^\dagger_p a^\dagger_q a_s a_r e^T \ .  
\]  
once the $T$-amplitudes are known. Their computation follows the same
diagrammatic rules as the computation of $\overline{H}$, see
\cite{shavittbartlett2009} for details.

Let $\la\phi|L_\alpha$ and $R_\beta|\phi\ra$ be the left and right
eigenvector corresponding to the energies $\omega_\alpha$ and
$\omega_\beta$, respectively. Then the transition matrix elements are
\[
\la\phi|L_\alpha \overline{\rho}^p_q R_\beta|\phi\ra \ \mbox{and} \:  \la\phi|L_\alpha \overline{\rho}^{pq}_{rs} R_\beta|\phi\ra \ ,
\]
and any one-body or two-body matrix element of interest can be
computed this way.

An alternative approach for the computation of expectation values
consists of using the Hellmann-Feynman theorem. In this approach one
computes the expectation of the operator $\hat{O}$ from the
energy expectation $E(\lambda)$ for the Hamiltonian $H+\lambda
\hat{O}$ and evaluates 
\[
\la\hat{O}\ra\approx \left.{\partial
E(\lambda)\over\partial\lambda}\right|_{\lambda=0} 
\] numerically.  Here, we have used the approximate sign instead of
the equality sign because the coupled-cluster method is not
variational but rather bi-variational. Strictly speaking, it does not
fulfill the Hellmann-Feynman theorem. This is less of a concern in
practical applications \cite{noga1988}. For a detailed discussion of
this aspect, and for an extension of the method that fulfill the
Hellmann-Feynman theorem, we refer the reader to the review
by~\citeasnoun{bishop1991}.

\subsubsection{Beyond the CCSD approximation}

The full inclusion of 3$p$-3$h$ or triples (T) excitations (i.e. $T_3$
amplitudes ), in coupled-cluster theory is numerically expensive and
scales as $A^3n^6$ per iteration in the solution of the CCSDT
equations. It is $An^2$ times more expensive than the solution of the
CCSD equations. This high computational cost [and the fact that
non-iterative triples approximations often yield results that are
closer to exact solutions than CCSDT~\cite{kutzelnigg1991}] made
approximative triples the standard in quantum chemistry. Approximative
triples are motivated by arguments from perturbation theory, and there are
several implementations that differ in
complexity and computational cost, see \cite{bartlett2007}. Generally
speaking, the computationally more expensive iterative triples
corrections are more accurate and also useful in non-perturbative
nuclear physics applications~\cite{kowalski2004,hagen2007b}.

The non-iterative CCSD(T)
approximation~\cite{raghavachari1989,bartlett1990} has become ``the
gold standard'' in quantum chemistry and presents a very good
compromise between affordability and accuracy. The more recent
$\Lambda$-triples approximation [denoted as $\Lambda$-CCSD(T) in
\cite{kucharski1998,taube2008}] is similar in spirit to the CCSD(T)
approximation but employs the nonperturbative (and computationally
somewhat more expensive) left eigenvector solution of the
similarity-transformed Hamiltonian~(\ref{hsim}). The energy correction
in the Hartree-Fock basis is
\[
\Delta E_3 = \frac{1}{(3!)^2}\sum_{ijkabc}\frac{\langle\phi |L V_N|\phi_{ijk}^{abc}\rangle\langle\phi_{ijk}^{abc}|\left(V_N T_2\right)_C|\phi\rangle}{\varepsilon_{ijk}^{abc}} \ .
\]
Here $\phi_{ijk}^{abc}$ is a 3$p$-3$h$
excitation of the reference $|\phi\rangle$, and 
\[
\varepsilon_{ijk}^{abc}=f_{ii}+f_{jj}+f_{kk}-f_{aa}-f_{bb}-f_{cc}
\]
is computed from the diagonal elements of the Fock matrix.  For
``soft'' interactions there is very little difference between the
accuracy of both approximations, but the nonperturbative
$\Lambda$-CCSD(T) approximation is more accurate for ``harder''
interactions. 

The ``completely renormalized'' non-iterative triples
corrections~\cite{kowalski2000,piecuch2002} are more sophisticated
than the $\Lambda$-CCSD(T) approximation (but also computationally
somewhat more expensive). They also perform very well in nuclear
structure
calculations~\cite{kowalski2004,wloch2005,roth2009,binder2013}.

\subsubsection{Time-dependent coupled-cluster method}
\citeasnoun{HooNeg78,HooNeg79} and \citeasnoun{SchGun78} developed
time-dependent coupled-cluster theory. For time-dependent phenomena
with small amplitudes, this approach leads to linear-response theory
or to coupled-cluster EOM~\cite{DalMon83,Mon87,TakPal86}. However,
large-amplitude time-dependent phenomena were not well understood
until very recently. If one attempts to base time-dependent phenomena
on the similarity-transformed Hamiltonian~(\ref{hsim}), one does not
obtain a real energy, and the energy is not time independent for
conservative systems~\cite{HubKla11}. Instead \citeasnoun{kvaal2012}
showed that time-dependent coupled-cluster theory must be based on the
energy functional~(\ref{bi}), and that this formulation yields a real
energy, which is constant for the time evolution of conservative
systems.  Here, the similarity-transformed Hamiltonian and the
de-excitation operator $L$ in Eq.~(\ref{bi}) both evolve in time. In
this formulation, observables that commute with the Hamiltonian are
also conserved quantities under time evolution~\cite{pigg2012}.

\citeasnoun{pigg2012} studied \citeasnoun{kvaal2012} formulation of
time-dependent CCSD in simple nuclear Hamiltonians. They found that
the imaginary time evolution of the similarity-transformed
Hamiltonian~(\ref{hsim}) drives $\overline{H}$ into a decoupled form
such that the coupled-cluster equations~(\ref{ccsd}) are fulfilled.
     
\subsubsection{Computational aspects of the nuclear coupled-cluster
  method} Let us briefly discuss some computational aspects regarding
the nuclear coupled-cluster method. The solution of the CCSD
equations~(\ref{ccsd}) yields the amplitudes $t^a_i$ and
$t^{ab}_{ij}$. These enter the CCSD energy~(\ref{erg}), and they are
used to construct the similarity-transformed
Hamiltonian~(\ref{bch}). In order to arrive at equations that are
suitable for numerical implementation, one can use Wicks' theorem or a
diagrammatic approach (see for example \cite{kucharski1986},
\cite{crawford2007}, \cite{bartlett2007} for details) to arrive at a
set of coupled non-linear equations in the $T$ amplitudes. The number
of coupled non-linear equations depends on the cluster truncation
level. For example, in the CCSD approximation there are $An$ +
$A^2n^2$ number of coupled equations. In the $m$-scheme CCSD
calculation of $^{40}$Ca in 9 major oscillator shells one deals with
40 occupied orbitals and 660 unoccupied orbitals, resulting in $\sim
10^{9}$ non-linear coupled equations. While this is a very large
number of non-linear equations, one can only use soft interactions to
obtain reasonably converged ground-state energies for medium mass
nuclei in model spaces of such size \cite{hagen2007b}. In order to
obtain the solution for such a large number of non-linear equations,
one needs to implement the coupled-cluster equations in a numerical
efficient scheme that aims at minimizing the number of computational
cycles and the memory in terms of storage. For a given nucleus one can
rewrite the non-linear CCSD equations in a quasi-linearized form by
defining appropriate intermediates (see for example \cite{gour2006}
and \cite{hagen2007a} for Hamiltonians with three-nucleon
forces). There are many ways to define these intermediates, and in
order to obtain the most efficient numerical scheme, one needs to take
into account the memory requirements in storing the various
intermediates and the number of computational cycles involved
\cite{hagen2007a}. When rewriting the CCSD equations in the
quasi-linear form, it is easy to see that
$\sum_{cd}\chi^{ab}_{cd}t^{cd}_{ij} $ is the most expensive term with
a cost of $A^2n^4$ computational cycles. However, this term can be
efficiently computed as a matrix-matrix product. 

The vast number of coupled-cluster equations does not allow for direct
inversion techniques. Rather, one uses iterative procedures to obtain
the solution. In nuclear physics, \citeasnoun{dean2008b} and
\citeasnoun{baran2008} explored different Krylov sub-space methods to
increase the convergence rate of iterative methods. They found that
the Broyden method~\cite{broyden1965} or DIIS (Direct Inversion in the
Iterative Subspace) \cite{pulay1980} achieve convergence of the CCSD
equations already with 20-30 iterations.

In order to use ``bare'' chiral interactions \cite{entem2003} in
coupled-cluster calculations of medium-mass nuclei like $^{40,48}$Ca,
one needs to handle model-space sizes of about 14-18 major oscillator
shells to obtain converged results (see for example
\cite{hagen2010b}). Such model-space sizes are too large for the
$m-$scheme representation. For example, the computation of $^{40}$Ca
in 15 major oscillator shells in the $m$-scheme uses 2720
single-particle orbitals, and would require the solution of $10^{10}$
non-linear equations. The memory requirements for the $t_{ij}^{ab}$
amplitudes alone would amount to about $100$~GBytes.
\citeasnoun{hagen2008} derived and implemented the CCSD equations in
an angular-momentum-coupled scheme. The Hamiltonian is a scalar under
rotation and so are the similarity transformed Hamiltonian and the $T$
amplitudes. From this property, and the fact that the degeneracy of
single-particle levels $nlj$ near and above the Fermi surface in
medium mass nuclei becomes rapidly very large, there are huge
computational savings in going from the uncoupled $m$-scheme to the
angular-momentum-coupled ($j$-coupled) scheme. For example, 15 major
oscillator shells only consist of 240 single-particle orbitals in the
$j$-coupled scheme. Further the number of non-linear equations to be
solved in the CCSD approximation would be reduced from $10^{10}$ in
the $m$-scheme to about $10^6$ in the $j$-coupled scheme. To estimate
the computational savings of the $j$-coupled scheme, we note that
there are about $n^{2/3}$ single $j$-shells in an oscillator space
with $n$ orbitals in the $m$-scheme.

\begin{figure}[h]
  \begin{center}
    \includegraphics[width=0.5\textwidth,clip=]{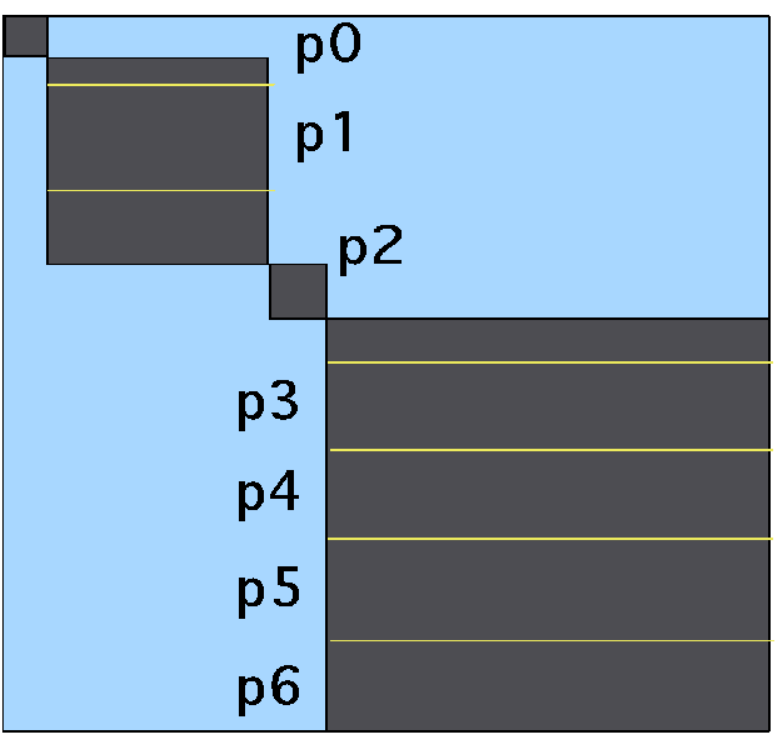}
  \end{center}
  \caption{Block diagonal structure of the interaction
    matrix in angular-momentum-coupled representation and the
    parallel distribution scheme used in the numerical
    implementation. The matrix elements of the block-diagonal matrix
    is distributed over the processes $p0 \ldots p6$ as shown. Taken
    from arXiv:1203.3765 with permission, see also \cite{hagen2012d}.}
  \label{fig:ccm_par_dist}
\end{figure}

The numerical implementation of the $j$-scheme coupled-cluster
equations utilizes MPI and OpenMP environments. Special care has to be
taken regarding load-balancing and memory distribution of the
interaction matrix elements, see \cite{hagen2012d} for details. In the
$j$-coupled representation the interaction matrix is sparse and block
diagonal in the quantum numbers ($J^\pi, T_z$). In order to utilize
optimized linear algebra libraries such as {\sc BLAS} and {\sc
  LAPACK}, \citeasnoun{hagen2009b} developed a load balancing scheme
with an optimal balance between memory distribution and computational
cycles as depicted in Fig.~\ref{fig:ccm_par_dist}. In this scheme one
adds successive rows of the $j$-coupled matrix to a given processor
until the optimal load-balancing criterion is reached.  In this way
one can utilize {\sc BLAS} and {\sc LAPACK} routines to compute
contractions between the interaction matrix with the $T$-amplitudes.
One needs to construct an optimal distribution for each separate part
of the full interaction matrix that enters in the various diagrams and
contractions with the $T$-amplitudes (see App.~\ref{app1}). This makes
it a challenge to obtain a numerical implementation that scales to
large number of processors, and examples are presented in
\cite{hagen2012d}.

Implementing CCSD in the $j$-coupled scheme, allowed for the first
ab-initio calculations of medium mass nuclei starting from ``bare''
chiral interactions \cite{hagen2008}. In Appendices ~\ref{app1} and
\ref{app2} we give the diagrammatic, uncoupled $m$-scheme, and
angular-momentum-coupled expressions for the CCSD approximation and
various equations of motion techniques used for ground- and excited
states in and around closed (sub)-shell nuclei.

%% file: TechDetails_Berggren.tex
\subsubsection{Harmonic-oscillator basis}
The coupled cluster method is a wave-function-based approach, and it is
thus convenient to take the three-dimensional spherical harmonic
oscillator as a basis for atomic nuclei. The matrix elements of the
nuclear interaction are computed numerically and transformed from the
center-of-mass system to the laboratory system. The single-particle
model space consists of $N+1$ oscillator shells with frequency
$\omega$ (and oscillator length $b\equiv \sqrt{\hbar/(m\omega) }$)
Thus, the maximum oscillator energy is $(N+3/2)\hbar\omega$. The
parameters $N$ and $\hbar\omega$ have to be chosen such that the
radial extent $\sqrt{2(N+3/2)} b$ of the basis is large enough to
accommodate the nucleus in position space, while the ultraviolet
momentum $\sqrt{2(N+3/2)}/b$ of the basis has to be larger than the
momentum cutoff of the employed interaction to accommodate the nucleus
in momentum space~\cite{stetcu2007,hagen2010b,jurgenson2011}.

Phenomenological extrapolation schemes have long been used in nuclear
structure
calculations~\cite{hagen2007b,bogner2008,forssen2008,maris2009}.  Very
recently,~\citeasnoun{furnstahl2012} built on \cite{coon2012} and
proposed a theoretical basis for the convergence properties of nuclear
energies and radii in the oscillator basis. The main idea is that the
finite spatial extent of the oscillator basis essentially forces the
wave function to fulfill Dirichlet boundary conditions at
$L_2\equiv\sqrt{2(N+3/2+2)} b$~\cite{more2013,furnstahl2013b}, and
this allows one to derive expressions similar to those
that \citeasnoun{luscher1985} derived for finite lattices.  For
quantum dots, i.e.  electrons with Coulomb interactions confined by a
harmonic oscillator potential, \citeasnoun{kvaal2009} gave closed-form
mathematical relations for the error in the energy due to truncation
in the harmonic oscillator basis. These relations are expected to
apply for screened Coulomb interactions as well, and could provide
useful insights to truncations made in nuclear many-body calculations.

In coupled-cluster calculations it is often convenient to start from a
Hartree-Fock basis. In this basis the coefficients $t_i^a$ of the
singles amplitude become very small~\cite{kowalski2004}. One also
finds that this approach reduces the $\hbar\omega$-dependence of the
computed energies. In practical computations one aims at increasing
the number $N+1$ of employed oscillator shells until the results
become virtually independent on the model-space parameters.  Another
advantage of the Hartree-Fock basis is that the Fock matrix is
diagonal. Thus, one can compute several triples approximations such as
CCSD(T) and $\Lambda$-CCSD(T) (see Section
\ref{subsec:techdetails_scheme} for more details) in a single
non-iterative step.

\subsubsection{Berggren basis} 
Due to the Gaussian falloff of the harmonic oscillator functions, this
basis is not able to capture the physics of unbound nuclei, or excited
resonant states above the particle threshold. Similar comments apply
to the description of extended objects such as halo nuclei. Nuclei and
nuclear states near and above the particle emission thresholds are
open quantum systems, and the coupling to open decay channels and the
particle continuum is essential. We note that the coupling to the
continuum is relevant for the understanding of shell structure at the
driplines~\cite{dobaczewski1994,hagen2012b}, and the parity inversion
of the ground state in $^{11}$Be~\cite{forssen2005,quaglioni2008}.

For open systems it is of advantage to employ the Berggren
basis~\cite{berggren1968,berggren1971,lind1993}. This basis is a
generalization of the standard completeness relation from the real
energy axis to the complex energy plane.  Completeness is given by a
finite set of bound- and resonant states together with a non-resonant
continuum. A finite-range potential that supports a finite set of
bound and resonant states is always accompanied by a scattering
continuum. Thus, such potentials have a much richer spectrum than the
standard harmonic oscillator potential which only supports
bound-states.

\begin{figure}[h]
  \begin{center}
    \includegraphics[width=0.9\textwidth,clip=]{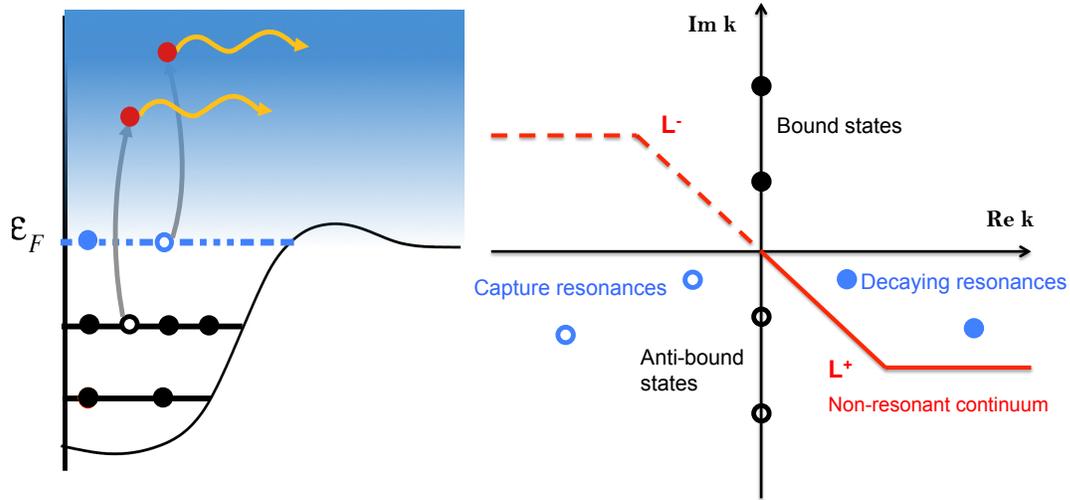}
  \end{center}
  \caption{Left: A schematic picture of a finite range potential
    together with the spectrum of bound, resonant and non-resonant
    continuum states. Right: Corresponding distribution of bound
    (black filled circles) states and resonant states (blue filled
    circles) in the complex momentum plane together with the
    non-resonant continuum (red line). }
  \label{smat_poles}
\end{figure}

Figure~\ref{smat_poles} shows an example of a finite-range potential
and its spectrum. The bound and resonant states are poles of the
scattering matrix in the complex momentum plane. This is shown to the
right half of Fig.~\ref{smat_poles}. Here the non-resonant continuum
is shown as the red colored contour. In practice one can choose any
contour for the non-resonant continuum as long as the potential has an
analytic continuation in the complex plane, see for example
\cite{hagen2004}.

Over the last decade the Berggren basis has seen many applications in
calculations of weakly bound and unbound nuclei.
\citeasnoun{michel2002} and \citeasnoun{idbetan2002} employed the
Berggren basis in shell model computations of nuclei and introduced
the Gamow shell model~\cite{michel2009}. Very recently,
\citeasnoun{papadimitriou2013} combined the no-core shell model with a
Berggren basis and applied it to computations of the unbound nucleus
$^5$He. The coupled-cluster method is well suited to describe unbound
nuclei with the Gamow basis because it can handle the increased size
of the model space and it is size extensive. \citeasnoun{hagen2007d}
generalized the coupled-cluster method to the complex and
bi-orthogonal Berggren basis, and computed the ground-state energies
and decay widths of the $^{3-10}$He isotopes, see
Subsection~\ref{subsec:results_he} for details.

In coupled-cluster calculations with Berggren bases, we find it
convenient to compute the Berggren basis from the analytically
continued Schr\"odinger equation in momentum space
\begin{equation}
\label{eq:neweq1}
{\hbar^{2}\over 2\mu}k^2\psi_{nl}(k) + \int_{L^{+}} 
dq {q}^{2}V_{l}(k,q)\psi_{nl}(q) = E_{nl}\psi_{nl}(k),
\end{equation}
see \cite{hagen2004} for details. Here both $k$ and $q$ are defined
on an inversion symmetric contour $L^+$ in the lower half complex
$k$-plane (see Fig.~\ref{smat_poles}), resulting in a closed integral
equation.  The eigenfunctions constitute a complete bi-orthogonal set,
normalized according to the Berggren completeness relation
\cite{berggren1968,berggren1971,lind1993}
\begin{equation}
\label{eq:unity2}
{\bf 1} = \sum _{n\in \bf{C}}\vert\psi_{nl}\rangle\langle\psi_{nl}^{*}\vert + 
\int_{L^{+}} dk k^2\vert\psi_{l}(k)\rangle\langle\psi_{l}^{*}(k)\vert.   
\end{equation} 
In Fig.~\ref{smat_poles} the contour $L^+$ is given by a rotation
followed by a translation in the lower half complex $k$-plane.
$V_l(k,q)$ is the Fourier-Bessel transformation of a suitable finite
range potential (e.g. Woods-Saxon).  An advantage of solving the
Schr\"odinger equation in momentum space is that we do not need to
impose boundary conditions for the bound-, resonant-, and scattering
states. The momentum-space Schr\"odinger equation is solved as a
matrix equation by discretizing the contour $L^+$ by some suitable
quadrature rule, (e.g. Gauss-Legendre quadrature). For non-singular,
finite-range potentials that fall off faster than $1/r$, one typically
finds convergence for both narrow and wide resonant states with
$n_{\rm max} \approx 40$ integration points and with the integration
limits $k \in [0,4-5]$~fm$^{-1}$, see e.g. \cite{hagen2006}. The
maximal radial extent of the wave function in position space can then
be estimated by the formula $R_{\rm max} \approx \pi/\Delta k$, where
$\Delta k$ is the step length of the $k$-space integration. For the
typical example above this corresponds to a maximal radial extent of
$R_{\rm max} \approx 30$~fm for the wave function in position space.

Solving Eq.~(\ref{eq:neweq1}) for protons with the Coulomb potential
in momentum space is not trivial due to the singular behavior of the
Legendre function of the second kind, $Q_\ell(k,q)$, on the diagonal
$k=q$.  There are various methods that deal with this problem such as
the Land{\'e} subtraction method or by introducing a cut in the
$r$-space integration. However, neither of these methods works for
resonances as the integration kernel is no longer analytical in the
complex $k$-plane. \citeasnoun{michel2011} developed an off-diagonal
method that works very well for both resonances and scattering states.
To illustrate the method we follow \cite{hagen2012c} and write
the Coulomb potential in momentum space as
\begin{equation}
  U_{\rm Coul}(k,q) =  \langle k | U_{\rm Coul}(r) - \frac{(Z-1) e^2}{r} | q \rangle + 
  \frac{(Z-1) e^2}{\pi} Q_\ell \left( \frac{k^2 + {q}^2}{2 k q} \right)  \label{Uc_k_kp}.
\end{equation}
The first term of Eq.~(\ref{Uc_k_kp}) decreases very quickly for $r
\rightarrow +\infty$ and can be calculated by numerical
integration. However, the second term has a logarithmic singularity at
$k = q$. The off-diagonal method consists of replacing the infinite
value $Q_\ell(1)$ in Eq.~(\ref{Uc_k_kp}) occurring at $k = q$ by a
finite value depending on the discretization used.

Table~\ref{tab2} shows results for the $s_{1/2},d_{3/2},d_{5/2}$
resonant states from a Woods-Saxon plus Coulomb potential for an
increasing number of integration points. We used a rotated and
translated contour as shown in Fig.~\ref{smat_poles} with $N_R$
($N_T$) Gauss-Legendre quadrature points along the rotated
(translated) line.  We see that $N_R+N_T \approx 40$ is sufficient to
reach a convergence at the level of a few keV, proving the accuracy
and efficiency of the off-diagonal method applied to the solution of
Eq.~(\ref{eq:neweq1}).
\begin{table}[h]
\begin{center}
\begin{tabular}{|l|l|l|l|l|l|l|l|l|}\hline
\multicolumn{2}{|c|}{} & \multicolumn{2}{c|}{$s_{1/2}$ }  & \multicolumn{2}{c|}{$d_{3/2}$}  & \multicolumn{2}{c|}{$d_{5/2}$} \\ 
\hline
 $N_R$ & $N_T$   &Re[$E$] & $\Gamma$&Re[$E$] & $\Gamma$ & Re[$E$]  & $\Gamma$ \\ \hline
 5 & 15    &    1.1054    &  0.1446 & 5.0832 & 1.3519   & 1.4923   &  0.0038  \\ 
 5 & 20    &    1.1033    &  0.1483 & 5.0785 & 1.3525   & 1.4873   &  0.0079  \\
 10& 25    &    1.0989    &  0.1360 & 5.0765 & 1.3525   & 1.4858   &  0.0093  \\
 10& 30    &    1.0986    &  0.1366 & 5.0757 & 1.3529   & 1.4849   &  0.0103  \\  
 15& 40    &    1.0978    &  0.1351 & 5.0749 & 1.3531   & 1.4842   &  0.0111  \\              
 15& 50    &    1.0978    &  0.1353 & 5.0746 & 1.3533   & 1.4838   &  0.0114  \\
 20& 60    &    1.0976    &  0.1349 & 5.0745 & 1.3533   & 1.4837   &  0.0116  \\
 30& 70    &    1.0975    &  0.1346 & 5.0744 & 1.3534   & 1.4837   &  0.0117  \\
\hline \hline 
\multicolumn{2}{|c|}{\cite{michel2011}} &   1.0975    &  0.1346 & 5.0744 & 1.3535   & 1.4836   &  0.0119  \\
\hline 
\end{tabular}
\end{center} 
\caption{Convergence of the $s_{1/2}, d_{3/2} $ and $ d_{5/2}$ proton resonant states with number of integration points 
  for a Woods-Saxon potential compared to results of \cite{michel2011}. 
  The width is $\Gamma = -2{\rm Im}[E]$ (in MeV).}
\label{tab2}
\end{table}

Let us briefly discuss the computation of the $NN$ interaction in the
Berggren basis. \citeasnoun{hagen2006b} expressed the $NN$ interaction
in a Berggren basis by introducing an intermediate expansion over a
finite number of harmonic oscillator states,
\begin{equation}
  \langle ab\vert V_{NN}\vert cd\rangle \approx
  \sum_{\alpha \beta \gamma \delta}^{n_{\rm max}} \langle ab\vert \alpha \beta \rangle
  \langle \alpha \beta \vert V_{NN}\vert \gamma \delta \rangle \langle \gamma \delta \vert cd \rangle,
\end{equation}
here $n_{\rm max}$ is the maximum number of radial harmonic oscillator
functions for a given partial wave $lj$. The two-particle overlap
integrals $ \langle ab\vert \alpha \beta \rangle $ are given in terms
of the single-particle overlaps $ \langle a \vert \alpha \rangle
\langle b \vert \beta \rangle $. Here Roman letters label the Berggren
states and Greek letters label the harmonic oscillator states. Because
of the Gaussian falloff of the harmonic oscillator functions the
single-particle overlap integrals are always finite.
\citeasnoun{hagen2006b} showed that a good convergence of both narrow
and wide resonances can be obtained already with $n_{\rm max} =
4$. Having represented the Hamiltonian in the Berggren basis, one can
then perform Gamow-Hartree-Fock calculations which gives the input to
coupled-cluster calculations.  Recently, this approach was used by
\citeasnoun{caprio2012} in the context of the no-core shell model and
using a Coulomb Sturmian basis.  We finally note that
\citeasnoun{mihaila2003} proposed a continuum coupled-cluster
expansion based on real momentum states, and reported first
applications for a single-particle problem.

%% file: TechDetails_Interact.tex
The coupled-cluster method can employ a variety of interactions. In
this Subsection, we briefly describe some of the interactions that
found applications within coupled-cluster theory.

\subsubsection{Interactions from chiral EFT}

The spontaneous breaking of chiral symmetry at low energies is one of
the hallmarks of quantum chromodynamics, the theory of the strong
interaction. The pion is the Nambu-Goldstone boson of the broken
symmetry (with corrections to this picture due to an explicit breaking
of chiral symmetry), and the long-ranged part of $NN$ interactions is
therefore due to pion exchange. The short-ranged contributions of the
$NN$ interaction is treated either using one-boson-exchange inspired
models, or in a model-independent approach via chiral effective field
theory. Popular examples for the former are the Argonne
interaction~\cite{wiringa1995} and the CD-Bonn
interaction~\cite{machleidt2001}, and examples for the latter are
interactions from chiral EFT
\cite{weinberg1990,weinberg1991,ordonez1992,ordonez1994,ordonez1996,vankolck1999,darocha1994,darocha1995,kaiser1997,epelbaoum1998,epelbaum2000,entem2003,machleidt2011,epelbaum2009,ekstrom2013}. We
note that high-precision interaction models depend on several
parameters that are adjusted to $NN$ scattering data; the optimization
yields a $\chi^2\approx 1$ per degree of freedom.

Interactions derived from EFT have theoretical advantages over
optimized interaction models. First, their currents are consistently
formulated with the Lagrangian (or Hamiltonian), and this is important
for the correct description of observables other than the
energy. Second, a power counting (in the ratio of the probed momentum
scale $Q$ over the cutoff scale $\Lambda$) exists for systematic
improvements of the interaction and observables. Third, the hierarchy
of $NN$ forces, three-nucleon forces, and forces of even higher rank
is explained by the power counting. Three-nucleon forces enter at
next-to-next-to-leading order (N$^2$LO)
order~\cite{vankolck1994,epelbaum2002,navratil2007}, and four-nucleon
forces at N$^3$LO.

Pion-based three-nucleon interactions were first derived
by~\citeasnoun{fujita1957}, and they have been known to improve the
agreement between empirical data and nuclear structure calculations
for a long time (see, e.g., ~\cite{kuemmel1978}). More recently, the
{\it ab initio} Green's function Monte Carlo calculations demonstrated
convincingly that a realistic computation of light nuclei requires
three-nucleon forces~\cite{pudliner1997,pieper2001}.  On physical
grounds, the appearance of three-nucleon forces is not surprising:
nucleons are not point particles and their substructure must lead to
effective three-nucleon forces when three nucleons are sufficiently
close to each other. We note, however, that three-nucleon forces can
only consistently be formulated for a given $NN$
interaction. \citeasnoun{polyzou1990} showed that a two-body
Hamiltonian alone might be equivalent to a different two-body
Hamiltonian plus three-body forces. A nice illustration of this
finding in the context of renormalization scheme dependence was given
by \citeasnoun{jurgenson2009}. It is thus not surprising, that the
character of three-nucleon forces depend on the $NN$ interaction they
accompany. Three-nucleon forces employed with the Argonne $NN$
interactions are mainly attractive in light nuclei~\cite{pieper2001},
while three-nucleon forces for low-momentum interactions effectively
act repulsive in shell-model calculations~\cite{otsuka2010}.

In recent years, the important role of chiral three-nucleon forces has
been explored and confirmed in the lightest nuclei, $p$-shell nuclei,
and neutron and nuclear matter. We note that these calculations often
combine $NN$ forces at N$^3$LO with three-nucleon forces at N$^2$LO,
see~\cite{hammer2013} for a recent review. Here, we briefly summarize
some important results. In few-nucleon systems,
\citeasnoun{kalantar2012} reviewed the present status of the field. In
light nuclei, chiral three-nucleon forces affect the binding energy,
radii and transitions~\cite{navratil2009,maris2012}.  They are
responsible for the correct level ordering in
$^{10}$B~\cite{navratil2007b}, and the anomalous long half life of
$^{14}$C~\cite{holtjw2008,holtjw2009,maris2011}. In oxygen isotopes,
chiral three-nucleon forces determine the position of the neutron drip
line and the structure of neutron-rich
isotopes~\cite{otsuka2010,hagen2012a,hergert2013,cipollone2013}. In
calcium isotopes, chiral three-nucleon forces are pivotal for our
understanding of shell evolution and (sub)shell
closures~\cite{holt2012,hagen2012b,holt2013,wienholtz2013}. Three-nucleon
forces also play an important role in neutron
matter~\cite{hebeler2010b} and the saturation of nuclear
matter~\cite{holtjw2010,hebeler2011}.

Very recently, this picture about the decisive role of three-nucleon
forces in chiral effective field has been questioned to some
extent. \citeasnoun{ekstrom2013} optimized the chiral $NN$
interactions at NNLO with the optimization tool {\sc
POUNDerS}~\cite{kortelainen2010}. The resulting $NN$ interaction
NNLO$_{\rm opt}$ exhibits $\chi^2\approx 1$ per degree of freedom for
laboratory energies below 125~MeV and is thus a high-precision
potential. For NNLO$_{\rm opt}$ the adopted value for the pion-nucleon
constant $c_4$ falls outside the range expected from pion-nucleon
scattering (but is similar in size as for the chiral N$^3$LO
interaction by~\citeasnoun{entem2003}); its spin-orbit force exhibits
deficiencies above laboratory energies of about 80~MeV, as is evident
in the $p$-wave phase shifts. Remarkably, the chiral $NN$
interaction NNLO$_{\rm opt}$ alone is able to reproduce the experimental
binding energies and the dripline in neutron-rich oxygen isotopes. In
isotopes of calcium, NNLO$_{\rm opt}$ overbinds but energy
differences and shell closures are well reproduced -- again hinting at
a possibly less complex role of the omitted three-nucleon forces. This
matter is still very much fluid, and it is too early to draw
conclusions. In particular, the role of three-nucleon forces
corresponding to NNLO$_{\rm opt}$, and the effects of higher orders of
the power counting are subjects of ongoing studies.

Let us turn to how three-nucleon forces are used in practical
computations.  \citeasnoun{hagen2007a} derived the coupled-cluster
equations for three-body Hamiltonians in the CCSD approximation. This
approach adds 68 diagrams to the coupled-cluster equations, and
significantly increases the computational cost compared to $NN$ forces
alone. \citeasnoun{binder2013} computed $\Lambda$-CCSD(T) corrections 
in the presence of three-nucleon forces. The inclusion of
three-nucleon forces in nuclear structure calculations is also
challenging due to the sheer number of three-body matrix elements that
are input to such calculations~\cite{vary2009}.

To deal with these challenges, nuclear theorists have employed
approximations that essentially reduce three-nucleon forces to
in-medium two-body forces by averaging the effect of the third
particle over the density.  The coupled-cluster
\cite{emrich1977,coon1977} and shell-model
calculations~\cite{coon1978} of light nuclei, for instance, used the
effective two-body force by \citeasnoun{blatt1975} that resulted from
averaging a two-pion exchange over the third particle in nuclear
matter. Similar approaches are employed
today~\cite{mihaila2000a,hagen2007a,holtjw2009,hebeler2010b,otsuka2010,hagen2012a,roth2012}.
These approximations assume that the most important contributions of
three-nucleon forces can be written as density-dependent two-nucleon
forces~\cite{zuker2003}. Let us describe this idea here in the
coupled-cluster framework~\cite{hagen2007a}. In its normal-ordered
form, the three-body Hamiltonian is \ba
\label{normal3}
\hat{H}_3 &=& {1\over 6} \sum_{ijk} \langle ijk||ijk\rangle
+ {1\over 2} \sum_{ijpq} \langle ijp||ijq\rangle
\{\hat{a}^\dagger_p \hat{a}_q \} \nonumber\\
&+& {1\over 4} \sum_{ipqrs} \langle ipq||irs\rangle
\{\hat{a}^\dagger_p \hat{a}^\dagger_q \hat{a}_s \hat{a}_r \} + \hat{h}_3 \ . 
\ea
Thus, the contributions of the three-body forces consist of an 
energy shift, of normal-ordered one-body and two-body terms, and of a  
``residual'' three-body Hamiltonian
\be
\label{h3}
\hat{h}_3\equiv {1\over 36} \sum_{pqrstu} \langle pqr||stu\rangle
\{\hat{a}^\dagger_p \hat{a}^\dagger_q \hat{a}^\dagger_r \hat{a}_u
\hat{a}_t \hat{a}_s \} \ . 
\ee 
In these expressions, the brackets $\{\ldots\}$ denote normal ordering
with respect to the coupled-cluster reference state. Note that the
normal-ordered Hamiltonian $\hat{H}_3-\hat{h}_3$ already contains
important contributions of three-nucleon forces though it is of
two-body form.  This is a tremendous technical
simplification. \citeasnoun{hagen2007a} found that the omission of
the three-body terms~(\ref{h3}) is a very good approximation for
$^4$He and low-momentum $NN$ interactions. The comparison between
coupled-cluster and no-core shell model results showed that this
approximation improves with increasing mass of the
nucleus~\cite{roth2012,binder2013}. In nuclear matter, the quality of the
normal-ordered approximation can be understood through arguments from
Fermi-liquid theory~\cite{friman2011}.

In large model spaces, the normal-ordered three-body contribution to
the two-body part of the Hamiltonian of Eq.~(\ref{normal3}), is still
expensive since it requires a significant number of three-body matrix
elements to be computed only to be summed over subsequently [see
Eq.~(\ref{normal3})]. A further simplification has been proposed
recently in the computation of neutron-rich isotopes of
oxygen~\cite{hagen2012a}, see Sect.~\ref{subsec:results_o} for
results. In this approach, one directly employs the two-body force
that results from normal-ordering of the three-nucleon force in
symmetric nuclear matter as a schematic correction to the chiral $NN$
interaction. This correction depends on the Fermi momentum and the
low-energy constants of the short-ranged contribution of the chiral
three-nucleon force, which are treated as adjustable parameters. This
approximation is a poor-man's solution to a challenging problem and
useful in practical calculations.

\subsubsection{Low-momentum and renormalization group 
interactions}

Often, it is not convenient to employ very large model spaces. In such
a case one has to properly renormalize the ``bare'' interaction. Popular
examples for renormalization procedures are low-momentum potentials
$\vlowk$~\cite{bogner2003}, SRG
transformations~\cite{bogner2007,furnstahl2013}, or the
$G$-matrix~\cite{hjorthjensen1995}. In what follows, we briefly
summarize these procedures because they entered in coupled-cluster
computations~\cite{kowalski2004,hagen2007b,roth2012,binder2013}.

The low-momentum potentials $\vlowk$ result from ``bare'' $NN$
potentials by integrating out high-momentum modes above a cutoff
$\lambda$ in a renormalization group procedure. This method works by
``decimation'' of degrees of freedom in the sense of the
Kadanoff-Wilson renormalization group (RG), and it removes high-momentum
modes from the interaction. Up to the cutoff, phase shifts of the $NN$
interaction are unchanged.  Note that the corresponding
renormalization flow of three-nucleon forces has not yet been worked out,
yet. Instead, three-nucleon forces are taken in the form of chiral
three-nucleon forces. The corresponding low-energy constants $c_D$ and
$c_E$ of the short-ranged parts of the three-nucleon force are fit to
data in the $A=3,4$ nuclei and are available for different
cutoffs~\cite{nogga2004}.

The SRG transformation~\cite{bogner2007} does not remove any modes
from the interaction but instead decouples low-momentum modes from
high-momentum modes via a similarity
transformation~\cite{glazek1993,wegner1994}, see \cite{furnstahl2013}
for a recent review. The transformation is governed by a generator and
implemented as a flow equation, i.e. as a first-order differential
equation in the space of matrix elements or coupling constants. The
flow parameter $\lambda$ has dimensions of a momentum cutoff. The
generator determines which modes are decoupled, and the flow parameter
determines the degree of the decoupling. In this approach, $NN$ and
three-nucleon forces are consistently transformed, and the
transformations have been worked out in the oscillator
basis~\cite{jurgenson2009} and in momentum
space~\cite{hebeler2012}. In contrast to the $\vlowk$ potentials which
are phase-shift equivalent only up to the cutoff $\lambda$, the SRG
potentials are unitary transformations and thus phase-shift
equivalent.

\begin{figure}[thbp]
  \begin{center}
    \includegraphics[width=0.6\textwidth,clip=]{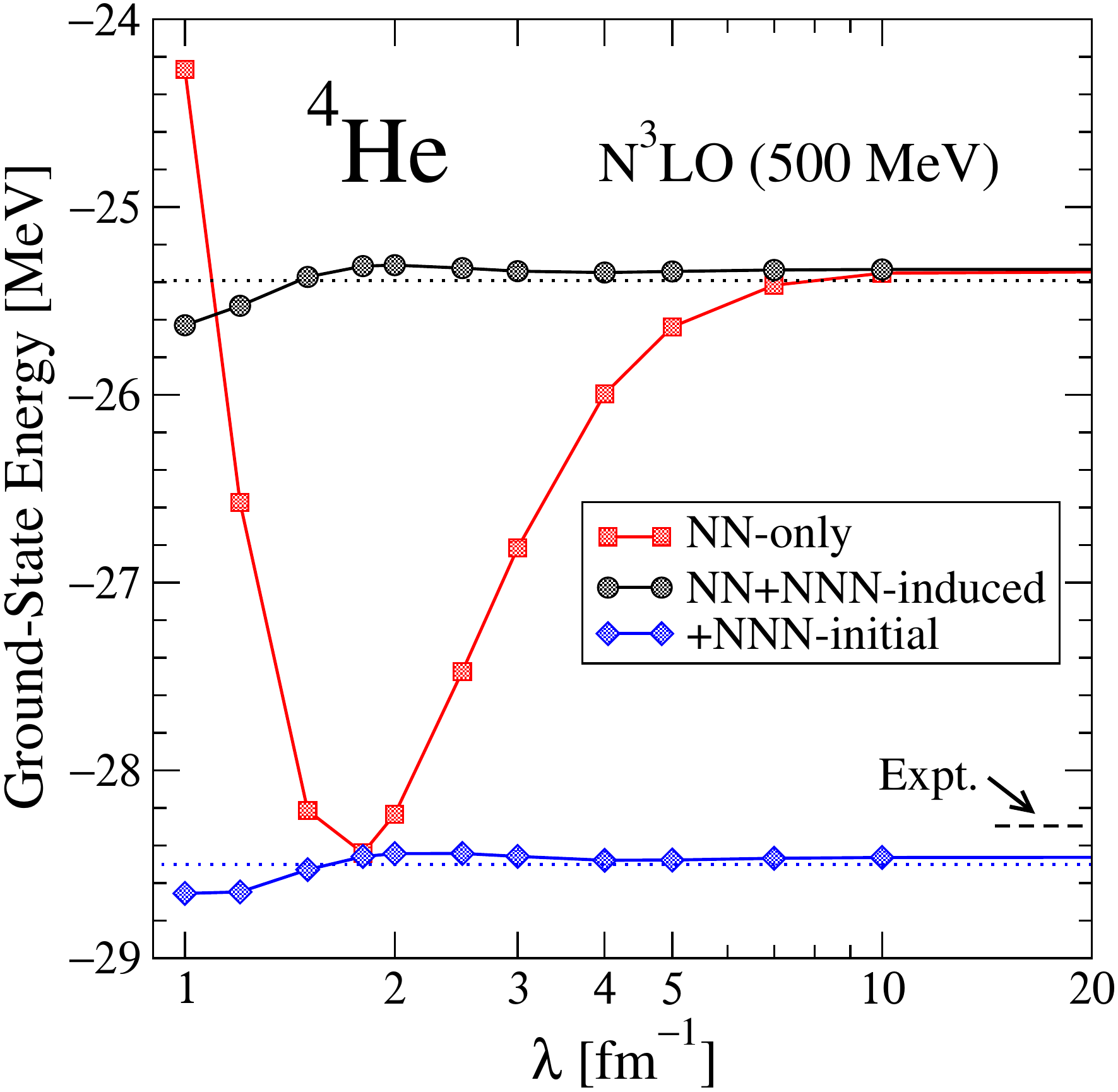}
  \end{center}
  \caption{Ground-state energy of
$^4$He as a function of the flow parameter $\lambda$ computed using an
$NN$ interaction only, an induced three-body interaction starting from
an $NN$ interaction but evolving both two- and three-body operators,
and finally a calculations which includes both $NN$ and three-nucleon
interactions. See text for more details. Taken from arXiv:0905.1873
with permission, see also~\cite{jurgenson2009}.}
  \label{fig_jurgenson}
\end{figure}

It is important to note that the SRG transformation is only a
similarity transformation in the $A$-body system if one generates and
follows the flow of up to $A$-body operators. A nice illustration of
this concept by \citeasnoun{jurgenson2009} is shown in
Fig.~\ref{fig_jurgenson}. The figure shows the ground-state energy of
$^4$He as a function of the flow parameter $\lambda$ computed in three
different ways.  The red line starts (at large values of the flow
parameter $\lambda$) from a ``bare'' $NN$ interaction from chiral EFT
and shows how the ground-state energy varies under the evolution when
only $NN$ forces are kept and evolved. The grey line again shows the
results if one starts (at high values of $\lambda$) from a ``bare''
interaction but evolves $NN$ and three-nucleon operators
simultaneously. The resulting ground-state energy is almost
independent of the evolution parameter, and at $\lambda\approx 2\fmi$
one sees a variation that is presumably due to missing four-body
operators. The blue line starts (at high values of $\lambda$) from
``bare'' $NN$ and three-nucleon interactions from chiral EFT, and in
the SRG transformation one evolves two-body and three-body
operators. Again, the results are almost independent of $\lambda$, and
a small variation develops for $\lambda\approx 2~\fmi$.  Note also
that the variation of the red curve gives an estimate for the
contribution of the omitted three-nucleon forces, and the variation of
the blue curve gives and estimate for the missing contributions of
four-nucleon forces. Thus, any variation of the results with the
cutoff or flow parameter measure contributions from omitted
short-ranged forces that are of higher rank. This turns the study of
cutoff dependencies into a useful tool. For more details on
low-momentum interactions, we refer the reader to the review
by~\citeasnoun{bogner2010}.

The following simple argument might be useful for understanding why
the renormalization group transformation induces forces of higher
rank, and at which scale $\lambda$ the RG-induced $a$-body forces
might become relevant. We note that the cutoff $\lambda$ induces the
shortest length scale $2\pi/\lambda$, and $a$-body forces become
relevant once the $a^{\rm th}$ particle ``sees'' about $a-1$ particles
in the volume of a sphere with diameter $2\pi/\lambda$. Let $\rho$
denote the nuclear saturation density. Thus, $a$-body forces become
relevant for cutoffs \be
\label{guesstimate}
\lambda \approx \left({4\pi^4\rho \over 3(a-1)}\right)^{1/3} \ .  
\ee 
Inserting  $a\approx 2.5$ ($a\approx 3.5$) into Eq.~(\ref{guesstimate}), 
one would thus expect that three-nucleon forces
(four-nucleon forces) become relevant below a cutoff of
$\lambda\approx 2.4\fmi$ ($\lambda\approx 2.0\fmi$). These estimates are
in reasonable agreement with actual calculations, see
e.g. Fig.~\ref{fig_jurgenson}.

The reduction of the size of the model space via SRG appears
particularly attractive for computations of matrix elements of
three-nucleon forces. \citeasnoun{roth2011a} employed SRG evolved
interactions from chiral effective field theory and computed matrix
elements of $NN$ and three-nucleon forces in oscillator spaces up to energies
of 12 oscillator spacings for two and three nucleons,
respectively. (The maximum average energy per nucleon is smaller for
the three-nucleon forces than for two-nucleon
forces). \citeasnoun{roth2011a} found that the binding energies exhibit
a dependence on the SRG cutoff $\lambda$ that increases with
increasing mass number. We recall that a renormalization-scale
dependence of observables indicates that the SRG is not unitary in the
many-nucleon system. The origin of this behavior is subject of ongoing
research~\cite{hebeler2012,jurgenson2013}.  A practical solution to
this problem is to start the SRG with a three-nucleon force that
exhibits a lower chiral cutoff than the corresponding two-nucleon
force~\cite{roth2011a,roth2012}.  This approach is a better controlled
approximation than the treatment of three-nucleon forces as in-medium
corrections to nucleon-nucleon forces discussed in the previous
Subsubsection.

SRG transformations of nuclear interactions can also be performed in
the presence of a non-trivial vacuum state. The resulting in-medium
SRG~\cite{tsukiyama2011} is a means to directly compute the
ground-state energy of an $A$-body nucleus and also yields the nuclear
interaction with respect to the non-trivial vacuum state. Here,
three-nucleon forces and forces of higher rank are included in the
normal-ordered approximation. Recently, this approach was extended to
open-shell nuclei~\cite{hergert2013,hergert2013b}.

The in-medium SRG and the coupled-cluster method have much in common.
Both are built on similarity-transformed Hamiltonians with respect to
a reference state.  The former employs a unitary similarity
transformation that keeps a Hermitian Hamiltonian, while the latter
yields a non-Hermitian Hamiltonian. In the in-medium SRG, the
similarity transform results from the solution of a differential
equation while the coupled-cluster method seeks an iterative solution
of the coupled-cluster equations.  The imaginary-time evolution of the
coupled-cluster equations also yields a continuous similarity
transformation of the Hamiltonian that converges towards the solution
of the coupled-cluster equations~\cite{pigg2012}. It is clear that
both methods can be used in the derivation of effective interactions
with respect to a non-trivial vacuum state. This aspect of
coupled-cluster theory has been emphasized in several recent
works~\cite{suzuki1992,suzuki1994,kohno2012}.

\citeasnoun{hergert2007} pointed out that the earlier developed
unitary correlation operator method
(UCOM)~\cite{feldmeier1998,roth2010} can also be viewed as a SRG
transformation. However, in {\it ab initio} calculations, the UCOM
potential exhibits a slower convergence than other SRG
interactions~\cite{roth2009,hagen2010b}.

\subsubsection{$G$ matrix}

The Brueckner $G$ matrix was developed to employ $NN$ potentials with
a strong short-range repulsion in the description of nuclear
matter~\cite{brueckner1954,brueckner1955,day1967}. This method is also
the basis for the derivation of effective shell-model
interactions~\cite{kuo1968}, see the review by
\citeasnoun{hjorthjensen1995}. It is illuminating to study the
convergence properties of the $G$ matrix, and we briefly summarize the
results from ~\cite{hagen2010b}. The $G(\overline{\omega})$ matrix
depends on the ``starting energy'' $\overline{\omega}$, and in the
limit of an infinite model space (i) becomes independent of
$\overline{\omega}$ and (ii) becomes identical to the interaction used
in its construction.  For increasing number of oscillator shells
($N+1$), the ground-state energy approaches the value obtained for the
``bare'' chiral $NN$ interaction, and the dependence on
$\overline{\omega}$ weakens. The $G$ matrix converges slowly (and from
below) to the ``bare'' Hamiltonian, and this makes it less is ideal
for {\it ab initio} computations. In the practical construction of
effective interactions for the nuclear shell model, however, the $G$
matrix is a most useful tool~\cite{poves1981,honma2002}.

%% file: TechDetails_CoM.tex
In this Subsection we discuss the treatment of the center of
mass.  The nuclear interaction is invariant under translations, and as
a result, the $A$-body wave function $\psi$ of the atomic nucleus is a
product of an intrinsic wave function $\psi_{\rm in}$ that depends on
$3(A-1)$ coordinates and the wave function $\psi_{\rm cm}$ of the
center of mass
\ba 
\label{fac}
\psi=\psi_{\rm in} \psi_{\rm cm} \ .  
\ea 
The factorization~(\ref{fac}) is manifestly expressed in the $A$-body
basis states if one chooses a single-particle basis of plane wave
states or a single-particle basis of harmonic-oscillator states. In
the latter case (and denoting the oscillator spacing by
$\hbar\omega$), the $A$-body Hilbert space must be a complete
$N\hbar\omega$ space, i.e. the space that consists of all Slater
determinants with excitation energies up to and including
$N\hbar\omega$. The no-core shell model~\cite{navratil2009} employs
such a model space, and the factorization~(\ref{fac}) is thus
guaranteed through the choice of the basis. The coupled-cluster
method, however, does not employ a complete $N\hbar\omega$ space, and
one has to seek alternatives. We discuss three alternative approaches.

The first approach consists of solving the two-body-cluster equations
in the center-of-mass system~\cite{bishop1990b}. In this approach, no
reference is made to the center of mass, and the coupled-cluster
equations are solved in the intrinsic coordinates of the
harmonic-oscillator basis. \citeasnoun{bishop1990} computed $^4$He in
very large model spaces and found a complicated dependence on the
oscillator frequency and a slow convergence with respect to the size
of the oscillator space. Variances of this approach with
translationally invariant cluster functions in position space were
developed in a series of
papers~\cite{guardiola1996,bishop1998,guardiola1998}.  The recent
calculations~\cite{moliner2002} of He, Be, C, and O isotopes with this
approach exhibits a much-improved convergence with respect to the
model space.

The second approach, consists of application of the ``Lawson
method''~\cite{gloeckner1974}, i.e. one employs an oscillator basis of
frequency $\omega$ and adds a harmonic-oscillator Hamiltonian of the
center of mass coordinate (with the same frequency $\omega$) to the
Hamiltonian and varies the corresponding Lagrange multiplier. Within
the NCSM and a full $N\hbar\omega$ model space, the Lawson method
indeed pushes up all spurious states. Within the traditional
shell-model (which does not employ a full $N\hbar\omega$ space), the
Lawson method was criticized by \citeasnoun{mcgrory1975} because the
Lagrangian constraint distorts relevant correlations in the intrinsic
wave function. In coupled-cluster theory, the Lawson method was
applied in several
works~\cite{heisenberg1999,mihaila2000b,dean2004,kowalski2004,wloch2005,gour2006},
and the computed observables depend mildly on the Lagrange
multiplier~\cite{mihaila2000a,roth2009b}. We believe that this
approach, which is tailored to a complete $N\hbar\omega$ oscillator
space, is not the most suitable for other model spaces.

The third approach
\cite{fink1974,zabolitzky1974,kuemmel1978,hagen2007d} employs the
intrinsic Hamiltonian
\ba
\label{ham_in}
H_{\rm in}\equiv  H-T_{\rm cm} = T-T_{\rm cm} +V \ ,
\ea
and appears to be most attractive. Here, $T_{\rm cm}$ denotes the
kinetic energy of the center of mass, and the interaction $V$ is
invariant under translations. The Hamiltonian~(\ref{ham_in}) clearly
does not reference the center of mass, and one finds empirically that
its ground-state wave function factorizes in sufficiently large model
spaces~\cite{hagen2009a}. To quantify this statement, one expands the
ground-state wave function of the $A$-body problem
\be
\label{svd}
\psi = \sum_i s_i \psi_{\rm in}^{(i)} \psi_{\rm cm}^{(i)}
\ee
into sums of products of intrinsic and center-of-mass wave
functions. Such an expansion can always be achieved and amounts to a
singular value decomposition of $\psi$. Here, the singular values are
ordered and non-negative, i.e.  $1\ge s_1\ge s_2\ge s_3 \ge 0$, and
the intrinsic and center-of-mass wave functions $\psi_{\rm in}^{(i)}$ and $\psi_{\rm
cm}^{(i)}$, respectively, are orthonormalized, and thus $\sum_i s_i^2=1$ for
proper normalization of $\psi$. Note that the expansion~(\ref{svd}) corresponds
to the factorization~(\ref{fac}) for $s_1=1$ (and all other singular
values vanish), and $s_1^2$ thus is a good measure of the
factorization achieved in practice. For a toy model of two
interacting particles in one dimension, \citeasnoun{hagen2010b} showed
that $s_1$ quickly approaches $s_1=1$ in sufficiently large oscillator
spaces (containing about 10 oscillator shells), the difference
$1-s_1^2$ assuming tiny values, with $1-s_1^2< 10^{-7}$ or smaller. 

In coupled-cluster calculations, it is impractical to verify the
factorization via the singular value decomposition~(\ref{svd}) as one
avoids expanding the cluster wave function in terms of Slater
determinants. Instead one can demonstrate that the coupled-cluster
ground-state $|\psi\rangle$ fulfills to a good approximation
$\langle\psi|H_{\rm cm}(\tilde{\omega})|\psi\rangle \approx
0$~\cite{hagen2009a,hagen2010b}. Here 
\be
\label{Hcm}
H_{\rm cm}(\tilde{\omega}) \equiv T_{\rm cm} + {1\over 2}Am\tilde{\omega}^2
\vec{R}^2-{3\over 2}\hbar\tilde{\omega} 
\ee 
is the oscillator Hamiltonian of the center-of-mass coordinate
$\vec{R}$, and $\tilde\omega$ a suitably chosen frequency that usually
differs from the frequency $\omega$ of the underlying oscillator
basis. Note that the Hamiltonian~(\ref{Hcm}) is non-negative, and has
a vanishing ground-state energy. Thus, a vanishing expectation value
of this operator indicates that one deals with its ground-state. The
computation of $\tilde\omega$ is described in~\cite{hagen2009a}. In
sufficiently large model spaces, $\tilde\omega$ exhibits only a weak
dependence on the parameter $\omega$. The exact vanishing of the
expectation value $\langle\psi|H_{\rm cm}(\tilde{\omega})|\psi\rangle$
indicates that the wave function $\psi$ factorizes as in
Eq.~(\ref{fac}), and that $\psi_{\rm cm}$ is the Gaussian ground-state
wave function of the center-of-mass oscillator
Hamiltonian~(\ref{Hcm}). In practice, one finds that the expectation
value $\langle\psi|H_{\rm cm}(\tilde{\omega})|\psi\rangle$ is small
compared to the scale $\hbar\tilde{\omega}$ of spurious center-of-mass
excitations, and the smallness of the ratio $\langle\psi|H_{\rm
cm}(\tilde{\omega})|\psi\rangle / (\hbar\tilde\omega)$ is thus a
measure of the quality of the factorization. 

\begin{figure}[thbp]
   \begin{center}
    \includegraphics[width=0.32\textwidth,clip=]{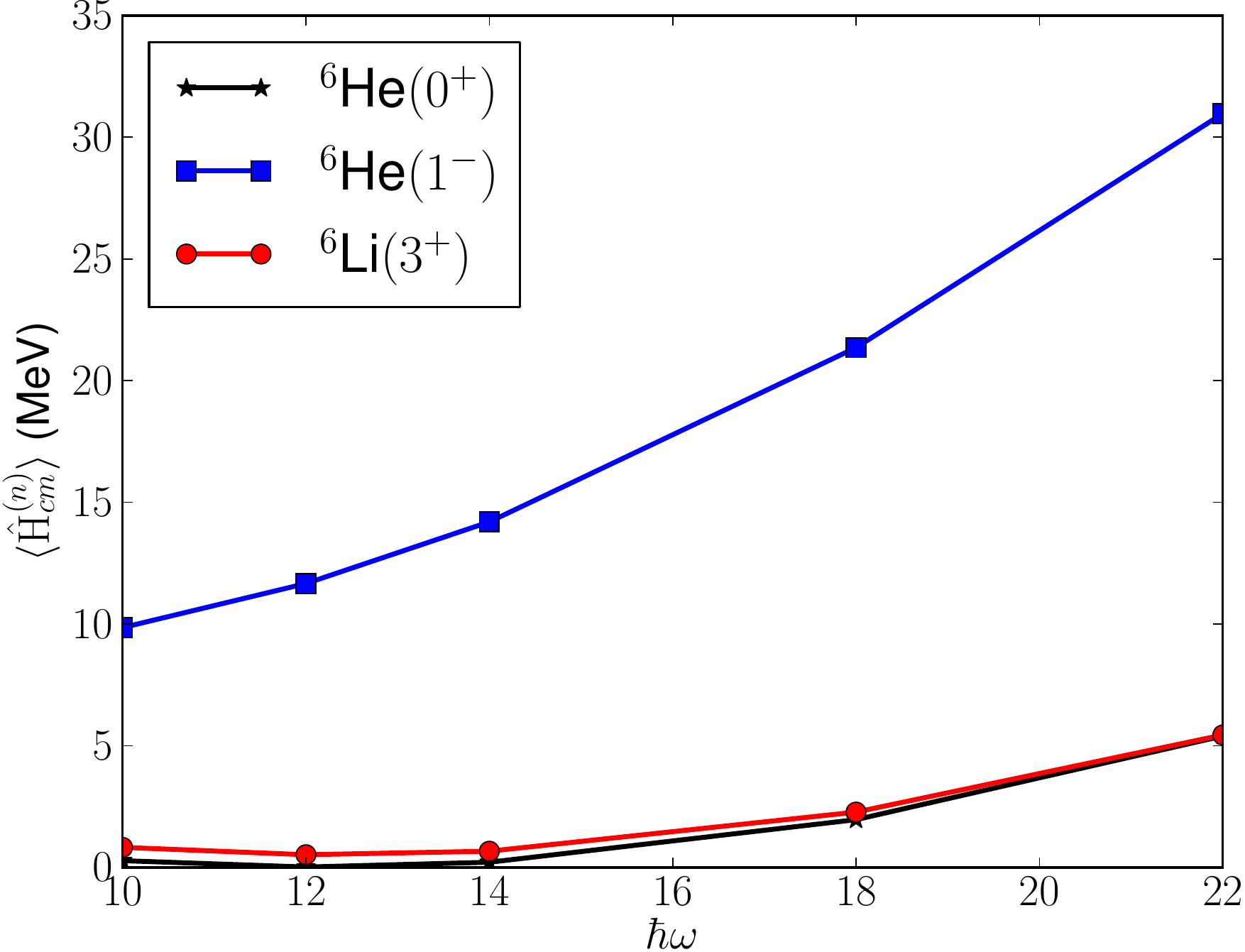}
    \includegraphics[width=0.32\textwidth,clip=]{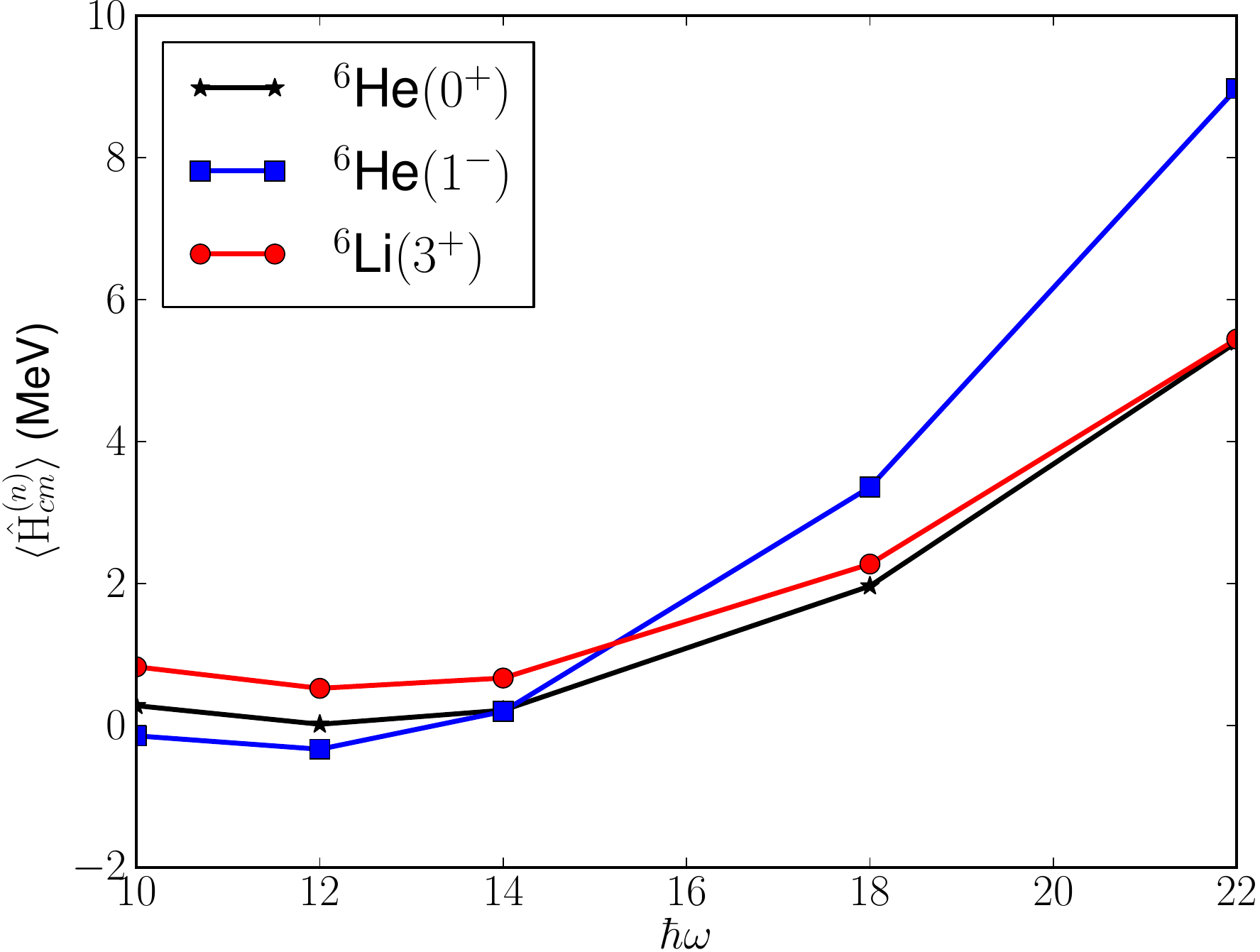}
    \includegraphics[width=0.32\textwidth,clip=]{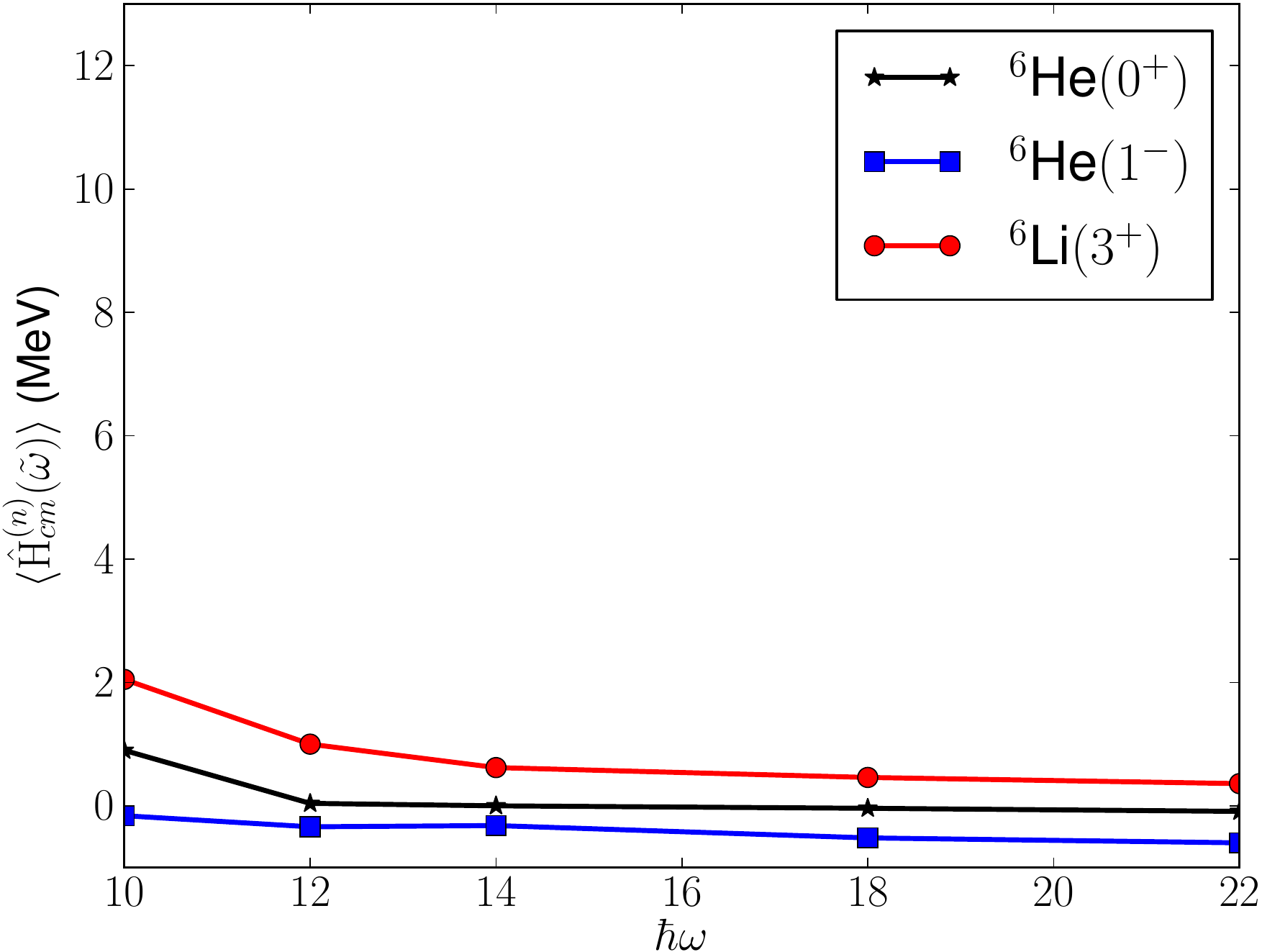}  
  \end{center}
  \caption{Left panel: expectation value of the center of mass
    Hamiltonian (\ref{Hcm}) for $\tilde{\omega}=\omega$ in selected
    states in $^6$He and $^6$Li.  Middle panel: same as in left panel
    except that an additional unit of $\hbar\tilde{\omega}$ is
    subtracted for the $1^-$ state.  Right panel: expectation value
    of the center of mass Hamiltonian evaluated at the appropriate
    $\tilde{\omega}$. Taken from arXiv:1207.7099 with permission, see
    also~\cite{jansen2012}}
        \label{jansen_fig}
\end{figure}

This approach has been extended to neighbors of closed shell nuclei
\cite{hagen2010b,jansen2012}, and one can also identify and remove
spurious states that exhibit excitations of the center-of-mass
coordinate. \citeasnoun{jansen2012} computed states in $^6$He and
$^6$Li using the two-particle attached coupled-cluster method. The
left panel of Fig.~\ref{jansen_fig} shows the expectation value of
Eq.~(\ref{Hcm}) evaluated for $\tilde{\omega} = \omega$ as a function
of the oscillator spacing of the model space. For the $0^+$ state in
$^6$He and $3^+$ state in $^6$Li these expectation values are small
for $\omega \sim 10-14$~MeV. However, the $1^-$ state in $^6$He
exhibits a large expectation value because the center of mass wave
function is in an excited state.  Subtracting another unit of
$\hbar\tilde{\omega}$ for this state in Eq.~(\ref{Hcm}) yields the
expectation values shown in the middle panel of
Fig.~\ref{jansen_fig}. The fact that the expectation values are still
non-negligible demonstrates that the Lawson method should not be
applied, particularly at higher frequency $\omega$. The center of mass
wave functions are approximately oscillator wave functions, but with a
frequency $\tilde{\omega}$ that is in general different from the
frequency $\omega$ of the model space.  Applying the appropriate
formula to compute $\tilde{\omega}$, and computing the expectation
values (\ref{Hcm}), shows that the center of mass wave function
approximately factorizes as depicted in the right panel of
Fig.~\ref{jansen_fig}. The expectation values are not zero, but much
smaller than $\hbar\tilde{\omega} \approx 10$~MeV. This is a measure
of the quality of the factorization.

Note that the factorization argument based on the
Hamiltonian~(\ref{Hcm}) is not limited to coupled-cluster
computations. Recently, the usefulness of this approach was reported
in in-medium SRG computations~\cite{tsukiyama2012}, and studied for the NCSM
with a Sturmian basis~\cite{caprio2012}. It seems that the use
of the intrinsic Hamiltonian~(\ref{ham_in}) in sufficiently large
model spaces indeed yields approximately factorized wave functions, and this solves the
center-of-mass problem in practical computations. Unfortunately,
however, we are still lacking a theoretical understanding of the
Gaussian nature of the factorized center-of-mass wave function.

For a Gaussian center-of-mass wave function it is straightforward to
compute the intrinsic density (i.e. the density in the coordinate of
the $A^{\rm th}$ particle with respect to the center of mass of the
remaining $A-1$ particles) from the density in the laboratory
system~\cite{elliott1955,gartenhaus1957,navratil2004,giraud2008}. The
density in the laboratory system is a convolution of the Gaussian
center-of-mass wave function with the intrinsic density, and the
de-convolution can easily be performed in Fourier space. This procedure
has been employed within the coupled-cluster method for extracting the
intrinsic density of $^{23}$O~\cite{kanungo2011}, see
Subsection~\ref{subsec:results_o}.

An alternative approach was followed by~\citeasnoun{mihaila2000b} in
their computation of the structure function for electron scattering
off $^{16}$O. \citeasnoun{mihaila1999} expand the form factor (a
one-body operator when written with respect to the center of mass) in
terms on one-body, two-body,..., $A$-body density matrices in the
laboratory system.  After a truncation at the two-body density, an
impressive agreement between theory and electron scattering data was
obtained, see Subsection~\ref{subsec:results_o16}.

%% file: Results.tex
In this Section, we review coupled-cluster results for finite nuclei
and focus attention on isotopes of helium, oxygen, calcium, and some
of their neighbors.

\subsection{Helium isotopes} 
\label{subsec:results_he}
\input{Results_He}

\subsection{Oxygen-16 and its neighbors} 
\label{subsec:results_o16}
\input{Results_O16}

\subsection{Neutron-rich isotopes of oxygen}
\label{subsec:results_o}
\input{Results_Oxygens}

\subsection{Neutron-rich isotopes of calcium}
\label{subsec:results_ca}
\input{Results_Calciums}

%% file: Results_He.tex
Bound isotopes of helium display the most extreme ratios of
proton-to-neutron numbers: $^{8}$He is the isotope with the maximum
binding energy, and $^{10}$He is bound with respect to one-neutron
emission but unbound with respect to two-neutron emission (see
Ref.~\cite{tanihata2013} for a recent review on neutron rich halo
nuclei). As few-nucleon systems, these isotopes are an excellent
testing ground for precision comparisons between experiment and
theory. For such light nuclei close to the dripline, the role of the
continuum is particularly important and affects the entire nucleus.

\citeasnoun{hagen2007d} developed the complex coupled-cluster method
to describe structure of loosely bound and unbound neutron-rich
nuclei.  Utilizing a Berggren basis~\cite{berggren1968,berggren1971}
that treats bound, unbound and scattering states on equal footing, a
Gamow-Hartree-Fock basis~\cite{hagen2006b} was constructed and
employed in ab-initio coupled-cluster calculations of very
neutron-rich isotopes of helium ranging from $^{3-10}$He. These
calculations were performed in the $m$-scheme and employed a
low-momentum nucleon-nucleon interaction $V_{{\rm low} k}$ generated
from the N$^3$LO interaction of Entem and
Machleidt~\cite{entem2003}. The cutoff of the interaction was
$\lambda=1.9$~fm$^{-1}$, a value that for this particular interaction
minimizes the expectation value of three-nucleon forces to the binding
energy of $^3$H and $^{4}$He~\cite{nogga2004}. Nevertheless, the
heavier isotopes of helium are underbound for this value of the
cutoff. By utilizing a low-momentum interaction and a truncation in
partial waves, well converged results were obtained in the largest
employed model-space consisting of ~850 active single-particle
orbitals.

\begin{figure}[h]
  \begin{center}
    \includegraphics[width=0.5\textwidth,clip=]{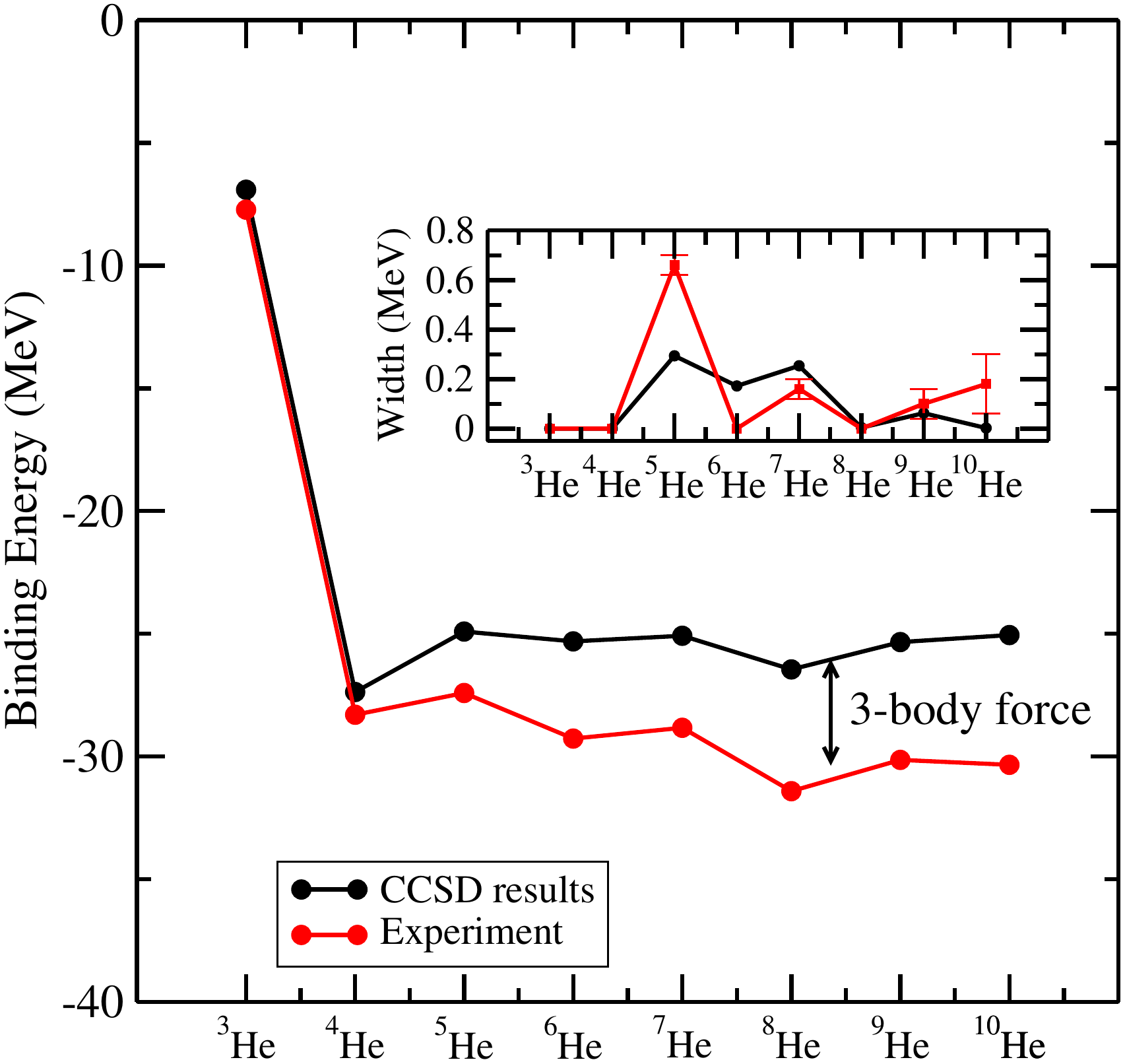}
  \end{center}
  \caption{Binding energies of helium isotopes computed
    with a $\vlowk$ nucleon-nucleon interaction, and compared to data.
    The inset shows the decay widths due to neutron emission. 
    Adapted from \cite{dean2008b}.}
  \label{helium_chain}
\end{figure}

The paper~\cite{hagen2007d} also benchmarked coupled-cluster results
with various triples excitations by comparing them to results from
exact diagonalizations. In particular it was found that CCSDT is
needed to restore spherical symmetry of the open-shell nucleus $^6$He
starting from a deformed reference state within the $m$-scheme
coupled-cluster approach. The computation of particle decay widths in
the complex coupled-cluster framework using a Gamow-Hartree-Fock basis
made predictions for the entire isotopic chain that are in
semi-quantitative agreement with data, see Fig.~\ref{helium_chain}.
For $^{5}$He (i.e. the unbound $\alpha +n$ system) in particular, the
computed resonance energy and width are in reasonable agreement with
the results from the combination of the resonating group method and
the NCSM~\cite{quaglioni2008} and the recent no-core Gamow-shell
model~\cite{papadimitriou2013}. The {\it ab initio} GFMC
calculation~\cite{nollett2007} and the recent NCSM with
continuum~\cite{hupin2013} with two-nucleon and three-nucleon forces
yield a quantitative description of this nucleus. Recently
\citeasnoun{baroni2013a,baroni2013b} developed the ab-initio NCSM with
continuum to compute the lifetimes of the resonance spin-orbit
partners ${1/2}^-$ and ${3/2}^-$ in the unbound nucleus $^7$He. These
calculations addressed a long standing controversy regarding the
lifetime of the excited ${1/2}^-$ resonance state and showed that
theory favors experiments giving a broad ${1/2}^-$ state.

\citeasnoun{bacca2009} studied the cutoff-dependence for low-momentum
$NN$ interactions based on chiral EFT potentials in neutron-rich
isotopes of helium and performed benchmarks between the hyperspherical
harmonics method and the coupled-cluster method in the CCSD and
$\Lambda-$CCSD(T) approximations for $^4$He. They found that
low-momentum $NN$ interactions alone bind $^{6,8}$He with respect to
$^4$He at sufficiently low cutoffs, and estimated omitted
contributions of three-nucleon forces by variation of the cutoff.
There are still open questions regarding the structure of very neutron
rich Helium isotopes. For example, it is still not settled, neither
experimentally nor theoretically, what the spin and parity of the
unbound $^9$He is, and whether a parity inversion of the ground-state
similar to that of $^{11}$Be is the case~\cite{tanihata2013}. An
inversion between the levels ${1/2}^-$ and ${1/2}^+$ in $^9$He would
also impact the structure of the unbound nucleus $^{10}$He.

%% file: Results_O16.tex
Coupled-cluster calculations of $^{16}$O date back decades
ago~\cite{kuemmel1978}, but it is only recently that calculations are
based on high precision nucleon-nucleon interactions. In this Subsection
we describe the results of coupled-cluster calculations for nuclei
around $^{16}$O.

\begin{figure}[thbp]
  \begin{center}
    \includegraphics[width=0.6\textwidth,clip=]{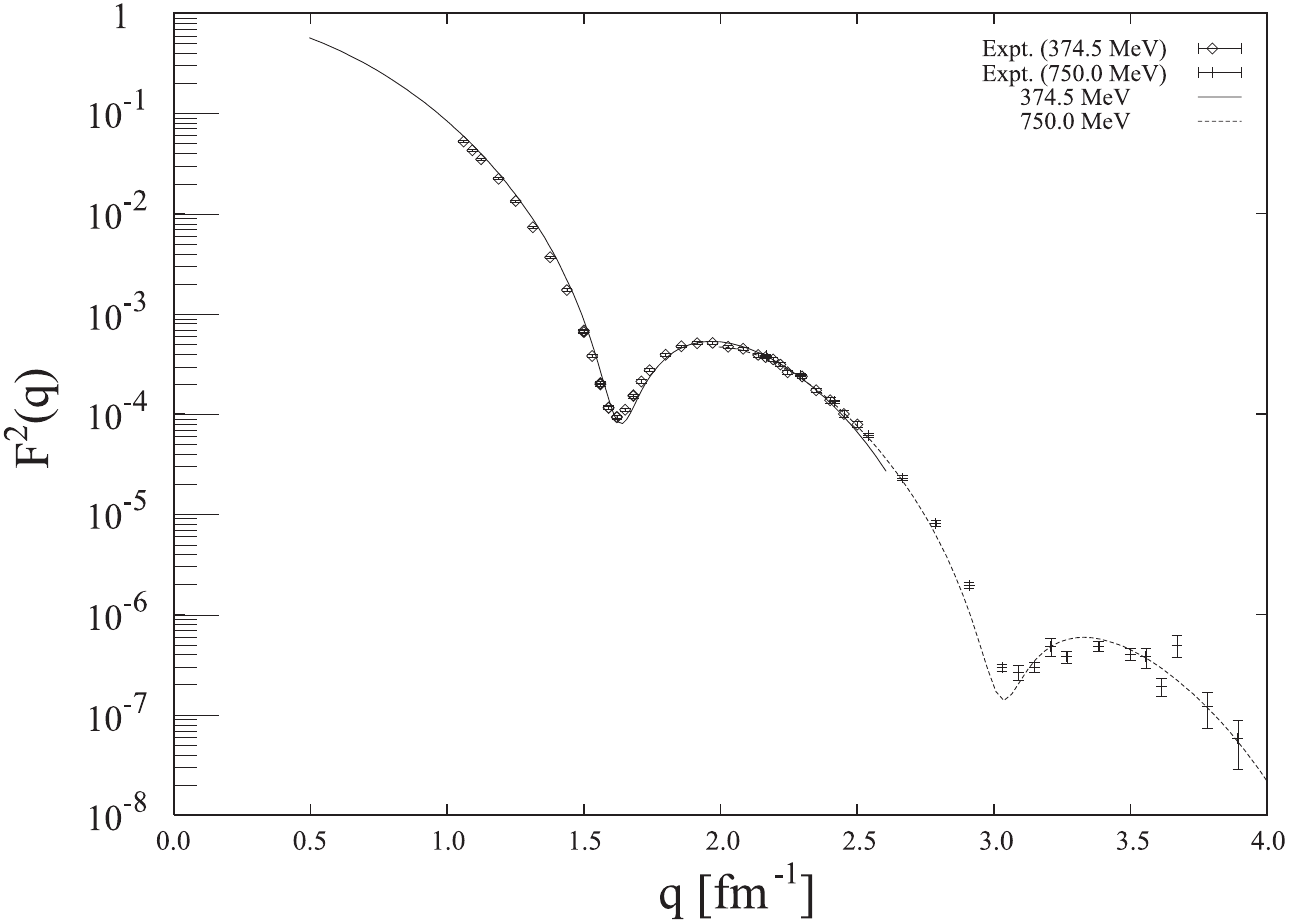}
  \end{center}
  \caption{Elastic scattering form factor as a function of the
    momentum transfer. Taken from arXiv:nucl-th/9910007 with
    permission, see also~\cite{mihaila2000b}.}
  \label{mihaila_fig}
\end{figure}

In a series of papers, Mihaila and Heisenberg developed the
coupled-cluster method for their computation of the elastic scattering
form factor in $^{16}$O. They followed the Bochum approach to deal
with the hard core of the Argonne interactions, and obtained a binding
energy of about 5.9~MeV and 7.0~MeV per nucleon in $^{16}$O for the
Argonne $v_{18}$ and the $v_8$ interaction,
respectively~\cite{heisenberg1999}. Three-nucleon forces were included
by using a two-body force that resulted from summation over the third
particle~\cite{mihaila2000a}. To mitigate the center-of-mass problem,
they used an intrinsic Hamiltonian and the Lawson method; the
calculation of the intrinsic scattering form factor itself included
the necessary two-body corrections in the laboratory
system~\cite{mihaila1999}. The results of these calculations are in
impressive agreement with data~\cite{mihaila2000b}, see
Fig.~\ref{mihaila_fig}.

\citeasnoun{dean2004} followed the standard approach from quantum
chemistry. In contrast to the Bochum approach where the
coupled-cluster method itself tames the hard-core interaction, they
employed a $G$-matrix~\cite{hjorthjensen1995}. The initial studies
computed the nuclei $^4$He and $^{16}$O and focused on conceptual
questions such as convergence in finite model spaces, inclusion of
three-body cluster excitations, treatment of the center of mass, and
the computation of excited states~\cite{kowalski2004}. These papers
demonstrated that the standard coupled-cluster techniques from quantum
chemistry can indeed be used for the description of atomic nuclei.
The computation of ground and excited states in
$^{16}$O~\cite{wloch2005} extended {\it ab initio} nuclear structure
computation to $^{16}$O. Of particular interest was the computation of
the excited $J^\pi=3^-$ state. This state is thought to be a low-lying
1$p$-1$h$ excitation and should thus be captured well by the
coupled-cluster method. However, the coupled-cluster calculation gave
this state at almost twice its expected energy and this pointed to
deficiencies in the employed Hamiltonian and missing many-body
correlations.

\citeasnoun{gour2006} computed excited states in the mass $A=15,17$
neighbors of $^{16}$O by considering them as generalized excited
states of the $^{16}$O closed-shell reference. This again showed that
techniques from quantum chemistry could be transferred to nuclear
structure. They employed $G$-matrices based on the CD-Bonn
interaction, the Argonne $v_{18}$ interaction and the chiral EFT
interaction~\cite{entem2003}. The relative binding energies were
reproduced well, but the spin-orbit splittings exhibited larger
deviations from data.

$^{17}$F is a particularly interesting nucleus because it
exhibits a proton halo as an excited state. The $J^\pi=1/2^+$ halo
state is bound by merely 105~keV, and the computation of such a
fragile state is a challenge. \citeasnoun{hagen2010a} employed $NN$
interactions from chiral EFT, a Gamow basis and the spherical
implementation of the coupled-cluster method in their computation of
the proton halo state. In lieu of three-nucleon forces, they varied
the cutoff of the $NN$ interaction with a similarity-renormalization
group transformation~\cite{bogner2007} and thereby gauged the
dependence of the results on short-ranged three-nucleon forces. They
found a very weakly bound halo state that is insensitive to
variation of the cutoff and a reduced ($3/2^+-5/2^+$) spin-orbit
splitting that exhibits considerable dependence on the cutoff. The
computed $J^\pi=3/2^+$ states in $^{17}$O and $^{17}$F are resonances, 
and the corresponding widths were in reasonable agreement with data.

The binding energy of $^{16}$O has also been computed with
interactions from chiral EFT. Coupled-cluster
calculations~\cite{hagen2010b} based on ``bare'' $NN$ interactions
alone yielded a binding energy of about 7.6~MeV per nucleon when
approximate triples clusters are included, compared to 6.7~MeV per
nucleon in the CCSD approximation~\cite{hagen2008}. The
coupled-cluster energies were confirmed by the unitary model operator
approach~\cite{fujii2009} and the no-core shell
model~\cite{roth2011a}.

%% file: Results_Oxygens.tex
Neutron-rich isotopes of oxygen are very interesting nuclei for
several reasons. First, $^{22}$O and $^{24}$O are closed-shell
nuclei~\cite{thirolf2000,ozawa2000,hoffman2009,kanungo2009}, making
$N=14$ and $N=16$ magic numbers for neutrons in these
isotopes. Second, the doubly magic nucleus $^{24}$O is the heaviest
bound isotope of oxygen. This nucleus has been discovered a long time
ago~\cite{artukh1970}, but only recently did \citeasnoun{hoffman2008}
and \citeasnoun{lunderberg2012} and \citeasnoun{caesar2013} establish
that $^{25}$O and $^{26}$O are unbound resonances in their ground
states. Thus, the dripline in oxygen extends only to neutron number
$N=16$. For the fluorine isotopes, adding one proton shifts the
dripline by six neutrons to $^{31}$F~\cite{sakurai1999}. For
shell-model calculations in this region of the nuclear chart we refer
the reader to ~\cite{caurier1998}.

\begin{figure}[thbp]
  \begin{center}
    \includegraphics[width=0.8\textwidth,clip=]{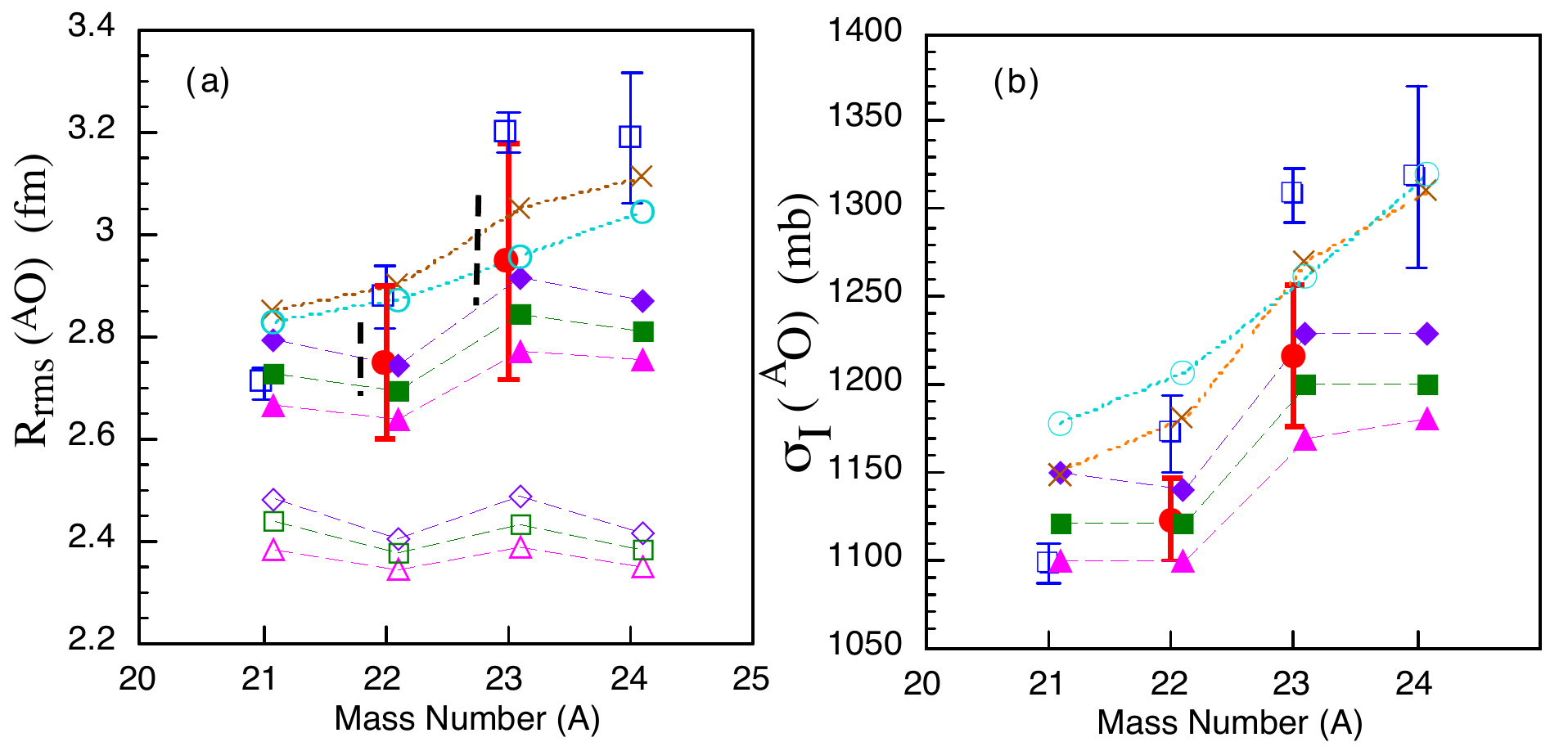}
  \end{center}
  \caption{Right: Interaction cross section of $^{23}$O upon
    scattering off $^{12}$C. Left: Inferred matter radius of $^{23}$O.
    Full circle with error bars: data. Diamonds, squares, triangles:
    results from coupled-cluster calculations with a cutoff parameter =
    4.0, 3.8, 3.6~fm$^{-1}$, respectively. Filled symbols are matter
    radii, and corresponding open symbols are charge radii.  Taken
    from arXiv:1112.3282 with permission, see
    also~\cite{kanungo2011}.}
  \label{kanungo_fig}
\end{figure}

The structure of $^{23}$O has not been without controversy. Early
indications that $^{23}$O could be a halo nucleus~\cite{ozawa2001}
were difficult to reconcile with the sub-shell closures of its
neighbors.  Recently, \citeasnoun{kanungo2011} remeasured the
interaction cross section of $^{23}$O upon scattering off $^{12}$C. In
the framework of the Glauber model, the interaction cross section is
related to the density of the nucleus. Coupled-cluster computations,
based on SRG nucleon-nucleon interactions from chiral EFT, were used
to compute the intrinsic densities and radii of isotopes $^{21-24}$O.
The results are shown in Fig.~\ref{kanungo_fig} for different values
of the SRG cutoff $\lambda$ and compared to the measurements. Overall,
the theoretical calculations very well reproduce the staggering. The
cutoff dependence probes contributions from omitted short-ranged
three-nucleon forces.  The results confirm that $^{23}$O does not
exhibit a halo.

\begin{figure}[thbp]
  \begin{center}
    \includegraphics[width=0.6\textwidth,clip=]{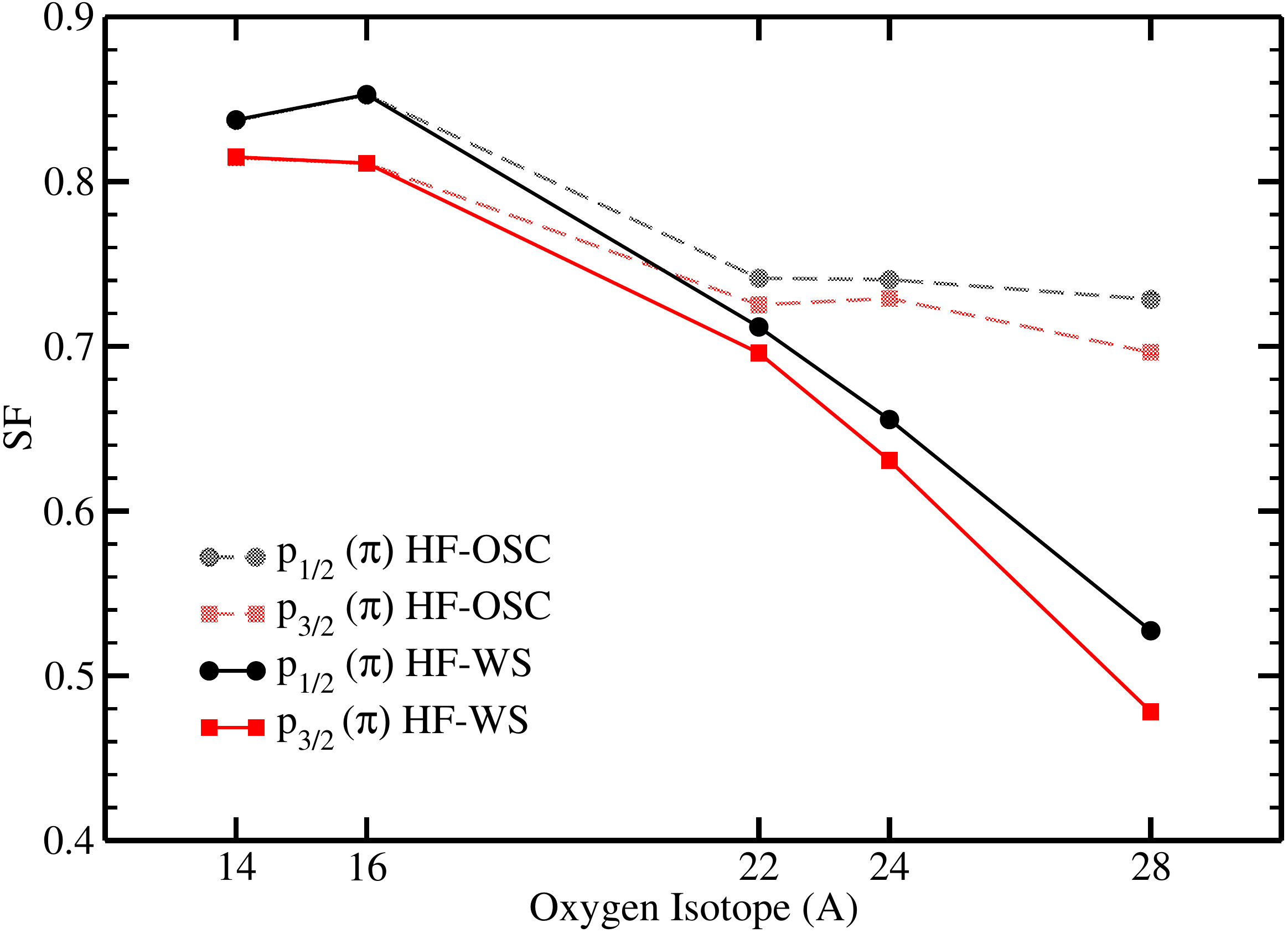}
  \end{center}
  \caption{Theoretical spectroscopic factors for proton removal of
    neutron-rich isotopes of oxygen based on $NN$ interactions from
    chiral EFT. Taken from arXiv:1104.1552 with permission, see
    also~\cite{jensen2011b}.}
  \label{jensen_fig}
\end{figure}

\citeasnoun{jensen2011b} computed spectroscopic factors for proton
removal of neutron-rich isotopes of oxygen based on $NN$ interactions
from chiral EFT. These authors found that the neutron continuum yields
a quenching of the spectroscopic factors. Computations in a Berggren
basis yield reduced spectroscopic factors compared with corresponding
results obtained with a harmonic oscillator
basis. Figure~\ref{jensen_fig} shows that the differences are small
close to the valley of $\beta$ stability but significant close to the
neutron drip line. This demonstrates the importance of coupling to the
continuum, and correlations in very neutron-rich nuclei.

Let us turn to the dripline in oxygen.  The location of the neutron
dripline is a challenging scientific problem, and one needs a very
good understanding of the nuclear interaction, an inclusion of
continuum effects, and an accurate many-body solver to address the
problem.  Thus, it is not surprising that several theoretical
calculations predicted $^{26}$O to be bound, see, e.g., references
cited in the work by \citeasnoun{lunderberg2012}.

\citeasnoun{volya2005} performed a shell-model calculation of
neutron-rich oxygen isotopes. They considered a model space with
$^{16}$O as a closed core and included the scattering continuum. The
empirical two-body interaction was adjusted to data of $sd$-shell
nuclei. Among the key results are the predictions of unbound nuclei
$^{25,26}$O, and the predicted $Q$ values are in good agreement with
the recent data by \citeasnoun{hoffman2008} and
\citeasnoun{lunderberg2012}.

The role of short-ranged three-nucleon forces in the location of the
neutron-drip line was studied by \citeasnoun{hagen2009b}. Their
coupled-cluster calculations of the closed-shell isotopes
$^{22,24,28}$O are based on $NN$ interactions from chiral
EFT~\cite{entem2003}. The $NN$ interactions derived with a momentum
cutoff of $\Lambda=500$~MeV$c^{-1}$ resulted in $^{28}$O being bound
with respect to $^{24}$O, while a ``harder'' $NN$ interaction with a
momentum cutoff of $\Lambda=600$~MeV$c^{-1}$ yielded $^{28}$O to be
unbound. In a renormalization group picture, the removal of
(integrating out) high momentum modes of the $NN$ interaction
generates short-ranged three-nucleon forces. Thus, short-ranged
three-nucleon forces are already relevant for the location of the
dripline in oxygen. Long-ranged three-nucleon forces were also
expected to be most relevant, as both employed $NN$ forces lacked
overall binding of the computed oxygen isotopes.

\citeasnoun{otsuka2010} first studied the role of three-nucleon forces
in the neutron-rich isotopes of oxygen. These authors kept $^{16}$O as
a closed core, employed the $sd$-shell with an appropriate oscillator
frequency as a model space, and employed non-empirical low-momentum
$NN$ forces and three-nucleon forces from chiral EFT. Core
polarization effects were included employing many-body perturbation
theory, see for example \cite{hjorthjensen1995}.  They found that
three-nucleon forces act repulsively in the employed framework, making
$^{24}$O the drip line nucleus. This picture was confirmed in an
enlarged model space that also contains the $f_{7/2}$ and $p_{3/2}$
orbitals~\cite{holt2012b}.

\begin{figure}[thbp]
  \begin{center}
    \includegraphics[width=0.9\textwidth,clip=]{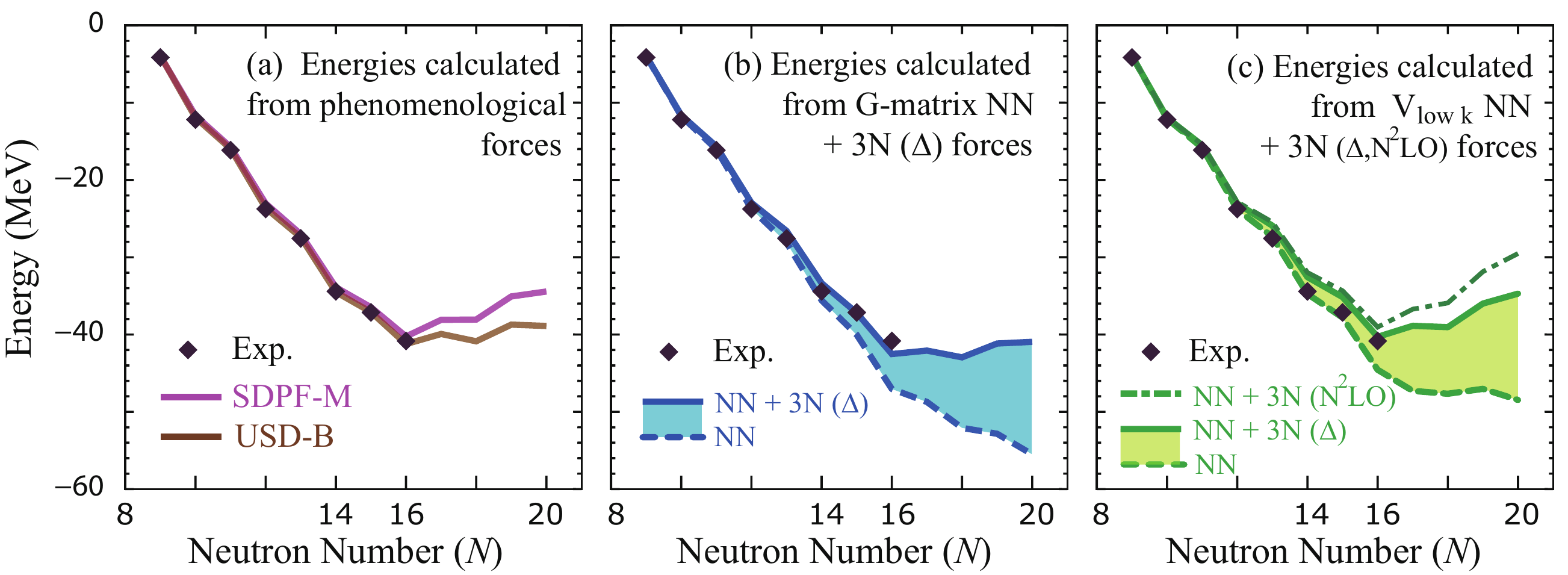}
  \end{center}
  \caption{Shell-model calculations of oxygen isotopes using $^{16}$O
    as closed-shell core and the $sd$-shell as model space.  Both
    phenomenological (SDPF-M and USD-B \cite{utsuno2004,brown2006})
    and microscopic effective interactions were employed, with the
    latter including two different three-body forces as well. Taken
    from arXiv:0908.2607 with permission, see also~\cite{otsuka2010}.}
  \label{subsec_ox_fig0}
\end{figure}
\begin{figure}[thbp]
  \begin{center}
    \includegraphics[width=0.6\textwidth,clip=]{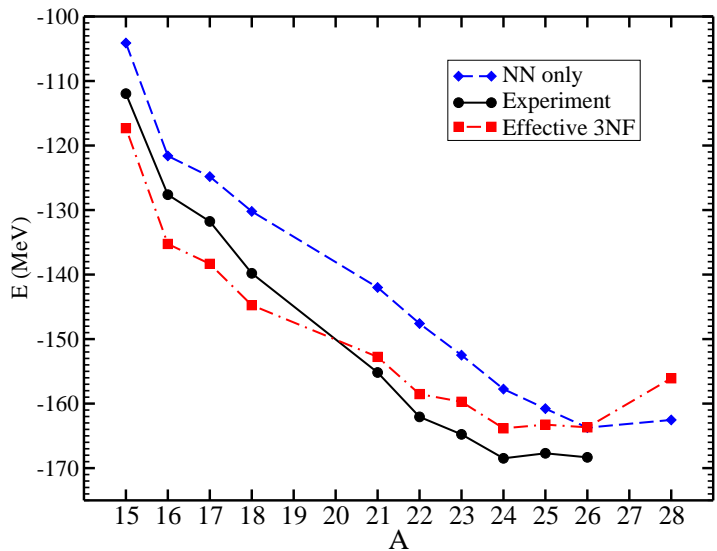}
  \end{center}
  \caption{Ground-state energy of the oxygen isotope $^A$O as a
    function of the mass number $A$. Black circles: experimental data;
    blue diamonds: results from nucleon-nucleon interactions; red
    squares: results including the effects of three-nucleon forces.
    Taken from arXiv:1202.2839 with permission, see
    also~\cite{hagen2012a}.}
  \label{subsec_ox_fig1}
\end{figure}

\citeasnoun{hagen2012a} performed a technically more sophisticated
computation of neutron-rich oxygen isotopes. They used the
coupled-cluster method and addressed continuum effects with a Berggren
basis. Effects of three-nucleon forces were included as in-medium
corrections to $NN$ forces by employing the two-body potential by
\citeasnoun{holtjw2009} that results from taking three-nucleon forces
from chiral EFT and averaging the third nucleon over the Fermi sea of
symmetric nuclear matter. The Fermi momentum and the low-energy
constant $c_E$ of the three-body contact interaction were adjusted to
the binding energies of $^{16,22,24}$O.  Figure~\ref{subsec_ox_fig1}
shows the ground-state energies of oxygen isotopes computed from
chiral $NN$ interactions (diamonds), from the inclusion of effects of
three-nucleon forces (squares) and data (circles). In isotopes of
oxygen, the employed chiral three-nucleon forces act mainly
attractive, but with subtle effects on separation energies. While the
inclusion of three-nucleon forces yields a significant improvement
over $NN$ interactions alone, the employed approximation could still
benefit from further improvements. The computed nuclei are limited to
nuclei that differ by $\pm 1$ or $+2$ mass numbers from references
with closed subshells.  The ground-state energies of nuclei with
closed references were computed in the $\Lambda$-triples
approximation, and the separation energies of their neighbors with
equation-of-motion methods.

\begin{figure}[thbp]
  \begin{center}
    \includegraphics[width=0.6\textwidth,clip=]{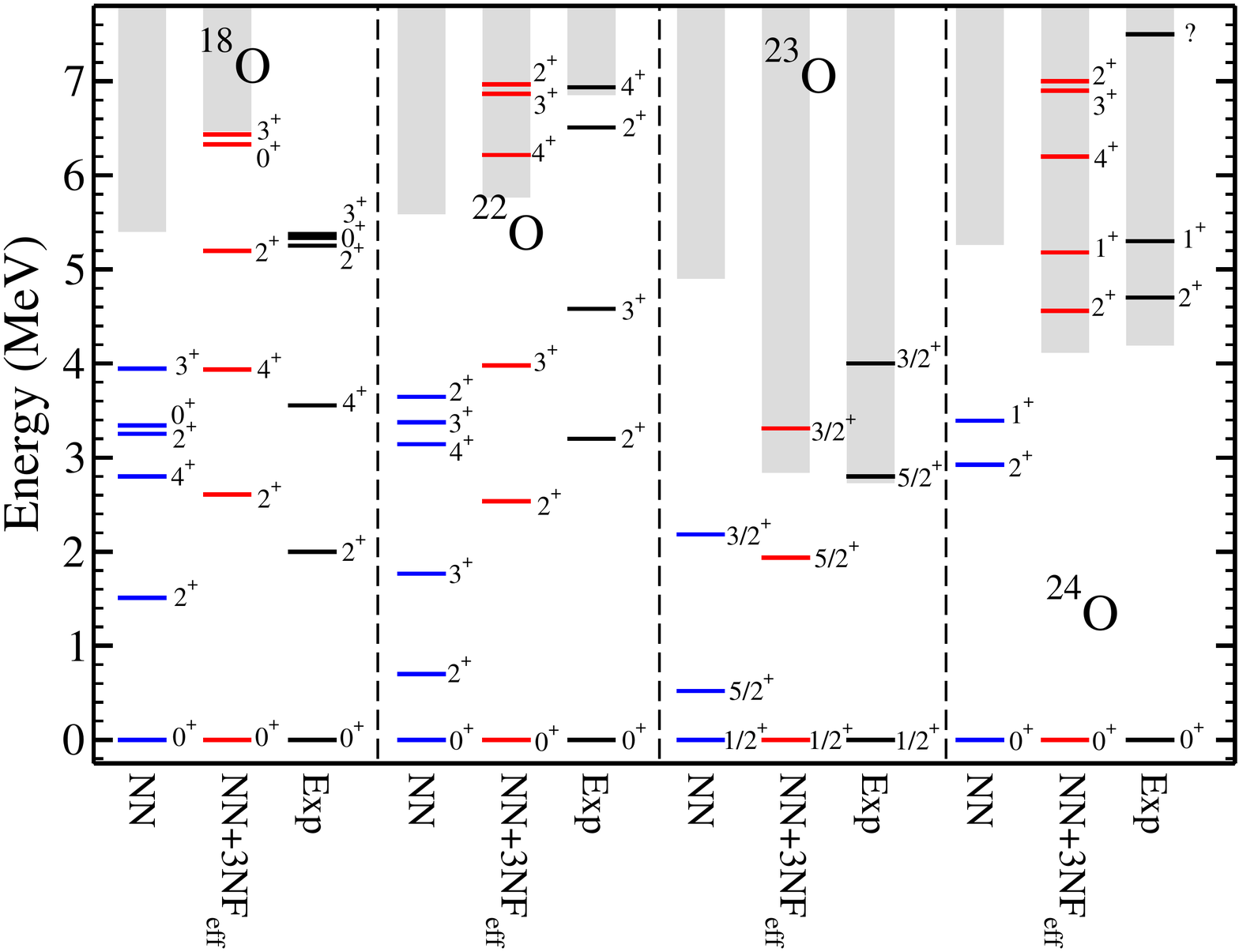}
  \end{center}
  \caption{Excitation spectra of oxygen isotopes computed from chiral
    nucleon-nucleon interactions, with inclusion of the effects of
    three-nucleon forces, and compared to data. Taken from
    arXiv:1202.2839 with permission, see also~\cite{hagen2012a}.}
  \label{subsec_ox_fig2}
\end{figure}

The spectra of neutron-rich oxygen isotopes were computed with
equation-of-motion methods starting from references with closed
subshells. Figure~\ref{subsec_ox_fig2} shows the results. Again, the
inclusion of effects of three-nucleon forces significantly improves
the agreement between computations and data. The Berggren basis lowers
the energy of resonances by about 0.3~MeV. For $^{24}$O, the
coupled-cluster calculations suggest spin assignments for the recently
observed resonance~\cite{hoffman2011,tshoo2012}.

The interaction~\cite{hagen2012a} that includes three-nucleon forces
as in-medium corrections to $NN$ interactions was also employed in the
theoretical description of neutron-rich isotopes of
fluorine~\cite{lepailleur2013}. The parameter-free results for
$^{25}$F ($^{26}$F), described as one (two) neutrons attached to
$^{24}$O, agreed very well with the experimental data. 

\begin{figure}[thbp]
  \begin{center}
    \includegraphics[width=0.6\textwidth,clip=]{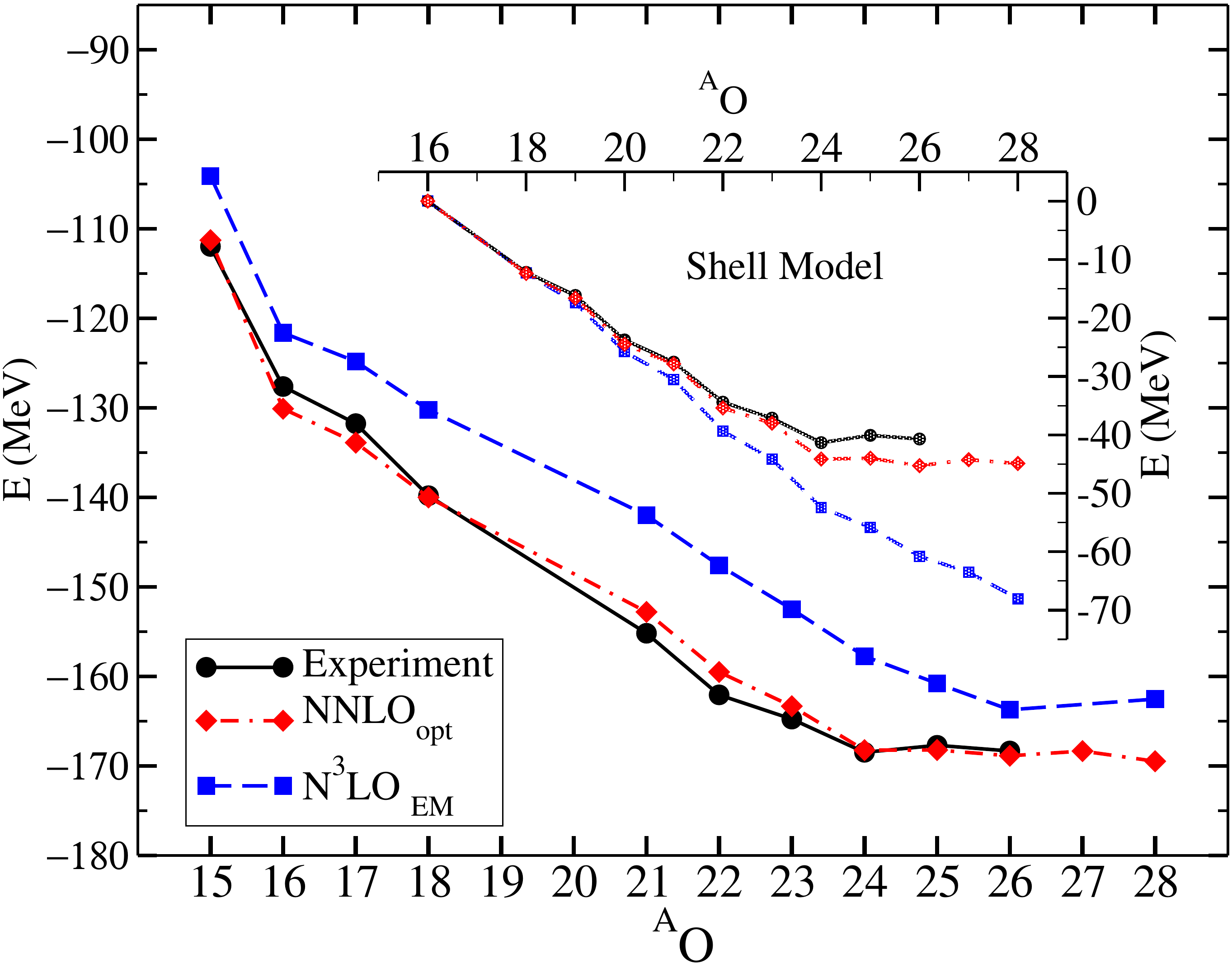}
  \end{center}
  \caption{Ground-state energy of the oxygen isotope $^A$O as a
    function of the mass number $A$. Black circles: experimental data;
    blue squares: results from the N$^3$LO chiral nucleon-nucleon
    interaction of \cite{entem2003}; red diamonds: results from the
    NNLO$_{\rm opt}$ chiral nucleon-nucleon interaction. Taken from
    arXiv:1303.4674 with permission, see also~\cite{ekstrom2013}.}
  \label{subsec_ox_fig3}
\end{figure}

We also note that an accurate description of the dripline in oxygen
can be achieved with chiral $NN$ forces alone~\cite{ekstrom2013}. By
optimizing the chiral $NN$ interaction at next-to-next-to leading
order (NNLO) to $NN$ phaseshifts, a $\chi^2 \sim 1$ per degree of
freedom was obtained for laboratory energies below $\sim
125$~MeV. With NNLO$_{\rm opt}$ it was shown that many aspects of
nuclei could be understood without invoking 3NFs explicitly.  As seen
in Fig.~\ref{subsec_ox_fig3}, there is an overall good agreement
between both coupled-cluster and shell-model calculations with
experimental binding energies of oxygen isotopes using the NNLO$_{\rm
  opt}$ $NN$ interaction.

It will be interesting to study the role three-nucleon forces in
combination with this newly optimized chiral $NN$ interaction at
NNLO. Furthermore, it has been recently pointed out by
\cite{baardsen2013} that the main difference between the N$^3$LO $NN$
interaction of \cite{entem2003} and NNLO$_{opt}$ is a poorer
reproduction of the $p$-wave phaseshifts at energies above $\sim
100$~MeV with NNLO$_{opt}$. Only at N$^3$LO can a good fit to
phaseshifts be obtained for higher energies, and it remains to be seen
how this will impact the results.

Very recently, the in-medium SRG has also been extended to deal with
reference states that are not simple product states.
\citeasnoun{hergert2013} employed a number-projected quasi-particle
state as a reference and computed the binding energies for even oxygen
isotopes with SRG-evolved $NN$ and $NNN$ interactions. Using
consistently evolved chiral $NN$ and 3NFs they obtained a good
agreement with data. For the same Hamiltonian, the self-consistent
Green's function method was applied by ~\citeasnoun{cipollone2013} to
the ground-states of isotopic chains around oxygen.  They obtained a
very good agreement with data, showing the predictive power of
consistently evolved chiral $NN$ interactions and 3NFs in this region
of the nuclear chart.

%% file: Results_Calciums.tex
Neutron-rich isotopes of calcium are of particular interest for
experiment and theory. Recent advances in ab-initio many-body methods
allow for a systematic investigation of structure and reaction
properties of calcium isotopes starting from modern chiral
interactions \cite{hagen2012b,hagen2012c,soma2013,hergert2013}.
Important questions concern magic neutron numbers beyond the well
established $N=20$ and $N=28$, and the evolution of shell structure in
heavy isotopes of calcium. $^{58}$Ca is the heaviest isotope of
calcium produced so far~\cite{tarasov2009}, but precise masses are
only known up to $^{54}$Ca~\cite{wienholtz2013}.  Mean-field
calculations predict that the neutron drip line is around
$^{70}$Ca~\cite{nazarewicz1996,fayans2000,meng2002,erler2012}. These
calculations report a near-degeneracy of the orbitals $g_{9/2}$,
$d_{5/2}$, and $s_{1/2}$, and that these orbitals are very close to
the continuum. Of course, shell-structure is expected to be modified
at the driplines~\cite{dobaczewski1994}.  \citeasnoun{hamamoto2012}
recently discussed the near-degeneracy of the $gds$ shell at the
neutron dripline due to weak binding and deformation effects. 

In atomic nuclei, a robust indication of shell closures is based on
the combined signatures of several observables, such as enhanced
nucleon separation energies, enhanced $\alpha$-particle separation
energies, high excitation energy of low-lying $J^\pi=2^+$ states, and
small quadrupole transition strengths $B(E2;2^+\to 0^+)$. In practice,
one often has to infer information from just a few available
observables.  The $N=32$ sub-shell closure is well established
experimentally for isotopes of Cr~\cite{prisciandaro2001},
Ti~\cite{janssens2002}, and Ca~\cite{huck1985,gade2006}. These nuclei
exhibit -- compared to their neighbors -- an increase in the first
excited $J^\pi=2^+$ state. This state is at about 2.5~MeV of
excitation energy in $^{52}$Ca, compared to 3.8~MeV in $^{48}$Ca. For
the $N=34$ neutron number, no sub-shell closure is found
experimentally in isotopes of chromium~\cite{marginean2006} and
titanium~\cite{liddick2004,dinca2005}, and doubts have been raised
regarding a possible shell closure in
calcium~\cite{rejmund2007,rodriguez2007,coraggio2009,hagen2012b}.  The
theoretical results exhibit a considerable scatter.
\citeasnoun{honma2002} predicted a strong shell gap in $^{54}$Ca based
on the empirical GXPF1 interaction in the $0f1p$ model space. This
result is in contrast to the monopole corrected KB3
interaction~\cite{poves1981} which yields no shell
gap~\cite{caurier2005}. In the same model space,
\citeasnoun{coraggio2009} employed a low-momentum interaction (with a
fixed cutoff) and adjusted single-particle energies to reproduce the
binding of $^{49}$Ca relative to the $^{40}$Ca core. These authors
then find a soft sub-shell closure in $^{54}$Ca, with the $J^\pi=2^+$
state at an excitation of about 2~MeV.

\citeasnoun{holt2012} investigated the role of three-nucleon forces in
isotopes of calcium. Their calculations are based on low-momentum
interactions with contributions from three-nucleon forces in the
normal ordered approximation. The model space consisted of the $0f1p$
shell, and an enhanced model space including the $g_{9/2}$ orbital was
also considered. These authors found that $^{48}$Ca is magic due to
three-nucleon forces, and they predict a shell gap for $^{54}$Ca that
is larger than for $^{52}$Ca. In the enhanced model space, the shell
closure is reduced, and the drip line is predicted to be around
$^{60}$Ca. Very recently \citeasnoun{soma2013b} computed the masses of
isotopic chains around the calcium region with SRG evolved chiral $NN$
and three-nucleon forces. They find good systematics of
separation energies, and an overbinding for all isotopes (see also the
very recent calculations by \citeasnoun{binder2013b}).

The theoretical calculations clearly show that the prediction of the
shell evolution in isotopes of calcium is a challenging task. Small
changes in the effective interaction (or the model space) impact the
calculated shell evolution of isotopes of calcium. The effects of
three-nucleon forces have to be included in the description. Based on
the mean-field calculations~\cite{fayans2000,meng2002}, the full
$0g1d2s$ shell plays a role in the location of the drip line, and it
thus seems that model spaces with a considerable size have to be
considered in computing the dripline. This challenge is compounded by
the center-of-mass problem in model spaces consisting of a few
oscillator shells, and by the need to include the continuum.

\begin{figure}[thbp]
  \begin{center}
    \includegraphics[width=0.6\textwidth,clip=]{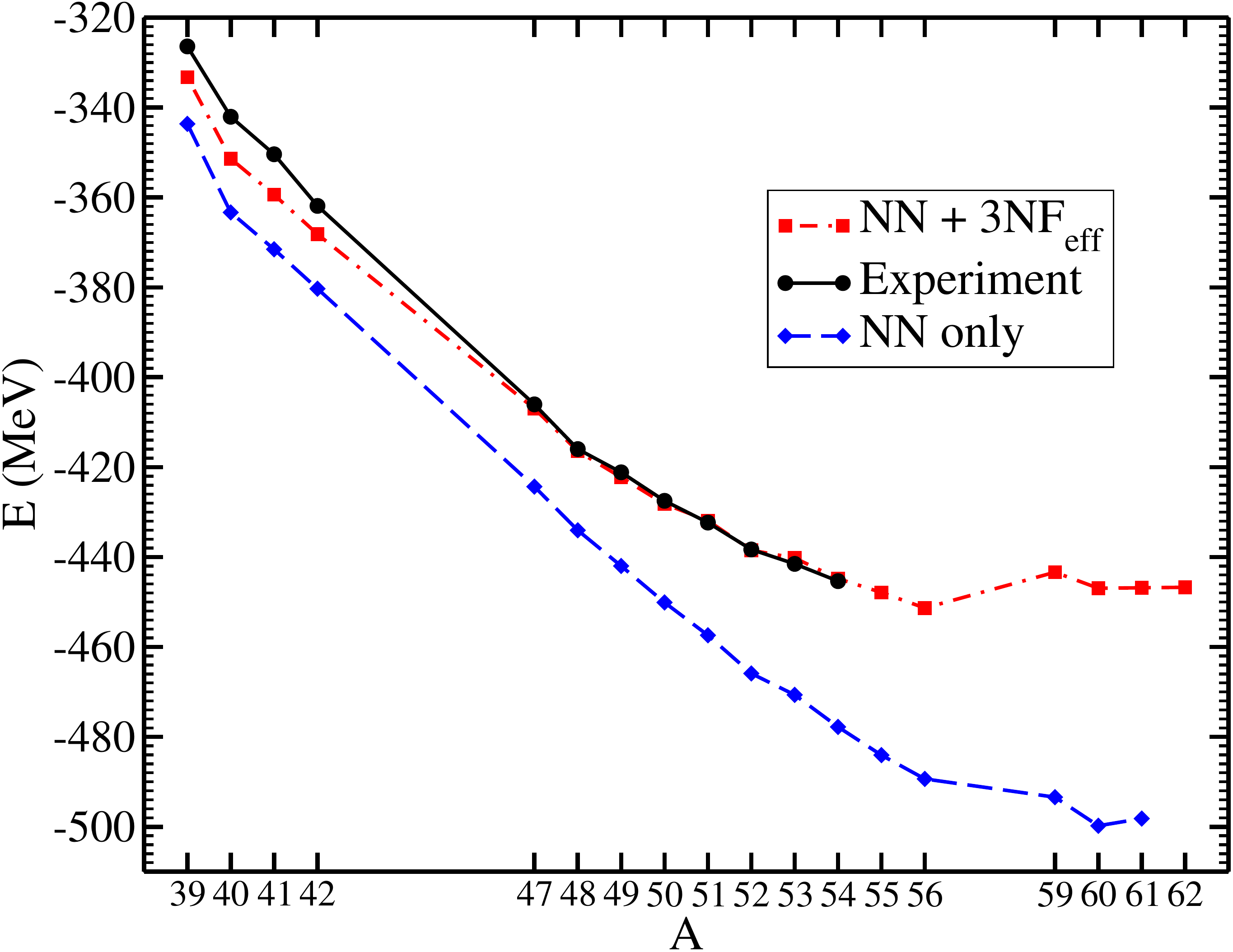}
  \end{center}
  \caption{Binding energies of calcium isotopes as function of mass
    number $A$. In addition to the experimental results, we include
    theoretical estimates using $NN$ forces and effective three-body
    forces as discussed in the text. Adapted from \cite{hagen2012b}.}
  \label{hagen2012b_fig1}
\end{figure}

\citeasnoun{hagen2012b} aimed at addressing several of the above
challenges in their calculation of the shell evolution in neutron-rich
isotopes of calcium. Their coupled-cluster calculation has all
nucleons as active degrees of freedom, the employed Gamow basis is
suitable for the description of weakly bound nuclei, and effects of
chiral three-nucleon forces were included as schematic corrections to
two-nucleon forces. The employed parameters are $k_f=0.95\fmi$,
$c_D=-0.2$, and $c_E=0.735$, and they are determined by adjustment to
the binding energies around $^{40,48}$Ca. The resulting binding energies
are shown in Fig.~\ref{hagen2012b_fig1}. As in the oxygen isotopes
(compare with Fig.~\ref{subsec_ox_fig1}), the inclusion of the effects
of three-nucleon forces as in-medium corrections to nucleon-nucleon
forces significantly improves the overall binding.  Very recently the
mass of $^{53,54}$Ca was measured~\cite{wienholtz2013}, and the
predicted binding energies and separation energies from the
coupled-cluster calculations are in good agreement with these data.

\begin{figure}[thbp]
  \begin{center}
    \includegraphics[width=0.6\textwidth,clip=]{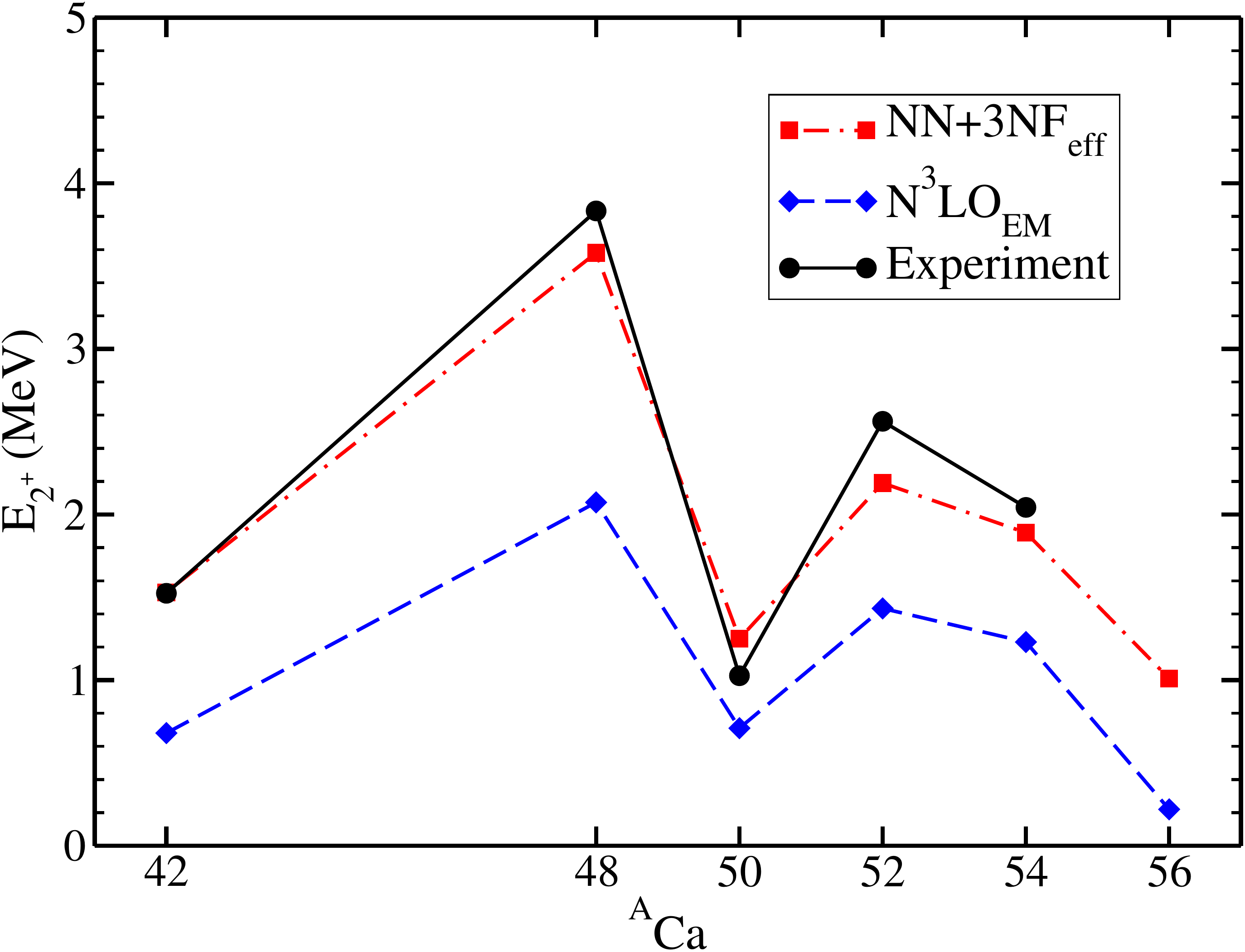}
  \end{center}
  \caption{First excited $2^+$ states of selected calcium isotopes. In
    addition to the experimental results, we include theoretical
    estimates using $NN$ forces and effective three-body forces. See
    text for further details. Adapted from from \cite{hagen2012b}.}
  \label{hagen2012b_fig2}
\end{figure}

\begin{figure}[thbp]
  \begin{center}
    \includegraphics[width=0.6\textwidth,clip=]{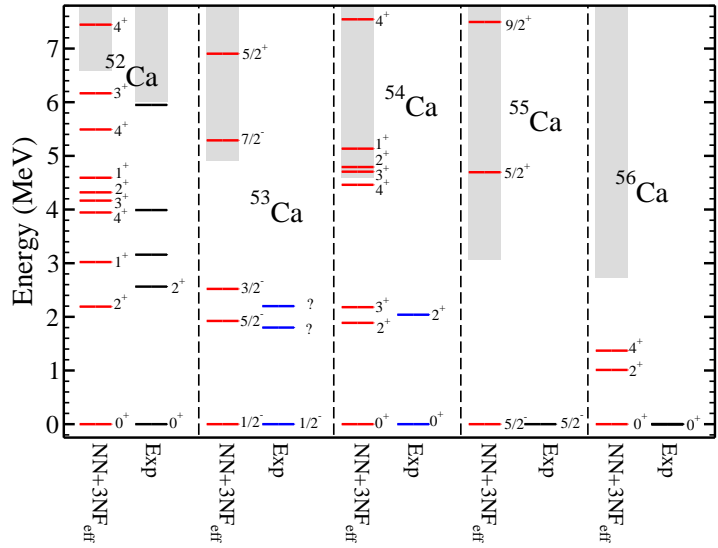}
  \end{center}
  \caption{Excited states in neutron rich calcium isotopes. The black
    lines show the known experimental levels in $^{52}$Ca, while the
    blue lines show the recently measured excitation levels in
    $^{53}$Ca and $^{54}$Ca at RIKEN \cite{steppenbeck2013b}. See text
    for further details. Adapted from \cite{hagen2012b}.}
  \label{hagen2012b_fig3}
\end{figure}

Figure~\ref{hagen2012b_fig2} shows the energies of the excited
$J^\pi=2^+$ states in $^{48,50,52,54}$Ca from coupled-cluster
calculations and compared with experimental values. Again, it is seen
that with the N$^3$LO $NN$ interaction alone, results deviate strongly
from experimental values and fail to reproduce magicity of
$^{48}$Ca. Inclusion of effective 3NFs improves the picture
considerably, and an overall good agreement with experiment is
achieved. In $^{54}$Ca, the coupled-cluster calculation yields
$E_{2^+}\approx 2$~MeV, and this suggests that $^{54}$Ca exhibits only
a soft subshell closure. This picture is confirmed by the computation
of neutron-separation energies [see Table 1 in \cite{hagen2012b}] and
the $4^+/2^+$ ratio (see Fig.~\ref{hagen2012b_fig3}). The separation
energy of the magic nucleus $^{52}$Ca is an interesting example.
Extrapolations based on atomic mass table evaluations yield
$S_n\approx 4.7$~MeV for this nucleus, while the recent measurement by
\citeasnoun{gallant2012} is $S_n\approx 6$~MeV. Calculations
by~\citeasnoun{hagen2012b} show that three-nucleon forces play an
important role in determining this separation energy.  The measured
value of $S_n$ is close to the coupled-cluster prediction $S_n\approx
6.6$~MeV. The coupled-cluster prediction for the excitation energy of
the $J^\pi=2^+$ state in $^{54}$Ca was recently confirmed
experimentally by~\citeasnoun{steppenbeck2013}, see
\cite{steppenbeck2013b} for details.

The evolution of shell structure is very interesting in neutron-rich
nuclei. In the naive shell model, the $gds$ orbitals (in this order)
are getting filled beyond $^{60}$Ca. However, the large-scale shell
model calculations by~\citeasnoun{sieja2012} suggest that in $^{60}$Ca
the $d_{5/2}$ orbital is lower in energy than the $g_{9/2}$ orbital.
This result is supported by coupled-cluster
calculations~\cite{hagen2012b}.  States in the odd isotopes
$^{53,55}$Ca are computed via one-particle attached/removed from its
``closed-shell'' neighbors. Figure~\ref{hagen2012b_fig3} shows the
computed and known experimental energies of excited states in
$^{52,53,54,55,56}$Ca. The recent experimental data on excitation
levels in $^{53,54}$Ca from \cite{steppenbeck2013b} are shown as blue
lines in the figure. As can be seen the coupled-cluster predictions
for levels in $^{53}$Ca are in good agreement with measured levels,
and they assign the spins and parities ${5/2}^{-}$ and
${3/2}^{-}$. The calculations also show that the excited $J^\pi=5/2^+$
and $J^\pi=9/2^+$ states are dominated by single-particle excitations,
and that the former is lower in energy than the latter. Furthermore,
it was found that the ground state of $^{61}$Ca is very close to
threshold, with spin and parity ${1/2}^+$ and dominated by
$s$-waves. The ${5/2}^+$ and ${9/2}^+$ states were found to be
resonances at $\sim 1~$~MeV and $\sim 2$~MeV above threshold. Using a
harmonic oscillator basis the order of states are inverted, giving
${9/2}^+$ as the ground-state in accordance with the naive shell model
filling. This demonstrates the importance of coupling to the continuum
near the particle threshold.

The very large value of the $s$-wave scattering length in the
$^{60}$Ca $+n$ system led~\citeasnoun{hagen2013} to discuss the
possibility of Efimov physics in $^{62}$Ca. This work was the first to
use input from ab-initio coupled-cluster calculations to determine
the low-energy constants of Halo-Effective-Field-Theory (Halo-EFT)
\cite{bertulani2002,bedaque2003}. Starting from the same Hamiltonian
as used in \cite{hagen2012b} and using tools developed by
\cite{hagen2012c}, the separation energy and scattering length was
computed very accurately for $^{61}$Ca. These observables was then
used as input in solving for $^{62}$Ca as a three-body cluster using
tools of Halo-EFT. The authors then explored correlations between the
scattering length of the $^{61}$Ca $+n$ system, the radius of
$^{62}$Ca and its two-neutron separation energy. Given the
uncertainties of the method and input Hamiltonian, the authors
concluded that $^{62}$Ca could be bound and have a second excited
Efimov state close to threshold. This would imply that $^{62}$Ca could
potentially be the largest halo nucleus in the chart of nuclei so far.

%% file: Misc.tex
\subsection{Shell model studies}

One strength of the coupled cluster method is its ability to treat the
$A$ body system fully microscopically. There are, however, also
applications of the method to the traditional shell-model problem with
a closed core. \citeasnoun{horoi2007} and \citeasnoun{gour2008}
compared coupled-cluster results to large-scale shell-model
computations of $fp$-shell nuclei around $^{56}$Ni. In these
calculations, $^{40}$Ca is a closed core, the $fp$ shell model space
consist of only 40 single-particle states, but wave functions can be
very correlated. In such situations, the accuracy of the
coupled-cluster method depends on a sufficient shell gap between the
$f_{7/2}$ orbital and the remaining orbitals of the $fp$ shell. This
is an interesting study in model spaces that are solvable by exact
diagonalization~\cite{caurier1999}.  

For the exactly solvable pairing model, \citeasnoun{dukelsky2003}
compared the CCSD approximation to the BCS method and the
self-consistent random phase approximation (RPA). The CCSD
approximation performs very well below a critical value of the pairing
strengths wich indicates the onset of a phase transition (and a new
reference state). In a very recent study~\citeasnoun{jemai2013} showed
that the CCSD ground state is annihilated by a generalized RPA
operator $\hat{Q}$, that also includes a two-body component. The
generalized RPA is successful tested with various solvable
models. This is an interesting finding because the adjoint
$\hat{Q}^\dagger$ creates collective excitations that are by
construction orthogonal to the CCSD ground state. (Recall that excited
states in the coupled-cluster method fulfill bi-orthogonality
relations.)

\subsection{Nuclear reactions}
\begin{figure}[htb]
  \begin{center}
    \includegraphics[scale=0.5,clip=]{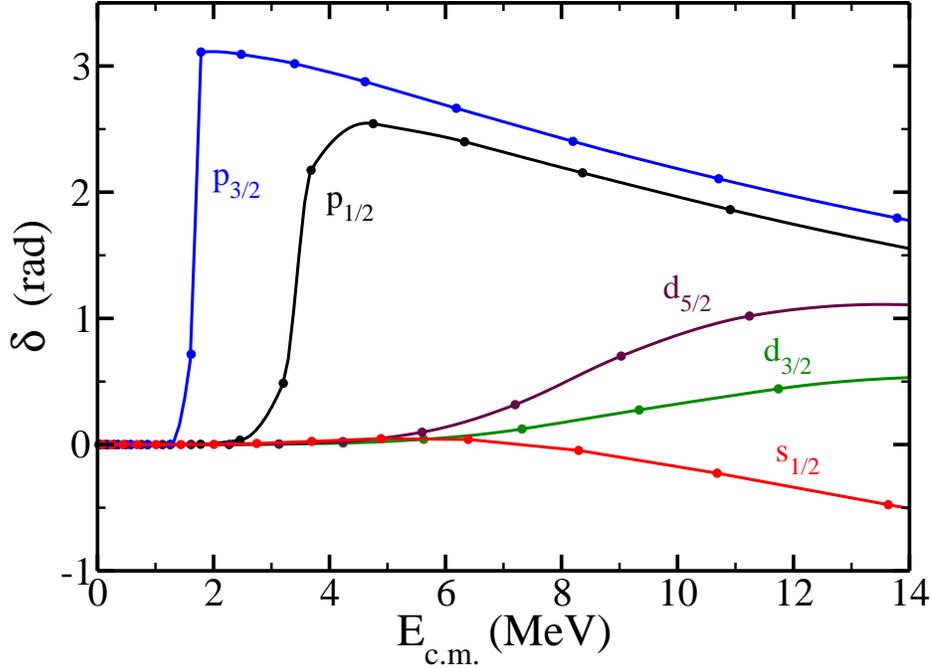}
  \end{center}
  \caption{Computed phase-shifts for elastic proton
    scattering on $^{40}$Ca for low-lying partial waves and energies
    below $14$~MeV. Taken from arXiv:1206.2336 with permission, see
    also~\cite{hagen2012c}.}
  \label{fig:40CaScatt}
\end{figure}

First steps have been made to employ the coupled-cluster method for
the description of nuclear reactions.  \citeasnoun{jensen2010}
developed the formalism for computing spectroscopic factors and
one-nucleon overlap functions from ab-initio coupled-cluster
theory. From one-nucleon overlap functions one can in principle
compute transfer and knockout reactions, as well as elastic and
inelastic nucleon-nucleus scattering. \citeasnoun{hagen2012c} showed
that accurate solutions for both resonances and scattering states with
a Coulomb interaction can be obtained in momentum space by utilizing
the off-diagonal method developed in~\cite{michel2011}. By utilizing a
single-particle basis defined along the real energy axis in ab-initio
coupled-cluster calculations, it was shown that one-nucleon overlap
functions with correct asymptotic behavior can be obtained. In
\cite{hagen2012c} this real energy continuum basis was used in
combination with the formalism of \cite{jensen2010} for the
computation of the one-proton overlap functions of $^{41}$Sc with
$^{40}$Ca. Elastic scattering phase-shifts was obtained by matching
the radial one-proton overlap functions to the known regular and
irregular Coloumb functions. Figure~\ref{fig:40CaScatt} shows the 
computed phaseshifts for elastic proton scattering on $^{40}$Ca. 

From the computed elastic scattering
phase-shifts the differential cross sections were computed at energies
$9.6~$MeV and $12.44~$MeV, respectively. The proof-of-principle
calculation are in fair agreement with data. The computed cross
section minima were in good agreement with data, while the
calculations overestimated the cross sections at large angles.

Very recently, \citeasnoun{bacca2013} combined the Lorentz integral
transform~\cite{efros1994,efros2007} -- a method for the computation
of continuum response functions -- with the coupled-cluster method for
the computation of the giant dipole resonance in $^{16}$O. Of course
this giant dipole resonance has been described in many works, see
e.g. \cite{shlomo1975,nakatsukasa2012,lyutorovich2012} and references
therein. The dipole response function of the ground state
$|\psi\rangle$ with spin $J_0$ is
\begin{equation}
\label{resp}
S(\omega)=\frac{1}{2J_0+1}\sum_f |\langle \psi
|{\hat{D}_0}|\psi_f\rangle|^2 \delta (E_f -E_0 -\omega ) \ .
\end{equation}  
Here, $\hat{D}_0$ is the component of the (translationally invariant)
dipole operator in the direction of the photon emission. The sum is
over all final states, most of which are in the continuum. This makes
the direct evaluation of Eq.~(\ref{resp}) very difficult, and instead one
considers its Lorentz integral transform (LIT)
\begin{eqnarray} 
\label{LIT}
  {\cal L}(\omega_0,\Gamma )=\int_{\omega_{\rm th}}^{\infty} d\omega
  \frac{S(\omega)}{(\omega -\omega_0) ^2+\Gamma^2} \ ,
\end{eqnarray}   
which can be computed directly.  In Eq.~(\ref{LIT}) $\omega_{\rm th}$
is the threshold energy and $\Gamma > 0$.  The closure relation yields
\be
\label{lorenzog} 
{\cal L}(z)= \langle \psi | {\hat{D}_0}^{\dagger}\frac{1}{\hat
  {H}-z^*}\frac{1}{\hat{H}-z}\hat{D}_0|\psi \rangle = \langle
\widetilde{\psi} | \widetilde{\psi} \rangle \ , \ee
with the complex energy $z=E_0+\omega_0+i\Gamma$. In coupled-cluster
theory the LIT of the dipole response function is obtained by
employing similarity-transformed operators, i.e.  Eq.~(\ref{lorenzog})
becomes
\be
\label{cc_lorentz0bar}
  {\cal L}(z)= \langle 0_L | {\bar{D}_0}^{\dagger}\frac{1}{\bar
    {H}-z^*}\frac{1}{\bar{H} - z}\bar{D}_0 |0_R \rangle = \langle
  \widetilde{\psi}_R | \widetilde{\psi}_L \rangle \ .  \ee
Here, $\langle 0_L|\equiv \langle \psi| L$ and $|0_R\rangle\equiv
R|\psi\rangle$ are the left and right ground states of $\overline{H}$,
see Subsect.~\ref{subsec:techdetails_scheme} for details.  The states
$|\widetilde{\psi}_R\rangle$ and $\langle\widetilde{\psi}_L|$ are the
solutions of a right and left Schr{\"o}dinger-like equation
\begin{eqnarray} 
\label{cc_psi1}
  (\bar{H}-z)|\widetilde{\psi}_R(z)\rangle &=& \bar{D}_0 | 0_R\rangle
\ , \nonumber\\ \langle\widetilde{\psi}_L(z)\vert(\bar{H}-z^*) &=&
\langle 0_L\vert \bar{D}_0^\dagger \ ,
\end{eqnarray}
whose right-hand-side is known.  The inversion of the LIT yields the
response function itself. In principle, the response function is
independent of the employed width $\Gamma$. In practice, however, the
inversion of the LIT is an ill-posed problem (i.e. the LIT kernel has
zero modes), and a nonzero width is necessary for obtaining results that
are stable under the inversion.

\begin{figure}[htb]
\begin{center}
  \includegraphics[scale=0.5,clip=]{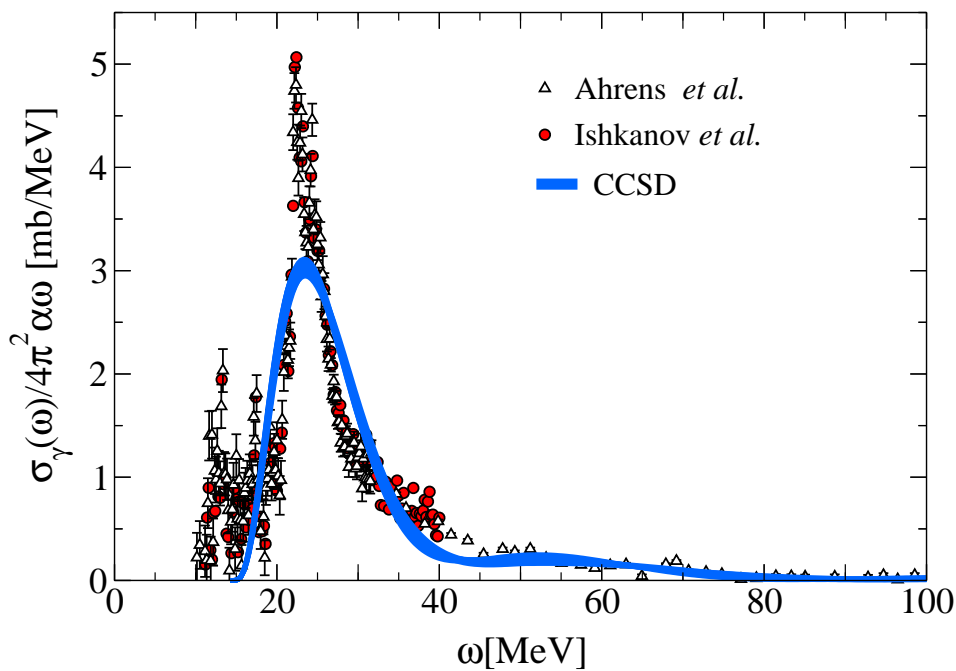}
\end{center}
\caption{(Color online) Comparison of the $^{16}$O dipole response
  calculated in the CCSD scheme against experimental data by
  \citeasnoun{ahrens1975} (triangles with error bars),
  \citeasnoun{ishkhanov2002} (red circles). Figure taken from
  arXiv:1303.7446 with permission, see also~\cite{bacca2013}.}
\label{fig_resp_O16}
\end{figure}

Figure~\ref{fig_resp_O16} shows the response function (the relation
between cross section and response is $\sigma=4\pi^2\alpha S$ with
fine structure constant $\alpha$) for $^{16}$O computed from the
chiral nucleon-nucleon interaction by \citeasnoun{entem2003} and
compare to data, see \cite{bacca2013} for details. The position of the
giant dipole resonance is well reproduced by the theory.

This method in computing giant dipole resonances can also be applied
to heavier nuclei. \citeasnoun{orlandini2013} very recently reported
corresponding calculations for $^{40}$Ca. 

\subsection{Neutron matter and nuclear matter}
\label{nucleonic}
Calculations of neutron matter and nuclear matter connect the atomic
nucleus with astrophysical objects such as neutron stars.  Of
particular interest is the equation of state and its isospin
dependence.  This subject is too vast to be reviewed here, and we
refer the reader to the recent reviews
\cite{hh2000,dickhoff2004,sammarruca2010,vandalen2010,tsang2012,hebeler2013b}.

Coupled-cluster calculations of nuclear matter were reported
by~\citeasnoun{kuemmel1978}. More recently, the method was employed in
calculations of neutron matter and nuclear matter with chiral
interactions \cite{ekstrom2013,baardsen2013,hagen2013b}. The
calculations ~\cite{ekstrom2013,baardsen2013} employ a spherical
implementation of the coupled-cluster method and work in the relative
and center-of-mass frame employing a partial wave basis. This approach
is similar to Brueckner-Hartree-Fock calculations of nuclear
matter~\cite{day1981b,haftel1970,suzuki2000}.  However,
\citeasnoun{ekstrom2013} and \citeasnoun{baardsen2013} sum
particle-particle and hole-hole ladder diagrams to infinite order,
while treating the Pauli operator exactly with angle-averaged
single-particle energies.  The equations of state for nuclear matter
with $NN$ forces obtained by \citeasnoun{hagen2013b} are in good
agreement with results from the self-consistent Green's function
method \cite{carbone2013}.

Alternatively, one might compute nucleonic matter directly in a
discrete momentum-space basis in the laboratory frame using periodic
boundary conditions. Here, the conservation of momentum implies the
absence of singles excitations for closed-shell references (i.e. all
single-particle states of the reference are doubly occupied by
neutrons and/or protons). The resulting coupled-clusters with doubles
(CCD) approximation very much reduces the numerical
effort~\cite{bishop1978}.  However, in this approach one does not work
in the thermodynamic limit and has to average over Bloch states to
mitigate finite-size effects~\cite{gros1992,gros1996,lin2001}. We also
note that the computation of matrix elements in the laboratory system
is much simpler in momentum space than in a basis that exhibits good
angular momentum, and this is particularly relevant for three-nucleon
forces.

\begin{figure}[htb]
  \begin{center}
    \includegraphics[scale=0.5,clip=]{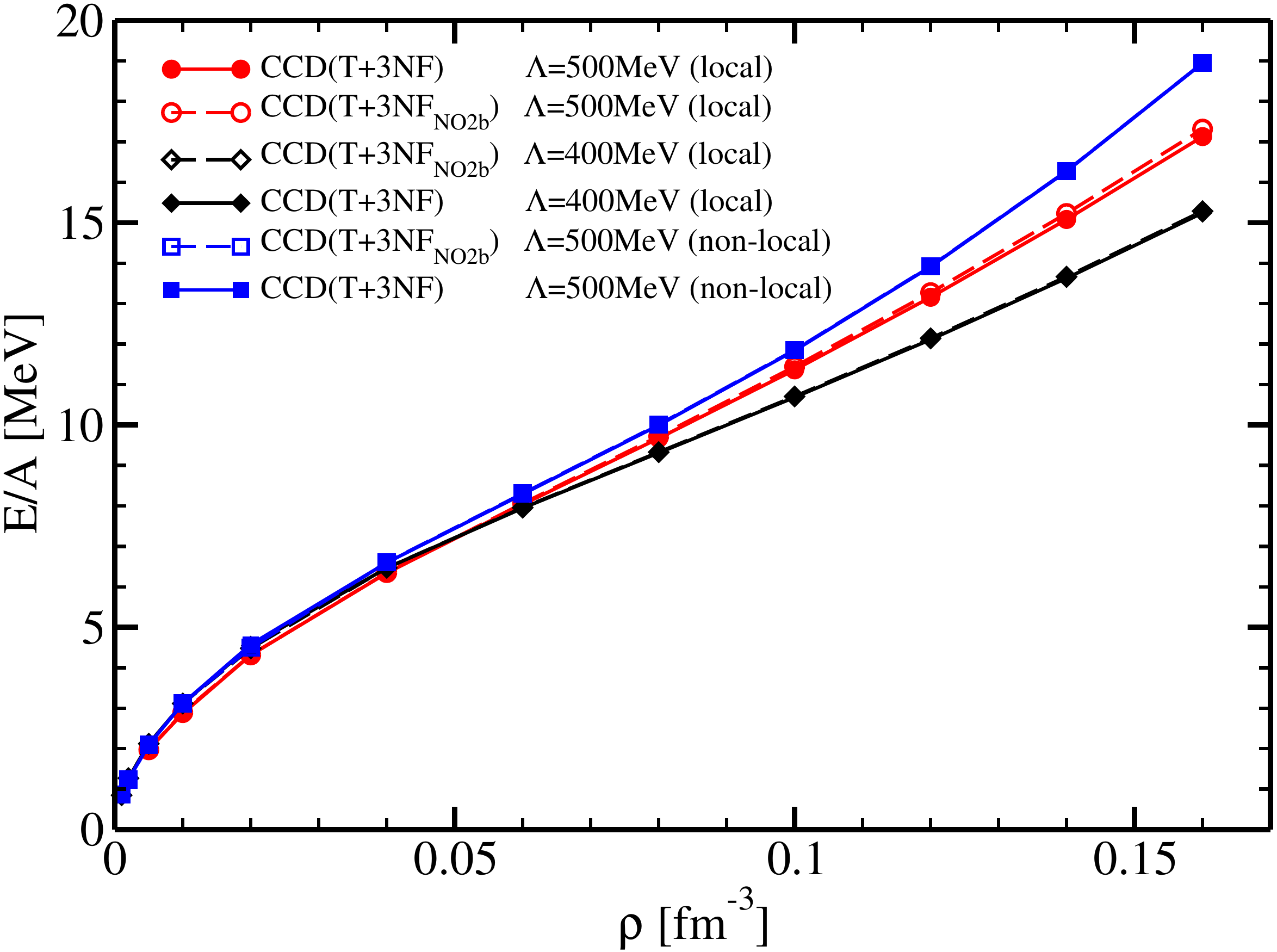}
  \end{center}
  \caption{Energy per particle in pure neutron matter with NNLO$_{\rm
      opt}$ and three-nucleon forces. For the three-nucleon force we
    used a local regulator with cutoffs $\Lambda = 400$ and $\Lambda =
    500$~ MeV.  The LECs of the three-nucleon force are given by $c_E
    = -0.389$ and $c_D = -0.39$ for the $\Lambda = 500$~MeV local
    regulator, while for the $\Lambda = 400$~MeV local regulator we
    used $ c_E = -0.27$ and $c_D = -0.39$ with $c_E$ adjusted to the
    $^4$He binding energy. For the non-local regulator with $\Lambda =
    500$~MeV cutoff we used $c_E = -0.791$ and $c_D = -2$, adjusted to
    the triton and $^3$He binding energies.  The calculations used 66
    neutrons, $n_{\rm max} = 4$, and periodic boundary conditions. Taken from
    arXiv:1311.2925 with permission, see also~\cite{hagen2013b}.}
  \label{fig:neutronmatter}
\end{figure}

Very recently, \citeasnoun{hagen2013b} presented momentum-space
coupled-cluster results of nucleonic matter based on the chiral
nucleon-nucleon interaction NNLO$_{\rm opt}$ and three-nucleon forces
with local and non-local regulators. The LECs that entered the
three-nucleon force was adjusted to the triton binding energy and
halflife. The main results can be summarized as follows.  Neutron
matter is perturbative. The coupled-cluster CCD results are close to
results from second-order many-body perturbation theory, and triples
corrections are small. Likewise, the role of three-nucleon forces is
small, and they act repulsively in neutron matter. The normal ordered
two-body approximation for the three-nucleon force works very well in
pure neutron matter. The results for pure neutron matter is summarized
in Fig.~\ref{fig:neutronmatter}. Note that the band obtained for
different regulators in the three-nucleon force are within the band
for neutron matter obtained by \cite{krueger2013}.

In contrast to neutron matter, symmetric nuclear matter is not
perturbative and therefore technically more challenging. Here, the
inclusion of all excitations in CCD (opposed to just $p$-$p$ ladders
and $h$-$h$ ladders) yields relevant corrections to the correlation
energy.  Somewhat surprisingly, local three-nucleon forces yield
considerable corrections to $NN$ forces, particularly at densities
beyond saturation density.  Three-nucleon forces with a reduced cutoff
or with a non-local regulator do not exhibit this unexpected
behavior. It seems difficult to adjust the low-energy coefficients
$c_D$ and $c_E$ of the short-ranged contributions to the three-nucleon
force, such that the binding energies of both light nuclei and nuclear
matter are acceptable. Thus, one might consider to include the
saturation point of symmetric nuclear matter in the optimization of
the chiral nuclear force.

\subsection{Quantum dots}

Quantum dots, artificial two-dimensional atoms made by confining
electrons in semiconductor heterostructures, are of considerable
interest in condensed matter physics, see for example the review of
\citeasnoun{reimann2002}.  The confining potential of these objects is
often modeled by a harmonic oscillator. These objects can be
interesting for nuclear physicists due to the prominent role of the
oscillator basis, commonalities in the treatment of the Coulomb
interaction, and as a test bed for various many-body methods. We note
that the correlation energy (i.e. the difference between the exact
energy and Hartree-Fock energy) is usually a small fraction of the
Hartree-Fock energy in electronic systems, and this is very different
for atomic nuclei. Of course, the accurate and precise computation of
the correlation energy is similarly challenging for electronic and
nuclear systems.  Recent applications of coupled-cluster theory to
quantum dot systems can be found in
\cite{bartlett2003,heidari2007,pedersen2011,waltersson2013}.

\citeasnoun{bartlett2003} computed ground and excited states quantum
dots and found that the usual EOM techniques work well also here. This
picture was confirmed by \citeasnoun{heidari2007} who also employed
multi-reference coupled-cluster methods. \citeasnoun{pedersen2011}
compared results from coupled-cluster calculations and diffusion Monte
Carlo for closed-shell systems with $2$, $6$, $12$ and $20$ electrons,
and some of their odd-numbered neighbors.  The calculations were
performed for several oscillator frequencies in the CCSD, the
non-iterative triples CCSD(T) and the $\Lambda$-CCSD(T) approximation.
To mitigate the slow convergence of the Coulomb interaction as a
function of the number of harmonic oscillator shells, a similarity
transformed (or effective) Coulomb interaction defined for a specific
model space was employed.  The effective Coulomb interaction yielded
essentially converged results for the ground-state energies in 20
major oscillator shells. The $\Lambda$-CCSD(T) calculations with an
effective two-body interaction resulted in an excellent agreement with
diffusion Monte Carlo calculations, with relative errors between
$10^{-5}$ and $10^{-4}$.  Table \ref{qdcorrelation} shows the
contribution to the total correlation energy for various
coupled-cluster approaches.

\begin{table}[hbtp]
\begin{center}
  \begin{tabular}{|c|c|r|r|r|r|r|}\hline
    \multicolumn{1}{|c}{} & \multicolumn{2}{|c}{$\omega =0.28$}&
    \multicolumn{2}{|c}{$\omega =0.5$ } & \multicolumn{2}{|c|}{
      $\omega =1.0$ } \\
    \hline
    \multicolumn{1}{|c}{$N$} & \multicolumn{1}{|c}{$\Delta E_2$}&
    \multicolumn{1}{|c}{$\Delta E_3$ } & \multicolumn{1}{|c}{
      $\Delta E_2$ }& \multicolumn{1}{|c}{ $\Delta E_3$ }& 
    \multicolumn{1}{|c}{ $\Delta E_2$}&
    \multicolumn{1}{|c|} {$\Delta E_3$} \\
    \hline
6  & 94$\%$ & 99$\%$ & 96$\%$ & 100$\%$ & 97$\%$ & 100$\%$  \\
12 & 91$\%$ & 99$\%$ & 94$\%$ & 100$\%$ & 96$\%$ & 100$\%$  \\
20 & 90$\%$ & 99$\%$ & 93$\%$ & 100$\%$ & 95$\%$ & 100$\%$  \\\hline
  \end{tabular}
\end{center}
  \caption{Percentage of correlation energy at the CCSD level ($\Delta
    E_2$) and at the $\Lambda$-CCSD(T) level ($\Delta E_3$), for
    different numbers of electrons $N$ and values of the confining
    harmonic potential $\omega$ in atomic units. All numbers are for
    $20$ major oscillator shells.  A Hartree-Fock basis and an
    effective two-body interaction were employed. Taken from
    \cite{pedersen2011}.}
  \label{qdcorrelation}
\end{table}

Here, the correlation energy is the difference between the Monte Carlo
results and the Hartree-Fock reference energy.  We see that at the
CCSD level ($\Delta E_2$ in the table) of approximation to
Eq.~(\ref{t_expan}), we obtain approximately $90\%$ or more of the
correlation energy.  At the $\Lambda$-CCSD(T) level ($\Delta E_3$), we
are close to $100\%$ of the correlation energy. We note that the CCSD
approximation becomes less accurate for smaller oscillator frequencies
(in atomic units here).  This behavior can be understood form the
following observation: the electron density decreases with decreasing
oscillator frequencies, and correlations become increasingly important
as the regime of a Wigner crystal is approached.  As a consequence,
the contributions from $\Lambda$-CCSD(T) become more important as the
frequency is reduced. It is however rewarding to see that the
$\Lambda$-CCSD(T) approximation recovers almost the benchmark result
of the diffusion Monte Carlo.  For frequencies below $0.05$ atomic
units, however, correlations tend to become more important and the
discrepancy between CCSD and diffusion Monte Carlo calculations tend
to become larger, as demonstrated in \cite{reimann2013}.

%% file: Summary.tex
We have reviewed the recent advances of the coupled-cluster method in
nuclear physics. Over the last decade, coupled-cluster theory has been
exploring ``bare'' chiral interactions in medium-mass nuclei. Relevant
steps for these calculations were (i) the formulation in an
angular-momentum coupled scheme that permits calculations with
``bare'' interactions from chiral effective field theory without the
need for secondary renormalizations, (ii) the practical solution of
the center-of-mass problem, (iii) the use of a Gamow basis for the
computation of weakly bound and unbound nuclei, (iv) the development
of nucleon attached/removed methods for the description of neighbors
of nuclei with closed subshells, and (v) a computational
implementation of the numerical methods that is suited for super
computers. A culmination of these developments was the prediction of
the structure of the exotic nucleus $^{54}$Ca. Many challenges
remain. In what follows we present a few open problems.

The coupled-cluster method is computationally most efficient for the
description of closed-shell nuclei and their neighbors. While this is
a limitation, the properties of the doubly-magic isotopes of oxygen,
calcium, nickel, and tin are relevant for entire regions of the
nuclear chart. Interesting future applications concern predictions for
very neutron-rich isotopes of calcium, the isotopes around $^{78}$Ni,
the neighborhood of proton-deficient $^{100}$Sn~\cite{darby2010} and
nuclei around neutron-rich $^{132}$Sn~\cite{jones2010}. Apart from
necessary computational avances, the main challenge consists of the
availability of nuclear interactions that reasonably accurately
describe such heavy nuclei~\cite{binder2013b}.

There are now several methods that aim at the {\it ab initio}
description of medium-mass nuclei, and some of these have been
extended to open-shell nuclei~\cite{soma2013,hergert2013b}.  It would
be interesting to see whether some of these ideas can also be used for
coupled-cluster calculations of superfluid or deformed nuclei.

First steps have been undertaken to describe elastic and inelastic
reactions with the coupled-cluster method. Of particular interest are
nucleon knockout and transfer reactions, and it would be interesting
to extend bound-state methods such as coupled-cluster for the
description of these experimentally relevant reactions in medium-mass
nuclei~\cite{carbonell2014}.

All experience with the coupled-cluster method shows that there is a
quickly converging hierarchy of approximations (singles and doubles,
triples, quadruplets, etc. ...), but there is only little theoretical
work on error estimates~\cite{kutzelnigg1991}. It would be desirable
to better quantify this hierarchy, and to give reliable error
estimates of the truncation scheme. The underlying question is about a
power counting for closed-shell nuclei. We note that the empirical
hierarchy of the coupled-cluster method is a good match for
interactions from effective field theory because it is not necessary
to solve an approximate Hamiltonian more precisely than demanded by
its power counting.

%% file: app1.tex
 \parindent=0pt

 \def\correction#1{%
     \abovedisplayshortskip=#1\baselineskip\relax\belowdisplayshortskip=#1\baselineskip\relax%
     \abovedisplayskip=#1\baselineskip\relax\belowdisplayskip=#1\baselineskip\relax}

 \newcolumntype{A}[2]{%
     >{\minipage{\dimexpr#1\linewidth-2\tabcolsep-#2\arrayrulewidth\relax}\vspace\tabcolsep}%
     c<{\vspace\tabcolsep\endminipage}}

 \arrayrulewidth=1pt\relax
 \tabcolsep=5pt\relax
 \arrayrulecolor{black}
 \fboxsep=\tabcolsep\relax
 \fboxrule=\arrayrulewidth\relax

\section{CCSD in angular momentum coupled representation}
\label{app1}
  
In this Section we present the equations for the $T_1$ and $T_2$
amplitudes in the CCSD approximation using an angular momentum coupled
scheme. Recall that the CCSD equations can be written in compact
form as 
\ba
\label{ccsd2}
\la\phi_{i}^{a}|\overline{H}|\phi\ra  & =&  0 \ , \nonumber\\
\la\phi_{ij}^{ab}|\overline{H}|\phi\ra &=& 0 \ .  
\ea 

Here, $|\phi_i^a\ra\equiv \adag_a a_i|\phi\ra$,
$|\phi_{ij}^{ab}\ra\equiv \adag_a \adag_b a_j a_i|\phi\ra$, and
$\overline{H} = e^{-T}H_Ne^{T}$ is the normal-ordered
similarity-transformed Hamiltonian. The diagrams representing the
$T_1$ and $T_2$ amplitudes, together with the uncoupled and coupled
algebraic expressions are given in Table \ref{tab:t_amps}. Here,
$t^{jm}$ and $t^j$ denote the cluster amplitudes in $m$-scheme and
$j$-scheme, respectively. Clebsch-Gordan coefficients are given by
$C_{j_1m_1j_2m_2}^{JM}$ with the coupling order $[ j_1 \to j_2]J$.

\begin{table*}[!htbp]
  \begin{Table_app}{0.25}{0.33}{0.42}
    Diagram & \mathrm{Uncoupled~expression} & \mathrm{Coupled~expression} 
    \tabularnewline\hline
    \includegraphics[scale=0.2]{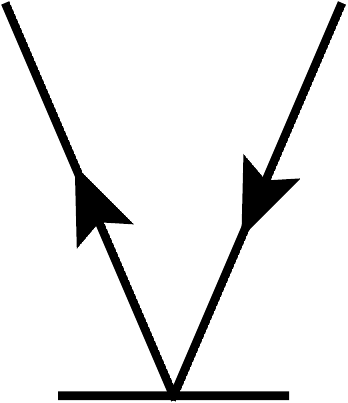} & 
    \langle a\vert t^{00} \vert i\rangle   & 
    \langle a\vert\vert t^{0} \vert\vert i\rangle\delta_{j_{i},j_{a}} 
    \tabularnewline\hline
    \includegraphics[scale=0.2]{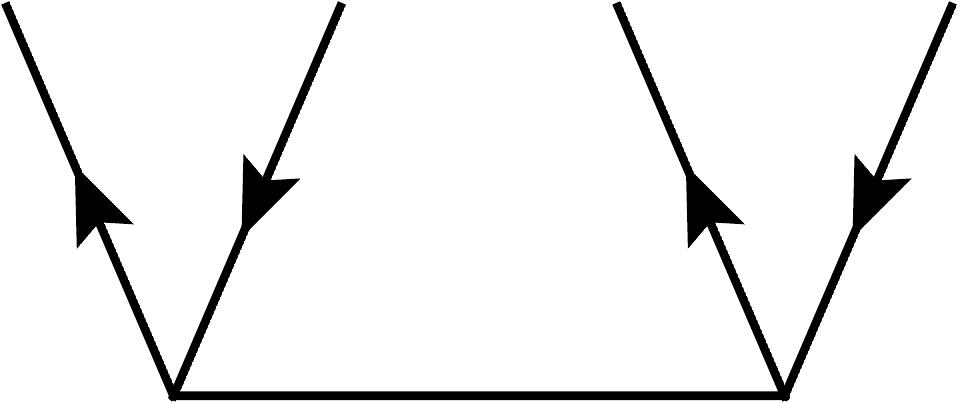} & 
    \langle ab\vert t^{00} \vert ij\rangle  &
    C_{j_a m_a j_b m_b}^{J_{ab}M_{ab}} C_{j_im_i j_jm_j}^{J_{ij}M_{ij}} 
    \langle ab\vert\vert t^{0} \vert\vert ij\rangle 
    \tabularnewline\hline
    \includegraphics[scale=0.2]{T2_diag} & \langle ab\vert t^{00}
    \vert ij\rangle & - (-1)^{j_i -m_i}(-1)^{j_b-m_b} \times C_{j_a m_a j_i
      -m_i}^{J_{ai}M_{ai}} C_{j_j m_j j_b -m_b}^{J_{ai}M_{ai}} \langle
    ai^{-1}\vert\vert t^{0} \vert\vert jb^{-1}\rangle
    \tabularnewline\hline
  \end{Table_app}
  \caption{The diagrams representing the $T_1$ and $T_2$ amplitudes,
    together with their uncoupled ($m$-scheme) and coupled
    ($j$-scheme) algebraic expressions. Both $T_1$ and $T_2$ are
    scalar under rotations.}
  \label{tab:t_amps}
\end{table*}

The last row in Table~\ref{tab:t_amps} give the so-called
cross-coupled representation of the $T_2$ amplitude.  Note that in the
cross-coupled reduced matrix element $ \langle ai^{-1} \vert\vert
t^{0} \vert\vert jb^{-1}\rangle $, the coupling order is
$[j_a\rightarrow j_i]J_{ai}$ and $[j_j\rightarrow j_b]J_{ai}$, and
that we define reduced matrix elements by the coupling order $\langle
p \vert T^{\mu \nu}\vert q\rangle = C_{j_q m_q \mu \nu}^{j_p m_p}
\langle p \vert\vert T^{\mu }\vert\vert q\rangle$.  We can express the
angular momentum coupled matrix element of $T_2$ in terms of
cross-coupled matrix elements of $T_2$ where the coupling order goes
across $T_2$ by the following recoupling
\cite{pandya1956,baranger1960,kuo1968},
\begin{equation}
  \langle ab \vert\vert t^{0} \vert\vert ij \rangle = 
  - \sum_{J_{ai}}(-1)^{j_i+j_j+J_{ab}}\hat{J}_{ai}^2
  \left\{\begin{array}{ccc}
      j_a & j_b & J_{ab} \\
      j_j & j_i & J_{ai} 
    \end{array}\right\} \\ 
  \langle ai^{-1} \vert\vert t^{0} \vert\vert jb^{-1}\rangle. 
  \label{eq:pandya1}
\end{equation}

Similarly we can express the cross coupled matrix
elements in terms of the normal coupled matrix elements by the
recoupling,
\begin{equation}
  \langle ai^{-1} \vert\vert t^{0} \vert\vert jb^{-1}\rangle =
  - \sum_{J_{ab}}(-1)^{j_i+j_j+J_{ab}}\hat{J}_{ab}^2
  \left\{\begin{array}{ccc}
      j_a & j_b & J_{ab} \\
      j_j & j_i & J_{ai} 
    \end{array}\right\} \\ 
   \langle ab \vert\vert t^{0} \vert\vert ij \rangle.
  \label{eq:pandya2}
\end{equation}

As will be seen below, some of the diagrams that involve intermediate
summations over particles and holes can be much more efficiently
computed using the cross-coupled representation. This avoids the
computation of complicated recoupling coefficients (e.g. $9j$ and
$12j$ symbols), and permits us to use matrix-matrix and matrix-vector
multiplication routines.

The $T_1$ and $T_2$ amplitude equations can be written in quasi-linear
form by the use of intermediates (see e.g.
\cite{gour2006,bartlett2007}). In terms of diagrams the $T_1$
amplitude equations can be written as
\begin{eqnarray}
  \nonumber
    0 &= 
    \parbox{18mm}{\includegraphics[scale=0.20]{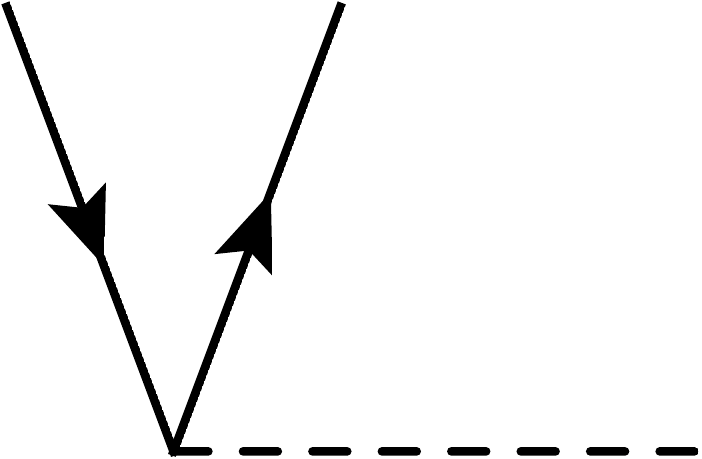}}
    + \parbox{28mm}{\includegraphics[scale=0.20]{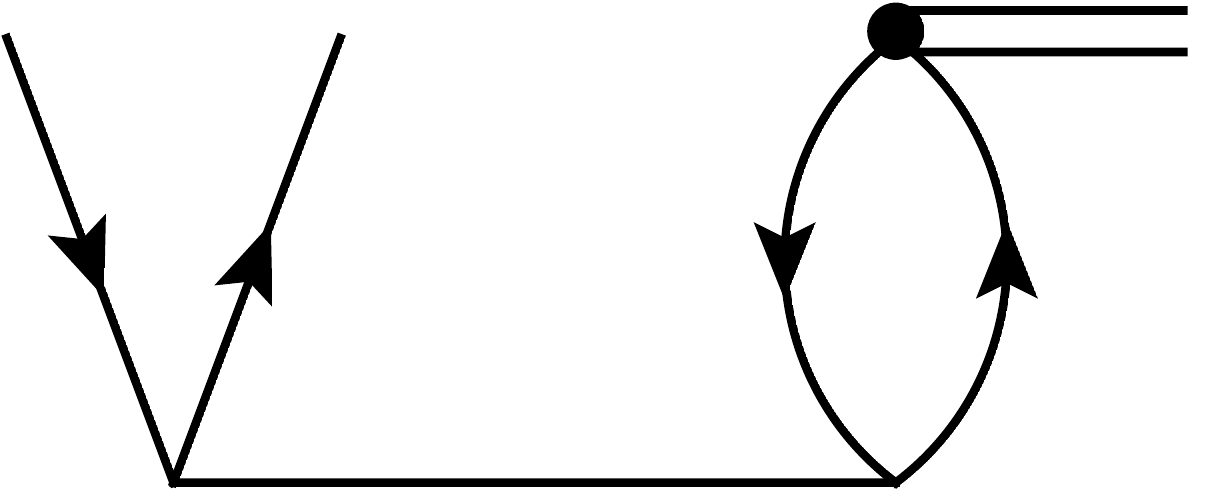}} 
    + \parbox{18mm}{\includegraphics[scale=0.20]{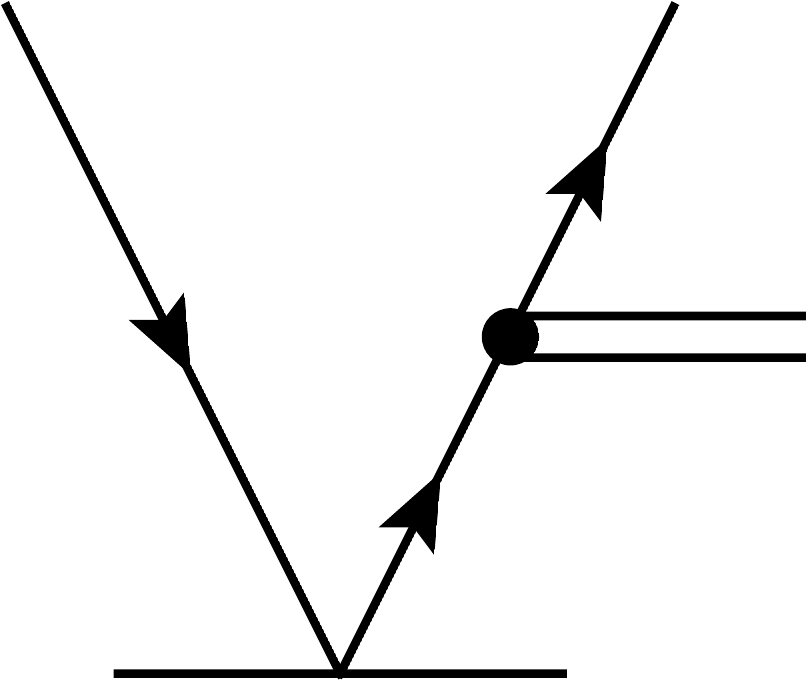}}
    + \parbox{18mm}{\includegraphics[scale=0.20]{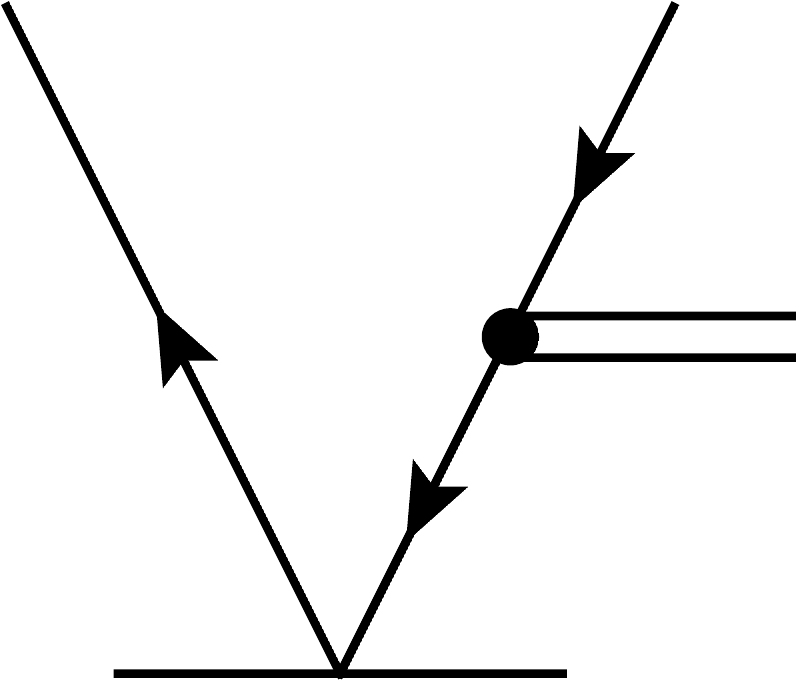}}\\
    & \quad + \parbox{28mm}{\includegraphics[scale=0.20]{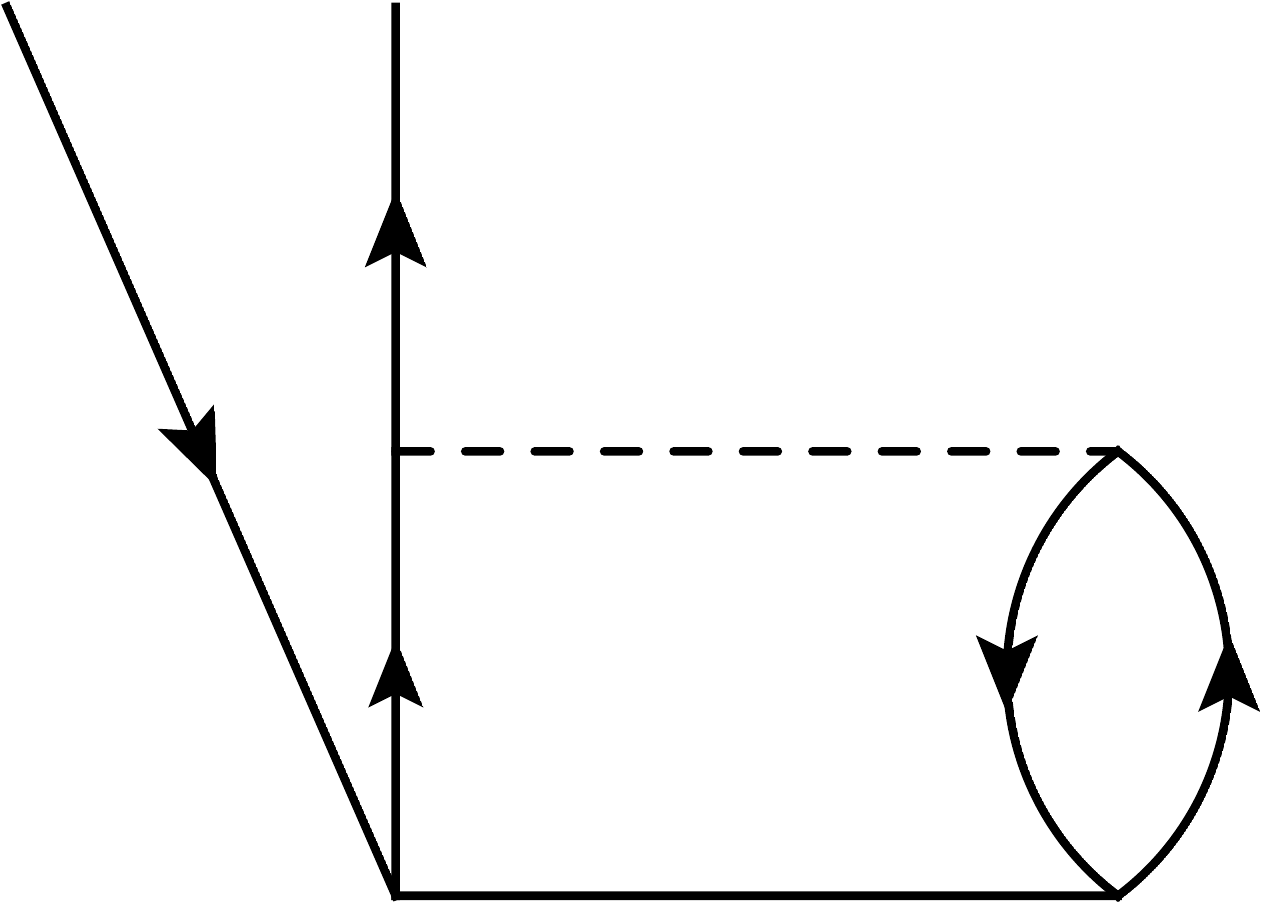}}
    + \parbox{28mm}{\includegraphics[scale=0.20]{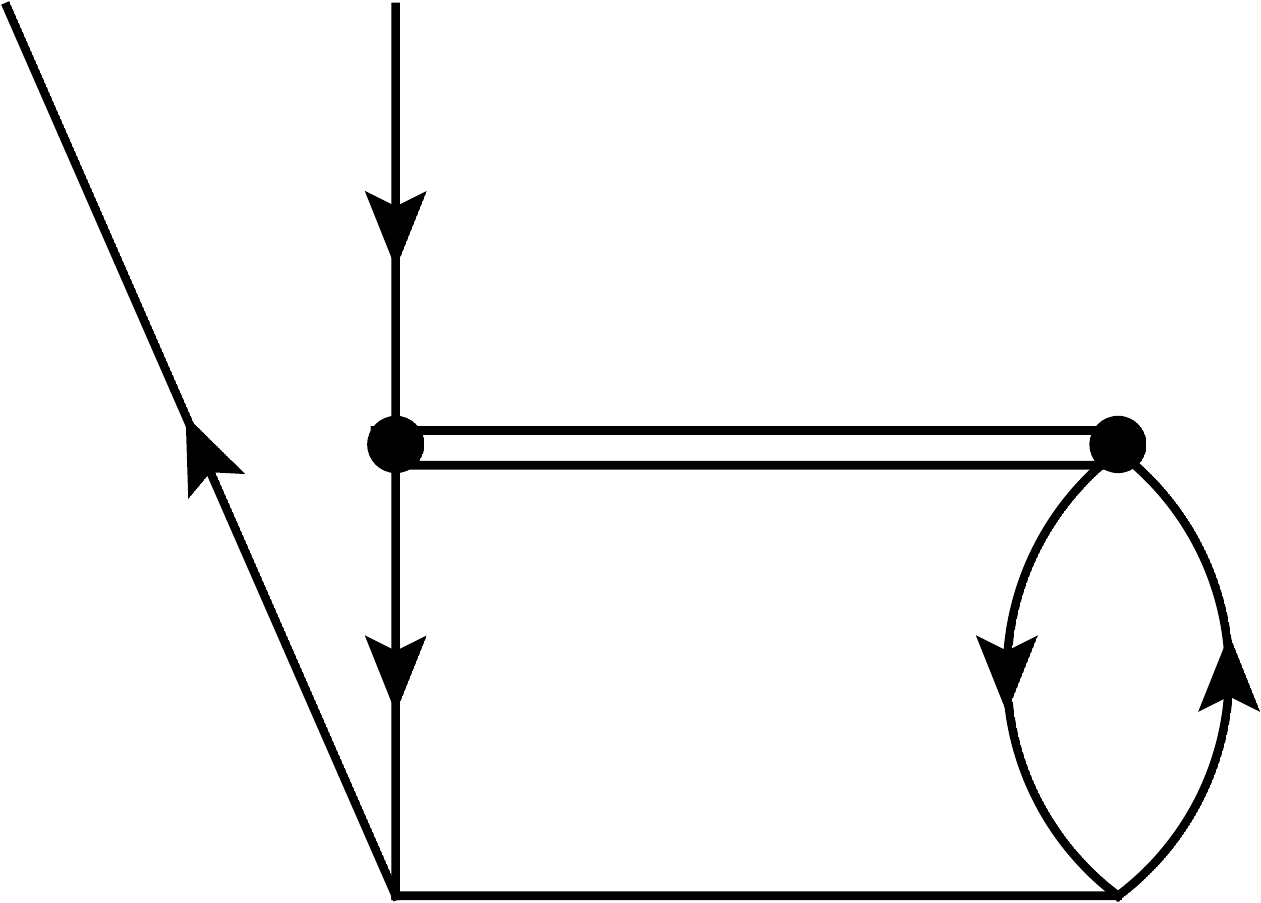}} 
    + \parbox{28mm}{\includegraphics[scale=0.20]{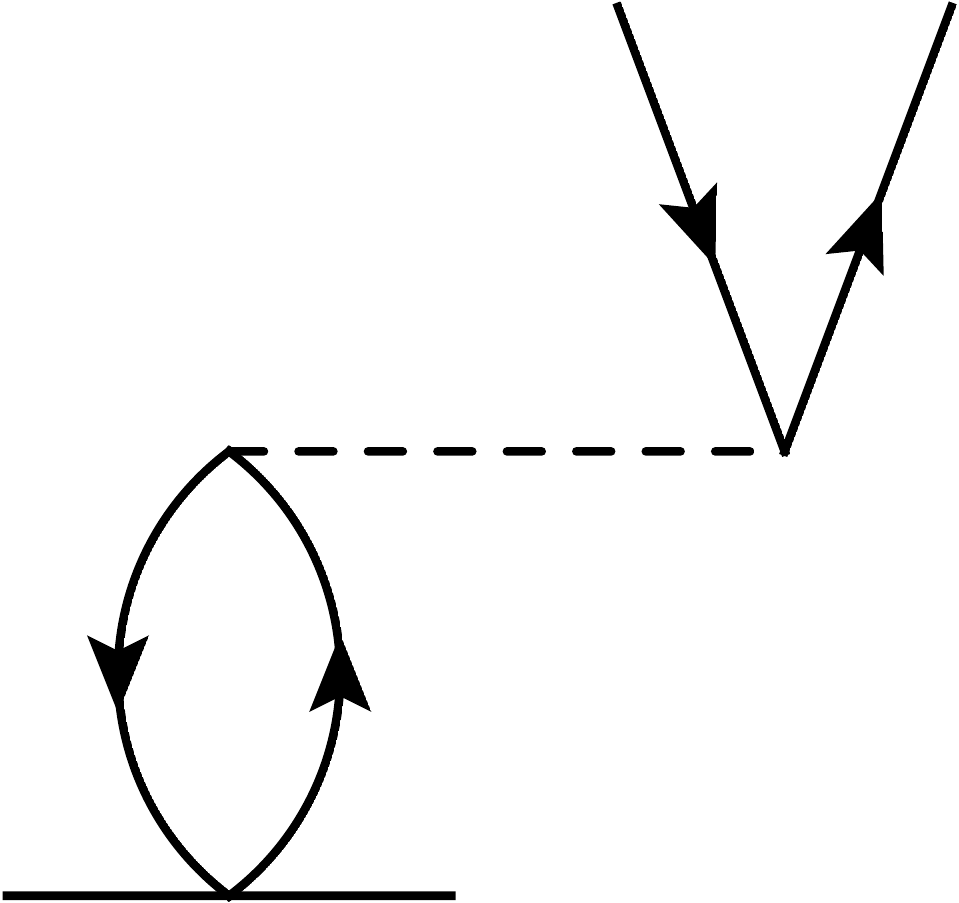}}.
    \label{eq:t1_eqns}
\end{eqnarray}

Similarly the $T_2$ amplitude equations can be written as
\begin{eqnarray}
  \nonumber
    0 &= 
    \parbox{35mm}{\includegraphics[scale=0.20]{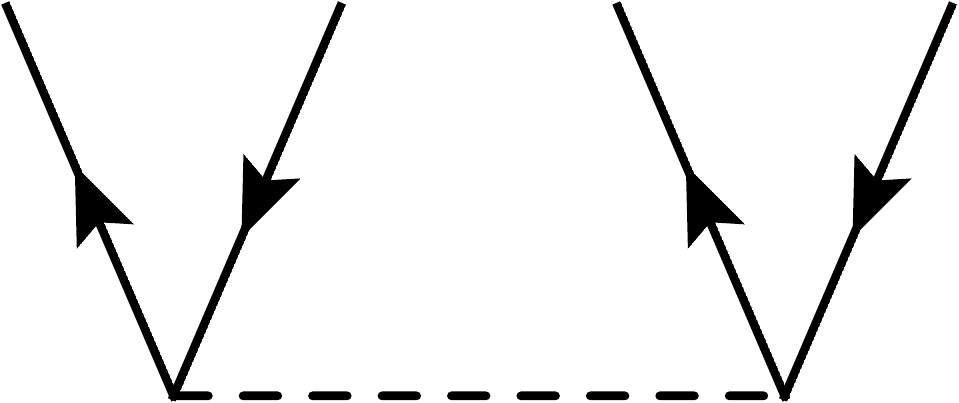}}
    + \parbox{35mm}{\includegraphics[scale=0.20]{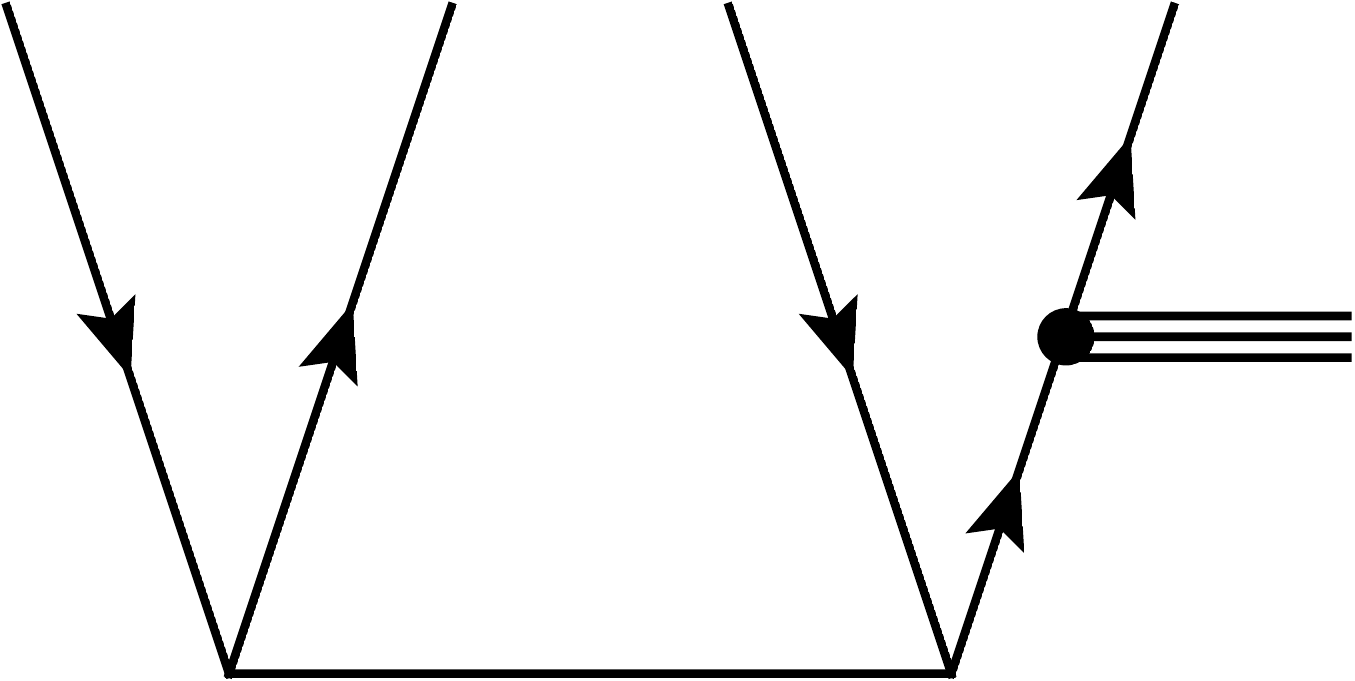}} 
    + \parbox{35mm}{\includegraphics[scale=0.20]{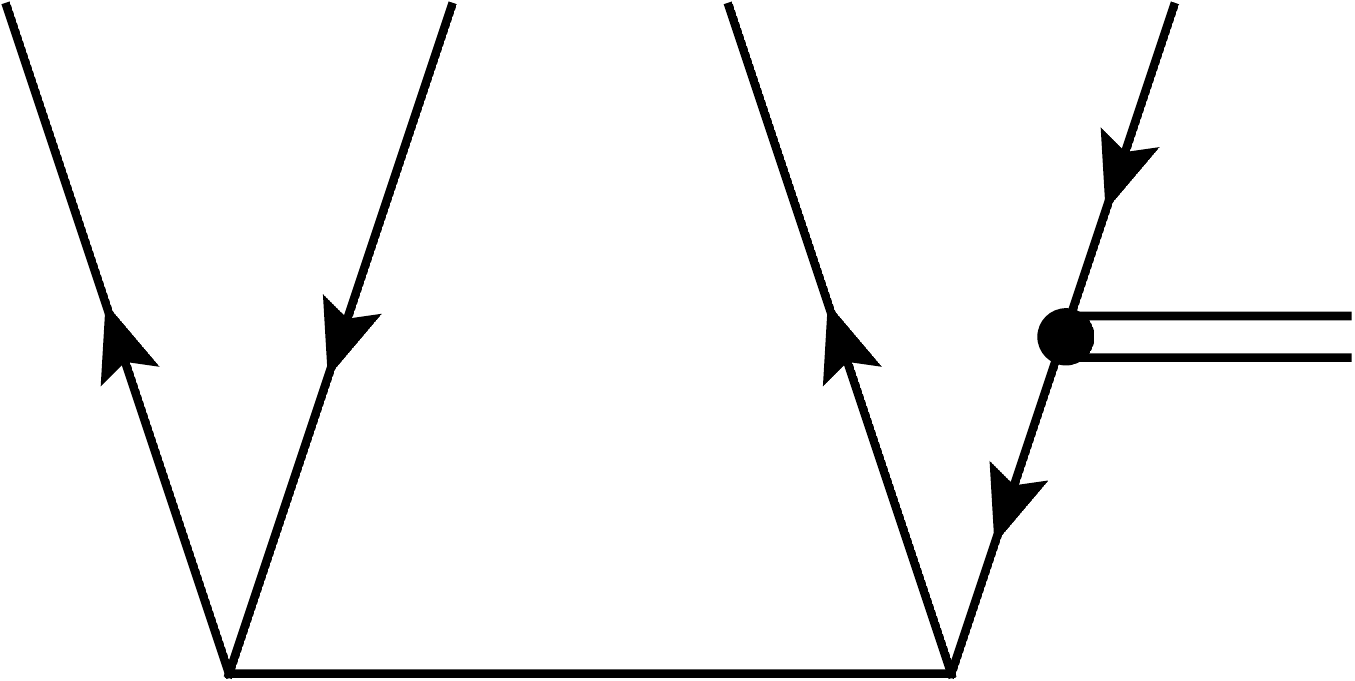}}\\
    \nonumber
    & \quad + \parbox{35mm}{\includegraphics[scale=0.20]{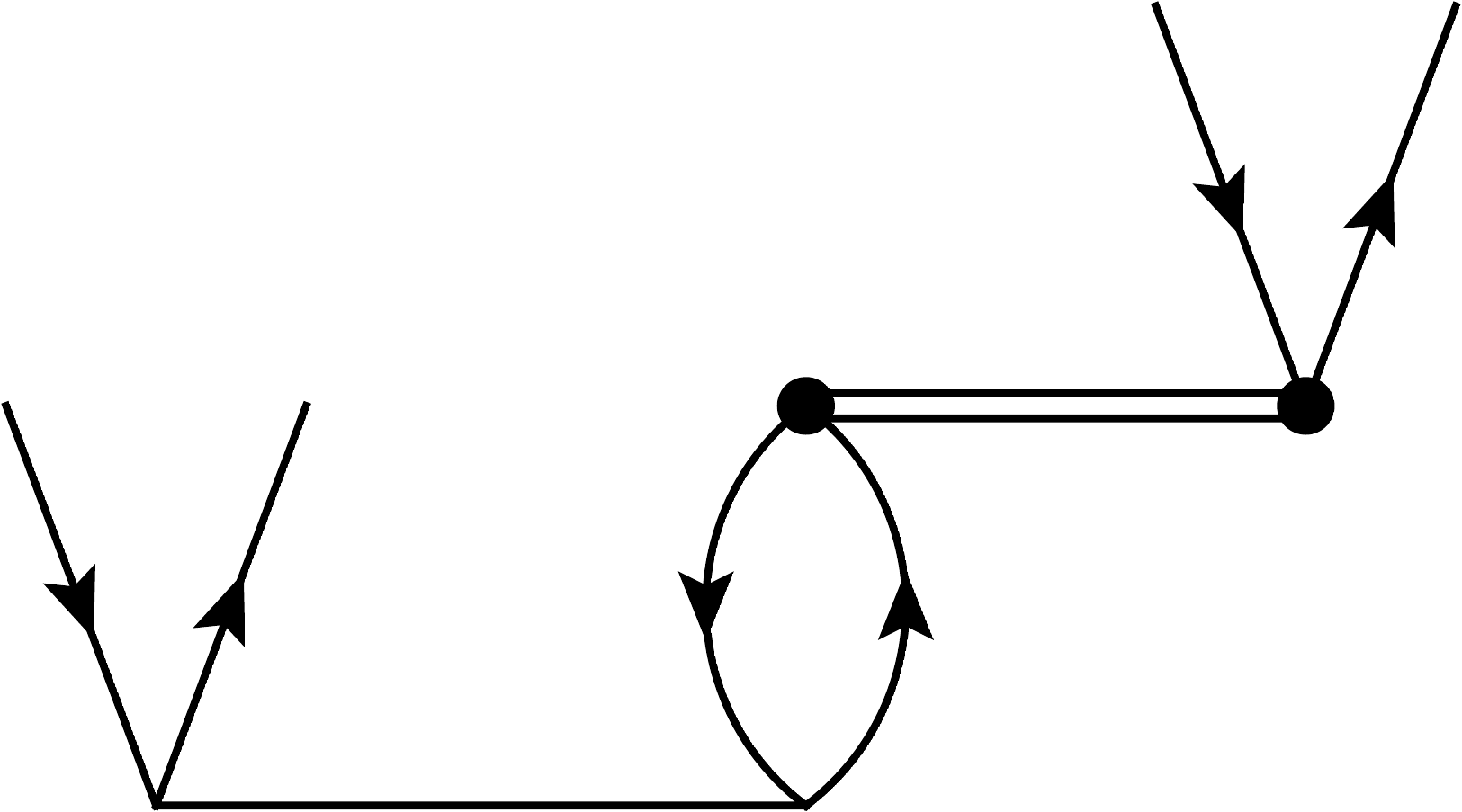}}
    + \parbox{35mm}{\includegraphics[scale=0.20]{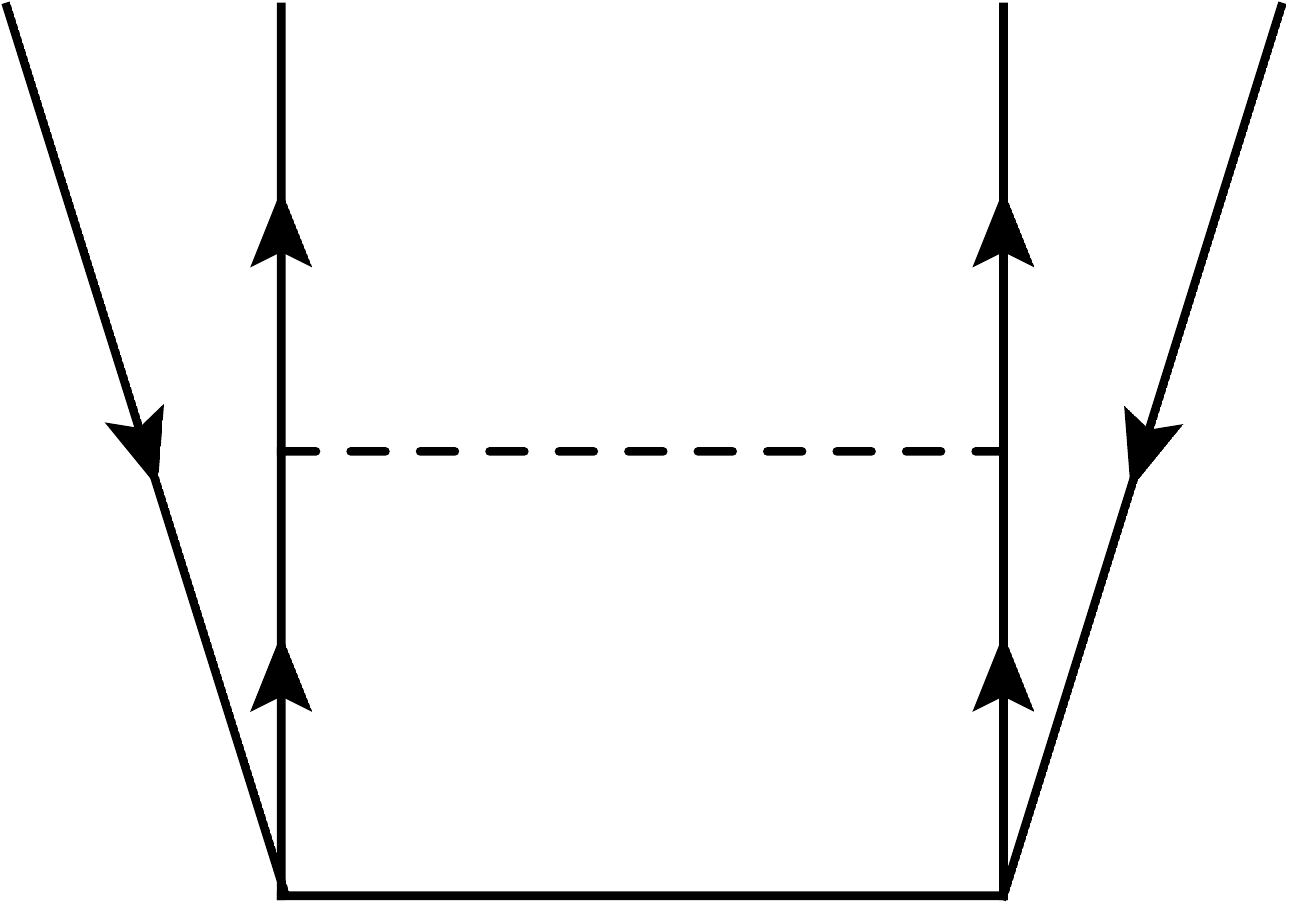}}
    + \parbox{35mm}{\includegraphics[scale=0.20]{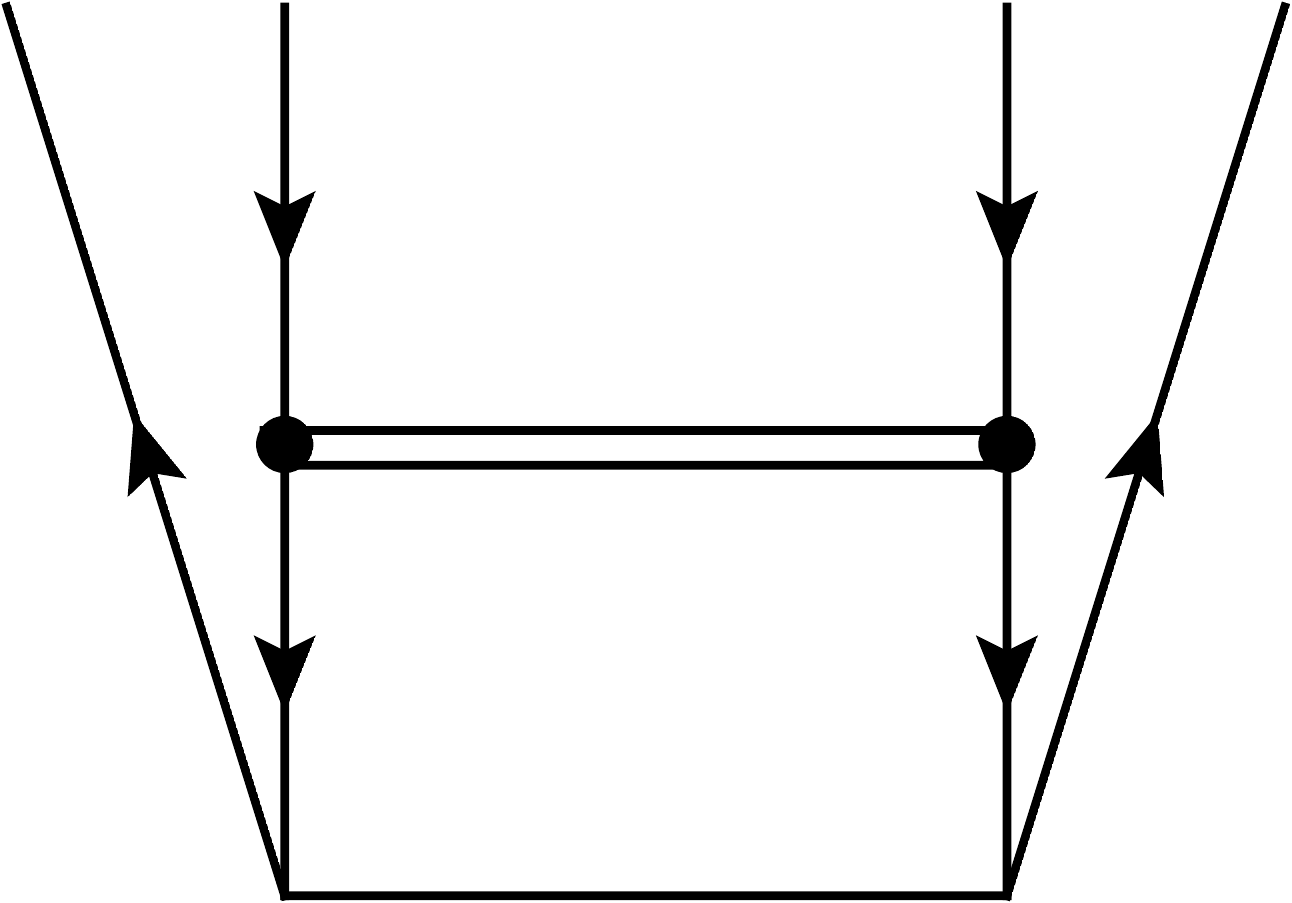}}\\ 
    & \quad + \parbox{35mm}{\includegraphics[scale=0.20]{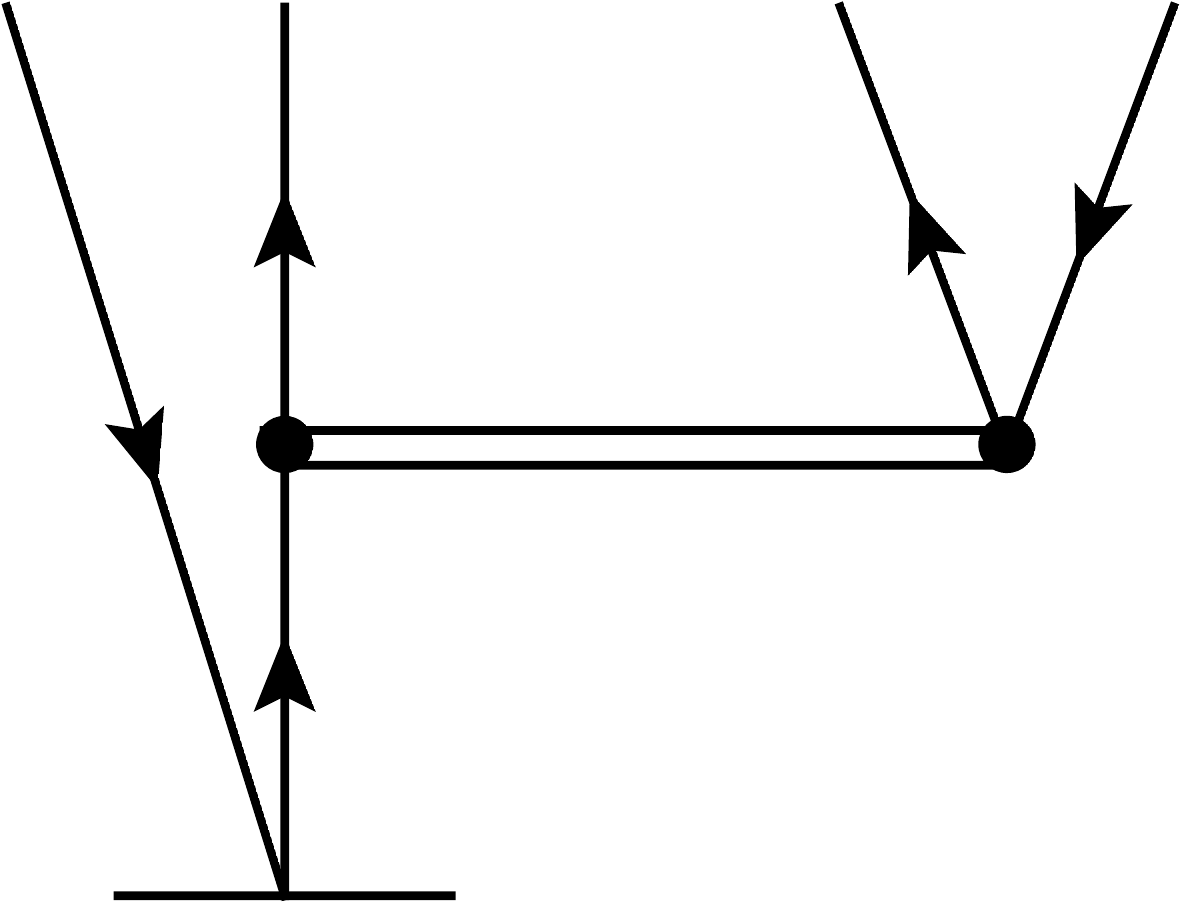}}
    + \parbox{35mm}{\includegraphics[scale=0.20]{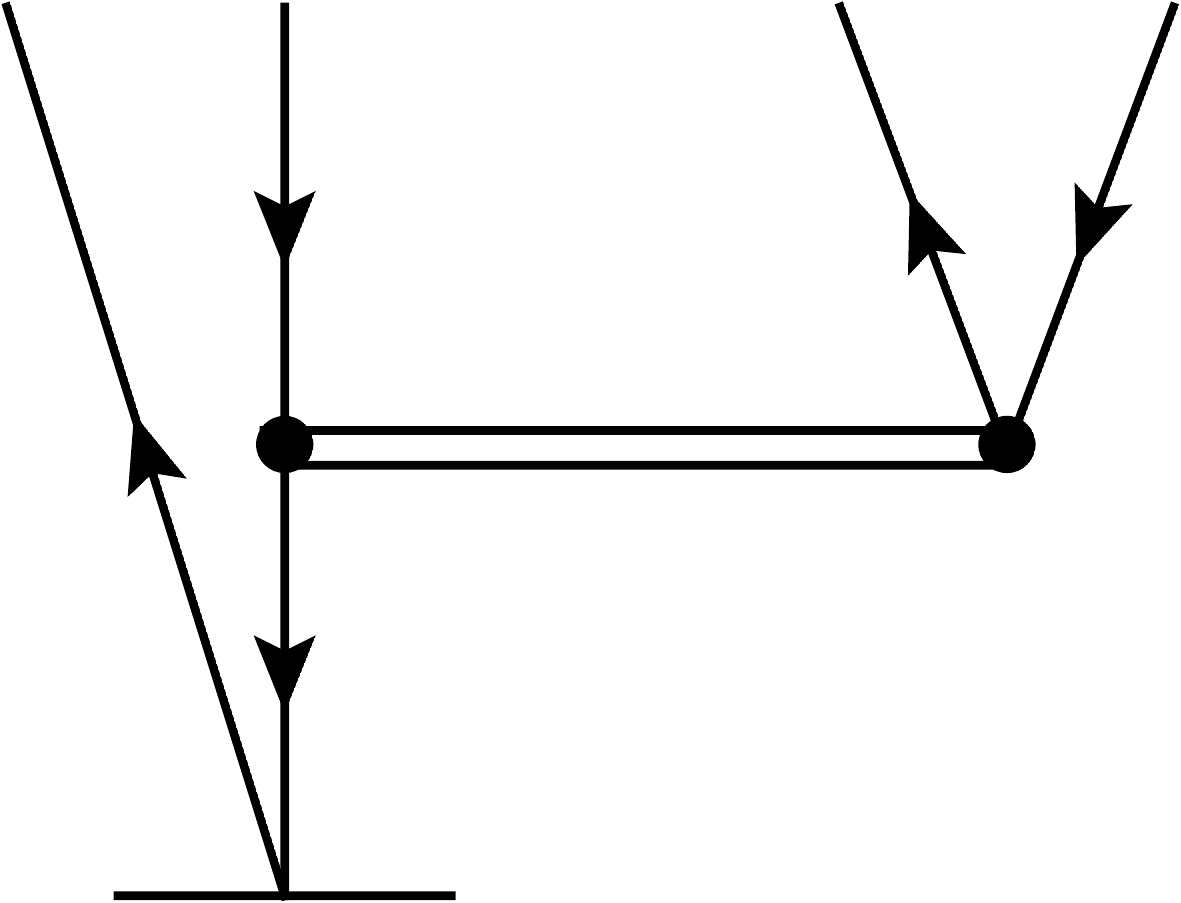}}.
    \label{eq:t2_eqns}
\end{eqnarray}

The $T_1$ and $T_2$ amplitude equations as presented in
Eqs.~\ref{eq:t1_eqns} and \ref{eq:t2_eqns} are ``quasi''-linear in the
$T_1$ and $T_2$ amplitudes, and the the non-linearity is hidden in
appropriately chosen intermediates. Dashed horizontal lines denote the
normal-ordered Hamiltonian (Fock matrix and $NN$ interaction), while
all other horizontal lines define intermediates.  The uncoupled and
coupled expressions for each of the diagrams defining the $T_1$ and
$T_2$ amplitude equations are given in Tables \ref{tab:int1} and
\ref{tab:int2}, respectively.

\begin{table*}[htbp!]
  \begin{Table_app}{0.25}{0.33}{0.42}
    Diagram & \mathrm{Uncoupled~expression} & \mathrm{Coupled~expression} 
    \tabularnewline\hline
    \includegraphics[scale=0.2]{T1_FockMat_diag} & \langle a\vert f^{00}\vert i\rangle  & 
    \langle a\vert\vert f^{0} \vert\vert i\rangle\delta_{j_{i},j_{a}}  \tabularnewline\hline
    \includegraphics[scale=0.2]{T1_I5_diag} & \langle ac\vert t^{00}\vert ik\rangle
    \langle k \vert\chi^{00} \vert c\rangle  &   { \hat{J}_{ac}^2\over \hat{j}_a^2} 
    \langle ac\vert\vert t^{0}\vert\vert ik\rangle
    \langle k \vert\vert\chi^{0} \vert\vert c\rangle  \tabularnewline\hline
    \includegraphics[scale=0.2]{T1_I7_diag} & \langle c\vert t^{00}\vert i\rangle
    \langle a\vert\chi^{00}\vert c\rangle  &  \langle c\vert\vert t^{0}\vert\vert i\rangle
    \langle a\vert\vert\chi^{0}\vert\vert c\rangle  \tabularnewline\hline
    \includegraphics[scale=0.2]{T1_I6_diag} & \langle a\vert t^{00}\vert k\rangle
    \langle k\vert\chi^{00}\vert i\rangle  & \langle a\vert\vert t^{0}\vert\vert k\rangle
    \langle k\vert\vert\chi^{0}\vert\vert i\rangle  \tabularnewline\hline
    \includegraphics[scale=0.2]{T1_I2_diag} & {1\over 2}\langle ak \vert v^{00} \vert cd\rangle
    \langle cd\vert t^{00}\vert ik\rangle  &  {1\over
      2} {\hat{J}_{cd}^2\over \hat{j}_a^2}
    \langle ak \vert\vert v^{0} \vert\vert cd\rangle
    \langle cd\vert\vert t^{0}\vert\vert ik\rangle  \tabularnewline\hline
    \includegraphics[scale=0.2]{T1_I3_diag} & -{1\over 2}\langle ac\vert t^{00}\vert kl\rangle
    \langle kl \vert\chi^{00} \vert ic\rangle  &  -{1\over 2}{\hat{J}_{cd}^2\over \hat{j}_a^2}
    \langle ac\vert\vert t^{0}\vert\vert kl\rangle
    \langle kl \vert\vert \chi^{0} \vert\vert ic\rangle  \tabularnewline\hline
    \includegraphics[scale=0.2]{T2_I4_diag} & \langle ka \vert v^{00} \vert ci\rangle
    \langle c\vert t^{00}\vert k\rangle  &  { \hat{J}_{ka}^2\over \hat{j}_a^2} 
    \langle ka \vert\vert v^{0} \vert\vert ci\rangle
    \langle c\vert\vert t^{0}\vert\vert k\rangle  
    \tabularnewline\hline
  \end{Table_app}
  \caption{Coupled and uncoupled algebraic expressions for the
    diagrams of the quasi-linearized $ T_1$ equation given in
    Eq.(\ref{eq:t1_eqns}). Repeated indices are summed over.
    $\langle p\vert f^{00}\vert q\rangle $ and $ \langle pq \vert
    v^{00} \vert rs \rangle$ are matrix elements of the Fock-matrix
    and the nucleon-nucleon interaction, respectively. }
  \label{tab:int1}
\end{table*}

\begin{table*}[htbp!]
  \begin{Table_app}{0.25}{0.33}{0.42}
    Diagram & \mathrm{Uncoupled~expression} & \mathrm{Coupled~expression} 
    \tabularnewline\hline
    \includegraphics[scale=0.2]{T2_gmat_diag} & \langle ab \vert v^{00} \vert ij\rangle & 
       \langle ab \vert\vert v^{0} \vert\vert ij\rangle \tabularnewline\hline 
      \includegraphics[scale=0.2]{T2_I6_diag} & 
      P(ab)\langle c \vert\ \tilde{\chi}^{00} \vert b\rangle
      \langle ac \vert t^{00} \vert ij\rangle  &  P(ab)\langle c \vert\vert\chi^{0} \vert\vert b\rangle
      \langle ac \vert\vert t^{0} \vert\vert ij\rangle  \tabularnewline\hline
      \includegraphics[scale=0.2]{T2_I7_diag} &  -P(ij)\langle k \vert\chi^{00} \vert j\rangle
      \langle ab \vert t^{00} \vert ik\rangle  &  -P(ij)\langle k \vert\vert\chi^{0} \vert\vert j\rangle
      \langle ab \vert\vert t^{0} \vert\vert ik\rangle  \tabularnewline\hline
      \includegraphics[scale=0.2]{T1_I4_diag} &  P(ab)P(ij)\times\langle kb \vert\chi^{00} \vert cj\rangle
      \langle ac\vert t^{00}\vert ik\rangle  &  P(ab)P(ij) 
      \times\langle kc^{-1} \vert\vert \chi^{0} \vert\vert jb^{-1}\rangle
      \langle ai^{-1} \vert\vert t^{0} \vert\vert kc^{-1}\rangle \tabularnewline\hline
      \includegraphics[scale=0.2]{T2_I8_diag} & {1\over 2}\langle ab \vert\chi^{00} \vert cd\rangle
      \langle cd\vert t^{00}\vert ij\rangle  & {1\over 2}\langle ab \vert\vert\chi^{0} \vert\vert cd\rangle
      \langle cd\vert\vert t^{0}\vert\vert ij\rangle  \tabularnewline\hline
      \includegraphics[scale=0.2]{T2_I9_diag} & {1\over 2}\langle kl \vert\chi^{00} \vert ij\rangle
      \langle ab\vert t^{00}\vert kl\rangle   & {1\over 2}\langle kl \vert\vert\chi^{0} \vert\vert ij\rangle
      \langle ab\vert\vert t^{0}\vert\vert kl\rangle  \tabularnewline\hline
      \includegraphics[scale=0.2]{T2_I10_diag} & P(ij)\langle ab \vert\chi^{00} \vert cj\rangle
      \langle c\vert t^{00}\vert i\rangle   & P(ij)\langle ab \vert\vert \chi^{0} \vert\vert cj\rangle
      \langle c\vert\vert t^{0}\vert\vert i\rangle   \tabularnewline\hline
      \includegraphics[scale=0.2]{T2_I11_diag} & -P(ab)\langle kb \vert\chi^{00} \vert ij\rangle
      \langle a\vert t^{00}\vert k\rangle   & -P(ab)\langle kb \vert\vert\chi^{0} \vert\vert ij\rangle
      \langle a\vert\vert t^{0}\vert\vert k\rangle   \tabularnewline\hline
    \end{Table_app}
    \caption{Coupled and uncoupled algebraic expressions for the
      diagrams of the quasi-linearized $ T_2$ equation given in
      Eq.(\ref{eq:t2_eqns}). Repeated indices are summed
      over. $\langle p\vert f^{00}\vert q\rangle $ and $ \langle pq
      \vert v^{00} \vert rs \rangle$ are matrix elements of the
      Fock-matrix and the nucleon-nucleon interaction, respectively.
      Note the coupled expressions are divided by
      $\sum_{J_{ab},M_{ab}}
      C_{j_am_aj_bm_m}^{J_{ab}M_{ab}}C_{j_im_ij_jm_j}^{J_{ij}M_{ij}}\delta_{J_{ab},J_{ij}}\delta_{M_{ab},M_{ij}}$
      to give the equation for the reduced amplitudes $\langle
      ab\vert\vert t^0\vert\vert ij\rangle$. }
    \label{tab:int2}  
\end{table*}

The intermediates given in Tables~\ref{tab:int1} and \ref{tab:int2}
are sums of various contractions between the Fock-matrix and
nucleon-nucleon interaction with the $T_1$ and $T_2$ amplitudes. In
Eqs.~\ref{eq:I5_eqns}, \ref{eq:I6_eqns}, \ref{eq:I6_eqns2},
\ref{eq:I7_eqns}, \ref{eq:I3_eqns}, \ref{eq:I4_eqns},
\ref{eq:I9_eqns}, \ref{eq:I10_eqns} and \ref{eq:I11_eqns} the angular
momentum coupled algebraic expressions for the intermediates are given
in terms of contractions between the normal-ordered Hamiltonian with
the $T_1$ and $T_2$ amplitudes.

\begin{eqnarray}
  \nonumber
  \parbox{20mm}{\includegraphics[scale=0.20]{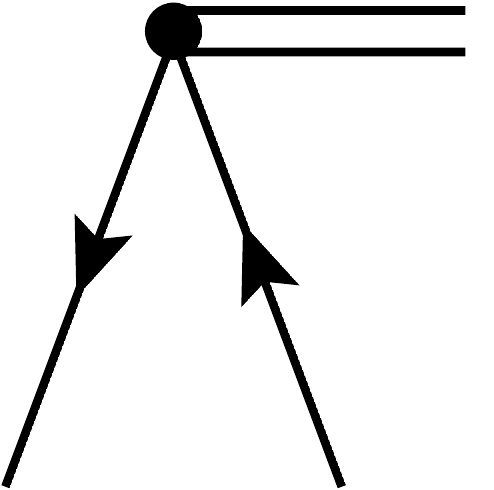}}  &
  = &  
  \parbox{20mm}{\includegraphics[scale=0.20]{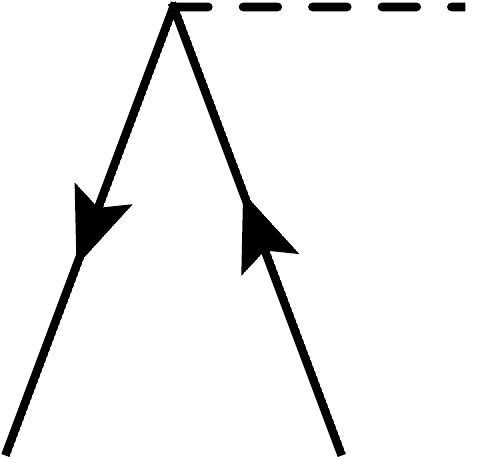}} 
  + \parbox{20mm}{\includegraphics[scale=0.20]{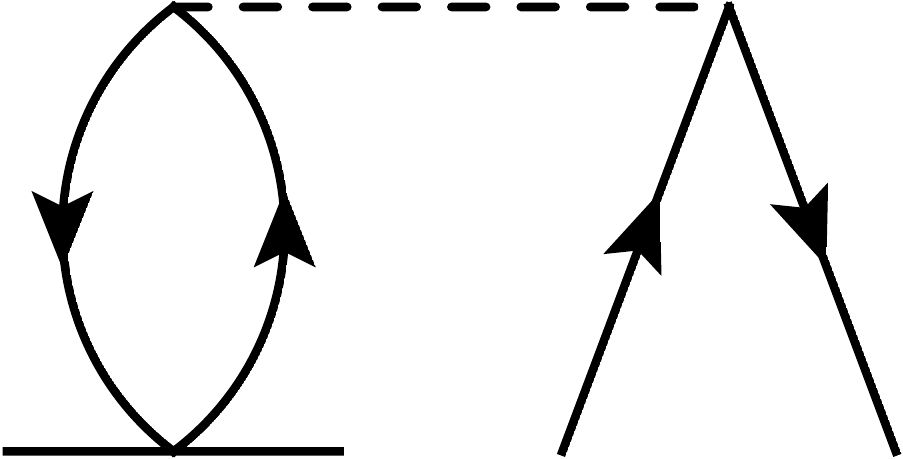}} \\
  & = & \langle k\vert\vert f^{0} \vert\vert c \rangle + {
    \hat{J}_{cd}^2\over \hat{j}_k^2} \langle
  lk \vert\vert v^{0} \vert\vert
  dc \rangle\langle d \vert\vert t^{0} \vert\vert l\rangle.
  \label{eq:I5_eqns}
\end{eqnarray}

\begin{eqnarray}
  \nonumber
  \parbox{20mm}{\includegraphics[scale=0.20]{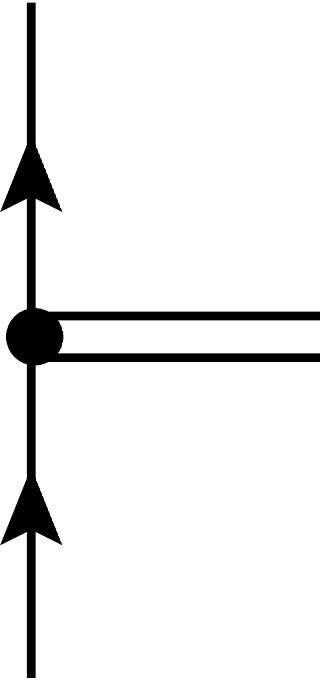}}  &
  = &  
  \parbox{20mm}{\includegraphics[scale=0.20]{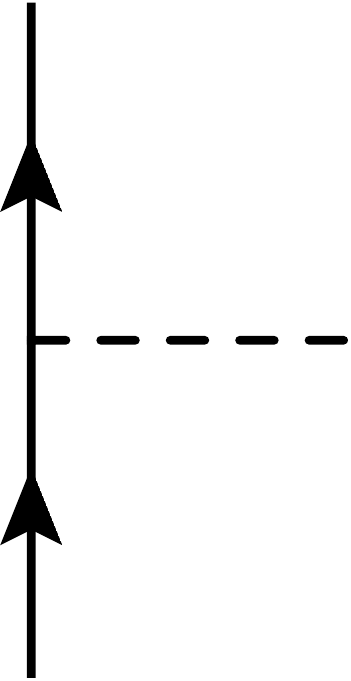}} 
  + \parbox{20mm}{\includegraphics[scale=0.20]{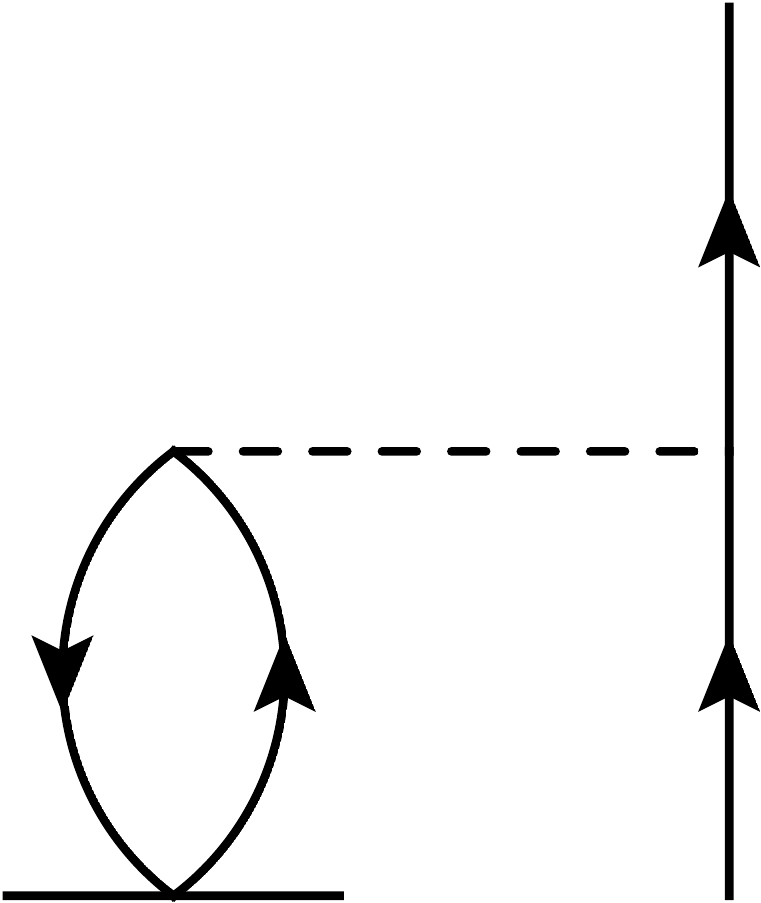}}  \\
    & = & \langle a\vert\vert f^{0} \vert\vert c \rangle + {
    \hat{J}_{cd}^2\over \hat{j}_a^2} \langle
  ka \vert\vert v^{0} \vert\vert
  dc \rangle\langle d \vert\vert t^{0} \vert\vert k\rangle.
  \label{eq:I6_eqns}
\end{eqnarray}

\begin{eqnarray}
  \nonumber
  \parbox{20mm}{\includegraphics[scale=0.20]{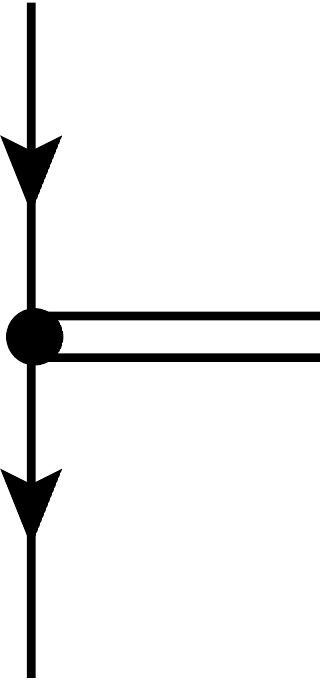}}  &
  = &   
  \parbox{20mm}{\includegraphics[scale=0.20]{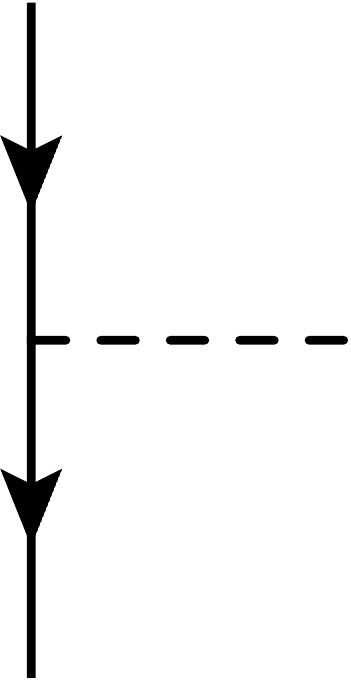}} 
  + \parbox{20mm}{\includegraphics[scale=0.20]{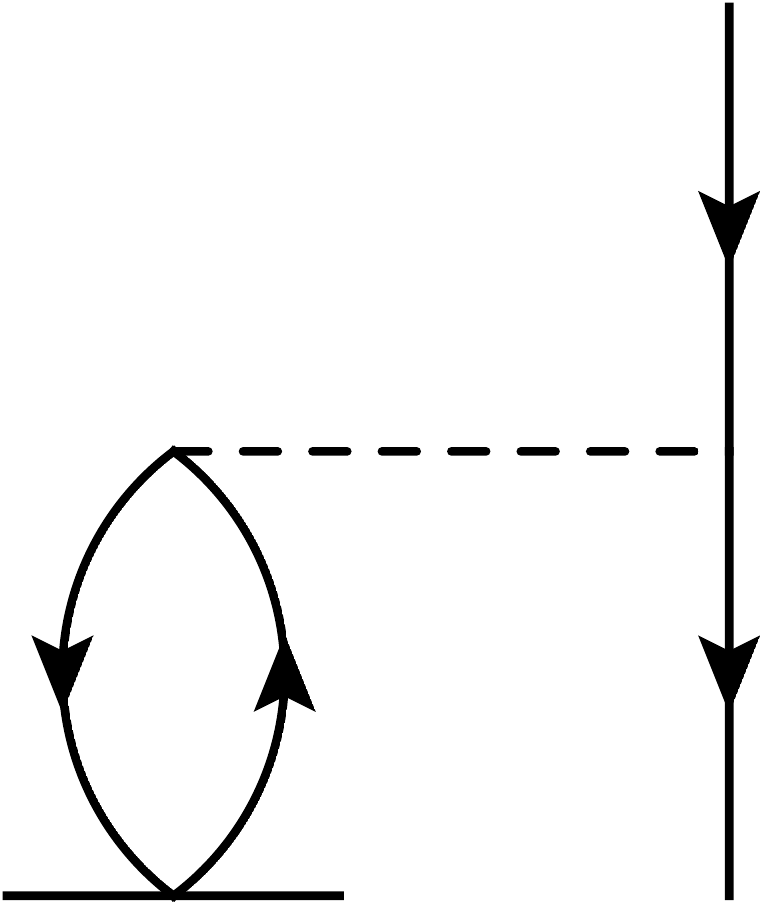}} 
  + \parbox{20mm}{\includegraphics[scale=0.20]{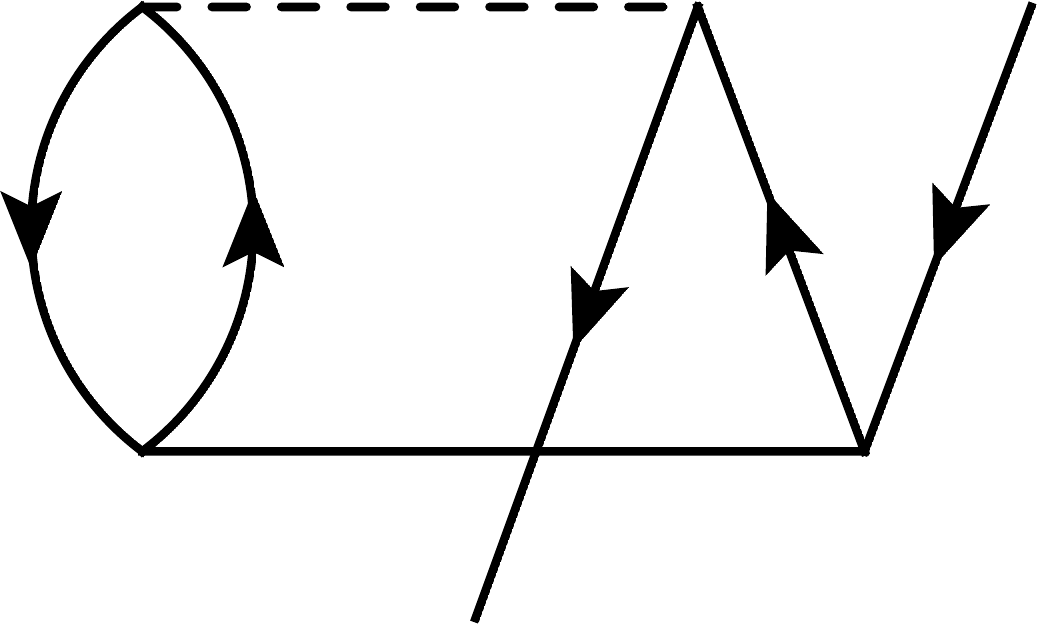}} 
  + \parbox{20mm}{\includegraphics[scale=0.20]{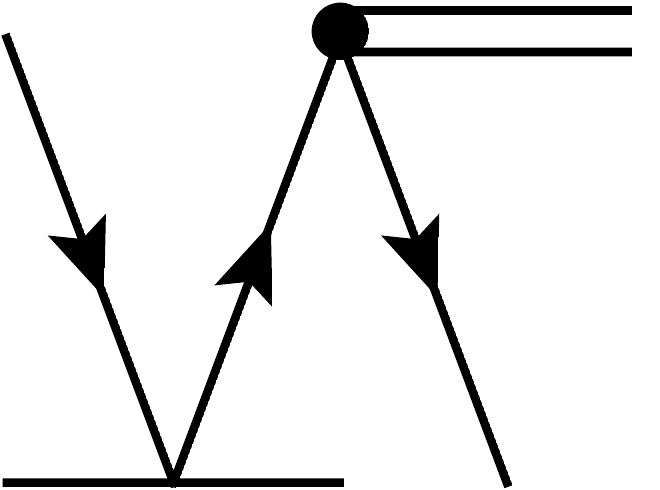}} \\
\nonumber
& = & \langle k\vert\vert f^{0} \vert\vert i \rangle + 
{ \hat{J}_{kl}^2\over \hat{j}_i^2} \langle
  lk \vert\vert v^{0} \vert\vert
  ci \rangle\langle c \vert\vert t^{0} \vert\vert l\rangle  \\ 
& + &
   {1\over 2}{ \hat{J}_{kl}^2\over \hat{j}_i^2} \langle
  lk \vert\vert v^{0} \vert\vert
  cd \rangle\langle cd \vert\vert t^{0} \vert\vert li\rangle + 
  \langle
  c \vert\vert t^{0} \vert\vert
  i\rangle\langle k \vert\vert \chi^{0} \vert\vert c\rangle.
  \label{eq:I7_eqns}
\end{eqnarray}

\begin{eqnarray}
  \nonumber
  \parbox{20mm}{\includegraphics[scale=0.20]{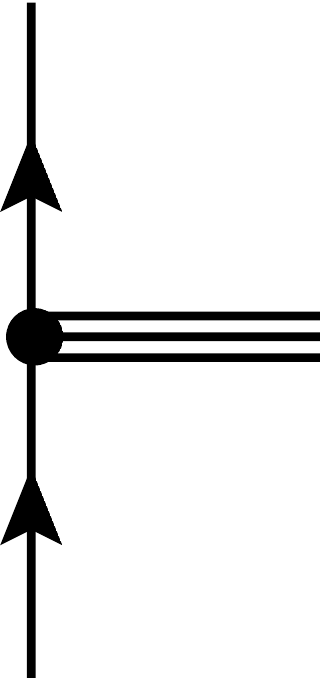}}  &
  = &   
  \parbox{20mm}{\includegraphics[scale=0.20]{I6int1_diag}} 
  + \parbox{20mm}{\includegraphics[scale=0.20]{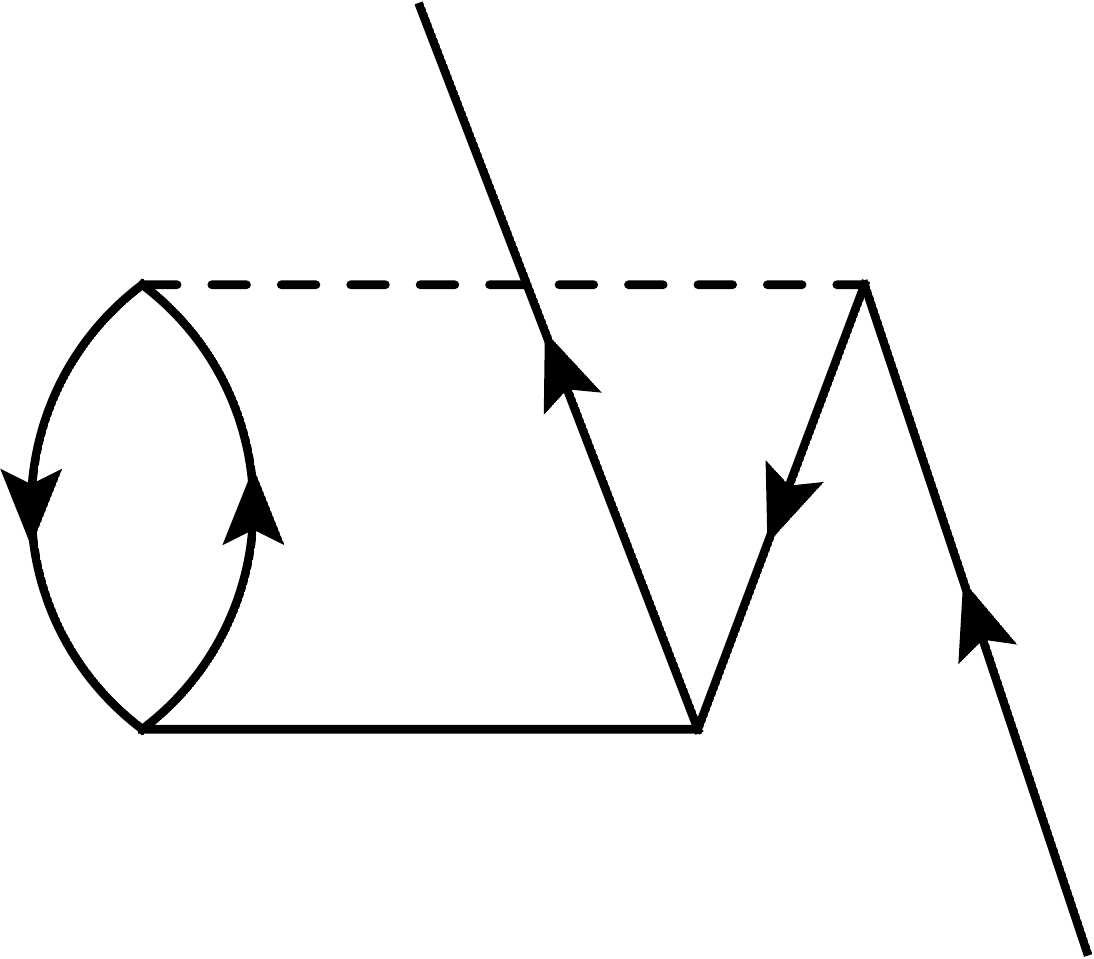}} \\
  & = & \langle k\vert\vert f^{0} \vert\vert i \rangle -
  {1\over 2}{ \hat{J}_{kl}^2\over \hat{j}_a^2} \langle
  da \vert\vert t^{0} \vert\vert
  kl \rangle\langle kl \vert\vert v^{0} \vert\vert dc\rangle.
  \label{eq:I6_eqns2}
\end{eqnarray}

\begin{eqnarray}
  \nonumber
  \parbox{20mm}{\includegraphics[scale=0.20]{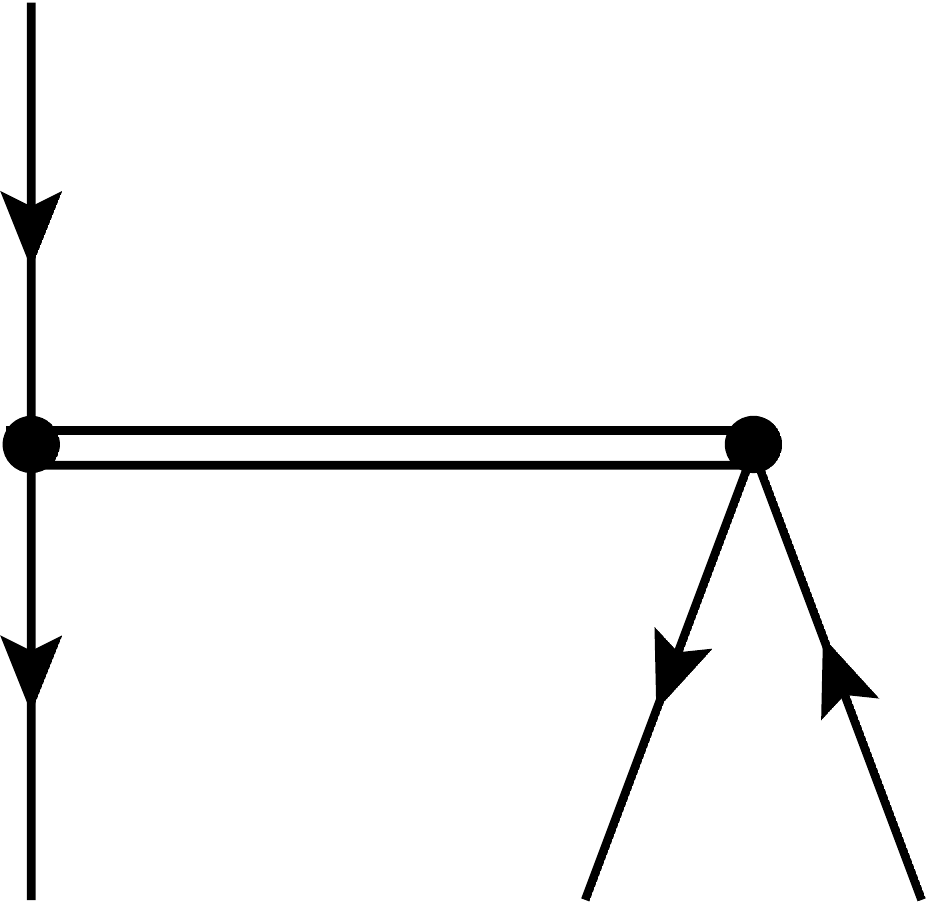}} &
  = &   
  \parbox{20mm}{\includegraphics[scale=0.20]{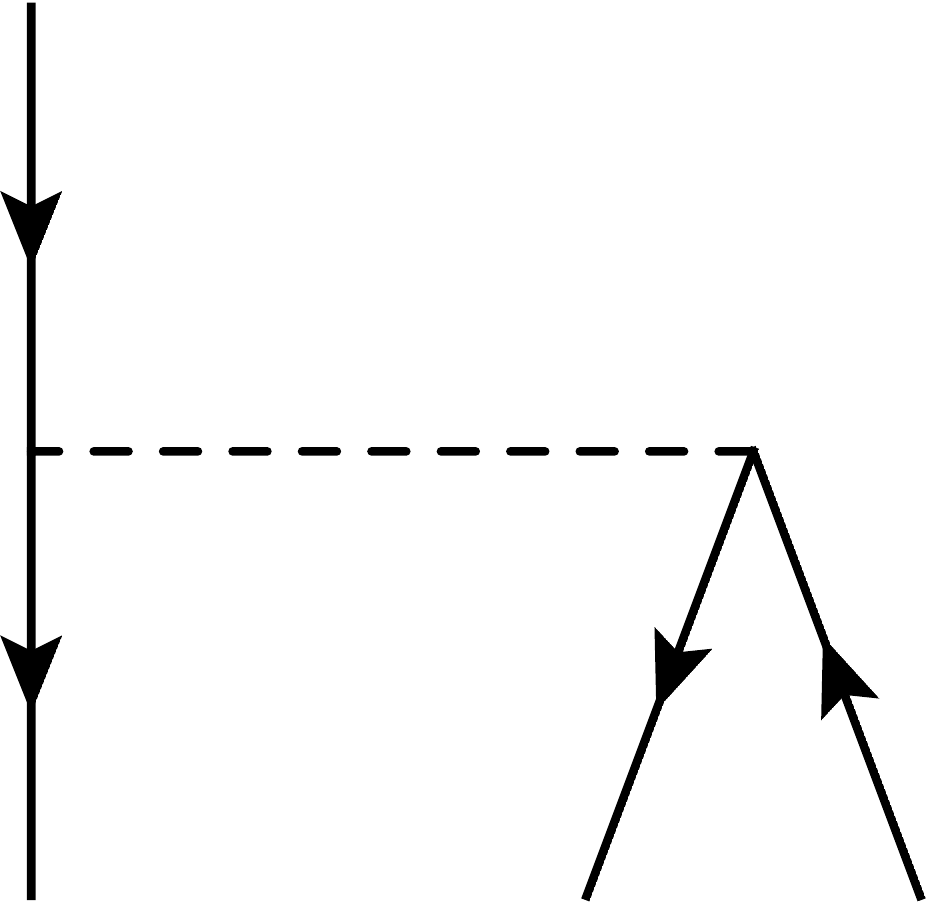}} 
  + \parbox{20mm}{\includegraphics[scale=0.20]{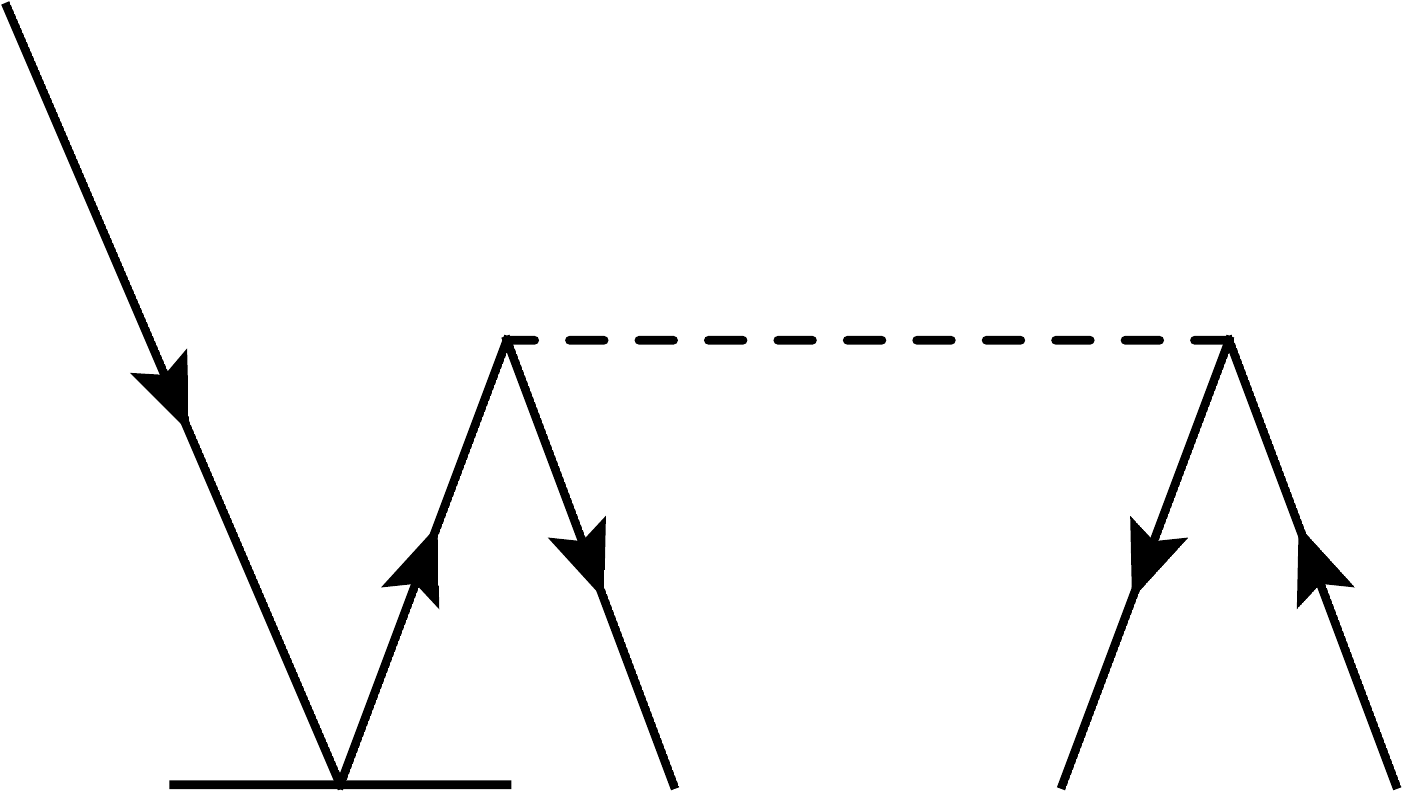}} \\ 
  & = & \langle kl \vert\vert v^{0} \vert\vert
  ic \rangle + \langle kl \vert\vert v^{0} \vert\vert dc\rangle \langle d \vert\vert t^{0} \vert\vert i\rangle.
  \label{eq:I3_eqns}
\end{eqnarray}

\begin{eqnarray}
  \nonumber
  \parbox{20mm}{\includegraphics[scale=0.20]{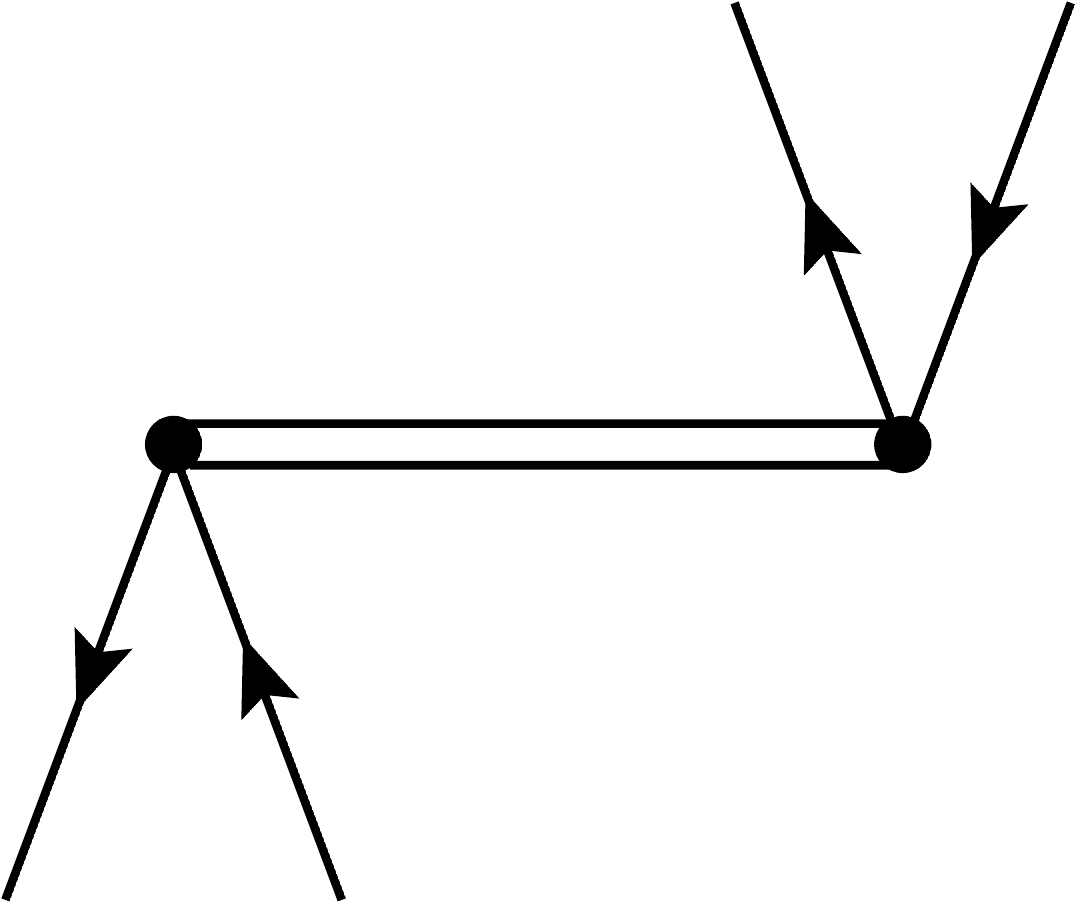}}  &
   = &  
  \parbox{30mm}{\includegraphics[scale=0.20]{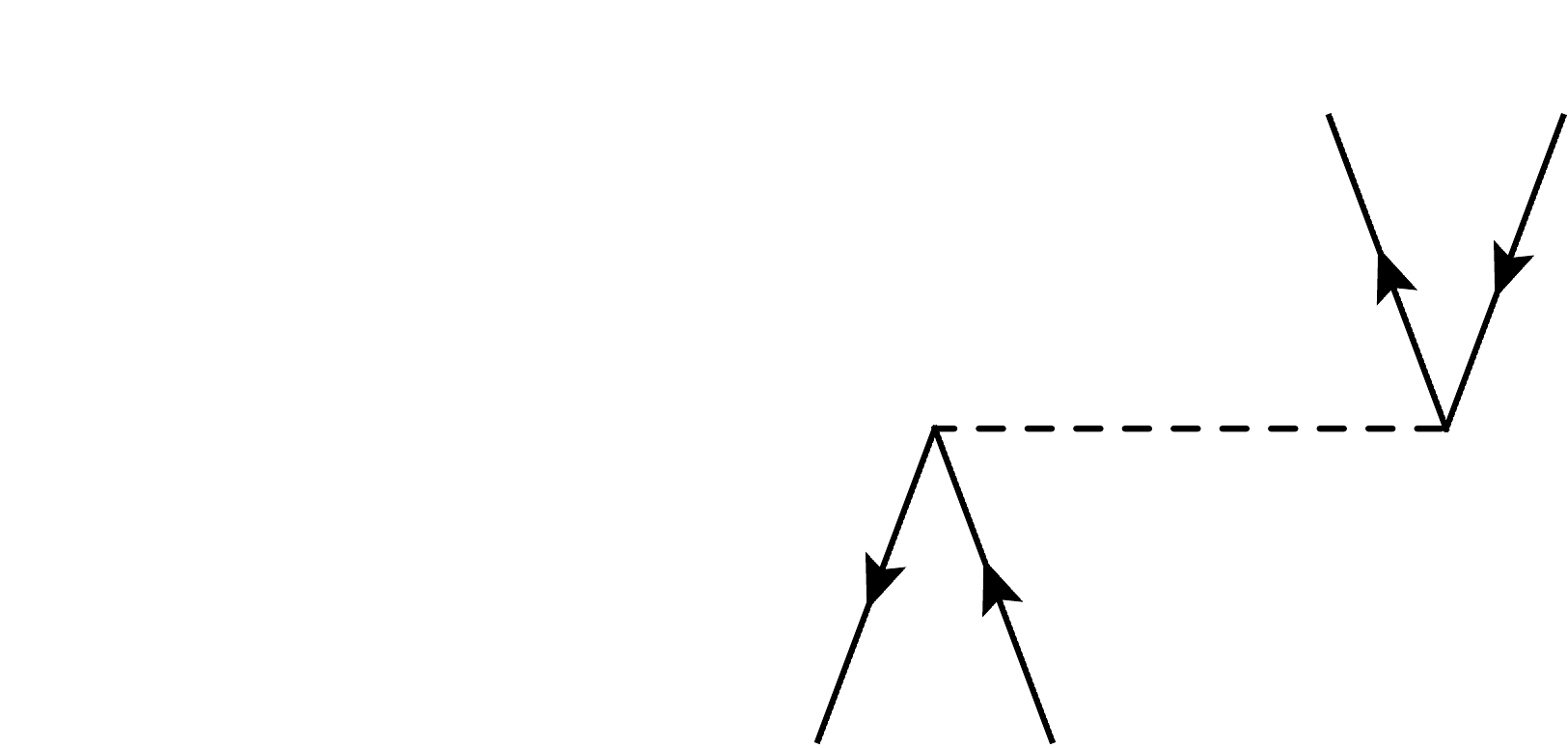}} 
  + \parbox{30mm}{\includegraphics[scale=0.20]{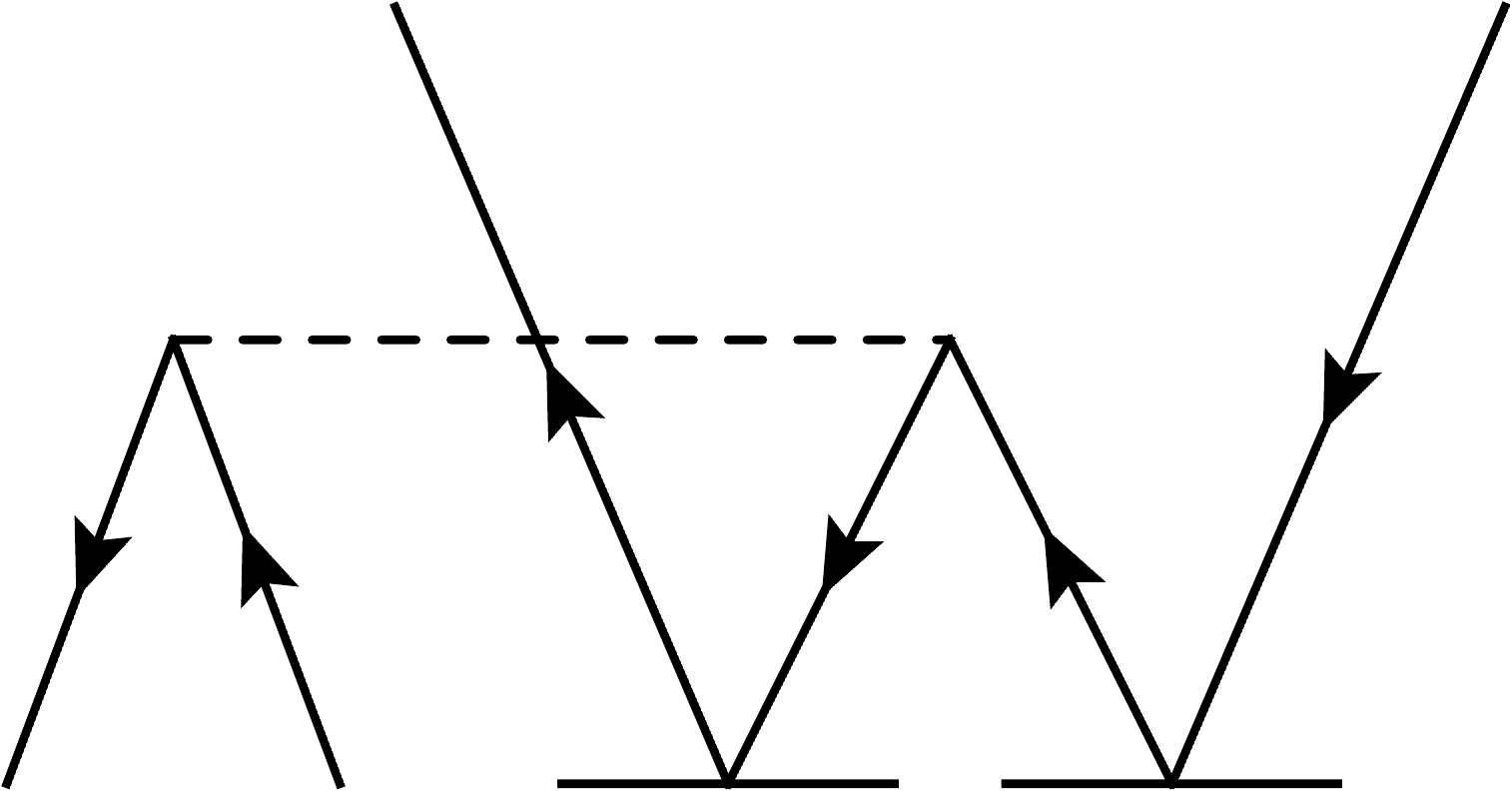}} 
  + \parbox{30mm}{\includegraphics[scale=0.20]{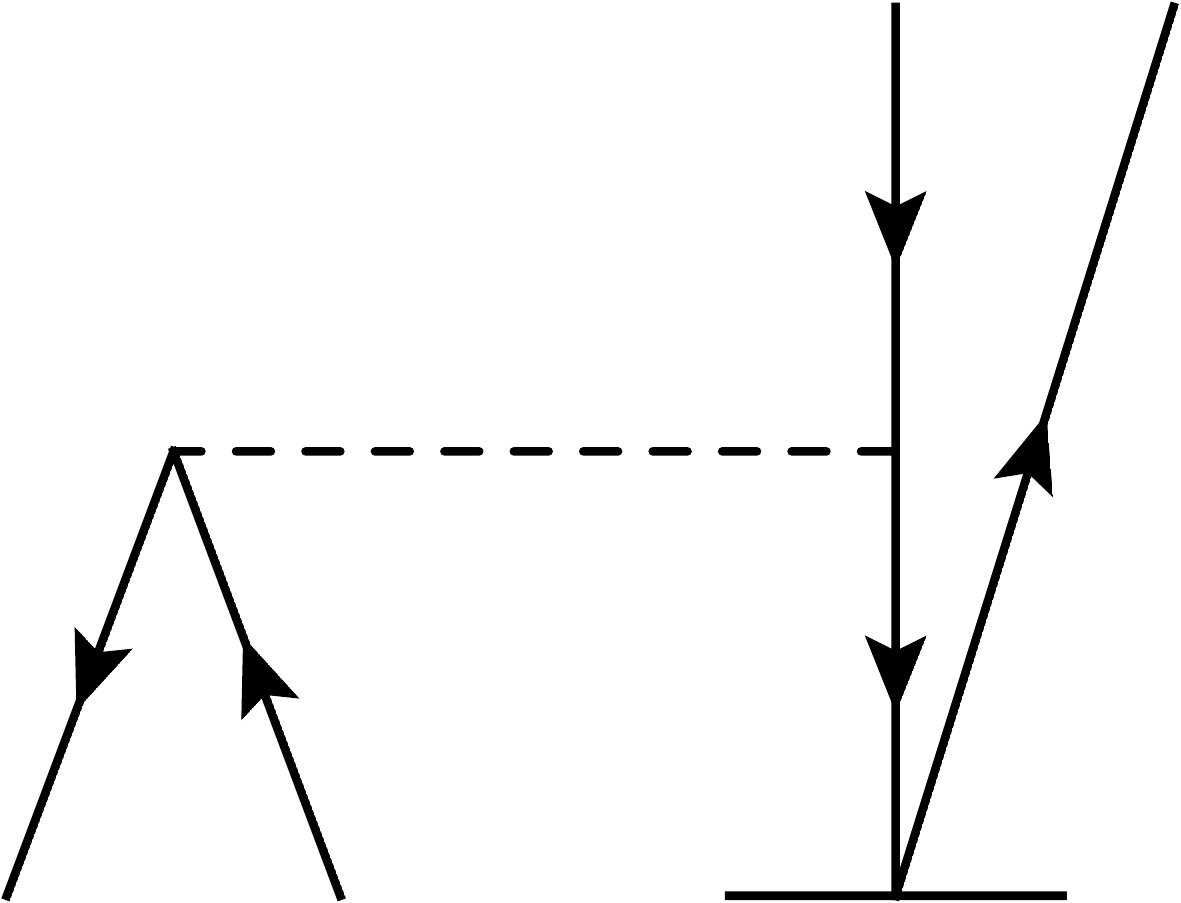}}  \\
\nonumber
  & + & \parbox{30mm}{\includegraphics[scale=0.20]{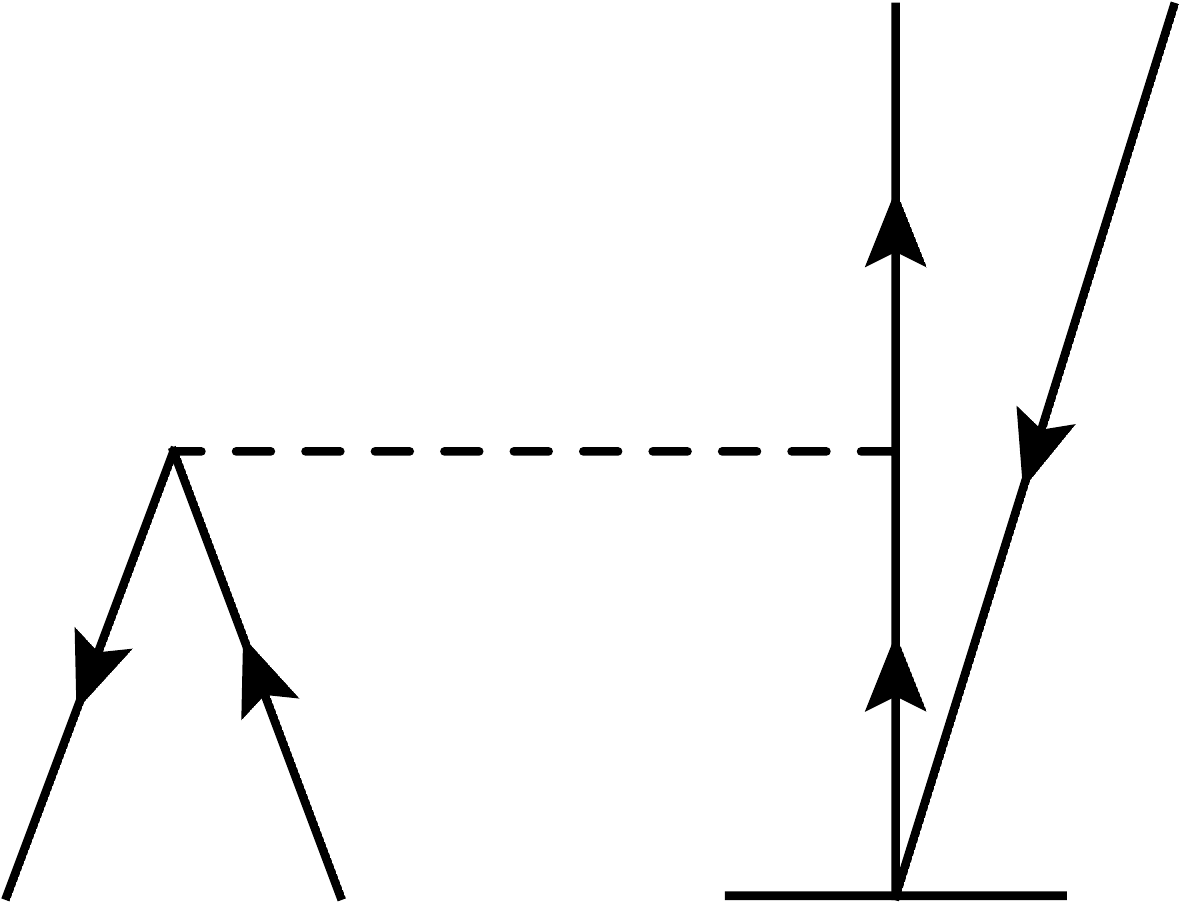}} 
  + \parbox{30mm}{\includegraphics[scale=0.20]{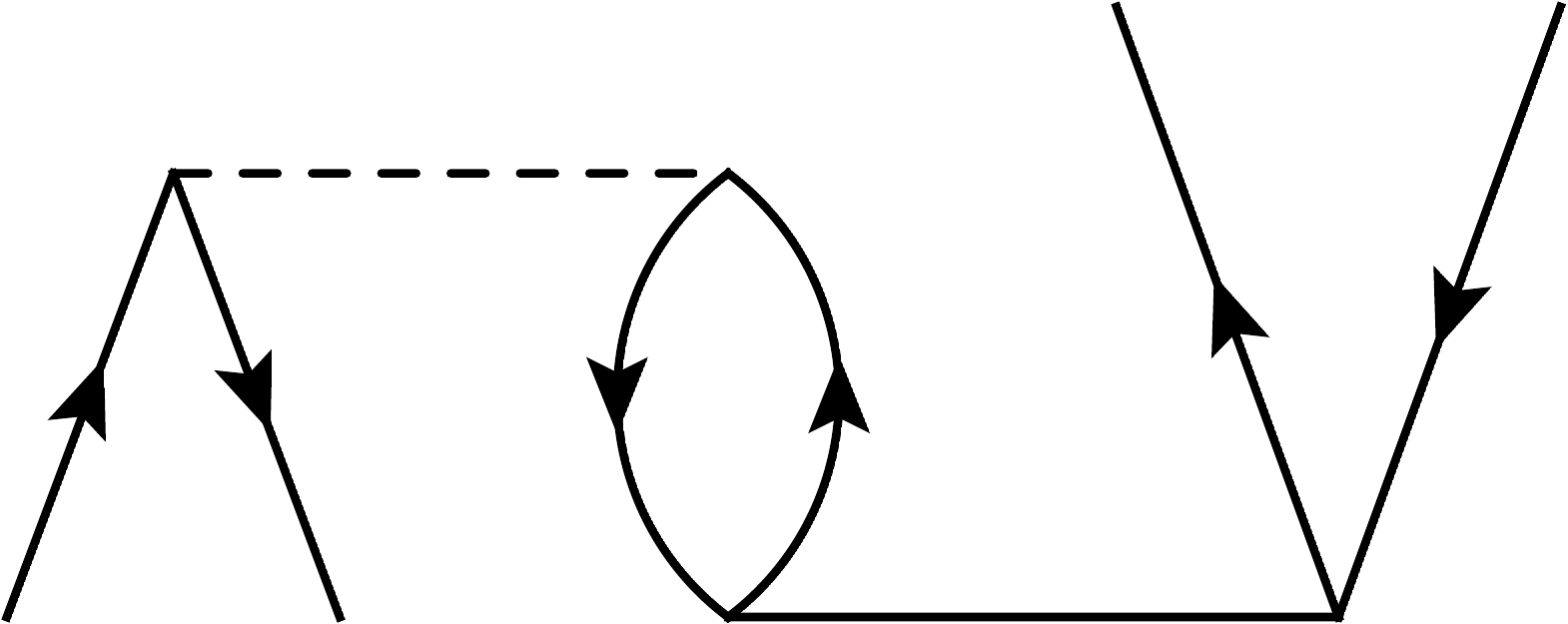}} \\ 
\nonumber
 &= & \langle kc^{-1} \vert\vert v^{0} \vert\vert jb^{-1} \rangle +  \langle kc^{-1}
 \vert\vert v^{0} \vert\vert dl^{-1} \rangle \langle b
 \vert\vert t^{0} \vert\vert l \rangle \langle d
 \vert\vert t^{0} \vert\vert j \rangle \\
\nonumber
&+ &
 \langle kc^{-1}
 \vert\vert v^{0} \vert\vert jl^{-1} \rangle \langle b
 \vert\vert t^{0} \vert\vert l \rangle
 + 
 \langle kc^{-1}
 \vert\vert v^{0} \vert\vert db^{-1} \rangle \langle d
 \vert\vert t^{0} \vert\vert j \rangle \\ 
& + &
\langle kc^{-1} \vert\vert v^{0} \vert\vert dl^{-1}\rangle
\langle dl^{-1} \vert\vert t^{0} \vert\vert jb^{-1}\rangle 
  \label{eq:I4_eqns}
\end{eqnarray}

\begin{eqnarray}
  \nonumber
  \parbox{20mm}{\includegraphics[scale=0.20]{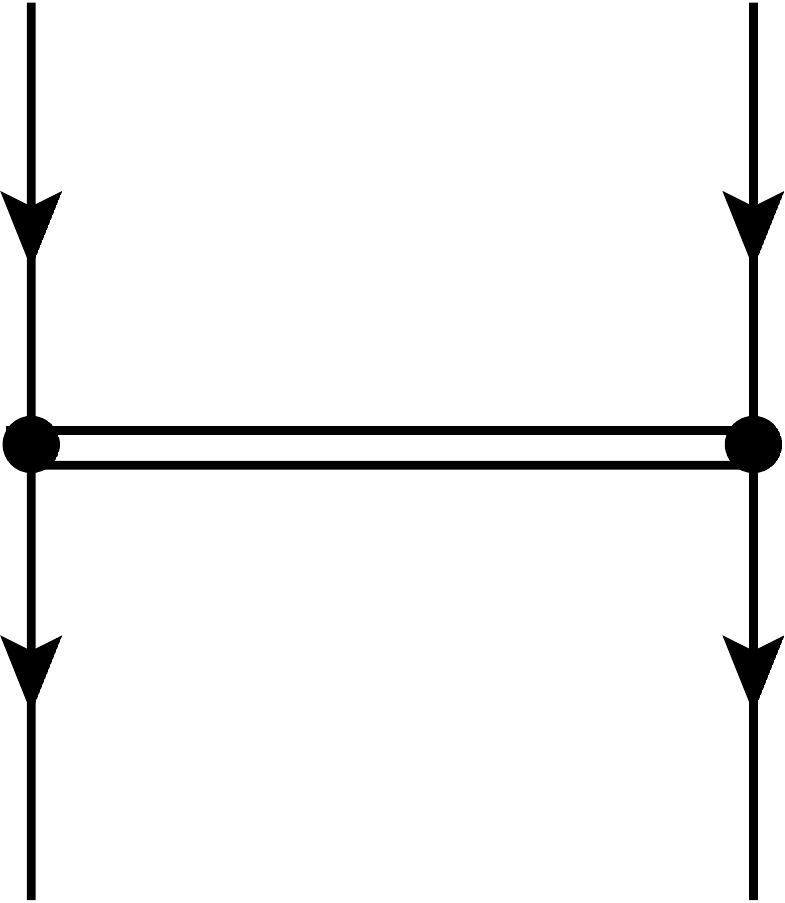}}  &
   = &  
  \parbox{30mm}{\includegraphics[scale=0.20]{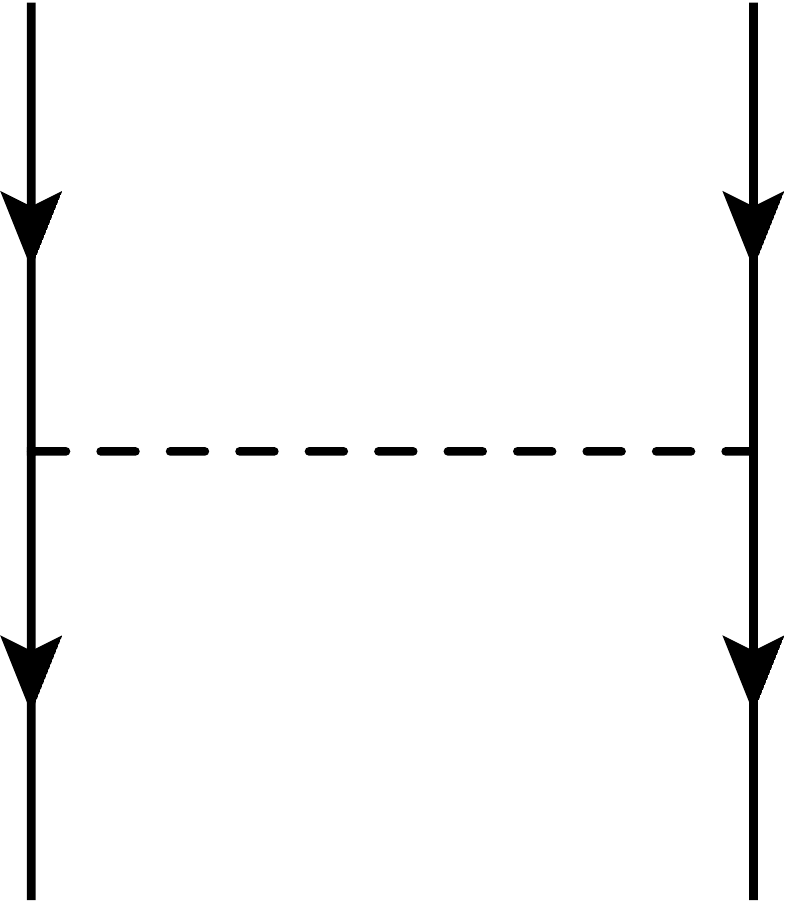}} 
  + \parbox{30mm}{\includegraphics[scale=0.20]{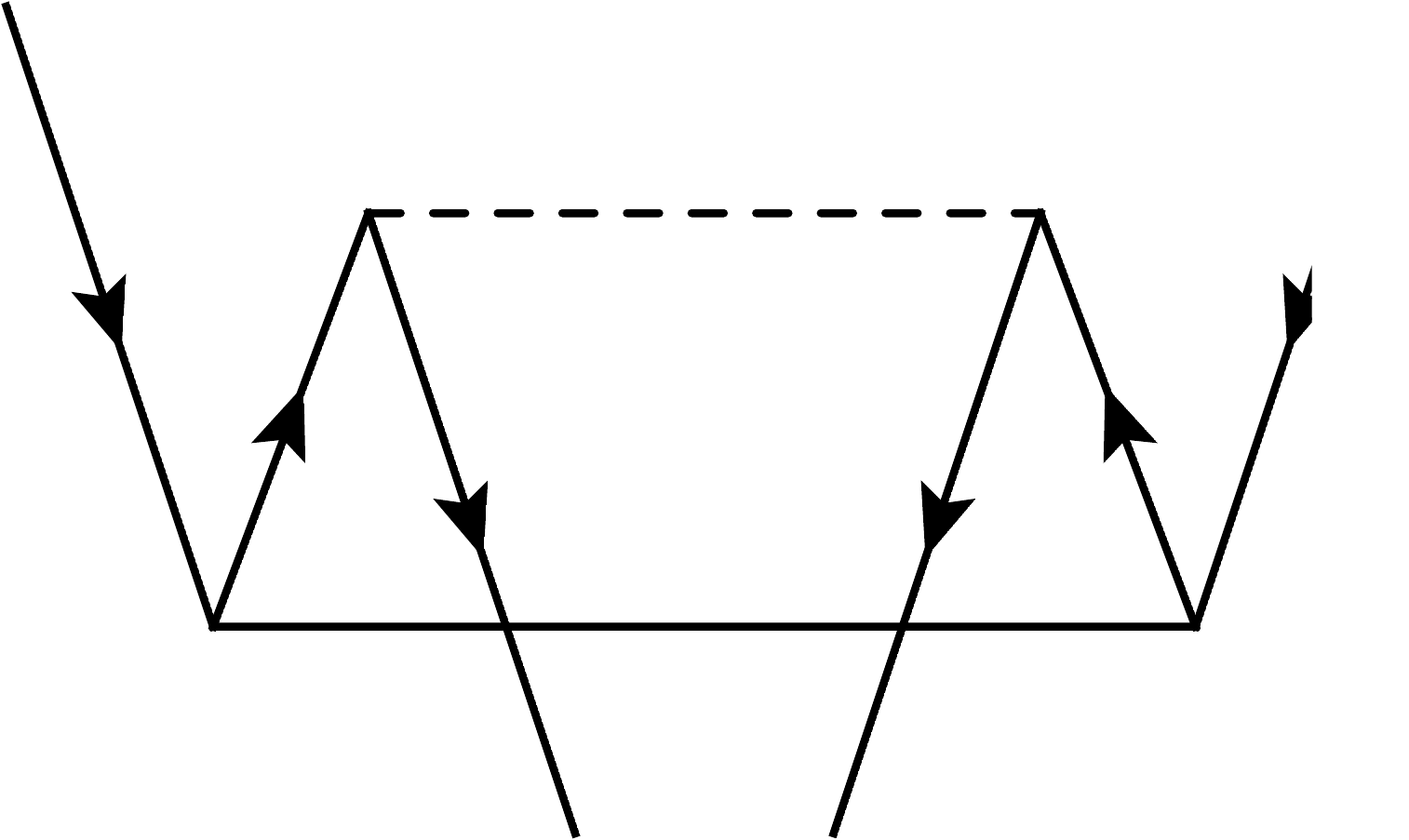}}  \\
\nonumber
  & + & {1\over 2} \:\: \parbox{30mm}{\includegraphics[scale=0.20]{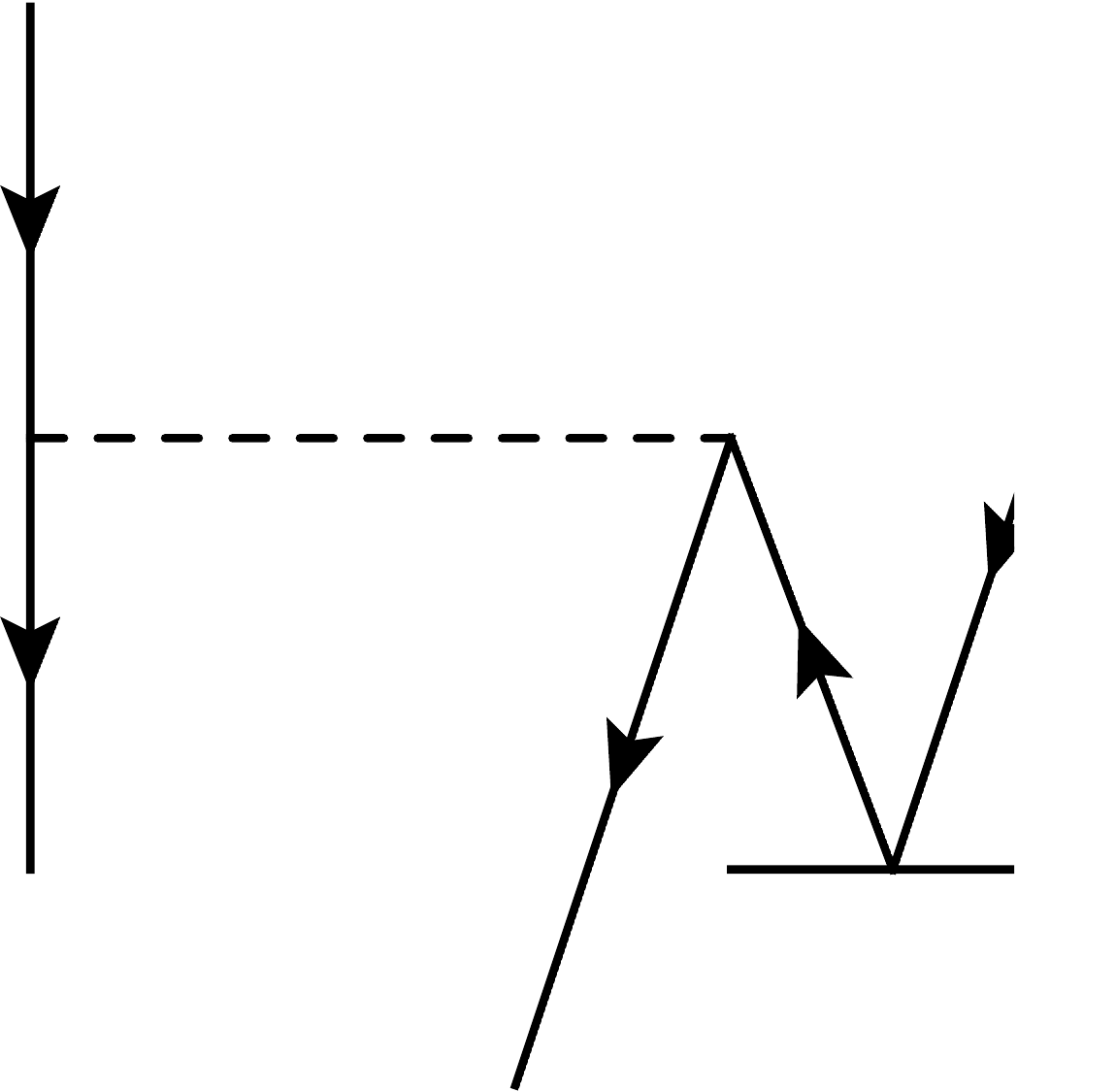}} 
  + {1\over 2}
  \:\: \parbox{30mm}{\includegraphics[scale=0.20]{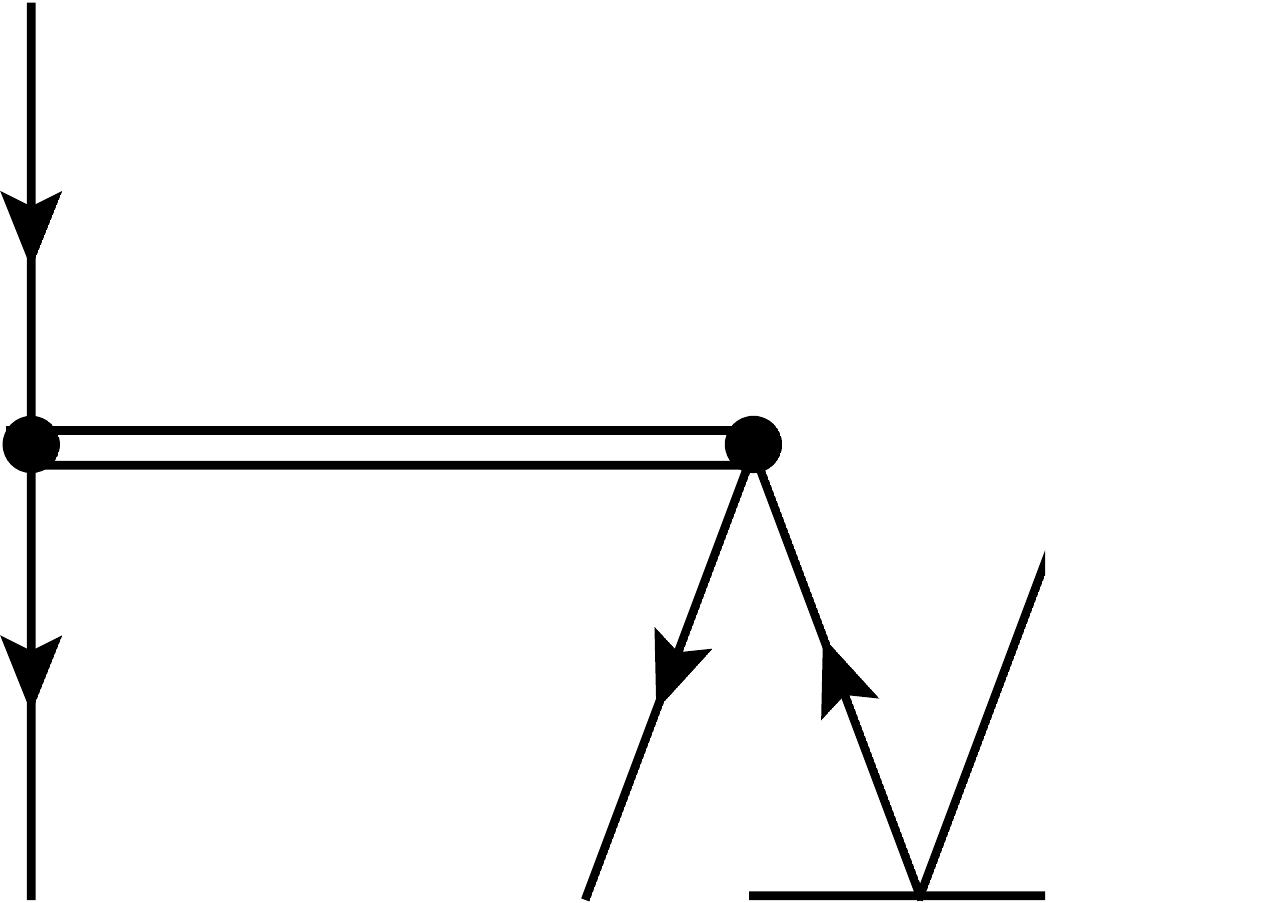}}  \\
\nonumber
 &= & \langle kl \vert\vert v^{0} \vert\vert ij \rangle +  {1\over 2} \langle kl
 \vert\vert v^{0} \vert\vert cd \rangle \langle cd
 \vert\vert t^{0} \vert\vert ij \rangle\\ 
& + & 
{1\over 2} P(ij)\langle kl
 \vert\vert v^{0} \vert\vert ic \rangle \langle c
 \vert\vert t^{0} \vert\vert j \rangle \\
\nonumber 
& + &  
{1\over 2} P(ij)\langle kl
 \vert\vert \chi^{0} \vert\vert ic \rangle \langle c
 \vert\vert t^{0} \vert\vert j \rangle.
  \label{eq:I9_eqns}
\end{eqnarray}

\begin{eqnarray}
  \nonumber
  \parbox{20mm}{\includegraphics[scale=0.20]{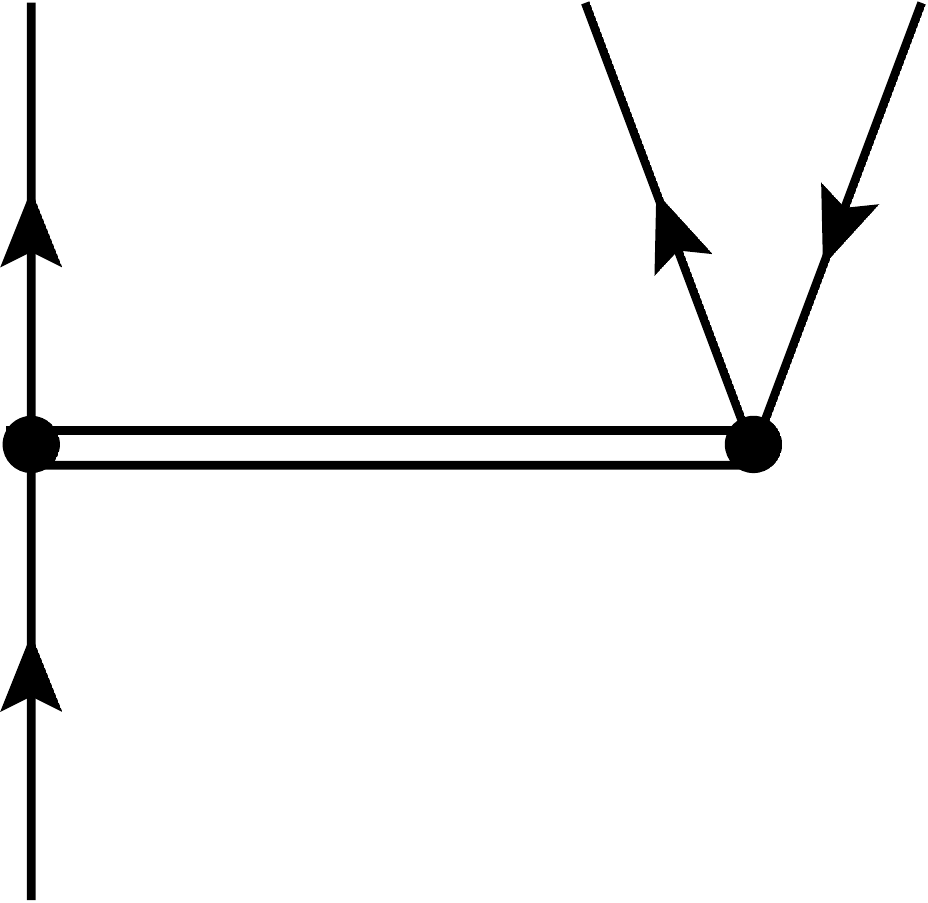}} &
  = &   
  \parbox{20mm}{\includegraphics[scale=0.20]{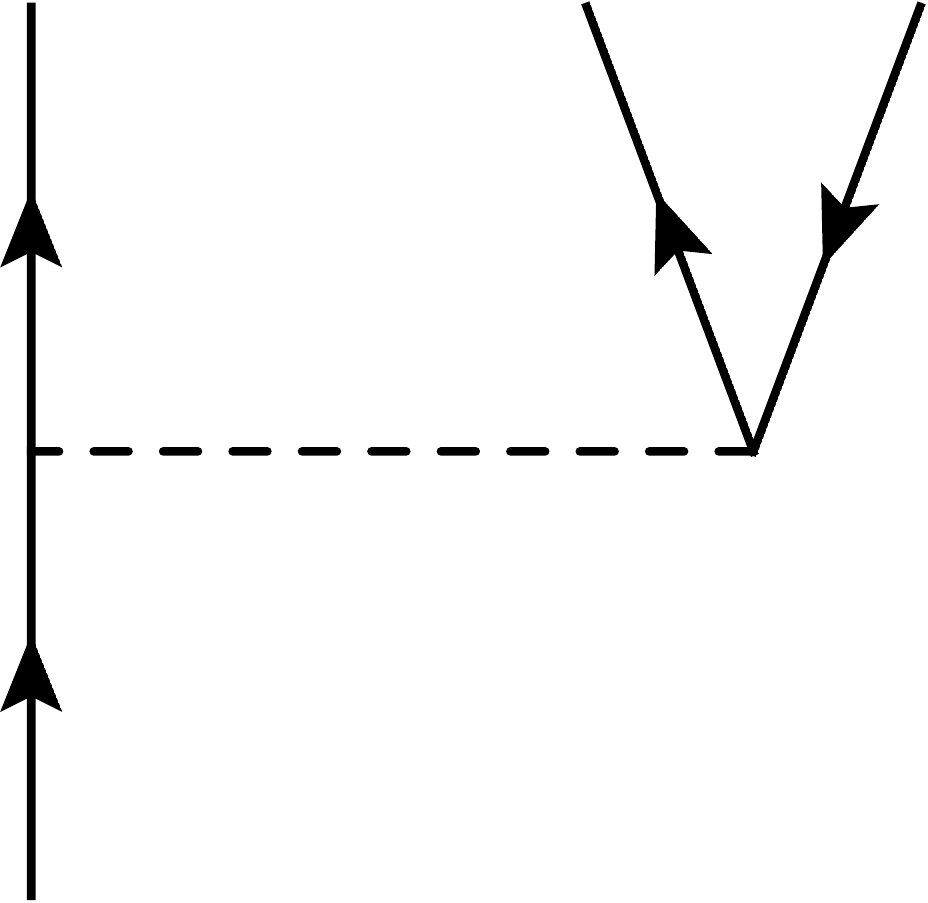}} 
  + \parbox{20mm}{\includegraphics[scale=0.20]{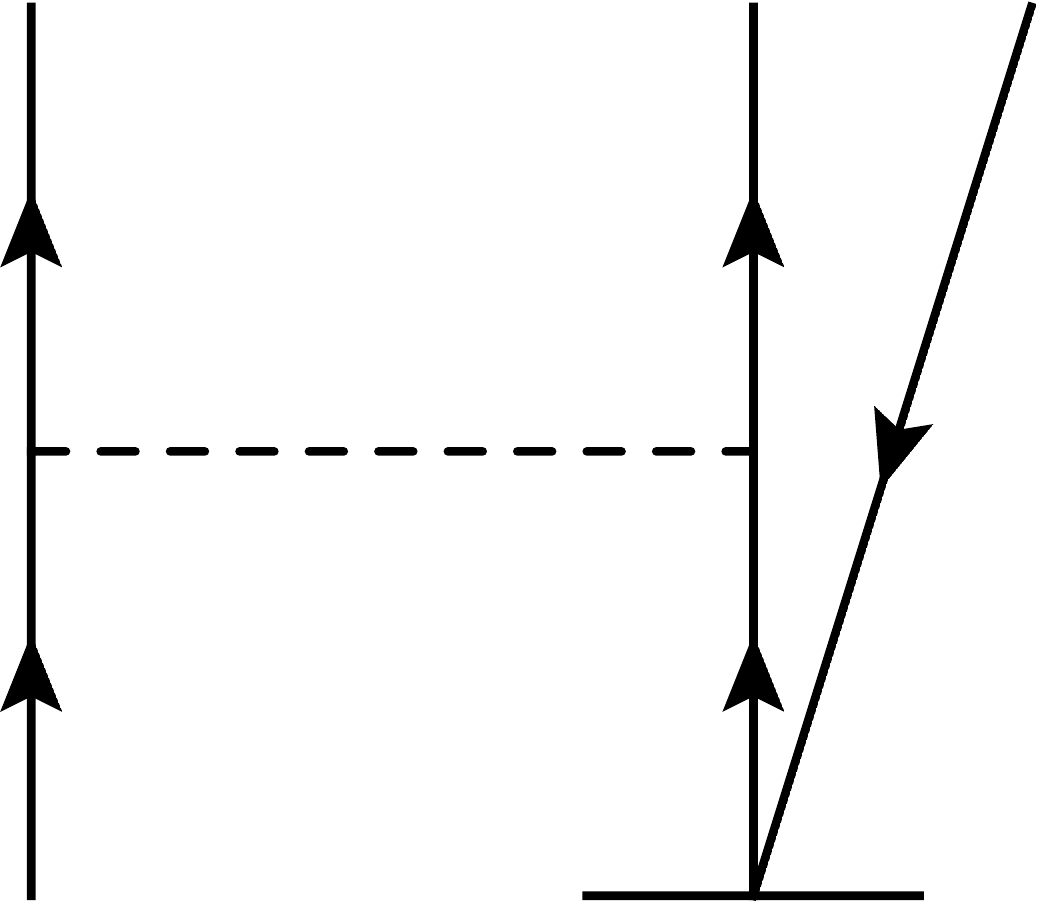}} \\
& = &  \langle ab \vert\vert v^{0} \vert\vert cj \rangle +  \langle ab
\vert\vert v^{0} \vert\vert cd \rangle  \langle d \vert\vert t^{0}
\vert\vert j \rangle.
  \label{eq:I10_eqns}
\end{eqnarray}

\begin{eqnarray}
  \nonumber
  \parbox{20mm}{\includegraphics[scale=0.20]{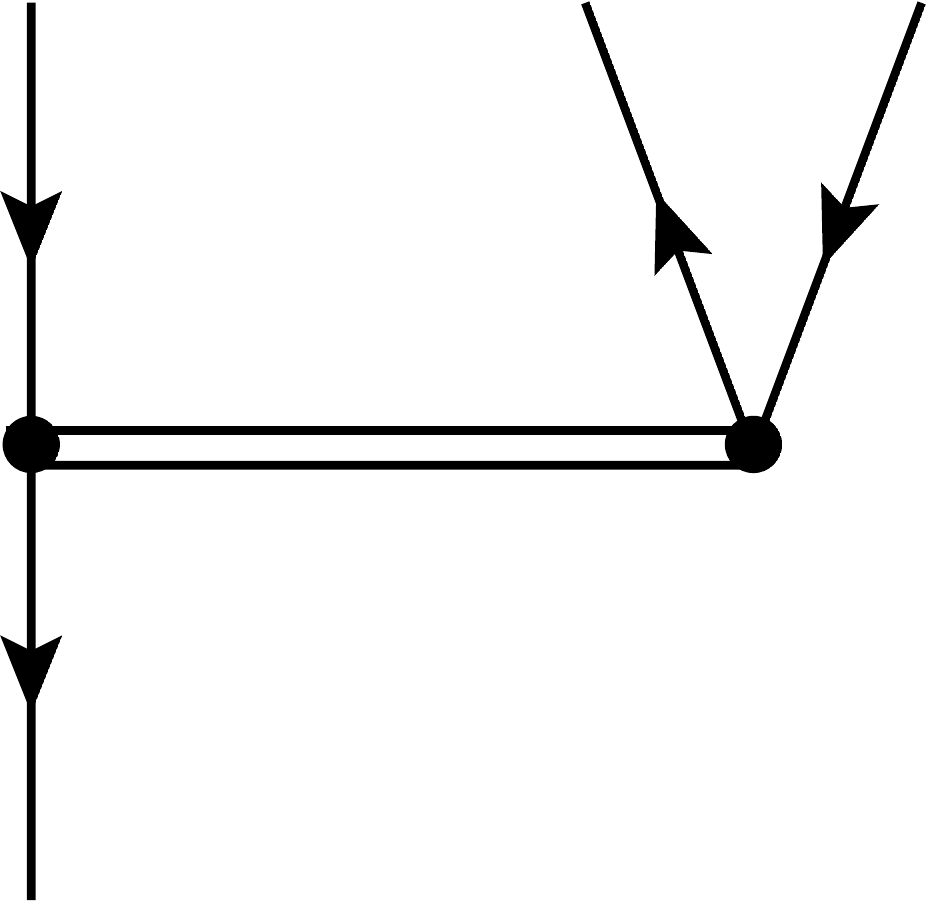}}  &
   = &  
  \parbox{30mm}{\includegraphics[scale=0.20]{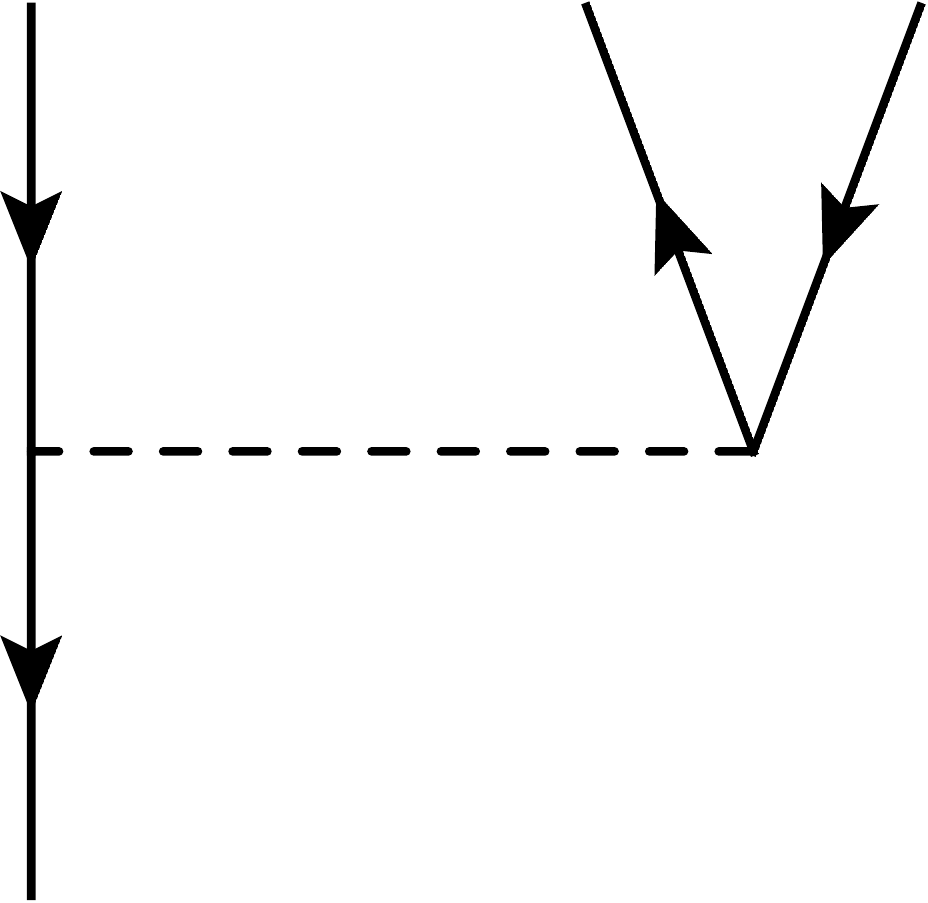}} 
  + \parbox{30mm}{\includegraphics[scale=0.20]{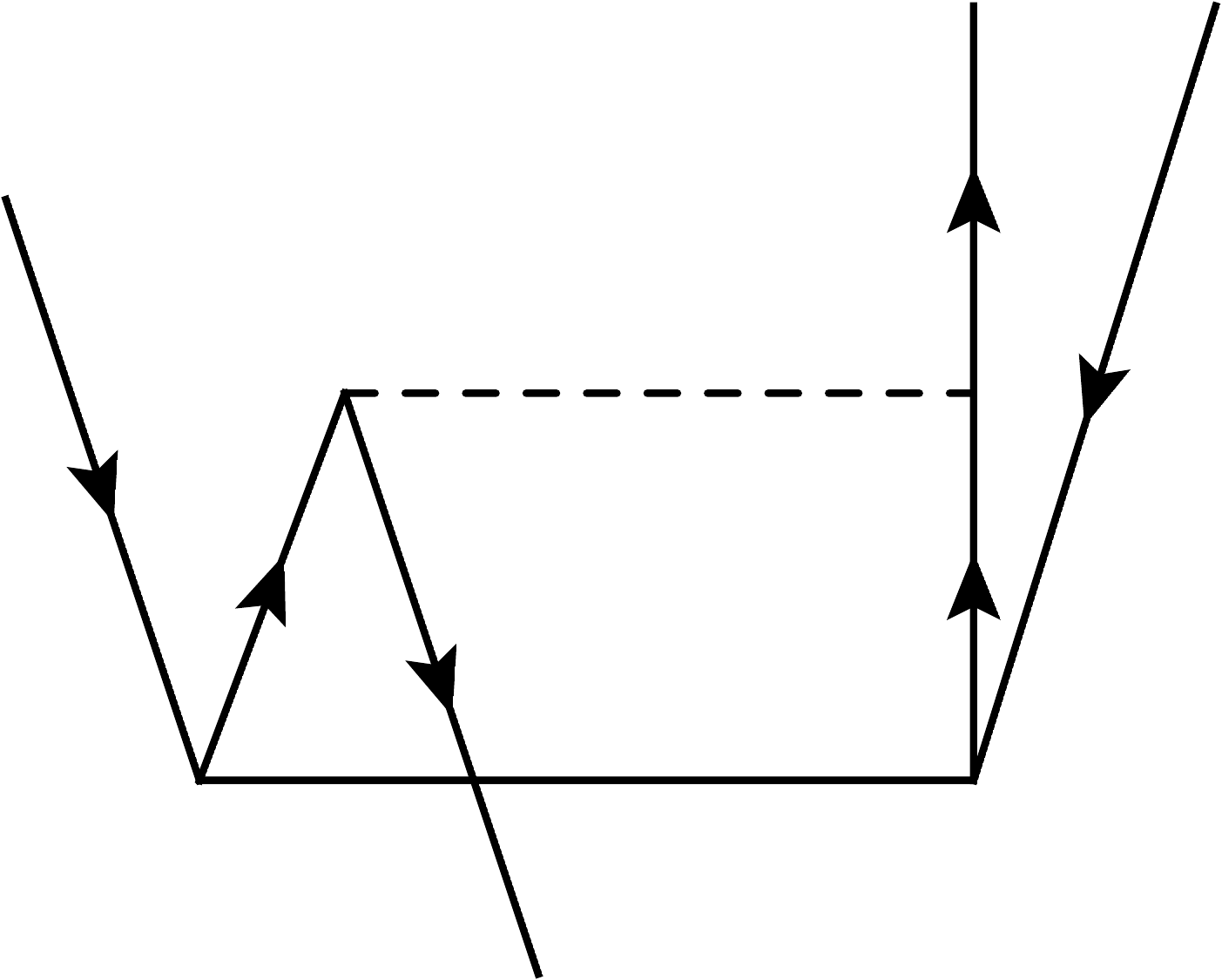}}
  + \parbox{30mm}{\includegraphics[scale=0.20]{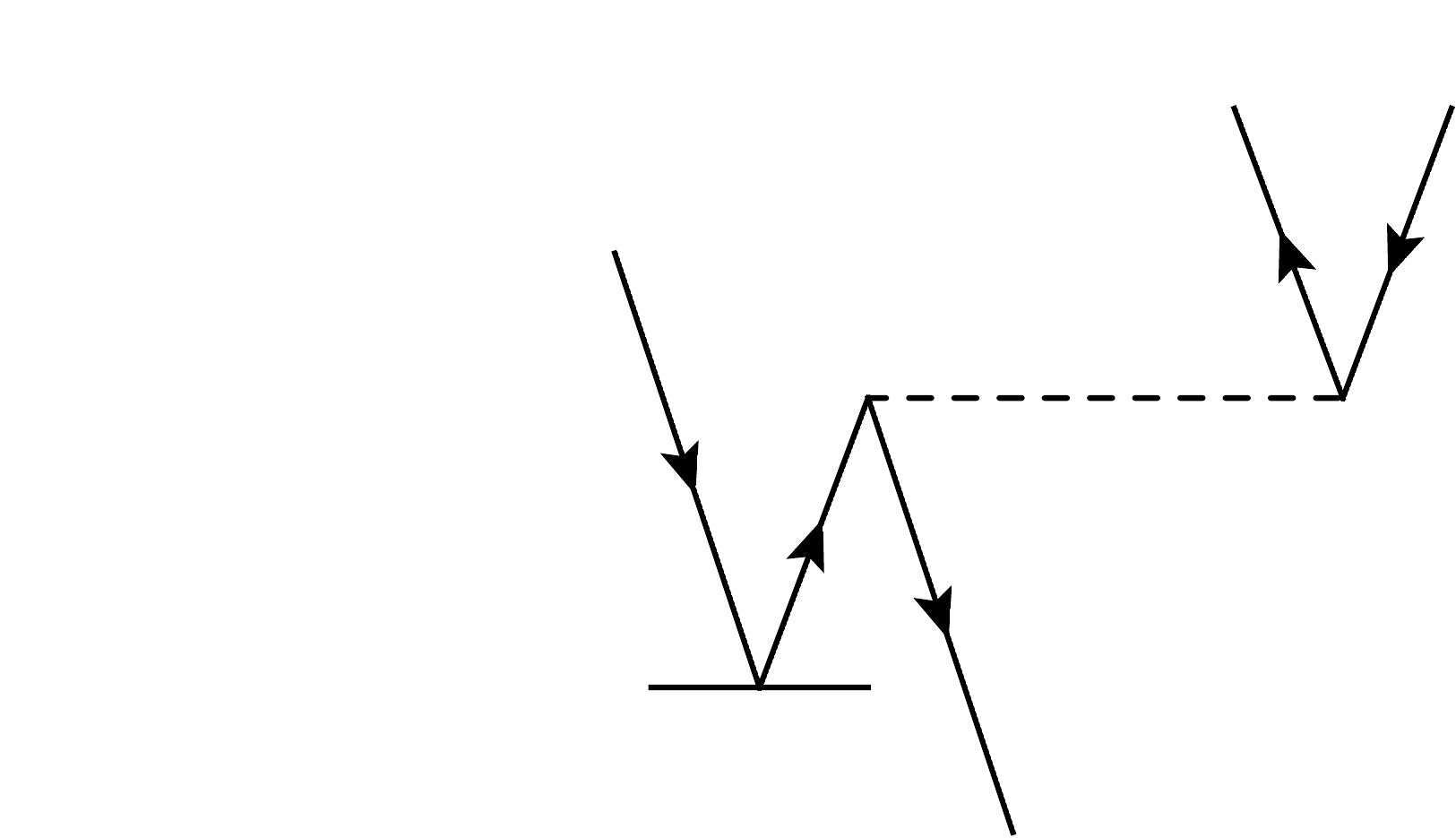}}  \\
\nonumber
  & + & \parbox{30mm}{\includegraphics[scale=0.20]{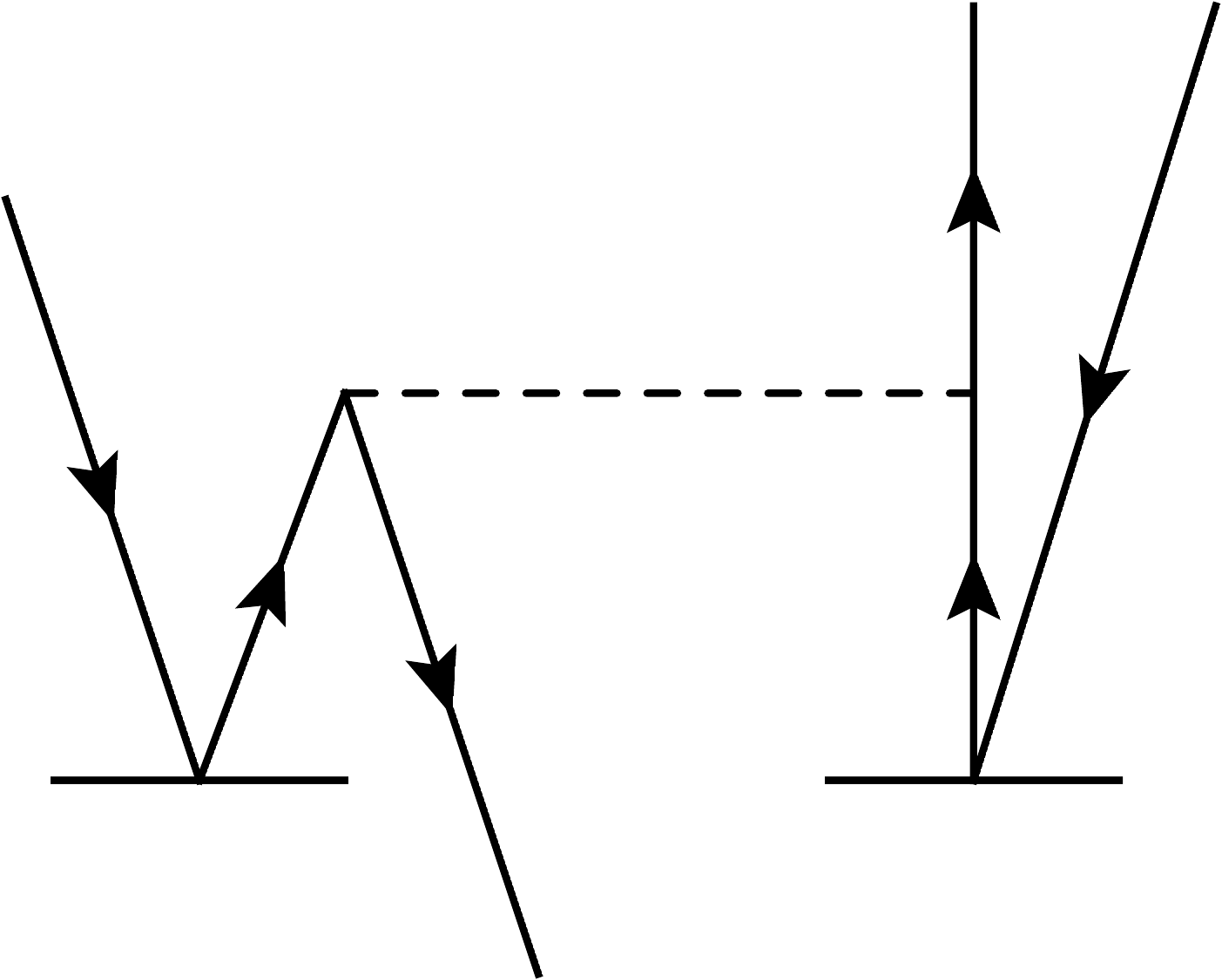}} 
  + \parbox{30mm}{\includegraphics[scale=0.20]{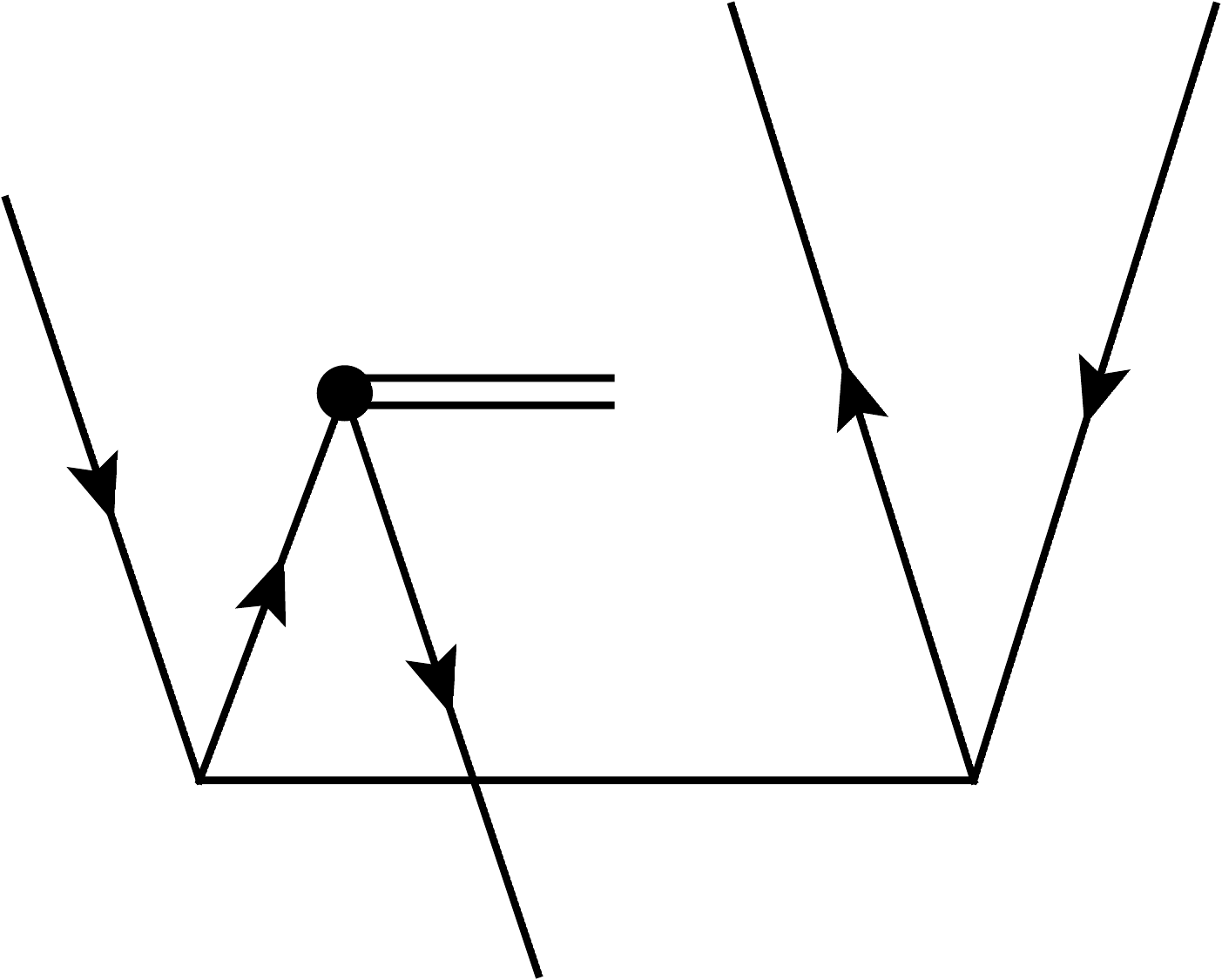}}  
 + {1\over 2}
 \:\: \parbox{30mm}{\includegraphics[scale=0.20]{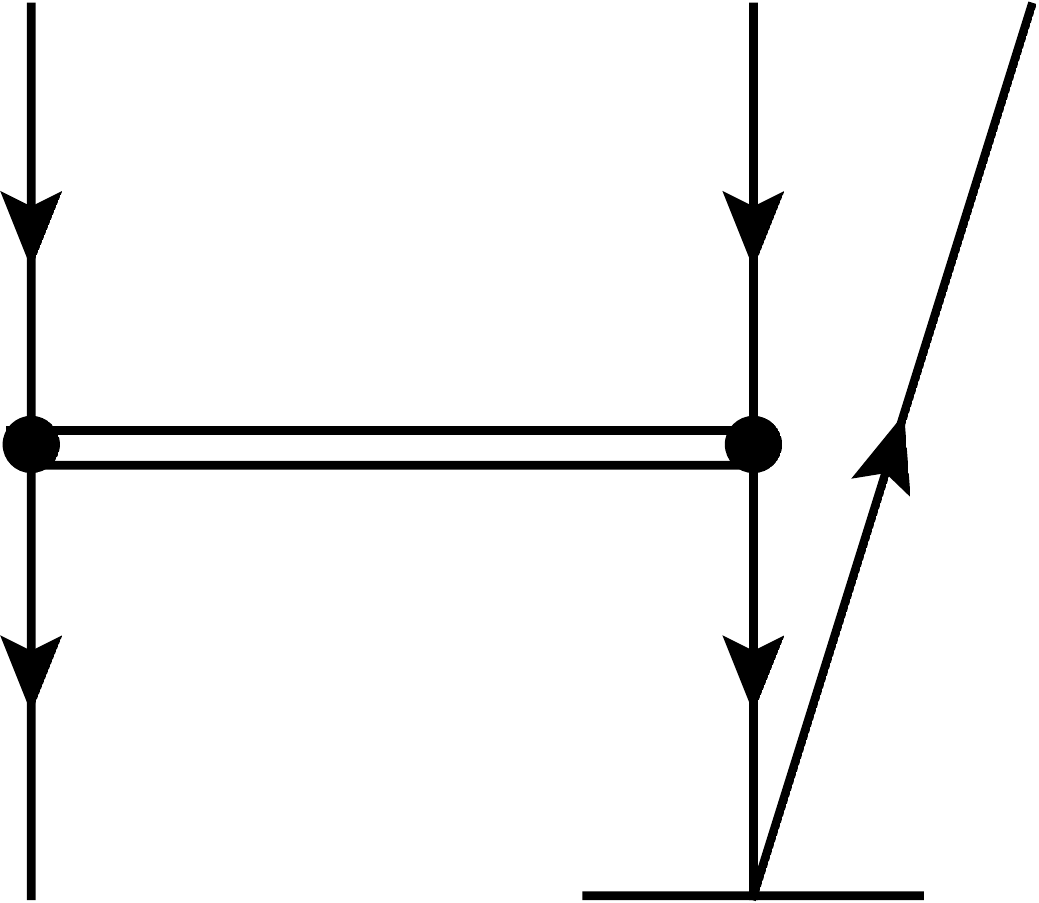}}
 \\
\nonumber
& = & \langle kb \vert\vert v^{0} \vert\vert ij \rangle +  {1\over 2} \langle kb
 \vert\vert v^{0} \vert\vert cd \rangle \langle cd
 \vert\vert t^{0} \vert\vert ij \rangle \\
\nonumber
& + &
 P(ij)\langle kb
 \vert\vert v^{0} \vert\vert cj \rangle \langle c
 \vert\vert t^{0} \vert\vert i \rangle \\
\nonumber
& - &  P(ij)\langle kl
 \vert\vert v^{0} \vert\vert cj \rangle \langle c
 \vert\vert t^{0} \vert\vert i \rangle \langle b
 \vert\vert t^{0} \vert\vert l \rangle \\
& + &  
 \langle cb \vert\vert t^{0} \vert\vert ij \rangle \langle k
 \vert\vert \chi^{0} \vert\vert c \rangle 
 +  {1\over 2} \langle kl \vert\vert \chi^{0} \vert\vert ij \rangle \langle b
 \vert\vert t^{0} \vert\vert l \rangle.
 \label{eq:I11_eqns}
\end{eqnarray}

\section{Angular-momentum-coupled equations of motion}
\label{app2}
In this Section we present the equations and diagrams for excited
states equation-of-motion (EOM), particle-attached EOM, and
particle-removed EOM coupled-cluster theory in the CCSD approximation
using an angular momentum coupled scheme. The EOM solutions results
from diagonalizing the similarity transformed Hamiltonian,
$\overline{H} = e^{-T}H_Ne^T$, in sub-space of $n$-particle-$m$-hole
excited reference states, 
\be
\label{this}
\left[ \overline{H},R_\mu \right] |\phi\ra = \omega_\mu R_\mu |\phi\ra \ .  
\ee 
Here, $R_\mu$ is the excitation operator acting on the reference state
$\vert\phi\ra$, and $\omega_\mu $ is the excitation energy with respect to
the CCSD ground-state energy.  The diagrammatic representation of the
various excitation amplitudes $R$, and their uncoupled and coupled
representations are given in Table~\ref{tab:r_amps},

\begin{table*}[htbp!]
  \begin{Table_app}{0.25}{0.33}{0.42}
    Diagram & \mathrm{Uncoupled~expression} & \mathrm{Coupled~expression} 
    \tabularnewline\hline
    \includegraphics[scale=0.2]{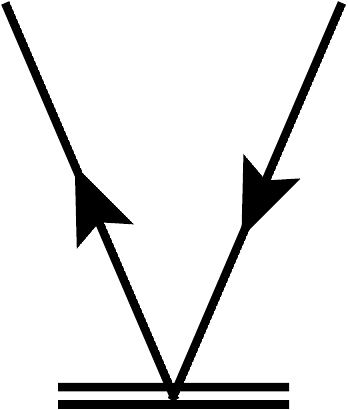} & 
    \langle a\vert r^{JM} \vert i \rangle   & 
    C_{j_i m_i JM}^{j_{a}m_{a}} \langle a\vert\vert r^{J} \vert\vert i \rangle  
    \tabularnewline\hline
    \includegraphics[scale=0.2]{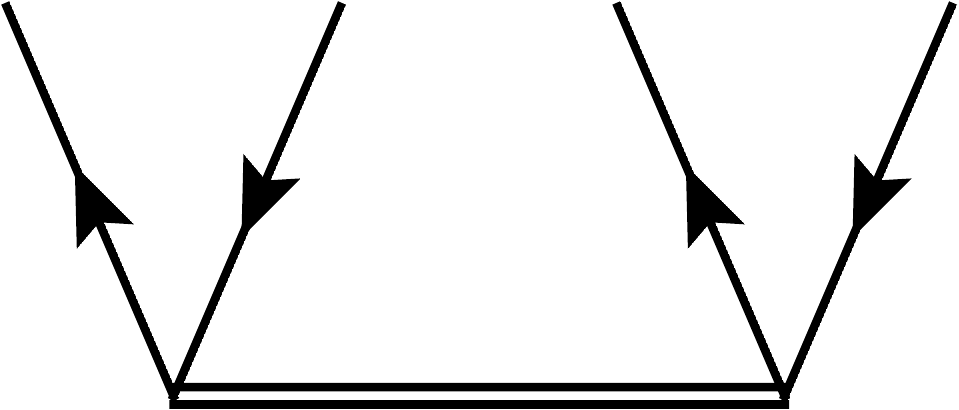} & \langle ab\vert r^{JM}
    \vert ij\rangle & C_{j_a m_a j_b m_b}^{J_{ab}M_{ab}} C_{j_i m_i
      j_j m_j}^{J_{ij}M_{ij}} C_{J_{ij} M_{ij} JM}^{J_{ab}M_{ab}}
    \times\langle ab\vert\vert r^{J} \vert\vert ij\rangle
    \tabularnewline\hline
    \includegraphics[scale=0.2]{R2p2h_diag} & \langle ab\vert r^{JM}
    \vert ij\rangle & -(-1)^{j_i-m_i}(-1)^{j_b-m_b}C_{j_a m_a j_i
      -m_i}^{J_{ai}M_{ai}}\times C_{j_j m_j j_b -m_b}^{J_{bj}M_{bj}}
    C_{J_{bj} M_{bj} JM}^{J_{ai}M_{ai}} \times\langle
    ai^{-1}\vert\vert r^{J} \vert\vert jb^{-1}\rangle
    \tabularnewline\hline
    \includegraphics[scale=0.2]{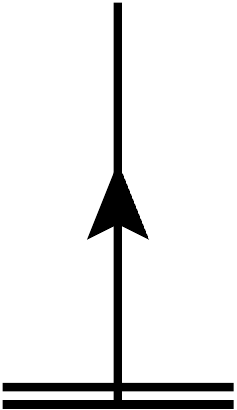} & 
    \langle a\vert r^{JM} \vert \rangle   & 
    \langle a\vert\vert r^{J} \vert\vert \rangle\delta_{J,j_{a}} 
    \tabularnewline\hline
    \includegraphics[scale=0.2]{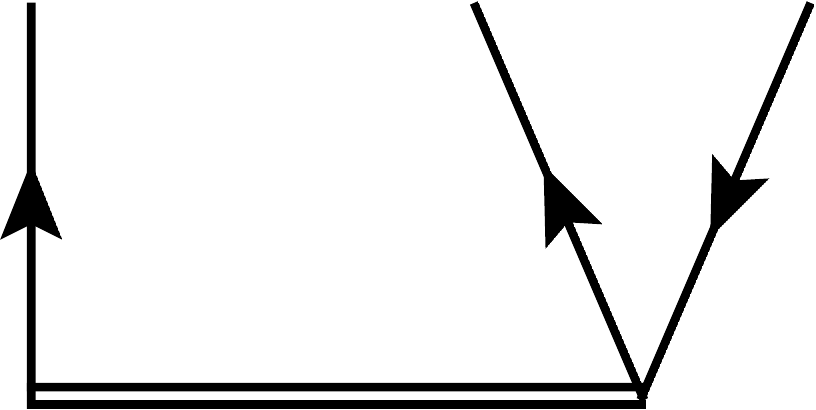} & 
    \langle ab\vert r^{JM} \vert j\rangle  &
    C_{j_a m_a j_b m_b}^{J_{ab}M_{ab}} C_{J M j_j m_j}^{J_{ab}M_{ab}} 
    \langle ab\vert\vert r^{J} \vert\vert j\rangle 
    \tabularnewline\hline
    \includegraphics[scale=0.2]{R1h2p_diag} & \langle ab\vert r^{JM}
    \vert j\rangle & - (-1)^{J -M}(-1)^{j_b-m_b} \times C_{j_a m_a J
      -M}^{J_{ai}M_{ai}} C_{j_j m_j j_b -m_b}^{J_{ai}M_{ai}} \langle a
    \vert\vert r^{J} \vert\vert jb^{-1}\rangle \tabularnewline\hline
    \includegraphics[scale=0.2]{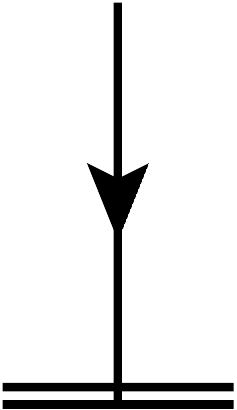} & 
    \langle \vert r^{JM} \vert i \rangle   & 
    \langle \vert\vert r^{J} \vert\vert i \rangle\delta_{J,j_{i}}  
    \tabularnewline\hline
    \includegraphics[scale=0.2]{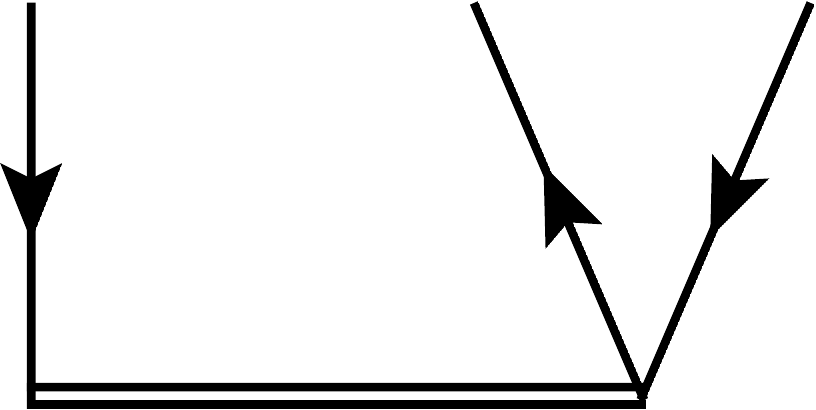} & 
    \langle b \vert r^{JM} \vert ij\rangle  &
    C_{j_i m_i j_j m_j}^{J_{ij}M_{ij}} C_{j_b m_b JM}^{J_{ij}M_{ij}} 
    \langle b \vert\vert r^{J} \vert\vert ij\rangle  
    \tabularnewline\hline
    \includegraphics[scale=0.2]{R2h1p_diag} & \langle b \vert r^{JM}
    \vert ij\rangle & -(-1)^{j_b-m_b}(-1)^{j_i-m_i} C_{j_j m_j j_b
      -m_b}^{J_{bj}M_{bj}} \times C_{J M j_i-m_i}^{J_{bj}M_{bj}}
    \langle i^{-1} \vert\vert r^{J} \vert\vert jb^{-1}\rangle 
    \tabularnewline\hline
  \end{Table_app}
  \caption{The diagrams representing the one-particle-one-hole $R(1p$-$1h)$, the two-particle-two-hole 
    $R(2p$-$2h)$, the one-particle $R(1p)$, the two-particle-one-hole $R(2p$-$1h)$, the one-hole $R(1h)$, and 
    finally the one-particle-two-hole $R(1h$-$2p)$ excitation amplitudes, together with their
    uncoupled ($m$-scheme) and coupled ($j$-scheme) algebraic
    expressions. Note, that the coupling order of the one-hole
    excitation operator is $\langle \vert r^{JM} \vert i \rangle =
    C_{0 0 J M}^{j_i m_i} \langle \vert\vert r^{J} \vert\vert i
    \rangle $. Repeated indices are summed over. }
  \label{tab:r_amps}
\end{table*}
In Table~\ref{tab:r_amps} we have given both the normal-coupled and
cross-coupled representations of the excitation amplitude $R_2$. Note
that the coupling order in the reduced matrix elements of the
two-particle-one-hole and one-particle-two-hole amplitudes follow a
different coupling order than for the two-particle-two-hole amplitude.

In
Eqs. ~\ref{eq:pandya3},\ref{eq:pandya4},\ref{eq:pandya5},\ref{eq:pandya6},\ref{eq:pandya7}
and \ref{eq:pandya8} we give the recouplings of the
two-particle-two-hole, two-particle-one-hole and one-particle-two-hole
amplitudes in terms of normal-coupled and cross-coupled matrix elements. Note
that the coupling order in the bra and ket is always from left to
right, 

\begin{eqnarray}
  \nonumber
  \langle ab \vert\vert R^{J} \vert\vert ij \rangle = \\
  - \sum_{J_{ai},J_{bj}}(-1)^{j_j+j_b-J_{bj}}\hat{J}_{bj}\hat{J}_{ij}\hat{J}_{ai}^2
  \left\{\begin{array}{ccc}
      J & J_{ab} & J_{ij} \\
      J_{ai} & j_a & j_i \\
      J_{bj} & j_b & j_j 
    \end{array}\right\} 
  \langle ai^{-1} \vert\vert R^{J} \vert\vert jb^{-1}\rangle,
  \label{eq:pandya3}
  \\  
  \nonumber
  \langle ai^{-1} \vert\vert R^{J} \vert\vert jb^{-1}\rangle = \\ 
  - \sum_{J_{ab},J_{ij}}(-1)^{j_j+j_b-J_{bj}}\hat{J}_{bj}\hat{J}_{ij}\hat{J}_{ab}^2 
  \left\{\begin{array}{ccc}
      J & J_{ai} & J_{bj} \\
      J_{ab} & j_a & j_b \\
      J_{ij} & j_i & j_j 
    \end{array}\right\} 
  \langle ab \vert\vert R^{J} \vert\vert ij \rangle,
  \label{eq:pandya4} 
  \\
  \label{eq:pandya5}
  \langle ab \vert\vert r^{J} \vert\vert j \rangle = 
  - \sum_{J_{bj}} (-1)^{J+j_j+J_{ab}}\hat{J}_{bj}^2
  \left\{\begin{array}{ccc}
      j_a & j_b & J_{ab} \\
      j_j & J & J_{bj} 
    \end{array}\right\} 
  \langle a \vert\vert r^{J} \vert\vert jb^{-1}\rangle, \\
  \label{eq:pandya6}
  \langle a \vert\vert r^{J} \vert\vert jb^{-1}\rangle =
  -\sum_{J_{ab}}(-1)^{J+j_j+J_{ab}}\hat{J}_{ab}^2
  \left\{\begin{array}{ccc}
      j_a & j_b & J_{ab} \\
      j_j & J & J_{bj} 
    \end{array}\right\}  
  \langle ab \vert\vert r^{J} \vert\vert j \rangle, \\
  \nonumber  
  \langle b \vert\vert r^{J} \vert\vert ij \rangle = 
  - \sum_{J_{bj}}(-1)^{j_i+j_j+J_{ij}}\hat{J}_{bj}^2
  \left\{\begin{array}{ccc}
      J & j_b & J_{ij} \\
      j_j & j_i & J_{bj} 
    \end{array}\right\}  
  \langle i^{-1} \vert\vert r^{J} \vert\vert jb^{-1}\rangle,
  \label{eq:pandya7}
  \\
  \langle i^{-1} \vert\vert r^{J} \vert\vert jb^{-1}\rangle =
  - \sum_{J_{ij}}(-1)^{j_i+j_j+J_{ij}}\hat{J}_{ij}^2
  \left\{\begin{array}{ccc}
      J & j_b & J_{ij} \\
      j_j & j_i & J_{bj} 
    \end{array}\right\} 
  \langle b \vert\vert r^{J} \vert\vert ij \rangle.
  \label{eq:pandya8}
\end{eqnarray}

\subsection{Excited states equation of motion}
Below we give the diagrammatic representation and algebraic
expressions in an angular momentum coupled scheme for the
coupled-cluster Equation-of-Motion (EOM) method in the
singles-and-doubles approximation. The EOM-CCSD method results from
diagonalizing the similarity transformed Hamiltonian in a sub-space of
one-particle-one-hole and two-particle-two-hole excitations. This
approximation has been shown to work particularly well for low-lying
states that are dominated by one-particle-one-hole excitations.  The
diagrammatic representation of the left-hand side of Eq.~\ref{this}
gives the one-particle-one-hole excitation amplitude

\begin{eqnarray}
  \nonumber
  \parbox{20mm}{\includegraphics[scale=0.20]{R1p1h_diag}} & 
  = &  
  \parbox{30mm}{\includegraphics[scale=0.20]{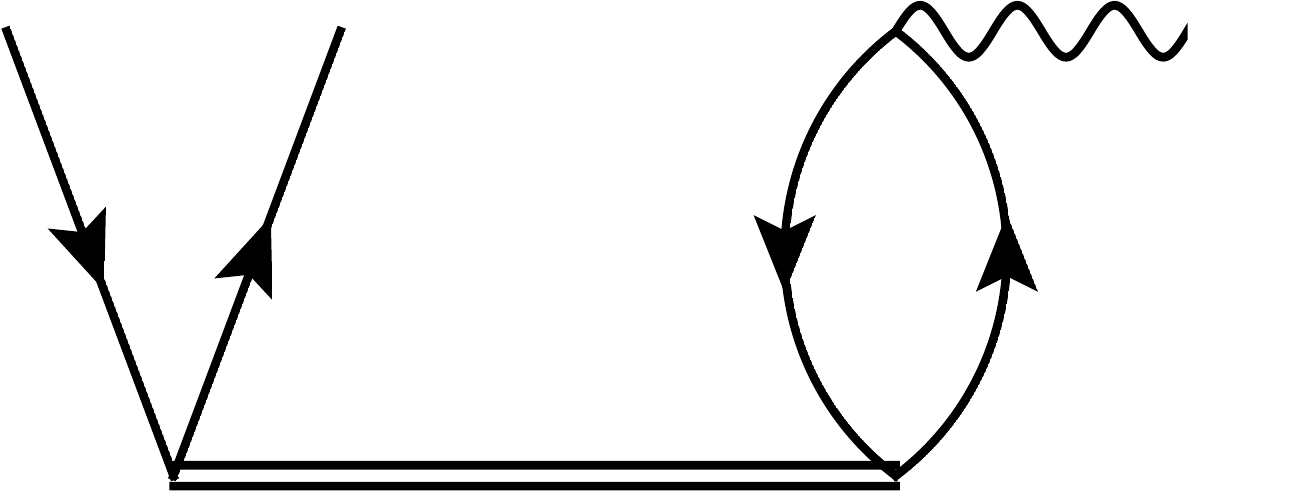}} 
  + \parbox{30mm}{\includegraphics[scale=0.20]{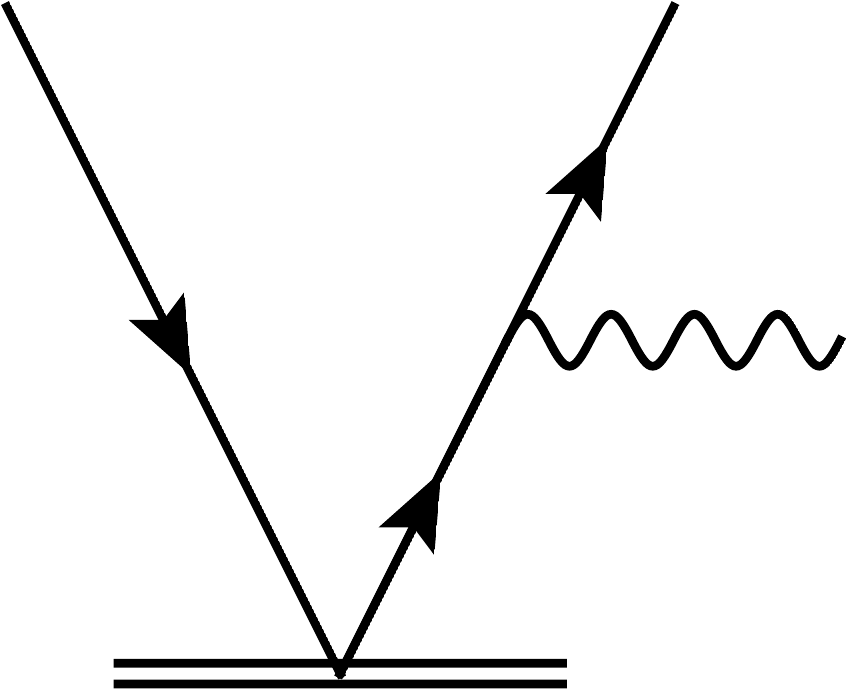}}
  + \parbox{30mm}{\includegraphics[scale=0.20]{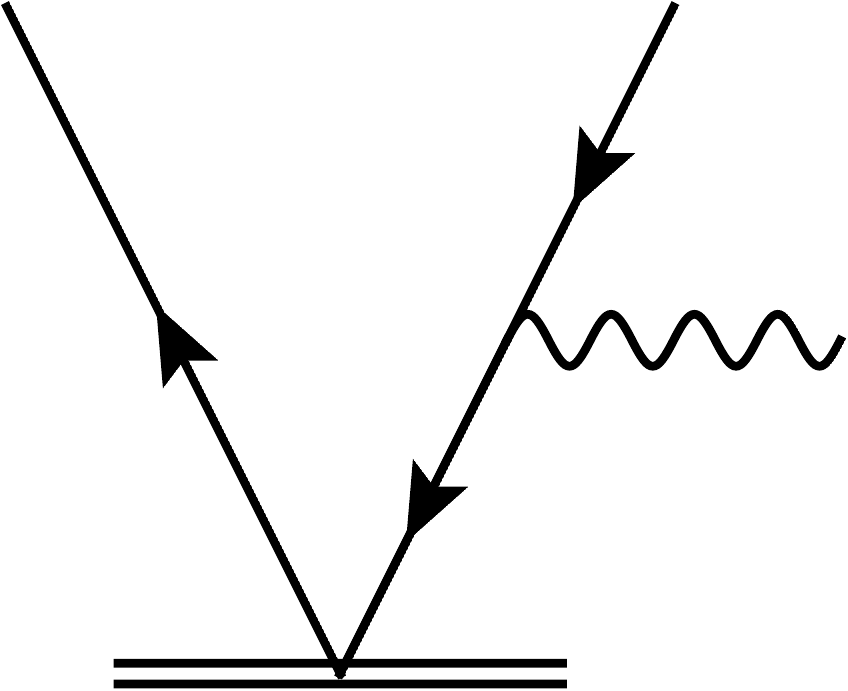}}\\
  &  + & \parbox{30mm}{\includegraphics[scale=0.20]{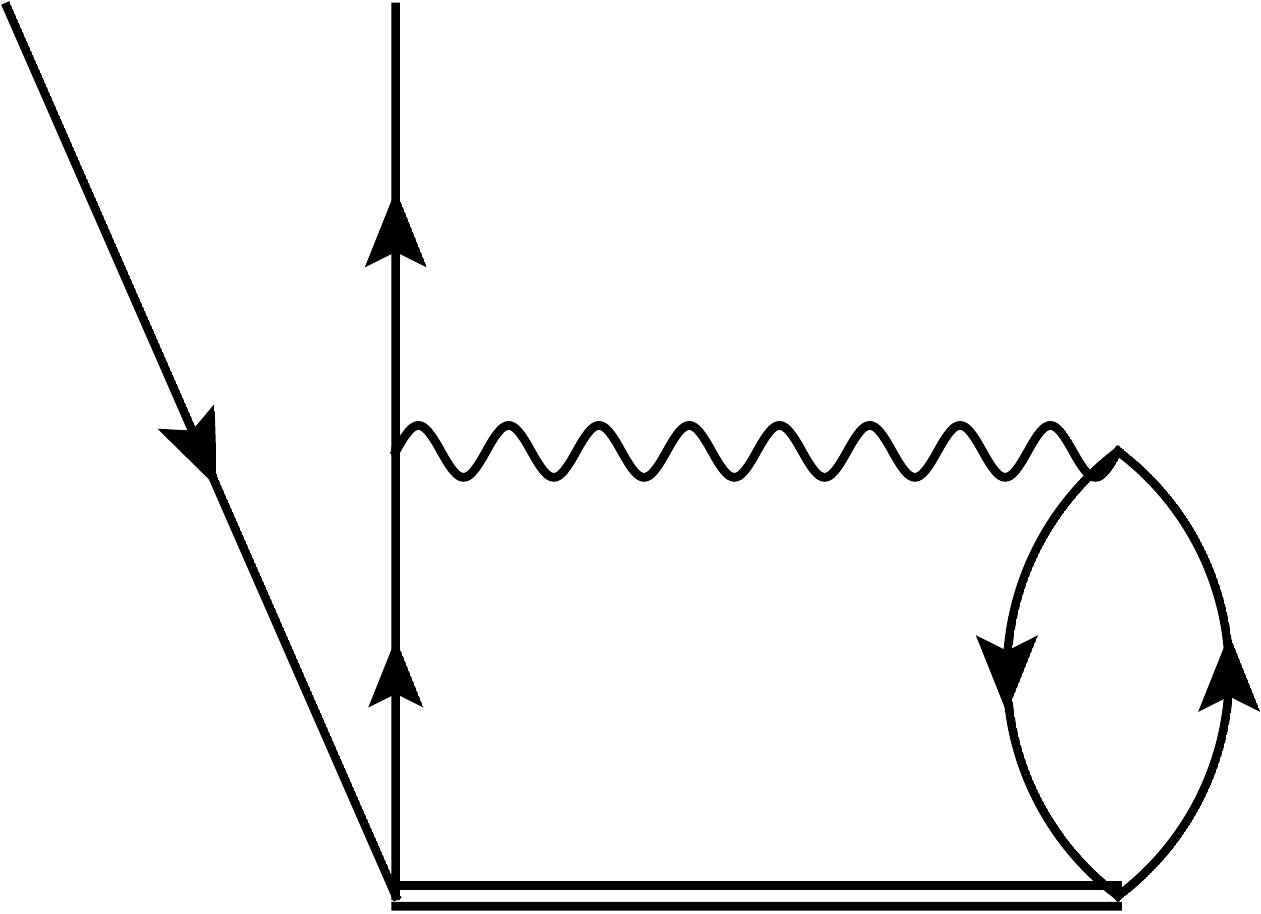}}
  + \parbox{30mm}{\includegraphics[scale=0.20]{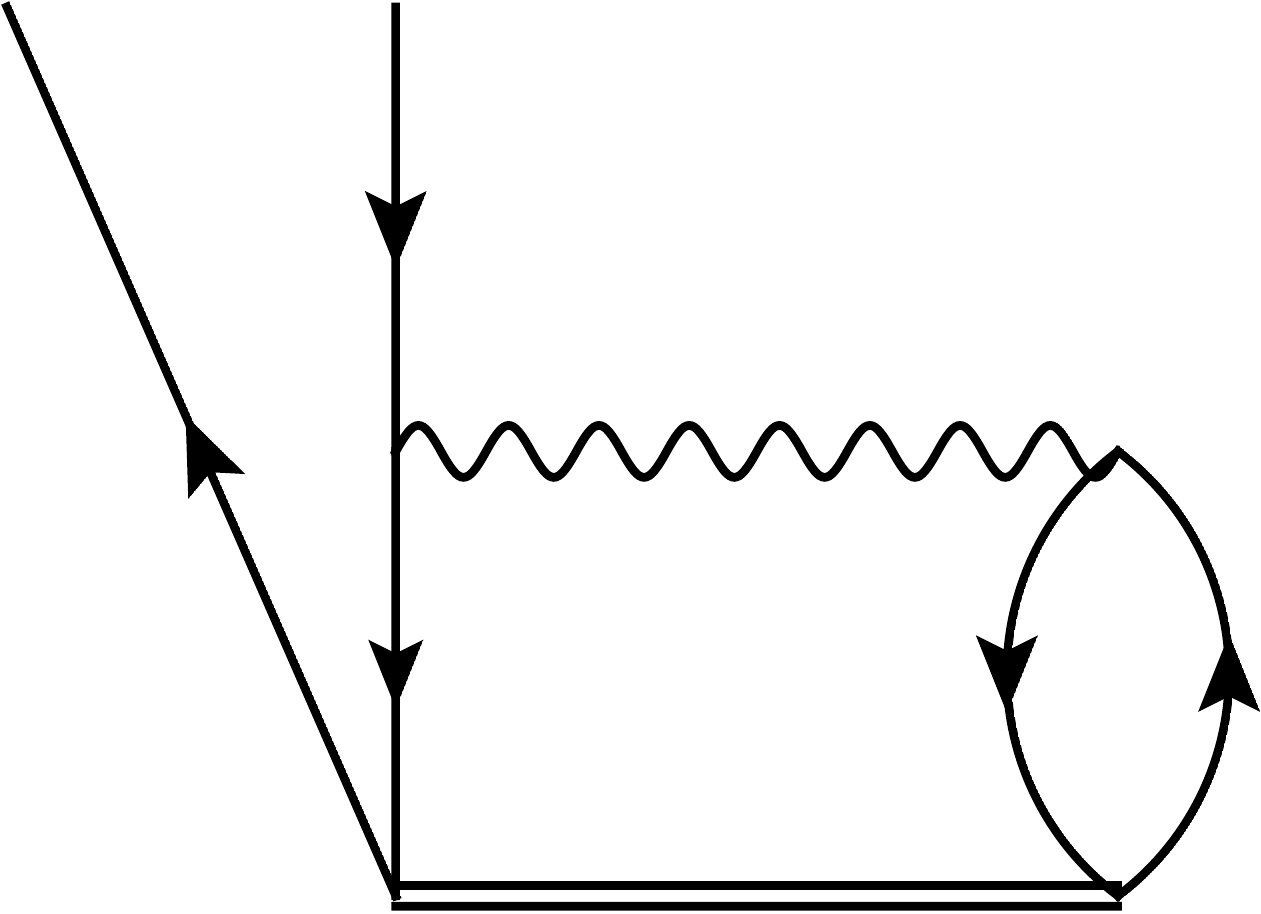}} 
  + \parbox{30mm}{\includegraphics[scale=0.20]{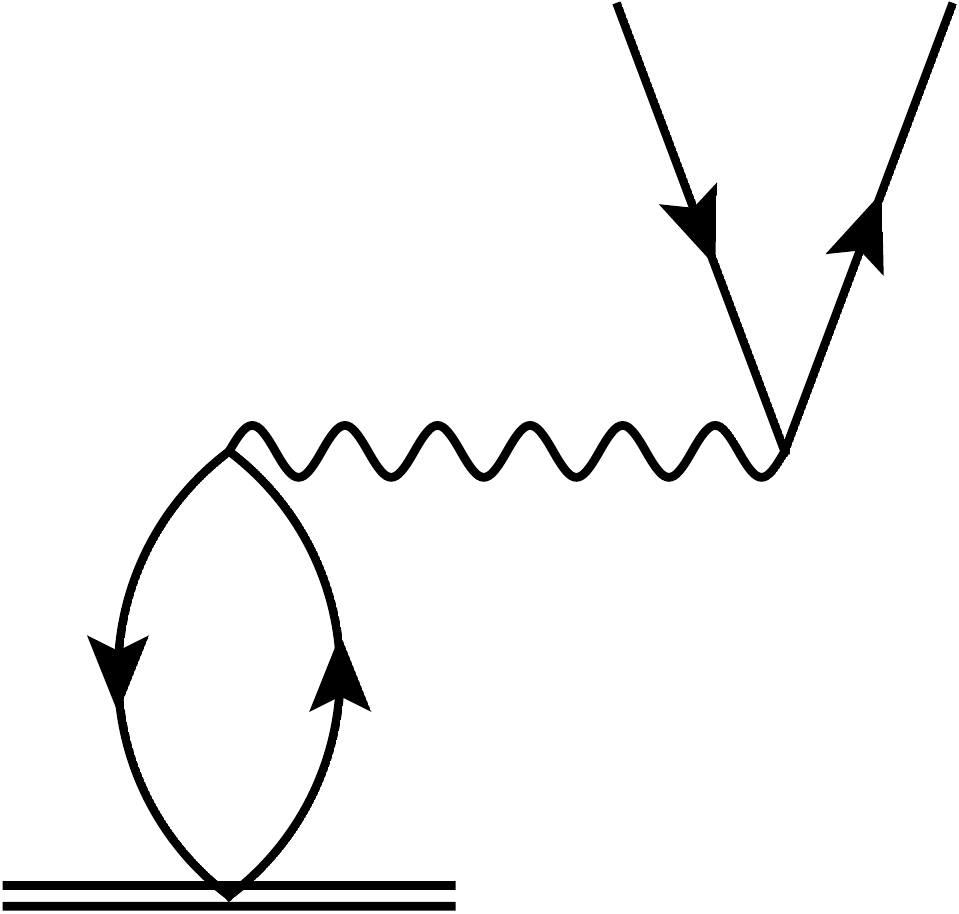}} .
  \label{eq:r1h1p_eqns}
\end{eqnarray}
Here the wavy lines represent the vertices of the similarity
transformed Hamiltonian. The corresponding diagrammatic representation
of the two-particle-two-hole excitation amplitudes are
\begin{eqnarray}
  \nonumber
  \parbox{20mm}{\includegraphics[scale=0.20]{R2p2h_diag}}  & 
  = &
  \parbox{30mm}{\includegraphics[scale=0.20]{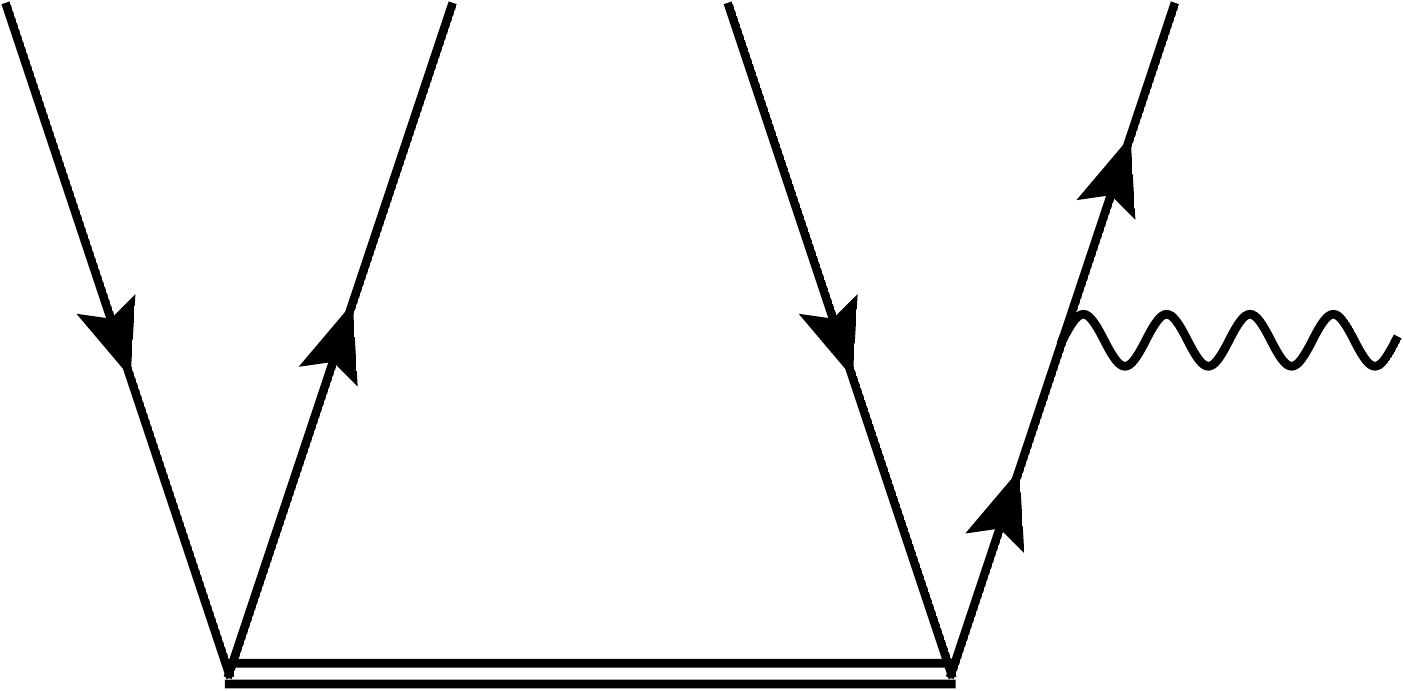}} 
  + \parbox{30mm}{\includegraphics[scale=0.20]{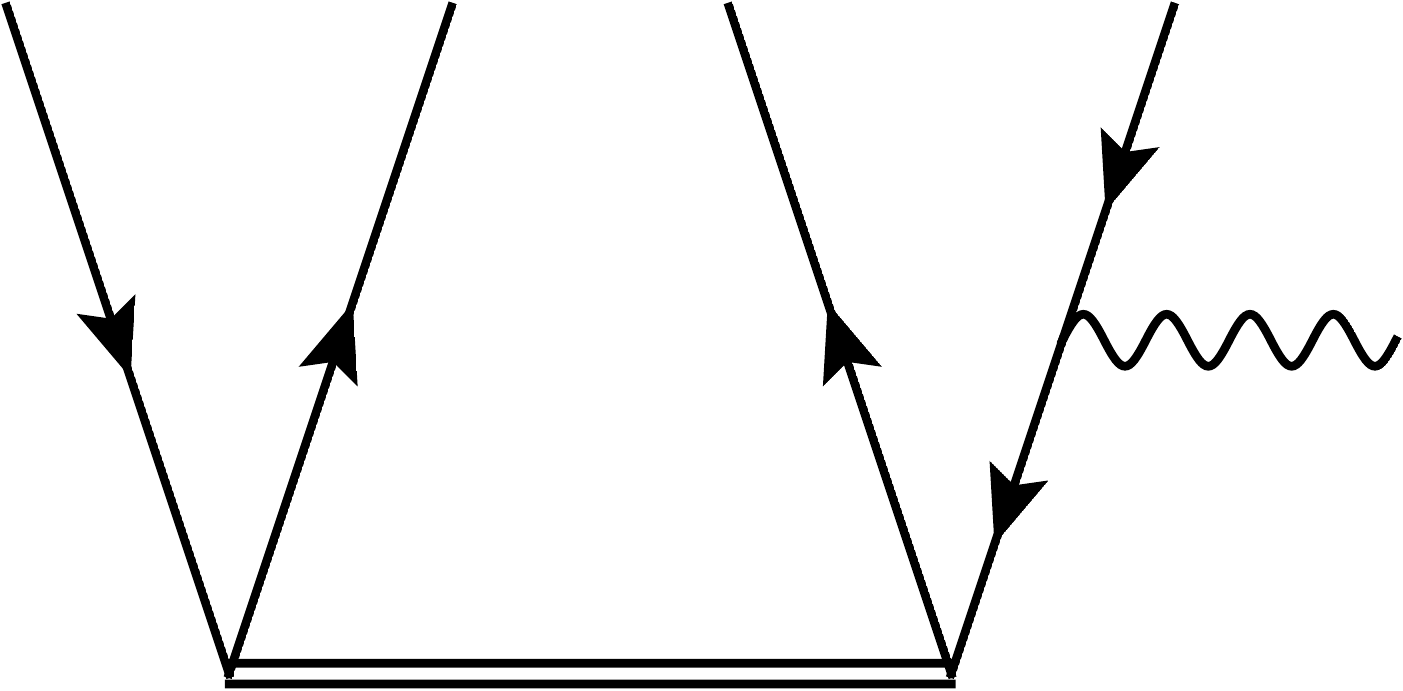}}\\
  \nonumber
  &  + & \parbox{30mm}{\includegraphics[scale=0.20]{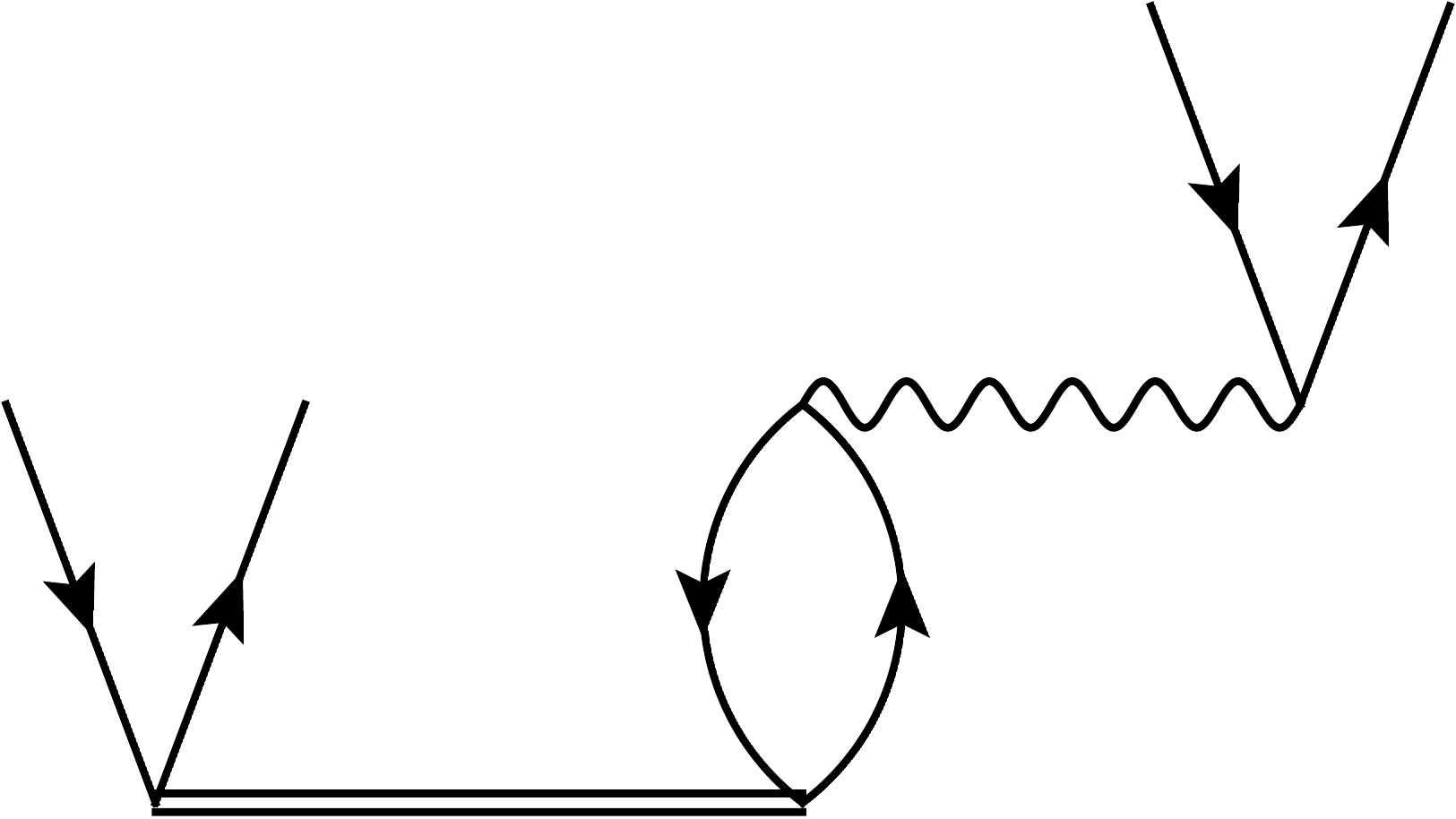}}
  + \parbox{30mm}{\includegraphics[scale=0.20]{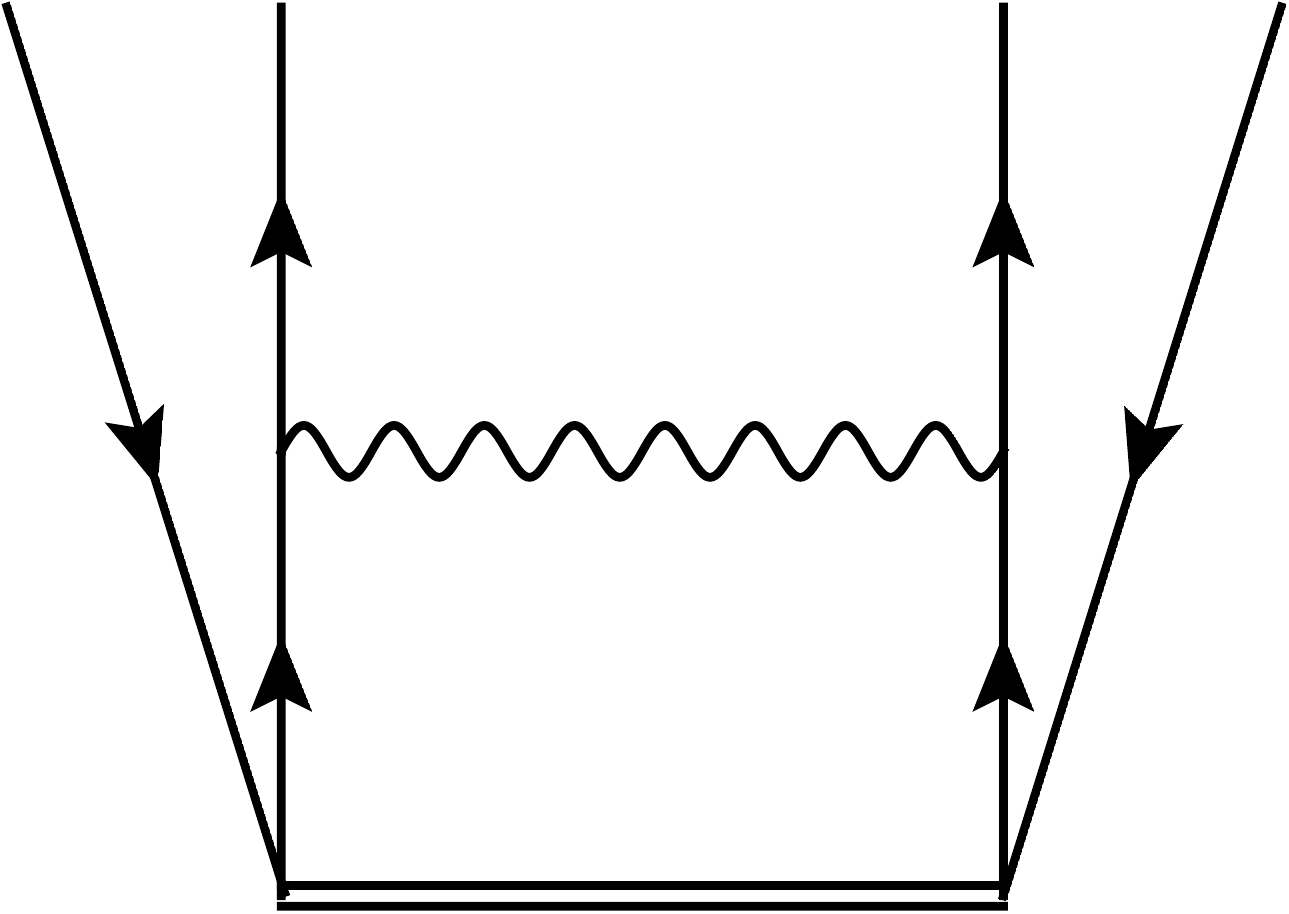}}
  + \parbox{30mm}{\includegraphics[scale=0.20]{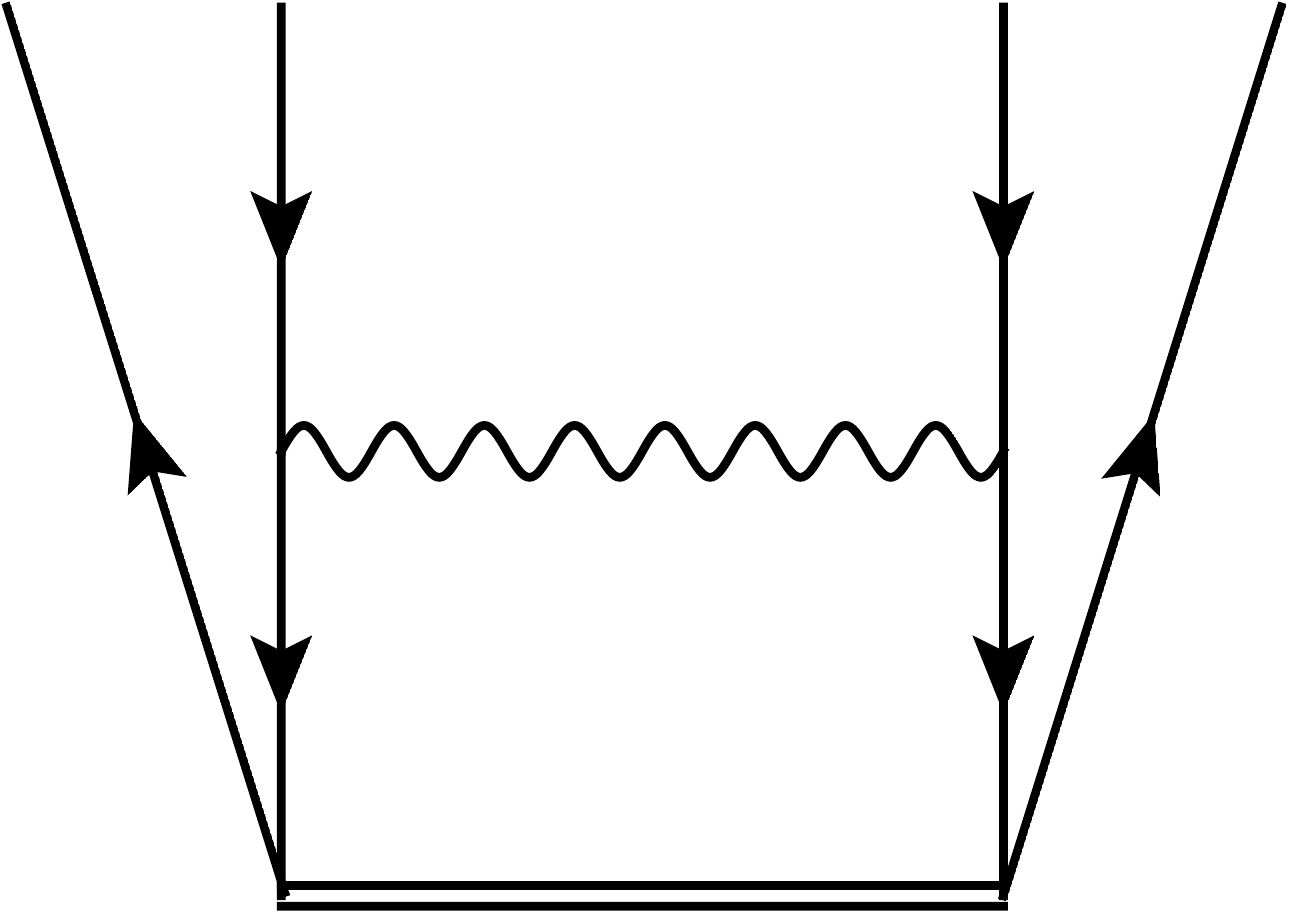}}\\ 
  \nonumber
  & + & \parbox{30mm}{\includegraphics[scale=0.20]{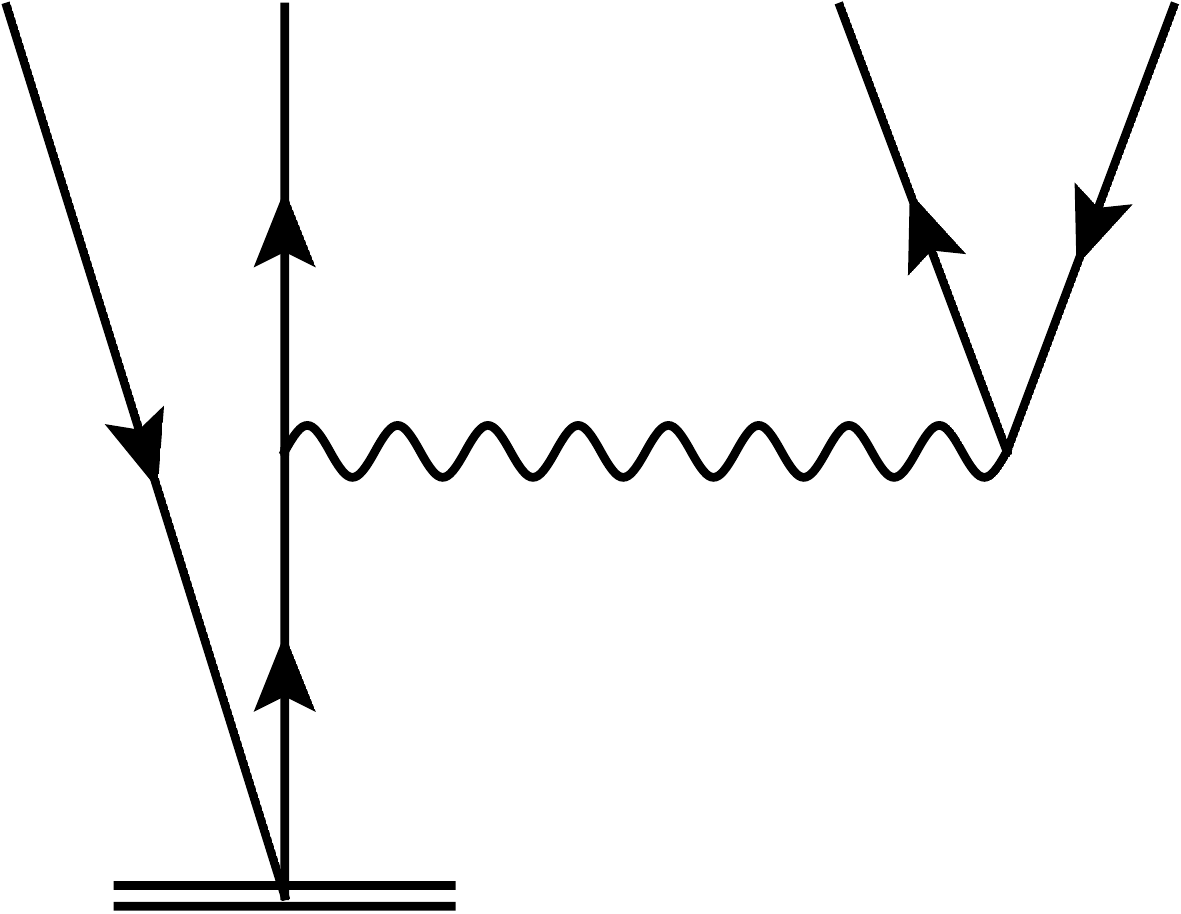}}
  + \parbox{30mm}{\includegraphics[scale=0.20]{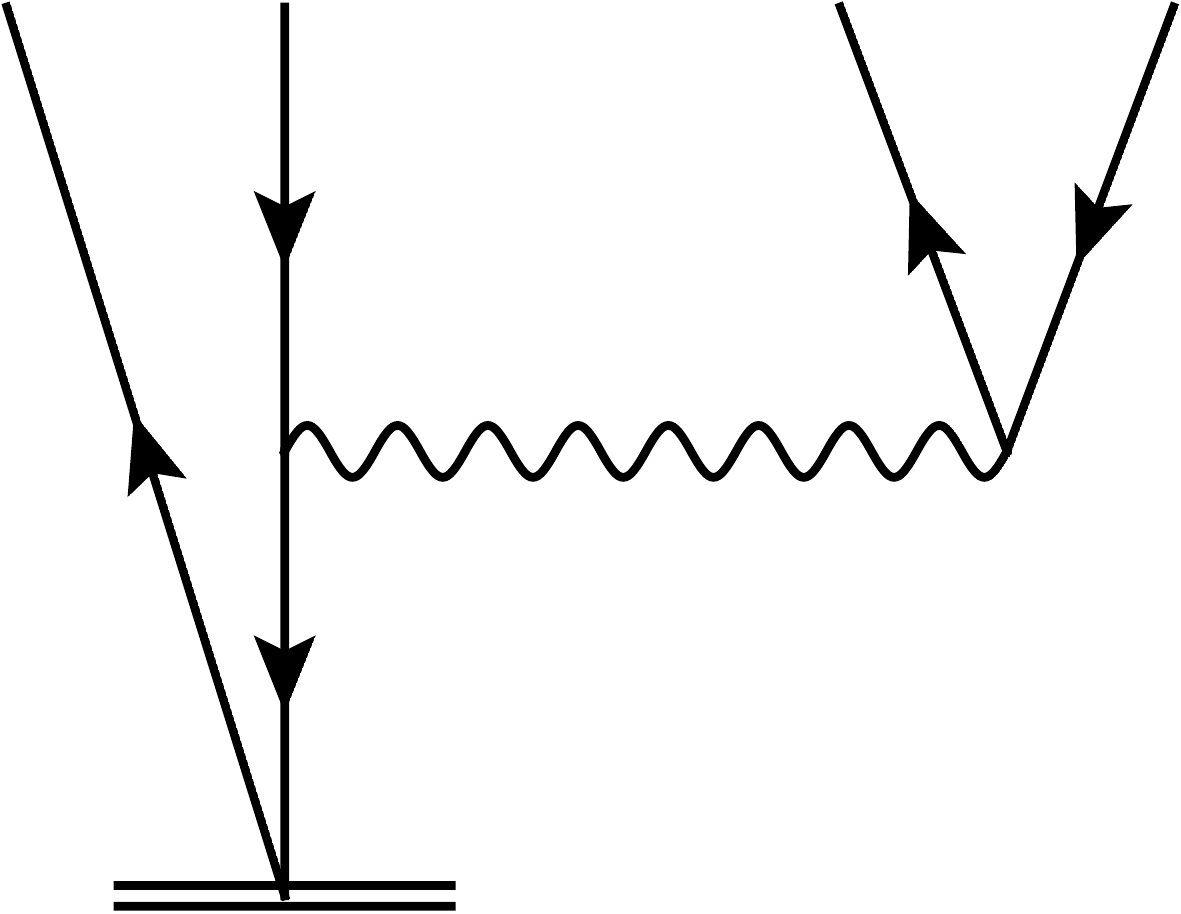}} \\
  \nonumber
  & + & \parbox{40mm}{\includegraphics[scale=0.20]{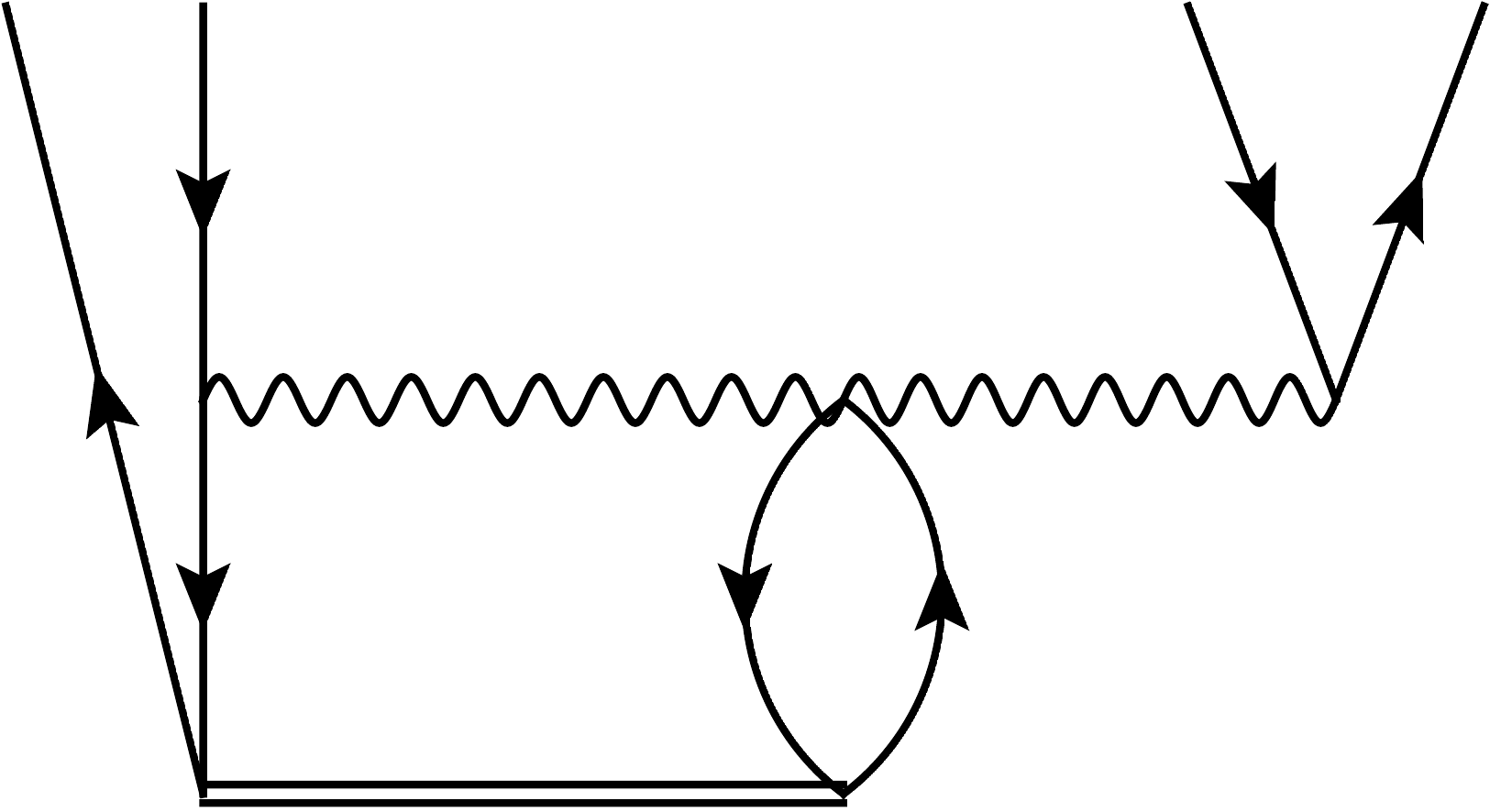}} 
    + \parbox{40mm}{\includegraphics[scale=0.20]{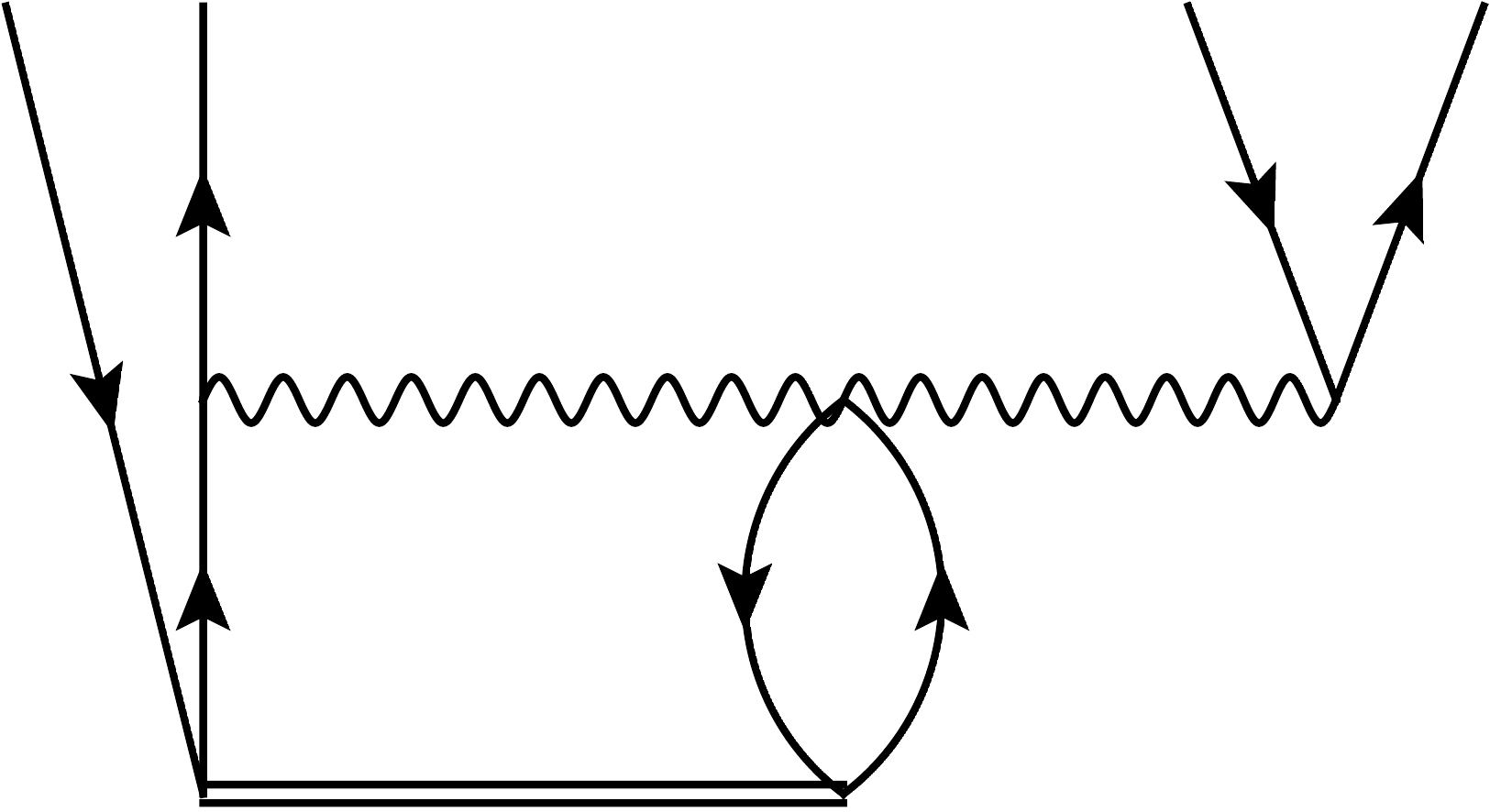}} \\
  & + &  \parbox{40mm}{\includegraphics[scale=0.20]{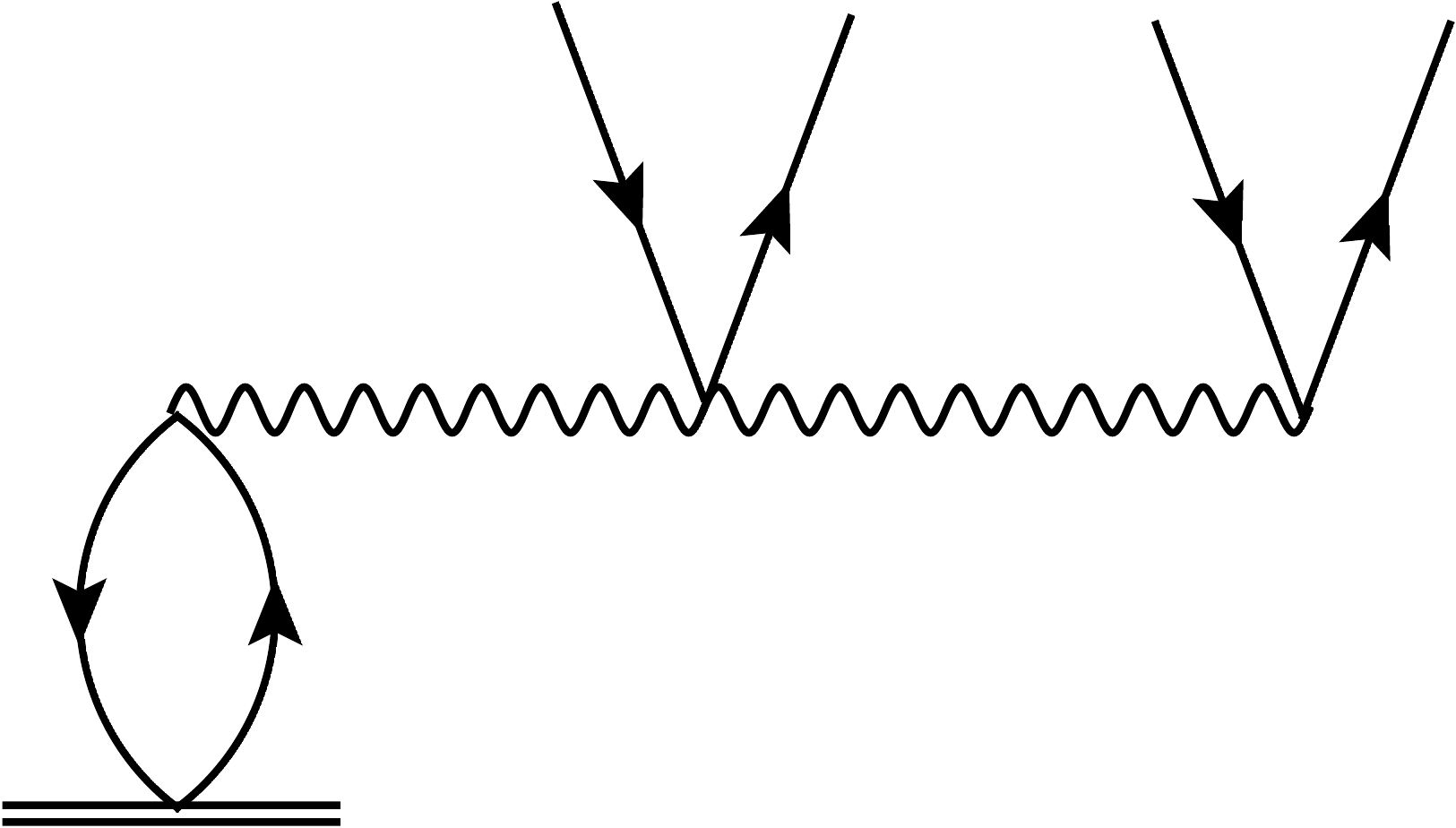}} .
  \label{eq:r2h2p_eqns}
\end{eqnarray}

The three last diagrams that enter in Eq.~\ref{eq:r2h2p_eqns} involve
three-body terms of the similarity transformed Hamiltonian. These
terms can be very memory expensive in numerical implementations, and
it is therefore more convenient (from a computational point of view)
to rewrite these terms using intermediates that involves only one- and
two-body terms. By defining the following intermediates,
\begin{eqnarray} \nonumber
  \parbox{20mm}{\includegraphics[scale=0.25]{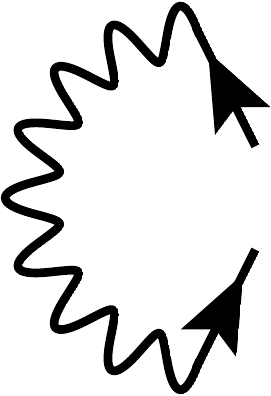}}  &
  = &  
  \parbox{30mm}{\includegraphics[scale=0.20]{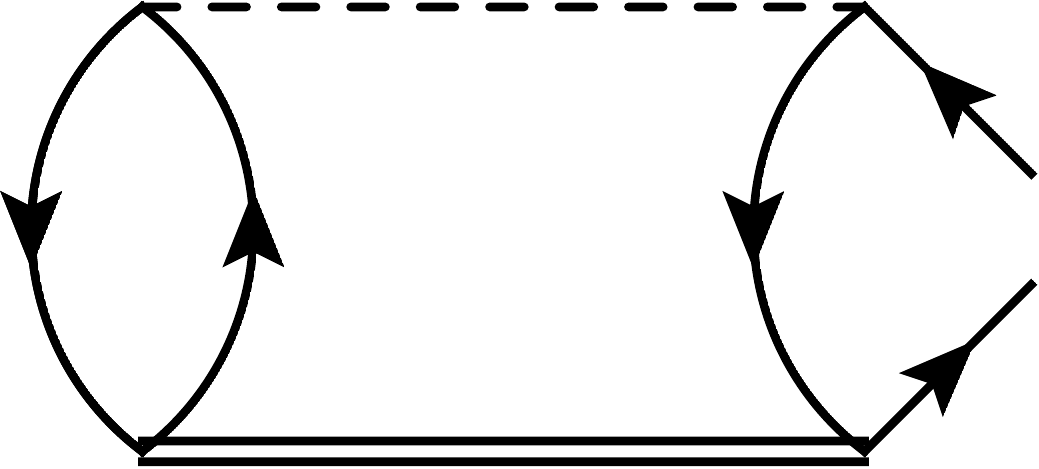}} + 
  \parbox{30mm}{\includegraphics[scale=0.20]{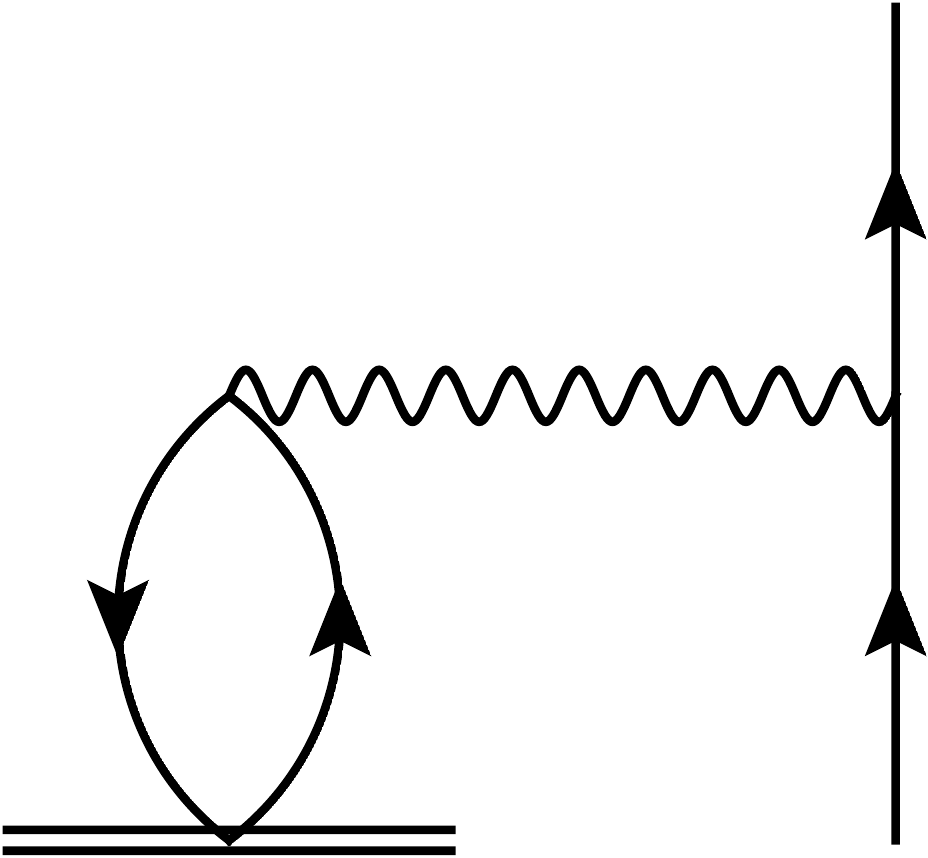}} \\ 
  \parbox{20mm}{\includegraphics[scale=0.25]{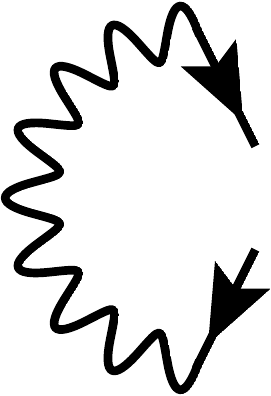}}  &
  = &  
  \parbox{30mm}{\includegraphics[scale=0.20]{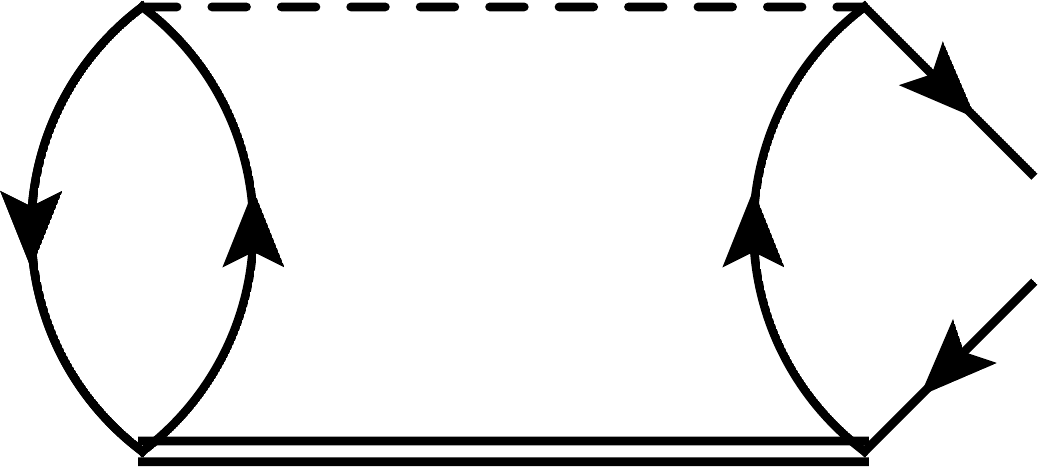}} + 
  \parbox{30mm}{\includegraphics[scale=0.20]{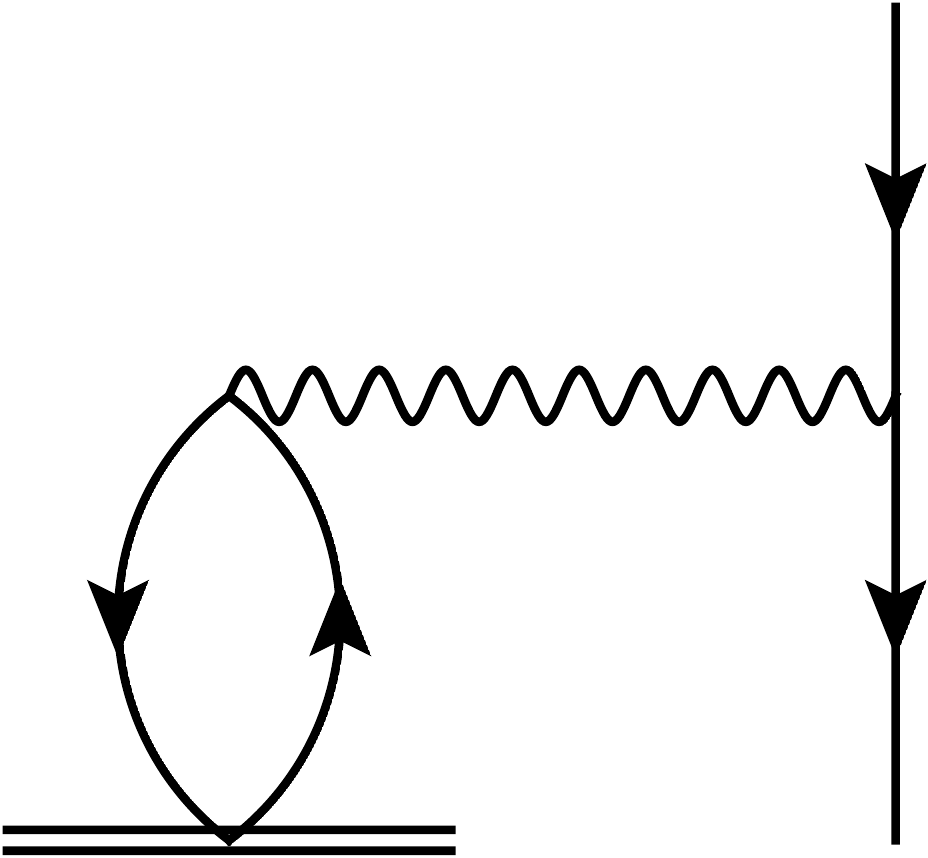}} \\ 
  \label{eq:threebody_eqns}
\end{eqnarray}
we can rewrite the last three diagrams in Eq.~\ref{eq:r2h2p_eqns} in
the following way,
\begin{equation}
  \parbox{30mm}{\includegraphics[scale=0.30]{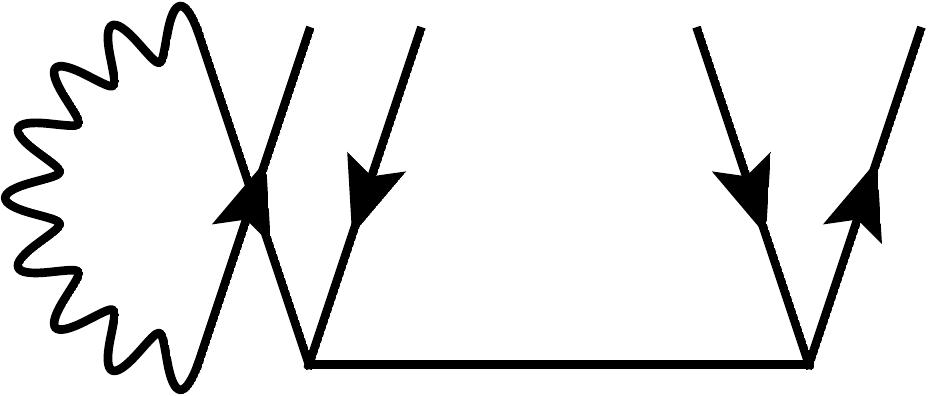}} + 
  \parbox{30mm}{\includegraphics[scale=0.30]{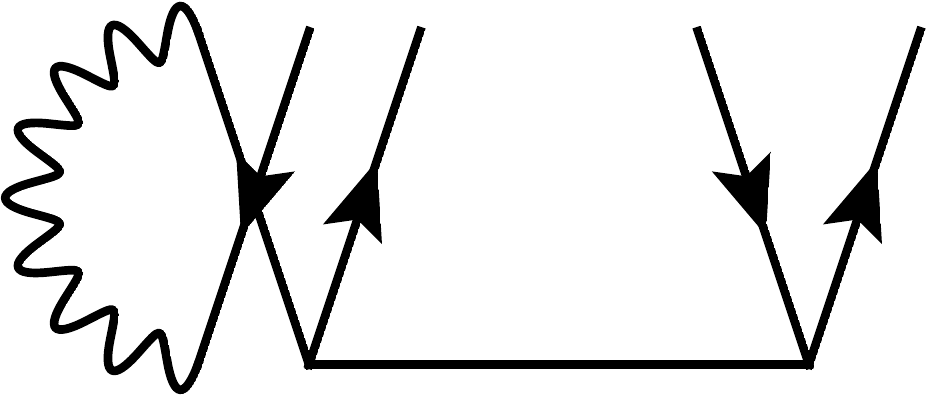}},
  \label{eq:threebody_int_eqns}
\end{equation}
which only involves one- and two-body terms.  In Table~\ref{tab:r1h1p}
we give the diagrams for the one-particle-one-hole excitation
amplitudes, and their corresponding algebraic expression in both
uncoupled ($m$-mscheme) and angular momentum coupled representations,
\begin{table*}[htbp!]
  \begin{Table_app}{0.25}{0.33}{0.42}
    Diagram & \mathrm{Uncoupled~expression} & \mathrm{Coupled~expression} 
    \tabularnewline\hline
    \includegraphics[scale=0.2]{R1h1p_HbarI7_diag} & \langle c\vert r^{JM}\vert i\rangle
    \langle a\vert\bar{h}^{00}\vert c\rangle  &  \langle c\vert\vert r^{J}\vert\vert i\rangle
    \langle a\vert\vert\bar{h}^{0}\vert\vert c\rangle \tabularnewline\hline
    \includegraphics[scale=0.2]{R1h1p_HbarI6_diag}  & \langle a\vert r^{JM}\vert k\rangle
    \langle k\vert\bar{h}^{00}\vert i\rangle  & \langle a\vert\vert r^{J}\vert\vert k\rangle
    \langle k\vert\vert\bar{h}^{0}\vert\vert i\rangle  \tabularnewline\hline
    \includegraphics[scale=0.2]{R1h1p_HbarI5_diag} & \langle ac\vert r^{JM}\vert ik\rangle
    \langle k \vert\bar{h}^{00} \vert c\rangle   & 
    { \hat{J}_{ac}^2\hat{J}_{ik}\over \hat{j}_a}
    \left\{\begin{array}{ccc}
        J   & J_{ac}   & J_{ik} \\
        j_c & j_i & j_a
      \end{array}\right\}  
    \times (-1)^{J-j_a-j_c+J_{ik}} \times \langle k \vert\vert \bar{h}^{0} \vert\vert c \rangle 
    \langle ac \vert\vert r^{J} \vert\vert ik \rangle  \tabularnewline\hline
    \includegraphics[scale=0.2]{R1h1p_HbarI2_diag} & {1\over 2}\langle ak \vert \bar{h}^{00} \vert cd\rangle
    \langle cd\vert r^{JM}\vert ik\rangle  &  {1\over 2}{ \hat{J}_{cd}^2\hat{J}_{ik}\over \hat{j}_a}
    \left\{\begin{array}{ccc}
        j_i  & J   & j_a \\
        J_{cd} & j_k & J_{ik}
      \end{array}\right\}  
    \times (-1)^{j_a+j_k+J_{ik}+J} \times \langle ak \vert\vert \bar{h}^{0} \vert\vert cd \rangle 
    \langle cd \vert\vert r^{J} \vert\vert ik \rangle \tabularnewline\hline 
    \includegraphics[scale=0.2]{R1h1p_HbarI3_diag}  & -{1\over 2}\langle ac\vert r^{JM}\vert kl\rangle
    \langle kl \vert\bar{h}^{00} \vert ic\rangle  &  -{1\over 2}{ \hat{J}_{ac}^2\hat{J}_{ic}\over \hat{j}_a}
    \left\{\begin{array}{ccc}
        j_i   & J   & j_a \\
        J_{ac} & j_c & J_{ic}
      \end{array}\right\} 
    \times (-1)^{j_a+j_c+J+J_{ic}} \times \langle kl \vert\vert \bar{h}^{0} \vert\vert ic \rangle 
    \langle ac \vert\vert r^{J} \vert\vert kl \rangle  \tabularnewline\hline
    \includegraphics[scale=0.2]{R1h1p_HbarI4_diag}  & \langle ka \vert \bar{h}^{00} \vert ci\rangle
    \langle c\vert r^{JM}\vert k\rangle & (-1)^{j_c + j_a + j_i+ j_k} { \hat{j}_c\over \hat{j}_a}
    \times \langle kc^{-1}
    \vert\vert\bar{h}^{0} \vert\vert ia^{-1}\rangle \langle c\vert\vert
    r^{J}\vert\vert k\rangle \tabularnewline\hline
  \end{Table_app}
  \caption{Coupled and uncoupled algebraic expressions for the
    diagrams of the one-particle-one-hole excitation amplitude $R_1$ given in
    Eq.(\ref{eq:r1h1p_eqns}). Repeated indices are summed over.}
  \label{tab:r1h1p}
\end{table*}
In Table~\ref{tab:r2h2p} we give the diagrams for the
two-particle-two-hole excitation amplitudes, and their corresponding
algebraic expression in both uncoupled ($m$-mscheme) and angular
momentum coupled representations,
\begin{table*}[htbp!]
  \begin{Table_app}{0.25}{0.33}{0.43}
    Diagram & \mathrm{Uncoupled~expression} & \mathrm{Coupled~expression} 
    \tabularnewline\hline
    \includegraphics[scale=0.2]{R2h2p_HbarI6_diag} & 
    P(ab)\langle c \vert\ \bar{h}^{00} \vert b\rangle
    \langle ac \vert r^{JM} \vert ij\rangle  &  P(ab)\langle c \vert\vert\bar{h}^{0} \vert\vert b\rangle
    \langle ac \vert\vert r^{J} \vert\vert ij\rangle  \tabularnewline\hline 
    \includegraphics[scale=0.2]{R2h2p_HbarI7_diag} &  -P(ij)\langle k \vert\bar{h}^{00} \vert j\rangle
    \langle ab \vert r^{JM} \vert ik\rangle  &  -P(ij)\langle k \vert\vert\bar{h}^{0} \vert\vert j\rangle
    \langle ab \vert\vert r^{J} \vert\vert ik\rangle  \tabularnewline\hline
    \includegraphics[scale=0.2]{R2h2p_HbarI4_diag} &
    P(ab)P(ij)\times \langle ac\vert r^{JM}\vert ik\rangle \langle kb
    \vert\bar{h}^{00} \vert cj\rangle & P(ab)P(ij)\times \langle
    ai^{-1}\vert\vert r^{J}\vert\vert kc^{-1}\rangle \langle kc^{-1}
    \vert\vert\bar{h}^{0} \vert\vert jb^{-1}\rangle
    \tabularnewline\hline
    \includegraphics[scale=0.2]{R2h2p_HbarI8_diag} & {1\over 2}\langle ab \vert\bar{h}^{00} \vert cd\rangle
    \langle cd\vert r^{JM}\vert ij\rangle  & {1\over 2}\langle ab \vert\vert\bar{h}^{0} \vert\vert cd\rangle
    \langle cd\vert\vert r^{J}\vert\vert ij\rangle  \tabularnewline\hline
    \includegraphics[scale=0.2]{R2h2p_HbarI9_diag} & {1\over 2}\langle kl \vert\bar{h}^{00} \vert ij\rangle
    \langle ab\vert r^{JM}\vert kl\rangle   & {1\over 2}\langle kl \vert\vert\bar{h}^{0} \vert\vert ij\rangle
    \langle ab\vert\vert r^{J}\vert\vert kl\rangle  \tabularnewline\hline
    \includegraphics[scale=0.2]{R2h2p_HbarI10_diag} & P(ij)\langle ab \vert\bar{h}^{00} \vert cj\rangle
    \langle c\vert r^{JM}\vert i\rangle   & P(ij) \hat{J}_{ij}\hat{j}_{c}
    \left\{\begin{array}{ccc}
        J   & J_{ab}   & J_{ij} \\
        j_j & j_i & j_c
      \end{array}\right\} 
    \times (-1)^{j_j+j_c+J+J_{ij}} \langle ab \vert\vert \bar{h}^{0} \vert\vert cj\rangle
    \langle c\vert\vert r^{J}\vert\vert i\rangle   \tabularnewline\hline
    \includegraphics[scale=0.2]{R2h2p_HbarI11_diag} & -P(ab)\langle kb \vert\bar{h}^{00} \vert ij\rangle
    \langle a\vert r^{JM}\vert k\rangle   & P(ab){1\over 2} \hat{J}_{ij}\hat{j}_{c}
    \left\{\begin{array}{ccc}
        J   & J_{ab}   & J_{ij} \\
        j_b & j_k & j_a
      \end{array}\right\}  
    \times (-1)^{j_k-j_b+J+J_{ab}} \langle kb \vert\vert \bar{h}^{0} \vert\vert ij\rangle
    \langle a\vert\vert r^{J}\vert\vert k\rangle  \tabularnewline\hline
    \includegraphics[scale=0.30]{R2h2p_Hbar_threebody_int1_diag}
    & P(ab)\langle cb\vert t^{00}\vert ij\rangle\langle a\vert
    \chi^{JM}\vert c\rangle & P(ab) (-1)^{1+j_b-j_a+J_{ab}} \times 
    \hat{J}_{ij}\hat{j}_a  \left\{\begin{array}{ccc}
        j_c   & j_{b}   & J_{ij} \\
        J_{ab} & J & j_a
      \end{array}\right\}  
    \times \langle cb\vert\vert t^0\vert\vert ij\rangle\langle
    a\vert\vert \chi^{J}\vert\vert c\rangle \tabularnewline\hline
    \includegraphics[scale=0.30]{R2h2p_Hbar_threebody_int2_diag}
    & -P(ij)\langle ab\vert t^{00}\vert kj\rangle\langle k\vert
    \chi^{JM}\vert i\rangle & -P(ij)(-1)^{j_k+j_j+J+J_{ij}} \times
    \hat{J}_{ij}\hat{j}_k \left\{\begin{array}{ccc}
        j_i   & j_{j}   & J_{ij} \\
        J_{ab} & J & j_k
      \end{array}\right\}  
    \times \langle ab \vert\vert t^0 \vert\vert
    kj\rangle\langle k\vert\vert \chi^{J}\vert \vert i\rangle
    \tabularnewline\hline
  \end{Table_app}
  \caption{Coupled and uncoupled algebraic expressions for the
    diagrams of the two-particle-two-hole excitation amplitudes $ R_2$ given in
    Eq.(\ref{eq:r2h2p_eqns}). Repeated indices are summed
    over. }
  \label{tab:r2h2p}  
\end{table*}
The diagrammatic representation and algebraic expressions for the
various matrix elements of the similarity transformed Hamiltonian that
enter in Tables \ref{tab:r1h1p} and \ref{tab:r2h2p} can be found in
e.g. \cite{bartlett2007,gour2006}.
 
\newpage
\subsection{Particle-attached equation of motion}
Below we give the diagrammatic representation and algebraic
expressions in an angular momentum coupled scheme for
Particle-Attached Equation-of-Motion method in the singles-and-doubles
approximation (PA-EOM-CCSD). The PA-EOM-CCSD results from
diagonalizing the similarity transformed Hamiltonian in a sub-space of
one-particle and two-particle-one-hole excitations. This approximation
has been shown to work particularly well for low-lying states that are
dominated by one-particle excitations
\cite{bartlett2007,gour2006}. The diagrammatic representation of the
one-particle excitation amplitude equations are given by, 
\begin{eqnarray}
  \parbox{20mm}{\includegraphics[scale=0.20]{R1p_diag}}  &
  = &  
  \parbox{30mm}{\includegraphics[scale=0.20]{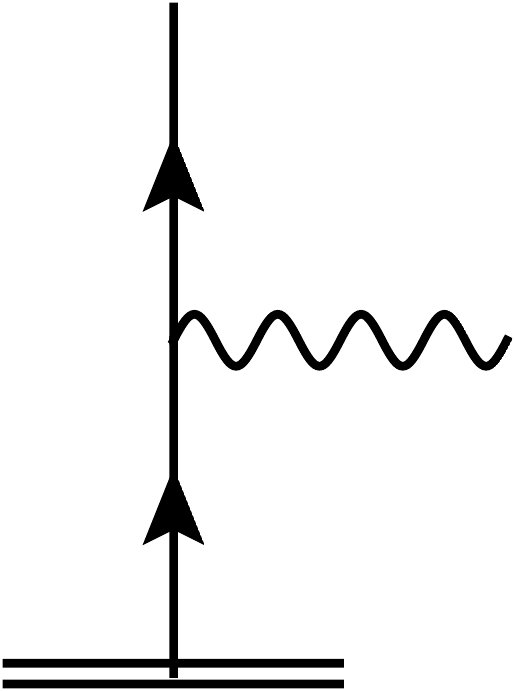}} 
  + \parbox{30mm}{\includegraphics[scale=0.20]{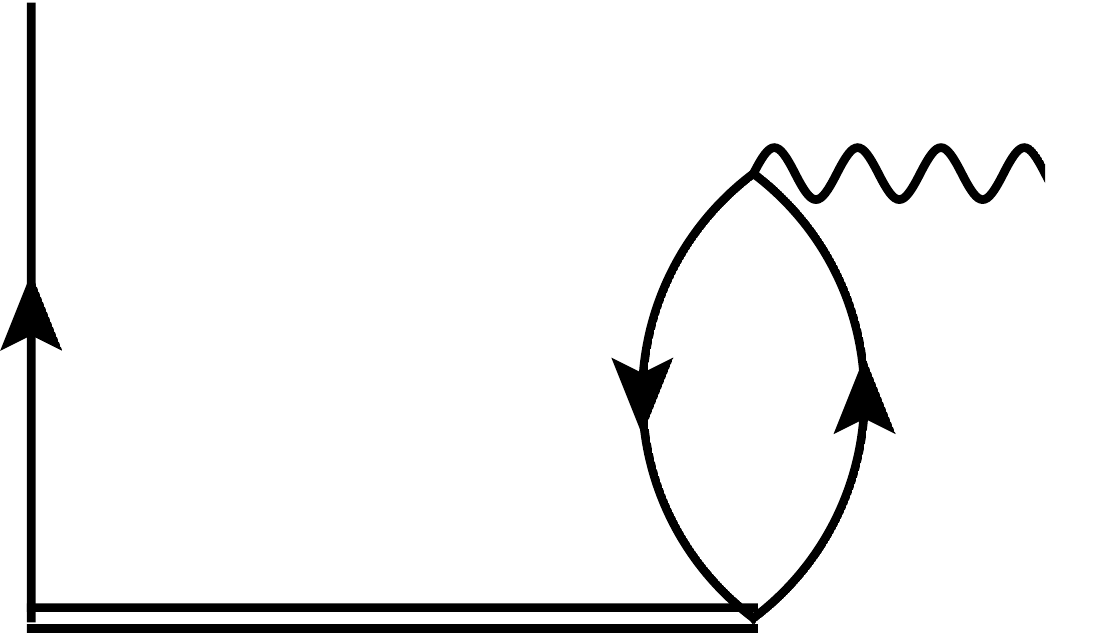}} 
  + \parbox{30mm}{\includegraphics[scale=0.20]{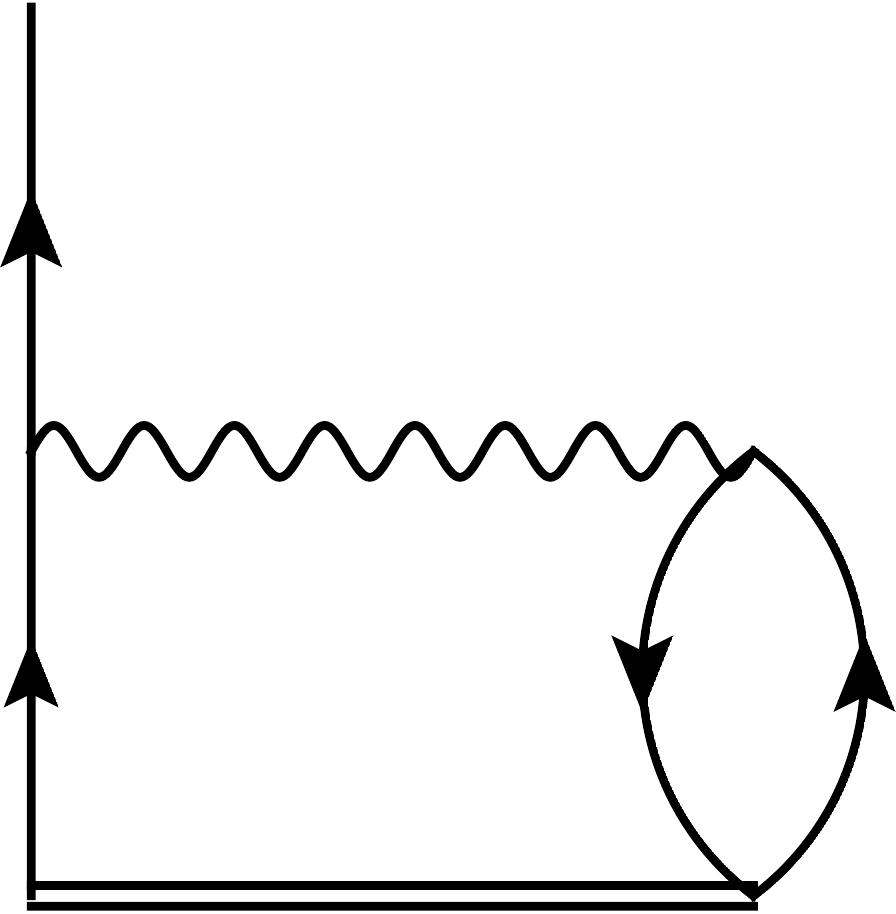}},
  \label{eq:R1p_eqns}
\end{eqnarray}
and the diagrammatic representation of the
two-particle-one-hole excitation amplitude equations are given by, 
\begin{eqnarray}
  \nonumber
  \parbox{20mm}{\includegraphics[scale=0.20]{R1h2p_diag}}  &
   = &  
   \parbox{30mm}{\includegraphics[scale=0.20]{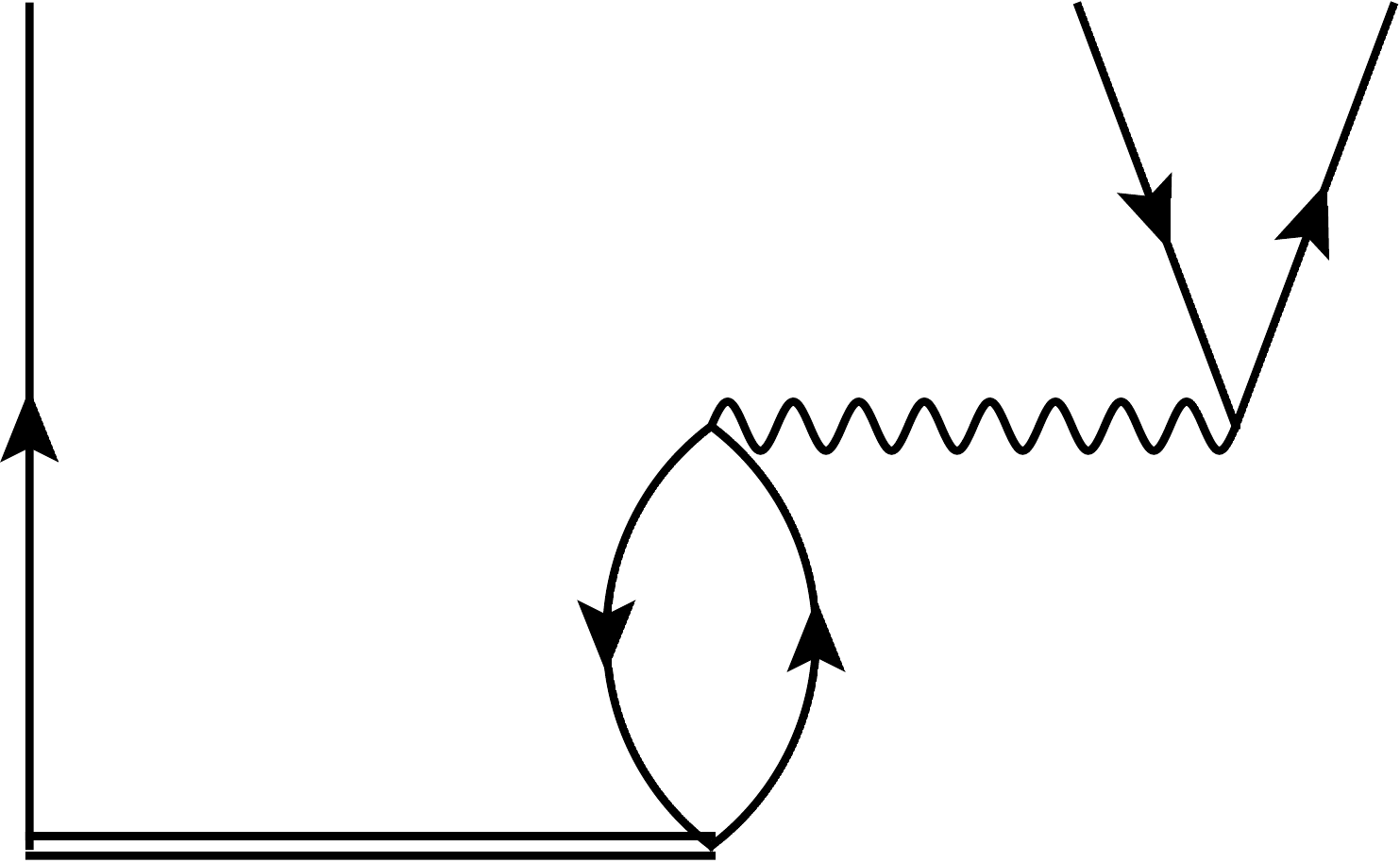}} 
   + \parbox{30mm}{\includegraphics[scale=0.20]{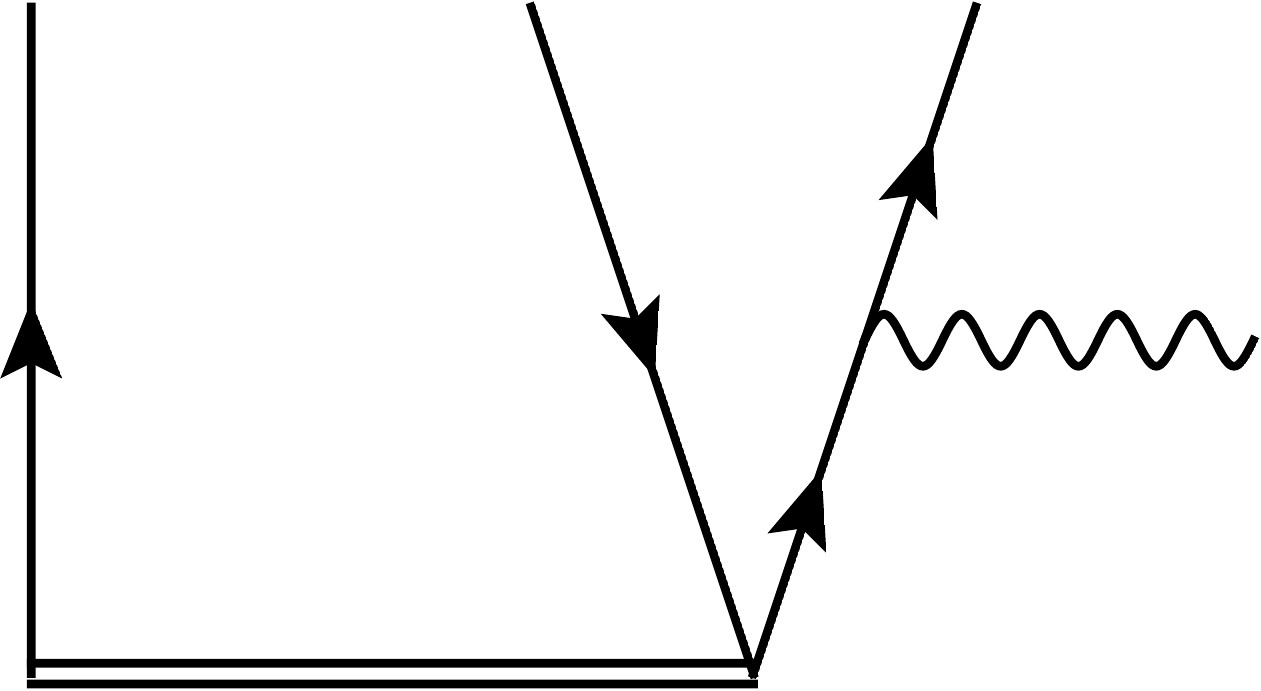}} 
   + \parbox{30mm}{\includegraphics[scale=0.20]{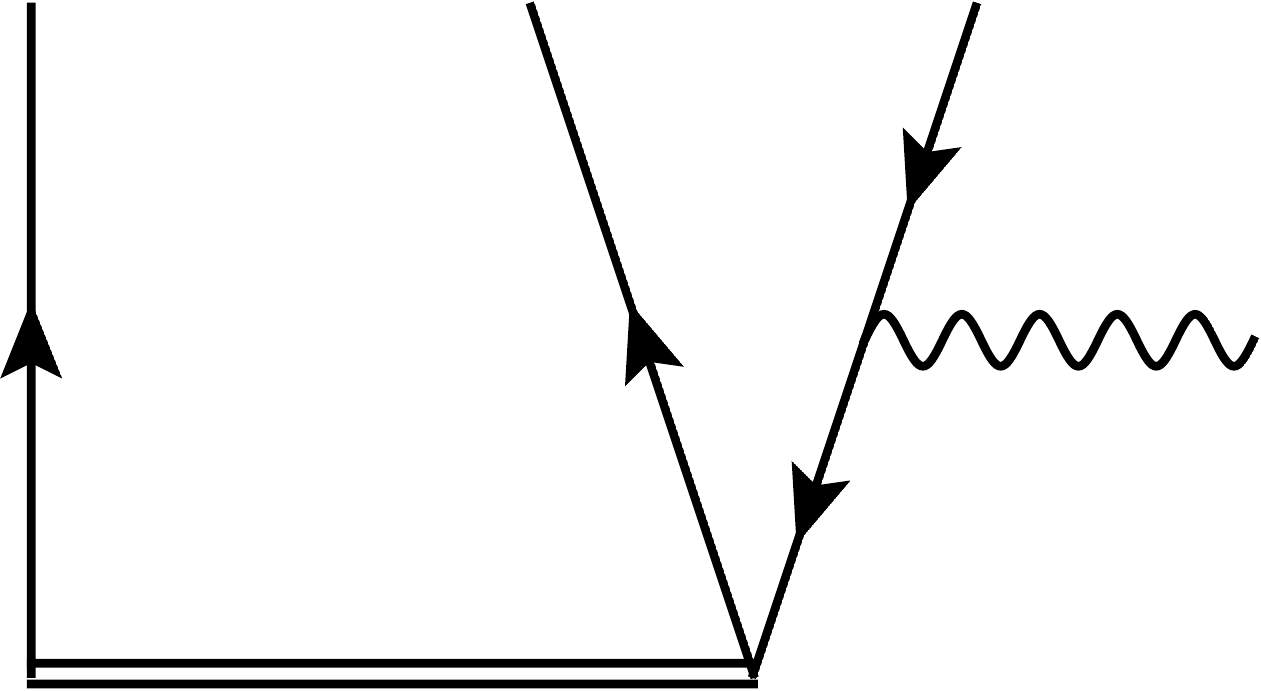}} \\
   & + & \parbox{30mm}{\includegraphics[scale=0.20]{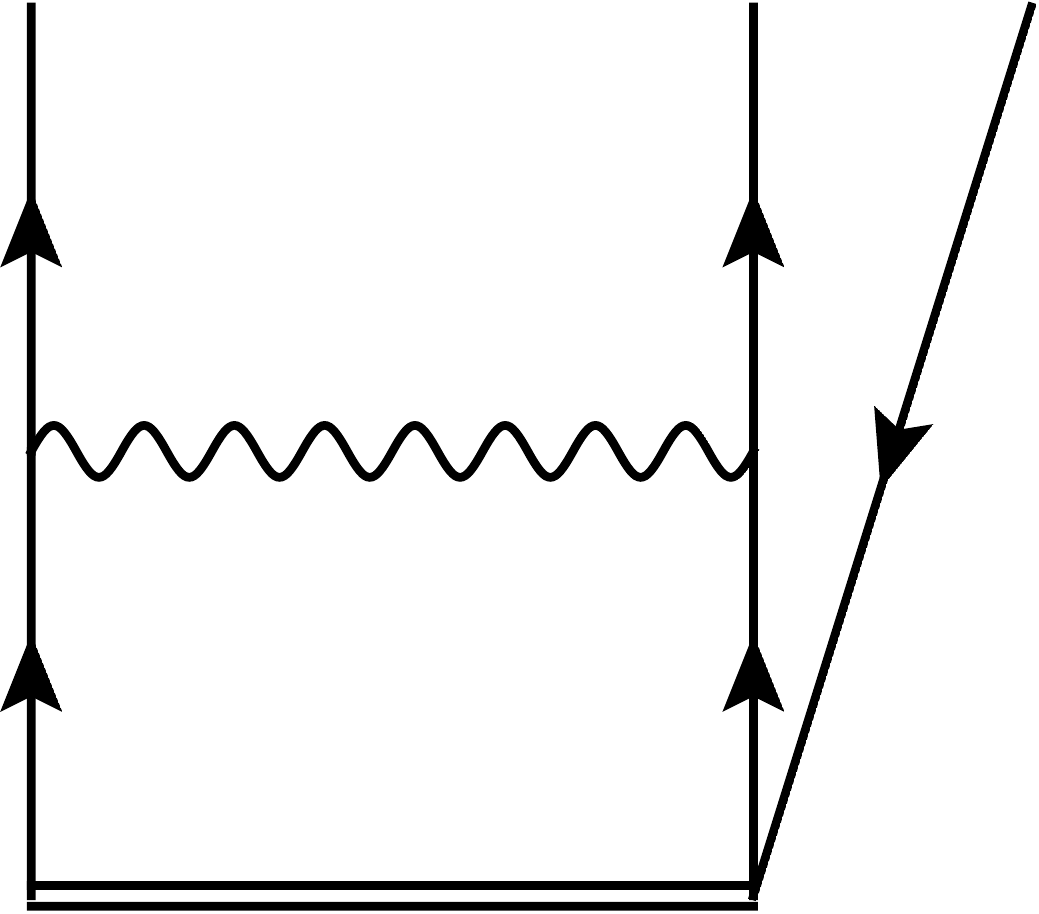}} 
   + \parbox{30mm}{\includegraphics[scale=0.20]{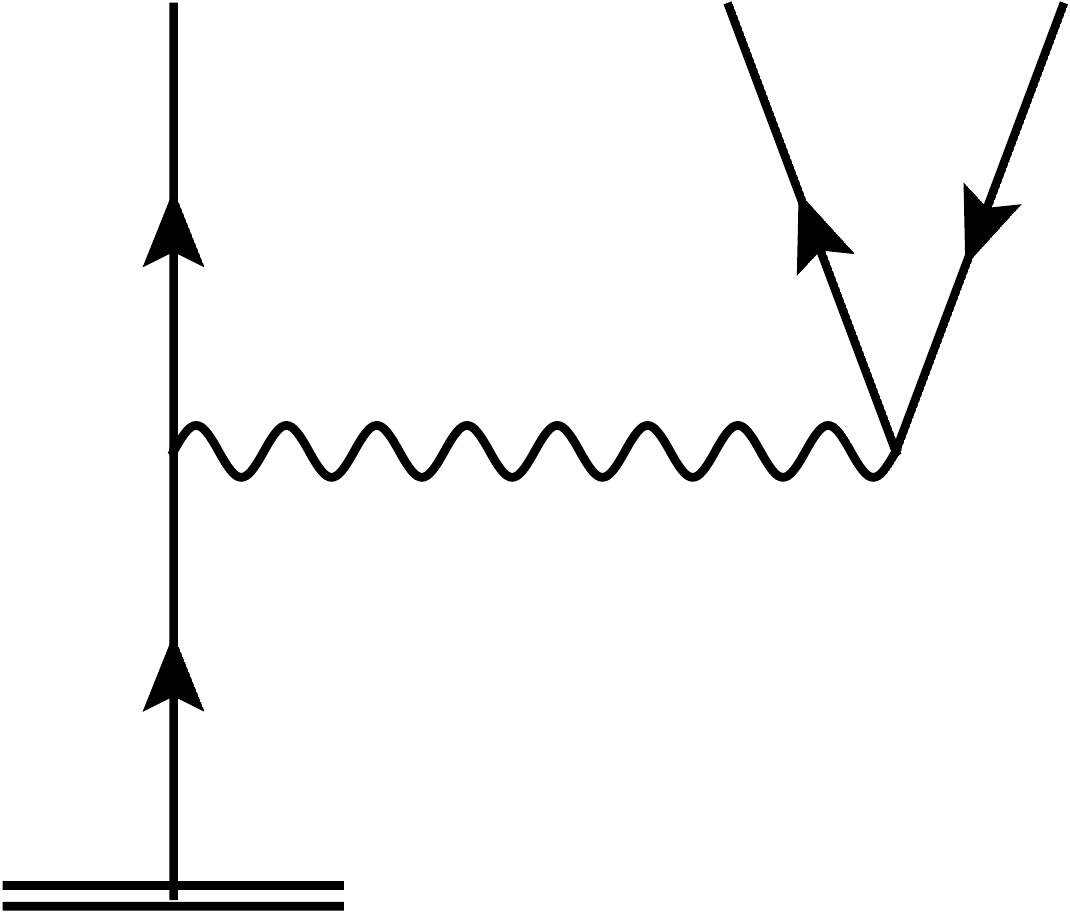}} \\
   & + & \parbox{30mm}{\includegraphics[scale=0.20]{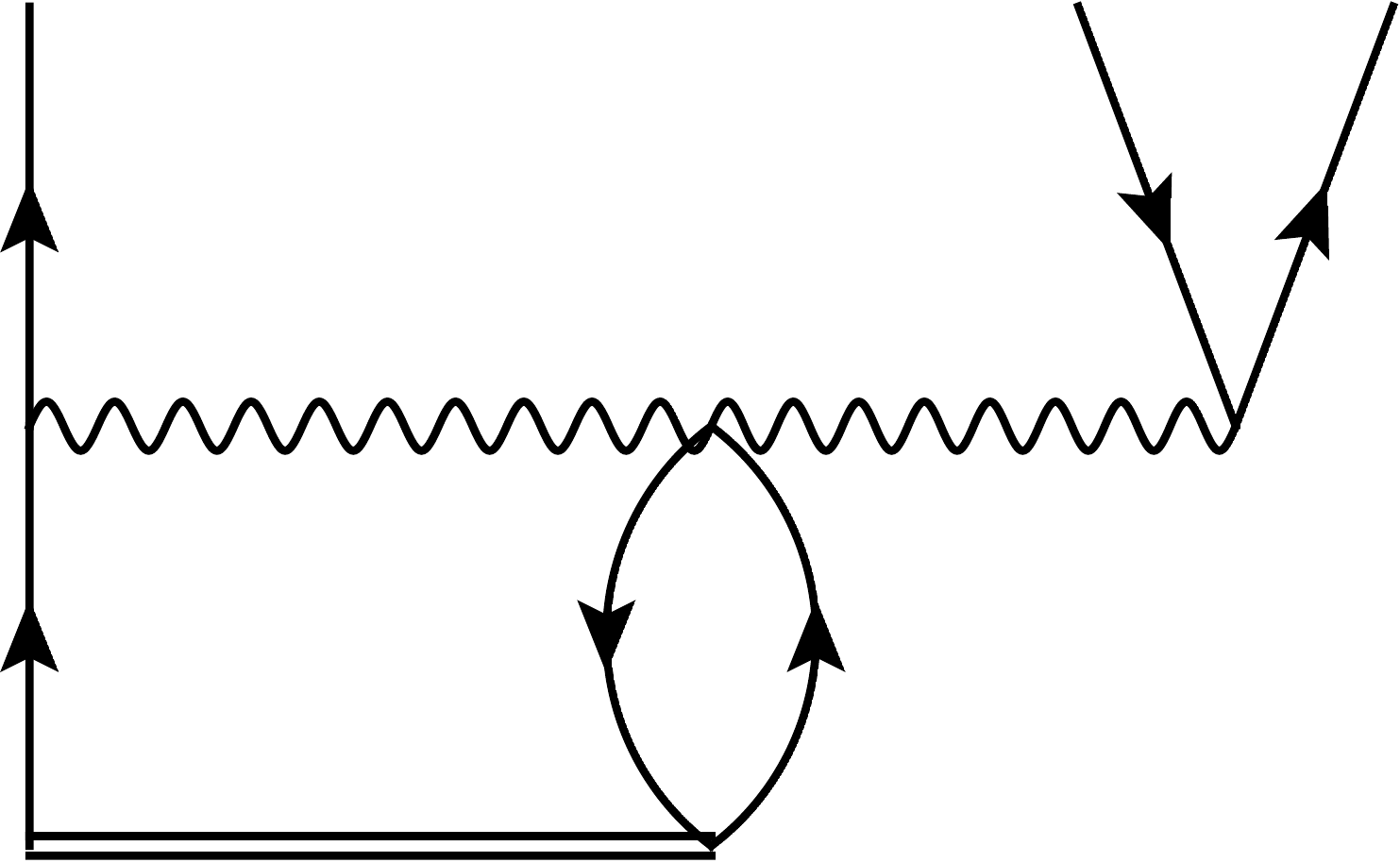}} \ .
   \label{eq:R1h2p_eqns}
\end{eqnarray}
The last diagram in Eq.~\ref{eq:R1h2p_eqns} involves a three-body term of
the similarity transformed Hamiltonian, and it is therefore convenient
to define the following intermediate,
\begin{equation} 
  \parbox{20mm}{\includegraphics[scale=0.25]{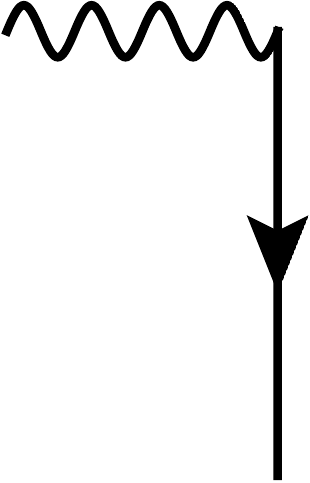}}  
  =   
  \parbox{30mm}{\includegraphics[scale=0.20]{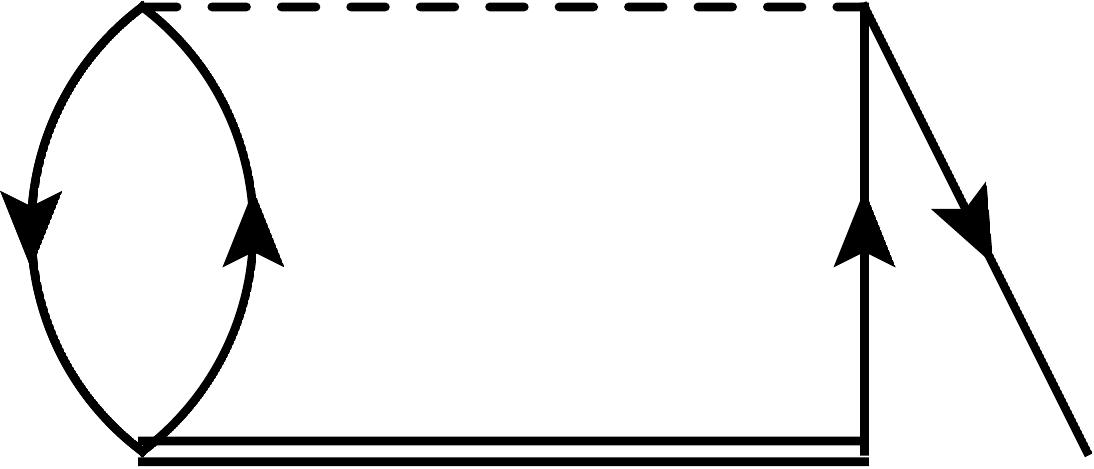}} + 
  \label{eq:threebody_eqns2}
\end{equation}
we can then rewrite the last diagram in Eq.~\ref{eq:R1h2p_eqns} in
the following way,
\begin{equation}
  \parbox{30mm}{\includegraphics[scale=0.20]{Rabi_Hbar_threebody_diag}}
  =  \parbox{30mm}{\includegraphics[scale=0.30]{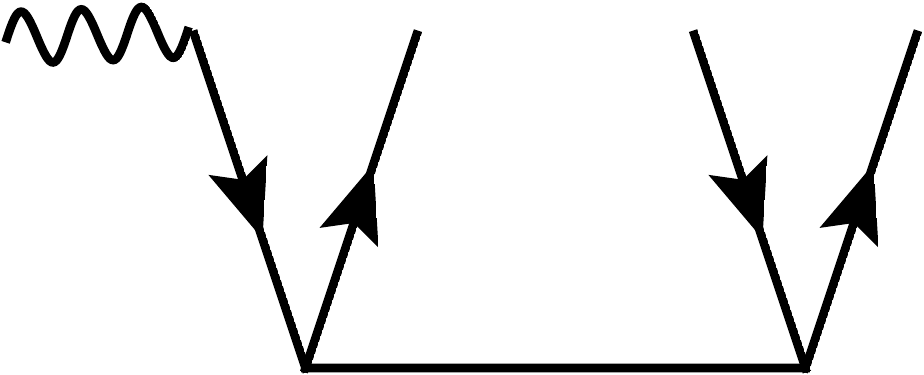}}.
  \label{eq:threebody_int_eqns2}
\end{equation}

In Table~\ref{tab:r2p1h} the algebraic expressions for the one-particle
and two-particle-one-hole excitation amplitudes in an uncoupled and
coupled angular momentum scheme are given,
\begin{table*}[htbp!]
  \begin{Table_app}{0.25}{0.33}{0.42}
    Diagram & \mathrm{Uncoupled~expression} & \mathrm{Coupled~expression} 
    \tabularnewline\hline
    \includegraphics[scale=0.2]{Ra_HbarI6_diag} &   \langle a \vert \bar{h}^{00} \vert c \rangle 
    \langle c\vert r^{JM}\vert\rangle  & 
     \langle a \vert\vert \bar{h}^{0} \vert\vert c \rangle \langle c\vert\vert r^J\vert\vert\rangle   
     \tabularnewline\hline
     \includegraphics[scale=0.2]{Ra_HbarI5_diag} &  -\langle k \vert \bar{h}^{00} \vert c \rangle 
    \langle ac \vert r^{JM} \vert k \rangle  & 
     -(-1)^{j_c+J_{ac}-J}{\hat{J}_{ac}^2\over \hat{J}^2} \times \langle k \vert\vert \bar{h}^{0} \vert\vert c \rangle 
    \langle ac \vert\vert r^{J} \vert\vert k \rangle  
    \tabularnewline\hline
    \includegraphics[scale=0.2]{Ra_HbarI2_diag} &  - \langle ka \vert \bar{h}^{00} \vert cd \rangle 
    \langle cd \vert r^{JM}\vert k \rangle  & 
     - {\hat{J}_{cd}^2\over \hat{J}^2} \langle ka \vert\vert \bar{h}^0 \vert\vert cd \rangle 
     \langle cd\vert\vert r^J\vert\vert k \rangle \tabularnewline\hline
    \includegraphics[scale=0.2]{Rabi_HbarI4_diag} &  P(ab)\langle kb \vert \bar{h}^{00} \vert cj\rangle
    \langle ac\vert r^{JM}\vert k\rangle  &  P(ab) \langle a \vert\vert r^{J} \vert\vert kc^{-1}\rangle 
    \langle kc^{-1} \vert\vert \bar{h}^{0} \vert\vert jb^{-1}\rangle
    \tabularnewline\hline
    \includegraphics[scale=0.2]{Rabi_HbarI6_diag} & P(ab)\langle ac\vert r^{JM}\vert j\rangle
    \langle b \vert\bar{h}^{00} \vert c\rangle  &   P(ab){ \hat{J}_{ac}^2\over \hat{j}_a^2} 
    \langle ac\vert\vert r^{J}\vert\vert j\rangle
    \langle b \vert\vert\bar{h}^{0} \vert\vert c\rangle  
    \tabularnewline\hline
    \includegraphics[scale=0.2]{Rabi_HbarI7_diag} &  -\langle k \vert\bar{h}^{00} \vert j\rangle
    \langle ab \vert r^{JM} \vert k\rangle  &  -\langle k \vert\vert\bar{h}^{0} \vert\vert j\rangle
    \langle ab \vert\vert r^{J} \vert\vert k\rangle  
    \tabularnewline\hline
    \includegraphics[scale=0.2]{Rabi_HbarI8_diag}  & {1\over 2}\langle ab \vert\bar{h}^{00} \vert cd\rangle
    \langle cd\vert r^{JM}\vert j\rangle   & {1\over 2}\langle ab \vert\vert\bar{h}^{0} \vert\vert cd\rangle
    \langle cd\vert\vert r^{J}\vert\vert j\rangle 
    \tabularnewline\hline
    \includegraphics[scale=0.2]{Rabi_HbarI10_diag} & \langle ab \vert\bar{h}^{00} \vert cj\rangle
    \langle c\vert r^{JM}\vert \rangle   & \langle ab \vert\vert \bar{h}^{0} \vert\vert cj\rangle
    \langle c\vert\vert r^{J}\vert\vert \rangle   
    \tabularnewline\hline
    \includegraphics[scale=0.2]{R1h2p_Hbar_threebody_int1_diag}
    & -\langle k\vert \chi^{JM}\vert\rangle \langle ab\vert
    t^{00}\vert kj\rangle & -\langle k\vert\vert
    \chi^{J}\vert\vert\rangle \langle ab\vert\vert t^{0}\vert\vert
    kj\rangle \tabularnewline\hline
  \end{Table_app}
  \caption{Coupled and uncoupled algebraic expressions for the
    diagrams of the one-particle $R_1$, and two-particle-one-hole amplitude $R_2$ given in
    Eqs.~(\ref{eq:R1p_eqns}) and ~(\ref{eq:R1h2p_eqns}). Repeated indices are summed over.}
  \label{tab:r2p1h}
\end{table*}
The diagrammatic representation and algebraic expressions for the
various matrix elements of the similarity transformed Hamiltonian that
enter in Table~\ref{tab:r2p1h} can be found in
e.g. \cite{bartlett2007,gour2006}.

\subsection{Particle-removed equation of motion}
Below we give the diagrammatic representation and algebraic
expressions in an angular momentum coupled scheme for
Particle-Removed Equation-of-Motion method in the singles-and-doubles
approximation (PR-EOM-CCSD). The PR-EOM-CCSD results from
diagonalizing the similarity transformed Hamiltonian in a sub-space of
one-hole and one-particle-two-hole excitations. This approximation
has been shown to work particularly well for low-lying states that are
dominated by one-hole excitations
\cite{bartlett2007,gour2006}. The diagrammatic representation of the
one-hole excitation amplitude equations are given by, 

\begin{eqnarray}
  \parbox{20mm}{\includegraphics[scale=0.20]{R1h_diag}}  &
   = &  
   \parbox{30mm}{\includegraphics[scale=0.20]{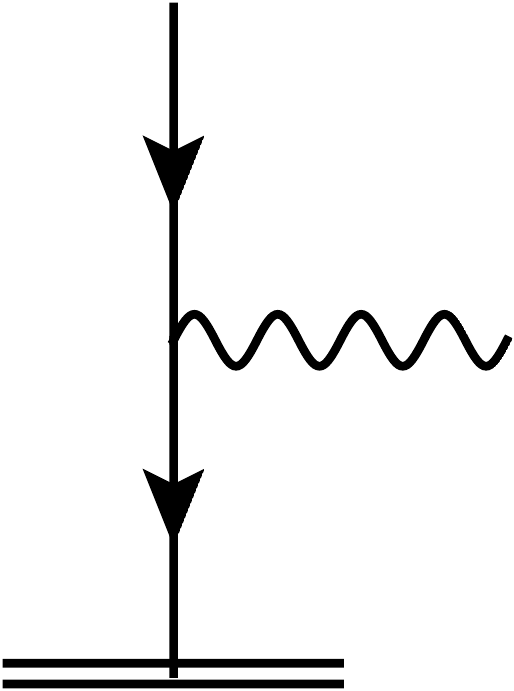}} 
   + \parbox{30mm}{\includegraphics[scale=0.20]{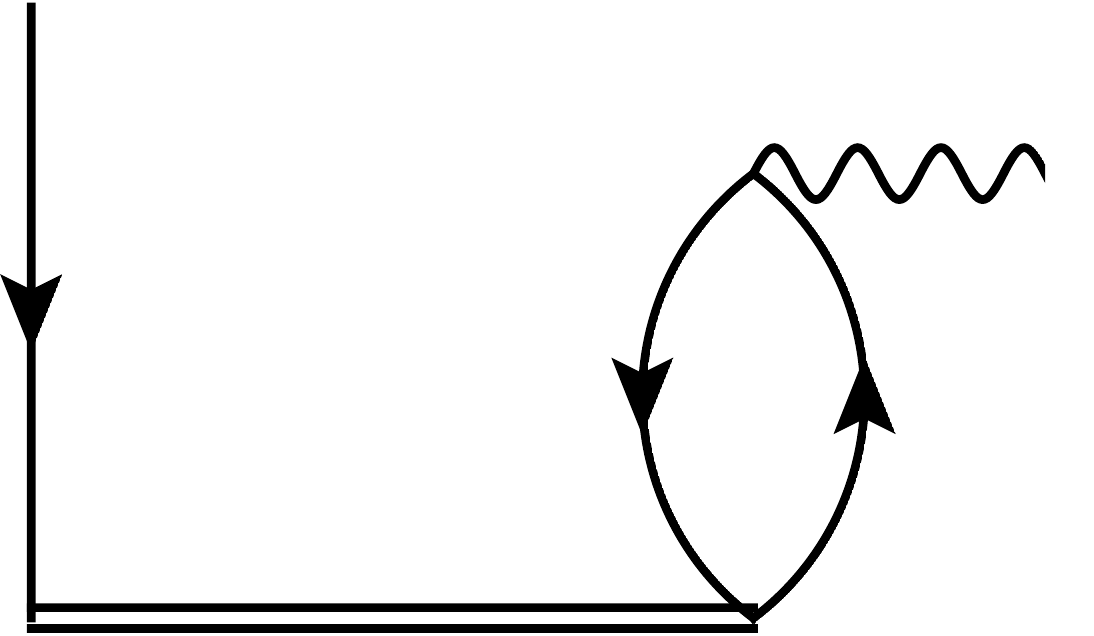}} 
   + \parbox{30mm}{\includegraphics[scale=0.20]{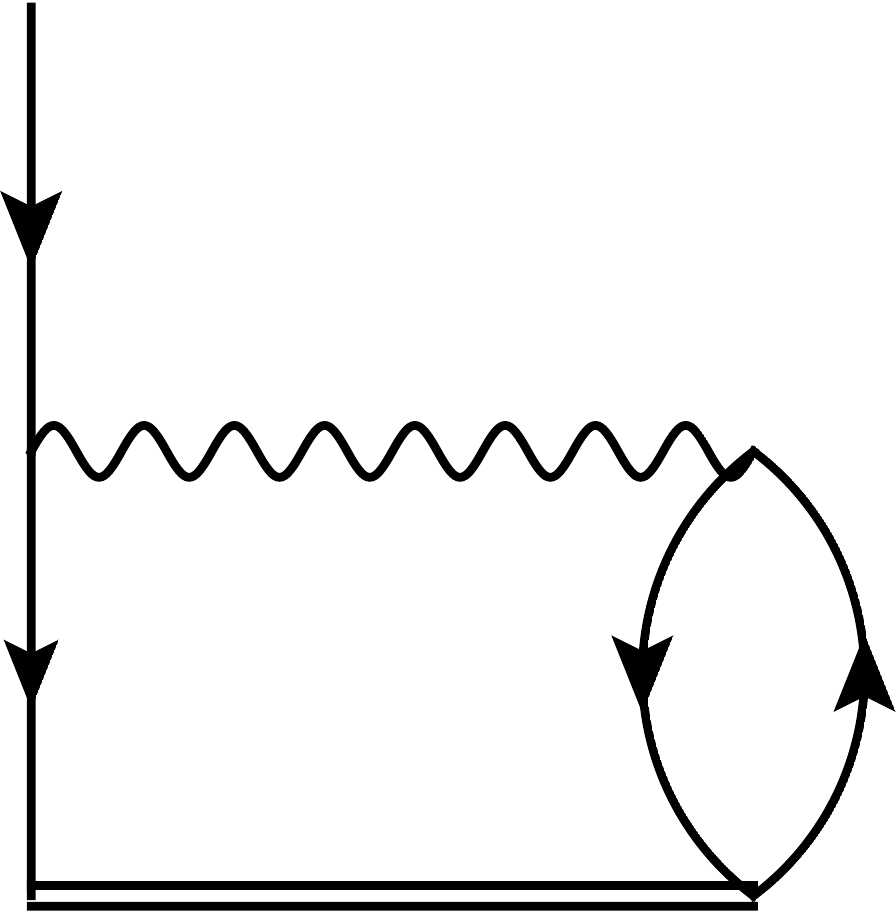}}, 
   \label{eq:R1h_eqns}
\end{eqnarray}

and the diagrammatic representation of the one-particle-two-hole
excitation amplitude equations are given by,

\begin{eqnarray}
  \nonumber
  \parbox{20mm}{\includegraphics[scale=0.20]{R2h1p_diag}}  &
   = &  
   \parbox{30mm}{\includegraphics[scale=0.20]{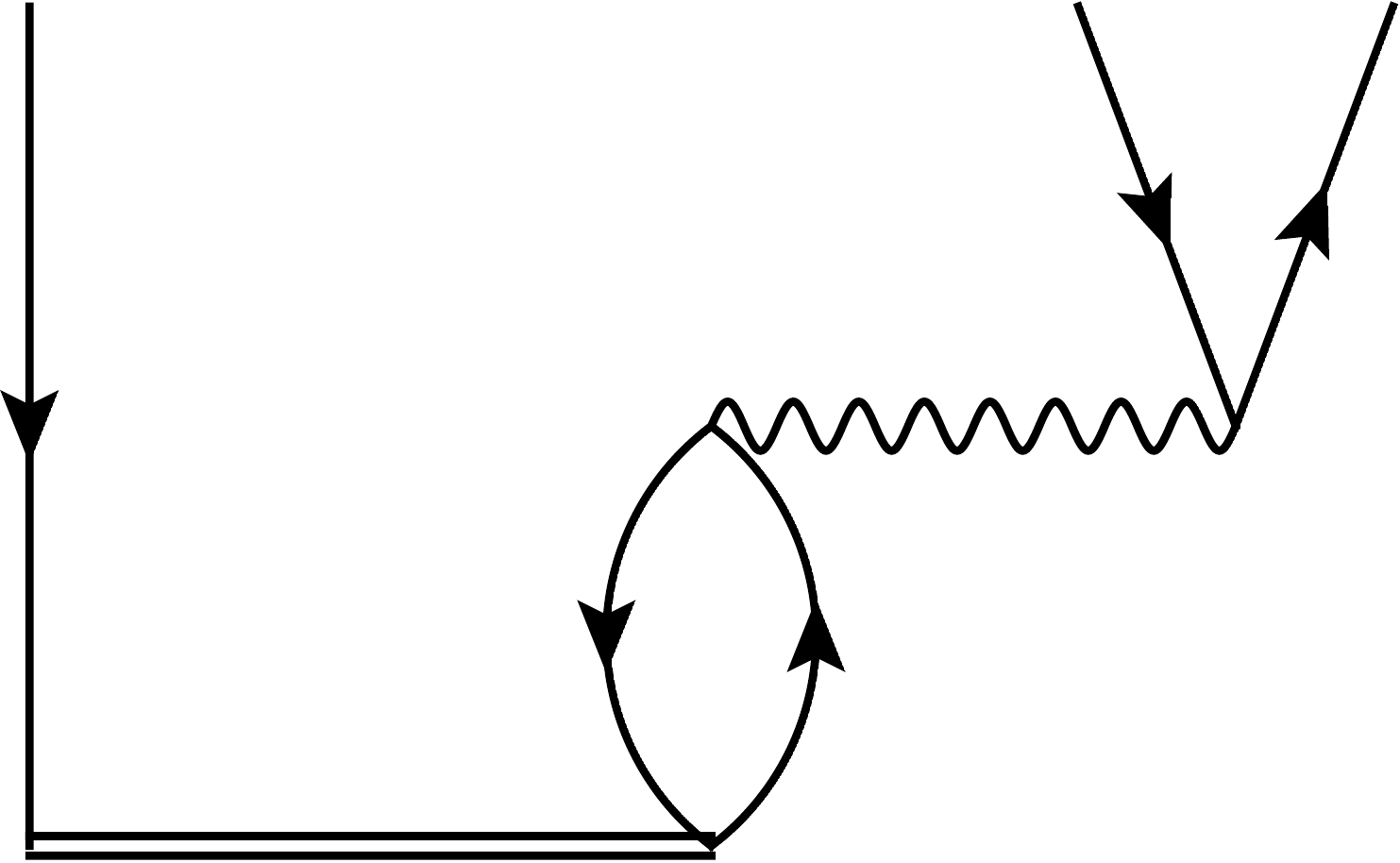}} 
   + \parbox{30mm}{\includegraphics[scale=0.20]{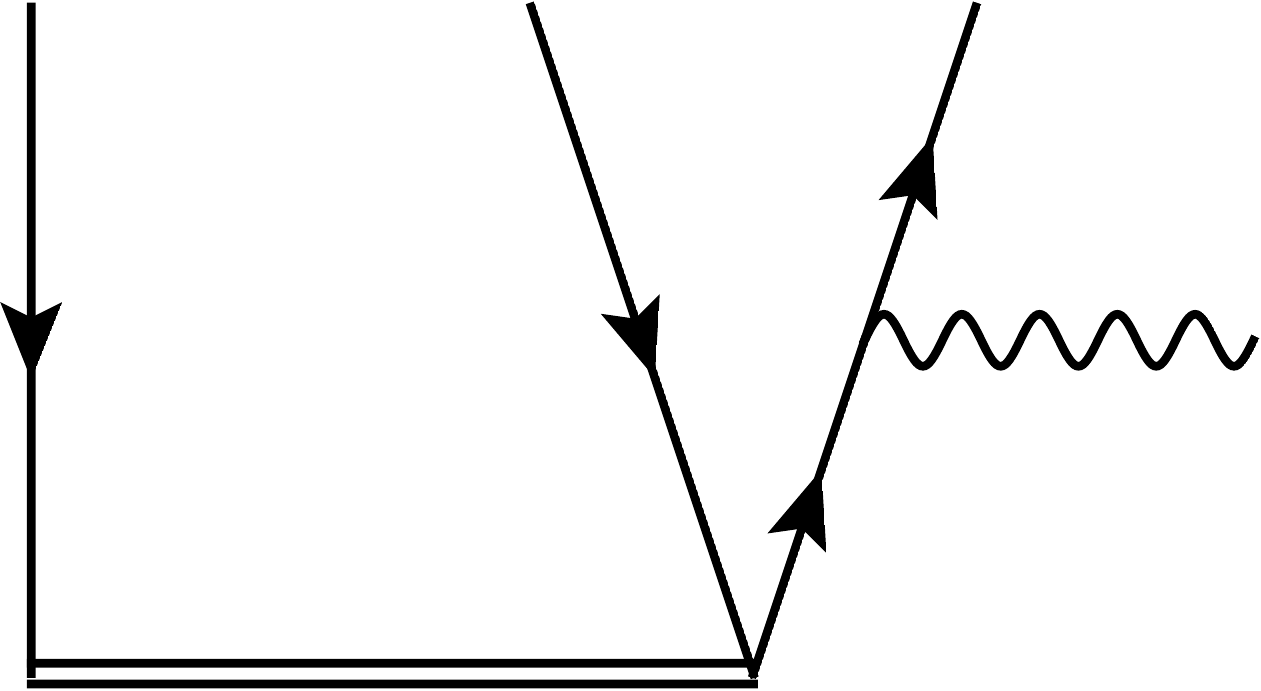}} 
   + \parbox{30mm}{\includegraphics[scale=0.20]{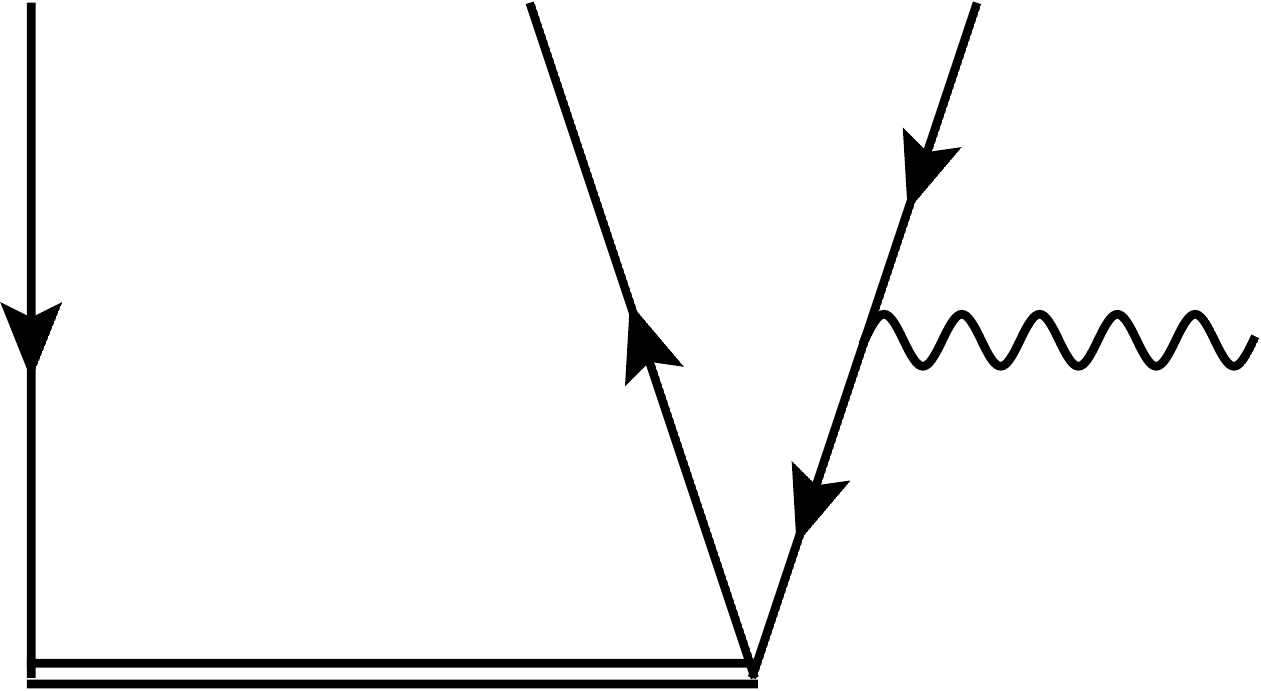}} \\
   & + & \parbox{30mm}{\includegraphics[scale=0.20]{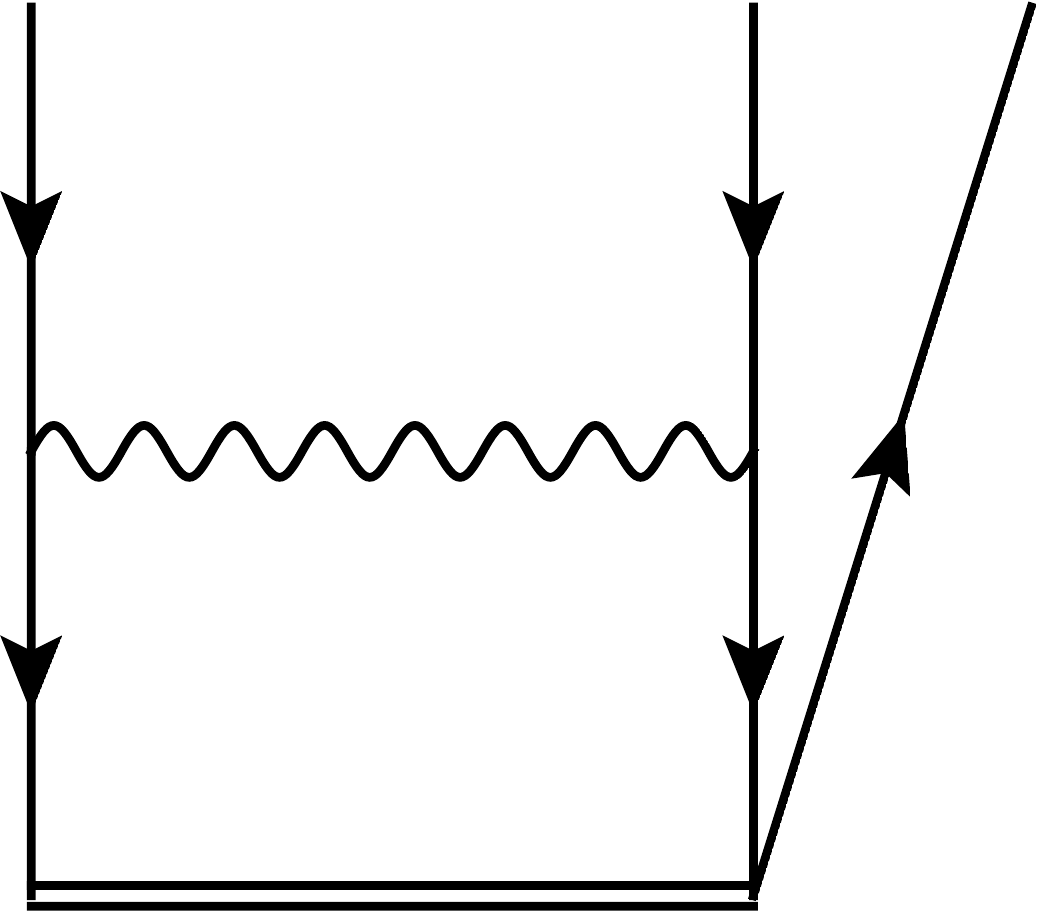}} 
   + \parbox{30mm}{\includegraphics[scale=0.20]{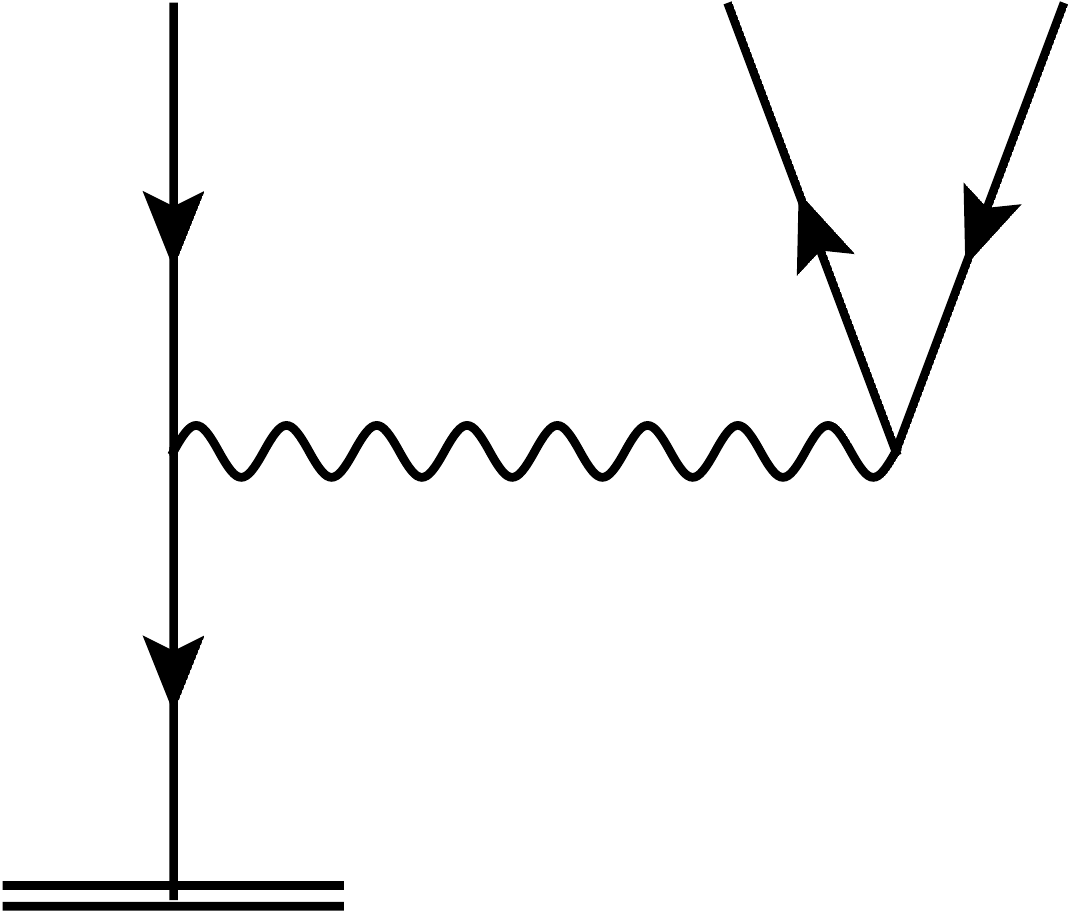}} \\
   & + & \parbox{30mm}{\includegraphics[scale=0.20]{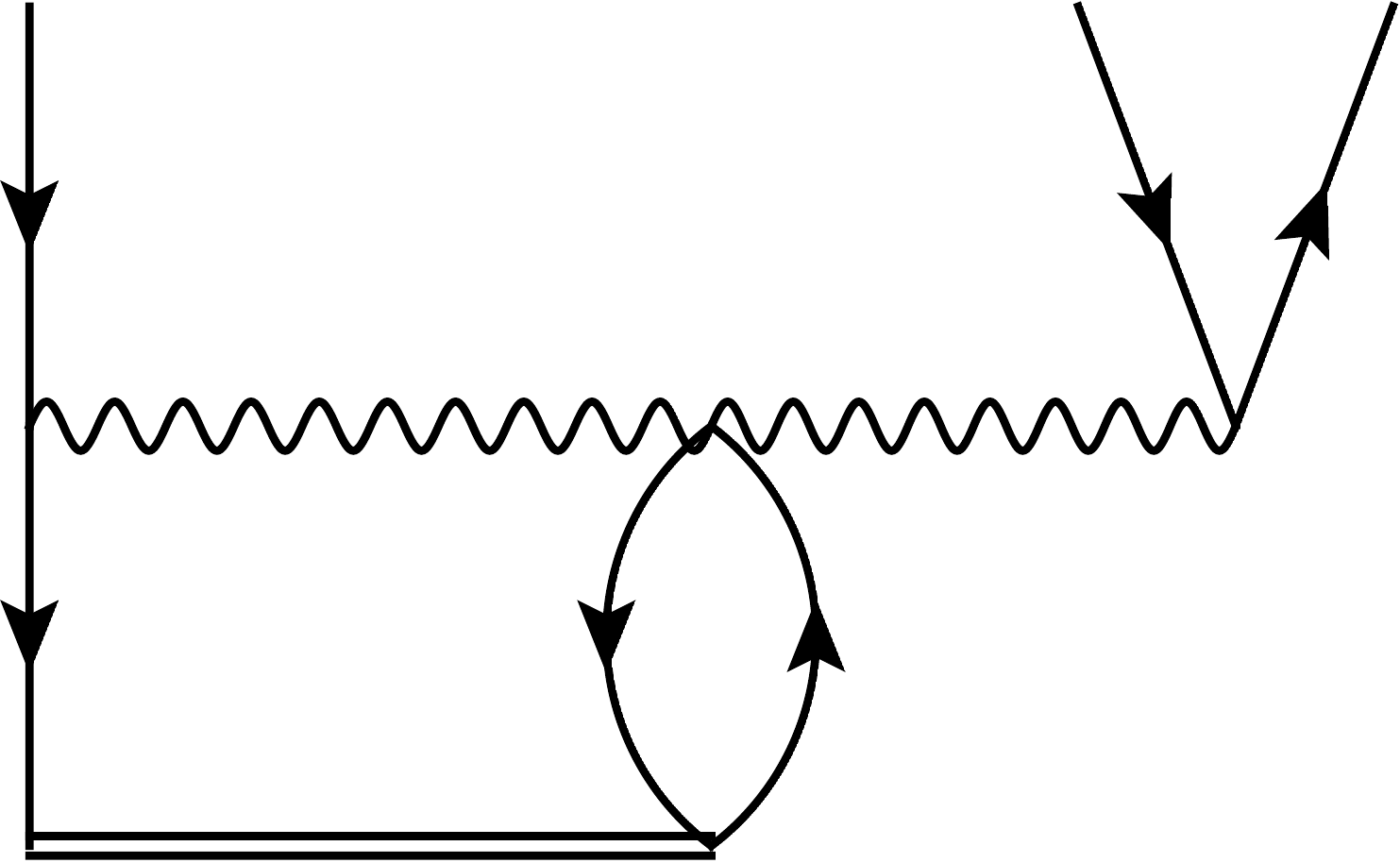}} \ .
   \label{eq:R2h1p_eqns}
\end{eqnarray}
Again, the last diagram involves a three-body term of the
similarity transformed Hamiltonian, and it is therefore convenient to
define the following intermediate,
\begin{equation} 
  \parbox{20mm}{\includegraphics[scale=0.25]{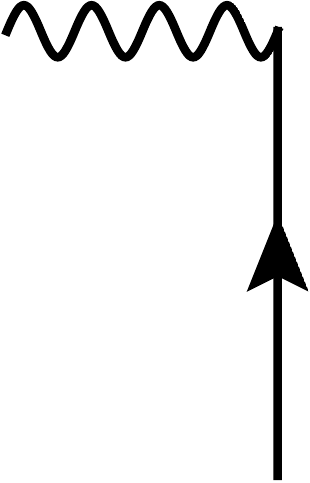}}  
  =   
  \parbox{30mm}{\includegraphics[scale=0.20]{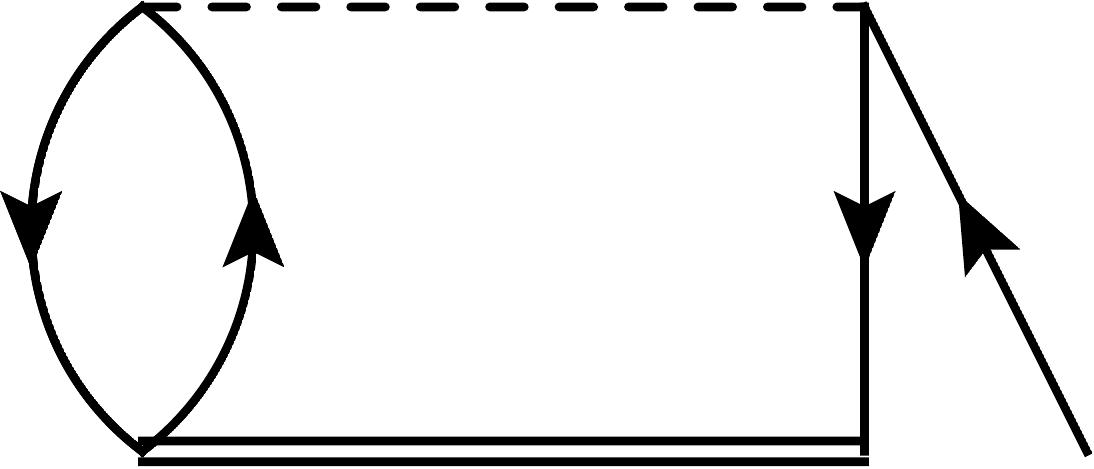}}, 
  \label{eq:threebody_eqns3}
\end{equation}
we can then rewrite the last diagram in Eq.~\ref{eq:R2h1p_eqns} in
the following way,
\begin{equation}
  \parbox{30mm}{\includegraphics[scale=0.20]{Raij_Hbar_threebody_diag}}
  =  \parbox{30mm}{\includegraphics[scale=0.30]{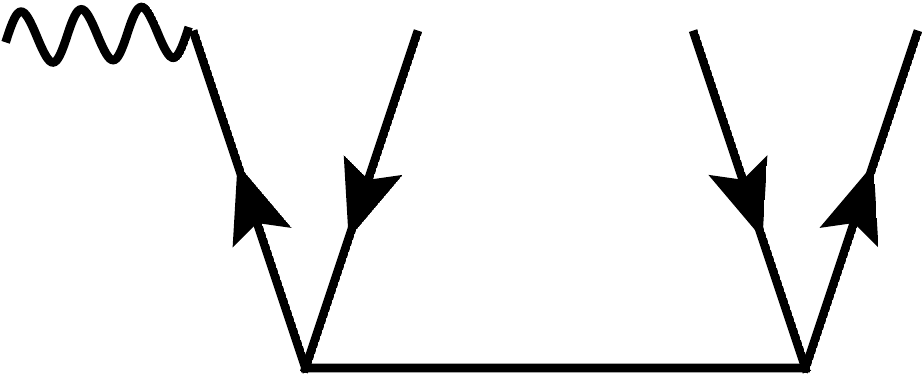}}.
  \label{eq:threebody_int_eqns3}
\end{equation}
In Table~\ref{tab:r1p2h} we give the algebraic expressions for the
one-hole and one-particle-two-hole excitation amplitudes in an
uncoupled and coupled angular momentum scheme,
\begin{table*}[htbp!]
  \begin{Table_app}{0.25}{0.33}{0.42}
    Diagram & \mathrm{Uncoupled~expression} & \mathrm{Coupled~expression} 
    \tabularnewline\hline
    \includegraphics[scale=0.2]{Ri_HbarI7_diag}  & \langle \vert r^{JM}\vert k\rangle
    \langle k\vert\bar{h}^{00}\vert i\rangle  &  \langle \vert\vert r^{J}\vert\vert k\rangle
    \langle k\vert\vert\bar{h}^{0}\vert\vert i\rangle  
    \tabularnewline\hline
    \includegraphics[scale=0.2]{Ri_HbarI5_diag} & \langle c\vert r^{JM}\vert ik\rangle
    \langle k \vert\bar{h}^{00} \vert c\rangle  &   { \hat{J}_{ac}^2\over \hat{j}_a^2} 
    \langle c\vert\vert r^{J}\vert\vert ik\rangle
    \langle k \vert\vert\bar{h}^{0} \vert\vert c\rangle  
    \tabularnewline\hline
    \includegraphics[scale=0.2]{Ri_HbarI3_diag} & -{1\over 2}\langle c\vert r^{JM}\vert kl\rangle
    \langle kl \vert\bar{h}^{00} \vert ic\rangle  &  -{1\over 2}{\hat{J}_{cd}^2\over \hat{j}_a^2}
    \langle c\vert\vert r^{J}\vert\vert kl\rangle
    \langle kl \vert\vert \bar{h}^{0} \vert\vert ic\rangle  \tabularnewline\hline
    \includegraphics[scale=0.2]{Raij_HbarI4_diag} &
    P(ij)\langle kb \vert\bar{h}^{00} \vert cj\rangle \langle c\vert
    r^{JM}\vert ik\rangle & P(ij) \langle i^{-1} \vert\vert r^{J}
    \vert\vert kc^{-1}\rangle \langle kc^{-1} \vert\vert \bar{h}^{0}
    \vert\vert jb^{-1}\rangle \tabularnewline\hline
    \includegraphics[scale=0.2]{Raij_HbarI6_diag}  & 
    \langle b \vert\bar{h}^{00} \vert c\rangle
    \langle c \vert r^{JM} \vert ij\rangle  &  \langle b \vert\vert\bar{h}^{0} \vert\vert c\rangle
    \langle c \vert\vert r^{J} \vert\vert ij\rangle  
    \tabularnewline\hline
    \includegraphics[scale=0.2]{Raij_HbarI7_diag} &  -P(ij)\langle k \vert\bar{h}^{00} \vert j\rangle
    \langle b \vert r^{JM} \vert ik\rangle  &  -P(ij)\langle k \vert\vert\bar{h}^{0} \vert\vert j\rangle
    \langle b \vert\vert r^{J} \vert\vert ik\rangle  
    \tabularnewline\hline
    \includegraphics[scale=0.2]{Raij_HbarI9_diag}  & {1\over 2}\langle kl \vert\bar{h}^{00} \vert ij\rangle
    \langle b\vert r^{JM}\vert kl\rangle   & {1\over 2}\langle kl \vert\vert\bar{h}^{0} \vert\vert ij\rangle
    \langle b\vert\vert r^{J}\vert\vert kl\rangle  
    \tabularnewline\hline
    \includegraphics[scale=0.2]{Raij_HbarI11_diag} & -\langle kb \vert\bar{h}^{00} \vert ij\rangle
    \langle \vert r^{JM}\vert k\rangle   & -\langle kb \vert\vert\bar{h}^{0} \vert\vert ij\rangle
    \langle \vert\vert r^{J}\vert\vert k\rangle   
    \tabularnewline\hline
    \includegraphics[scale=0.2]{R2h1p_Hbar_threebody_int1_diag}
    & -\langle \vert \chi^{JM}\vert c \rangle \langle cb\vert
    t^{00}\vert ij\rangle & -\langle \vert\vert \chi^{J}\vert\vert c \rangle \langle cb\vert\vert
    t^{0}\vert\vert ij\rangle 
    \tabularnewline\hline
  \end{Table_app}
  \caption{Coupled and uncoupled algebraic expressions for the
    diagrams of the one-particle $R_1$, and one-particle-two-hole amplitude $R_2$ given in
    Eqs.~(\ref{eq:R1h_eqns}), and~(\ref{eq:R2h1p_eqns}). Repeated indices are summed over.}
  \label{tab:r1p2h}
\end{table*}
The diagrammatic representation and algebraic expressions for the
various matrix elements of the similarity transformed Hamiltonian that
enter in Table \ref{tab:r1p2h} can be found in
e.g. \cite{bartlett2007,gour2006}.

\newpage

%% file: rop2.bbl
\begin{thebibliography}{300}%
\makeatletter
\providecommand \@ifxundefined [1]{%
 \@ifx{#1\undefined}
}%
\providecommand \@ifnum [1]{%
 \ifnum #1\expandafter \@firstoftwo
 \else \expandafter \@secondoftwo
 \fi
}%
\providecommand \@ifx [1]{%
 \ifx #1\expandafter \@firstoftwo
 \else \expandafter \@secondoftwo
 \fi
}%
\providecommand \natexlab [1]{#1}%
\providecommand \enquote  [1]{``#1''}%
\providecommand \bibnamefont  [1]{#1}%
\providecommand \bibfnamefont [1]{#1}%
\providecommand \citenamefont [1]{#1}%
\providecommand \href@noop [0]{\@secondoftwo}%
\providecommand \href [0]{\begingroup \@sanitize@url \@href}%
\providecommand \@href[1]{\@@startlink{#1}\@@href}%
\providecommand \@@href[1]{\endgroup#1\@@endlink}%
\providecommand \@sanitize@url [0]{\catcode `\\12\catcode `\$12\catcode
  `\&12\catcode `\#12\catcode `\^12\catcode `\_12\catcode `\%12\relax}%
\providecommand \@@startlink[1]{}%
\providecommand \@@endlink[0]{}%
\providecommand \url  [0]{\begingroup\@sanitize@url \@url }%
\providecommand \@url [1]{\endgroup\@href {#1}{\urlprefix }}%
\providecommand \urlprefix  [0]{URL }%
\providecommand \Eprint [0]{\href }%
\providecommand \doibase [0]{http://dx.doi.org/}%
\providecommand \selectlanguage [0]{\@gobble}%
\providecommand \bibinfo  [0]{\@secondoftwo}%
\providecommand \bibfield  [0]{\@secondoftwo}%
\providecommand \translation [1]{[#1]}%
\providecommand \BibitemOpen [0]{}%
\providecommand \bibitemStop [0]{}%
\providecommand \bibitemNoStop [0]{.\EOS\space}%
\providecommand \EOS [0]{\spacefactor3000\relax}%
\providecommand \BibitemShut  [1]{\csname bibitem#1\endcsname}%
\let\auto@bib@innerbib\@empty
\bibitem [{\citenamefont {Ahrens}\ \emph {et~al.}(1975)\citenamefont {Ahrens},
  \citenamefont {Borchert}, \citenamefont {Czock}, \citenamefont {Eppler},
  \citenamefont {Gimm}, \citenamefont {Gundrum}, \citenamefont {Kr{\"o}ning},
  \citenamefont {Riehn}, \citenamefont {Ram}, \citenamefont {Zieger},\ and\
  \citenamefont {Ziegler}}]{ahrens1975}%
  \BibitemOpen
  \bibfield  {author} {\bibinfo {author} {\bibnamefont {Ahrens}, \bibfnamefont
  {J.}}, \bibinfo {author} {\bibfnamefont {H.}~\bibnamefont {Borchert}},
  \bibinfo {author} {\bibfnamefont {K.}~\bibnamefont {Czock}}, \bibinfo
  {author} {\bibfnamefont {H.}~\bibnamefont {Eppler}}, \bibinfo {author}
  {\bibfnamefont {H.}~\bibnamefont {Gimm}}, \bibinfo {author} {\bibfnamefont
  {H.}~\bibnamefont {Gundrum}}, \bibinfo {author} {\bibfnamefont
  {M.}~\bibnamefont {Kr{\"o}ning}}, \bibinfo {author} {\bibfnamefont
  {P.}~\bibnamefont {Riehn}}, \bibinfo {author} {\bibfnamefont {G.~S.}\
  \bibnamefont {Ram}}, \bibinfo {author} {\bibfnamefont {A.}~\bibnamefont
  {Zieger}}, \ and\ \bibinfo {author} {\bibfnamefont {B.}~\bibnamefont
  {Ziegler}}} (\bibinfo {year} {1975}),\ \href {\doibase
  10.1016/0375-9474(75)90543-6} {\bibfield  {journal} {\bibinfo  {journal}
  {Nuclear Physics A}\ }\textbf {\bibinfo {volume} {251}}~(\bibinfo {number}
  {3}),\ \bibinfo {pages} {479 }}\BibitemShut {NoStop}%
\bibitem [{\citenamefont {Arponen}(1983)}]{arponen1983}%
  \BibitemOpen
  \bibfield  {author} {\bibinfo {author} {\bibnamefont {Arponen}, \bibfnamefont
  {J.}}} (\bibinfo {year} {1983}),\ \href {\doibase
  10.1016/0003-4916(83)90284-1} {\bibfield  {journal} {\bibinfo  {journal}
  {Annals of Physics}\ }\textbf {\bibinfo {volume} {151}}~(\bibinfo {number}
  {2}),\ \bibinfo {pages} {311 }}\BibitemShut {NoStop}%
\bibitem [{\citenamefont {Artukh}\ \emph {et~al.}(1970)\citenamefont {Artukh},
  \citenamefont {Avdeichikov}, \citenamefont {Chelnokov}, \citenamefont
  {Gridnev}, \citenamefont {Mikheev}, \citenamefont {Vakatov}, \citenamefont
  {Volkov},\ and\ \citenamefont {Wilczynski}}]{artukh1970}%
  \BibitemOpen
  \bibfield  {author} {\bibinfo {author} {\bibnamefont {Artukh}, \bibfnamefont
  {A.~G.}}, \bibinfo {author} {\bibfnamefont {V.~V.}\ \bibnamefont
  {Avdeichikov}}, \bibinfo {author} {\bibfnamefont {L.~P.}\ \bibnamefont
  {Chelnokov}}, \bibinfo {author} {\bibfnamefont {G.~F.}\ \bibnamefont
  {Gridnev}}, \bibinfo {author} {\bibfnamefont {V.~L.}\ \bibnamefont
  {Mikheev}}, \bibinfo {author} {\bibfnamefont {V.~I.}\ \bibnamefont
  {Vakatov}}, \bibinfo {author} {\bibfnamefont {V.~V.}\ \bibnamefont {Volkov}},
  \ and\ \bibinfo {author} {\bibfnamefont {J.}~\bibnamefont {Wilczynski}}}
  (\bibinfo {year} {1970}),\ \href {\doibase 10.1016/0370-2693(70)90332-1}
  {\bibfield  {journal} {\bibinfo  {journal} {Physics Letters B}\ }\textbf
  {\bibinfo {volume} {32}}~(\bibinfo {number} {1}),\ \bibinfo {pages} {43
  }}\BibitemShut {NoStop}%
\bibitem [{\citenamefont {Baardsen}\ \emph {et~al.}(2013)\citenamefont
  {Baardsen}, \citenamefont {Ekstr\"om}, \citenamefont {Hagen},\ and\
  \citenamefont {Hjorth-Jensen}}]{baardsen2013}%
  \BibitemOpen
  \bibfield  {author} {\bibinfo {author} {\bibnamefont {Baardsen},
  \bibfnamefont {G.}}, \bibinfo {author} {\bibfnamefont {A.}~\bibnamefont
  {Ekstr\"om}}, \bibinfo {author} {\bibfnamefont {G.}~\bibnamefont {Hagen}}, \
  and\ \bibinfo {author} {\bibfnamefont {M.}~\bibnamefont {Hjorth-Jensen}}}
  (\bibinfo {year} {2013}),\ \href {\doibase 10.1103/PhysRevC.88.054312}
  {\bibfield  {journal} {\bibinfo  {journal} {Phys. Rev. C}\ }\textbf {\bibinfo
  {volume} {88}},\ \bibinfo {pages} {054312}}\BibitemShut {NoStop}%
\bibitem [{\citenamefont {Bacca}\ \emph {et~al.}(2013)\citenamefont {Bacca},
  \citenamefont {Barnea}, \citenamefont {Hagen}, \citenamefont {Orlandini},\
  and\ \citenamefont {Papenbrock}}]{bacca2013}%
  \BibitemOpen
  \bibfield  {author} {\bibinfo {author} {\bibnamefont {Bacca}, \bibfnamefont
  {S.}}, \bibinfo {author} {\bibfnamefont {N.}~\bibnamefont {Barnea}}, \bibinfo
  {author} {\bibfnamefont {G.}~\bibnamefont {Hagen}}, \bibinfo {author}
  {\bibfnamefont {G.}~\bibnamefont {Orlandini}}, \ and\ \bibinfo {author}
  {\bibfnamefont {T.}~\bibnamefont {Papenbrock}}} (\bibinfo {year} {2013}),\
  \href {\doibase 10.1103/PhysRevLett.111.122502} {\bibfield  {journal}
  {\bibinfo  {journal} {Phys. Rev. Lett.}\ }\textbf {\bibinfo {volume} {111}},\
  \bibinfo {pages} {122502}}\BibitemShut {NoStop}%
\bibitem [{\citenamefont {Bacca}\ \emph {et~al.}(2009)\citenamefont {Bacca},
  \citenamefont {Schwenk}, \citenamefont {Hagen},\ and\ \citenamefont
  {Papenbrock}}]{bacca2009}%
  \BibitemOpen
  \bibfield  {author} {\bibinfo {author} {\bibnamefont {Bacca}, \bibfnamefont
  {S.}}, \bibinfo {author} {\bibfnamefont {A.}~\bibnamefont {Schwenk}},
  \bibinfo {author} {\bibfnamefont {G.}~\bibnamefont {Hagen}}, \ and\ \bibinfo
  {author} {\bibfnamefont {T.}~\bibnamefont {Papenbrock}}} (\bibinfo {year}
  {2009}),\ \href {\doibase 10.1140/epja/i2009-10815-5} {\bibfield  {journal}
  {\bibinfo  {journal} {The European Physical Journal A}\ }\textbf {\bibinfo
  {volume} {42}}~(\bibinfo {number} {3}),\ \bibinfo {pages} {553}}\BibitemShut
  {NoStop}%
\bibitem [{\citenamefont {Baran}\ \emph {et~al.}(2008)\citenamefont {Baran},
  \citenamefont {Bulgac}, \citenamefont {Forbes}, \citenamefont {Hagen},
  \citenamefont {Nazarewicz}, \citenamefont {Schunck},\ and\ \citenamefont
  {Stoitsov}}]{baran2008}%
  \BibitemOpen
  \bibfield  {author} {\bibinfo {author} {\bibnamefont {Baran}, \bibfnamefont
  {A.}}, \bibinfo {author} {\bibfnamefont {A.}~\bibnamefont {Bulgac}}, \bibinfo
  {author} {\bibfnamefont {M.~M.}\ \bibnamefont {Forbes}}, \bibinfo {author}
  {\bibfnamefont {G.}~\bibnamefont {Hagen}}, \bibinfo {author} {\bibfnamefont
  {W.}~\bibnamefont {Nazarewicz}}, \bibinfo {author} {\bibfnamefont
  {N.}~\bibnamefont {Schunck}}, \ and\ \bibinfo {author} {\bibfnamefont
  {M.~V.}\ \bibnamefont {Stoitsov}}} (\bibinfo {year} {2008}),\ \href {\doibase
  10.1103/PhysRevC.78.014318} {\bibfield  {journal} {\bibinfo  {journal} {Phys.
  Rev. C}\ }\textbf {\bibinfo {volume} {78}},\ \bibinfo {pages}
  {014318}}\BibitemShut {NoStop}%
\bibitem [{\citenamefont {Baranger}(1960)}]{baranger1960}%
  \BibitemOpen
  \bibfield  {author} {\bibinfo {author} {\bibnamefont {Baranger},
  \bibfnamefont {M.}}} (\bibinfo {year} {1960}),\ \href {\doibase
  10.1103/PhysRev.120.957} {\bibfield  {journal} {\bibinfo  {journal} {Phys.
  Rev.}\ }\textbf {\bibinfo {volume} {120}},\ \bibinfo {pages}
  {957}}\BibitemShut {NoStop}%
\bibitem [{\citenamefont {Barbieri}\ and\ \citenamefont
  {Hjorth-Jensen}(2009)}]{barbieri2009}%
  \BibitemOpen
  \bibfield  {author} {\bibinfo {author} {\bibnamefont {Barbieri},
  \bibfnamefont {C.}}, \ and\ \bibinfo {author} {\bibfnamefont
  {M.}~\bibnamefont {Hjorth-Jensen}}} (\bibinfo {year} {2009}),\ \href
  {\doibase 10.1103/PhysRevC.79.064313} {\bibfield  {journal} {\bibinfo
  {journal} {Phys. Rev. C}\ }\textbf {\bibinfo {volume} {79}},\ \bibinfo
  {pages} {064313}}\BibitemShut {NoStop}%
\bibitem [{\citenamefont {Baroni}\ \emph
  {et~al.}(2013{\natexlab{a}})\citenamefont {Baroni}, \citenamefont
  {Navr\'atil},\ and\ \citenamefont {Quaglioni}}]{baroni2013a}%
  \BibitemOpen
  \bibfield  {author} {\bibinfo {author} {\bibnamefont {Baroni}, \bibfnamefont
  {S.}}, \bibinfo {author} {\bibfnamefont {P.}~\bibnamefont {Navr\'atil}}, \
  and\ \bibinfo {author} {\bibfnamefont {S.}~\bibnamefont {Quaglioni}}}
  (\bibinfo {year} {2013}{\natexlab{a}}),\ \href {\doibase
  10.1103/PhysRevLett.110.022505} {\bibfield  {journal} {\bibinfo  {journal}
  {Phys. Rev. Lett.}\ }\textbf {\bibinfo {volume} {110}},\ \bibinfo {pages}
  {022505}}\BibitemShut {NoStop}%
\bibitem [{\citenamefont {Baroni}\ \emph
  {et~al.}(2013{\natexlab{b}})\citenamefont {Baroni}, \citenamefont
  {Navr\'atil},\ and\ \citenamefont {Quaglioni}}]{baroni2013b}%
  \BibitemOpen
  \bibfield  {author} {\bibinfo {author} {\bibnamefont {Baroni}, \bibfnamefont
  {S.}}, \bibinfo {author} {\bibfnamefont {P.}~\bibnamefont {Navr\'atil}}, \
  and\ \bibinfo {author} {\bibfnamefont {S.}~\bibnamefont {Quaglioni}}}
  (\bibinfo {year} {2013}{\natexlab{b}}),\ \href {\doibase
  10.1103/PhysRevC.87.034326} {\bibfield  {journal} {\bibinfo  {journal} {Phys.
  Rev. C}\ }\textbf {\bibinfo {volume} {87}},\ \bibinfo {pages}
  {034326}}\BibitemShut {NoStop}%
\bibitem [{\citenamefont {Barrett}\ \emph {et~al.}(2013)\citenamefont
  {Barrett}, \citenamefont {Navr{\'a}til},\ and\ \citenamefont
  {Vary}}]{barrett2013}%
  \BibitemOpen
  \bibfield  {author} {\bibinfo {author} {\bibnamefont {Barrett}, \bibfnamefont
  {B.~R.}}, \bibinfo {author} {\bibfnamefont {P.}~\bibnamefont {Navr{\'a}til}},
  \ and\ \bibinfo {author} {\bibfnamefont {J.~P.}\ \bibnamefont {Vary}}}
  (\bibinfo {year} {2013}),\ \href {\doibase 10.1016/j.ppnp.2012.10.003}
  {\bibfield  {journal} {\bibinfo  {journal} {Progress in Particle and Nuclear
  Physics}\ }\textbf {\bibinfo {volume} {69}}~(\bibinfo {number} {0}),\
  \bibinfo {pages} {131 }}\BibitemShut {NoStop}%
\bibitem [{\citenamefont {Bartlett}(1981)}]{bartlett1981}%
  \BibitemOpen
  \bibfield  {author} {\bibinfo {author} {\bibnamefont {Bartlett},
  \bibfnamefont {R.~J.}}} (\bibinfo {year} {1981}),\ \href {\doibase
  10.1146/annurev.pc.32.100181.002043} {\bibfield  {journal} {\bibinfo
  {journal} {Annual Review of Physical Chemistry}\ }\textbf {\bibinfo {volume}
  {32}}~(\bibinfo {number} {1}),\ \bibinfo {pages} {359}}\BibitemShut {NoStop}%
\bibitem [{\citenamefont {Bartlett}\ and\ \citenamefont
  {Musia\l{}}(2007)}]{bartlett2007}%
  \BibitemOpen
  \bibfield  {author} {\bibinfo {author} {\bibnamefont {Bartlett},
  \bibfnamefont {R.~J.}}, \ and\ \bibinfo {author} {\bibfnamefont
  {M.}~\bibnamefont {Musia\l{}}}} (\bibinfo {year} {2007}),\ \href {\doibase
  10.1103/RevModPhys.79.291} {\bibfield  {journal} {\bibinfo  {journal} {Rev.
  Mod. Phys.}\ }\textbf {\bibinfo {volume} {79}},\ \bibinfo {pages}
  {291}}\BibitemShut {NoStop}%
\bibitem [{\citenamefont {Bartlett}\ and\ \citenamefont
  {Purvis}(1978)}]{bartlett1978}%
  \BibitemOpen
  \bibfield  {author} {\bibinfo {author} {\bibnamefont {Bartlett},
  \bibfnamefont {R.~J.}}, \ and\ \bibinfo {author} {\bibfnamefont {G.~D.}\
  \bibnamefont {Purvis}}} (\bibinfo {year} {1978}),\ \href {\doibase
  10.1002/qua.560140504} {\bibfield  {journal} {\bibinfo  {journal}
  {International Journal of Quantum Chemistry}\ }\textbf {\bibinfo {volume}
  {14}}~(\bibinfo {number} {5}),\ \bibinfo {pages} {561}}\BibitemShut {NoStop}%
\bibitem [{\citenamefont {Bartlett}\ \emph {et~al.}(1990)\citenamefont
  {Bartlett}, \citenamefont {Watts}, \citenamefont {Kucharski},\ and\
  \citenamefont {Noga}}]{bartlett1990}%
  \BibitemOpen
  \bibfield  {author} {\bibinfo {author} {\bibnamefont {Bartlett},
  \bibfnamefont {R.~J.}}, \bibinfo {author} {\bibfnamefont {J.}~\bibnamefont
  {Watts}}, \bibinfo {author} {\bibfnamefont {S.}~\bibnamefont {Kucharski}}, \
  and\ \bibinfo {author} {\bibfnamefont {J.}~\bibnamefont {Noga}}} (\bibinfo
  {year} {1990}),\ \href {\doibase 10.1016/0009-2614(90)87031-L} {\bibfield
  {journal} {\bibinfo  {journal} {Chemical Physics Letters}\ }\textbf {\bibinfo
  {volume} {165}}~(\bibinfo {number} {6}),\ \bibinfo {pages} {513
  }}\BibitemShut {NoStop}%
\bibitem [{\citenamefont {Bedaque}\ \emph {et~al.}(2003)\citenamefont
  {Bedaque}, \citenamefont {Hammer},\ and\ \citenamefont {van
  Kolck}}]{bedaque2003}%
  \BibitemOpen
  \bibfield  {author} {\bibinfo {author} {\bibnamefont {Bedaque}, \bibfnamefont
  {P.}}, \bibinfo {author} {\bibfnamefont {H.-W.}\ \bibnamefont {Hammer}}, \
  and\ \bibinfo {author} {\bibfnamefont {U.}~\bibnamefont {van Kolck}}}
  (\bibinfo {year} {2003}),\ \href {\doibase 10.1016/j.physletb.2003.07.049}
  {\bibfield  {journal} {\bibinfo  {journal} {Physics Letters B}\ }\textbf
  {\bibinfo {volume} {569}}~(\bibinfo {number} {3–4}),\ \bibinfo {pages} {159
  }}\BibitemShut {NoStop}%
\bibitem [{\citenamefont {Bender}\ \emph {et~al.}(2003)\citenamefont {Bender},
  \citenamefont {Heenen},\ and\ \citenamefont {Reinhard}}]{bender2003}%
  \BibitemOpen
  \bibfield  {author} {\bibinfo {author} {\bibnamefont {Bender}, \bibfnamefont
  {M.}}, \bibinfo {author} {\bibfnamefont {P.-H.}\ \bibnamefont {Heenen}}, \
  and\ \bibinfo {author} {\bibfnamefont {P.-G.}\ \bibnamefont {Reinhard}}}
  (\bibinfo {year} {2003}),\ \href {\doibase 10.1103/RevModPhys.75.121}
  {\bibfield  {journal} {\bibinfo  {journal} {Rev. Mod. Phys.}\ }\textbf
  {\bibinfo {volume} {75}},\ \bibinfo {pages} {121}}\BibitemShut {NoStop}%
\bibitem [{\citenamefont {Berggren}(1968)}]{berggren1968}%
  \BibitemOpen
  \bibfield  {author} {\bibinfo {author} {\bibnamefont {Berggren},
  \bibfnamefont {T.}}} (\bibinfo {year} {1968}),\ \href {\doibase
  10.1016/0375-9474(68)90593-9} {\bibfield  {journal} {\bibinfo  {journal}
  {Nuclear Physics A}\ }\textbf {\bibinfo {volume} {109}}~(\bibinfo {number}
  {2}),\ \bibinfo {pages} {265 }}\BibitemShut {NoStop}%
\bibitem [{\citenamefont {Berggren}(1971)}]{berggren1971}%
  \BibitemOpen
  \bibfield  {author} {\bibinfo {author} {\bibnamefont {Berggren},
  \bibfnamefont {T.}}} (\bibinfo {year} {1971}),\ \href {\doibase
  10.1016/0375-9474(71)90889-X} {\bibfield  {journal} {\bibinfo  {journal}
  {Nuclear Physics A}\ }\textbf {\bibinfo {volume} {169}}~(\bibinfo {number}
  {2}),\ \bibinfo {pages} {353 }}\BibitemShut {NoStop}%
\bibitem [{\citenamefont {Bertulani}\ \emph {et~al.}(2002)\citenamefont
  {Bertulani}, \citenamefont {Hammer},\ and\ \citenamefont {van
  Kolck}}]{bertulani2002}%
  \BibitemOpen
  \bibfield  {author} {\bibinfo {author} {\bibnamefont {Bertulani},
  \bibfnamefont {C.}}, \bibinfo {author} {\bibfnamefont {H.-W.}\ \bibnamefont
  {Hammer}}, \ and\ \bibinfo {author} {\bibfnamefont {U.}~\bibnamefont {van
  Kolck}}} (\bibinfo {year} {2002}),\ \href {\doibase
  10.1016/S0375-9474(02)01270-8} {\bibfield  {journal} {\bibinfo  {journal}
  {Nuclear Physics A}\ }\textbf {\bibinfo {volume} {712}}~(\bibinfo {number}
  {1–2}),\ \bibinfo {pages} {37 }}\BibitemShut {NoStop}%
\bibitem [{\citenamefont {{Binder}}\ \emph {et~al.}(2013)\citenamefont
  {{Binder}}, \citenamefont {{Langhammer}}, \citenamefont {{Calci}},\ and\
  \citenamefont {{Roth}}}]{binder2013b}%
  \BibitemOpen
  \bibfield  {author} {\bibinfo {author} {\bibnamefont {{Binder}},
  \bibfnamefont {S.}}, \bibinfo {author} {\bibfnamefont {J.}~\bibnamefont
  {{Langhammer}}}, \bibinfo {author} {\bibfnamefont {A.}~\bibnamefont
  {{Calci}}}, \ and\ \bibinfo {author} {\bibfnamefont {R.}~\bibnamefont
  {{Roth}}}} (\bibinfo {year} {2013}),\ \href@noop {} {\bibfield  {journal}
  {\bibinfo  {journal} {ArXiv e-prints}\ }}\Eprint
  {http://arxiv.org/abs/1312.5685} {arXiv:1312.5685 [nucl-th]} \BibitemShut
  {NoStop}%
\bibitem [{\citenamefont {Binder}\ \emph {et~al.}(2013)\citenamefont {Binder},
  \citenamefont {Piecuch}, \citenamefont {Calci}, \citenamefont {Langhammer},
  \citenamefont {Navr\'atil},\ and\ \citenamefont {Roth}}]{binder2013}%
  \BibitemOpen
  \bibfield  {author} {\bibinfo {author} {\bibnamefont {Binder}, \bibfnamefont
  {S.}}, \bibinfo {author} {\bibfnamefont {P.}~\bibnamefont {Piecuch}},
  \bibinfo {author} {\bibfnamefont {A.}~\bibnamefont {Calci}}, \bibinfo
  {author} {\bibfnamefont {J.}~\bibnamefont {Langhammer}}, \bibinfo {author}
  {\bibfnamefont {P.}~\bibnamefont {Navr\'atil}}, \ and\ \bibinfo {author}
  {\bibfnamefont {R.}~\bibnamefont {Roth}}} (\bibinfo {year} {2013}),\ \href
  {\doibase 10.1103/PhysRevC.88.054319} {\bibfield  {journal} {\bibinfo
  {journal} {Phys. Rev. C}\ }\textbf {\bibinfo {volume} {88}},\ \bibinfo
  {pages} {054319}}\BibitemShut {NoStop}%
\bibitem [{\citenamefont {Bishop}\ \emph {et~al.}(1998)\citenamefont {Bishop},
  \citenamefont {Guardiola}, \citenamefont {Moliner}, \citenamefont {Navarro},
  \citenamefont {Portesi}, \citenamefont {Puente},\ and\ \citenamefont
  {Walet}}]{bishop1998}%
  \BibitemOpen
  \bibfield  {author} {\bibinfo {author} {\bibnamefont {Bishop}, \bibfnamefont
  {R.}}, \bibinfo {author} {\bibfnamefont {R.}~\bibnamefont {Guardiola}},
  \bibinfo {author} {\bibfnamefont {I.}~\bibnamefont {Moliner}}, \bibinfo
  {author} {\bibfnamefont {J.}~\bibnamefont {Navarro}}, \bibinfo {author}
  {\bibfnamefont {M.}~\bibnamefont {Portesi}}, \bibinfo {author} {\bibfnamefont
  {A.}~\bibnamefont {Puente}}, \ and\ \bibinfo {author} {\bibfnamefont
  {N.}~\bibnamefont {Walet}}} (\bibinfo {year} {1998}),\ \href {\doibase
  10.1016/S0375-9474(98)00562-4} {\bibfield  {journal} {\bibinfo  {journal}
  {Nuclear Physics A}\ }\textbf {\bibinfo {volume} {643}}~(\bibinfo {number}
  {3}),\ \bibinfo {pages} {243 }}\BibitemShut {NoStop}%
\bibitem [{\citenamefont {Bishop}(1991)}]{bishop1991}%
  \BibitemOpen
  \bibfield  {author} {\bibinfo {author} {\bibnamefont {Bishop}, \bibfnamefont
  {R.~F.}}} (\bibinfo {year} {1991}),\ \href
  {http://dx.doi.org/10.1007/BF01119617} {\bibfield  {journal} {\bibinfo
  {journal} {Theoretical Chemistry Accounts: Theory, Computation, and Modeling
  (Theoretica Chimica Acta)}\ }\textbf {\bibinfo {volume} {80}},\ \bibinfo
  {pages} {95}},\ \bibinfo {note} {10.1007/BF01119617}\BibitemShut {NoStop}%
\bibitem [{\citenamefont {Bishop}\ \emph {et~al.}(1992)\citenamefont {Bishop},
  \citenamefont {Buend{\'i}a}, \citenamefont {Flynn},\ and\ \citenamefont
  {Guardiola}}]{bishop1992}%
  \BibitemOpen
  \bibfield  {author} {\bibinfo {author} {\bibnamefont {Bishop}, \bibfnamefont
  {R.~F.}}, \bibinfo {author} {\bibfnamefont {E.}~\bibnamefont {Buend{\'i}a}},
  \bibinfo {author} {\bibfnamefont {M.~F.}\ \bibnamefont {Flynn}}, \ and\
  \bibinfo {author} {\bibfnamefont {R.}~\bibnamefont {Guardiola}}} (\bibinfo
  {year} {1992}),\ \href {http://stacks.iop.org/0954-3899/18/i=7/a=007}
  {\bibfield  {journal} {\bibinfo  {journal} {Journal of Physics G: Nuclear and
  Particle Physics}\ }\textbf {\bibinfo {volume} {18}}~(\bibinfo {number}
  {7}),\ \bibinfo {pages} {1157}}\BibitemShut {NoStop}%
\bibitem [{\citenamefont {Bishop}\ \emph
  {et~al.}(1990{\natexlab{a}})\citenamefont {Bishop}, \citenamefont {Flynn},
  \citenamefont {Bosca}, \citenamefont {Buendia},\ and\ \citenamefont
  {Guardiola}}]{bishop1990b}%
  \BibitemOpen
  \bibfield  {author} {\bibinfo {author} {\bibnamefont {Bishop}, \bibfnamefont
  {R.~F.}}, \bibinfo {author} {\bibfnamefont {M.~F.}\ \bibnamefont {Flynn}},
  \bibinfo {author} {\bibfnamefont {M.~C.}\ \bibnamefont {Bosca}}, \bibinfo
  {author} {\bibfnamefont {E.}~\bibnamefont {Buendia}}, \ and\ \bibinfo
  {author} {\bibfnamefont {R.}~\bibnamefont {Guardiola}}} (\bibinfo {year}
  {1990}{\natexlab{a}}),\ \href {http://stacks.iop.org/0954-3899/16/i=3/a=005}
  {\bibfield  {journal} {\bibinfo  {journal} {Journal of Physics G: Nuclear and
  Particle Physics}\ }\textbf {\bibinfo {volume} {16}}~(\bibinfo {number}
  {3}),\ \bibinfo {pages} {L61}}\BibitemShut {NoStop}%
\bibitem [{\citenamefont {Bishop}\ \emph
  {et~al.}(1990{\natexlab{b}})\citenamefont {Bishop}, \citenamefont {Flynn},
  \citenamefont {Bosc{\'a}}, \citenamefont {Buend{\'i}a},\ and\ \citenamefont
  {Guardiola}}]{bishop1990}%
  \BibitemOpen
  \bibfield  {author} {\bibinfo {author} {\bibnamefont {Bishop}, \bibfnamefont
  {R.~F.}}, \bibinfo {author} {\bibfnamefont {M.~F.}\ \bibnamefont {Flynn}},
  \bibinfo {author} {\bibfnamefont {M.~C.}\ \bibnamefont {Bosc{\'a}}}, \bibinfo
  {author} {\bibfnamefont {E.}~\bibnamefont {Buend{\'i}a}}, \ and\ \bibinfo
  {author} {\bibfnamefont {R.}~\bibnamefont {Guardiola}}} (\bibinfo {year}
  {1990}{\natexlab{b}}),\ \href {\doibase 10.1103/PhysRevC.42.1341} {\bibfield
  {journal} {\bibinfo  {journal} {Phys. Rev. C}\ }\textbf {\bibinfo {volume}
  {42}},\ \bibinfo {pages} {1341}}\BibitemShut {NoStop}%
\bibitem [{\citenamefont {Bishop}\ and\ \citenamefont
  {L\"uhrmann}(1978)}]{bishop1978}%
  \BibitemOpen
  \bibfield  {author} {\bibinfo {author} {\bibnamefont {Bishop}, \bibfnamefont
  {R.~F.}}, \ and\ \bibinfo {author} {\bibfnamefont {K.~H.}\ \bibnamefont
  {L\"uhrmann}}} (\bibinfo {year} {1978}),\ \href {\doibase
  10.1103/PhysRevB.17.3757} {\bibfield  {journal} {\bibinfo  {journal} {Phys.
  Rev. B}\ }\textbf {\bibinfo {volume} {17}},\ \bibinfo {pages}
  {3757}}\BibitemShut {NoStop}%
\bibitem [{\citenamefont {Blatt}\ and\ \citenamefont
  {McKellar}(1975)}]{blatt1975}%
  \BibitemOpen
  \bibfield  {author} {\bibinfo {author} {\bibnamefont {Blatt}, \bibfnamefont
  {D.~W.~E.}}, \ and\ \bibinfo {author} {\bibfnamefont {B.~H.~J.}\ \bibnamefont
  {McKellar}}} (\bibinfo {year} {1975}),\ \href {\doibase
  10.1103/PhysRevC.11.614} {\bibfield  {journal} {\bibinfo  {journal} {Phys.
  Rev. C}\ }\textbf {\bibinfo {volume} {11}},\ \bibinfo {pages}
  {614}}\BibitemShut {NoStop}%
\bibitem [{\citenamefont {Bogner}\ \emph {et~al.}(2013)\citenamefont {Bogner},
  \citenamefont {Bulgac}, \citenamefont {Carlson}, \citenamefont {Engel},
  \citenamefont {Fann}, \citenamefont {Furnstahl}, \citenamefont {Gandolfi},
  \citenamefont {Hagen}, \citenamefont {Horoi}, \citenamefont {Johnson},
  \citenamefont {Kortelainen}, \citenamefont {Lusk}, \citenamefont {Maris},
  \citenamefont {Nam}, \citenamefont {Navratil}, \citenamefont {Nazarewicz},
  \citenamefont {Ng}, \citenamefont {Nobre}, \citenamefont {Ormand},
  \citenamefont {Papenbrock}, \citenamefont {Pei}, \citenamefont {Pieper},
  \citenamefont {Quaglioni}, \citenamefont {Roche}, \citenamefont {Sarich},
  \citenamefont {Schunck}, \citenamefont {Sosonkina}, \citenamefont {Terasaki},
  \citenamefont {Thompson}, \citenamefont {Vary},\ and\ \citenamefont
  {Wild}}]{bogner2013}%
  \BibitemOpen
  \bibfield  {author} {\bibinfo {author} {\bibnamefont {Bogner}, \bibfnamefont
  {S.}}, \bibinfo {author} {\bibfnamefont {A.}~\bibnamefont {Bulgac}}, \bibinfo
  {author} {\bibfnamefont {J.}~\bibnamefont {Carlson}}, \bibinfo {author}
  {\bibfnamefont {J.}~\bibnamefont {Engel}}, \bibinfo {author} {\bibfnamefont
  {G.}~\bibnamefont {Fann}}, \bibinfo {author} {\bibfnamefont {R.}~\bibnamefont
  {Furnstahl}}, \bibinfo {author} {\bibfnamefont {S.}~\bibnamefont {Gandolfi}},
  \bibinfo {author} {\bibfnamefont {G.}~\bibnamefont {Hagen}}, \bibinfo
  {author} {\bibfnamefont {M.}~\bibnamefont {Horoi}}, \bibinfo {author}
  {\bibfnamefont {C.}~\bibnamefont {Johnson}}, \bibinfo {author} {\bibfnamefont
  {M.}~\bibnamefont {Kortelainen}}, \bibinfo {author} {\bibfnamefont
  {E.}~\bibnamefont {Lusk}}, \bibinfo {author} {\bibfnamefont {P.}~\bibnamefont
  {Maris}}, \bibinfo {author} {\bibfnamefont {H.}~\bibnamefont {Nam}}, \bibinfo
  {author} {\bibfnamefont {P.}~\bibnamefont {Navratil}}, \bibinfo {author}
  {\bibfnamefont {W.}~\bibnamefont {Nazarewicz}}, \bibinfo {author}
  {\bibfnamefont {E.}~\bibnamefont {Ng}}, \bibinfo {author} {\bibfnamefont
  {G.}~\bibnamefont {Nobre}}, \bibinfo {author} {\bibfnamefont
  {E.}~\bibnamefont {Ormand}}, \bibinfo {author} {\bibfnamefont
  {T.}~\bibnamefont {Papenbrock}}, \bibinfo {author} {\bibfnamefont
  {J.}~\bibnamefont {Pei}}, \bibinfo {author} {\bibfnamefont {S.}~\bibnamefont
  {Pieper}}, \bibinfo {author} {\bibfnamefont {S.}~\bibnamefont {Quaglioni}},
  \bibinfo {author} {\bibfnamefont {K.}~\bibnamefont {Roche}}, \bibinfo
  {author} {\bibfnamefont {J.}~\bibnamefont {Sarich}}, \bibinfo {author}
  {\bibfnamefont {N.}~\bibnamefont {Schunck}}, \bibinfo {author} {\bibfnamefont
  {M.}~\bibnamefont {Sosonkina}}, \bibinfo {author} {\bibfnamefont
  {J.}~\bibnamefont {Terasaki}}, \bibinfo {author} {\bibfnamefont
  {I.}~\bibnamefont {Thompson}}, \bibinfo {author} {\bibfnamefont
  {J.}~\bibnamefont {Vary}}, \ and\ \bibinfo {author} {\bibfnamefont
  {S.}~\bibnamefont {Wild}}} (\bibinfo {year} {2013}),\ \href {\doibase
  10.1016/j.cpc.2013.05.020} {\bibfield  {journal} {\bibinfo  {journal}
  {Computer Physics Communications}\ }\textbf {\bibinfo {volume}
  {184}}~(\bibinfo {number} {10}),\ \bibinfo {pages} {2235 }}\BibitemShut
  {NoStop}%
\bibitem [{\citenamefont {Bogner}\ \emph {et~al.}(2008)\citenamefont {Bogner},
  \citenamefont {Furnstahl}, \citenamefont {Maris}, \citenamefont {Perry},
  \citenamefont {Schwenk},\ and\ \citenamefont {Vary}}]{bogner2008}%
  \BibitemOpen
  \bibfield  {author} {\bibinfo {author} {\bibnamefont {Bogner}, \bibfnamefont
  {S.}}, \bibinfo {author} {\bibfnamefont {R.}~\bibnamefont {Furnstahl}},
  \bibinfo {author} {\bibfnamefont {P.}~\bibnamefont {Maris}}, \bibinfo
  {author} {\bibfnamefont {R.}~\bibnamefont {Perry}}, \bibinfo {author}
  {\bibfnamefont {A.}~\bibnamefont {Schwenk}}, \ and\ \bibinfo {author}
  {\bibfnamefont {J.}~\bibnamefont {Vary}}} (\bibinfo {year} {2008}),\ \href
  {\doibase 10.1016/j.nuclphysa.2007.12.008} {\bibfield  {journal} {\bibinfo
  {journal} {Nuclear Physics A}\ }\textbf {\bibinfo {volume} {801}}~(\bibinfo
  {number} {1–2}),\ \bibinfo {pages} {21 }}\BibitemShut {NoStop}%
\bibitem [{\citenamefont {Bogner}\ \emph {et~al.}(2010)\citenamefont {Bogner},
  \citenamefont {Furnstahl},\ and\ \citenamefont {Schwenk}}]{bogner2010}%
  \BibitemOpen
  \bibfield  {author} {\bibinfo {author} {\bibnamefont {Bogner}, \bibfnamefont
  {S.}}, \bibinfo {author} {\bibfnamefont {R.}~\bibnamefont {Furnstahl}}, \
  and\ \bibinfo {author} {\bibfnamefont {A.}~\bibnamefont {Schwenk}}} (\bibinfo
  {year} {2010}),\ \href {\doibase 10.1016/j.ppnp.2010.03.001} {\bibfield
  {journal} {\bibinfo  {journal} {Progress in Particle and Nuclear Physics}\
  }\textbf {\bibinfo {volume} {65}}~(\bibinfo {number} {1}),\ \bibinfo {pages}
  {94 }}\BibitemShut {NoStop}%
\bibitem [{\citenamefont {Bogner}\ \emph {et~al.}(2007)\citenamefont {Bogner},
  \citenamefont {Furnstahl},\ and\ \citenamefont {Perry}}]{bogner2007}%
  \BibitemOpen
  \bibfield  {author} {\bibinfo {author} {\bibnamefont {Bogner}, \bibfnamefont
  {S.~K.}}, \bibinfo {author} {\bibfnamefont {R.~J.}\ \bibnamefont
  {Furnstahl}}, \ and\ \bibinfo {author} {\bibfnamefont {R.~J.}\ \bibnamefont
  {Perry}}} (\bibinfo {year} {2007}),\ \href {\doibase
  10.1103/PhysRevC.75.061001} {\bibfield  {journal} {\bibinfo  {journal} {Phys.
  Rev. C}\ }\textbf {\bibinfo {volume} {75}},\ \bibinfo {pages}
  {061001}}\BibitemShut {NoStop}%
\bibitem [{\citenamefont {Bogner}\ \emph {et~al.}(2003)\citenamefont {Bogner},
  \citenamefont {Kuo},\ and\ \citenamefont {Schwenk}}]{bogner2003}%
  \BibitemOpen
  \bibfield  {author} {\bibinfo {author} {\bibnamefont {Bogner}, \bibfnamefont
  {S.~K.}}, \bibinfo {author} {\bibfnamefont {T.~T.~S.}\ \bibnamefont {Kuo}}, \
  and\ \bibinfo {author} {\bibfnamefont {A.}~\bibnamefont {Schwenk}}} (\bibinfo
  {year} {2003}),\ \href {\doibase 10.1016/j.physrep.2003.07.001} {\bibfield
  {journal} {\bibinfo  {journal} {Physics Reports}\ }\textbf {\bibinfo {volume}
  {386}}~(\bibinfo {number} {1}),\ \bibinfo {pages} {1 }}\BibitemShut {NoStop}%
\bibitem [{\citenamefont {Brown}\ and\ \citenamefont
  {Richter}(2006)}]{brown2006}%
  \BibitemOpen
  \bibfield  {author} {\bibinfo {author} {\bibnamefont {Brown}, \bibfnamefont
  {B.~A.}}, \ and\ \bibinfo {author} {\bibfnamefont {W.~A.}\ \bibnamefont
  {Richter}}} (\bibinfo {year} {2006}),\ \href {\doibase
  10.1103/PhysRevC.74.034315} {\bibfield  {journal} {\bibinfo  {journal}
  {Physical Review C (Nuclear Physics)}\ }\textbf {\bibinfo {volume}
  {74}}~(\bibinfo {number} {3}),\ \bibinfo {eid} {034315}}\BibitemShut
  {NoStop}%
\bibitem [{\citenamefont {Brown}\ and\ \citenamefont
  {Green}(1966)}]{brown1966}%
  \BibitemOpen
  \bibfield  {author} {\bibinfo {author} {\bibnamefont {Brown}, \bibfnamefont
  {G.}}, \ and\ \bibinfo {author} {\bibfnamefont {A.}~\bibnamefont {Green}}}
  (\bibinfo {year} {1966}),\ \href {\doibase 10.1016/0029-5582(66)90771-1}
  {\bibfield  {journal} {\bibinfo  {journal} {Nuclear Physics}\ }\textbf
  {\bibinfo {volume} {75}}~(\bibinfo {number} {2}),\ \bibinfo {pages} {401
  }}\BibitemShut {NoStop}%
\bibitem [{\citenamefont {Broyden}(1965)}]{broyden1965}%
  \BibitemOpen
  \bibfield  {author} {\bibinfo {author} {\bibnamefont {Broyden}, \bibfnamefont
  {C.~G.}}} (\bibinfo {year} {1965}),\ \href
  {http://dx.doi.org/10.1090/S0025-5718-1965-0198670-6} {\bibfield  {journal}
  {\bibinfo  {journal} {Math. Comp.}\ }\textbf {\bibinfo {volume} {19}},\
  \bibinfo {pages} {577}}\BibitemShut {NoStop}%
\bibitem [{\citenamefont {Brueckner}(1955)}]{brueckner1955}%
  \BibitemOpen
  \bibfield  {author} {\bibinfo {author} {\bibnamefont {Brueckner},
  \bibfnamefont {K.~A.}}} (\bibinfo {year} {1955}),\ \href {\doibase
  10.1103/PhysRev.100.36} {\bibfield  {journal} {\bibinfo  {journal} {Phys.
  Rev.}\ }\textbf {\bibinfo {volume} {100}},\ \bibinfo {pages}
  {36}}\BibitemShut {NoStop}%
\bibitem [{\citenamefont {Brueckner}\ \emph {et~al.}(1954)\citenamefont
  {Brueckner}, \citenamefont {Levinson},\ and\ \citenamefont
  {Mahmoud}}]{brueckner1954}%
  \BibitemOpen
  \bibfield  {author} {\bibinfo {author} {\bibnamefont {Brueckner},
  \bibfnamefont {K.~A.}}, \bibinfo {author} {\bibfnamefont {C.~A.}\
  \bibnamefont {Levinson}}, \ and\ \bibinfo {author} {\bibfnamefont {H.~M.}\
  \bibnamefont {Mahmoud}}} (\bibinfo {year} {1954}),\ \href {\doibase
  10.1103/PhysRev.95.217} {\bibfield  {journal} {\bibinfo  {journal} {Phys.
  Rev.}\ }\textbf {\bibinfo {volume} {95}},\ \bibinfo {pages}
  {217}}\BibitemShut {NoStop}%
\bibitem [{\citenamefont {Caesar}\ \emph {et~al.}(2013)\citenamefont {Caesar},
  \citenamefont {Simonis}, \citenamefont {Adachi}, \citenamefont {Aksyutina},
  \citenamefont {Alcantara}, \citenamefont {Altstadt}, \citenamefont
  {Alvarez-Pol}, \citenamefont {Ashwood}, \citenamefont {Aumann}, \citenamefont
  {Avdeichikov}, \citenamefont {Barr}, \citenamefont {Beceiro}, \citenamefont
  {Bemmerer}, \citenamefont {Benlliure}, \citenamefont {Bertulani},
  \citenamefont {Boretzky}, \citenamefont {Borge}, \citenamefont {Burgunder},
  \citenamefont {Caamano}, \citenamefont {Casarejos}, \citenamefont {Catford},
  \citenamefont {Cederk\"all}, \citenamefont {Chakraborty}, \citenamefont
  {Chartier}, \citenamefont {Chulkov}, \citenamefont {Cortina-Gil},
  \citenamefont {Datta~Pramanik}, \citenamefont {Diaz~Fernandez}, \citenamefont
  {Dillmann}, \citenamefont {Elekes}, \citenamefont {Enders}, \citenamefont
  {Ershova}, \citenamefont {Estrade}, \citenamefont {Farinon}, \citenamefont
  {Fraile}, \citenamefont {Freer}, \citenamefont {Freudenberger}, \citenamefont
  {Fynbo}, \citenamefont {Galaviz}, \citenamefont {Geissel}, \citenamefont
  {Gernh\"auser}, \citenamefont {Golubev}, \citenamefont {Gonzalez~Diaz},
  \citenamefont {Hagdahl}, \citenamefont {Heftrich}, \citenamefont {Heil},
  \citenamefont {Heine}, \citenamefont {Heinz}, \citenamefont {Henriques},
  \citenamefont {Holl}, \citenamefont {Holt}, \citenamefont {Ickert},
  \citenamefont {Ignatov}, \citenamefont {Jakobsson}, \citenamefont
  {Johansson}, \citenamefont {Jonson}, \citenamefont {Kalantar-Nayestanaki},
  \citenamefont {Kanungo}, \citenamefont {Kelic-Heil}, \citenamefont
  {Kn\"obel}, \citenamefont {Kr\"oll}, \citenamefont {Kr\"ucken}, \citenamefont
  {Kurcewicz}, \citenamefont {Labiche}, \citenamefont {Langer}, \citenamefont
  {Le~Bleis}, \citenamefont {Lemmon}, \citenamefont {Lepyoshkina},
  \citenamefont {Lindberg}, \citenamefont {Machado}, \citenamefont {Marganiec},
  \citenamefont {Maroussov}, \citenamefont {Men\'endez}, \citenamefont
  {Mostazo}, \citenamefont {Movsesyan}, \citenamefont {Najafi}, \citenamefont
  {Nilsson}, \citenamefont {Nociforo}, \citenamefont {Panin}, \citenamefont
  {Perea}, \citenamefont {Pietri}, \citenamefont {Plag}, \citenamefont
  {Prochazka}, \citenamefont {Rahaman}, \citenamefont {Rastrepina},
  \citenamefont {Reifarth}, \citenamefont {Ribeiro}, \citenamefont {Ricciardi},
  \citenamefont {Rigollet}, \citenamefont {Riisager}, \citenamefont {R\"oder},
  \citenamefont {Rossi}, \citenamefont {Sanchez~del Rio}, \citenamefont
  {Savran}, \citenamefont {Scheit}, \citenamefont {Schwenk}, \citenamefont
  {Simon}, \citenamefont {Sorlin}, \citenamefont {Stoica}, \citenamefont
  {Streicher}, \citenamefont {Taylor}, \citenamefont {Tengblad}, \citenamefont
  {Terashima}, \citenamefont {Thies}, \citenamefont {Togano}, \citenamefont
  {Uberseder}, \citenamefont {Van~de Walle}, \citenamefont {Velho},
  \citenamefont {Volkov}, \citenamefont {Wagner}, \citenamefont {Wamers},
  \citenamefont {Weick}, \citenamefont {Weigand}, \citenamefont {Wheldon},
  \citenamefont {Wilson}, \citenamefont {Wimmer}, \citenamefont {Winfield},
  \citenamefont {Woods}, \citenamefont {Yakorev}, \citenamefont {Zhukov},
  \citenamefont {Zilges}, \citenamefont {Zoric},\ and\ \citenamefont
  {Zuber}}]{caesar2013}%
  \BibitemOpen
  \bibfield  {author} {\bibinfo {author} {\bibnamefont {Caesar}, \bibfnamefont
  {C.}}, \bibinfo {author} {\bibfnamefont {J.}~\bibnamefont {Simonis}},
  \bibinfo {author} {\bibfnamefont {T.}~\bibnamefont {Adachi}}, \bibinfo
  {author} {\bibfnamefont {Y.}~\bibnamefont {Aksyutina}}, \bibinfo {author}
  {\bibfnamefont {J.}~\bibnamefont {Alcantara}}, \bibinfo {author}
  {\bibfnamefont {S.}~\bibnamefont {Altstadt}}, \bibinfo {author}
  {\bibfnamefont {H.}~\bibnamefont {Alvarez-Pol}}, \bibinfo {author}
  {\bibfnamefont {N.}~\bibnamefont {Ashwood}}, \bibinfo {author} {\bibfnamefont
  {T.}~\bibnamefont {Aumann}}, \bibinfo {author} {\bibfnamefont
  {V.}~\bibnamefont {Avdeichikov}}, \bibinfo {author} {\bibfnamefont
  {M.}~\bibnamefont {Barr}}, \bibinfo {author} {\bibfnamefont {S.}~\bibnamefont
  {Beceiro}}, \bibinfo {author} {\bibfnamefont {D.}~\bibnamefont {Bemmerer}},
  \bibinfo {author} {\bibfnamefont {J.}~\bibnamefont {Benlliure}}, \bibinfo
  {author} {\bibfnamefont {C.~A.}\ \bibnamefont {Bertulani}}, \bibinfo {author}
  {\bibfnamefont {K.}~\bibnamefont {Boretzky}}, \bibinfo {author}
  {\bibfnamefont {M.~J.~G.}\ \bibnamefont {Borge}}, \bibinfo {author}
  {\bibfnamefont {G.}~\bibnamefont {Burgunder}}, \bibinfo {author}
  {\bibfnamefont {M.}~\bibnamefont {Caamano}}, \bibinfo {author} {\bibfnamefont
  {E.}~\bibnamefont {Casarejos}}, \bibinfo {author} {\bibfnamefont
  {W.}~\bibnamefont {Catford}}, \bibinfo {author} {\bibfnamefont
  {J.}~\bibnamefont {Cederk\"all}}, \bibinfo {author} {\bibfnamefont
  {S.}~\bibnamefont {Chakraborty}}, \bibinfo {author} {\bibfnamefont
  {M.}~\bibnamefont {Chartier}}, \bibinfo {author} {\bibfnamefont
  {L.}~\bibnamefont {Chulkov}}, \bibinfo {author} {\bibfnamefont
  {D.}~\bibnamefont {Cortina-Gil}}, \bibinfo {author} {\bibfnamefont
  {U.}~\bibnamefont {Datta~Pramanik}}, \bibinfo {author} {\bibfnamefont
  {P.}~\bibnamefont {Diaz~Fernandez}}, \bibinfo {author} {\bibfnamefont
  {I.}~\bibnamefont {Dillmann}}, \bibinfo {author} {\bibfnamefont
  {Z.}~\bibnamefont {Elekes}}, \bibinfo {author} {\bibfnamefont
  {J.}~\bibnamefont {Enders}}, \bibinfo {author} {\bibfnamefont
  {O.}~\bibnamefont {Ershova}}, \bibinfo {author} {\bibfnamefont
  {A.}~\bibnamefont {Estrade}}, \bibinfo {author} {\bibfnamefont
  {F.}~\bibnamefont {Farinon}}, \bibinfo {author} {\bibfnamefont {L.~M.}\
  \bibnamefont {Fraile}}, \bibinfo {author} {\bibfnamefont {M.}~\bibnamefont
  {Freer}}, \bibinfo {author} {\bibfnamefont {M.}~\bibnamefont
  {Freudenberger}}, \bibinfo {author} {\bibfnamefont {H.~O.~U.}\ \bibnamefont
  {Fynbo}}, \bibinfo {author} {\bibfnamefont {D.}~\bibnamefont {Galaviz}},
  \bibinfo {author} {\bibfnamefont {H.}~\bibnamefont {Geissel}}, \bibinfo
  {author} {\bibfnamefont {R.}~\bibnamefont {Gernh\"auser}}, \bibinfo {author}
  {\bibfnamefont {P.}~\bibnamefont {Golubev}}, \bibinfo {author} {\bibfnamefont
  {D.}~\bibnamefont {Gonzalez~Diaz}}, \bibinfo {author} {\bibfnamefont
  {J.}~\bibnamefont {Hagdahl}}, \bibinfo {author} {\bibfnamefont
  {T.}~\bibnamefont {Heftrich}}, \bibinfo {author} {\bibfnamefont
  {M.}~\bibnamefont {Heil}}, \bibinfo {author} {\bibfnamefont {M.}~\bibnamefont
  {Heine}}, \bibinfo {author} {\bibfnamefont {A.}~\bibnamefont {Heinz}},
  \bibinfo {author} {\bibfnamefont {A.}~\bibnamefont {Henriques}}, \bibinfo
  {author} {\bibfnamefont {M.}~\bibnamefont {Holl}}, \bibinfo {author}
  {\bibfnamefont {J.~D.}\ \bibnamefont {Holt}}, \bibinfo {author}
  {\bibfnamefont {G.}~\bibnamefont {Ickert}}, \bibinfo {author} {\bibfnamefont
  {A.}~\bibnamefont {Ignatov}}, \bibinfo {author} {\bibfnamefont
  {B.}~\bibnamefont {Jakobsson}}, \bibinfo {author} {\bibfnamefont {H.~T.}\
  \bibnamefont {Johansson}}, \bibinfo {author} {\bibfnamefont {B.}~\bibnamefont
  {Jonson}}, \bibinfo {author} {\bibfnamefont {N.}~\bibnamefont
  {Kalantar-Nayestanaki}}, \bibinfo {author} {\bibfnamefont {R.}~\bibnamefont
  {Kanungo}}, \bibinfo {author} {\bibfnamefont {A.}~\bibnamefont {Kelic-Heil}},
  \bibinfo {author} {\bibfnamefont {R.}~\bibnamefont {Kn\"obel}}, \bibinfo
  {author} {\bibfnamefont {T.}~\bibnamefont {Kr\"oll}}, \bibinfo {author}
  {\bibfnamefont {R.}~\bibnamefont {Kr\"ucken}}, \bibinfo {author}
  {\bibfnamefont {J.}~\bibnamefont {Kurcewicz}}, \bibinfo {author}
  {\bibfnamefont {M.}~\bibnamefont {Labiche}}, \bibinfo {author} {\bibfnamefont
  {C.}~\bibnamefont {Langer}}, \bibinfo {author} {\bibfnamefont
  {T.}~\bibnamefont {Le~Bleis}}, \bibinfo {author} {\bibfnamefont
  {R.}~\bibnamefont {Lemmon}}, \bibinfo {author} {\bibfnamefont
  {O.}~\bibnamefont {Lepyoshkina}}, \bibinfo {author} {\bibfnamefont
  {S.}~\bibnamefont {Lindberg}}, \bibinfo {author} {\bibfnamefont
  {J.}~\bibnamefont {Machado}}, \bibinfo {author} {\bibfnamefont
  {J.}~\bibnamefont {Marganiec}}, \bibinfo {author} {\bibfnamefont
  {V.}~\bibnamefont {Maroussov}}, \bibinfo {author} {\bibfnamefont
  {J.}~\bibnamefont {Men\'endez}}, \bibinfo {author} {\bibfnamefont
  {M.}~\bibnamefont {Mostazo}}, \bibinfo {author} {\bibfnamefont
  {A.}~\bibnamefont {Movsesyan}}, \bibinfo {author} {\bibfnamefont
  {A.}~\bibnamefont {Najafi}}, \bibinfo {author} {\bibfnamefont
  {T.}~\bibnamefont {Nilsson}}, \bibinfo {author} {\bibfnamefont
  {C.}~\bibnamefont {Nociforo}}, \bibinfo {author} {\bibfnamefont
  {V.}~\bibnamefont {Panin}}, \bibinfo {author} {\bibfnamefont
  {A.}~\bibnamefont {Perea}}, \bibinfo {author} {\bibfnamefont
  {S.}~\bibnamefont {Pietri}}, \bibinfo {author} {\bibfnamefont
  {R.}~\bibnamefont {Plag}}, \bibinfo {author} {\bibfnamefont {A.}~\bibnamefont
  {Prochazka}}, \bibinfo {author} {\bibfnamefont {A.}~\bibnamefont {Rahaman}},
  \bibinfo {author} {\bibfnamefont {G.}~\bibnamefont {Rastrepina}}, \bibinfo
  {author} {\bibfnamefont {R.}~\bibnamefont {Reifarth}}, \bibinfo {author}
  {\bibfnamefont {G.}~\bibnamefont {Ribeiro}}, \bibinfo {author} {\bibfnamefont
  {M.~V.}\ \bibnamefont {Ricciardi}}, \bibinfo {author} {\bibfnamefont
  {C.}~\bibnamefont {Rigollet}}, \bibinfo {author} {\bibfnamefont
  {K.}~\bibnamefont {Riisager}}, \bibinfo {author} {\bibfnamefont
  {M.}~\bibnamefont {R\"oder}}, \bibinfo {author} {\bibfnamefont
  {D.}~\bibnamefont {Rossi}}, \bibinfo {author} {\bibfnamefont
  {J.}~\bibnamefont {Sanchez~del Rio}}, \bibinfo {author} {\bibfnamefont
  {D.}~\bibnamefont {Savran}}, \bibinfo {author} {\bibfnamefont
  {H.}~\bibnamefont {Scheit}}, \bibinfo {author} {\bibfnamefont
  {A.}~\bibnamefont {Schwenk}}, \bibinfo {author} {\bibfnamefont
  {H.}~\bibnamefont {Simon}}, \bibinfo {author} {\bibfnamefont
  {O.}~\bibnamefont {Sorlin}}, \bibinfo {author} {\bibfnamefont
  {V.}~\bibnamefont {Stoica}}, \bibinfo {author} {\bibfnamefont
  {B.}~\bibnamefont {Streicher}}, \bibinfo {author} {\bibfnamefont
  {J.}~\bibnamefont {Taylor}}, \bibinfo {author} {\bibfnamefont
  {O.}~\bibnamefont {Tengblad}}, \bibinfo {author} {\bibfnamefont
  {S.}~\bibnamefont {Terashima}}, \bibinfo {author} {\bibfnamefont
  {R.}~\bibnamefont {Thies}}, \bibinfo {author} {\bibfnamefont
  {Y.}~\bibnamefont {Togano}}, \bibinfo {author} {\bibfnamefont
  {E.}~\bibnamefont {Uberseder}}, \bibinfo {author} {\bibfnamefont
  {J.}~\bibnamefont {Van~de Walle}}, \bibinfo {author} {\bibfnamefont
  {P.}~\bibnamefont {Velho}}, \bibinfo {author} {\bibfnamefont
  {V.}~\bibnamefont {Volkov}}, \bibinfo {author} {\bibfnamefont
  {A.}~\bibnamefont {Wagner}}, \bibinfo {author} {\bibfnamefont
  {F.}~\bibnamefont {Wamers}}, \bibinfo {author} {\bibfnamefont
  {H.}~\bibnamefont {Weick}}, \bibinfo {author} {\bibfnamefont
  {M.}~\bibnamefont {Weigand}}, \bibinfo {author} {\bibfnamefont
  {C.}~\bibnamefont {Wheldon}}, \bibinfo {author} {\bibfnamefont
  {G.}~\bibnamefont {Wilson}}, \bibinfo {author} {\bibfnamefont
  {C.}~\bibnamefont {Wimmer}}, \bibinfo {author} {\bibfnamefont {J.~S.}\
  \bibnamefont {Winfield}}, \bibinfo {author} {\bibfnamefont {P.}~\bibnamefont
  {Woods}}, \bibinfo {author} {\bibfnamefont {D.}~\bibnamefont {Yakorev}},
  \bibinfo {author} {\bibfnamefont {M.~V.}\ \bibnamefont {Zhukov}}, \bibinfo
  {author} {\bibfnamefont {A.}~\bibnamefont {Zilges}}, \bibinfo {author}
  {\bibfnamefont {M.}~\bibnamefont {Zoric}}, \ and\ \bibinfo {author}
  {\bibfnamefont {K.}~\bibnamefont {Zuber}} (\bibinfo {collaboration} {R3B
  collaboration})} (\bibinfo {year} {2013}),\ \href {\doibase
  10.1103/PhysRevC.88.034313} {\bibfield  {journal} {\bibinfo  {journal} {Phys.
  Rev. C}\ }\textbf {\bibinfo {volume} {88}},\ \bibinfo {pages}
  {034313}}\BibitemShut {NoStop}%
\bibitem [{\citenamefont {Caprio}\ \emph {et~al.}(2012)\citenamefont {Caprio},
  \citenamefont {Maris},\ and\ \citenamefont {Vary}}]{caprio2012}%
  \BibitemOpen
  \bibfield  {author} {\bibinfo {author} {\bibnamefont {Caprio}, \bibfnamefont
  {M.~A.}}, \bibinfo {author} {\bibfnamefont {P.}~\bibnamefont {Maris}}, \ and\
  \bibinfo {author} {\bibfnamefont {J.~P.}\ \bibnamefont {Vary}}} (\bibinfo
  {year} {2012}),\ \href {\doibase 10.1103/PhysRevC.86.034312} {\bibfield
  {journal} {\bibinfo  {journal} {Phys. Rev. C}\ }\textbf {\bibinfo {volume}
  {86}},\ \bibinfo {pages} {034312}}\BibitemShut {NoStop}%
\bibitem [{\citenamefont {Carbone}\ \emph {et~al.}(2013)\citenamefont
  {Carbone}, \citenamefont {Polls},\ and\ \citenamefont {Rios}}]{carbone2013}%
  \BibitemOpen
  \bibfield  {author} {\bibinfo {author} {\bibnamefont {Carbone}, \bibfnamefont
  {A.}}, \bibinfo {author} {\bibfnamefont {A.}~\bibnamefont {Polls}}, \ and\
  \bibinfo {author} {\bibfnamefont {A.}~\bibnamefont {Rios}}} (\bibinfo {year}
  {2013}),\ \href {\doibase 10.1103/PhysRevC.88.044302} {\bibfield  {journal}
  {\bibinfo  {journal} {Phys. Rev. C}\ }\textbf {\bibinfo {volume} {88}},\
  \bibinfo {pages} {044302}}\BibitemShut {NoStop}%
\bibitem [{\citenamefont {Carbonell}\ \emph {et~al.}(2014)\citenamefont
  {Carbonell}, \citenamefont {Deltuva}, \citenamefont {Fonseca},\ and\
  \citenamefont {Lazauskas}}]{carbonell2014}%
  \BibitemOpen
  \bibfield  {author} {\bibinfo {author} {\bibnamefont {Carbonell},
  \bibfnamefont {J.}}, \bibinfo {author} {\bibfnamefont {A.}~\bibnamefont
  {Deltuva}}, \bibinfo {author} {\bibfnamefont {A.}~\bibnamefont {Fonseca}}, \
  and\ \bibinfo {author} {\bibfnamefont {R.}~\bibnamefont {Lazauskas}}}
  (\bibinfo {year} {2014}),\ \href {\doibase 10.1016/j.ppnp.2013.10.003}
  {\bibfield  {journal} {\bibinfo  {journal} {Progress in Particle and Nuclear
  Physics}\ }\textbf {\bibinfo {volume} {74}}~(\bibinfo {number} {0}),\
  \bibinfo {pages} {55 }}\BibitemShut {NoStop}%
\bibitem [{\citenamefont {Carlson}(1987)}]{carlson1987}%
  \BibitemOpen
  \bibfield  {author} {\bibinfo {author} {\bibnamefont {Carlson}, \bibfnamefont
  {J.}}} (\bibinfo {year} {1987}),\ \href {\doibase 10.1103/PhysRevC.36.2026}
  {\bibfield  {journal} {\bibinfo  {journal} {Phys. Rev. C}\ }\textbf {\bibinfo
  {volume} {36}},\ \bibinfo {pages} {2026}}\BibitemShut {NoStop}%
\bibitem [{\citenamefont {Caurier}\ \emph {et~al.}(1999)\citenamefont
  {Caurier}, \citenamefont {Mart{\'i}nez-Pinedo}, \citenamefont {Nowacki},
  \citenamefont {Poves}, \citenamefont {Retamosa},\ and\ \citenamefont
  {Zuker}}]{caurier1999}%
  \BibitemOpen
  \bibfield  {author} {\bibinfo {author} {\bibnamefont {Caurier}, \bibfnamefont
  {E.}}, \bibinfo {author} {\bibfnamefont {G.}~\bibnamefont
  {Mart{\'i}nez-Pinedo}}, \bibinfo {author} {\bibfnamefont {F.}~\bibnamefont
  {Nowacki}}, \bibinfo {author} {\bibfnamefont {A.}~\bibnamefont {Poves}},
  \bibinfo {author} {\bibfnamefont {J.}~\bibnamefont {Retamosa}}, \ and\
  \bibinfo {author} {\bibfnamefont {A.~P.}\ \bibnamefont {Zuker}}} (\bibinfo
  {year} {1999}),\ \href {\doibase 10.1103/PhysRevC.59.2033} {\bibfield
  {journal} {\bibinfo  {journal} {Phys. Rev. C}\ }\textbf {\bibinfo {volume}
  {59}},\ \bibinfo {pages} {2033}}\BibitemShut {NoStop}%
\bibitem [{\citenamefont {Caurier}\ \emph {et~al.}(2005)\citenamefont
  {Caurier}, \citenamefont {Mart{\'i}nez-Pinedo}, \citenamefont {Nowacki},
  \citenamefont {Poves},\ and\ \citenamefont {Zuker}}]{caurier2005}%
  \BibitemOpen
  \bibfield  {author} {\bibinfo {author} {\bibnamefont {Caurier}, \bibfnamefont
  {E.}}, \bibinfo {author} {\bibfnamefont {G.}~\bibnamefont
  {Mart{\'i}nez-Pinedo}}, \bibinfo {author} {\bibfnamefont {F.}~\bibnamefont
  {Nowacki}}, \bibinfo {author} {\bibfnamefont {A.}~\bibnamefont {Poves}}, \
  and\ \bibinfo {author} {\bibfnamefont {A.~P.}\ \bibnamefont {Zuker}}}
  (\bibinfo {year} {2005}),\ \href {\doibase 10.1103/RevModPhys.77.427}
  {\bibfield  {journal} {\bibinfo  {journal} {Rev. Mod. Phys.}\ }\textbf
  {\bibinfo {volume} {77}},\ \bibinfo {pages} {427}}\BibitemShut {NoStop}%
\bibitem [{\citenamefont {Caurier}\ and\ \citenamefont
  {Navr\'atil}(2006)}]{caurier2006}%
  \BibitemOpen
  \bibfield  {author} {\bibinfo {author} {\bibnamefont {Caurier}, \bibfnamefont
  {E.}}, \ and\ \bibinfo {author} {\bibfnamefont {P.}~\bibnamefont
  {Navr\'atil}}} (\bibinfo {year} {2006}),\ \href {\doibase
  10.1103/PhysRevC.73.021302} {\bibfield  {journal} {\bibinfo  {journal} {Phys.
  Rev. C}\ }\textbf {\bibinfo {volume} {73}},\ \bibinfo {pages}
  {021302}}\BibitemShut {NoStop}%
\bibitem [{\citenamefont {Caurier}\ \emph {et~al.}(1998)\citenamefont
  {Caurier}, \citenamefont {Nowacki}, \citenamefont {Poves},\ and\
  \citenamefont {Retamosa}}]{caurier1998}%
  \BibitemOpen
  \bibfield  {author} {\bibinfo {author} {\bibnamefont {Caurier}, \bibfnamefont
  {E.}}, \bibinfo {author} {\bibfnamefont {F.}~\bibnamefont {Nowacki}},
  \bibinfo {author} {\bibfnamefont {A.}~\bibnamefont {Poves}}, \ and\ \bibinfo
  {author} {\bibfnamefont {J.}~\bibnamefont {Retamosa}}} (\bibinfo {year}
  {1998}),\ \href {\doibase 10.1103/PhysRevC.58.2033} {\bibfield  {journal}
  {\bibinfo  {journal} {Phys. Rev. C}\ }\textbf {\bibinfo {volume} {58}},\
  \bibinfo {pages} {2033}}\BibitemShut {NoStop}%
\bibitem [{\citenamefont {Cipollone}\ \emph {et~al.}(2013)\citenamefont
  {Cipollone}, \citenamefont {Barbieri},\ and\ \citenamefont
  {Navr\'atil}}]{cipollone2013}%
  \BibitemOpen
  \bibfield  {author} {\bibinfo {author} {\bibnamefont {Cipollone},
  \bibfnamefont {A.}}, \bibinfo {author} {\bibfnamefont {C.}~\bibnamefont
  {Barbieri}}, \ and\ \bibinfo {author} {\bibfnamefont {P.}~\bibnamefont
  {Navr\'atil}}} (\bibinfo {year} {2013}),\ \href {\doibase
  10.1103/PhysRevLett.111.062501} {\bibfield  {journal} {\bibinfo  {journal}
  {Phys. Rev. Lett.}\ }\textbf {\bibinfo {volume} {111}},\ \bibinfo {pages}
  {062501}}\BibitemShut {NoStop}%
\bibitem [{\citenamefont {{\v {C}}{\'\i}{\v z}ek}(1966)}]{cizek1966}%
  \BibitemOpen
  \bibfield  {author} {\bibinfo {author} {\bibnamefont {{\v {C}}{\'\i}{\v
  z}ek}, \bibfnamefont {J.}}} (\bibinfo {year} {1966}),\ \href {\doibase
  10.1063/1.1727484} {\bibfield  {journal} {\bibinfo  {journal} {The Journal of
  Chemical Physics}\ }\textbf {\bibinfo {volume} {45}}~(\bibinfo {number}
  {11}),\ \bibinfo {pages} {4256}}\BibitemShut {NoStop}%
\bibitem [{\citenamefont {{\v {C}}{\'\i}{\v z}ek}\ and\ \citenamefont
  {Paldus}(1971)}]{cizek1971}%
  \BibitemOpen
  \bibfield  {author} {\bibinfo {author} {\bibnamefont {{\v {C}}{\'\i}{\v
  z}ek}, \bibfnamefont {J.}}, \ and\ \bibinfo {author} {\bibfnamefont
  {J.}~\bibnamefont {Paldus}}} (\bibinfo {year} {1971}),\ \href {\doibase
  10.1002/qua.560050402} {\bibfield  {journal} {\bibinfo  {journal}
  {International Journal of Quantum Chemistry}\ }\textbf {\bibinfo {volume}
  {5}}~(\bibinfo {number} {4}),\ \bibinfo {pages} {359}}\BibitemShut {NoStop}%
\bibitem [{\citenamefont {Coester}(1958)}]{coester1958}%
  \BibitemOpen
  \bibfield  {author} {\bibinfo {author} {\bibnamefont {Coester}, \bibfnamefont
  {F.}}} (\bibinfo {year} {1958}),\ \href {\doibase
  10.1016/0029-5582(58)90280-3} {\bibfield  {journal} {\bibinfo  {journal}
  {Nuclear Physics}\ }\textbf {\bibinfo {volume} {7}}~(\bibinfo {number} {0}),\
  \bibinfo {pages} {421 }}\BibitemShut {NoStop}%
\bibitem [{\citenamefont {Coester}\ and\ \citenamefont
  {K{\"u}mmel}(1960)}]{coester1960}%
  \BibitemOpen
  \bibfield  {author} {\bibinfo {author} {\bibnamefont {Coester}, \bibfnamefont
  {F.}}, \ and\ \bibinfo {author} {\bibfnamefont {H.}~\bibnamefont
  {K{\"u}mmel}}} (\bibinfo {year} {1960}),\ \href {\doibase
  10.1016/0029-5582(60)90140-1} {\bibfield  {journal} {\bibinfo  {journal}
  {Nuclear Physics}\ }\textbf {\bibinfo {volume} {17}}~(\bibinfo {number}
  {0}),\ \bibinfo {pages} {477 }}\BibitemShut {NoStop}%
\bibitem [{\citenamefont {Coon}\ \emph {et~al.}(1977)\citenamefont {Coon},
  \citenamefont {Zabolitzky},\ and\ \citenamefont {Blatt}}]{coon1977}%
  \BibitemOpen
  \bibfield  {author} {\bibinfo {author} {\bibnamefont {Coon}, \bibfnamefont
  {S.}}, \bibinfo {author} {\bibfnamefont {J.}~\bibnamefont {Zabolitzky}}, \
  and\ \bibinfo {author} {\bibfnamefont {D.}~\bibnamefont {Blatt}}} (\bibinfo
  {year} {1977}),\ \href {\doibase 10.1007/BF01408625} {\bibfield  {journal}
  {\bibinfo  {journal} {Zeitschrift f{\"u}r Physik A Atoms and Nuclei}\
  }\textbf {\bibinfo {volume} {281}}~(\bibinfo {number} {1-2}),\ \bibinfo
  {pages} {137}}\BibitemShut {NoStop}%
\bibitem [{\citenamefont {Coon}\ \emph {et~al.}(2012)\citenamefont {Coon},
  \citenamefont {Avetian}, \citenamefont {Kruse}, \citenamefont {van Kolck},
  \citenamefont {Maris},\ and\ \citenamefont {Vary}}]{coon2012}%
  \BibitemOpen
  \bibfield  {author} {\bibinfo {author} {\bibnamefont {Coon}, \bibfnamefont
  {S.~A.}}, \bibinfo {author} {\bibfnamefont {M.~I.}\ \bibnamefont {Avetian}},
  \bibinfo {author} {\bibfnamefont {M.~K.~G.}\ \bibnamefont {Kruse}}, \bibinfo
  {author} {\bibfnamefont {U.}~\bibnamefont {van Kolck}}, \bibinfo {author}
  {\bibfnamefont {P.}~\bibnamefont {Maris}}, \ and\ \bibinfo {author}
  {\bibfnamefont {J.~P.}\ \bibnamefont {Vary}}} (\bibinfo {year} {2012}),\
  \href {\doibase 10.1103/PhysRevC.86.054002} {\bibfield  {journal} {\bibinfo
  {journal} {Phys. Rev. C}\ }\textbf {\bibinfo {volume} {86}},\ \bibinfo
  {pages} {054002}}\BibitemShut {NoStop}%
\bibitem [{\citenamefont {Coon}\ \emph {et~al.}(1978)\citenamefont {Coon},
  \citenamefont {McCarthy},\ and\ \citenamefont {Malta}}]{coon1978}%
  \BibitemOpen
  \bibfield  {author} {\bibinfo {author} {\bibnamefont {Coon}, \bibfnamefont
  {S.~A.}}, \bibinfo {author} {\bibfnamefont {R.~J.}\ \bibnamefont {McCarthy}},
  \ and\ \bibinfo {author} {\bibfnamefont {C.~P.}\ \bibnamefont {Malta}}}
  (\bibinfo {year} {1978}),\ \href
  {http://stacks.iop.org/0305-4616/4/i=2/a=009} {\bibfield  {journal} {\bibinfo
   {journal} {Journal of Physics G: Nuclear Physics}\ }\textbf {\bibinfo
  {volume} {4}}~(\bibinfo {number} {2}),\ \bibinfo {pages} {183}}\BibitemShut
  {NoStop}%
\bibitem [{\citenamefont {Coraggio}\ \emph {et~al.}(2009)\citenamefont
  {Coraggio}, \citenamefont {Covello}, \citenamefont {Gargano},\ and\
  \citenamefont {Itaco}}]{coraggio2009}%
  \BibitemOpen
  \bibfield  {author} {\bibinfo {author} {\bibnamefont {Coraggio},
  \bibfnamefont {L.}}, \bibinfo {author} {\bibfnamefont {A.}~\bibnamefont
  {Covello}}, \bibinfo {author} {\bibfnamefont {A.}~\bibnamefont {Gargano}}, \
  and\ \bibinfo {author} {\bibfnamefont {N.}~\bibnamefont {Itaco}}} (\bibinfo
  {year} {2009}),\ \href {\doibase 10.1103/PhysRevC.80.044311} {\bibfield
  {journal} {\bibinfo  {journal} {Phys. Rev. C}\ }\textbf {\bibinfo {volume}
  {80}},\ \bibinfo {pages} {044311}}\BibitemShut {NoStop}%
\bibitem [{\citenamefont {Crawford}\ and\ \citenamefont
  {Schaefer}(2007)}]{crawford2007}%
  \BibitemOpen
  \bibfield  {author} {\bibinfo {author} {\bibnamefont {Crawford},
  \bibfnamefont {T.~D.}}, \ and\ \bibinfo {author} {\bibfnamefont {H.~F.}\
  \bibnamefont {Schaefer}}} (\bibinfo {year} {2007}),\ \href {\doibase
  10.1002/9780470125915.ch2} {\bibfield  {journal} {\bibinfo  {journal}
  {Reviews in Computational Chemistry}\ }\textbf {\bibinfo {volume} {14}},\
  \bibinfo {pages} {33}}\BibitemShut {NoStop}%
\bibitem [{\citenamefont {van Dalen}\ and\ \citenamefont
  {M{\"u}ther}(2010)}]{vandalen2010}%
  \BibitemOpen
  \bibfield  {author} {\bibinfo {author} {\bibnamefont {van Dalen},
  \bibfnamefont {E.}}, \ and\ \bibinfo {author} {\bibfnamefont
  {H.}~\bibnamefont {M{\"u}ther}}} (\bibinfo {year} {2010}),\ \href {\doibase
  10.1142/S0218301310016533} {\bibfield  {journal} {\bibinfo  {journal}
  {International Journal of Modern Physics E}\ }\textbf {\bibinfo {volume}
  {19}}~(\bibinfo {number} {11}),\ \bibinfo {pages} {2077}}\BibitemShut
  {NoStop}%
\bibitem [{\citenamefont {Dalgaard}\ and\ \citenamefont
  {Monkhorst}(1983)}]{DalMon83}%
  \BibitemOpen
  \bibfield  {author} {\bibinfo {author} {\bibnamefont {Dalgaard},
  \bibfnamefont {E.}}, \ and\ \bibinfo {author} {\bibfnamefont {H.~J.}\
  \bibnamefont {Monkhorst}}} (\bibinfo {year} {1983}),\ \href {\doibase
  10.1103/PhysRevA.28.1217} {\bibfield  {journal} {\bibinfo  {journal} {Phys.
  Rev. A}\ }\textbf {\bibinfo {volume} {28}},\ \bibinfo {pages}
  {1217}}\BibitemShut {NoStop}%
\bibitem [{\citenamefont {Darby}\ \emph {et~al.}(2010)\citenamefont {Darby},
  \citenamefont {Grzywacz}, \citenamefont {Batchelder}, \citenamefont
  {Bingham}, \citenamefont {Cartegni}, \citenamefont {Gross}, \citenamefont
  {Hjorth-Jensen}, \citenamefont {Joss}, \citenamefont {Liddick}, \citenamefont
  {Nazarewicz}, \citenamefont {Padgett}, \citenamefont {Page}, \citenamefont
  {Papenbrock}, \citenamefont {Rajabali}, \citenamefont {Rotureau},\ and\
  \citenamefont {Rykaczewski}}]{darby2010}%
  \BibitemOpen
  \bibfield  {author} {\bibinfo {author} {\bibnamefont {Darby}, \bibfnamefont
  {I.~G.}}, \bibinfo {author} {\bibfnamefont {R.~K.}\ \bibnamefont {Grzywacz}},
  \bibinfo {author} {\bibfnamefont {J.~C.}\ \bibnamefont {Batchelder}},
  \bibinfo {author} {\bibfnamefont {C.~R.}\ \bibnamefont {Bingham}}, \bibinfo
  {author} {\bibfnamefont {L.}~\bibnamefont {Cartegni}}, \bibinfo {author}
  {\bibfnamefont {C.~J.}\ \bibnamefont {Gross}}, \bibinfo {author}
  {\bibfnamefont {M.}~\bibnamefont {Hjorth-Jensen}}, \bibinfo {author}
  {\bibfnamefont {D.~T.}\ \bibnamefont {Joss}}, \bibinfo {author}
  {\bibfnamefont {S.~N.}\ \bibnamefont {Liddick}}, \bibinfo {author}
  {\bibfnamefont {W.}~\bibnamefont {Nazarewicz}}, \bibinfo {author}
  {\bibfnamefont {S.}~\bibnamefont {Padgett}}, \bibinfo {author} {\bibfnamefont
  {R.~D.}\ \bibnamefont {Page}}, \bibinfo {author} {\bibfnamefont
  {T.}~\bibnamefont {Papenbrock}}, \bibinfo {author} {\bibfnamefont {M.~M.}\
  \bibnamefont {Rajabali}}, \bibinfo {author} {\bibfnamefont {J.}~\bibnamefont
  {Rotureau}}, \ and\ \bibinfo {author} {\bibfnamefont {K.~P.}\ \bibnamefont
  {Rykaczewski}}} (\bibinfo {year} {2010}),\ \href {\doibase
  10.1103/PhysRevLett.105.162502} {\bibfield  {journal} {\bibinfo  {journal}
  {Phys. Rev. Lett.}\ }\textbf {\bibinfo {volume} {105}},\ \bibinfo {pages}
  {162502}}\BibitemShut {NoStop}%
\bibitem [{\citenamefont {Day}(1967)}]{day1967}%
  \BibitemOpen
  \bibfield  {author} {\bibinfo {author} {\bibnamefont {Day}, \bibfnamefont
  {B.~D.}}} (\bibinfo {year} {1967}),\ \href {\doibase
  10.1103/RevModPhys.39.719} {\bibfield  {journal} {\bibinfo  {journal} {Rev.
  Mod. Phys.}\ }\textbf {\bibinfo {volume} {39}},\ \bibinfo {pages}
  {719}}\BibitemShut {NoStop}%
\bibitem [{\citenamefont {Day}(1981)}]{day1981b}%
  \BibitemOpen
  \bibfield  {author} {\bibinfo {author} {\bibnamefont {Day}, \bibfnamefont
  {B.~D.}}} (\bibinfo {year} {1981}),\ \href {\doibase
  10.1103/PhysRevC.24.1203} {\bibfield  {journal} {\bibinfo  {journal} {Phys.
  Rev. C}\ }\textbf {\bibinfo {volume} {24}},\ \bibinfo {pages}
  {1203}}\BibitemShut {NoStop}%
\bibitem [{\citenamefont {Dean}\ \emph {et~al.}(2008)\citenamefont {Dean},
  \citenamefont {Hagen}, \citenamefont {Hjorth-Jensen},\ and\ \citenamefont
  {Papenbrock}}]{dean2008b}%
  \BibitemOpen
  \bibfield  {author} {\bibinfo {author} {\bibnamefont {Dean}, \bibfnamefont
  {D.~J.}}, \bibinfo {author} {\bibfnamefont {G.}~\bibnamefont {Hagen}},
  \bibinfo {author} {\bibfnamefont {M.}~\bibnamefont {Hjorth-Jensen}}, \ and\
  \bibinfo {author} {\bibfnamefont {T.}~\bibnamefont {Papenbrock}}} (\bibinfo
  {year} {2008}),\ \href {http://stacks.iop.org/1749-4699/1/i=1/a=015008}
  {\bibfield  {journal} {\bibinfo  {journal} {Computational Science \&
  Discovery}\ }\textbf {\bibinfo {volume} {1}}~(\bibinfo {number} {1}),\
  \bibinfo {pages} {015008}}\BibitemShut {NoStop}%
\bibitem [{\citenamefont {Dean}\ and\ \citenamefont
  {Hjorth-Jensen}(2004)}]{dean2004}%
  \BibitemOpen
  \bibfield  {author} {\bibinfo {author} {\bibnamefont {Dean}, \bibfnamefont
  {D.~J.}}, \ and\ \bibinfo {author} {\bibfnamefont {M.}~\bibnamefont
  {Hjorth-Jensen}}} (\bibinfo {year} {2004}),\ \href {\doibase
  10.1103/PhysRevC.69.054320} {\bibfield  {journal} {\bibinfo  {journal} {Phys.
  Rev. C}\ }\textbf {\bibinfo {volume} {69}},\ \bibinfo {pages}
  {054320}}\BibitemShut {NoStop}%
\bibitem [{\citenamefont {Dickhoff}\ and\ \citenamefont
  {Barbieri}(2004)}]{dickhoff2004}%
  \BibitemOpen
  \bibfield  {author} {\bibinfo {author} {\bibnamefont {Dickhoff},
  \bibfnamefont {W.}}, \ and\ \bibinfo {author} {\bibfnamefont
  {C.}~\bibnamefont {Barbieri}}} (\bibinfo {year} {2004}),\ \href {\doibase
  10.1016/j.ppnp.2004.02.038} {\bibfield  {journal} {\bibinfo  {journal}
  {Progress in Particle and Nuclear Physics}\ }\textbf {\bibinfo {volume}
  {52}}~(\bibinfo {number} {2}),\ \bibinfo {pages} {377 }}\BibitemShut
  {NoStop}%
\bibitem [{\citenamefont {Dinca}\ \emph {et~al.}(2005)\citenamefont {Dinca},
  \citenamefont {Janssens}, \citenamefont {Gade}, \citenamefont {Bazin},
  \citenamefont {Broda}, \citenamefont {Brown}, \citenamefont {Campbell},
  \citenamefont {Carpenter}, \citenamefont {Chowdhury}, \citenamefont {Cook},
  \citenamefont {Deacon}, \citenamefont {Fornal}, \citenamefont {Freeman},
  \citenamefont {Glasmacher}, \citenamefont {Honma}, \citenamefont {Kondev},
  \citenamefont {Lecouey}, \citenamefont {Liddick}, \citenamefont {Mantica},
  \citenamefont {Mueller}, \citenamefont {Olliver}, \citenamefont {Otsuka},
  \citenamefont {Terry}, \citenamefont {Tomlin},\ and\ \citenamefont
  {Yoneda}}]{dinca2005}%
  \BibitemOpen
  \bibfield  {author} {\bibinfo {author} {\bibnamefont {Dinca}, \bibfnamefont
  {D.-C.}}, \bibinfo {author} {\bibfnamefont {R.~V.~F.}\ \bibnamefont
  {Janssens}}, \bibinfo {author} {\bibfnamefont {A.}~\bibnamefont {Gade}},
  \bibinfo {author} {\bibfnamefont {D.}~\bibnamefont {Bazin}}, \bibinfo
  {author} {\bibfnamefont {R.}~\bibnamefont {Broda}}, \bibinfo {author}
  {\bibfnamefont {B.~A.}\ \bibnamefont {Brown}}, \bibinfo {author}
  {\bibfnamefont {C.~M.}\ \bibnamefont {Campbell}}, \bibinfo {author}
  {\bibfnamefont {M.~P.}\ \bibnamefont {Carpenter}}, \bibinfo {author}
  {\bibfnamefont {P.}~\bibnamefont {Chowdhury}}, \bibinfo {author}
  {\bibfnamefont {J.~M.}\ \bibnamefont {Cook}}, \bibinfo {author}
  {\bibfnamefont {A.~N.}\ \bibnamefont {Deacon}}, \bibinfo {author}
  {\bibfnamefont {B.}~\bibnamefont {Fornal}}, \bibinfo {author} {\bibfnamefont
  {S.~J.}\ \bibnamefont {Freeman}}, \bibinfo {author} {\bibfnamefont
  {T.}~\bibnamefont {Glasmacher}}, \bibinfo {author} {\bibfnamefont
  {M.}~\bibnamefont {Honma}}, \bibinfo {author} {\bibfnamefont {F.~G.}\
  \bibnamefont {Kondev}}, \bibinfo {author} {\bibfnamefont {J.-L.}\
  \bibnamefont {Lecouey}}, \bibinfo {author} {\bibfnamefont {S.~N.}\
  \bibnamefont {Liddick}}, \bibinfo {author} {\bibfnamefont {P.~F.}\
  \bibnamefont {Mantica}}, \bibinfo {author} {\bibfnamefont {W.~F.}\
  \bibnamefont {Mueller}}, \bibinfo {author} {\bibfnamefont {H.}~\bibnamefont
  {Olliver}}, \bibinfo {author} {\bibfnamefont {T.}~\bibnamefont {Otsuka}},
  \bibinfo {author} {\bibfnamefont {J.~R.}\ \bibnamefont {Terry}}, \bibinfo
  {author} {\bibfnamefont {B.~A.}\ \bibnamefont {Tomlin}}, \ and\ \bibinfo
  {author} {\bibfnamefont {K.}~\bibnamefont {Yoneda}}} (\bibinfo {year}
  {2005}),\ \href {\doibase 10.1103/PhysRevC.71.041302} {\bibfield  {journal}
  {\bibinfo  {journal} {Phys. Rev. C}\ }\textbf {\bibinfo {volume} {71}},\
  \bibinfo {pages} {041302}}\BibitemShut {NoStop}%
\bibitem [{\citenamefont {Dobaczewski}\ \emph {et~al.}(1994)\citenamefont
  {Dobaczewski}, \citenamefont {Hamamoto}, \citenamefont {Nazarewicz},\ and\
  \citenamefont {Sheikh}}]{dobaczewski1994}%
  \BibitemOpen
  \bibfield  {author} {\bibinfo {author} {\bibnamefont {Dobaczewski},
  \bibfnamefont {J.}}, \bibinfo {author} {\bibfnamefont {I.}~\bibnamefont
  {Hamamoto}}, \bibinfo {author} {\bibfnamefont {W.}~\bibnamefont
  {Nazarewicz}}, \ and\ \bibinfo {author} {\bibfnamefont {J.~A.}\ \bibnamefont
  {Sheikh}}} (\bibinfo {year} {1994}),\ \href {\doibase
  10.1103/PhysRevLett.72.981} {\bibfield  {journal} {\bibinfo  {journal} {Phys.
  Rev. Lett.}\ }\textbf {\bibinfo {volume} {72}},\ \bibinfo {pages}
  {981}}\BibitemShut {NoStop}%
\bibitem [{\citenamefont {Dukelsky}\ \emph {et~al.}(2003)\citenamefont
  {Dukelsky}, \citenamefont {Dussel}, \citenamefont {Hirsch},\ and\
  \citenamefont {Schuck}}]{dukelsky2003}%
  \BibitemOpen
  \bibfield  {author} {\bibinfo {author} {\bibnamefont {Dukelsky},
  \bibfnamefont {J.}}, \bibinfo {author} {\bibfnamefont {G.}~\bibnamefont
  {Dussel}}, \bibinfo {author} {\bibfnamefont {J.}~\bibnamefont {Hirsch}}, \
  and\ \bibinfo {author} {\bibfnamefont {P.}~\bibnamefont {Schuck}}} (\bibinfo
  {year} {2003}),\ \href {\doibase 10.1016/S0375-9474(02)01361-1} {\bibfield
  {journal} {\bibinfo  {journal} {Nuclear Physics A}\ }\textbf {\bibinfo
  {volume} {714}}~(\bibinfo {number} {1–2}),\ \bibinfo {pages} {63
  }}\BibitemShut {NoStop}%
\bibitem [{\citenamefont {Efros}\ \emph {et~al.}(1994)\citenamefont {Efros},
  \citenamefont {Leidemann},\ and\ \citenamefont {Orlandini}}]{efros1994}%
  \BibitemOpen
  \bibfield  {author} {\bibinfo {author} {\bibnamefont {Efros}, \bibfnamefont
  {V.~D.}}, \bibinfo {author} {\bibfnamefont {W.}~\bibnamefont {Leidemann}}, \
  and\ \bibinfo {author} {\bibfnamefont {G.}~\bibnamefont {Orlandini}}}
  (\bibinfo {year} {1994}),\ \href {\doibase 10.1016/0370-2693(94)91355-2}
  {\bibfield  {journal} {\bibinfo  {journal} {Physics Letters B}\ }\textbf
  {\bibinfo {volume} {338}}~(\bibinfo {number} {2–3}),\ \bibinfo {pages} {130
  }}\BibitemShut {NoStop}%
\bibitem [{\citenamefont {Efros}\ \emph {et~al.}(2007)\citenamefont {Efros},
  \citenamefont {Leidemann}, \citenamefont {Orlandini},\ and\ \citenamefont
  {Barnea}}]{efros2007}%
  \BibitemOpen
  \bibfield  {author} {\bibinfo {author} {\bibnamefont {Efros}, \bibfnamefont
  {V.~D.}}, \bibinfo {author} {\bibfnamefont {W.}~\bibnamefont {Leidemann}},
  \bibinfo {author} {\bibfnamefont {G.}~\bibnamefont {Orlandini}}, \ and\
  \bibinfo {author} {\bibfnamefont {N.}~\bibnamefont {Barnea}}} (\bibinfo
  {year} {2007}),\ \href {http://stacks.iop.org/0954-3899/34/i=12/a=R02}
  {\bibfield  {journal} {\bibinfo  {journal} {Journal of Physics G: Nuclear and
  Particle Physics}\ }\textbf {\bibinfo {volume} {34}}~(\bibinfo {number}
  {12}),\ \bibinfo {pages} {R459}}\BibitemShut {NoStop}%
\bibitem [{\citenamefont {Ekstr\"om}\ \emph {et~al.}(2013)\citenamefont
  {Ekstr\"om}, \citenamefont {Baardsen}, \citenamefont {Forss\'en},
  \citenamefont {Hagen}, \citenamefont {Hjorth-Jensen}, \citenamefont {Jansen},
  \citenamefont {Machleidt}, \citenamefont {Nazarewicz}, \citenamefont
  {Papenbrock}, \citenamefont {Sarich},\ and\ \citenamefont
  {Wild}}]{ekstrom2013}%
  \BibitemOpen
  \bibfield  {author} {\bibinfo {author} {\bibnamefont {Ekstr\"om},
  \bibfnamefont {A.}}, \bibinfo {author} {\bibfnamefont {G.}~\bibnamefont
  {Baardsen}}, \bibinfo {author} {\bibfnamefont {C.}~\bibnamefont {Forss\'en}},
  \bibinfo {author} {\bibfnamefont {G.}~\bibnamefont {Hagen}}, \bibinfo
  {author} {\bibfnamefont {M.}~\bibnamefont {Hjorth-Jensen}}, \bibinfo {author}
  {\bibfnamefont {G.~R.}\ \bibnamefont {Jansen}}, \bibinfo {author}
  {\bibfnamefont {R.}~\bibnamefont {Machleidt}}, \bibinfo {author}
  {\bibfnamefont {W.}~\bibnamefont {Nazarewicz}}, \bibinfo {author}
  {\bibfnamefont {T.}~\bibnamefont {Papenbrock}}, \bibinfo {author}
  {\bibfnamefont {J.}~\bibnamefont {Sarich}}, \ and\ \bibinfo {author}
  {\bibfnamefont {S.~M.}\ \bibnamefont {Wild}}} (\bibinfo {year} {2013}),\
  \href {\doibase 10.1103/PhysRevLett.110.192502} {\bibfield  {journal}
  {\bibinfo  {journal} {Phys. Rev. Lett.}\ }\textbf {\bibinfo {volume} {110}},\
  \bibinfo {pages} {192502}}\BibitemShut {NoStop}%
\bibitem [{\citenamefont {Elliott}\ and\ \citenamefont
  {Skyrme}(1955)}]{elliott1955}%
  \BibitemOpen
  \bibfield  {author} {\bibinfo {author} {\bibnamefont {Elliott}, \bibfnamefont
  {J.~P.}}, \ and\ \bibinfo {author} {\bibfnamefont {T.~H.~R.}\ \bibnamefont
  {Skyrme}}} (\bibinfo {year} {1955}),\ \href {\doibase 10.1098/rspa.1955.0239}
  {\bibfield  {journal} {\bibinfo  {journal} {Proc. Roy. Soc. (London) A}\
  }\textbf {\bibinfo {volume} {232}},\ \bibinfo {pages} {561}}\BibitemShut
  {NoStop}%
\bibitem [{\citenamefont {Emrich}\ \emph {et~al.}(1977)\citenamefont {Emrich},
  \citenamefont {Zabolitzky},\ and\ \citenamefont {L\"uhrmann}}]{emrich1977}%
  \BibitemOpen
  \bibfield  {author} {\bibinfo {author} {\bibnamefont {Emrich}, \bibfnamefont
  {K.}}, \bibinfo {author} {\bibfnamefont {J.~G.}\ \bibnamefont {Zabolitzky}},
  \ and\ \bibinfo {author} {\bibfnamefont {K.~H.}\ \bibnamefont {L\"uhrmann}}}
  (\bibinfo {year} {1977}),\ \href {\doibase 10.1103/PhysRevC.16.1650}
  {\bibfield  {journal} {\bibinfo  {journal} {Phys. Rev. C}\ }\textbf {\bibinfo
  {volume} {16}},\ \bibinfo {pages} {1650}}\BibitemShut {NoStop}%
\bibitem [{\citenamefont {Entem}\ and\ \citenamefont
  {Machleidt}(2003)}]{entem2003}%
  \BibitemOpen
  \bibfield  {author} {\bibinfo {author} {\bibnamefont {Entem}, \bibfnamefont
  {D.~R.}}, \ and\ \bibinfo {author} {\bibfnamefont {R.}~\bibnamefont
  {Machleidt}}} (\bibinfo {year} {2003}),\ \href {\doibase
  10.1103/PhysRevC.68.041001} {\bibfield  {journal} {\bibinfo  {journal} {Phys.
  Rev. C}\ }\textbf {\bibinfo {volume} {68}},\ \bibinfo {pages}
  {041001}}\BibitemShut {NoStop}%
\bibitem [{\citenamefont {Epelbaoum}\ \emph {et~al.}(1998)\citenamefont
  {Epelbaoum}, \citenamefont {Gl{\"o}ckle},\ and\ \citenamefont
  {Mei{\ss}ner}}]{epelbaoum1998}%
  \BibitemOpen
  \bibfield  {author} {\bibinfo {author} {\bibnamefont {Epelbaoum},
  \bibfnamefont {E.}}, \bibinfo {author} {\bibfnamefont {W.}~\bibnamefont
  {Gl{\"o}ckle}}, \ and\ \bibinfo {author} {\bibfnamefont {U.-G.}\ \bibnamefont
  {Mei{\ss}ner}}} (\bibinfo {year} {1998}),\ \href {\doibase
  10.1016/S0375-9474(98)00220-6} {\bibfield  {journal} {\bibinfo  {journal}
  {Nuclear Physics A}\ }\textbf {\bibinfo {volume} {637}}~(\bibinfo {number}
  {1}),\ \bibinfo {pages} {107 }}\BibitemShut {NoStop}%
\bibitem [{\citenamefont {Epelbaum}\ \emph {et~al.}(2000)\citenamefont
  {Epelbaum}, \citenamefont {Gl{\"o}ckle},\ and\ \citenamefont
  {Mei{\ss}ner}}]{epelbaum2000}%
  \BibitemOpen
  \bibfield  {author} {\bibinfo {author} {\bibnamefont {Epelbaum},
  \bibfnamefont {E.}}, \bibinfo {author} {\bibfnamefont {W.}~\bibnamefont
  {Gl{\"o}ckle}}, \ and\ \bibinfo {author} {\bibfnamefont {U.-G.}\ \bibnamefont
  {Mei{\ss}ner}}} (\bibinfo {year} {2000}),\ \href {\doibase
  10.1016/S0375-9474(99)00821-0} {\bibfield  {journal} {\bibinfo  {journal}
  {Nuclear Physics A}\ }\textbf {\bibinfo {volume} {671}}~(\bibinfo {number}
  {1–4}),\ \bibinfo {pages} {295 }}\BibitemShut {NoStop}%
\bibitem [{\citenamefont {Epelbaum}\ \emph {et~al.}(2009)\citenamefont
  {Epelbaum}, \citenamefont {Hammer},\ and\ \citenamefont
  {Mei\ss{}ner}}]{epelbaum2009}%
  \BibitemOpen
  \bibfield  {author} {\bibinfo {author} {\bibnamefont {Epelbaum},
  \bibfnamefont {E.}}, \bibinfo {author} {\bibfnamefont {H.-W.}\ \bibnamefont
  {Hammer}}, \ and\ \bibinfo {author} {\bibfnamefont {U.-G.}\ \bibnamefont
  {Mei\ss{}ner}}} (\bibinfo {year} {2009}),\ \href {\doibase
  10.1103/RevModPhys.81.1773} {\bibfield  {journal} {\bibinfo  {journal} {Rev.
  Mod. Phys.}\ }\textbf {\bibinfo {volume} {81}},\ \bibinfo {pages}
  {1773}}\BibitemShut {NoStop}%
\bibitem [{\citenamefont {Epelbaum}\ \emph {et~al.}(2011)\citenamefont
  {Epelbaum}, \citenamefont {Krebs}, \citenamefont {Lee},\ and\ \citenamefont
  {Mei\ss{}ner}}]{epelbaum2011}%
  \BibitemOpen
  \bibfield  {author} {\bibinfo {author} {\bibnamefont {Epelbaum},
  \bibfnamefont {E.}}, \bibinfo {author} {\bibfnamefont {H.}~\bibnamefont
  {Krebs}}, \bibinfo {author} {\bibfnamefont {D.}~\bibnamefont {Lee}}, \ and\
  \bibinfo {author} {\bibfnamefont {U.-G.}\ \bibnamefont {Mei\ss{}ner}}}
  (\bibinfo {year} {2011}),\ \href {\doibase 10.1103/PhysRevLett.106.192501}
  {\bibfield  {journal} {\bibinfo  {journal} {Phys. Rev. Lett.}\ }\textbf
  {\bibinfo {volume} {106}},\ \bibinfo {pages} {192501}}\BibitemShut {NoStop}%
\bibitem [{\citenamefont {Epelbaum}\ \emph {et~al.}(2002)\citenamefont
  {Epelbaum}, \citenamefont {Nogga}, \citenamefont {Gl\"ockle}, \citenamefont
  {Kamada}, \citenamefont {Mei\ss{}ner},\ and\ \citenamefont
  {Wita\l{}a}}]{epelbaum2002}%
  \BibitemOpen
  \bibfield  {author} {\bibinfo {author} {\bibnamefont {Epelbaum},
  \bibfnamefont {E.}}, \bibinfo {author} {\bibfnamefont {A.}~\bibnamefont
  {Nogga}}, \bibinfo {author} {\bibfnamefont {W.}~\bibnamefont {Gl\"ockle}},
  \bibinfo {author} {\bibfnamefont {H.}~\bibnamefont {Kamada}}, \bibinfo
  {author} {\bibfnamefont {U.-G.}\ \bibnamefont {Mei\ss{}ner}}, \ and\ \bibinfo
  {author} {\bibfnamefont {H.}~\bibnamefont {Wita\l{}a}}} (\bibinfo {year}
  {2002}),\ \href {\doibase 10.1103/PhysRevC.66.064001} {\bibfield  {journal}
  {\bibinfo  {journal} {Phys. Rev. C}\ }\textbf {\bibinfo {volume} {66}},\
  \bibinfo {pages} {064001}}\BibitemShut {NoStop}%
\bibitem [{\citenamefont {Erler}\ \emph {et~al.}(2012)\citenamefont {Erler},
  \citenamefont {Birge}, \citenamefont {Kortelainen}, \citenamefont
  {Nazarewicz}, \citenamefont {Olsen}, \citenamefont {Perhac},\ and\
  \citenamefont {Stoitsov}}]{erler2012}%
  \BibitemOpen
  \bibfield  {author} {\bibinfo {author} {\bibnamefont {Erler}, \bibfnamefont
  {J.}}, \bibinfo {author} {\bibfnamefont {N.}~\bibnamefont {Birge}}, \bibinfo
  {author} {\bibfnamefont {M.}~\bibnamefont {Kortelainen}}, \bibinfo {author}
  {\bibfnamefont {W.}~\bibnamefont {Nazarewicz}}, \bibinfo {author}
  {\bibfnamefont {E.}~\bibnamefont {Olsen}}, \bibinfo {author} {\bibfnamefont
  {A.~M.}\ \bibnamefont {Perhac}}, \ and\ \bibinfo {author} {\bibfnamefont
  {M.}~\bibnamefont {Stoitsov}}} (\bibinfo {year} {2012}),\ \href {\doibase
  10.1038/nature11188} {\bibfield  {journal} {\bibinfo  {journal} {Nature}\
  }\textbf {\bibinfo {volume} {486}},\ \bibinfo {pages} {509 }}\BibitemShut
  {NoStop}%
\bibitem [{\citenamefont {Fayans}\ \emph {et~al.}(2000)\citenamefont {Fayans},
  \citenamefont {Tolokonnikov},\ and\ \citenamefont {Zawischa}}]{fayans2000}%
  \BibitemOpen
  \bibfield  {author} {\bibinfo {author} {\bibnamefont {Fayans}, \bibfnamefont
  {S.}}, \bibinfo {author} {\bibfnamefont {S.}~\bibnamefont {Tolokonnikov}}, \
  and\ \bibinfo {author} {\bibfnamefont {D.}~\bibnamefont {Zawischa}}}
  (\bibinfo {year} {2000}),\ \href {\doibase 10.1016/S0370-2693(00)01053-4}
  {\bibfield  {journal} {\bibinfo  {journal} {Physics Letters B}\ }\textbf
  {\bibinfo {volume} {491}}~(\bibinfo {number} {3 - 4}),\ \bibinfo {pages} {245
  }}\BibitemShut {NoStop}%
\bibitem [{\citenamefont {Feldmeier}\ \emph {et~al.}(1998)\citenamefont
  {Feldmeier}, \citenamefont {Neff}, \citenamefont {Roth},\ and\ \citenamefont
  {Schnack}}]{feldmeier1998}%
  \BibitemOpen
  \bibfield  {author} {\bibinfo {author} {\bibnamefont {Feldmeier},
  \bibfnamefont {H.}}, \bibinfo {author} {\bibfnamefont {T.}~\bibnamefont
  {Neff}}, \bibinfo {author} {\bibfnamefont {R.}~\bibnamefont {Roth}}, \ and\
  \bibinfo {author} {\bibfnamefont {J.}~\bibnamefont {Schnack}}} (\bibinfo
  {year} {1998}),\ \href {\doibase 10.1016/S0375-9474(97)00805-1} {\bibfield
  {journal} {\bibinfo  {journal} {Nuclear Physics A}\ }\textbf {\bibinfo
  {volume} {632}}~(\bibinfo {number} {1}),\ \bibinfo {pages} {61 }}\BibitemShut
  {NoStop}%
\bibitem [{\citenamefont {Fink}(1974)}]{fink1974}%
  \BibitemOpen
  \bibfield  {author} {\bibinfo {author} {\bibnamefont {Fink}, \bibfnamefont
  {M.}}} (\bibinfo {year} {1974}),\ \href {\doibase
  10.1016/0375-9474(74)90105-5} {\bibfield  {journal} {\bibinfo  {journal}
  {Nuclear Physics A}\ }\textbf {\bibinfo {volume} {221}}~(\bibinfo {number}
  {1}),\ \bibinfo {pages} {163 }}\BibitemShut {NoStop}%
\bibitem [{\citenamefont {Forss\'en}\ \emph {et~al.}(2005)\citenamefont
  {Forss\'en}, \citenamefont {Navr\'atil}, \citenamefont {Ormand},\ and\
  \citenamefont {Caurier}}]{forssen2005}%
  \BibitemOpen
  \bibfield  {author} {\bibinfo {author} {\bibnamefont {Forss\'en},
  \bibfnamefont {C.}}, \bibinfo {author} {\bibfnamefont {P.}~\bibnamefont
  {Navr\'atil}}, \bibinfo {author} {\bibfnamefont {W.~E.}\ \bibnamefont
  {Ormand}}, \ and\ \bibinfo {author} {\bibfnamefont {E.}~\bibnamefont
  {Caurier}}} (\bibinfo {year} {2005}),\ \href {\doibase
  10.1103/PhysRevC.71.044312} {\bibfield  {journal} {\bibinfo  {journal} {Phys.
  Rev. C}\ }\textbf {\bibinfo {volume} {71}},\ \bibinfo {pages}
  {044312}}\BibitemShut {NoStop}%
\bibitem [{\citenamefont {Forss\'en}\ \emph {et~al.}(2008)\citenamefont
  {Forss\'en}, \citenamefont {Vary}, \citenamefont {Caurier},\ and\
  \citenamefont {Navr\'atil}}]{forssen2008}%
  \BibitemOpen
  \bibfield  {author} {\bibinfo {author} {\bibnamefont {Forss\'en},
  \bibfnamefont {C.}}, \bibinfo {author} {\bibfnamefont {J.~P.}\ \bibnamefont
  {Vary}}, \bibinfo {author} {\bibfnamefont {E.}~\bibnamefont {Caurier}}, \
  and\ \bibinfo {author} {\bibfnamefont {P.}~\bibnamefont {Navr\'atil}}}
  (\bibinfo {year} {2008}),\ \href {\doibase 10.1103/PhysRevC.77.024301}
  {\bibfield  {journal} {\bibinfo  {journal} {Phys. Rev. C}\ }\textbf {\bibinfo
  {volume} {77}},\ \bibinfo {pages} {024301}}\BibitemShut {NoStop}%
\bibitem [{\citenamefont {{Friman}}\ and\ \citenamefont
  {{Schwenk}}(2011)}]{friman2011}%
  \BibitemOpen
  \bibfield  {author} {\bibinfo {author} {\bibnamefont {{Friman}},
  \bibfnamefont {B.}}, \ and\ \bibinfo {author} {\bibfnamefont
  {A.}~\bibnamefont {{Schwenk}}}} (\bibinfo {year} {2011}),\ \href
  {http://adsabs.harvard.edu/abs/2011arXiv1101.4858F} {\bibfield  {journal}
  {\bibinfo  {journal} {ArXiv e-prints}\ }}\Eprint
  {http://arxiv.org/abs/1101.4858} {arXiv:1101.4858 [nucl-th]} \BibitemShut
  {NoStop}%
\bibitem [{\citenamefont {Fujii}\ \emph {et~al.}(2009)\citenamefont {Fujii},
  \citenamefont {Okamoto},\ and\ \citenamefont {Suzuki}}]{fujii2009}%
  \BibitemOpen
  \bibfield  {author} {\bibinfo {author} {\bibnamefont {Fujii}, \bibfnamefont
  {S.}}, \bibinfo {author} {\bibfnamefont {R.}~\bibnamefont {Okamoto}}, \ and\
  \bibinfo {author} {\bibfnamefont {K.}~\bibnamefont {Suzuki}}} (\bibinfo
  {year} {2009}),\ \href {\doibase 10.1103/PhysRevLett.103.182501} {\bibfield
  {journal} {\bibinfo  {journal} {Phys. Rev. Lett.}\ }\textbf {\bibinfo
  {volume} {103}},\ \bibinfo {pages} {182501}}\BibitemShut {NoStop}%
\bibitem [{\citenamefont {Fujita}\ and\ \citenamefont
  {Miyazawa}(1957)}]{fujita1957}%
  \BibitemOpen
  \bibfield  {author} {\bibinfo {author} {\bibnamefont {Fujita}, \bibfnamefont
  {J.}}, \ and\ \bibinfo {author} {\bibfnamefont {H.}~\bibnamefont {Miyazawa}}}
  (\bibinfo {year} {1957}),\ \href {\doibase 10.1143/PTP.17.360} {\bibfield
  {journal} {\bibinfo  {journal} {Progress of Theoretical Physics}\ }\textbf
  {\bibinfo {volume} {17}}~(\bibinfo {number} {3}),\ \bibinfo {pages}
  {360}}\BibitemShut {NoStop}%
\bibitem [{\citenamefont {Furnstahl}\ \emph {et~al.}(2012)\citenamefont
  {Furnstahl}, \citenamefont {Hagen},\ and\ \citenamefont
  {Papenbrock}}]{furnstahl2012}%
  \BibitemOpen
  \bibfield  {author} {\bibinfo {author} {\bibnamefont {Furnstahl},
  \bibfnamefont {R.~J.}}, \bibinfo {author} {\bibfnamefont {G.}~\bibnamefont
  {Hagen}}, \ and\ \bibinfo {author} {\bibfnamefont {T.}~\bibnamefont
  {Papenbrock}}} (\bibinfo {year} {2012}),\ \href {\doibase
  10.1103/PhysRevC.86.031301} {\bibfield  {journal} {\bibinfo  {journal} {Phys.
  Rev. C}\ }\textbf {\bibinfo {volume} {86}},\ \bibinfo {pages}
  {031301}}\BibitemShut {NoStop}%
\bibitem [{\citenamefont {{Furnstahl}}\ and\ \citenamefont
  {{Hebeler}}(2013)}]{furnstahl2013}%
  \BibitemOpen
  \bibfield  {author} {\bibinfo {author} {\bibnamefont {{Furnstahl}},
  \bibfnamefont {R.~J.}}, \ and\ \bibinfo {author} {\bibfnamefont
  {K.}~\bibnamefont {{Hebeler}}}} (\bibinfo {year} {2013}),\ \href
  {http://stacks.iop.org/0034-4885/76/i=12/a=126301} {\bibfield  {journal}
  {\bibinfo  {journal} {Reports on Progress in Physics}\ }\textbf {\bibinfo
  {volume} {76}}~(\bibinfo {number} {12}),\ \bibinfo {pages}
  {126301}}\BibitemShut {NoStop}%
\bibitem [{\citenamefont {{Furnstahl}}\ \emph {et~al.}(2013)\citenamefont
  {{Furnstahl}}, \citenamefont {{Papenbrock}},\ and\ \citenamefont
  {{More}}}]{furnstahl2013b}%
  \BibitemOpen
  \bibfield  {author} {\bibinfo {author} {\bibnamefont {{Furnstahl}},
  \bibfnamefont {R.~J.}}, \bibinfo {author} {\bibfnamefont {T.}~\bibnamefont
  {{Papenbrock}}}, \ and\ \bibinfo {author} {\bibfnamefont {S.~N.}\
  \bibnamefont {{More}}}} (\bibinfo {year} {2013}),\ \href@noop {} {\bibfield
  {journal} {\bibinfo  {journal} {ArXiv e-prints}\ }}\Eprint
  {http://arxiv.org/abs/1312.6876} {arXiv:1312.6876 [nucl-th]} \BibitemShut
  {NoStop}%
\bibitem [{\citenamefont {Gade}\ \emph {et~al.}(2006)\citenamefont {Gade},
  \citenamefont {Janssens}, \citenamefont {Bazin}, \citenamefont {Broda},
  \citenamefont {Brown}, \citenamefont {Campbell}, \citenamefont {Carpenter},
  \citenamefont {Cook}, \citenamefont {Deacon}, \citenamefont {Dinca},
  \citenamefont {Fornal}, \citenamefont {Freeman}, \citenamefont {Glasmacher},
  \citenamefont {Hansen}, \citenamefont {Kay}, \citenamefont {Mantica},
  \citenamefont {Mueller}, \citenamefont {Terry}, \citenamefont {Tostevin},\
  and\ \citenamefont {Zhu}}]{gade2006}%
  \BibitemOpen
  \bibfield  {author} {\bibinfo {author} {\bibnamefont {Gade}, \bibfnamefont
  {A.}}, \bibinfo {author} {\bibfnamefont {R.~V.~F.}\ \bibnamefont {Janssens}},
  \bibinfo {author} {\bibfnamefont {D.}~\bibnamefont {Bazin}}, \bibinfo
  {author} {\bibfnamefont {R.}~\bibnamefont {Broda}}, \bibinfo {author}
  {\bibfnamefont {B.~A.}\ \bibnamefont {Brown}}, \bibinfo {author}
  {\bibfnamefont {C.~M.}\ \bibnamefont {Campbell}}, \bibinfo {author}
  {\bibfnamefont {M.~P.}\ \bibnamefont {Carpenter}}, \bibinfo {author}
  {\bibfnamefont {J.~M.}\ \bibnamefont {Cook}}, \bibinfo {author}
  {\bibfnamefont {A.~N.}\ \bibnamefont {Deacon}}, \bibinfo {author}
  {\bibfnamefont {D.-C.}\ \bibnamefont {Dinca}}, \bibinfo {author}
  {\bibfnamefont {B.}~\bibnamefont {Fornal}}, \bibinfo {author} {\bibfnamefont
  {S.~J.}\ \bibnamefont {Freeman}}, \bibinfo {author} {\bibfnamefont
  {T.}~\bibnamefont {Glasmacher}}, \bibinfo {author} {\bibfnamefont {P.~G.}\
  \bibnamefont {Hansen}}, \bibinfo {author} {\bibfnamefont {B.~P.}\
  \bibnamefont {Kay}}, \bibinfo {author} {\bibfnamefont {P.~F.}\ \bibnamefont
  {Mantica}}, \bibinfo {author} {\bibfnamefont {W.~F.}\ \bibnamefont
  {Mueller}}, \bibinfo {author} {\bibfnamefont {J.~R.}\ \bibnamefont {Terry}},
  \bibinfo {author} {\bibfnamefont {J.~A.}\ \bibnamefont {Tostevin}}, \ and\
  \bibinfo {author} {\bibfnamefont {S.}~\bibnamefont {Zhu}}} (\bibinfo {year}
  {2006}),\ \href {\doibase 10.1103/PhysRevC.74.021302} {\bibfield  {journal}
  {\bibinfo  {journal} {Phys. Rev. C}\ }\textbf {\bibinfo {volume} {74}},\
  \bibinfo {pages} {021302}}\BibitemShut {NoStop}%
\bibitem [{\citenamefont {Gallant}\ \emph {et~al.}(2012)\citenamefont
  {Gallant}, \citenamefont {Bale}, \citenamefont {Brunner}, \citenamefont
  {Chowdhury}, \citenamefont {Ettenauer}, \citenamefont {Lennarz},
  \citenamefont {Robertson}, \citenamefont {Simon}, \citenamefont {Chaudhuri},
  \citenamefont {Holt}, \citenamefont {Kwiatkowski}, \citenamefont {Man\'e},
  \citenamefont {Men\'endez}, \citenamefont {Schultz}, \citenamefont {Simon},
  \citenamefont {Andreoiu}, \citenamefont {Delheij}, \citenamefont {Pearson},
  \citenamefont {Savajols}, \citenamefont {Schwenk},\ and\ \citenamefont
  {Dilling}}]{gallant2012}%
  \BibitemOpen
  \bibfield  {author} {\bibinfo {author} {\bibnamefont {Gallant}, \bibfnamefont
  {A.~T.}}, \bibinfo {author} {\bibfnamefont {J.~C.}\ \bibnamefont {Bale}},
  \bibinfo {author} {\bibfnamefont {T.}~\bibnamefont {Brunner}}, \bibinfo
  {author} {\bibfnamefont {U.}~\bibnamefont {Chowdhury}}, \bibinfo {author}
  {\bibfnamefont {S.}~\bibnamefont {Ettenauer}}, \bibinfo {author}
  {\bibfnamefont {A.}~\bibnamefont {Lennarz}}, \bibinfo {author} {\bibfnamefont
  {D.}~\bibnamefont {Robertson}}, \bibinfo {author} {\bibfnamefont {V.~V.}\
  \bibnamefont {Simon}}, \bibinfo {author} {\bibfnamefont {A.}~\bibnamefont
  {Chaudhuri}}, \bibinfo {author} {\bibfnamefont {J.~D.}\ \bibnamefont {Holt}},
  \bibinfo {author} {\bibfnamefont {A.~A.}\ \bibnamefont {Kwiatkowski}},
  \bibinfo {author} {\bibfnamefont {E.}~\bibnamefont {Man\'e}}, \bibinfo
  {author} {\bibfnamefont {J.}~\bibnamefont {Men\'endez}}, \bibinfo {author}
  {\bibfnamefont {B.~E.}\ \bibnamefont {Schultz}}, \bibinfo {author}
  {\bibfnamefont {M.~C.}\ \bibnamefont {Simon}}, \bibinfo {author}
  {\bibfnamefont {C.}~\bibnamefont {Andreoiu}}, \bibinfo {author}
  {\bibfnamefont {P.}~\bibnamefont {Delheij}}, \bibinfo {author} {\bibfnamefont
  {M.~R.}\ \bibnamefont {Pearson}}, \bibinfo {author} {\bibfnamefont
  {H.}~\bibnamefont {Savajols}}, \bibinfo {author} {\bibfnamefont
  {A.}~\bibnamefont {Schwenk}}, \ and\ \bibinfo {author} {\bibfnamefont
  {J.}~\bibnamefont {Dilling}}} (\bibinfo {year} {2012}),\ \href {\doibase
  10.1103/PhysRevLett.109.032506} {\bibfield  {journal} {\bibinfo  {journal}
  {Phys. Rev. Lett.}\ }\textbf {\bibinfo {volume} {109}},\ \bibinfo {pages}
  {032506}}\BibitemShut {NoStop}%
\bibitem [{\citenamefont {Gartenhaus}\ and\ \citenamefont
  {Schwartz}(1957)}]{gartenhaus1957}%
  \BibitemOpen
  \bibfield  {author} {\bibinfo {author} {\bibnamefont {Gartenhaus},
  \bibfnamefont {S.}}, \ and\ \bibinfo {author} {\bibfnamefont
  {C.}~\bibnamefont {Schwartz}}} (\bibinfo {year} {1957}),\ \href {\doibase
  10.1103/PhysRev.108.482} {\bibfield  {journal} {\bibinfo  {journal} {Phys.
  Rev.}\ }\textbf {\bibinfo {volume} {108}},\ \bibinfo {pages}
  {482}}\BibitemShut {NoStop}%
\bibitem [{\citenamefont {Gazit}\ \emph {et~al.}(2006)\citenamefont {Gazit},
  \citenamefont {Bacca}, \citenamefont {Barnea}, \citenamefont {Leidemann},\
  and\ \citenamefont {Orlandini}}]{gazit2006}%
  \BibitemOpen
  \bibfield  {author} {\bibinfo {author} {\bibnamefont {Gazit}, \bibfnamefont
  {D.}}, \bibinfo {author} {\bibfnamefont {S.}~\bibnamefont {Bacca}}, \bibinfo
  {author} {\bibfnamefont {N.}~\bibnamefont {Barnea}}, \bibinfo {author}
  {\bibfnamefont {W.}~\bibnamefont {Leidemann}}, \ and\ \bibinfo {author}
  {\bibfnamefont {G.}~\bibnamefont {Orlandini}}} (\bibinfo {year} {2006}),\
  \href {\doibase 10.1103/PhysRevLett.96.112301} {\bibfield  {journal}
  {\bibinfo  {journal} {Phys. Rev. Lett.}\ }\textbf {\bibinfo {volume} {96}},\
  \bibinfo {pages} {112301}}\BibitemShut {NoStop}%
\bibitem [{\citenamefont {Giraud}(2008)}]{giraud2008}%
  \BibitemOpen
  \bibfield  {author} {\bibinfo {author} {\bibnamefont {Giraud}, \bibfnamefont
  {B.~G.}}} (\bibinfo {year} {2008}),\ \href {\doibase
  10.1103/PhysRevC.77.014311} {\bibfield  {journal} {\bibinfo  {journal} {Phys.
  Rev. C}\ }\textbf {\bibinfo {volume} {77}},\ \bibinfo {pages}
  {014311}}\BibitemShut {NoStop}%
\bibitem [{\citenamefont {G\l{}azek}\ and\ \citenamefont
  {Wilson}(1993)}]{glazek1993}%
  \BibitemOpen
  \bibfield  {author} {\bibinfo {author} {\bibnamefont {G\l{}azek},
  \bibfnamefont {S.~D.}}, \ and\ \bibinfo {author} {\bibfnamefont {K.~G.}\
  \bibnamefont {Wilson}}} (\bibinfo {year} {1993}),\ \href {\doibase
  10.1103/PhysRevD.48.5863} {\bibfield  {journal} {\bibinfo  {journal} {Phys.
  Rev. D}\ }\textbf {\bibinfo {volume} {48}},\ \bibinfo {pages}
  {5863}}\BibitemShut {NoStop}%
\bibitem [{\citenamefont {Gloeckner}\ and\ \citenamefont
  {Lawson}(1974)}]{gloeckner1974}%
  \BibitemOpen
  \bibfield  {author} {\bibinfo {author} {\bibnamefont {Gloeckner},
  \bibfnamefont {D.}}, \ and\ \bibinfo {author} {\bibfnamefont
  {R.}~\bibnamefont {Lawson}}} (\bibinfo {year} {1974}),\ \href {\doibase
  10.1016/0370-2693(74)90390-6} {\bibfield  {journal} {\bibinfo  {journal}
  {Physics Letters B}\ }\textbf {\bibinfo {volume} {53}}~(\bibinfo {number}
  {4}),\ \bibinfo {pages} {313 }}\BibitemShut {NoStop}%
\bibitem [{\citenamefont {Goldstone}(1957)}]{goldstone1957}%
  \BibitemOpen
  \bibfield  {author} {\bibinfo {author} {\bibnamefont {Goldstone},
  \bibfnamefont {J.}}} (\bibinfo {year} {1957}),\ \href {\doibase
  10.1098/rspa.1957.0037} {\bibfield  {journal} {\bibinfo  {journal}
  {Proceedings of the Royal Society of London. Series A. Mathematical and
  Physical Sciences}\ }\textbf {\bibinfo {volume} {239}}~(\bibinfo {number}
  {1217}),\ \bibinfo {pages} {267}}\BibitemShut {NoStop}%
\bibitem [{\citenamefont {Goriely}\ \emph {et~al.}(2009)\citenamefont
  {Goriely}, \citenamefont {Chamel},\ and\ \citenamefont
  {Pearson}}]{goriely2009}%
  \BibitemOpen
  \bibfield  {author} {\bibinfo {author} {\bibnamefont {Goriely}, \bibfnamefont
  {S.}}, \bibinfo {author} {\bibfnamefont {N.}~\bibnamefont {Chamel}}, \ and\
  \bibinfo {author} {\bibfnamefont {J.~M.}\ \bibnamefont {Pearson}}} (\bibinfo
  {year} {2009}),\ \href {\doibase 10.1103/PhysRevLett.102.152503} {\bibfield
  {journal} {\bibinfo  {journal} {Phys. Rev. Lett.}\ }\textbf {\bibinfo
  {volume} {102}},\ \bibinfo {pages} {152503}}\BibitemShut {NoStop}%
\bibitem [{\citenamefont {Gour}\ \emph {et~al.}(2008)\citenamefont {Gour},
  \citenamefont {Horoi}, \citenamefont {Piecuch},\ and\ \citenamefont
  {Brown}}]{gour2008}%
  \BibitemOpen
  \bibfield  {author} {\bibinfo {author} {\bibnamefont {Gour}, \bibfnamefont
  {J.~R.}}, \bibinfo {author} {\bibfnamefont {M.}~\bibnamefont {Horoi}},
  \bibinfo {author} {\bibfnamefont {P.}~\bibnamefont {Piecuch}}, \ and\
  \bibinfo {author} {\bibfnamefont {B.~A.}\ \bibnamefont {Brown}}} (\bibinfo
  {year} {2008}),\ \href {\doibase 10.1103/PhysRevLett.101.052501} {\bibfield
  {journal} {\bibinfo  {journal} {Phys. Rev. Lett.}\ }\textbf {\bibinfo
  {volume} {101}},\ \bibinfo {pages} {052501}}\BibitemShut {NoStop}%
\bibitem [{\citenamefont {Gour}\ \emph {et~al.}(2006)\citenamefont {Gour},
  \citenamefont {Piecuch}, \citenamefont {Hjorth-Jensen}, \citenamefont
  {W\l{}och},\ and\ \citenamefont {Dean}}]{gour2006}%
  \BibitemOpen
  \bibfield  {author} {\bibinfo {author} {\bibnamefont {Gour}, \bibfnamefont
  {J.~R.}}, \bibinfo {author} {\bibfnamefont {P.}~\bibnamefont {Piecuch}},
  \bibinfo {author} {\bibfnamefont {M.}~\bibnamefont {Hjorth-Jensen}}, \bibinfo
  {author} {\bibfnamefont {M.}~\bibnamefont {W\l{}och}}, \ and\ \bibinfo
  {author} {\bibfnamefont {D.~J.}\ \bibnamefont {Dean}}} (\bibinfo {year}
  {2006}),\ \href {\doibase 10.1103/PhysRevC.74.024310} {\bibfield  {journal}
  {\bibinfo  {journal} {Phys. Rev. C}\ }\textbf {\bibinfo {volume} {74}},\
  \bibinfo {pages} {024310}}\BibitemShut {NoStop}%
\bibitem [{\citenamefont {Gros}(1992)}]{gros1992}%
  \BibitemOpen
  \bibfield  {author} {\bibinfo {author} {\bibnamefont {Gros}, \bibfnamefont
  {C.}}} (\bibinfo {year} {1992}),\ \href {\doibase 10.1007/BF01323728}
  {\bibfield  {journal} {\bibinfo  {journal} {Zeitschrift f{\"u}r Physik B
  Condensed Matter}\ }\textbf {\bibinfo {volume} {86}}~(\bibinfo {number}
  {3}),\ \bibinfo {pages} {359}}\BibitemShut {NoStop}%
\bibitem [{\citenamefont {Gros}(1996)}]{gros1996}%
  \BibitemOpen
  \bibfield  {author} {\bibinfo {author} {\bibnamefont {Gros}, \bibfnamefont
  {C.}}} (\bibinfo {year} {1996}),\ \href {\doibase 10.1103/PhysRevB.53.6865}
  {\bibfield  {journal} {\bibinfo  {journal} {Phys. Rev. B}\ }\textbf {\bibinfo
  {volume} {53}},\ \bibinfo {pages} {6865}}\BibitemShut {NoStop}%
\bibitem [{\citenamefont {Guardiola}\ \emph {et~al.}(1998)\citenamefont
  {Guardiola}, \citenamefont {Moliner}, \citenamefont {Navarro},\ and\
  \citenamefont {Portesi}}]{guardiola1998}%
  \BibitemOpen
  \bibfield  {author} {\bibinfo {author} {\bibnamefont {Guardiola},
  \bibfnamefont {R.}}, \bibinfo {author} {\bibfnamefont {I.}~\bibnamefont
  {Moliner}}, \bibinfo {author} {\bibfnamefont {J.}~\bibnamefont {Navarro}}, \
  and\ \bibinfo {author} {\bibfnamefont {M.}~\bibnamefont {Portesi}}} (\bibinfo
  {year} {1998}),\ \href {\doibase 10.1016/S0375-9474(97)00627-1} {\bibfield
  {journal} {\bibinfo  {journal} {Nuclear Physics A}\ }\textbf {\bibinfo
  {volume} {628}}~(\bibinfo {number} {2}),\ \bibinfo {pages} {187
  }}\BibitemShut {NoStop}%
\bibitem [{\citenamefont {Guardiola}\ \emph {et~al.}(1996)\citenamefont
  {Guardiola}, \citenamefont {Moliner}, \citenamefont {Navarro}, \citenamefont
  {Bishop}, \citenamefont {Puente},\ and\ \citenamefont
  {Walet}}]{guardiola1996}%
  \BibitemOpen
  \bibfield  {author} {\bibinfo {author} {\bibnamefont {Guardiola},
  \bibfnamefont {R.}}, \bibinfo {author} {\bibfnamefont {P.}~\bibnamefont
  {Moliner}}, \bibinfo {author} {\bibfnamefont {J.}~\bibnamefont {Navarro}},
  \bibinfo {author} {\bibfnamefont {R.}~\bibnamefont {Bishop}}, \bibinfo
  {author} {\bibfnamefont {A.}~\bibnamefont {Puente}}, \ and\ \bibinfo {author}
  {\bibfnamefont {N.~R.}\ \bibnamefont {Walet}}} (\bibinfo {year} {1996}),\
  \href {\doibase 10.1016/0375-9474(96)00315-6} {\bibfield  {journal} {\bibinfo
   {journal} {Nuclear Physics A}\ }\textbf {\bibinfo {volume} {609}}~(\bibinfo
  {number} {2}),\ \bibinfo {pages} {218 }}\BibitemShut {NoStop}%
\bibitem [{\citenamefont {Haftel}\ and\ \citenamefont
  {Tabakin}(1970)}]{haftel1970}%
  \BibitemOpen
  \bibfield  {author} {\bibinfo {author} {\bibnamefont {Haftel}, \bibfnamefont
  {M.~I.}}, \ and\ \bibinfo {author} {\bibfnamefont {F.}~\bibnamefont
  {Tabakin}}} (\bibinfo {year} {1970}),\ \href {\doibase
  10.1016/0375-9474(70)90047-3} {\bibfield  {journal} {\bibinfo  {journal}
  {Nuclear Physics A}\ }\textbf {\bibinfo {volume} {158}}~(\bibinfo {number}
  {1}),\ \bibinfo {pages} {1 }}\BibitemShut {NoStop}%
\bibitem [{\citenamefont {Hagen}\ \emph
  {et~al.}(2007{\natexlab{a}})\citenamefont {Hagen}, \citenamefont {Dean},
  \citenamefont {Hjorth-Jensen},\ and\ \citenamefont
  {Papenbrock}}]{hagen2007d}%
  \BibitemOpen
  \bibfield  {author} {\bibinfo {author} {\bibnamefont {Hagen}, \bibfnamefont
  {G.}}, \bibinfo {author} {\bibfnamefont {D.~J.}\ \bibnamefont {Dean}},
  \bibinfo {author} {\bibfnamefont {M.}~\bibnamefont {Hjorth-Jensen}}, \ and\
  \bibinfo {author} {\bibfnamefont {T.}~\bibnamefont {Papenbrock}}} (\bibinfo
  {year} {2007}{\natexlab{a}}),\ \href {\doibase
  10.1016/j.physletb.2007.07.072} {\bibfield  {journal} {\bibinfo  {journal}
  {Physics Letters B}\ }\textbf {\bibinfo {volume} {656}}~(\bibinfo {number}
  {4}),\ \bibinfo {pages} {169 }}\BibitemShut {NoStop}%
\bibitem [{\citenamefont {Hagen}\ \emph
  {et~al.}(2007{\natexlab{b}})\citenamefont {Hagen}, \citenamefont {Dean},
  \citenamefont {Hjorth-Jensen}, \citenamefont {Papenbrock},\ and\
  \citenamefont {Schwenk}}]{hagen2007b}%
  \BibitemOpen
  \bibfield  {author} {\bibinfo {author} {\bibnamefont {Hagen}, \bibfnamefont
  {G.}}, \bibinfo {author} {\bibfnamefont {D.~J.}\ \bibnamefont {Dean}},
  \bibinfo {author} {\bibfnamefont {M.}~\bibnamefont {Hjorth-Jensen}}, \bibinfo
  {author} {\bibfnamefont {T.}~\bibnamefont {Papenbrock}}, \ and\ \bibinfo
  {author} {\bibfnamefont {A.}~\bibnamefont {Schwenk}}} (\bibinfo {year}
  {2007}{\natexlab{b}}),\ \href {\doibase 10.1103/PhysRevC.76.044305}
  {\bibfield  {journal} {\bibinfo  {journal} {Phys. Rev. C}\ }\textbf {\bibinfo
  {volume} {76}},\ \bibinfo {pages} {044305}}\BibitemShut {NoStop}%
\bibitem [{\citenamefont {Hagen}\ \emph {et~al.}(2013)\citenamefont {Hagen},
  \citenamefont {Hagen}, \citenamefont {Hammer},\ and\ \citenamefont
  {Platter}}]{hagen2013}%
  \BibitemOpen
  \bibfield  {author} {\bibinfo {author} {\bibnamefont {Hagen}, \bibfnamefont
  {G.}}, \bibinfo {author} {\bibfnamefont {P.}~\bibnamefont {Hagen}}, \bibinfo
  {author} {\bibfnamefont {H.-W.}\ \bibnamefont {Hammer}}, \ and\ \bibinfo
  {author} {\bibfnamefont {L.}~\bibnamefont {Platter}}} (\bibinfo {year}
  {2013}),\ \href {\doibase 10.1103/PhysRevLett.111.132501} {\bibfield
  {journal} {\bibinfo  {journal} {Phys. Rev. Lett.}\ }\textbf {\bibinfo
  {volume} {111}},\ \bibinfo {pages} {132501}}\BibitemShut {NoStop}%
\bibitem [{\citenamefont {Hagen}\ \emph
  {et~al.}(2012{\natexlab{a}})\citenamefont {Hagen}, \citenamefont
  {Hjorth-Jensen}, \citenamefont {Jansen}, \citenamefont {Machleidt},\ and\
  \citenamefont {Papenbrock}}]{hagen2012a}%
  \BibitemOpen
  \bibfield  {author} {\bibinfo {author} {\bibnamefont {Hagen}, \bibfnamefont
  {G.}}, \bibinfo {author} {\bibfnamefont {M.}~\bibnamefont {Hjorth-Jensen}},
  \bibinfo {author} {\bibfnamefont {G.~R.}\ \bibnamefont {Jansen}}, \bibinfo
  {author} {\bibfnamefont {R.}~\bibnamefont {Machleidt}}, \ and\ \bibinfo
  {author} {\bibfnamefont {T.}~\bibnamefont {Papenbrock}}} (\bibinfo {year}
  {2012}{\natexlab{a}}),\ \href {\doibase 10.1103/PhysRevLett.108.242501}
  {\bibfield  {journal} {\bibinfo  {journal} {Phys. Rev. Lett.}\ }\textbf
  {\bibinfo {volume} {108}},\ \bibinfo {pages} {242501}}\BibitemShut {NoStop}%
\bibitem [{\citenamefont {Hagen}\ \emph
  {et~al.}(2012{\natexlab{b}})\citenamefont {Hagen}, \citenamefont
  {Hjorth-Jensen}, \citenamefont {Jansen}, \citenamefont {Machleidt},\ and\
  \citenamefont {Papenbrock}}]{hagen2012b}%
  \BibitemOpen
  \bibfield  {author} {\bibinfo {author} {\bibnamefont {Hagen}, \bibfnamefont
  {G.}}, \bibinfo {author} {\bibfnamefont {M.}~\bibnamefont {Hjorth-Jensen}},
  \bibinfo {author} {\bibfnamefont {G.~R.}\ \bibnamefont {Jansen}}, \bibinfo
  {author} {\bibfnamefont {R.}~\bibnamefont {Machleidt}}, \ and\ \bibinfo
  {author} {\bibfnamefont {T.}~\bibnamefont {Papenbrock}}} (\bibinfo {year}
  {2012}{\natexlab{b}}),\ \href {\doibase 10.1103/PhysRevLett.109.032502}
  {\bibfield  {journal} {\bibinfo  {journal} {Phys. Rev. Lett.}\ }\textbf
  {\bibinfo {volume} {109}},\ \bibinfo {pages} {032502}}\BibitemShut {NoStop}%
\bibitem [{\citenamefont {Hagen}\ \emph {et~al.}(2006)\citenamefont {Hagen},
  \citenamefont {Hjorth-Jensen},\ and\ \citenamefont {Michel}}]{hagen2006b}%
  \BibitemOpen
  \bibfield  {author} {\bibinfo {author} {\bibnamefont {Hagen}, \bibfnamefont
  {G.}}, \bibinfo {author} {\bibfnamefont {M.}~\bibnamefont {Hjorth-Jensen}}, \
  and\ \bibinfo {author} {\bibfnamefont {N.}~\bibnamefont {Michel}}} (\bibinfo
  {year} {2006}),\ \href {\doibase 10.1103/PhysRevC.73.064307} {\bibfield
  {journal} {\bibinfo  {journal} {Phys. Rev. C}\ }\textbf {\bibinfo {volume}
  {73}},\ \bibinfo {pages} {064307}}\BibitemShut {NoStop}%
\bibitem [{\citenamefont {Hagen}\ and\ \citenamefont
  {Michel}(2012)}]{hagen2012c}%
  \BibitemOpen
  \bibfield  {author} {\bibinfo {author} {\bibnamefont {Hagen}, \bibfnamefont
  {G.}}, \ and\ \bibinfo {author} {\bibfnamefont {N.}~\bibnamefont {Michel}}}
  (\bibinfo {year} {2012}),\ \href {\doibase 10.1103/PhysRevC.86.021602}
  {\bibfield  {journal} {\bibinfo  {journal} {Phys. Rev. C}\ }\textbf {\bibinfo
  {volume} {86}},\ \bibinfo {pages} {021602}}\BibitemShut {NoStop}%
\bibitem [{\citenamefont {Hagen}\ and\ \citenamefont {Nam}(2012)}]{hagen2012d}%
  \BibitemOpen
  \bibfield  {author} {\bibinfo {author} {\bibnamefont {Hagen}, \bibfnamefont
  {G.}}, \ and\ \bibinfo {author} {\bibfnamefont {H.~A.}\ \bibnamefont {Nam}}}
  (\bibinfo {year} {2012}),\ \href {\doibase 10.1143/PTPS.196.102} {\bibfield
  {journal} {\bibinfo  {journal} {Progress of Theoretical Physics Supplement}\
  }\textbf {\bibinfo {volume} {196}},\ \bibinfo {pages} {102}}\BibitemShut
  {NoStop}%
\bibitem [{\citenamefont {Hagen}\ \emph
  {et~al.}(2009{\natexlab{a}})\citenamefont {Hagen}, \citenamefont
  {Papenbrock},\ and\ \citenamefont {Dean}}]{hagen2009a}%
  \BibitemOpen
  \bibfield  {author} {\bibinfo {author} {\bibnamefont {Hagen}, \bibfnamefont
  {G.}}, \bibinfo {author} {\bibfnamefont {T.}~\bibnamefont {Papenbrock}}, \
  and\ \bibinfo {author} {\bibfnamefont {D.~J.}\ \bibnamefont {Dean}}}
  (\bibinfo {year} {2009}{\natexlab{a}}),\ \href {\doibase
  10.1103/PhysRevLett.103.062503} {\bibfield  {journal} {\bibinfo  {journal}
  {Phys. Rev. Lett.}\ }\textbf {\bibinfo {volume} {103}},\ \bibinfo {pages}
  {062503}}\BibitemShut {NoStop}%
\bibitem [{\citenamefont {Hagen}\ \emph {et~al.}(2008)\citenamefont {Hagen},
  \citenamefont {Papenbrock}, \citenamefont {Dean},\ and\ \citenamefont
  {Hjorth-Jensen}}]{hagen2008}%
  \BibitemOpen
  \bibfield  {author} {\bibinfo {author} {\bibnamefont {Hagen}, \bibfnamefont
  {G.}}, \bibinfo {author} {\bibfnamefont {T.}~\bibnamefont {Papenbrock}},
  \bibinfo {author} {\bibfnamefont {D.~J.}\ \bibnamefont {Dean}}, \ and\
  \bibinfo {author} {\bibfnamefont {M.}~\bibnamefont {Hjorth-Jensen}}}
  (\bibinfo {year} {2008}),\ \href {\doibase 10.1103/PhysRevLett.101.092502}
  {\bibfield  {journal} {\bibinfo  {journal} {Phys. Rev. Lett.}\ }\textbf
  {\bibinfo {volume} {101}},\ \bibinfo {pages} {092502}}\BibitemShut {NoStop}%
\bibitem [{\citenamefont {Hagen}\ \emph
  {et~al.}(2010{\natexlab{a}})\citenamefont {Hagen}, \citenamefont
  {Papenbrock}, \citenamefont {Dean},\ and\ \citenamefont
  {Hjorth-Jensen}}]{hagen2010b}%
  \BibitemOpen
  \bibfield  {author} {\bibinfo {author} {\bibnamefont {Hagen}, \bibfnamefont
  {G.}}, \bibinfo {author} {\bibfnamefont {T.}~\bibnamefont {Papenbrock}},
  \bibinfo {author} {\bibfnamefont {D.~J.}\ \bibnamefont {Dean}}, \ and\
  \bibinfo {author} {\bibfnamefont {M.}~\bibnamefont {Hjorth-Jensen}}}
  (\bibinfo {year} {2010}{\natexlab{a}}),\ \href {\doibase
  10.1103/PhysRevC.82.034330} {\bibfield  {journal} {\bibinfo  {journal} {Phys.
  Rev. C}\ }\textbf {\bibinfo {volume} {82}},\ \bibinfo {pages}
  {034330}}\BibitemShut {NoStop}%
\bibitem [{\citenamefont {Hagen}\ \emph
  {et~al.}(2009{\natexlab{b}})\citenamefont {Hagen}, \citenamefont
  {Papenbrock}, \citenamefont {Dean}, \citenamefont {Hjorth-Jensen},\ and\
  \citenamefont {Asokan}}]{hagen2009b}%
  \BibitemOpen
  \bibfield  {author} {\bibinfo {author} {\bibnamefont {Hagen}, \bibfnamefont
  {G.}}, \bibinfo {author} {\bibfnamefont {T.}~\bibnamefont {Papenbrock}},
  \bibinfo {author} {\bibfnamefont {D.~J.}\ \bibnamefont {Dean}}, \bibinfo
  {author} {\bibfnamefont {M.}~\bibnamefont {Hjorth-Jensen}}, \ and\ \bibinfo
  {author} {\bibfnamefont {B.~V.}\ \bibnamefont {Asokan}}} (\bibinfo {year}
  {2009}{\natexlab{b}}),\ \href {\doibase 10.1103/PhysRevC.80.021306}
  {\bibfield  {journal} {\bibinfo  {journal} {Phys. Rev. C}\ }\textbf {\bibinfo
  {volume} {80}},\ \bibinfo {pages} {021306}}\BibitemShut {NoStop}%
\bibitem [{\citenamefont {Hagen}\ \emph
  {et~al.}(2007{\natexlab{c}})\citenamefont {Hagen}, \citenamefont
  {Papenbrock}, \citenamefont {Dean}, \citenamefont {Schwenk}, \citenamefont
  {Nogga}, \citenamefont {W\l{}och},\ and\ \citenamefont
  {Piecuch}}]{hagen2007a}%
  \BibitemOpen
  \bibfield  {author} {\bibinfo {author} {\bibnamefont {Hagen}, \bibfnamefont
  {G.}}, \bibinfo {author} {\bibfnamefont {T.}~\bibnamefont {Papenbrock}},
  \bibinfo {author} {\bibfnamefont {D.~J.}\ \bibnamefont {Dean}}, \bibinfo
  {author} {\bibfnamefont {A.}~\bibnamefont {Schwenk}}, \bibinfo {author}
  {\bibfnamefont {A.}~\bibnamefont {Nogga}}, \bibinfo {author} {\bibfnamefont
  {M.}~\bibnamefont {W\l{}och}}, \ and\ \bibinfo {author} {\bibfnamefont
  {P.}~\bibnamefont {Piecuch}}} (\bibinfo {year} {2007}{\natexlab{c}}),\ \href
  {\doibase 10.1103/PhysRevC.76.034302} {\bibfield  {journal} {\bibinfo
  {journal} {Phys. Rev. C}\ }\textbf {\bibinfo {volume} {76}},\ \bibinfo
  {pages} {034302}}\BibitemShut {NoStop}%
\bibitem [{\citenamefont {{Hagen}}\ \emph {et~al.}(2013)\citenamefont
  {{Hagen}}, \citenamefont {{Papenbrock}}, \citenamefont {{Ekstr{\"o}m}},
  \citenamefont {{Wendt}}, \citenamefont {{Baardsen}}, \citenamefont
  {{Gandolfi}}, \citenamefont {{Hjorth-Jensen}},\ and\ \citenamefont
  {{Horowitz}}}]{hagen2013b}%
  \BibitemOpen
  \bibfield  {author} {\bibinfo {author} {\bibnamefont {{Hagen}}, \bibfnamefont
  {G.}}, \bibinfo {author} {\bibfnamefont {T.}~\bibnamefont {{Papenbrock}}},
  \bibinfo {author} {\bibfnamefont {A.}~\bibnamefont {{Ekstr{\"o}m}}}, \bibinfo
  {author} {\bibfnamefont {K.~A.}\ \bibnamefont {{Wendt}}}, \bibinfo {author}
  {\bibfnamefont {G.}~\bibnamefont {{Baardsen}}}, \bibinfo {author}
  {\bibfnamefont {S.}~\bibnamefont {{Gandolfi}}}, \bibinfo {author}
  {\bibfnamefont {M.}~\bibnamefont {{Hjorth-Jensen}}}, \ and\ \bibinfo {author}
  {\bibfnamefont {C.~J.}\ \bibnamefont {{Horowitz}}}} (\bibinfo {year}
  {2013}),\ \href@noop {} {\bibfield  {journal} {\bibinfo  {journal} {ArXiv
  e-prints}\ }}\Eprint {http://arxiv.org/abs/1311.2925} {arXiv:1311.2925
  [nucl-th]} \BibitemShut {NoStop}%
\bibitem [{\citenamefont {Hagen}\ \emph
  {et~al.}(2010{\natexlab{b}})\citenamefont {Hagen}, \citenamefont
  {Papenbrock},\ and\ \citenamefont {Hjorth-Jensen}}]{hagen2010a}%
  \BibitemOpen
  \bibfield  {author} {\bibinfo {author} {\bibnamefont {Hagen}, \bibfnamefont
  {G.}}, \bibinfo {author} {\bibfnamefont {T.}~\bibnamefont {Papenbrock}}, \
  and\ \bibinfo {author} {\bibfnamefont {M.}~\bibnamefont {Hjorth-Jensen}}}
  (\bibinfo {year} {2010}{\natexlab{b}}),\ \href {\doibase
  10.1103/PhysRevLett.104.182501} {\bibfield  {journal} {\bibinfo  {journal}
  {Phys. Rev. Lett.}\ }\textbf {\bibinfo {volume} {104}},\ \bibinfo {pages}
  {182501}}\BibitemShut {NoStop}%
\bibitem [{\citenamefont {Hagen}\ and\ \citenamefont
  {Vaagen}(2006)}]{hagen2006}%
  \BibitemOpen
  \bibfield  {author} {\bibinfo {author} {\bibnamefont {Hagen}, \bibfnamefont
  {G.}}, \ and\ \bibinfo {author} {\bibfnamefont {J.~S.}\ \bibnamefont
  {Vaagen}}} (\bibinfo {year} {2006}),\ \href {\doibase
  10.1103/PhysRevC.73.034321} {\bibfield  {journal} {\bibinfo  {journal} {Phys.
  Rev. C}\ }\textbf {\bibinfo {volume} {73}},\ \bibinfo {pages}
  {034321}}\BibitemShut {NoStop}%
\bibitem [{\citenamefont {Hagen}\ \emph {et~al.}(2004)\citenamefont {Hagen},
  \citenamefont {Vaagen},\ and\ \citenamefont {Hjorth-Jensen}}]{hagen2004}%
  \BibitemOpen
  \bibfield  {author} {\bibinfo {author} {\bibnamefont {Hagen}, \bibfnamefont
  {G.}}, \bibinfo {author} {\bibfnamefont {J.~S.}\ \bibnamefont {Vaagen}}, \
  and\ \bibinfo {author} {\bibfnamefont {M.}~\bibnamefont {Hjorth-Jensen}}}
  (\bibinfo {year} {2004}),\ \href
  {http://stacks.iop.org/0305-4470/37/i=38/a=006} {\bibfield  {journal}
  {\bibinfo  {journal} {Journal of Physics A: Mathematical and General}\
  }\textbf {\bibinfo {volume} {37}}~(\bibinfo {number} {38}),\ \bibinfo {pages}
  {8991}}\BibitemShut {NoStop}%
\bibitem [{\citenamefont {Hamamoto}(2012)}]{hamamoto2012}%
  \BibitemOpen
  \bibfield  {author} {\bibinfo {author} {\bibnamefont {Hamamoto},
  \bibfnamefont {I.}}} (\bibinfo {year} {2012}),\ \href {\doibase
  10.1103/PhysRevC.85.064329} {\bibfield  {journal} {\bibinfo  {journal} {Phys.
  Rev. C}\ }\textbf {\bibinfo {volume} {85}},\ \bibinfo {pages}
  {064329}}\BibitemShut {NoStop}%
\bibitem [{\citenamefont {Hammer}\ \emph {et~al.}(2013)\citenamefont {Hammer},
  \citenamefont {Nogga},\ and\ \citenamefont {Schwenk}}]{hammer2013}%
  \BibitemOpen
  \bibfield  {author} {\bibinfo {author} {\bibnamefont {Hammer}, \bibfnamefont
  {H.-W.}}, \bibinfo {author} {\bibfnamefont {A.}~\bibnamefont {Nogga}}, \ and\
  \bibinfo {author} {\bibfnamefont {A.}~\bibnamefont {Schwenk}}} (\bibinfo
  {year} {2013}),\ \href {\doibase 10.1103/RevModPhys.85.197} {\bibfield
  {journal} {\bibinfo  {journal} {Rev. Mod. Phys.}\ }\textbf {\bibinfo {volume}
  {85}},\ \bibinfo {pages} {197}}\BibitemShut {NoStop}%
\bibitem [{\citenamefont {Haxton}\ and\ \citenamefont
  {Johnson}(1990)}]{haxton1990}%
  \BibitemOpen
  \bibfield  {author} {\bibinfo {author} {\bibnamefont {Haxton}, \bibfnamefont
  {W.~C.}}, \ and\ \bibinfo {author} {\bibfnamefont {C.}~\bibnamefont
  {Johnson}}} (\bibinfo {year} {1990}),\ \href {\doibase
  10.1103/PhysRevLett.65.1325} {\bibfield  {journal} {\bibinfo  {journal}
  {Phys. Rev. Lett.}\ }\textbf {\bibinfo {volume} {65}},\ \bibinfo {pages}
  {1325}}\BibitemShut {NoStop}%
\bibitem [{\citenamefont {Hebeler}(2012)}]{hebeler2012}%
  \BibitemOpen
  \bibfield  {author} {\bibinfo {author} {\bibnamefont {Hebeler}, \bibfnamefont
  {K.}}} (\bibinfo {year} {2012}),\ \href {\doibase 10.1103/PhysRevC.85.021002}
  {\bibfield  {journal} {\bibinfo  {journal} {Phys. Rev. C}\ }\textbf {\bibinfo
  {volume} {85}},\ \bibinfo {pages} {021002}}\BibitemShut {NoStop}%
\bibitem [{\citenamefont {Hebeler}\ \emph {et~al.}(2011)\citenamefont
  {Hebeler}, \citenamefont {Bogner}, \citenamefont {Furnstahl}, \citenamefont
  {Nogga},\ and\ \citenamefont {Schwenk}}]{hebeler2011}%
  \BibitemOpen
  \bibfield  {author} {\bibinfo {author} {\bibnamefont {Hebeler}, \bibfnamefont
  {K.}}, \bibinfo {author} {\bibfnamefont {S.~K.}\ \bibnamefont {Bogner}},
  \bibinfo {author} {\bibfnamefont {R.~J.}\ \bibnamefont {Furnstahl}}, \bibinfo
  {author} {\bibfnamefont {A.}~\bibnamefont {Nogga}}, \ and\ \bibinfo {author}
  {\bibfnamefont {A.}~\bibnamefont {Schwenk}}} (\bibinfo {year} {2011}),\ \href
  {\doibase 10.1103/PhysRevC.83.031301} {\bibfield  {journal} {\bibinfo
  {journal} {Phys. Rev. C}\ }\textbf {\bibinfo {volume} {83}},\ \bibinfo
  {pages} {031301}}\BibitemShut {NoStop}%
\bibitem [{\citenamefont {Hebeler}\ \emph {et~al.}(2013)\citenamefont
  {Hebeler}, \citenamefont {Lattimer}, \citenamefont {Pethick},\ and\
  \citenamefont {Schwenk}}]{hebeler2013b}%
  \BibitemOpen
  \bibfield  {author} {\bibinfo {author} {\bibnamefont {Hebeler}, \bibfnamefont
  {K.}}, \bibinfo {author} {\bibfnamefont {J.~M.}\ \bibnamefont {Lattimer}},
  \bibinfo {author} {\bibfnamefont {C.~J.}\ \bibnamefont {Pethick}}, \ and\
  \bibinfo {author} {\bibfnamefont {A.}~\bibnamefont {Schwenk}}} (\bibinfo
  {year} {2013}),\ \href {http://stacks.iop.org/0004-637X/773/i=1/a=11}
  {\bibfield  {journal} {\bibinfo  {journal} {The Astrophysical Journal}\
  }\textbf {\bibinfo {volume} {773}}~(\bibinfo {number} {1}),\ \bibinfo {pages}
  {11}}\BibitemShut {NoStop}%
\bibitem [{\citenamefont {Hebeler}\ and\ \citenamefont
  {Schwenk}(2010)}]{hebeler2010b}%
  \BibitemOpen
  \bibfield  {author} {\bibinfo {author} {\bibnamefont {Hebeler}, \bibfnamefont
  {K.}}, \ and\ \bibinfo {author} {\bibfnamefont {A.}~\bibnamefont {Schwenk}}}
  (\bibinfo {year} {2010}),\ \href {\doibase 10.1103/PhysRevC.82.014314}
  {\bibfield  {journal} {\bibinfo  {journal} {Phys. Rev. C}\ }\textbf {\bibinfo
  {volume} {82}},\ \bibinfo {pages} {014314}}\BibitemShut {NoStop}%
\bibitem [{\citenamefont {Heidari}\ \emph {et~al.}(2007)\citenamefont
  {Heidari}, \citenamefont {Pal}, \citenamefont {Pujari},\ and\ \citenamefont
  {Kanhere}}]{heidari2007}%
  \BibitemOpen
  \bibfield  {author} {\bibinfo {author} {\bibnamefont {Heidari}, \bibfnamefont
  {I.}}, \bibinfo {author} {\bibfnamefont {S.}~\bibnamefont {Pal}}, \bibinfo
  {author} {\bibfnamefont {B.~S.}\ \bibnamefont {Pujari}}, \ and\ \bibinfo
  {author} {\bibfnamefont {D.~G.}\ \bibnamefont {Kanhere}}} (\bibinfo {year}
  {2007}),\ \href {\doibase 10.1063/1.2768523} {\bibfield  {journal} {\bibinfo
  {journal} {The Journal of Chemical Physics}\ }\textbf {\bibinfo {volume}
  {127}}~(\bibinfo {number} {11}),\ \bibinfo {eid} {114708}}\BibitemShut
  {NoStop}%
\bibitem [{\citenamefont {Heiselberg}\ and\ \citenamefont
  {Hjorth-Jensen}(2000)}]{hh2000}%
  \BibitemOpen
  \bibfield  {author} {\bibinfo {author} {\bibnamefont {Heiselberg},
  \bibfnamefont {H.}}, \ and\ \bibinfo {author} {\bibfnamefont
  {M.}~\bibnamefont {Hjorth-Jensen}}} (\bibinfo {year} {2000}),\ \href@noop {}
  {\bibfield  {journal} {\bibinfo  {journal} {Phys. Rep.}\ }\textbf {\bibinfo
  {volume} {328}},\ \bibinfo {pages} {237}}\BibitemShut {NoStop}%
\bibitem [{\citenamefont {Heisenberg}\ and\ \citenamefont
  {Mihaila}(1999)}]{heisenberg1999}%
  \BibitemOpen
  \bibfield  {author} {\bibinfo {author} {\bibnamefont {Heisenberg},
  \bibfnamefont {J.~H.}}, \ and\ \bibinfo {author} {\bibfnamefont
  {B.}~\bibnamefont {Mihaila}}} (\bibinfo {year} {1999}),\ \href {\doibase
  10.1103/PhysRevC.59.1440} {\bibfield  {journal} {\bibinfo  {journal} {Phys.
  Rev. C}\ }\textbf {\bibinfo {volume} {59}},\ \bibinfo {pages}
  {1440}}\BibitemShut {NoStop}%
\bibitem [{\citenamefont {Henderson}\ \emph {et~al.}(2003)\citenamefont
  {Henderson}, \citenamefont {Runge},\ and\ \citenamefont
  {Bartlett}}]{bartlett2003}%
  \BibitemOpen
  \bibfield  {author} {\bibinfo {author} {\bibnamefont {Henderson},
  \bibfnamefont {T.~M.}}, \bibinfo {author} {\bibfnamefont {K.}~\bibnamefont
  {Runge}}, \ and\ \bibinfo {author} {\bibfnamefont {R.~J.}\ \bibnamefont
  {Bartlett}}} (\bibinfo {year} {2003}),\ \href {\doibase
  10.1103/PhysRevB.67.045320} {\bibfield  {journal} {\bibinfo  {journal} {Phys.
  Rev. B}\ }\textbf {\bibinfo {volume} {67}},\ \bibinfo {pages}
  {045320}}\BibitemShut {NoStop}%
\bibitem [{\citenamefont {Hergert}\ \emph
  {et~al.}(2013{\natexlab{a}})\citenamefont {Hergert}, \citenamefont {Binder},
  \citenamefont {Calci}, \citenamefont {Langhammer},\ and\ \citenamefont
  {Roth}}]{hergert2013}%
  \BibitemOpen
  \bibfield  {author} {\bibinfo {author} {\bibnamefont {Hergert}, \bibfnamefont
  {H.}}, \bibinfo {author} {\bibfnamefont {S.}~\bibnamefont {Binder}}, \bibinfo
  {author} {\bibfnamefont {A.}~\bibnamefont {Calci}}, \bibinfo {author}
  {\bibfnamefont {J.}~\bibnamefont {Langhammer}}, \ and\ \bibinfo {author}
  {\bibfnamefont {R.}~\bibnamefont {Roth}}} (\bibinfo {year}
  {2013}{\natexlab{a}}),\ \href {\doibase 10.1103/PhysRevLett.110.242501}
  {\bibfield  {journal} {\bibinfo  {journal} {Phys. Rev. Lett.}\ }\textbf
  {\bibinfo {volume} {110}},\ \bibinfo {pages} {242501}}\BibitemShut {NoStop}%
\bibitem [{\citenamefont {Hergert}\ \emph
  {et~al.}(2013{\natexlab{b}})\citenamefont {Hergert}, \citenamefont {Bogner},
  \citenamefont {Binder}, \citenamefont {Calci}, \citenamefont {Langhammer},
  \citenamefont {Roth},\ and\ \citenamefont {Schwenk}}]{hergert2013b}%
  \BibitemOpen
  \bibfield  {author} {\bibinfo {author} {\bibnamefont {Hergert}, \bibfnamefont
  {H.}}, \bibinfo {author} {\bibfnamefont {S.~K.}\ \bibnamefont {Bogner}},
  \bibinfo {author} {\bibfnamefont {S.}~\bibnamefont {Binder}}, \bibinfo
  {author} {\bibfnamefont {A.}~\bibnamefont {Calci}}, \bibinfo {author}
  {\bibfnamefont {J.}~\bibnamefont {Langhammer}}, \bibinfo {author}
  {\bibfnamefont {R.}~\bibnamefont {Roth}}, \ and\ \bibinfo {author}
  {\bibfnamefont {A.}~\bibnamefont {Schwenk}}} (\bibinfo {year}
  {2013}{\natexlab{b}}),\ \href {\doibase 10.1103/PhysRevC.87.034307}
  {\bibfield  {journal} {\bibinfo  {journal} {Phys. Rev. C}\ }\textbf {\bibinfo
  {volume} {87}},\ \bibinfo {pages} {034307}}\BibitemShut {NoStop}%
\bibitem [{\citenamefont {Hergert}\ and\ \citenamefont
  {Roth}(2007)}]{hergert2007}%
  \BibitemOpen
  \bibfield  {author} {\bibinfo {author} {\bibnamefont {Hergert}, \bibfnamefont
  {H.}}, \ and\ \bibinfo {author} {\bibfnamefont {R.}~\bibnamefont {Roth}}}
  (\bibinfo {year} {2007}),\ \href {\doibase 10.1103/PhysRevC.75.051001}
  {\bibfield  {journal} {\bibinfo  {journal} {Phys. Rev. C}\ }\textbf {\bibinfo
  {volume} {75}},\ \bibinfo {pages} {051001}}\BibitemShut {NoStop}%
\bibitem [{\citenamefont {Hjorth-Jensen}\ \emph {et~al.}(1995)\citenamefont
  {Hjorth-Jensen}, \citenamefont {Kuo},\ and\ \citenamefont
  {Osnes}}]{hjorthjensen1995}%
  \BibitemOpen
  \bibfield  {author} {\bibinfo {author} {\bibnamefont {Hjorth-Jensen},
  \bibfnamefont {M.}}, \bibinfo {author} {\bibfnamefont {T.~T.}\ \bibnamefont
  {Kuo}}, \ and\ \bibinfo {author} {\bibfnamefont {E.}~\bibnamefont {Osnes}}}
  (\bibinfo {year} {1995}),\ \href {\doibase 10.1016/0370-1573(95)00012-6}
  {\bibfield  {journal} {\bibinfo  {journal} {Physics Reports}\ }\textbf
  {\bibinfo {volume} {261}}~(\bibinfo {number} {3}),\ \bibinfo {pages} {125
  }}\BibitemShut {NoStop}%
\bibitem [{\citenamefont {Hoffman}\ \emph {et~al.}(2009)\citenamefont
  {Hoffman}, \citenamefont {Baumann}, \citenamefont {Bazin}, \citenamefont
  {Brown}, \citenamefont {Christian}, \citenamefont {Denby}, \citenamefont
  {DeYoung}, \citenamefont {Finck}, \citenamefont {Frank}, \citenamefont
  {Hinnefeld}, \citenamefont {Mosby}, \citenamefont {Peters}, \citenamefont
  {Rogers}, \citenamefont {Schiller}, \citenamefont {Spyrou}, \citenamefont
  {Scott}, \citenamefont {Tabor}, \citenamefont {Thoennessen},\ and\
  \citenamefont {Voss}}]{hoffman2009}%
  \BibitemOpen
  \bibfield  {author} {\bibinfo {author} {\bibnamefont {Hoffman}, \bibfnamefont
  {C.~R.}}, \bibinfo {author} {\bibfnamefont {T.}~\bibnamefont {Baumann}},
  \bibinfo {author} {\bibfnamefont {D.}~\bibnamefont {Bazin}}, \bibinfo
  {author} {\bibfnamefont {J.}~\bibnamefont {Brown}}, \bibinfo {author}
  {\bibfnamefont {G.}~\bibnamefont {Christian}}, \bibinfo {author}
  {\bibfnamefont {D.~H.}\ \bibnamefont {Denby}}, \bibinfo {author}
  {\bibfnamefont {P.~A.}\ \bibnamefont {DeYoung}}, \bibinfo {author}
  {\bibfnamefont {J.~E.}\ \bibnamefont {Finck}}, \bibinfo {author}
  {\bibfnamefont {N.}~\bibnamefont {Frank}}, \bibinfo {author} {\bibfnamefont
  {J.}~\bibnamefont {Hinnefeld}}, \bibinfo {author} {\bibfnamefont
  {S.}~\bibnamefont {Mosby}}, \bibinfo {author} {\bibfnamefont {W.~A.}\
  \bibnamefont {Peters}}, \bibinfo {author} {\bibfnamefont {W.~F.}\
  \bibnamefont {Rogers}}, \bibinfo {author} {\bibfnamefont {A.}~\bibnamefont
  {Schiller}}, \bibinfo {author} {\bibfnamefont {A.}~\bibnamefont {Spyrou}},
  \bibinfo {author} {\bibfnamefont {M.~J.}\ \bibnamefont {Scott}}, \bibinfo
  {author} {\bibfnamefont {S.~L.}\ \bibnamefont {Tabor}}, \bibinfo {author}
  {\bibfnamefont {M.}~\bibnamefont {Thoennessen}}, \ and\ \bibinfo {author}
  {\bibfnamefont {P.}~\bibnamefont {Voss}}} (\bibinfo {year} {2009}),\ \href
  {\doibase 10.1016/j.physletb.2008.12.066} {\bibfield  {journal} {\bibinfo
  {journal} {Physics Letters B}\ }\textbf {\bibinfo {volume} {672}}~(\bibinfo
  {number} {1}),\ \bibinfo {pages} {17 }}\BibitemShut {NoStop}%
\bibitem [{\citenamefont {Hoffman}\ \emph {et~al.}(2008)\citenamefont
  {Hoffman}, \citenamefont {Baumann}, \citenamefont {Bazin}, \citenamefont
  {Brown}, \citenamefont {Christian}, \citenamefont {DeYoung}, \citenamefont
  {Finck}, \citenamefont {Frank}, \citenamefont {Hinnefeld}, \citenamefont
  {Howes}, \citenamefont {Mears}, \citenamefont {Mosby}, \citenamefont {Mosby},
  \citenamefont {Reith}, \citenamefont {Rizzo}, \citenamefont {Rogers},
  \citenamefont {Peaslee}, \citenamefont {Peters}, \citenamefont {Schiller},
  \citenamefont {Scott}, \citenamefont {Tabor}, \citenamefont {Thoennessen},
  \citenamefont {Voss},\ and\ \citenamefont {Williams}}]{hoffman2008}%
  \BibitemOpen
  \bibfield  {author} {\bibinfo {author} {\bibnamefont {Hoffman}, \bibfnamefont
  {C.~R.}}, \bibinfo {author} {\bibfnamefont {T.}~\bibnamefont {Baumann}},
  \bibinfo {author} {\bibfnamefont {D.}~\bibnamefont {Bazin}}, \bibinfo
  {author} {\bibfnamefont {J.}~\bibnamefont {Brown}}, \bibinfo {author}
  {\bibfnamefont {G.}~\bibnamefont {Christian}}, \bibinfo {author}
  {\bibfnamefont {P.~A.}\ \bibnamefont {DeYoung}}, \bibinfo {author}
  {\bibfnamefont {J.~E.}\ \bibnamefont {Finck}}, \bibinfo {author}
  {\bibfnamefont {N.}~\bibnamefont {Frank}}, \bibinfo {author} {\bibfnamefont
  {J.}~\bibnamefont {Hinnefeld}}, \bibinfo {author} {\bibfnamefont
  {R.}~\bibnamefont {Howes}}, \bibinfo {author} {\bibfnamefont
  {P.}~\bibnamefont {Mears}}, \bibinfo {author} {\bibfnamefont
  {E.}~\bibnamefont {Mosby}}, \bibinfo {author} {\bibfnamefont
  {S.}~\bibnamefont {Mosby}}, \bibinfo {author} {\bibfnamefont
  {J.}~\bibnamefont {Reith}}, \bibinfo {author} {\bibfnamefont
  {B.}~\bibnamefont {Rizzo}}, \bibinfo {author} {\bibfnamefont {W.~F.}\
  \bibnamefont {Rogers}}, \bibinfo {author} {\bibfnamefont {G.}~\bibnamefont
  {Peaslee}}, \bibinfo {author} {\bibfnamefont {W.~A.}\ \bibnamefont {Peters}},
  \bibinfo {author} {\bibfnamefont {A.}~\bibnamefont {Schiller}}, \bibinfo
  {author} {\bibfnamefont {M.~J.}\ \bibnamefont {Scott}}, \bibinfo {author}
  {\bibfnamefont {S.~L.}\ \bibnamefont {Tabor}}, \bibinfo {author}
  {\bibfnamefont {M.}~\bibnamefont {Thoennessen}}, \bibinfo {author}
  {\bibfnamefont {P.~J.}\ \bibnamefont {Voss}}, \ and\ \bibinfo {author}
  {\bibfnamefont {T.}~\bibnamefont {Williams}}} (\bibinfo {year} {2008}),\
  \href {\doibase 10.1103/PhysRevLett.100.152502} {\bibfield  {journal}
  {\bibinfo  {journal} {Phys. Rev. Lett.}\ }\textbf {\bibinfo {volume} {100}},\
  \bibinfo {pages} {152502}}\BibitemShut {NoStop}%
\bibitem [{\citenamefont {Hoffman}\ \emph {et~al.}(2011)\citenamefont
  {Hoffman}, \citenamefont {Baumann}, \citenamefont {Brown}, \citenamefont
  {DeYoung}, \citenamefont {Finck}, \citenamefont {Frank}, \citenamefont
  {Hinnefeld}, \citenamefont {Mosby}, \citenamefont {Peters}, \citenamefont
  {Rogers}, \citenamefont {Schiller}, \citenamefont {Snyder}, \citenamefont
  {Spyrou}, \citenamefont {Tabor},\ and\ \citenamefont
  {Thoennessen}}]{hoffman2011}%
  \BibitemOpen
  \bibfield  {author} {\bibinfo {author} {\bibnamefont {Hoffman}, \bibfnamefont
  {C.~R.}}, \bibinfo {author} {\bibfnamefont {T.}~\bibnamefont {Baumann}},
  \bibinfo {author} {\bibfnamefont {J.}~\bibnamefont {Brown}}, \bibinfo
  {author} {\bibfnamefont {P.~A.}\ \bibnamefont {DeYoung}}, \bibinfo {author}
  {\bibfnamefont {J.~E.}\ \bibnamefont {Finck}}, \bibinfo {author}
  {\bibfnamefont {N.}~\bibnamefont {Frank}}, \bibinfo {author} {\bibfnamefont
  {J.~D.}\ \bibnamefont {Hinnefeld}}, \bibinfo {author} {\bibfnamefont
  {S.}~\bibnamefont {Mosby}}, \bibinfo {author} {\bibfnamefont {W.~A.}\
  \bibnamefont {Peters}}, \bibinfo {author} {\bibfnamefont {W.~F.}\
  \bibnamefont {Rogers}}, \bibinfo {author} {\bibfnamefont {A.}~\bibnamefont
  {Schiller}}, \bibinfo {author} {\bibfnamefont {J.}~\bibnamefont {Snyder}},
  \bibinfo {author} {\bibfnamefont {A.}~\bibnamefont {Spyrou}}, \bibinfo
  {author} {\bibfnamefont {S.~L.}\ \bibnamefont {Tabor}}, \ and\ \bibinfo
  {author} {\bibfnamefont {M.}~\bibnamefont {Thoennessen}}} (\bibinfo {year}
  {2011}),\ \href {\doibase 10.1103/PhysRevC.83.031303} {\bibfield  {journal}
  {\bibinfo  {journal} {Phys. Rev. C}\ }\textbf {\bibinfo {volume} {83}},\
  \bibinfo {pages} {031303}}\BibitemShut {NoStop}%
\bibitem [{\citenamefont {Holt}\ \emph
  {et~al.}(2013{\natexlab{a}})\citenamefont {Holt}, \citenamefont
  {Men{\'e}ndez},\ and\ \citenamefont {Schwenk}}]{holt2012b}%
  \BibitemOpen
  \bibfield  {author} {\bibinfo {author} {\bibnamefont {Holt}, \bibfnamefont
  {J.~D.}}, \bibinfo {author} {\bibfnamefont {J.}~\bibnamefont {Men{\'e}ndez}},
  \ and\ \bibinfo {author} {\bibfnamefont {A.}~\bibnamefont {Schwenk}}}
  (\bibinfo {year} {2013}{\natexlab{a}}),\ \href {\doibase
  10.1140/epja/i2013-13039-2} {\bibfield  {journal} {\bibinfo  {journal} {The
  European Physical Journal A}\ }\textbf {\bibinfo {volume} {49}}~(\bibinfo
  {number} {3}),\ \bibinfo {pages} {1}}\BibitemShut {NoStop}%
\bibitem [{\citenamefont {Holt}\ \emph
  {et~al.}(2013{\natexlab{b}})\citenamefont {Holt}, \citenamefont
  {Men{\'e}ndez},\ and\ \citenamefont {Schwenk}}]{holt2013}%
  \BibitemOpen
  \bibfield  {author} {\bibinfo {author} {\bibnamefont {Holt}, \bibfnamefont
  {J.~D.}}, \bibinfo {author} {\bibfnamefont {J.}~\bibnamefont {Men{\'e}ndez}},
  \ and\ \bibinfo {author} {\bibfnamefont {A.}~\bibnamefont {Schwenk}}}
  (\bibinfo {year} {2013}{\natexlab{b}}),\ \href
  {http://stacks.iop.org/0954-3899/40/i=7/a=075105} {\bibfield  {journal}
  {\bibinfo  {journal} {Journal of Physics G: Nuclear and Particle Physics}\
  }\textbf {\bibinfo {volume} {40}}~(\bibinfo {number} {7}),\ \bibinfo {pages}
  {075105}}\BibitemShut {NoStop}%
\bibitem [{\citenamefont {Holt}\ \emph {et~al.}(2012)\citenamefont {Holt},
  \citenamefont {Otsuka}, \citenamefont {Schwenk},\ and\ \citenamefont
  {Suzuki}}]{holt2012}%
  \BibitemOpen
  \bibfield  {author} {\bibinfo {author} {\bibnamefont {Holt}, \bibfnamefont
  {J.~D.}}, \bibinfo {author} {\bibfnamefont {T.}~\bibnamefont {Otsuka}},
  \bibinfo {author} {\bibfnamefont {A.}~\bibnamefont {Schwenk}}, \ and\
  \bibinfo {author} {\bibfnamefont {T.}~\bibnamefont {Suzuki}}} (\bibinfo
  {year} {2012}),\ \href {http://stacks.iop.org/0954-3899/39/i=8/a=085111}
  {\bibfield  {journal} {\bibinfo  {journal} {Journal of Physics G: Nuclear and
  Particle Physics}\ }\textbf {\bibinfo {volume} {39}}~(\bibinfo {number}
  {8}),\ \bibinfo {pages} {085111}}\BibitemShut {NoStop}%
\bibitem [{\citenamefont {Holt}\ \emph {et~al.}(2008)\citenamefont {Holt},
  \citenamefont {Brown}, \citenamefont {Kuo}, \citenamefont {Holt},\ and\
  \citenamefont {Machleidt}}]{holtjw2008}%
  \BibitemOpen
  \bibfield  {author} {\bibinfo {author} {\bibnamefont {Holt}, \bibfnamefont
  {J.~W.}}, \bibinfo {author} {\bibfnamefont {G.~E.}\ \bibnamefont {Brown}},
  \bibinfo {author} {\bibfnamefont {T.~T.~S.}\ \bibnamefont {Kuo}}, \bibinfo
  {author} {\bibfnamefont {J.~D.}\ \bibnamefont {Holt}}, \ and\ \bibinfo
  {author} {\bibfnamefont {R.}~\bibnamefont {Machleidt}}} (\bibinfo {year}
  {2008}),\ \href {\doibase 10.1103/PhysRevLett.100.062501} {\bibfield
  {journal} {\bibinfo  {journal} {Phys. Rev. Lett.}\ }\textbf {\bibinfo
  {volume} {100}},\ \bibinfo {pages} {062501}}\BibitemShut {NoStop}%
\bibitem [{\citenamefont {Holt}\ \emph {et~al.}(2009)\citenamefont {Holt},
  \citenamefont {Kaiser},\ and\ \citenamefont {Weise}}]{holtjw2009}%
  \BibitemOpen
  \bibfield  {author} {\bibinfo {author} {\bibnamefont {Holt}, \bibfnamefont
  {J.~W.}}, \bibinfo {author} {\bibfnamefont {N.}~\bibnamefont {Kaiser}}, \
  and\ \bibinfo {author} {\bibfnamefont {W.}~\bibnamefont {Weise}}} (\bibinfo
  {year} {2009}),\ \href {\doibase 10.1103/PhysRevC.79.054331} {\bibfield
  {journal} {\bibinfo  {journal} {Phys. Rev. C}\ }\textbf {\bibinfo {volume}
  {79}},\ \bibinfo {pages} {054331}}\BibitemShut {NoStop}%
\bibitem [{\citenamefont {Holt}\ \emph {et~al.}(2010)\citenamefont {Holt},
  \citenamefont {Kaiser},\ and\ \citenamefont {Weise}}]{holtjw2010}%
  \BibitemOpen
  \bibfield  {author} {\bibinfo {author} {\bibnamefont {Holt}, \bibfnamefont
  {J.~W.}}, \bibinfo {author} {\bibfnamefont {N.}~\bibnamefont {Kaiser}}, \
  and\ \bibinfo {author} {\bibfnamefont {W.}~\bibnamefont {Weise}}} (\bibinfo
  {year} {2010}),\ \href {\doibase 10.1103/PhysRevC.81.024002} {\bibfield
  {journal} {\bibinfo  {journal} {Phys. Rev. C}\ }\textbf {\bibinfo {volume}
  {81}},\ \bibinfo {pages} {024002}}\BibitemShut {NoStop}%
\bibitem [{\citenamefont {Honma}\ \emph {et~al.}(2002)\citenamefont {Honma},
  \citenamefont {Otsuka}, \citenamefont {Brown},\ and\ \citenamefont
  {Mizusaki}}]{honma2002}%
  \BibitemOpen
  \bibfield  {author} {\bibinfo {author} {\bibnamefont {Honma}, \bibfnamefont
  {M.}}, \bibinfo {author} {\bibfnamefont {T.}~\bibnamefont {Otsuka}}, \bibinfo
  {author} {\bibfnamefont {B.~A.}\ \bibnamefont {Brown}}, \ and\ \bibinfo
  {author} {\bibfnamefont {T.}~\bibnamefont {Mizusaki}}} (\bibinfo {year}
  {2002}),\ \href {\doibase 10.1103/PhysRevC.65.061301} {\bibfield  {journal}
  {\bibinfo  {journal} {Phys. Rev. C}\ }\textbf {\bibinfo {volume} {65}},\
  \bibinfo {pages} {061301}}\BibitemShut {NoStop}%
\bibitem [{\citenamefont {Hoodbhoy}\ and\ \citenamefont
  {Negele}(1978)}]{HooNeg78}%
  \BibitemOpen
  \bibfield  {author} {\bibinfo {author} {\bibnamefont {Hoodbhoy},
  \bibfnamefont {P.}}, \ and\ \bibinfo {author} {\bibfnamefont {J.~W.}\
  \bibnamefont {Negele}}} (\bibinfo {year} {1978}),\ \href {\doibase
  10.1103/PhysRevC.18.2380} {\bibfield  {journal} {\bibinfo  {journal} {Phys.
  Rev. C}\ }\textbf {\bibinfo {volume} {18}},\ \bibinfo {pages}
  {2380}}\BibitemShut {NoStop}%
\bibitem [{\citenamefont {Hoodbhoy}\ and\ \citenamefont
  {Negele}(1979)}]{HooNeg79}%
  \BibitemOpen
  \bibfield  {author} {\bibinfo {author} {\bibnamefont {Hoodbhoy},
  \bibfnamefont {P.}}, \ and\ \bibinfo {author} {\bibfnamefont {J.~W.}\
  \bibnamefont {Negele}}} (\bibinfo {year} {1979}),\ \href {\doibase
  10.1103/PhysRevC.19.1971} {\bibfield  {journal} {\bibinfo  {journal} {Phys.
  Rev. C}\ }\textbf {\bibinfo {volume} {19}},\ \bibinfo {pages}
  {1971}}\BibitemShut {NoStop}%
\bibitem [{\citenamefont {Horoi}\ \emph {et~al.}(2007)\citenamefont {Horoi},
  \citenamefont {Gour}, \citenamefont {W\l{}och}, \citenamefont {Lodriguito},
  \citenamefont {Brown},\ and\ \citenamefont {Piecuch}}]{horoi2007}%
  \BibitemOpen
  \bibfield  {author} {\bibinfo {author} {\bibnamefont {Horoi}, \bibfnamefont
  {M.}}, \bibinfo {author} {\bibfnamefont {J.~R.}\ \bibnamefont {Gour}},
  \bibinfo {author} {\bibfnamefont {M.}~\bibnamefont {W\l{}och}}, \bibinfo
  {author} {\bibfnamefont {M.~D.}\ \bibnamefont {Lodriguito}}, \bibinfo
  {author} {\bibfnamefont {B.~A.}\ \bibnamefont {Brown}}, \ and\ \bibinfo
  {author} {\bibfnamefont {P.}~\bibnamefont {Piecuch}}} (\bibinfo {year}
  {2007}),\ \href {\doibase 10.1103/PhysRevLett.98.112501} {\bibfield
  {journal} {\bibinfo  {journal} {Phys. Rev. Lett.}\ }\textbf {\bibinfo
  {volume} {98}},\ \bibinfo {pages} {112501}}\BibitemShut {NoStop}%
\bibitem [{\citenamefont {Huber}\ and\ \citenamefont
  {Klamroth}(2011)}]{HubKla11}%
  \BibitemOpen
  \bibfield  {author} {\bibinfo {author} {\bibnamefont {Huber}, \bibfnamefont
  {C.}}, \ and\ \bibinfo {author} {\bibfnamefont {T.}~\bibnamefont {Klamroth}}}
  (\bibinfo {year} {2011}),\ \href {\doibase 10.1063/1.3530807} {\bibfield
  {journal} {\bibinfo  {journal} {The Journal of Chemical Physics}\ }\textbf
  {\bibinfo {volume} {134}}~(\bibinfo {number} {5}),\ \bibinfo {eid}
  {054113}}\BibitemShut {NoStop}%
\bibitem [{\citenamefont {Huck}\ \emph {et~al.}(1985)\citenamefont {Huck},
  \citenamefont {Klotz}, \citenamefont {Knipper}, \citenamefont {Mieh\'e},
  \citenamefont {Richard-Serre}, \citenamefont {Walter}, \citenamefont {Poves},
  \citenamefont {Ravn},\ and\ \citenamefont {Marguier}}]{huck1985}%
  \BibitemOpen
  \bibfield  {author} {\bibinfo {author} {\bibnamefont {Huck}, \bibfnamefont
  {A.}}, \bibinfo {author} {\bibfnamefont {G.}~\bibnamefont {Klotz}}, \bibinfo
  {author} {\bibfnamefont {A.}~\bibnamefont {Knipper}}, \bibinfo {author}
  {\bibfnamefont {C.}~\bibnamefont {Mieh\'e}}, \bibinfo {author} {\bibfnamefont
  {C.}~\bibnamefont {Richard-Serre}}, \bibinfo {author} {\bibfnamefont
  {G.}~\bibnamefont {Walter}}, \bibinfo {author} {\bibfnamefont
  {A.}~\bibnamefont {Poves}}, \bibinfo {author} {\bibfnamefont {H.~L.}\
  \bibnamefont {Ravn}}, \ and\ \bibinfo {author} {\bibfnamefont
  {G.}~\bibnamefont {Marguier}}} (\bibinfo {year} {1985}),\ \href {\doibase
  10.1103/PhysRevC.31.2226} {\bibfield  {journal} {\bibinfo  {journal} {Phys.
  Rev. C}\ }\textbf {\bibinfo {volume} {31}},\ \bibinfo {pages}
  {2226}}\BibitemShut {NoStop}%
\bibitem [{\citenamefont {Hupin}\ \emph {et~al.}(2013)\citenamefont {Hupin},
  \citenamefont {Langhammer}, \citenamefont {Navr\'atil}, \citenamefont
  {Quaglioni}, \citenamefont {Calci},\ and\ \citenamefont {Roth}}]{hupin2013}%
  \BibitemOpen
  \bibfield  {author} {\bibinfo {author} {\bibnamefont {Hupin}, \bibfnamefont
  {G.}}, \bibinfo {author} {\bibfnamefont {J.}~\bibnamefont {Langhammer}},
  \bibinfo {author} {\bibfnamefont {P.}~\bibnamefont {Navr\'atil}}, \bibinfo
  {author} {\bibfnamefont {S.}~\bibnamefont {Quaglioni}}, \bibinfo {author}
  {\bibfnamefont {A.}~\bibnamefont {Calci}}, \ and\ \bibinfo {author}
  {\bibfnamefont {R.}~\bibnamefont {Roth}}} (\bibinfo {year} {2013}),\ \href
  {\doibase 10.1103/PhysRevC.88.054622} {\bibfield  {journal} {\bibinfo
  {journal} {Phys. Rev. C}\ }\textbf {\bibinfo {volume} {88}},\ \bibinfo
  {pages} {054622}}\BibitemShut {NoStop}%
\bibitem [{\citenamefont {Id~Betan}\ \emph {et~al.}(2002)\citenamefont
  {Id~Betan}, \citenamefont {Liotta}, \citenamefont {Sandulescu},\ and\
  \citenamefont {Vertse}}]{idbetan2002}%
  \BibitemOpen
  \bibfield  {author} {\bibinfo {author} {\bibnamefont {Id~Betan},
  \bibfnamefont {R.}}, \bibinfo {author} {\bibfnamefont {R.~J.}\ \bibnamefont
  {Liotta}}, \bibinfo {author} {\bibfnamefont {N.}~\bibnamefont {Sandulescu}},
  \ and\ \bibinfo {author} {\bibfnamefont {T.}~\bibnamefont {Vertse}}}
  (\bibinfo {year} {2002}),\ \href {\doibase 10.1103/PhysRevLett.89.042501}
  {\bibfield  {journal} {\bibinfo  {journal} {Phys. Rev. Lett.}\ }\textbf
  {\bibinfo {volume} {89}},\ \bibinfo {pages} {042501}}\BibitemShut {NoStop}%
\bibitem [{\citenamefont {Ishkhanov}\ \emph {et~al.}(2002)\citenamefont
  {Ishkhanov}, \citenamefont {Kapitonov}, \citenamefont {Lileeva},
  \citenamefont {Shirokov}, \citenamefont {Erokhova}, \citenamefont {Elkin},\
  and\ \citenamefont {Izotova}}]{ishkhanov2002}%
  \BibitemOpen
  \bibfield  {author} {\bibinfo {author} {\bibnamefont {Ishkhanov},
  \bibfnamefont {B.~S.}}, \bibinfo {author} {\bibfnamefont {I.~M.}\
  \bibnamefont {Kapitonov}}, \bibinfo {author} {\bibfnamefont {E.~I.}\
  \bibnamefont {Lileeva}}, \bibinfo {author} {\bibfnamefont {E.~V.}\
  \bibnamefont {Shirokov}}, \bibinfo {author} {\bibfnamefont {V.~A.}\
  \bibnamefont {Erokhova}}, \bibinfo {author} {\bibfnamefont {M.~A.}\
  \bibnamefont {Elkin}}, \ and\ \bibinfo {author} {\bibfnamefont {A.~V.}\
  \bibnamefont {Izotova}}} (\bibinfo {year} {2002}),\ \href@noop {} {\emph
  {\bibinfo {title} {Cross sections of photon absorption by nuclei with nucleon
  numbers 12 - 65}}},\ \bibinfo {type} {Tech. Rep.}\ \bibinfo {number}
  {MSU-INP-2002-27/711}\ (\bibinfo  {institution} {Institute of Nuclear
  Physics},\ \bibinfo {address} {Moscow State University})\BibitemShut
  {NoStop}%
\bibitem [{\citenamefont {Jansen}(2013)}]{jansen2012}%
  \BibitemOpen
  \bibfield  {author} {\bibinfo {author} {\bibnamefont {Jansen}, \bibfnamefont
  {G.~R.}}} (\bibinfo {year} {2013}),\ \href {\doibase
  10.1103/PhysRevC.88.024305} {\bibfield  {journal} {\bibinfo  {journal} {Phys.
  Rev. C}\ }\textbf {\bibinfo {volume} {88}},\ \bibinfo {pages}
  {024305}}\BibitemShut {NoStop}%
\bibitem [{\citenamefont {Jansen}\ \emph {et~al.}(2011)\citenamefont {Jansen},
  \citenamefont {Hjorth-Jensen}, \citenamefont {Hagen},\ and\ \citenamefont
  {Papenbrock}}]{jansen2011}%
  \BibitemOpen
  \bibfield  {author} {\bibinfo {author} {\bibnamefont {Jansen}, \bibfnamefont
  {G.~R.}}, \bibinfo {author} {\bibfnamefont {M.}~\bibnamefont
  {Hjorth-Jensen}}, \bibinfo {author} {\bibfnamefont {G.}~\bibnamefont
  {Hagen}}, \ and\ \bibinfo {author} {\bibfnamefont {T.}~\bibnamefont
  {Papenbrock}}} (\bibinfo {year} {2011}),\ \href {\doibase
  10.1103/PhysRevC.83.054306} {\bibfield  {journal} {\bibinfo  {journal} {Phys.
  Rev. C}\ }\textbf {\bibinfo {volume} {83}},\ \bibinfo {pages}
  {054306}}\BibitemShut {NoStop}%
\bibitem [{\citenamefont {Janssens}\ \emph {et~al.}(2002)\citenamefont
  {Janssens}, \citenamefont {Fornal}, \citenamefont {Mantica}, \citenamefont
  {Brown}, \citenamefont {Broda}, \citenamefont {Bhattacharyya}, \citenamefont
  {Carpenter}, \citenamefont {Cinausero}, \citenamefont {Daly}, \citenamefont
  {Davies}, \citenamefont {Glasmacher}, \citenamefont {Grabowski},
  \citenamefont {Groh}, \citenamefont {Honma}, \citenamefont {Kondev},
  \citenamefont {Kr{\'o}las}, \citenamefont {Lauritsen}, \citenamefont
  {Liddick}, \citenamefont {Lunardi}, \citenamefont {Marginean}, \citenamefont
  {Mizusaki}, \citenamefont {Morrissey}, \citenamefont {Morton}, \citenamefont
  {Mueller}, \citenamefont {Otsuka}, \citenamefont {Pawlat}, \citenamefont
  {Seweryniak}, \citenamefont {Schatz}, \citenamefont {Stolz}, \citenamefont
  {Tabor}, \citenamefont {Ur}, \citenamefont {Viesti}, \citenamefont
  {Wiedenh{\"o}ver},\ and\ \citenamefont {Wrzesi{\'n}ski}}]{janssens2002}%
  \BibitemOpen
  \bibfield  {author} {\bibinfo {author} {\bibnamefont {Janssens},
  \bibfnamefont {R.}}, \bibinfo {author} {\bibfnamefont {B.}~\bibnamefont
  {Fornal}}, \bibinfo {author} {\bibfnamefont {P.}~\bibnamefont {Mantica}},
  \bibinfo {author} {\bibfnamefont {B.}~\bibnamefont {Brown}}, \bibinfo
  {author} {\bibfnamefont {R.}~\bibnamefont {Broda}}, \bibinfo {author}
  {\bibfnamefont {P.}~\bibnamefont {Bhattacharyya}}, \bibinfo {author}
  {\bibfnamefont {M.}~\bibnamefont {Carpenter}}, \bibinfo {author}
  {\bibfnamefont {M.}~\bibnamefont {Cinausero}}, \bibinfo {author}
  {\bibfnamefont {P.}~\bibnamefont {Daly}}, \bibinfo {author} {\bibfnamefont
  {A.}~\bibnamefont {Davies}}, \bibinfo {author} {\bibfnamefont
  {T.}~\bibnamefont {Glasmacher}}, \bibinfo {author} {\bibfnamefont
  {Z.}~\bibnamefont {Grabowski}}, \bibinfo {author} {\bibfnamefont
  {D.}~\bibnamefont {Groh}}, \bibinfo {author} {\bibfnamefont {M.}~\bibnamefont
  {Honma}}, \bibinfo {author} {\bibfnamefont {F.}~\bibnamefont {Kondev}},
  \bibinfo {author} {\bibfnamefont {W.}~\bibnamefont {Kr{\'o}las}}, \bibinfo
  {author} {\bibfnamefont {T.}~\bibnamefont {Lauritsen}}, \bibinfo {author}
  {\bibfnamefont {S.}~\bibnamefont {Liddick}}, \bibinfo {author} {\bibfnamefont
  {S.}~\bibnamefont {Lunardi}}, \bibinfo {author} {\bibfnamefont
  {N.}~\bibnamefont {Marginean}}, \bibinfo {author} {\bibfnamefont
  {T.}~\bibnamefont {Mizusaki}}, \bibinfo {author} {\bibfnamefont
  {D.}~\bibnamefont {Morrissey}}, \bibinfo {author} {\bibfnamefont
  {A.}~\bibnamefont {Morton}}, \bibinfo {author} {\bibfnamefont
  {W.}~\bibnamefont {Mueller}}, \bibinfo {author} {\bibfnamefont
  {T.}~\bibnamefont {Otsuka}}, \bibinfo {author} {\bibfnamefont
  {T.}~\bibnamefont {Pawlat}}, \bibinfo {author} {\bibfnamefont
  {D.}~\bibnamefont {Seweryniak}}, \bibinfo {author} {\bibfnamefont
  {H.}~\bibnamefont {Schatz}}, \bibinfo {author} {\bibfnamefont
  {A.}~\bibnamefont {Stolz}}, \bibinfo {author} {\bibfnamefont
  {S.}~\bibnamefont {Tabor}}, \bibinfo {author} {\bibfnamefont
  {C.}~\bibnamefont {Ur}}, \bibinfo {author} {\bibfnamefont {G.}~\bibnamefont
  {Viesti}}, \bibinfo {author} {\bibfnamefont {I.}~\bibnamefont
  {Wiedenh{\"o}ver}}, \ and\ \bibinfo {author} {\bibfnamefont {J.}~\bibnamefont
  {Wrzesi{\'n}ski}}} (\bibinfo {year} {2002}),\ \href {\doibase
  10.1016/S0370-2693(02)02682-5} {\bibfield  {journal} {\bibinfo  {journal}
  {Physics Letters B}\ }\textbf {\bibinfo {volume} {546}}~(\bibinfo {number}
  {1–2}),\ \bibinfo {pages} {55 }}\BibitemShut {NoStop}%
\bibitem [{\citenamefont {Jema\"\i}\ \emph {et~al.}(2013)\citenamefont
  {Jema\"\i}, \citenamefont {Delion},\ and\ \citenamefont
  {Schuck}}]{jemai2013}%
  \BibitemOpen
  \bibfield  {author} {\bibinfo {author} {\bibnamefont {Jema\"\i},
  \bibfnamefont {M.}}, \bibinfo {author} {\bibfnamefont {D.~S.}\ \bibnamefont
  {Delion}}, \ and\ \bibinfo {author} {\bibfnamefont {P.}~\bibnamefont
  {Schuck}}} (\bibinfo {year} {2013}),\ \href {\doibase
  10.1103/PhysRevC.88.044004} {\bibfield  {journal} {\bibinfo  {journal} {Phys.
  Rev. C}\ }\textbf {\bibinfo {volume} {88}},\ \bibinfo {pages}
  {044004}}\BibitemShut {NoStop}%
\bibitem [{\citenamefont {Jensen}\ \emph {et~al.}(2011)\citenamefont {Jensen},
  \citenamefont {Hagen}, \citenamefont {Hjorth-Jensen}, \citenamefont {Brown},\
  and\ \citenamefont {Gade}}]{jensen2011b}%
  \BibitemOpen
  \bibfield  {author} {\bibinfo {author} {\bibnamefont {Jensen}, \bibfnamefont
  {O.}}, \bibinfo {author} {\bibfnamefont {G.}~\bibnamefont {Hagen}}, \bibinfo
  {author} {\bibfnamefont {M.}~\bibnamefont {Hjorth-Jensen}}, \bibinfo {author}
  {\bibfnamefont {B.~A.}\ \bibnamefont {Brown}}, \ and\ \bibinfo {author}
  {\bibfnamefont {A.}~\bibnamefont {Gade}}} (\bibinfo {year} {2011}),\ \href
  {\doibase 10.1103/PhysRevLett.107.032501} {\bibfield  {journal} {\bibinfo
  {journal} {Phys. Rev. Lett.}\ }\textbf {\bibinfo {volume} {107}},\ \bibinfo
  {pages} {032501}}\BibitemShut {NoStop}%
\bibitem [{\citenamefont {Jensen}\ \emph {et~al.}(2010)\citenamefont {Jensen},
  \citenamefont {Hagen}, \citenamefont {Papenbrock}, \citenamefont {Dean},\
  and\ \citenamefont {Vaagen}}]{jensen2010}%
  \BibitemOpen
  \bibfield  {author} {\bibinfo {author} {\bibnamefont {Jensen}, \bibfnamefont
  {O.}}, \bibinfo {author} {\bibfnamefont {G.}~\bibnamefont {Hagen}}, \bibinfo
  {author} {\bibfnamefont {T.}~\bibnamefont {Papenbrock}}, \bibinfo {author}
  {\bibfnamefont {D.~J.}\ \bibnamefont {Dean}}, \ and\ \bibinfo {author}
  {\bibfnamefont {J.~S.}\ \bibnamefont {Vaagen}}} (\bibinfo {year} {2010}),\
  \href {\doibase 10.1103/PhysRevC.82.014310} {\bibfield  {journal} {\bibinfo
  {journal} {Phys. Rev. C}\ }\textbf {\bibinfo {volume} {82}},\ \bibinfo
  {pages} {014310}}\BibitemShut {NoStop}%
\bibitem [{\citenamefont {Jones}\ \emph {et~al.}(2010)\citenamefont {Jones},
  \citenamefont {Adekola}, \citenamefont {Bardayan}, \citenamefont {Blackmon},
  \citenamefont {Chae}, \citenamefont {Chipps}, \citenamefont {Cizewski},
  \citenamefont {Erikson}, \citenamefont {Harlin}, \citenamefont {Hatarik},
  \citenamefont {Kapler}, \citenamefont {Kozub}, \citenamefont {Liang},
  \citenamefont {Livesay}, \citenamefont {Ma}, \citenamefont {Moazen},
  \citenamefont {Nesaraja}, \citenamefont {Nunes}, \citenamefont {Pain},
  \citenamefont {Patterson}, \citenamefont {Shapira}, \citenamefont {Shriner},
  \citenamefont {Smith}, \citenamefont {Swan},\ and\ \citenamefont
  {Thomas}}]{jones2010}%
  \BibitemOpen
  \bibfield  {author} {\bibinfo {author} {\bibnamefont {Jones}, \bibfnamefont
  {K.~L.}}, \bibinfo {author} {\bibfnamefont {A.~S.}\ \bibnamefont {Adekola}},
  \bibinfo {author} {\bibfnamefont {D.~W.}\ \bibnamefont {Bardayan}}, \bibinfo
  {author} {\bibfnamefont {J.~C.}\ \bibnamefont {Blackmon}}, \bibinfo {author}
  {\bibfnamefont {K.~Y.}\ \bibnamefont {Chae}}, \bibinfo {author}
  {\bibfnamefont {K.~A.}\ \bibnamefont {Chipps}}, \bibinfo {author}
  {\bibfnamefont {J.~A.}\ \bibnamefont {Cizewski}}, \bibinfo {author}
  {\bibfnamefont {L.}~\bibnamefont {Erikson}}, \bibinfo {author} {\bibfnamefont
  {C.}~\bibnamefont {Harlin}}, \bibinfo {author} {\bibfnamefont
  {R.}~\bibnamefont {Hatarik}}, \bibinfo {author} {\bibfnamefont
  {R.}~\bibnamefont {Kapler}}, \bibinfo {author} {\bibfnamefont {R.~L.}\
  \bibnamefont {Kozub}}, \bibinfo {author} {\bibfnamefont {J.~F.}\ \bibnamefont
  {Liang}}, \bibinfo {author} {\bibfnamefont {R.}~\bibnamefont {Livesay}},
  \bibinfo {author} {\bibfnamefont {Z.}~\bibnamefont {Ma}}, \bibinfo {author}
  {\bibfnamefont {B.~H.}\ \bibnamefont {Moazen}}, \bibinfo {author}
  {\bibfnamefont {C.~D.}\ \bibnamefont {Nesaraja}}, \bibinfo {author}
  {\bibfnamefont {F.~M.}\ \bibnamefont {Nunes}}, \bibinfo {author}
  {\bibfnamefont {S.~D.}\ \bibnamefont {Pain}}, \bibinfo {author}
  {\bibfnamefont {N.~P.}\ \bibnamefont {Patterson}}, \bibinfo {author}
  {\bibfnamefont {D.}~\bibnamefont {Shapira}}, \bibinfo {author} {\bibfnamefont
  {J.~F.}\ \bibnamefont {Shriner}}, \bibinfo {author} {\bibfnamefont {M.~S.}\
  \bibnamefont {Smith}}, \bibinfo {author} {\bibfnamefont {T.~P.}\ \bibnamefont
  {Swan}}, \ and\ \bibinfo {author} {\bibfnamefont {J.~S.}\ \bibnamefont
  {Thomas}}} (\bibinfo {year} {2010}),\ \href {\doibase 10.1038/nature09048}
  {\bibfield  {journal} {\bibinfo  {journal} {Nature}\ }\textbf {\bibinfo
  {volume} {465}}~(\bibinfo {number} {7297}),\ \bibinfo {pages}
  {454}}\BibitemShut {NoStop}%
\bibitem [{\citenamefont {Jurgenson}\ \emph {et~al.}(2013)\citenamefont
  {Jurgenson}, \citenamefont {Maris}, \citenamefont {Furnstahl}, \citenamefont
  {Navr\'atil}, \citenamefont {Ormand},\ and\ \citenamefont
  {Vary}}]{jurgenson2013}%
  \BibitemOpen
  \bibfield  {author} {\bibinfo {author} {\bibnamefont {Jurgenson},
  \bibfnamefont {E.~D.}}, \bibinfo {author} {\bibfnamefont {P.}~\bibnamefont
  {Maris}}, \bibinfo {author} {\bibfnamefont {R.~J.}\ \bibnamefont
  {Furnstahl}}, \bibinfo {author} {\bibfnamefont {P.}~\bibnamefont
  {Navr\'atil}}, \bibinfo {author} {\bibfnamefont {W.~E.}\ \bibnamefont
  {Ormand}}, \ and\ \bibinfo {author} {\bibfnamefont {J.~P.}\ \bibnamefont
  {Vary}}} (\bibinfo {year} {2013}),\ \href {\doibase
  10.1103/PhysRevC.87.054312} {\bibfield  {journal} {\bibinfo  {journal} {Phys.
  Rev. C}\ }\textbf {\bibinfo {volume} {87}},\ \bibinfo {pages}
  {054312}}\BibitemShut {NoStop}%
\bibitem [{\citenamefont {Jurgenson}\ \emph {et~al.}(2009)\citenamefont
  {Jurgenson}, \citenamefont {Navr\'atil},\ and\ \citenamefont
  {Furnstahl}}]{jurgenson2009}%
  \BibitemOpen
  \bibfield  {author} {\bibinfo {author} {\bibnamefont {Jurgenson},
  \bibfnamefont {E.~D.}}, \bibinfo {author} {\bibfnamefont {P.}~\bibnamefont
  {Navr\'atil}}, \ and\ \bibinfo {author} {\bibfnamefont {R.~J.}\ \bibnamefont
  {Furnstahl}}} (\bibinfo {year} {2009}),\ \href {\doibase
  10.1103/PhysRevLett.103.082501} {\bibfield  {journal} {\bibinfo  {journal}
  {Phys. Rev. Lett.}\ }\textbf {\bibinfo {volume} {103}},\ \bibinfo {pages}
  {082501}}\BibitemShut {NoStop}%
\bibitem [{\citenamefont {Jurgenson}\ \emph {et~al.}(2011)\citenamefont
  {Jurgenson}, \citenamefont {Navr\'atil},\ and\ \citenamefont
  {Furnstahl}}]{jurgenson2011}%
  \BibitemOpen
  \bibfield  {author} {\bibinfo {author} {\bibnamefont {Jurgenson},
  \bibfnamefont {E.~D.}}, \bibinfo {author} {\bibfnamefont {P.}~\bibnamefont
  {Navr\'atil}}, \ and\ \bibinfo {author} {\bibfnamefont {R.~J.}\ \bibnamefont
  {Furnstahl}}} (\bibinfo {year} {2011}),\ \href {\doibase
  10.1103/PhysRevC.83.034301} {\bibfield  {journal} {\bibinfo  {journal} {Phys.
  Rev. C}\ }\textbf {\bibinfo {volume} {83}},\ \bibinfo {pages}
  {034301}}\BibitemShut {NoStop}%
\bibitem [{\citenamefont {Kaiser}\ \emph {et~al.}(1997)\citenamefont {Kaiser},
  \citenamefont {Brockmann},\ and\ \citenamefont {Weise}}]{kaiser1997}%
  \BibitemOpen
  \bibfield  {author} {\bibinfo {author} {\bibnamefont {Kaiser}, \bibfnamefont
  {N.}}, \bibinfo {author} {\bibfnamefont {R.}~\bibnamefont {Brockmann}}, \
  and\ \bibinfo {author} {\bibfnamefont {W.}~\bibnamefont {Weise}}} (\bibinfo
  {year} {1997}),\ \href {\doibase 10.1016/S0375-9474(97)00586-1} {\bibfield
  {journal} {\bibinfo  {journal} {Nuclear Physics A}\ }\textbf {\bibinfo
  {volume} {625}}~(\bibinfo {number} {4}),\ \bibinfo {pages} {758
  }}\BibitemShut {NoStop}%
\bibitem [{\citenamefont {Kalantar-Nayestanaki}\ \emph
  {et~al.}(2012)\citenamefont {Kalantar-Nayestanaki}, \citenamefont {Epelbaum},
  \citenamefont {Messchendorp},\ and\ \citenamefont {Nogga}}]{kalantar2012}%
  \BibitemOpen
  \bibfield  {author} {\bibinfo {author} {\bibnamefont {Kalantar-Nayestanaki},
  \bibfnamefont {N.}}, \bibinfo {author} {\bibfnamefont {E.}~\bibnamefont
  {Epelbaum}}, \bibinfo {author} {\bibfnamefont {J.~G.}\ \bibnamefont
  {Messchendorp}}, \ and\ \bibinfo {author} {\bibfnamefont {A.}~\bibnamefont
  {Nogga}}} (\bibinfo {year} {2012}),\ \href
  {http://stacks.iop.org/0034-4885/75/i=1/a=016301} {\bibfield  {journal}
  {\bibinfo  {journal} {Reports on Progress in Physics}\ }\textbf {\bibinfo
  {volume} {75}}~(\bibinfo {number} {1}),\ \bibinfo {pages}
  {016301}}\BibitemShut {NoStop}%
\bibitem [{\citenamefont {Kamada}\ \emph {et~al.}(2001)\citenamefont {Kamada},
  \citenamefont {Nogga}, \citenamefont {Gl\"ockle}, \citenamefont {Hiyama},
  \citenamefont {Kamimura}, \citenamefont {Varga}, \citenamefont {Suzuki},
  \citenamefont {Viviani}, \citenamefont {Kievsky}, \citenamefont {Rosati},
  \citenamefont {Carlson}, \citenamefont {Pieper}, \citenamefont {Wiringa},
  \citenamefont {Navr\'atil}, \citenamefont {Barrett}, \citenamefont {Barnea},
  \citenamefont {Leidemann},\ and\ \citenamefont {Orlandini}}]{kamada2001}%
  \BibitemOpen
  \bibfield  {author} {\bibinfo {author} {\bibnamefont {Kamada}, \bibfnamefont
  {H.}}, \bibinfo {author} {\bibfnamefont {A.}~\bibnamefont {Nogga}}, \bibinfo
  {author} {\bibfnamefont {W.}~\bibnamefont {Gl\"ockle}}, \bibinfo {author}
  {\bibfnamefont {E.}~\bibnamefont {Hiyama}}, \bibinfo {author} {\bibfnamefont
  {M.}~\bibnamefont {Kamimura}}, \bibinfo {author} {\bibfnamefont
  {K.}~\bibnamefont {Varga}}, \bibinfo {author} {\bibfnamefont
  {Y.}~\bibnamefont {Suzuki}}, \bibinfo {author} {\bibfnamefont
  {M.}~\bibnamefont {Viviani}}, \bibinfo {author} {\bibfnamefont
  {A.}~\bibnamefont {Kievsky}}, \bibinfo {author} {\bibfnamefont
  {S.}~\bibnamefont {Rosati}}, \bibinfo {author} {\bibfnamefont
  {J.}~\bibnamefont {Carlson}}, \bibinfo {author} {\bibfnamefont {S.~C.}\
  \bibnamefont {Pieper}}, \bibinfo {author} {\bibfnamefont {R.~B.}\
  \bibnamefont {Wiringa}}, \bibinfo {author} {\bibfnamefont {P.}~\bibnamefont
  {Navr\'atil}}, \bibinfo {author} {\bibfnamefont {B.~R.}\ \bibnamefont
  {Barrett}}, \bibinfo {author} {\bibfnamefont {N.}~\bibnamefont {Barnea}},
  \bibinfo {author} {\bibfnamefont {W.}~\bibnamefont {Leidemann}}, \ and\
  \bibinfo {author} {\bibfnamefont {G.}~\bibnamefont {Orlandini}}} (\bibinfo
  {year} {2001}),\ \href {\doibase 10.1103/PhysRevC.64.044001} {\bibfield
  {journal} {\bibinfo  {journal} {Phys. Rev. C}\ }\textbf {\bibinfo {volume}
  {64}},\ \bibinfo {pages} {044001}}\BibitemShut {NoStop}%
\bibitem [{\citenamefont {Kanungo}\ \emph {et~al.}(2009)\citenamefont
  {Kanungo}, \citenamefont {Nociforo}, \citenamefont {Prochazka}, \citenamefont
  {Aumann}, \citenamefont {Boutin}, \citenamefont {Cortina-Gil}, \citenamefont
  {Davids}, \citenamefont {Diakaki}, \citenamefont {Farinon}, \citenamefont
  {Geissel}, \citenamefont {Gernh\"auser}, \citenamefont {Gerl}, \citenamefont
  {Janik}, \citenamefont {Jonson}, \citenamefont {Kindler}, \citenamefont
  {Kn\"obel}, \citenamefont {Kr\"ucken}, \citenamefont {Lantz}, \citenamefont
  {Lenske}, \citenamefont {Litvinov}, \citenamefont {Lommel}, \citenamefont
  {Mahata}, \citenamefont {Maierbeck}, \citenamefont {Musumarra}, \citenamefont
  {Nilsson}, \citenamefont {Otsuka}, \citenamefont {Perro}, \citenamefont
  {Scheidenberger}, \citenamefont {Sitar}, \citenamefont {Strmen},
  \citenamefont {Sun}, \citenamefont {Szarka}, \citenamefont {Tanihata},
  \citenamefont {Utsuno}, \citenamefont {Weick},\ and\ \citenamefont
  {Winkler}}]{kanungo2009}%
  \BibitemOpen
  \bibfield  {author} {\bibinfo {author} {\bibnamefont {Kanungo}, \bibfnamefont
  {R.}}, \bibinfo {author} {\bibfnamefont {C.}~\bibnamefont {Nociforo}},
  \bibinfo {author} {\bibfnamefont {A.}~\bibnamefont {Prochazka}}, \bibinfo
  {author} {\bibfnamefont {T.}~\bibnamefont {Aumann}}, \bibinfo {author}
  {\bibfnamefont {D.}~\bibnamefont {Boutin}}, \bibinfo {author} {\bibfnamefont
  {D.}~\bibnamefont {Cortina-Gil}}, \bibinfo {author} {\bibfnamefont
  {B.}~\bibnamefont {Davids}}, \bibinfo {author} {\bibfnamefont
  {M.}~\bibnamefont {Diakaki}}, \bibinfo {author} {\bibfnamefont
  {F.}~\bibnamefont {Farinon}}, \bibinfo {author} {\bibfnamefont
  {H.}~\bibnamefont {Geissel}}, \bibinfo {author} {\bibfnamefont
  {R.}~\bibnamefont {Gernh\"auser}}, \bibinfo {author} {\bibfnamefont
  {J.}~\bibnamefont {Gerl}}, \bibinfo {author} {\bibfnamefont {R.}~\bibnamefont
  {Janik}}, \bibinfo {author} {\bibfnamefont {B.}~\bibnamefont {Jonson}},
  \bibinfo {author} {\bibfnamefont {B.}~\bibnamefont {Kindler}}, \bibinfo
  {author} {\bibfnamefont {R.}~\bibnamefont {Kn\"obel}}, \bibinfo {author}
  {\bibfnamefont {R.}~\bibnamefont {Kr\"ucken}}, \bibinfo {author}
  {\bibfnamefont {M.}~\bibnamefont {Lantz}}, \bibinfo {author} {\bibfnamefont
  {H.}~\bibnamefont {Lenske}}, \bibinfo {author} {\bibfnamefont
  {Y.}~\bibnamefont {Litvinov}}, \bibinfo {author} {\bibfnamefont
  {B.}~\bibnamefont {Lommel}}, \bibinfo {author} {\bibfnamefont
  {K.}~\bibnamefont {Mahata}}, \bibinfo {author} {\bibfnamefont
  {P.}~\bibnamefont {Maierbeck}}, \bibinfo {author} {\bibfnamefont
  {A.}~\bibnamefont {Musumarra}}, \bibinfo {author} {\bibfnamefont
  {T.}~\bibnamefont {Nilsson}}, \bibinfo {author} {\bibfnamefont
  {T.}~\bibnamefont {Otsuka}}, \bibinfo {author} {\bibfnamefont
  {C.}~\bibnamefont {Perro}}, \bibinfo {author} {\bibfnamefont
  {C.}~\bibnamefont {Scheidenberger}}, \bibinfo {author} {\bibfnamefont
  {B.}~\bibnamefont {Sitar}}, \bibinfo {author} {\bibfnamefont
  {P.}~\bibnamefont {Strmen}}, \bibinfo {author} {\bibfnamefont
  {B.}~\bibnamefont {Sun}}, \bibinfo {author} {\bibfnamefont {I.}~\bibnamefont
  {Szarka}}, \bibinfo {author} {\bibfnamefont {I.}~\bibnamefont {Tanihata}},
  \bibinfo {author} {\bibfnamefont {Y.}~\bibnamefont {Utsuno}}, \bibinfo
  {author} {\bibfnamefont {H.}~\bibnamefont {Weick}}, \ and\ \bibinfo {author}
  {\bibfnamefont {M.}~\bibnamefont {Winkler}}} (\bibinfo {year} {2009}),\ \href
  {\doibase 10.1103/PhysRevLett.102.152501} {\bibfield  {journal} {\bibinfo
  {journal} {Phys. Rev. Lett.}\ }\textbf {\bibinfo {volume} {102}},\ \bibinfo
  {pages} {152501}}\BibitemShut {NoStop}%
\bibitem [{\citenamefont {Kanungo}\ \emph {et~al.}(2011)\citenamefont
  {Kanungo}, \citenamefont {Prochazka}, \citenamefont {Uchida}, \citenamefont
  {Horiuchi}, \citenamefont {Hagen}, \citenamefont {Papenbrock}, \citenamefont
  {Nociforo}, \citenamefont {Aumann}, \citenamefont {Boutin}, \citenamefont
  {Cortina-Gil}, \citenamefont {Davids}, \citenamefont {Diakaki}, \citenamefont
  {Farinon}, \citenamefont {Geissel}, \citenamefont {Gernh\"auser},
  \citenamefont {Gerl}, \citenamefont {Janik}, \citenamefont {Jensen},
  \citenamefont {Jonson}, \citenamefont {Kindler}, \citenamefont {Kn\"obel},
  \citenamefont {Kr\"ucken}, \citenamefont {Lantz}, \citenamefont {Lenske},
  \citenamefont {Litvinov}, \citenamefont {Lommel}, \citenamefont {Mahata},
  \citenamefont {Maierbeck}, \citenamefont {Musumarra}, \citenamefont
  {Nilsson}, \citenamefont {Perro}, \citenamefont {Scheidenberger},
  \citenamefont {Sitar}, \citenamefont {Strmen}, \citenamefont {Sun},
  \citenamefont {Suzuki}, \citenamefont {Szarka}, \citenamefont {Tanihata},
  \citenamefont {Weick},\ and\ \citenamefont {Winkler}}]{kanungo2011}%
  \BibitemOpen
  \bibfield  {author} {\bibinfo {author} {\bibnamefont {Kanungo}, \bibfnamefont
  {R.}}, \bibinfo {author} {\bibfnamefont {A.}~\bibnamefont {Prochazka}},
  \bibinfo {author} {\bibfnamefont {M.}~\bibnamefont {Uchida}}, \bibinfo
  {author} {\bibfnamefont {W.}~\bibnamefont {Horiuchi}}, \bibinfo {author}
  {\bibfnamefont {G.}~\bibnamefont {Hagen}}, \bibinfo {author} {\bibfnamefont
  {T.}~\bibnamefont {Papenbrock}}, \bibinfo {author} {\bibfnamefont
  {C.}~\bibnamefont {Nociforo}}, \bibinfo {author} {\bibfnamefont
  {T.}~\bibnamefont {Aumann}}, \bibinfo {author} {\bibfnamefont
  {D.}~\bibnamefont {Boutin}}, \bibinfo {author} {\bibfnamefont
  {D.}~\bibnamefont {Cortina-Gil}}, \bibinfo {author} {\bibfnamefont
  {B.}~\bibnamefont {Davids}}, \bibinfo {author} {\bibfnamefont
  {M.}~\bibnamefont {Diakaki}}, \bibinfo {author} {\bibfnamefont
  {F.}~\bibnamefont {Farinon}}, \bibinfo {author} {\bibfnamefont
  {H.}~\bibnamefont {Geissel}}, \bibinfo {author} {\bibfnamefont
  {R.}~\bibnamefont {Gernh\"auser}}, \bibinfo {author} {\bibfnamefont
  {J.}~\bibnamefont {Gerl}}, \bibinfo {author} {\bibfnamefont {R.}~\bibnamefont
  {Janik}}, \bibinfo {author} {\bibfnamefont {O.}~\bibnamefont {Jensen}},
  \bibinfo {author} {\bibfnamefont {B.}~\bibnamefont {Jonson}}, \bibinfo
  {author} {\bibfnamefont {B.}~\bibnamefont {Kindler}}, \bibinfo {author}
  {\bibfnamefont {R.}~\bibnamefont {Kn\"obel}}, \bibinfo {author}
  {\bibfnamefont {R.}~\bibnamefont {Kr\"ucken}}, \bibinfo {author}
  {\bibfnamefont {M.}~\bibnamefont {Lantz}}, \bibinfo {author} {\bibfnamefont
  {H.}~\bibnamefont {Lenske}}, \bibinfo {author} {\bibfnamefont
  {Y.}~\bibnamefont {Litvinov}}, \bibinfo {author} {\bibfnamefont
  {B.}~\bibnamefont {Lommel}}, \bibinfo {author} {\bibfnamefont
  {K.}~\bibnamefont {Mahata}}, \bibinfo {author} {\bibfnamefont
  {P.}~\bibnamefont {Maierbeck}}, \bibinfo {author} {\bibfnamefont
  {A.}~\bibnamefont {Musumarra}}, \bibinfo {author} {\bibfnamefont
  {T.}~\bibnamefont {Nilsson}}, \bibinfo {author} {\bibfnamefont
  {C.}~\bibnamefont {Perro}}, \bibinfo {author} {\bibfnamefont
  {C.}~\bibnamefont {Scheidenberger}}, \bibinfo {author} {\bibfnamefont
  {B.}~\bibnamefont {Sitar}}, \bibinfo {author} {\bibfnamefont
  {P.}~\bibnamefont {Strmen}}, \bibinfo {author} {\bibfnamefont
  {B.}~\bibnamefont {Sun}}, \bibinfo {author} {\bibfnamefont {Y.}~\bibnamefont
  {Suzuki}}, \bibinfo {author} {\bibfnamefont {I.}~\bibnamefont {Szarka}},
  \bibinfo {author} {\bibfnamefont {I.}~\bibnamefont {Tanihata}}, \bibinfo
  {author} {\bibfnamefont {H.}~\bibnamefont {Weick}}, \ and\ \bibinfo {author}
  {\bibfnamefont {M.}~\bibnamefont {Winkler}}} (\bibinfo {year} {2011}),\ \href
  {\doibase 10.1103/PhysRevC.84.061304} {\bibfield  {journal} {\bibinfo
  {journal} {Phys. Rev. C}\ }\textbf {\bibinfo {volume} {84}},\ \bibinfo
  {pages} {061304}}\BibitemShut {NoStop}%
\bibitem [{\citenamefont {Kohno}\ and\ \citenamefont
  {Okamoto}(2012)}]{kohno2012}%
  \BibitemOpen
  \bibfield  {author} {\bibinfo {author} {\bibnamefont {Kohno}, \bibfnamefont
  {M.}}, \ and\ \bibinfo {author} {\bibfnamefont {R.}~\bibnamefont {Okamoto}}}
  (\bibinfo {year} {2012}),\ \href {\doibase 10.1103/PhysRevC.86.014317}
  {\bibfield  {journal} {\bibinfo  {journal} {Phys. Rev. C}\ }\textbf {\bibinfo
  {volume} {86}},\ \bibinfo {pages} {014317}}\BibitemShut {NoStop}%
\bibitem [{\citenamefont {van Kolck}(1994)}]{vankolck1994}%
  \BibitemOpen
  \bibfield  {author} {\bibinfo {author} {\bibnamefont {van Kolck},
  \bibfnamefont {U.}}} (\bibinfo {year} {1994}),\ \href {\doibase
  10.1103/PhysRevC.49.2932} {\bibfield  {journal} {\bibinfo  {journal} {Phys.
  Rev. C}\ }\textbf {\bibinfo {volume} {49}},\ \bibinfo {pages}
  {2932}}\BibitemShut {NoStop}%
\bibitem [{\citenamefont {Kolck}(1999)}]{vankolck1999}%
  \BibitemOpen
  \bibfield  {author} {\bibinfo {author} {\bibnamefont {Kolck}, \bibfnamefont
  {U.~V.}}} (\bibinfo {year} {1999}),\ \href {\doibase
  10.1016/S0146-6410(99)00097-6} {\bibfield  {journal} {\bibinfo  {journal}
  {Progress in Particle and Nuclear Physics}\ }\textbf {\bibinfo {volume}
  {43}}~(\bibinfo {number} {0}),\ \bibinfo {pages} {337 }}\BibitemShut
  {NoStop}%
\bibitem [{\citenamefont {Kortelainen}\ \emph {et~al.}(2010)\citenamefont
  {Kortelainen}, \citenamefont {Lesinski}, \citenamefont {Mor\'e},
  \citenamefont {Nazarewicz}, \citenamefont {Sarich}, \citenamefont {Schunck},
  \citenamefont {Stoitsov},\ and\ \citenamefont {Wild}}]{kortelainen2010}%
  \BibitemOpen
  \bibfield  {author} {\bibinfo {author} {\bibnamefont {Kortelainen},
  \bibfnamefont {M.}}, \bibinfo {author} {\bibfnamefont {T.}~\bibnamefont
  {Lesinski}}, \bibinfo {author} {\bibfnamefont {J.}~\bibnamefont {Mor\'e}},
  \bibinfo {author} {\bibfnamefont {W.}~\bibnamefont {Nazarewicz}}, \bibinfo
  {author} {\bibfnamefont {J.}~\bibnamefont {Sarich}}, \bibinfo {author}
  {\bibfnamefont {N.}~\bibnamefont {Schunck}}, \bibinfo {author} {\bibfnamefont
  {M.~V.}\ \bibnamefont {Stoitsov}}, \ and\ \bibinfo {author} {\bibfnamefont
  {S.}~\bibnamefont {Wild}}} (\bibinfo {year} {2010}),\ \href {\doibase
  10.1103/PhysRevC.82.024313} {\bibfield  {journal} {\bibinfo  {journal} {Phys.
  Rev. C}\ }\textbf {\bibinfo {volume} {82}},\ \bibinfo {pages}
  {024313}}\BibitemShut {NoStop}%
\bibitem [{\citenamefont {Kowalski}\ \emph {et~al.}(2004)\citenamefont
  {Kowalski}, \citenamefont {Dean}, \citenamefont {Hjorth-Jensen},
  \citenamefont {Papenbrock},\ and\ \citenamefont {Piecuch}}]{kowalski2004}%
  \BibitemOpen
  \bibfield  {author} {\bibinfo {author} {\bibnamefont {Kowalski},
  \bibfnamefont {K.}}, \bibinfo {author} {\bibfnamefont {D.~J.}\ \bibnamefont
  {Dean}}, \bibinfo {author} {\bibfnamefont {M.}~\bibnamefont {Hjorth-Jensen}},
  \bibinfo {author} {\bibfnamefont {T.}~\bibnamefont {Papenbrock}}, \ and\
  \bibinfo {author} {\bibfnamefont {P.}~\bibnamefont {Piecuch}}} (\bibinfo
  {year} {2004}),\ \href {\doibase 10.1103/PhysRevLett.92.132501} {\bibfield
  {journal} {\bibinfo  {journal} {Phys. Rev. Lett.}\ }\textbf {\bibinfo
  {volume} {92}},\ \bibinfo {pages} {132501}}\BibitemShut {NoStop}%
\bibitem [{\citenamefont {Kowalski}\ and\ \citenamefont
  {Piecuch}(2000)}]{kowalski2000}%
  \BibitemOpen
  \bibfield  {author} {\bibinfo {author} {\bibnamefont {Kowalski},
  \bibfnamefont {K.}}, \ and\ \bibinfo {author} {\bibfnamefont
  {P.}~\bibnamefont {Piecuch}}} (\bibinfo {year} {2000}),\ \href {\doibase
  10.1063/1.481769} {\bibfield  {journal} {\bibinfo  {journal} {The Journal of
  Chemical Physics}\ }\textbf {\bibinfo {volume} {113}}~(\bibinfo {number}
  {1}),\ \bibinfo {pages} {18}}\BibitemShut {NoStop}%
\bibitem [{\citenamefont {Kr\"uger}\ \emph {et~al.}(2013)\citenamefont
  {Kr\"uger}, \citenamefont {Tews}, \citenamefont {Hebeler},\ and\
  \citenamefont {Schwenk}}]{krueger2013}%
  \BibitemOpen
  \bibfield  {author} {\bibinfo {author} {\bibnamefont {Kr\"uger},
  \bibfnamefont {T.}}, \bibinfo {author} {\bibfnamefont {I.}~\bibnamefont
  {Tews}}, \bibinfo {author} {\bibfnamefont {K.}~\bibnamefont {Hebeler}}, \
  and\ \bibinfo {author} {\bibfnamefont {A.}~\bibnamefont {Schwenk}}} (\bibinfo
  {year} {2013}),\ \href {\doibase 10.1103/PhysRevC.88.025802} {\bibfield
  {journal} {\bibinfo  {journal} {Phys. Rev. C}\ }\textbf {\bibinfo {volume}
  {88}},\ \bibinfo {pages} {025802}}\BibitemShut {NoStop}%
\bibitem [{\citenamefont {Kucharski}\ and\ \citenamefont
  {Bartlett}(1991)}]{kucharski1991}%
  \BibitemOpen
  \bibfield  {author} {\bibinfo {author} {\bibnamefont {Kucharski},
  \bibfnamefont {S.}}, \ and\ \bibinfo {author} {\bibfnamefont
  {R.}~\bibnamefont {Bartlett}}} (\bibinfo {year} {1991}),\ \href {\doibase
  10.1007/BF01117419} {\bibfield  {journal} {\bibinfo  {journal} {Theoretica
  chimica acta}\ }\textbf {\bibinfo {volume} {80}}~(\bibinfo {number} {4-5}),\
  \bibinfo {pages} {387}}\BibitemShut {NoStop}%
\bibitem [{\citenamefont {Kucharski}\ and\ \citenamefont
  {Bartlett}(1986)}]{kucharski1986}%
  \BibitemOpen
  \bibfield  {author} {\bibinfo {author} {\bibnamefont {Kucharski},
  \bibfnamefont {S.~A.}}, \ and\ \bibinfo {author} {\bibfnamefont {R.~J.}\
  \bibnamefont {Bartlett}}} (\bibinfo {year} {1986}),\ \href {\doibase
  10.1016/S0065-3276(08)60051-9} {\bibfield  {journal} {\bibinfo  {journal}
  {Advances in Quantum Chemistry}\ }\textbf {\bibinfo {volume} {18}},\ \bibinfo
  {pages} {281 }}\BibitemShut {NoStop}%
\bibitem [{\citenamefont {Kucharski}\ and\ \citenamefont
  {Bartlett}(1998)}]{kucharski1998}%
  \BibitemOpen
  \bibfield  {author} {\bibinfo {author} {\bibnamefont {Kucharski},
  \bibfnamefont {S.~A.}}, \ and\ \bibinfo {author} {\bibfnamefont {R.~J.}\
  \bibnamefont {Bartlett}}} (\bibinfo {year} {1998}),\ \href {\doibase
  10.1063/1.475961} {\bibfield  {journal} {\bibinfo  {journal} {The Journal of
  Chemical Physics}\ }\textbf {\bibinfo {volume} {108}}~(\bibinfo {number}
  {13}),\ \bibinfo {pages} {5243}}\BibitemShut {NoStop}%
\bibitem [{\citenamefont {K{\"u}mmel}\ \emph {et~al.}(1978)\citenamefont
  {K{\"u}mmel}, \citenamefont {L{\"u}hrmann},\ and\ \citenamefont
  {Zabolitzky}}]{kuemmel1978}%
  \BibitemOpen
  \bibfield  {author} {\bibinfo {author} {\bibnamefont {K{\"u}mmel},
  \bibfnamefont {H.}}, \bibinfo {author} {\bibfnamefont {K.~H.}\ \bibnamefont
  {L{\"u}hrmann}}, \ and\ \bibinfo {author} {\bibfnamefont {J.~G.}\
  \bibnamefont {Zabolitzky}}} (\bibinfo {year} {1978}),\ \href {\doibase
  10.1016/0370-1573(78)90081-9} {\bibfield  {journal} {\bibinfo  {journal}
  {Physics Reports}\ }\textbf {\bibinfo {volume} {36}}~(\bibinfo {number}
  {1}),\ \bibinfo {pages} {1 }}\BibitemShut {NoStop}%
\bibitem [{\citenamefont {Kuo}\ and\ \citenamefont {Brown}(1968)}]{kuo1968}%
  \BibitemOpen
  \bibfield  {author} {\bibinfo {author} {\bibnamefont {Kuo}, \bibfnamefont
  {T.}}, \ and\ \bibinfo {author} {\bibfnamefont {G.}~\bibnamefont {Brown}}}
  (\bibinfo {year} {1968}),\ \href {\doibase 10.1016/0375-9474(68)90353-9}
  {\bibfield  {journal} {\bibinfo  {journal} {Nuclear Physics A}\ }\textbf
  {\bibinfo {volume} {114}}~(\bibinfo {number} {2}),\ \bibinfo {pages} {241
  }}\BibitemShut {NoStop}%
\bibitem [{\citenamefont {Kutzelnigg}(1991)}]{kutzelnigg1991}%
  \BibitemOpen
  \bibfield  {author} {\bibinfo {author} {\bibnamefont {Kutzelnigg},
  \bibfnamefont {W.}}} (\bibinfo {year} {1991}),\ \href
  {http://dx.doi.org/10.1007/BF01117418} {\bibfield  {journal} {\bibinfo
  {journal} {Theoretical Chemistry Accounts: Theory, Computation, and Modeling
  (Theoretica Chimica Acta)}\ }\textbf {\bibinfo {volume} {80}},\ \bibinfo
  {pages} {349}},\ \bibinfo {note} {10.1007/BF01117418}\BibitemShut {NoStop}%
\bibitem [{\citenamefont {Kvaal}(2009)}]{kvaal2009}%
  \BibitemOpen
  \bibfield  {author} {\bibinfo {author} {\bibnamefont {Kvaal}, \bibfnamefont
  {S.}}} (\bibinfo {year} {2009}),\ \href {\doibase 10.1103/PhysRevB.80.045321}
  {\bibfield  {journal} {\bibinfo  {journal} {Phys. Rev. B}\ }\textbf {\bibinfo
  {volume} {80}},\ \bibinfo {pages} {045321}}\BibitemShut {NoStop}%
\bibitem [{\citenamefont {Kvaal}(2012)}]{kvaal2012}%
  \BibitemOpen
  \bibfield  {author} {\bibinfo {author} {\bibnamefont {Kvaal}, \bibfnamefont
  {S.}}} (\bibinfo {year} {2012}),\ \href {\doibase 10.1063/1.4718427}
  {\bibfield  {journal} {\bibinfo  {journal} {The Journal of Chemical Physics}\
  }\textbf {\bibinfo {volume} {136}}~(\bibinfo {number} {19}),\ \bibinfo {eid}
  {194109}}\BibitemShut {NoStop}%
\bibitem [{\citenamefont {Lee}(2009)}]{lee2009}%
  \BibitemOpen
  \bibfield  {author} {\bibinfo {author} {\bibnamefont {Lee}, \bibfnamefont
  {D.}}} (\bibinfo {year} {2009}),\ \href {\doibase 10.1016/j.ppnp.2008.12.001}
  {\bibfield  {journal} {\bibinfo  {journal} {Progress in Particle and Nuclear
  Physics}\ }\textbf {\bibinfo {volume} {63}}~(\bibinfo {number} {1}),\
  \bibinfo {pages} {117 }}\BibitemShut {NoStop}%
\bibitem [{\citenamefont {Leidemann}\ and\ \citenamefont
  {Orlandini}(2013)}]{leidemann2013}%
  \BibitemOpen
  \bibfield  {author} {\bibinfo {author} {\bibnamefont {Leidemann},
  \bibfnamefont {W.}}, \ and\ \bibinfo {author} {\bibfnamefont
  {G.}~\bibnamefont {Orlandini}}} (\bibinfo {year} {2013}),\ \href {\doibase
  10.1016/j.ppnp.2012.09.001} {\bibfield  {journal} {\bibinfo  {journal}
  {Progress in Particle and Nuclear Physics}\ }\textbf {\bibinfo {volume}
  {68}}~(\bibinfo {number} {0}),\ \bibinfo {pages} {158 }}\BibitemShut
  {NoStop}%
\bibitem [{\citenamefont {Lepailleur}\ \emph {et~al.}(2013)\citenamefont
  {Lepailleur}, \citenamefont {Sorlin}, \citenamefont {Caceres}, \citenamefont
  {Bastin}, \citenamefont {Borcea}, \citenamefont {Borcea}, \citenamefont
  {Brown}, \citenamefont {Gaudefroy}, \citenamefont {Gr\'evy}, \citenamefont
  {Grinyer}, \citenamefont {Hagen}, \citenamefont {Hjorth-Jensen},
  \citenamefont {Jansen}, \citenamefont {Llidoo}, \citenamefont {Negoita},
  \citenamefont {de~Oliveira}, \citenamefont {Porquet}, \citenamefont {Rotaru},
  \citenamefont {Saint-Laurent}, \citenamefont {Sohler}, \citenamefont
  {Stanoiu},\ and\ \citenamefont {Thomas}}]{lepailleur2013}%
  \BibitemOpen
  \bibfield  {author} {\bibinfo {author} {\bibnamefont {Lepailleur},
  \bibfnamefont {A.}}, \bibinfo {author} {\bibfnamefont {O.}~\bibnamefont
  {Sorlin}}, \bibinfo {author} {\bibfnamefont {L.}~\bibnamefont {Caceres}},
  \bibinfo {author} {\bibfnamefont {B.}~\bibnamefont {Bastin}}, \bibinfo
  {author} {\bibfnamefont {C.}~\bibnamefont {Borcea}}, \bibinfo {author}
  {\bibfnamefont {R.}~\bibnamefont {Borcea}}, \bibinfo {author} {\bibfnamefont
  {B.~A.}\ \bibnamefont {Brown}}, \bibinfo {author} {\bibfnamefont
  {L.}~\bibnamefont {Gaudefroy}}, \bibinfo {author} {\bibfnamefont
  {S.}~\bibnamefont {Gr\'evy}}, \bibinfo {author} {\bibfnamefont {G.~F.}\
  \bibnamefont {Grinyer}}, \bibinfo {author} {\bibfnamefont {G.}~\bibnamefont
  {Hagen}}, \bibinfo {author} {\bibfnamefont {M.}~\bibnamefont
  {Hjorth-Jensen}}, \bibinfo {author} {\bibfnamefont {G.~R.}\ \bibnamefont
  {Jansen}}, \bibinfo {author} {\bibfnamefont {O.}~\bibnamefont {Llidoo}},
  \bibinfo {author} {\bibfnamefont {F.}~\bibnamefont {Negoita}}, \bibinfo
  {author} {\bibfnamefont {F.}~\bibnamefont {de~Oliveira}}, \bibinfo {author}
  {\bibfnamefont {M.-G.}\ \bibnamefont {Porquet}}, \bibinfo {author}
  {\bibfnamefont {F.}~\bibnamefont {Rotaru}}, \bibinfo {author} {\bibfnamefont
  {M.-G.}\ \bibnamefont {Saint-Laurent}}, \bibinfo {author} {\bibfnamefont
  {D.}~\bibnamefont {Sohler}}, \bibinfo {author} {\bibfnamefont
  {M.}~\bibnamefont {Stanoiu}}, \ and\ \bibinfo {author} {\bibfnamefont
  {J.~C.}\ \bibnamefont {Thomas}}} (\bibinfo {year} {2013}),\ \href {\doibase
  10.1103/PhysRevLett.110.082502} {\bibfield  {journal} {\bibinfo  {journal}
  {Phys. Rev. Lett.}\ }\textbf {\bibinfo {volume} {110}},\ \bibinfo {pages}
  {082502}}\BibitemShut {NoStop}%
\bibitem [{\citenamefont {Liddick}\ \emph {et~al.}(2004)\citenamefont
  {Liddick}, \citenamefont {Mantica}, \citenamefont {Janssens}, \citenamefont
  {Broda}, \citenamefont {Brown}, \citenamefont {Carpenter}, \citenamefont
  {Fornal}, \citenamefont {Honma}, \citenamefont {Mizusaki}, \citenamefont
  {Morton}, \citenamefont {Mueller}, \citenamefont {Otsuka}, \citenamefont
  {Pavan}, \citenamefont {Stolz}, \citenamefont {Tabor}, \citenamefont
  {Tomlin},\ and\ \citenamefont {Wiedeking}}]{liddick2004}%
  \BibitemOpen
  \bibfield  {author} {\bibinfo {author} {\bibnamefont {Liddick}, \bibfnamefont
  {S.~N.}}, \bibinfo {author} {\bibfnamefont {P.~F.}\ \bibnamefont {Mantica}},
  \bibinfo {author} {\bibfnamefont {R.~V.~F.}\ \bibnamefont {Janssens}},
  \bibinfo {author} {\bibfnamefont {R.}~\bibnamefont {Broda}}, \bibinfo
  {author} {\bibfnamefont {B.~A.}\ \bibnamefont {Brown}}, \bibinfo {author}
  {\bibfnamefont {M.~P.}\ \bibnamefont {Carpenter}}, \bibinfo {author}
  {\bibfnamefont {B.}~\bibnamefont {Fornal}}, \bibinfo {author} {\bibfnamefont
  {M.}~\bibnamefont {Honma}}, \bibinfo {author} {\bibfnamefont
  {T.}~\bibnamefont {Mizusaki}}, \bibinfo {author} {\bibfnamefont {A.~C.}\
  \bibnamefont {Morton}}, \bibinfo {author} {\bibfnamefont {W.~F.}\
  \bibnamefont {Mueller}}, \bibinfo {author} {\bibfnamefont {T.}~\bibnamefont
  {Otsuka}}, \bibinfo {author} {\bibfnamefont {J.}~\bibnamefont {Pavan}},
  \bibinfo {author} {\bibfnamefont {A.}~\bibnamefont {Stolz}}, \bibinfo
  {author} {\bibfnamefont {S.~L.}\ \bibnamefont {Tabor}}, \bibinfo {author}
  {\bibfnamefont {B.~E.}\ \bibnamefont {Tomlin}}, \ and\ \bibinfo {author}
  {\bibfnamefont {M.}~\bibnamefont {Wiedeking}}} (\bibinfo {year} {2004}),\
  \href {\doibase 10.1103/PhysRevLett.92.072502} {\bibfield  {journal}
  {\bibinfo  {journal} {Phys. Rev. Lett.}\ }\textbf {\bibinfo {volume} {92}},\
  \bibinfo {pages} {072502}}\BibitemShut {NoStop}%
\bibitem [{\citenamefont {Lin}\ \emph {et~al.}(2001)\citenamefont {Lin},
  \citenamefont {Zong},\ and\ \citenamefont {Ceperley}}]{lin2001}%
  \BibitemOpen
  \bibfield  {author} {\bibinfo {author} {\bibnamefont {Lin}, \bibfnamefont
  {C.}}, \bibinfo {author} {\bibfnamefont {F.~H.}\ \bibnamefont {Zong}}, \ and\
  \bibinfo {author} {\bibfnamefont {D.~M.}\ \bibnamefont {Ceperley}}} (\bibinfo
  {year} {2001}),\ \href {\doibase 10.1103/PhysRevE.64.016702} {\bibfield
  {journal} {\bibinfo  {journal} {Phys. Rev. E}\ }\textbf {\bibinfo {volume}
  {64}},\ \bibinfo {pages} {016702}}\BibitemShut {NoStop}%
\bibitem [{\citenamefont {Lind}(1993)}]{lind1993}%
  \BibitemOpen
  \bibfield  {author} {\bibinfo {author} {\bibnamefont {Lind}, \bibfnamefont
  {P.}}} (\bibinfo {year} {1993}),\ \href {\doibase 10.1103/PhysRevC.47.1903}
  {\bibfield  {journal} {\bibinfo  {journal} {Phys. Rev. C}\ }\textbf {\bibinfo
  {volume} {47}},\ \bibinfo {pages} {1903}}\BibitemShut {NoStop}%
\bibitem [{\citenamefont {Lunderberg}\ \emph {et~al.}(2012)\citenamefont
  {Lunderberg}, \citenamefont {DeYoung}, \citenamefont {Kohley}, \citenamefont
  {Attanayake}, \citenamefont {Baumann}, \citenamefont {Bazin}, \citenamefont
  {Christian}, \citenamefont {Divaratne}, \citenamefont {Grimes}, \citenamefont
  {Haagsma}, \citenamefont {Finck}, \citenamefont {Frank}, \citenamefont
  {Luther}, \citenamefont {Mosby}, \citenamefont {Nagi}, \citenamefont
  {Peaslee}, \citenamefont {Schiller}, \citenamefont {Snyder}, \citenamefont
  {Spyrou}, \citenamefont {Strongman},\ and\ \citenamefont
  {Thoennessen}}]{lunderberg2012}%
  \BibitemOpen
  \bibfield  {author} {\bibinfo {author} {\bibnamefont {Lunderberg},
  \bibfnamefont {E.}}, \bibinfo {author} {\bibfnamefont {P.~A.}\ \bibnamefont
  {DeYoung}}, \bibinfo {author} {\bibfnamefont {Z.}~\bibnamefont {Kohley}},
  \bibinfo {author} {\bibfnamefont {H.}~\bibnamefont {Attanayake}}, \bibinfo
  {author} {\bibfnamefont {T.}~\bibnamefont {Baumann}}, \bibinfo {author}
  {\bibfnamefont {D.}~\bibnamefont {Bazin}}, \bibinfo {author} {\bibfnamefont
  {G.}~\bibnamefont {Christian}}, \bibinfo {author} {\bibfnamefont
  {D.}~\bibnamefont {Divaratne}}, \bibinfo {author} {\bibfnamefont {S.~M.}\
  \bibnamefont {Grimes}}, \bibinfo {author} {\bibfnamefont {A.}~\bibnamefont
  {Haagsma}}, \bibinfo {author} {\bibfnamefont {J.~E.}\ \bibnamefont {Finck}},
  \bibinfo {author} {\bibfnamefont {N.}~\bibnamefont {Frank}}, \bibinfo
  {author} {\bibfnamefont {B.}~\bibnamefont {Luther}}, \bibinfo {author}
  {\bibfnamefont {S.}~\bibnamefont {Mosby}}, \bibinfo {author} {\bibfnamefont
  {T.}~\bibnamefont {Nagi}}, \bibinfo {author} {\bibfnamefont {G.~F.}\
  \bibnamefont {Peaslee}}, \bibinfo {author} {\bibfnamefont {A.}~\bibnamefont
  {Schiller}}, \bibinfo {author} {\bibfnamefont {J.}~\bibnamefont {Snyder}},
  \bibinfo {author} {\bibfnamefont {A.}~\bibnamefont {Spyrou}}, \bibinfo
  {author} {\bibfnamefont {M.~J.}\ \bibnamefont {Strongman}}, \ and\ \bibinfo
  {author} {\bibfnamefont {M.}~\bibnamefont {Thoennessen}}} (\bibinfo {year}
  {2012}),\ \href {\doibase 10.1103/PhysRevLett.108.142503} {\bibfield
  {journal} {\bibinfo  {journal} {Phys. Rev. Lett.}\ }\textbf {\bibinfo
  {volume} {108}},\ \bibinfo {pages} {142503}}\BibitemShut {NoStop}%
\bibitem [{\citenamefont {L{\"u}scher}(1986)}]{luscher1985}%
  \BibitemOpen
  \bibfield  {author} {\bibinfo {author} {\bibnamefont {L{\"u}scher},
  \bibfnamefont {M.}}} (\bibinfo {year} {1986}),\ \href {\doibase
  10.1007/BF01211589} {\bibfield  {journal} {\bibinfo  {journal} {Commun. Math.
  Phys.}\ }\textbf {\bibinfo {volume} {104}},\ \bibinfo {pages}
  {177}}\BibitemShut {NoStop}%
\bibitem [{\citenamefont {Lyutorovich}\ \emph {et~al.}(2012)\citenamefont
  {Lyutorovich}, \citenamefont {Tselyaev}, \citenamefont {Speth}, \citenamefont
  {Krewald}, \citenamefont {Gr\"ummer},\ and\ \citenamefont
  {Reinhard}}]{lyutorovich2012}%
  \BibitemOpen
  \bibfield  {author} {\bibinfo {author} {\bibnamefont {Lyutorovich},
  \bibfnamefont {N.}}, \bibinfo {author} {\bibfnamefont {V.~I.}\ \bibnamefont
  {Tselyaev}}, \bibinfo {author} {\bibfnamefont {J.}~\bibnamefont {Speth}},
  \bibinfo {author} {\bibfnamefont {S.}~\bibnamefont {Krewald}}, \bibinfo
  {author} {\bibfnamefont {F.}~\bibnamefont {Gr\"ummer}}, \ and\ \bibinfo
  {author} {\bibfnamefont {P.-G.}\ \bibnamefont {Reinhard}}} (\bibinfo {year}
  {2012}),\ \href {\doibase 10.1103/PhysRevLett.109.092502} {\bibfield
  {journal} {\bibinfo  {journal} {Phys. Rev. Lett.}\ }\textbf {\bibinfo
  {volume} {109}},\ \bibinfo {pages} {092502}}\BibitemShut {NoStop}%
\bibitem [{\citenamefont {Machleidt}(2001)}]{machleidt2001}%
  \BibitemOpen
  \bibfield  {author} {\bibinfo {author} {\bibnamefont {Machleidt},
  \bibfnamefont {R.}}} (\bibinfo {year} {2001}),\ \href {\doibase
  10.1103/PhysRevC.63.024001} {\bibfield  {journal} {\bibinfo  {journal} {Phys.
  Rev. C}\ }\textbf {\bibinfo {volume} {63}},\ \bibinfo {pages}
  {024001}}\BibitemShut {NoStop}%
\bibitem [{\citenamefont {Machleidt}\ and\ \citenamefont
  {Entem}(2011)}]{machleidt2011}%
  \BibitemOpen
  \bibfield  {author} {\bibinfo {author} {\bibnamefont {Machleidt},
  \bibfnamefont {R.}}, \ and\ \bibinfo {author} {\bibfnamefont
  {D.}~\bibnamefont {Entem}}} (\bibinfo {year} {2011}),\ \href {\doibase
  10.1016/j.physrep.2011.02.001} {\bibfield  {journal} {\bibinfo  {journal}
  {Physics Reports}\ }\textbf {\bibinfo {volume} {503}}~(\bibinfo {number}
  {1}),\ \bibinfo {pages} {1 }}\BibitemShut {NoStop}%
\bibitem [{\citenamefont {Marginean}\ \emph {et~al.}(2006)\citenamefont
  {Marginean}, \citenamefont {Lenzi}, \citenamefont {Gadea}, \citenamefont
  {Farnea}, \citenamefont {Freeman}, \citenamefont {Napoli}, \citenamefont
  {Bazzacco}, \citenamefont {Beghini}, \citenamefont {Behera}, \citenamefont
  {Bizzeti}, \citenamefont {Bizzeti-Sona}, \citenamefont {Bucurescu},
  \citenamefont {Chapman}, \citenamefont {Corradi}, \citenamefont {Deacon},
  \citenamefont {de~Angelis}, \citenamefont {Vedova}, \citenamefont {Fioretto},
  \citenamefont {Ionescu-Bujor}, \citenamefont {Iordachescu}, \citenamefont
  {Kr{\"o}ll}, \citenamefont {Latina}, \citenamefont {Liang}, \citenamefont
  {Lunardi}, \citenamefont {Montagnoli}, \citenamefont {Marginean},
  \citenamefont {Nespolo}, \citenamefont {Pollarolo}, \citenamefont {Rusu},
  \citenamefont {Scarlassara}, \citenamefont {Smith}, \citenamefont {Spohr},
  \citenamefont {Stefanini}, \citenamefont {Szilner}, \citenamefont {Trotta},
  \citenamefont {Ur}, \citenamefont {Varley},\ and\ \citenamefont
  {Zhimin}}]{marginean2006}%
  \BibitemOpen
  \bibfield  {author} {\bibinfo {author} {\bibnamefont {Marginean},
  \bibfnamefont {N.}}, \bibinfo {author} {\bibfnamefont {S.}~\bibnamefont
  {Lenzi}}, \bibinfo {author} {\bibfnamefont {A.}~\bibnamefont {Gadea}},
  \bibinfo {author} {\bibfnamefont {E.}~\bibnamefont {Farnea}}, \bibinfo
  {author} {\bibfnamefont {S.}~\bibnamefont {Freeman}}, \bibinfo {author}
  {\bibfnamefont {D.}~\bibnamefont {Napoli}}, \bibinfo {author} {\bibfnamefont
  {D.}~\bibnamefont {Bazzacco}}, \bibinfo {author} {\bibfnamefont
  {S.}~\bibnamefont {Beghini}}, \bibinfo {author} {\bibfnamefont
  {B.}~\bibnamefont {Behera}}, \bibinfo {author} {\bibfnamefont
  {P.}~\bibnamefont {Bizzeti}}, \bibinfo {author} {\bibfnamefont
  {A.}~\bibnamefont {Bizzeti-Sona}}, \bibinfo {author} {\bibfnamefont
  {D.}~\bibnamefont {Bucurescu}}, \bibinfo {author} {\bibfnamefont
  {R.}~\bibnamefont {Chapman}}, \bibinfo {author} {\bibfnamefont
  {L.}~\bibnamefont {Corradi}}, \bibinfo {author} {\bibfnamefont
  {A.}~\bibnamefont {Deacon}}, \bibinfo {author} {\bibfnamefont
  {G.}~\bibnamefont {de~Angelis}}, \bibinfo {author} {\bibfnamefont {F.~D.}\
  \bibnamefont {Vedova}}, \bibinfo {author} {\bibfnamefont {E.}~\bibnamefont
  {Fioretto}}, \bibinfo {author} {\bibfnamefont {M.}~\bibnamefont
  {Ionescu-Bujor}}, \bibinfo {author} {\bibfnamefont {A.}~\bibnamefont
  {Iordachescu}}, \bibinfo {author} {\bibfnamefont {T.}~\bibnamefont
  {Kr{\"o}ll}}, \bibinfo {author} {\bibfnamefont {A.}~\bibnamefont {Latina}},
  \bibinfo {author} {\bibfnamefont {X.}~\bibnamefont {Liang}}, \bibinfo
  {author} {\bibfnamefont {S.}~\bibnamefont {Lunardi}}, \bibinfo {author}
  {\bibfnamefont {G.}~\bibnamefont {Montagnoli}}, \bibinfo {author}
  {\bibfnamefont {R.}~\bibnamefont {Marginean}}, \bibinfo {author}
  {\bibfnamefont {M.}~\bibnamefont {Nespolo}}, \bibinfo {author} {\bibfnamefont
  {G.}~\bibnamefont {Pollarolo}}, \bibinfo {author} {\bibfnamefont
  {C.}~\bibnamefont {Rusu}}, \bibinfo {author} {\bibfnamefont {F.}~\bibnamefont
  {Scarlassara}}, \bibinfo {author} {\bibfnamefont {J.}~\bibnamefont {Smith}},
  \bibinfo {author} {\bibfnamefont {K.}~\bibnamefont {Spohr}}, \bibinfo
  {author} {\bibfnamefont {A.}~\bibnamefont {Stefanini}}, \bibinfo {author}
  {\bibfnamefont {S.}~\bibnamefont {Szilner}}, \bibinfo {author} {\bibfnamefont
  {M.}~\bibnamefont {Trotta}}, \bibinfo {author} {\bibfnamefont
  {C.}~\bibnamefont {Ur}}, \bibinfo {author} {\bibfnamefont {B.}~\bibnamefont
  {Varley}}, \ and\ \bibinfo {author} {\bibfnamefont {W.}~\bibnamefont
  {Zhimin}}} (\bibinfo {year} {2006}),\ \href {\doibase
  10.1016/j.physletb.2005.12.047} {\bibfield  {journal} {\bibinfo  {journal}
  {Physics Letters B}\ }\textbf {\bibinfo {volume} {633}}~(\bibinfo {number}
  {6}),\ \bibinfo {pages} {696 }}\BibitemShut {NoStop}%
\bibitem [{\citenamefont {Maris}\ \emph {et~al.}(2010)\citenamefont {Maris},
  \citenamefont {Shirokov},\ and\ \citenamefont {Vary}}]{maris2010}%
  \BibitemOpen
  \bibfield  {author} {\bibinfo {author} {\bibnamefont {Maris}, \bibfnamefont
  {P.}}, \bibinfo {author} {\bibfnamefont {A.~M.}\ \bibnamefont {Shirokov}}, \
  and\ \bibinfo {author} {\bibfnamefont {J.~P.}\ \bibnamefont {Vary}}}
  (\bibinfo {year} {2010}),\ \href {\doibase 10.1103/PhysRevC.81.021301}
  {\bibfield  {journal} {\bibinfo  {journal} {Phys. Rev. C}\ }\textbf {\bibinfo
  {volume} {81}},\ \bibinfo {pages} {021301}}\BibitemShut {NoStop}%
\bibitem [{\citenamefont {Maris}\ \emph {et~al.}(2013)\citenamefont {Maris},
  \citenamefont {Vary},\ and\ \citenamefont {Navr\'atil}}]{maris2012}%
  \BibitemOpen
  \bibfield  {author} {\bibinfo {author} {\bibnamefont {Maris}, \bibfnamefont
  {P.}}, \bibinfo {author} {\bibfnamefont {J.~P.}\ \bibnamefont {Vary}}, \ and\
  \bibinfo {author} {\bibfnamefont {P.}~\bibnamefont {Navr\'atil}}} (\bibinfo
  {year} {2013}),\ \href {\doibase 10.1103/PhysRevC.87.014327} {\bibfield
  {journal} {\bibinfo  {journal} {Phys. Rev. C}\ }\textbf {\bibinfo {volume}
  {87}},\ \bibinfo {pages} {014327}}\BibitemShut {NoStop}%
\bibitem [{\citenamefont {Maris}\ \emph {et~al.}(2011)\citenamefont {Maris},
  \citenamefont {Vary}, \citenamefont {Navr\'atil}, \citenamefont {Ormand},
  \citenamefont {Nam},\ and\ \citenamefont {Dean}}]{maris2011}%
  \BibitemOpen
  \bibfield  {author} {\bibinfo {author} {\bibnamefont {Maris}, \bibfnamefont
  {P.}}, \bibinfo {author} {\bibfnamefont {J.~P.}\ \bibnamefont {Vary}},
  \bibinfo {author} {\bibfnamefont {P.}~\bibnamefont {Navr\'atil}}, \bibinfo
  {author} {\bibfnamefont {W.~E.}\ \bibnamefont {Ormand}}, \bibinfo {author}
  {\bibfnamefont {H.}~\bibnamefont {Nam}}, \ and\ \bibinfo {author}
  {\bibfnamefont {D.~J.}\ \bibnamefont {Dean}}} (\bibinfo {year} {2011}),\
  \href {\doibase 10.1103/PhysRevLett.106.202502} {\bibfield  {journal}
  {\bibinfo  {journal} {Phys. Rev. Lett.}\ }\textbf {\bibinfo {volume} {106}},\
  \bibinfo {pages} {202502}}\BibitemShut {NoStop}%
\bibitem [{\citenamefont {Maris}\ \emph {et~al.}(2009)\citenamefont {Maris},
  \citenamefont {Vary},\ and\ \citenamefont {Shirokov}}]{maris2009}%
  \BibitemOpen
  \bibfield  {author} {\bibinfo {author} {\bibnamefont {Maris}, \bibfnamefont
  {P.}}, \bibinfo {author} {\bibfnamefont {J.~P.}\ \bibnamefont {Vary}}, \ and\
  \bibinfo {author} {\bibfnamefont {A.~M.}\ \bibnamefont {Shirokov}}} (\bibinfo
  {year} {2009}),\ \href {\doibase 10.1103/PhysRevC.79.014308} {\bibfield
  {journal} {\bibinfo  {journal} {Phys. Rev. C}\ }\textbf {\bibinfo {volume}
  {79}},\ \bibinfo {pages} {014308}}\BibitemShut {NoStop}%
\bibitem [{\citenamefont {McGrory}\ and\ \citenamefont
  {Wildenthal}(1975)}]{mcgrory1975}%
  \BibitemOpen
  \bibfield  {author} {\bibinfo {author} {\bibnamefont {McGrory}, \bibfnamefont
  {J.}}, \ and\ \bibinfo {author} {\bibfnamefont {B.}~\bibnamefont
  {Wildenthal}}} (\bibinfo {year} {1975}),\ \href {\doibase
  10.1016/0370-2693(75)90513-4} {\bibfield  {journal} {\bibinfo  {journal}
  {Physics Letters B}\ }\textbf {\bibinfo {volume} {60}}~(\bibinfo {number}
  {1}),\ \bibinfo {pages} {5 }}\BibitemShut {NoStop}%
\bibitem [{\citenamefont {Meng}\ \emph {et~al.}(2002)\citenamefont {Meng},
  \citenamefont {Toki}, \citenamefont {Zeng}, \citenamefont {Zhang},\ and\
  \citenamefont {Zhou}}]{meng2002}%
  \BibitemOpen
  \bibfield  {author} {\bibinfo {author} {\bibnamefont {Meng}, \bibfnamefont
  {J.}}, \bibinfo {author} {\bibfnamefont {H.}~\bibnamefont {Toki}}, \bibinfo
  {author} {\bibfnamefont {J.~Y.}\ \bibnamefont {Zeng}}, \bibinfo {author}
  {\bibfnamefont {S.~Q.}\ \bibnamefont {Zhang}}, \ and\ \bibinfo {author}
  {\bibfnamefont {S.-G.}\ \bibnamefont {Zhou}}} (\bibinfo {year} {2002}),\
  \href {\doibase 10.1103/PhysRevC.65.041302} {\bibfield  {journal} {\bibinfo
  {journal} {Phys. Rev. C}\ }\textbf {\bibinfo {volume} {65}},\ \bibinfo
  {pages} {041302}}\BibitemShut {NoStop}%
\bibitem [{\citenamefont {Michel}(2011)}]{michel2011}%
  \BibitemOpen
  \bibfield  {author} {\bibinfo {author} {\bibnamefont {Michel}, \bibfnamefont
  {N.}}} (\bibinfo {year} {2011}),\ \href {\doibase 10.1103/PhysRevC.83.034325}
  {\bibfield  {journal} {\bibinfo  {journal} {Phys. Rev. C}\ }\textbf {\bibinfo
  {volume} {83}},\ \bibinfo {pages} {034325}}\BibitemShut {NoStop}%
\bibitem [{\citenamefont {Michel}\ \emph {et~al.}(2002)\citenamefont {Michel},
  \citenamefont {Nazarewicz}, \citenamefont {P\l{}oszajczak},\ and\
  \citenamefont {Bennaceur}}]{michel2002}%
  \BibitemOpen
  \bibfield  {author} {\bibinfo {author} {\bibnamefont {Michel}, \bibfnamefont
  {N.}}, \bibinfo {author} {\bibfnamefont {W.}~\bibnamefont {Nazarewicz}},
  \bibinfo {author} {\bibfnamefont {M.}~\bibnamefont {P\l{}oszajczak}}, \ and\
  \bibinfo {author} {\bibfnamefont {K.}~\bibnamefont {Bennaceur}}} (\bibinfo
  {year} {2002}),\ \href {\doibase 10.1103/PhysRevLett.89.042502} {\bibfield
  {journal} {\bibinfo  {journal} {Phys. Rev. Lett.}\ }\textbf {\bibinfo
  {volume} {89}},\ \bibinfo {pages} {042502}}\BibitemShut {NoStop}%
\bibitem [{\citenamefont {Michel}\ \emph {et~al.}(2009)\citenamefont {Michel},
  \citenamefont {Nazarewicz}, \citenamefont {P\l{}oszajczak},\ and\
  \citenamefont {Vertse}}]{michel2009}%
  \BibitemOpen
  \bibfield  {author} {\bibinfo {author} {\bibnamefont {Michel}, \bibfnamefont
  {N.}}, \bibinfo {author} {\bibfnamefont {W.}~\bibnamefont {Nazarewicz}},
  \bibinfo {author} {\bibfnamefont {M.}~\bibnamefont {P\l{}oszajczak}}, \ and\
  \bibinfo {author} {\bibfnamefont {T.}~\bibnamefont {Vertse}}} (\bibinfo
  {year} {2009}),\ \href {http://stacks.iop.org/0954-3899/36/i=1/a=013101}
  {\bibfield  {journal} {\bibinfo  {journal} {Journal of Physics G: Nuclear and
  Particle Physics}\ }\textbf {\bibinfo {volume} {36}}~(\bibinfo {number}
  {1}),\ \bibinfo {pages} {013101}}\BibitemShut {NoStop}%
\bibitem [{\citenamefont {Mihaila}(2003)}]{mihaila2003}%
  \BibitemOpen
  \bibfield  {author} {\bibinfo {author} {\bibnamefont {Mihaila}, \bibfnamefont
  {B.}}} (\bibinfo {year} {2003}),\ \href {\doibase 10.1103/PhysRevC.68.054327}
  {\bibfield  {journal} {\bibinfo  {journal} {Phys. Rev. C}\ }\textbf {\bibinfo
  {volume} {68}},\ \bibinfo {pages} {054327}}\BibitemShut {NoStop}%
\bibitem [{\citenamefont {Mihaila}\ and\ \citenamefont
  {Heisenberg}(1999)}]{mihaila1999}%
  \BibitemOpen
  \bibfield  {author} {\bibinfo {author} {\bibnamefont {Mihaila}, \bibfnamefont
  {B.}}, \ and\ \bibinfo {author} {\bibfnamefont {J.~H.}\ \bibnamefont
  {Heisenberg}}} (\bibinfo {year} {1999}),\ \href {\doibase
  10.1103/PhysRevC.60.054303} {\bibfield  {journal} {\bibinfo  {journal} {Phys.
  Rev. C}\ }\textbf {\bibinfo {volume} {60}},\ \bibinfo {pages}
  {054303}}\BibitemShut {NoStop}%
\bibitem [{\citenamefont {Mihaila}\ and\ \citenamefont
  {Heisenberg}(2000{\natexlab{a}})}]{mihaila2000a}%
  \BibitemOpen
  \bibfield  {author} {\bibinfo {author} {\bibnamefont {Mihaila}, \bibfnamefont
  {B.}}, \ and\ \bibinfo {author} {\bibfnamefont {J.~H.}\ \bibnamefont
  {Heisenberg}}} (\bibinfo {year} {2000}{\natexlab{a}}),\ \href {\doibase
  10.1103/PhysRevC.61.054309} {\bibfield  {journal} {\bibinfo  {journal} {Phys.
  Rev. C}\ }\textbf {\bibinfo {volume} {61}},\ \bibinfo {pages}
  {054309}}\BibitemShut {NoStop}%
\bibitem [{\citenamefont {Mihaila}\ and\ \citenamefont
  {Heisenberg}(2000{\natexlab{b}})}]{mihaila2000b}%
  \BibitemOpen
  \bibfield  {author} {\bibinfo {author} {\bibnamefont {Mihaila}, \bibfnamefont
  {B.}}, \ and\ \bibinfo {author} {\bibfnamefont {J.~H.}\ \bibnamefont
  {Heisenberg}}} (\bibinfo {year} {2000}{\natexlab{b}}),\ \href {\doibase
  10.1103/PhysRevLett.84.1403} {\bibfield  {journal} {\bibinfo  {journal}
  {Phys. Rev. Lett.}\ }\textbf {\bibinfo {volume} {84}},\ \bibinfo {pages}
  {1403}}\BibitemShut {NoStop}%
\bibitem [{\citenamefont {Moliner}\ \emph {et~al.}(2002)\citenamefont
  {Moliner}, \citenamefont {Walet},\ and\ \citenamefont
  {Bishop}}]{moliner2002}%
  \BibitemOpen
  \bibfield  {author} {\bibinfo {author} {\bibnamefont {Moliner}, \bibfnamefont
  {I.}}, \bibinfo {author} {\bibfnamefont {N.~R.}\ \bibnamefont {Walet}}, \
  and\ \bibinfo {author} {\bibfnamefont {R.~F.}\ \bibnamefont {Bishop}}}
  (\bibinfo {year} {2002}),\ \href
  {http://stacks.iop.org/0954-3899/28/i=6/a=305} {\bibfield  {journal}
  {\bibinfo  {journal} {Journal of Physics G: Nuclear and Particle Physics}\
  }\textbf {\bibinfo {volume} {28}}~(\bibinfo {number} {6}),\ \bibinfo {pages}
  {1209}}\BibitemShut {NoStop}%
\bibitem [{\citenamefont {Monkhorst}(1987)}]{Mon87}%
  \BibitemOpen
  \bibfield  {author} {\bibinfo {author} {\bibnamefont {Monkhorst},
  \bibfnamefont {H.~J.}}} (\bibinfo {year} {1987}),\ \href {\doibase
  10.1103/PhysRevA.36.1544} {\bibfield  {journal} {\bibinfo  {journal} {Phys.
  Rev. A}\ }\textbf {\bibinfo {volume} {36}},\ \bibinfo {pages}
  {1544}}\BibitemShut {NoStop}%
\bibitem [{\citenamefont {More}\ \emph {et~al.}(2013)\citenamefont {More},
  \citenamefont {Ekstr\"om}, \citenamefont {Furnstahl}, \citenamefont {Hagen},\
  and\ \citenamefont {Papenbrock}}]{more2013}%
  \BibitemOpen
  \bibfield  {author} {\bibinfo {author} {\bibnamefont {More}, \bibfnamefont
  {S.~N.}}, \bibinfo {author} {\bibfnamefont {A.}~\bibnamefont {Ekstr\"om}},
  \bibinfo {author} {\bibfnamefont {R.~J.}\ \bibnamefont {Furnstahl}}, \bibinfo
  {author} {\bibfnamefont {G.}~\bibnamefont {Hagen}}, \ and\ \bibinfo {author}
  {\bibfnamefont {T.}~\bibnamefont {Papenbrock}}} (\bibinfo {year} {2013}),\
  \href {\doibase 10.1103/PhysRevC.87.044326} {\bibfield  {journal} {\bibinfo
  {journal} {Phys. Rev. C}\ }\textbf {\bibinfo {volume} {87}},\ \bibinfo
  {pages} {044326}}\BibitemShut {NoStop}%
\bibitem [{\citenamefont {Nakatsukasa}(2012)}]{nakatsukasa2012}%
  \BibitemOpen
  \bibfield  {author} {\bibinfo {author} {\bibnamefont {Nakatsukasa},
  \bibfnamefont {T.}}} (\bibinfo {year} {2012}),\ \href {\doibase
  10.1093/ptep/pts016} {\bibfield  {journal} {\bibinfo  {journal} {Progress of
  Theoretical and Experimental Physics}\ }\textbf {\bibinfo {volume}
  {2012}}~(\bibinfo {number} {1}),\ 10.1093/ptep/pts016}\BibitemShut {NoStop}%
\bibitem [{\citenamefont {{Nam}}\ \emph {et~al.}(2012)\citenamefont {{Nam}},
  \citenamefont {{Stoitsov}}, \citenamefont {{Nazarewicz}}, \citenamefont
  {{Bulgac}}, \citenamefont {{Hagen}}, \citenamefont {{Kortelainen}},
  \citenamefont {{Maris}}, \citenamefont {{Pei}}, \citenamefont {{Roche}},
  \citenamefont {{Schunck}}, \citenamefont {{Thompson}}, \citenamefont
  {{Vary}},\ and\ \citenamefont {{Wild}}}]{nam2012}%
  \BibitemOpen
  \bibfield  {author} {\bibinfo {author} {\bibnamefont {{Nam}}, \bibfnamefont
  {H.}}, \bibinfo {author} {\bibfnamefont {M.}~\bibnamefont {{Stoitsov}}},
  \bibinfo {author} {\bibfnamefont {W.}~\bibnamefont {{Nazarewicz}}}, \bibinfo
  {author} {\bibfnamefont {A.}~\bibnamefont {{Bulgac}}}, \bibinfo {author}
  {\bibfnamefont {G.}~\bibnamefont {{Hagen}}}, \bibinfo {author} {\bibfnamefont
  {M.}~\bibnamefont {{Kortelainen}}}, \bibinfo {author} {\bibfnamefont
  {P.}~\bibnamefont {{Maris}}}, \bibinfo {author} {\bibfnamefont {J.~C.}\
  \bibnamefont {{Pei}}}, \bibinfo {author} {\bibfnamefont {K.~J.}\ \bibnamefont
  {{Roche}}}, \bibinfo {author} {\bibfnamefont {N.}~\bibnamefont {{Schunck}}},
  \bibinfo {author} {\bibfnamefont {I.}~\bibnamefont {{Thompson}}}, \bibinfo
  {author} {\bibfnamefont {J.~P.}\ \bibnamefont {{Vary}}}, \ and\ \bibinfo
  {author} {\bibfnamefont {S.~M.}\ \bibnamefont {{Wild}}}} (\bibinfo {year}
  {2012}),\ \href {http://stacks.iop.org/1742-6596/402/i=1/a=012033} {\bibfield
   {journal} {\bibinfo  {journal} {Journal of Physics: Conference Series}\
  }\textbf {\bibinfo {volume} {402}}~(\bibinfo {number} {1}),\ \bibinfo {pages}
  {012033}}\BibitemShut {NoStop}%
\bibitem [{\citenamefont {Navr\'atil}(2004)}]{navratil2004}%
  \BibitemOpen
  \bibfield  {author} {\bibinfo {author} {\bibnamefont {Navr\'atil},
  \bibfnamefont {P.}}} (\bibinfo {year} {2004}),\ \href {\doibase
  10.1103/PhysRevC.70.014317} {\bibfield  {journal} {\bibinfo  {journal} {Phys.
  Rev. C}\ }\textbf {\bibinfo {volume} {70}},\ \bibinfo {pages}
  {014317}}\BibitemShut {NoStop}%
\bibitem [{\citenamefont {Navr{\'a}til}(2007)}]{navratil2007}%
  \BibitemOpen
  \bibfield  {author} {\bibinfo {author} {\bibnamefont {Navr{\'a}til},
  \bibfnamefont {P.}}} (\bibinfo {year} {2007}),\ \href {\doibase
  10.1007/s00601-007-0193-3} {\bibfield  {journal} {\bibinfo  {journal}
  {Few-Body Systems}\ }\textbf {\bibinfo {volume} {41}}~(\bibinfo {number}
  {3-4}),\ \bibinfo {pages} {117}}\BibitemShut {NoStop}%
\bibitem [{\citenamefont {Navr\'atil}\ \emph {et~al.}(2007)\citenamefont
  {Navr\'atil}, \citenamefont {Gueorguiev}, \citenamefont {Vary}, \citenamefont
  {Ormand},\ and\ \citenamefont {Nogga}}]{navratil2007b}%
  \BibitemOpen
  \bibfield  {author} {\bibinfo {author} {\bibnamefont {Navr\'atil},
  \bibfnamefont {P.}}, \bibinfo {author} {\bibfnamefont {V.~G.}\ \bibnamefont
  {Gueorguiev}}, \bibinfo {author} {\bibfnamefont {J.~P.}\ \bibnamefont
  {Vary}}, \bibinfo {author} {\bibfnamefont {W.~E.}\ \bibnamefont {Ormand}}, \
  and\ \bibinfo {author} {\bibfnamefont {A.}~\bibnamefont {Nogga}}} (\bibinfo
  {year} {2007}),\ \href {\doibase 10.1103/PhysRevLett.99.042501} {\bibfield
  {journal} {\bibinfo  {journal} {Phys. Rev. Lett.}\ }\textbf {\bibinfo
  {volume} {99}},\ \bibinfo {pages} {042501}}\BibitemShut {NoStop}%
\bibitem [{\citenamefont {Navr{\'a}til}\ \emph {et~al.}(2009)\citenamefont
  {Navr{\'a}til}, \citenamefont {Quaglioni}, \citenamefont {Stetcu},\ and\
  \citenamefont {Barrett}}]{navratil2009}%
  \BibitemOpen
  \bibfield  {author} {\bibinfo {author} {\bibnamefont {Navr{\'a}til},
  \bibfnamefont {P.}}, \bibinfo {author} {\bibfnamefont {S.}~\bibnamefont
  {Quaglioni}}, \bibinfo {author} {\bibfnamefont {I.}~\bibnamefont {Stetcu}}, \
  and\ \bibinfo {author} {\bibfnamefont {B.~R.}\ \bibnamefont {Barrett}}}
  (\bibinfo {year} {2009}),\ \href
  {http://stacks.iop.org/0954-3899/36/i=8/a=083101} {\bibfield  {journal}
  {\bibinfo  {journal} {Journal of Physics G: Nuclear and Particle Physics}\
  }\textbf {\bibinfo {volume} {36}}~(\bibinfo {number} {8}),\ \bibinfo {pages}
  {083101}}\BibitemShut {NoStop}%
\bibitem [{\citenamefont {Navr\'atil}\ \emph {et~al.}(2000)\citenamefont
  {Navr\'atil}, \citenamefont {Vary},\ and\ \citenamefont
  {Barrett}}]{navratil2000}%
  \BibitemOpen
  \bibfield  {author} {\bibinfo {author} {\bibnamefont {Navr\'atil},
  \bibfnamefont {P.}}, \bibinfo {author} {\bibfnamefont {J.~P.}\ \bibnamefont
  {Vary}}, \ and\ \bibinfo {author} {\bibfnamefont {B.~R.}\ \bibnamefont
  {Barrett}}} (\bibinfo {year} {2000}),\ \href {\doibase
  10.1103/PhysRevC.62.054311} {\bibfield  {journal} {\bibinfo  {journal} {Phys.
  Rev. C}\ }\textbf {\bibinfo {volume} {62}},\ \bibinfo {pages}
  {054311}}\BibitemShut {NoStop}%
\bibitem [{\citenamefont {Nazarewicz}\ \emph {et~al.}(1996)\citenamefont
  {Nazarewicz}, \citenamefont {Dobaczewski}, \citenamefont {Werner},
  \citenamefont {Maruhn}, \citenamefont {Reinhard}, \citenamefont {Rutz},
  \citenamefont {Chinn}, \citenamefont {Umar},\ and\ \citenamefont
  {Strayer}}]{nazarewicz1996}%
  \BibitemOpen
  \bibfield  {author} {\bibinfo {author} {\bibnamefont {Nazarewicz},
  \bibfnamefont {W.}}, \bibinfo {author} {\bibfnamefont {J.}~\bibnamefont
  {Dobaczewski}}, \bibinfo {author} {\bibfnamefont {T.~R.}\ \bibnamefont
  {Werner}}, \bibinfo {author} {\bibfnamefont {J.~A.}\ \bibnamefont {Maruhn}},
  \bibinfo {author} {\bibfnamefont {P.-G.}\ \bibnamefont {Reinhard}}, \bibinfo
  {author} {\bibfnamefont {K.}~\bibnamefont {Rutz}}, \bibinfo {author}
  {\bibfnamefont {C.~R.}\ \bibnamefont {Chinn}}, \bibinfo {author}
  {\bibfnamefont {A.~S.}\ \bibnamefont {Umar}}, \ and\ \bibinfo {author}
  {\bibfnamefont {M.~R.}\ \bibnamefont {Strayer}}} (\bibinfo {year} {1996}),\
  \href {\doibase 10.1103/PhysRevC.53.740} {\bibfield  {journal} {\bibinfo
  {journal} {Phys. Rev. C}\ }\textbf {\bibinfo {volume} {53}},\ \bibinfo
  {pages} {740}}\BibitemShut {NoStop}%
\bibitem [{\citenamefont {Nik{\v s}i{\'c}}\ \emph {et~al.}(2011)\citenamefont
  {Nik{\v s}i{\'c}}, \citenamefont {Vretenar},\ and\ \citenamefont
  {Ring}}]{niksic2011}%
  \BibitemOpen
  \bibfield  {author} {\bibinfo {author} {\bibnamefont {Nik{\v s}i{\'c}},
  \bibfnamefont {T.}}, \bibinfo {author} {\bibfnamefont {D.}~\bibnamefont
  {Vretenar}}, \ and\ \bibinfo {author} {\bibfnamefont {P.}~\bibnamefont
  {Ring}}} (\bibinfo {year} {2011}),\ \href {\doibase
  10.1016/j.ppnp.2011.01.055} {\bibfield  {journal} {\bibinfo  {journal}
  {Progress in Particle and Nuclear Physics}\ }\textbf {\bibinfo {volume}
  {66}}~(\bibinfo {number} {3}),\ \bibinfo {pages} {519 }}\BibitemShut
  {NoStop}%
\bibitem [{\citenamefont {Noga}\ and\ \citenamefont {Urban}(1988)}]{noga1988}%
  \BibitemOpen
  \bibfield  {author} {\bibinfo {author} {\bibnamefont {Noga}, \bibfnamefont
  {J.}}, \ and\ \bibinfo {author} {\bibfnamefont {M.}~\bibnamefont {Urban}}}
  (\bibinfo {year} {1988}),\ \href {\doibase 10.1007/BF00527416} {\bibfield
  {journal} {\bibinfo  {journal} {Theoretica chimica acta}\ }\textbf {\bibinfo
  {volume} {73}}~(\bibinfo {number} {4}),\ \bibinfo {pages} {291}}\BibitemShut
  {NoStop}%
\bibitem [{\citenamefont {Nogga}\ \emph {et~al.}(2004)\citenamefont {Nogga},
  \citenamefont {Bogner},\ and\ \citenamefont {Schwenk}}]{nogga2004}%
  \BibitemOpen
  \bibfield  {author} {\bibinfo {author} {\bibnamefont {Nogga}, \bibfnamefont
  {A.}}, \bibinfo {author} {\bibfnamefont {S.~K.}\ \bibnamefont {Bogner}}, \
  and\ \bibinfo {author} {\bibfnamefont {A.}~\bibnamefont {Schwenk}}} (\bibinfo
  {year} {2004}),\ \href {\doibase 10.1103/PhysRevC.70.061002} {\bibfield
  {journal} {\bibinfo  {journal} {Phys. Rev. C}\ }\textbf {\bibinfo {volume}
  {70}},\ \bibinfo {pages} {061002}}\BibitemShut {NoStop}%
\bibitem [{\citenamefont {Nollett}\ \emph {et~al.}(2007)\citenamefont
  {Nollett}, \citenamefont {Pieper}, \citenamefont {Wiringa}, \citenamefont
  {Carlson},\ and\ \citenamefont {Hale}}]{nollett2007}%
  \BibitemOpen
  \bibfield  {author} {\bibinfo {author} {\bibnamefont {Nollett}, \bibfnamefont
  {K.~M.}}, \bibinfo {author} {\bibfnamefont {S.~C.}\ \bibnamefont {Pieper}},
  \bibinfo {author} {\bibfnamefont {R.~B.}\ \bibnamefont {Wiringa}}, \bibinfo
  {author} {\bibfnamefont {J.}~\bibnamefont {Carlson}}, \ and\ \bibinfo
  {author} {\bibfnamefont {G.~M.}\ \bibnamefont {Hale}}} (\bibinfo {year}
  {2007}),\ \href {\doibase 10.1103/PhysRevLett.99.022502} {\bibfield
  {journal} {\bibinfo  {journal} {Phys. Rev. Lett.}\ }\textbf {\bibinfo
  {volume} {99}},\ \bibinfo {pages} {022502}}\BibitemShut {NoStop}%
\bibitem [{\citenamefont {Nooijen}\ \emph {et~al.}(2005)\citenamefont
  {Nooijen}, \citenamefont {Shamasundar},\ and\ \citenamefont
  {Mukherjee}}]{nooijen2005}%
  \BibitemOpen
  \bibfield  {author} {\bibinfo {author} {\bibnamefont {Nooijen}, \bibfnamefont
  {M.}}, \bibinfo {author} {\bibfnamefont {K.~R.}\ \bibnamefont {Shamasundar}},
  \ and\ \bibinfo {author} {\bibfnamefont {D.}~\bibnamefont {Mukherjee}}}
  (\bibinfo {year} {2005}),\ \href {\doibase 10.1080/00268970500083952}
  {\bibfield  {journal} {\bibinfo  {journal} {Molecular Physics}\ }\textbf
  {\bibinfo {volume} {103}}~(\bibinfo {number} {15-16}),\ \bibinfo {pages}
  {2277}}\BibitemShut {NoStop}%
\bibitem [{\citenamefont {Ord\'o\~nez}\ \emph {et~al.}(1994)\citenamefont
  {Ord\'o\~nez}, \citenamefont {Ray},\ and\ \citenamefont {van
  Kolck}}]{ordonez1994}%
  \BibitemOpen
  \bibfield  {author} {\bibinfo {author} {\bibnamefont {Ord\'o\~nez},
  \bibfnamefont {C.}}, \bibinfo {author} {\bibfnamefont {L.}~\bibnamefont
  {Ray}}, \ and\ \bibinfo {author} {\bibfnamefont {U.}~\bibnamefont {van
  Kolck}}} (\bibinfo {year} {1994}),\ \href {\doibase
  10.1103/PhysRevLett.72.1982} {\bibfield  {journal} {\bibinfo  {journal}
  {Phys. Rev. Lett.}\ }\textbf {\bibinfo {volume} {72}},\ \bibinfo {pages}
  {1982}}\BibitemShut {NoStop}%
\bibitem [{\citenamefont {Ord\'o\~nez}\ \emph {et~al.}(1996)\citenamefont
  {Ord\'o\~nez}, \citenamefont {Ray},\ and\ \citenamefont {van
  Kolck}}]{ordonez1996}%
  \BibitemOpen
  \bibfield  {author} {\bibinfo {author} {\bibnamefont {Ord\'o\~nez},
  \bibfnamefont {C.}}, \bibinfo {author} {\bibfnamefont {L.}~\bibnamefont
  {Ray}}, \ and\ \bibinfo {author} {\bibfnamefont {U.}~\bibnamefont {van
  Kolck}}} (\bibinfo {year} {1996}),\ \href {\doibase 10.1103/PhysRevC.53.2086}
  {\bibfield  {journal} {\bibinfo  {journal} {Phys. Rev. C}\ }\textbf {\bibinfo
  {volume} {53}},\ \bibinfo {pages} {2086}}\BibitemShut {NoStop}%
\bibitem [{\citenamefont {Ord{\'o}{\~n}ez}\ and\ \citenamefont {van
  Kolck}(1992)}]{ordonez1992}%
  \BibitemOpen
  \bibfield  {author} {\bibinfo {author} {\bibnamefont {Ord{\'o}{\~n}ez},
  \bibfnamefont {C.}}, \ and\ \bibinfo {author} {\bibfnamefont
  {U.}~\bibnamefont {van Kolck}}} (\bibinfo {year} {1992}),\ \href {\doibase
  10.1016/0370-2693(92)91404-W} {\bibfield  {journal} {\bibinfo  {journal}
  {Physics Letters B}\ }\textbf {\bibinfo {volume} {291}}~(\bibinfo {number}
  {4}),\ \bibinfo {pages} {459 }}\BibitemShut {NoStop}%
\bibitem [{\citenamefont {{Orlandini}}\ \emph {et~al.}(2013)\citenamefont
  {{Orlandini}}, \citenamefont {{Bacca}}, \citenamefont {{Barnea}},
  \citenamefont {{Hagen}}, \citenamefont {{Miorelli}},\ and\ \citenamefont
  {{Papenbrock}}}]{orlandini2013}%
  \BibitemOpen
  \bibfield  {author} {\bibinfo {author} {\bibnamefont {{Orlandini}},
  \bibfnamefont {G.}}, \bibinfo {author} {\bibfnamefont {S.}~\bibnamefont
  {{Bacca}}}, \bibinfo {author} {\bibfnamefont {N.}~\bibnamefont {{Barnea}}},
  \bibinfo {author} {\bibfnamefont {G.}~\bibnamefont {{Hagen}}}, \bibinfo
  {author} {\bibfnamefont {M.}~\bibnamefont {{Miorelli}}}, \ and\ \bibinfo
  {author} {\bibfnamefont {T.}~\bibnamefont {{Papenbrock}}}} (\bibinfo {year}
  {2013}),\ \href@noop {} {\bibfield  {journal} {\bibinfo  {journal} {ArXiv
  e-prints}\ }}\Eprint {http://arxiv.org/abs/1311.2141} {arXiv:1311.2141
  [nucl-th]} \BibitemShut {NoStop}%
\bibitem [{\citenamefont {Otsuka}\ \emph {et~al.}(2010)\citenamefont {Otsuka},
  \citenamefont {Suzuki}, \citenamefont {Holt}, \citenamefont {Schwenk},\ and\
  \citenamefont {Akaishi}}]{otsuka2010}%
  \BibitemOpen
  \bibfield  {author} {\bibinfo {author} {\bibnamefont {Otsuka}, \bibfnamefont
  {T.}}, \bibinfo {author} {\bibfnamefont {T.}~\bibnamefont {Suzuki}}, \bibinfo
  {author} {\bibfnamefont {J.~D.}\ \bibnamefont {Holt}}, \bibinfo {author}
  {\bibfnamefont {A.}~\bibnamefont {Schwenk}}, \ and\ \bibinfo {author}
  {\bibfnamefont {Y.}~\bibnamefont {Akaishi}}} (\bibinfo {year} {2010}),\ \href
  {\doibase 10.1103/PhysRevLett.105.032501} {\bibfield  {journal} {\bibinfo
  {journal} {Phys. Rev. Lett.}\ }\textbf {\bibinfo {volume} {105}},\ \bibinfo
  {pages} {032501}}\BibitemShut {NoStop}%
\bibitem [{\citenamefont {Ozawa}\ \emph {et~al.}(2001)\citenamefont {Ozawa},
  \citenamefont {Bochkarev}, \citenamefont {Chulkov}, \citenamefont {Cortina},
  \citenamefont {Geissel}, \citenamefont {Hellstr{\"o}m}, \citenamefont
  {Ivanov}, \citenamefont {Janik}, \citenamefont {Kimura}, \citenamefont
  {Kobayashi}, \citenamefont {Korsheninnikov}, \citenamefont {M{\"u}nzenberg},
  \citenamefont {Nickel}, \citenamefont {Ogawa}, \citenamefont {Ogloblin},
  \citenamefont {Pf{\"u}tzner}, \citenamefont {Pribora}, \citenamefont {Simon},
  \citenamefont {Sit{\'a}r}, \citenamefont {Strmen}, \citenamefont
  {S{\"u}mmerer}, \citenamefont {Suzuki}, \citenamefont {Tanihata},
  \citenamefont {Winkler},\ and\ \citenamefont {Yoshida}}]{ozawa2001}%
  \BibitemOpen
  \bibfield  {author} {\bibinfo {author} {\bibnamefont {Ozawa}, \bibfnamefont
  {A.}}, \bibinfo {author} {\bibfnamefont {O.}~\bibnamefont {Bochkarev}},
  \bibinfo {author} {\bibfnamefont {L.}~\bibnamefont {Chulkov}}, \bibinfo
  {author} {\bibfnamefont {D.}~\bibnamefont {Cortina}}, \bibinfo {author}
  {\bibfnamefont {H.}~\bibnamefont {Geissel}}, \bibinfo {author} {\bibfnamefont
  {M.}~\bibnamefont {Hellstr{\"o}m}}, \bibinfo {author} {\bibfnamefont
  {M.}~\bibnamefont {Ivanov}}, \bibinfo {author} {\bibfnamefont
  {R.}~\bibnamefont {Janik}}, \bibinfo {author} {\bibfnamefont
  {K.}~\bibnamefont {Kimura}}, \bibinfo {author} {\bibfnamefont
  {T.}~\bibnamefont {Kobayashi}}, \bibinfo {author} {\bibfnamefont {A.~A.}\
  \bibnamefont {Korsheninnikov}}, \bibinfo {author} {\bibfnamefont
  {G.}~\bibnamefont {M{\"u}nzenberg}}, \bibinfo {author} {\bibfnamefont
  {F.}~\bibnamefont {Nickel}}, \bibinfo {author} {\bibfnamefont
  {Y.}~\bibnamefont {Ogawa}}, \bibinfo {author} {\bibfnamefont {A.~A.}\
  \bibnamefont {Ogloblin}}, \bibinfo {author} {\bibfnamefont {M.}~\bibnamefont
  {Pf{\"u}tzner}}, \bibinfo {author} {\bibfnamefont {V.}~\bibnamefont
  {Pribora}}, \bibinfo {author} {\bibfnamefont {H.}~\bibnamefont {Simon}},
  \bibinfo {author} {\bibfnamefont {B.}~\bibnamefont {Sit{\'a}r}}, \bibinfo
  {author} {\bibfnamefont {P.}~\bibnamefont {Strmen}}, \bibinfo {author}
  {\bibfnamefont {K.}~\bibnamefont {S{\"u}mmerer}}, \bibinfo {author}
  {\bibfnamefont {T.}~\bibnamefont {Suzuki}}, \bibinfo {author} {\bibfnamefont
  {I.}~\bibnamefont {Tanihata}}, \bibinfo {author} {\bibfnamefont
  {M.}~\bibnamefont {Winkler}}, \ and\ \bibinfo {author} {\bibfnamefont
  {K.}~\bibnamefont {Yoshida}}} (\bibinfo {year} {2001}),\ \href {\doibase
  10.1016/S0375-9474(01)00563-2} {\bibfield  {journal} {\bibinfo  {journal}
  {Nuclear Physics A}\ }\textbf {\bibinfo {volume} {691}},\ \bibinfo {pages}
  {599 }}\BibitemShut {NoStop}%
\bibitem [{\citenamefont {Ozawa}\ \emph {et~al.}(2000)\citenamefont {Ozawa},
  \citenamefont {Kobayashi}, \citenamefont {Suzuki}, \citenamefont {Yoshida},\
  and\ \citenamefont {Tanihata}}]{ozawa2000}%
  \BibitemOpen
  \bibfield  {author} {\bibinfo {author} {\bibnamefont {Ozawa}, \bibfnamefont
  {A.}}, \bibinfo {author} {\bibfnamefont {T.}~\bibnamefont {Kobayashi}},
  \bibinfo {author} {\bibfnamefont {T.}~\bibnamefont {Suzuki}}, \bibinfo
  {author} {\bibfnamefont {K.}~\bibnamefont {Yoshida}}, \ and\ \bibinfo
  {author} {\bibfnamefont {I.}~\bibnamefont {Tanihata}}} (\bibinfo {year}
  {2000}),\ \href {\doibase 10.1103/PhysRevLett.84.5493} {\bibfield  {journal}
  {\bibinfo  {journal} {Phys. Rev. Lett.}\ }\textbf {\bibinfo {volume} {84}},\
  \bibinfo {pages} {5493}}\BibitemShut {NoStop}%
\bibitem [{\citenamefont {Pandya}(1956)}]{pandya1956}%
  \BibitemOpen
  \bibfield  {author} {\bibinfo {author} {\bibnamefont {Pandya}, \bibfnamefont
  {S.~P.}}} (\bibinfo {year} {1956}),\ \href {\doibase 10.1103/PhysRev.103.956}
  {\bibfield  {journal} {\bibinfo  {journal} {Phys. Rev.}\ }\textbf {\bibinfo
  {volume} {103}},\ \bibinfo {pages} {956}}\BibitemShut {NoStop}%
\bibitem [{\citenamefont {Papadimitriou}\ \emph {et~al.}(2013)\citenamefont
  {Papadimitriou}, \citenamefont {Rotureau}, \citenamefont {Michel},
  \citenamefont {P\l{}oszajczak},\ and\ \citenamefont
  {Barrett}}]{papadimitriou2013}%
  \BibitemOpen
  \bibfield  {author} {\bibinfo {author} {\bibnamefont {Papadimitriou},
  \bibfnamefont {G.}}, \bibinfo {author} {\bibfnamefont {J.}~\bibnamefont
  {Rotureau}}, \bibinfo {author} {\bibfnamefont {N.}~\bibnamefont {Michel}},
  \bibinfo {author} {\bibfnamefont {M.}~\bibnamefont {P\l{}oszajczak}}, \ and\
  \bibinfo {author} {\bibfnamefont {B.~R.}\ \bibnamefont {Barrett}}} (\bibinfo
  {year} {2013}),\ \href {\doibase 10.1103/PhysRevC.88.044318} {\bibfield
  {journal} {\bibinfo  {journal} {Phys. Rev. C}\ }\textbf {\bibinfo {volume}
  {88}},\ \bibinfo {pages} {044318}}\BibitemShut {NoStop}%
\bibitem [{\citenamefont {Pedersen~Lohne}\ \emph {et~al.}(2011)\citenamefont
  {Pedersen~Lohne}, \citenamefont {Hagen}, \citenamefont {Hjorth-Jensen},
  \citenamefont {Kvaal},\ and\ \citenamefont {Pederiva}}]{pedersen2011}%
  \BibitemOpen
  \bibfield  {author} {\bibinfo {author} {\bibnamefont {Pedersen~Lohne},
  \bibfnamefont {M.}}, \bibinfo {author} {\bibfnamefont {G.}~\bibnamefont
  {Hagen}}, \bibinfo {author} {\bibfnamefont {M.}~\bibnamefont
  {Hjorth-Jensen}}, \bibinfo {author} {\bibfnamefont {S.}~\bibnamefont
  {Kvaal}}, \ and\ \bibinfo {author} {\bibfnamefont {F.}~\bibnamefont
  {Pederiva}}} (\bibinfo {year} {2011}),\ \href {\doibase
  10.1103/PhysRevB.84.115302} {\bibfield  {journal} {\bibinfo  {journal} {Phys.
  Rev. B}\ }\textbf {\bibinfo {volume} {84}},\ \bibinfo {pages}
  {115302}}\BibitemShut {NoStop}%
\bibitem [{\citenamefont {Piecuch}\ \emph {et~al.}(2002)\citenamefont
  {Piecuch}, \citenamefont {Kowalski}, \citenamefont {Pimienta},\ and\
  \citenamefont {Mcguire}}]{piecuch2002}%
  \BibitemOpen
  \bibfield  {author} {\bibinfo {author} {\bibnamefont {Piecuch}, \bibfnamefont
  {P.}}, \bibinfo {author} {\bibfnamefont {K.}~\bibnamefont {Kowalski}},
  \bibinfo {author} {\bibfnamefont {I.~S.~O.}\ \bibnamefont {Pimienta}}, \ and\
  \bibinfo {author} {\bibfnamefont {M.~J.}\ \bibnamefont {Mcguire}}} (\bibinfo
  {year} {2002}),\ \href {\doibase 10.1080/0144235021000053811} {\bibfield
  {journal} {\bibinfo  {journal} {International Reviews in Physical Chemistry}\
  }\textbf {\bibinfo {volume} {21}}~(\bibinfo {number} {4}),\ \bibinfo {pages}
  {527}}\BibitemShut {NoStop}%
\bibitem [{\citenamefont {Pieper}\ and\ \citenamefont
  {Wiringa}(2001)}]{pieper2001}%
  \BibitemOpen
  \bibfield  {author} {\bibinfo {author} {\bibnamefont {Pieper}, \bibfnamefont
  {S.~C.}}, \ and\ \bibinfo {author} {\bibfnamefont {R.~B.}\ \bibnamefont
  {Wiringa}}} (\bibinfo {year} {2001}),\ \href {\doibase
  10.1146/annurev.nucl.51.101701.132506} {\bibfield  {journal} {\bibinfo
  {journal} {Annual Review of Nuclear and Particle Science}\ }\textbf {\bibinfo
  {volume} {51}}~(\bibinfo {number} {1}),\ \bibinfo {pages} {53}}\BibitemShut
  {NoStop}%
\bibitem [{\citenamefont {Pigg}\ \emph {et~al.}(2012)\citenamefont {Pigg},
  \citenamefont {Hagen}, \citenamefont {Nam},\ and\ \citenamefont
  {Papenbrock}}]{pigg2012}%
  \BibitemOpen
  \bibfield  {author} {\bibinfo {author} {\bibnamefont {Pigg}, \bibfnamefont
  {D.~A.}}, \bibinfo {author} {\bibfnamefont {G.}~\bibnamefont {Hagen}},
  \bibinfo {author} {\bibfnamefont {H.}~\bibnamefont {Nam}}, \ and\ \bibinfo
  {author} {\bibfnamefont {T.}~\bibnamefont {Papenbrock}}} (\bibinfo {year}
  {2012}),\ \href {\doibase 10.1103/PhysRevC.86.014308} {\bibfield  {journal}
  {\bibinfo  {journal} {Phys. Rev. C}\ }\textbf {\bibinfo {volume} {86}},\
  \bibinfo {pages} {014308}}\BibitemShut {NoStop}%
\bibitem [{\citenamefont {Polyzou}\ and\ \citenamefont
  {Gl{\"o}ckle}(1990)}]{polyzou1990}%
  \BibitemOpen
  \bibfield  {author} {\bibinfo {author} {\bibnamefont {Polyzou}, \bibfnamefont
  {W.}}, \ and\ \bibinfo {author} {\bibfnamefont {W.}~\bibnamefont
  {Gl{\"o}ckle}}} (\bibinfo {year} {1990}),\ \href {\doibase
  10.1007/BF01091701} {\bibfield  {journal} {\bibinfo  {journal} {Few-Body
  Systems}\ }\textbf {\bibinfo {volume} {9}}~(\bibinfo {number} {2-3}),\
  \bibinfo {pages} {97}}\BibitemShut {NoStop}%
\bibitem [{\citenamefont {Poves}\ and\ \citenamefont
  {Zuker}(1981)}]{poves1981}%
  \BibitemOpen
  \bibfield  {author} {\bibinfo {author} {\bibnamefont {Poves}, \bibfnamefont
  {A.}}, \ and\ \bibinfo {author} {\bibfnamefont {A.}~\bibnamefont {Zuker}}}
  (\bibinfo {year} {1981}),\ \href {\doibase 10.1016/0370-1573(81)90153-8}
  {\bibfield  {journal} {\bibinfo  {journal} {Physics Reports}\ }\textbf
  {\bibinfo {volume} {70}}~(\bibinfo {number} {4}),\ \bibinfo {pages} {235
  }}\BibitemShut {NoStop}%
\bibitem [{\citenamefont {Prisciandaro}\ \emph {et~al.}(2001)\citenamefont
  {Prisciandaro}, \citenamefont {Mantica}, \citenamefont {Brown}, \citenamefont
  {Anthony}, \citenamefont {Cooper}, \citenamefont {Garcia}, \citenamefont
  {Groh}, \citenamefont {Komives}, \citenamefont {Kumarasiri}, \citenamefont
  {Lofy}, \citenamefont {Oros-Peusquens}, \citenamefont {Tabor},\ and\
  \citenamefont {Wiedeking}}]{prisciandaro2001}%
  \BibitemOpen
  \bibfield  {author} {\bibinfo {author} {\bibnamefont {Prisciandaro},
  \bibfnamefont {J.}}, \bibinfo {author} {\bibfnamefont {P.}~\bibnamefont
  {Mantica}}, \bibinfo {author} {\bibfnamefont {B.}~\bibnamefont {Brown}},
  \bibinfo {author} {\bibfnamefont {D.}~\bibnamefont {Anthony}}, \bibinfo
  {author} {\bibfnamefont {M.}~\bibnamefont {Cooper}}, \bibinfo {author}
  {\bibfnamefont {A.}~\bibnamefont {Garcia}}, \bibinfo {author} {\bibfnamefont
  {D.}~\bibnamefont {Groh}}, \bibinfo {author} {\bibfnamefont {A.}~\bibnamefont
  {Komives}}, \bibinfo {author} {\bibfnamefont {W.}~\bibnamefont {Kumarasiri}},
  \bibinfo {author} {\bibfnamefont {P.}~\bibnamefont {Lofy}}, \bibinfo {author}
  {\bibfnamefont {A.}~\bibnamefont {Oros-Peusquens}}, \bibinfo {author}
  {\bibfnamefont {S.}~\bibnamefont {Tabor}}, \ and\ \bibinfo {author}
  {\bibfnamefont {M.}~\bibnamefont {Wiedeking}}} (\bibinfo {year} {2001}),\
  \href {\doibase 10.1016/S0370-2693(01)00565-2} {\bibfield  {journal}
  {\bibinfo  {journal} {Physics Letters B}\ }\textbf {\bibinfo {volume}
  {510}}~(\bibinfo {number} {14}),\ \bibinfo {pages} {17 }}\BibitemShut
  {NoStop}%
\bibitem [{\citenamefont {Pudliner}\ \emph {et~al.}(1997)\citenamefont
  {Pudliner}, \citenamefont {Pandharipande}, \citenamefont {Carlson},
  \citenamefont {Pieper},\ and\ \citenamefont {Wiringa}}]{pudliner1997}%
  \BibitemOpen
  \bibfield  {author} {\bibinfo {author} {\bibnamefont {Pudliner},
  \bibfnamefont {B.~S.}}, \bibinfo {author} {\bibfnamefont {V.~R.}\
  \bibnamefont {Pandharipande}}, \bibinfo {author} {\bibfnamefont
  {J.}~\bibnamefont {Carlson}}, \bibinfo {author} {\bibfnamefont {S.~C.}\
  \bibnamefont {Pieper}}, \ and\ \bibinfo {author} {\bibfnamefont {R.~B.}\
  \bibnamefont {Wiringa}}} (\bibinfo {year} {1997}),\ \href {\doibase
  10.1103/PhysRevC.56.1720} {\bibfield  {journal} {\bibinfo  {journal} {Phys.
  Rev. C}\ }\textbf {\bibinfo {volume} {56}},\ \bibinfo {pages}
  {1720}}\BibitemShut {NoStop}%
\bibitem [{\citenamefont {Pulay}(1980)}]{pulay1980}%
  \BibitemOpen
  \bibfield  {author} {\bibinfo {author} {\bibnamefont {Pulay}, \bibfnamefont
  {P.}}} (\bibinfo {year} {1980}),\ \href {\doibase
  10.1016/0009-2614(80)80396-4} {\bibfield  {journal} {\bibinfo  {journal}
  {Chemical Physics Letters}\ }\textbf {\bibinfo {volume} {73}}~(\bibinfo
  {number} {2}),\ \bibinfo {pages} {393 }}\BibitemShut {NoStop}%
\bibitem [{\citenamefont {Quaglioni}\ and\ \citenamefont
  {Navr\'atil}(2008)}]{quaglioni2008}%
  \BibitemOpen
  \bibfield  {author} {\bibinfo {author} {\bibnamefont {Quaglioni},
  \bibfnamefont {S.}}, \ and\ \bibinfo {author} {\bibfnamefont
  {P.}~\bibnamefont {Navr\'atil}}} (\bibinfo {year} {2008}),\ \href {\doibase
  10.1103/PhysRevLett.101.092501} {\bibfield  {journal} {\bibinfo  {journal}
  {Phys. Rev. Lett.}\ }\textbf {\bibinfo {volume} {101}},\ \bibinfo {pages}
  {092501}}\BibitemShut {NoStop}%
\bibitem [{\citenamefont {Raghavachari}\ \emph {et~al.}(1989)\citenamefont
  {Raghavachari}, \citenamefont {Trucks}, \citenamefont {Pople},\ and\
  \citenamefont {Head-Gordon}}]{raghavachari1989}%
  \BibitemOpen
  \bibfield  {author} {\bibinfo {author} {\bibnamefont {Raghavachari},
  \bibfnamefont {K.}}, \bibinfo {author} {\bibfnamefont {G.~W.}\ \bibnamefont
  {Trucks}}, \bibinfo {author} {\bibfnamefont {J.~A.}\ \bibnamefont {Pople}}, \
  and\ \bibinfo {author} {\bibfnamefont {M.}~\bibnamefont {Head-Gordon}}}
  (\bibinfo {year} {1989}),\ \href {\doibase 10.1016/S0009-2614(89)87395-6}
  {\bibfield  {journal} {\bibinfo  {journal} {Chemical Physics Letters}\
  }\textbf {\bibinfo {volume} {157}}~(\bibinfo {number} {6}),\ \bibinfo {pages}
  {479 }}\BibitemShut {NoStop}%
\bibitem [{\citenamefont {Reimann}\ \emph {et~al.}(2013)\citenamefont
  {Reimann}, \citenamefont {H{\o}gberget}, \citenamefont {Bogner},\ and\
  \citenamefont {Hjorth-Jensen}}]{reimann2013}%
  \BibitemOpen
  \bibfield  {author} {\bibinfo {author} {\bibnamefont {Reimann}, \bibfnamefont
  {S.}}, \bibinfo {author} {\bibfnamefont {J.}~\bibnamefont {H{\o}gberget}},
  \bibinfo {author} {\bibfnamefont {S.~K.}\ \bibnamefont {Bogner}}, \ and\
  \bibinfo {author} {\bibfnamefont {M.}~\bibnamefont {Hjorth-Jensen}}}
  (\bibinfo {year} {2013}),\ \href@noop {} {\bibinfo  {journal} {in
  preparation}\ }\BibitemShut {NoStop}%
\bibitem [{\citenamefont {Reimann}\ and\ \citenamefont
  {Manninen}(2002)}]{reimann2002}%
  \BibitemOpen
\bibfield  {journal} {  }\bibfield  {author} {\bibinfo {author} {\bibnamefont
  {Reimann}, \bibfnamefont {S.~M.}}, \ and\ \bibinfo {author} {\bibfnamefont
  {M.}~\bibnamefont {Manninen}}} (\bibinfo {year} {2002}),\ \href {\doibase
  10.1103/RevModPhys.74.1283} {\bibfield  {journal} {\bibinfo  {journal} {Rev.
  Mod. Phys.}\ }\textbf {\bibinfo {volume} {74}},\ \bibinfo {pages}
  {1283}}\BibitemShut {NoStop}%
\bibitem [{\citenamefont {Rejmund}\ \emph {et~al.}(2007)\citenamefont
  {Rejmund}, \citenamefont {Bhattacharyya}, \citenamefont {Navin},
  \citenamefont {Mittig}, \citenamefont {Gaudefroy}, \citenamefont {Gelin},
  \citenamefont {Mukherjee}, \citenamefont {Rejmund}, \citenamefont
  {Roussel-Chomaz},\ and\ \citenamefont {Theisen}}]{rejmund2007}%
  \BibitemOpen
  \bibfield  {author} {\bibinfo {author} {\bibnamefont {Rejmund}, \bibfnamefont
  {M.}}, \bibinfo {author} {\bibfnamefont {S.}~\bibnamefont {Bhattacharyya}},
  \bibinfo {author} {\bibfnamefont {A.}~\bibnamefont {Navin}}, \bibinfo
  {author} {\bibfnamefont {W.}~\bibnamefont {Mittig}}, \bibinfo {author}
  {\bibfnamefont {L.}~\bibnamefont {Gaudefroy}}, \bibinfo {author}
  {\bibfnamefont {M.}~\bibnamefont {Gelin}}, \bibinfo {author} {\bibfnamefont
  {G.}~\bibnamefont {Mukherjee}}, \bibinfo {author} {\bibfnamefont
  {F.}~\bibnamefont {Rejmund}}, \bibinfo {author} {\bibfnamefont
  {P.}~\bibnamefont {Roussel-Chomaz}}, \ and\ \bibinfo {author} {\bibfnamefont
  {C.}~\bibnamefont {Theisen}}} (\bibinfo {year} {2007}),\ \href {\doibase
  10.1103/PhysRevC.76.021304} {\bibfield  {journal} {\bibinfo  {journal} {Phys.
  Rev. C}\ }\textbf {\bibinfo {volume} {76}},\ \bibinfo {pages}
  {021304}}\BibitemShut {NoStop}%
\bibitem [{\citenamefont {da~Rocha}\ and\ \citenamefont
  {Robilotta}(1994)}]{darocha1994}%
  \BibitemOpen
  \bibfield  {author} {\bibinfo {author} {\bibnamefont {da~Rocha},
  \bibfnamefont {C.~A.}}, \ and\ \bibinfo {author} {\bibfnamefont {M.~R.}\
  \bibnamefont {Robilotta}}} (\bibinfo {year} {1994}),\ \href {\doibase
  10.1103/PhysRevC.49.1818} {\bibfield  {journal} {\bibinfo  {journal} {Phys.
  Rev. C}\ }\textbf {\bibinfo {volume} {49}},\ \bibinfo {pages}
  {1818}}\BibitemShut {NoStop}%
\bibitem [{\citenamefont {da~Rocha}\ and\ \citenamefont
  {Robilotta}(1995)}]{darocha1995}%
  \BibitemOpen
  \bibfield  {author} {\bibinfo {author} {\bibnamefont {da~Rocha},
  \bibfnamefont {C.~A.}}, \ and\ \bibinfo {author} {\bibfnamefont {M.~R.}\
  \bibnamefont {Robilotta}}} (\bibinfo {year} {1995}),\ \href {\doibase
  10.1103/PhysRevC.52.531} {\bibfield  {journal} {\bibinfo  {journal} {Phys.
  Rev. C}\ }\textbf {\bibinfo {volume} {52}},\ \bibinfo {pages}
  {531}}\BibitemShut {NoStop}%
\bibitem [{\citenamefont {Rodr\'iguez}\ and\ \citenamefont
  {Egido}(2007)}]{rodriguez2007}%
  \BibitemOpen
  \bibfield  {author} {\bibinfo {author} {\bibnamefont {Rodr\'iguez},
  \bibfnamefont {T.~R.}}, \ and\ \bibinfo {author} {\bibfnamefont {J.~L.}\
  \bibnamefont {Egido}}} (\bibinfo {year} {2007}),\ \href {\doibase
  10.1103/PhysRevLett.99.062501} {\bibfield  {journal} {\bibinfo  {journal}
  {Phys. Rev. Lett.}\ }\textbf {\bibinfo {volume} {99}},\ \bibinfo {pages}
  {062501}}\BibitemShut {NoStop}%
\bibitem [{\citenamefont {Roth}\ \emph {et~al.}(2012)\citenamefont {Roth},
  \citenamefont {Binder}, \citenamefont {Vobig}, \citenamefont {Calci},
  \citenamefont {Langhammer},\ and\ \citenamefont {Navr\'atil}}]{roth2012}%
  \BibitemOpen
  \bibfield  {author} {\bibinfo {author} {\bibnamefont {Roth}, \bibfnamefont
  {R.}}, \bibinfo {author} {\bibfnamefont {S.}~\bibnamefont {Binder}}, \bibinfo
  {author} {\bibfnamefont {K.}~\bibnamefont {Vobig}}, \bibinfo {author}
  {\bibfnamefont {A.}~\bibnamefont {Calci}}, \bibinfo {author} {\bibfnamefont
  {J.}~\bibnamefont {Langhammer}}, \ and\ \bibinfo {author} {\bibfnamefont
  {P.}~\bibnamefont {Navr\'atil}}} (\bibinfo {year} {2012}),\ \href {\doibase
  10.1103/PhysRevLett.109.052501} {\bibfield  {journal} {\bibinfo  {journal}
  {Phys. Rev. Lett.}\ }\textbf {\bibinfo {volume} {109}},\ \bibinfo {pages}
  {052501}}\BibitemShut {NoStop}%
\bibitem [{\citenamefont {Roth}\ \emph
  {et~al.}(2009{\natexlab{a}})\citenamefont {Roth}, \citenamefont {Gour},\ and\
  \citenamefont {Piecuch}}]{roth2009b}%
  \BibitemOpen
  \bibfield  {author} {\bibinfo {author} {\bibnamefont {Roth}, \bibfnamefont
  {R.}}, \bibinfo {author} {\bibfnamefont {J.~R.}\ \bibnamefont {Gour}}, \ and\
  \bibinfo {author} {\bibfnamefont {P.}~\bibnamefont {Piecuch}}} (\bibinfo
  {year} {2009}{\natexlab{a}}),\ \href {\doibase
  10.1016/j.physletb.2009.07.071} {\bibfield  {journal} {\bibinfo  {journal}
  {Physics Letters B}\ }\textbf {\bibinfo {volume} {679}}~(\bibinfo {number}
  {4}),\ \bibinfo {pages} {334 }}\BibitemShut {NoStop}%
\bibitem [{\citenamefont {Roth}\ \emph
  {et~al.}(2009{\natexlab{b}})\citenamefont {Roth}, \citenamefont {Gour},\ and\
  \citenamefont {Piecuch}}]{roth2009}%
  \BibitemOpen
  \bibfield  {author} {\bibinfo {author} {\bibnamefont {Roth}, \bibfnamefont
  {R.}}, \bibinfo {author} {\bibfnamefont {J.~R.}\ \bibnamefont {Gour}}, \ and\
  \bibinfo {author} {\bibfnamefont {P.}~\bibnamefont {Piecuch}}} (\bibinfo
  {year} {2009}{\natexlab{b}}),\ \href {\doibase 10.1103/PhysRevC.79.054325}
  {\bibfield  {journal} {\bibinfo  {journal} {Phys. Rev. C}\ }\textbf {\bibinfo
  {volume} {79}},\ \bibinfo {pages} {054325}}\BibitemShut {NoStop}%
\bibitem [{\citenamefont {Roth}\ \emph {et~al.}(2011)\citenamefont {Roth},
  \citenamefont {Langhammer}, \citenamefont {Calci}, \citenamefont {Binder},\
  and\ \citenamefont {Navr\'atil}}]{roth2011a}%
  \BibitemOpen
  \bibfield  {author} {\bibinfo {author} {\bibnamefont {Roth}, \bibfnamefont
  {R.}}, \bibinfo {author} {\bibfnamefont {J.}~\bibnamefont {Langhammer}},
  \bibinfo {author} {\bibfnamefont {A.}~\bibnamefont {Calci}}, \bibinfo
  {author} {\bibfnamefont {S.}~\bibnamefont {Binder}}, \ and\ \bibinfo {author}
  {\bibfnamefont {P.}~\bibnamefont {Navr\'atil}}} (\bibinfo {year} {2011}),\
  \href {\doibase 10.1103/PhysRevLett.107.072501} {\bibfield  {journal}
  {\bibinfo  {journal} {Phys. Rev. Lett.}\ }\textbf {\bibinfo {volume} {107}},\
  \bibinfo {pages} {072501}}\BibitemShut {NoStop}%
\bibitem [{\citenamefont {Roth}\ \emph {et~al.}(2010)\citenamefont {Roth},
  \citenamefont {Neff},\ and\ \citenamefont {Feldmeier}}]{roth2010}%
  \BibitemOpen
  \bibfield  {author} {\bibinfo {author} {\bibnamefont {Roth}, \bibfnamefont
  {R.}}, \bibinfo {author} {\bibfnamefont {T.}~\bibnamefont {Neff}}, \ and\
  \bibinfo {author} {\bibfnamefont {H.}~\bibnamefont {Feldmeier}}} (\bibinfo
  {year} {2010}),\ \href {\doibase 10.1016/j.ppnp.2010.02.003} {\bibfield
  {journal} {\bibinfo  {journal} {Progress in Particle and Nuclear Physics}\
  }\textbf {\bibinfo {volume} {65}}~(\bibinfo {number} {1}),\ \bibinfo {pages}
  {50 }}\BibitemShut {NoStop}%
\bibitem [{\citenamefont {Sakurai}\ \emph {et~al.}(1999)\citenamefont
  {Sakurai}, \citenamefont {Lukyanov}, \citenamefont {Notani}, \citenamefont
  {Aoi}, \citenamefont {Beaumel}, \citenamefont {Fukuda}, \citenamefont
  {Hirai}, \citenamefont {Ideguchi}, \citenamefont {Imai}, \citenamefont
  {Ishihara}, \citenamefont {Iwasaki}, \citenamefont {Kubo}, \citenamefont
  {Kusaka}, \citenamefont {Kumagai}, \citenamefont {Nakamura}, \citenamefont
  {Ogawa}, \citenamefont {Penionzhkevich}, \citenamefont {Teranishi},
  \citenamefont {Watanabe}, \citenamefont {Yoneda},\ and\ \citenamefont
  {Yoshida}}]{sakurai1999}%
  \BibitemOpen
  \bibfield  {author} {\bibinfo {author} {\bibnamefont {Sakurai}, \bibfnamefont
  {H.}}, \bibinfo {author} {\bibfnamefont {S.~M.}\ \bibnamefont {Lukyanov}},
  \bibinfo {author} {\bibfnamefont {M.}~\bibnamefont {Notani}}, \bibinfo
  {author} {\bibfnamefont {N.}~\bibnamefont {Aoi}}, \bibinfo {author}
  {\bibfnamefont {D.}~\bibnamefont {Beaumel}}, \bibinfo {author} {\bibfnamefont
  {N.}~\bibnamefont {Fukuda}}, \bibinfo {author} {\bibfnamefont
  {M.}~\bibnamefont {Hirai}}, \bibinfo {author} {\bibfnamefont
  {E.}~\bibnamefont {Ideguchi}}, \bibinfo {author} {\bibfnamefont
  {N.}~\bibnamefont {Imai}}, \bibinfo {author} {\bibfnamefont {M.}~\bibnamefont
  {Ishihara}}, \bibinfo {author} {\bibfnamefont {H.}~\bibnamefont {Iwasaki}},
  \bibinfo {author} {\bibfnamefont {T.}~\bibnamefont {Kubo}}, \bibinfo {author}
  {\bibfnamefont {K.}~\bibnamefont {Kusaka}}, \bibinfo {author} {\bibfnamefont
  {H.}~\bibnamefont {Kumagai}}, \bibinfo {author} {\bibfnamefont
  {T.}~\bibnamefont {Nakamura}}, \bibinfo {author} {\bibfnamefont
  {H.}~\bibnamefont {Ogawa}}, \bibinfo {author} {\bibfnamefont {Y.~E.}\
  \bibnamefont {Penionzhkevich}}, \bibinfo {author} {\bibfnamefont
  {T.}~\bibnamefont {Teranishi}}, \bibinfo {author} {\bibfnamefont {Y.~X.}\
  \bibnamefont {Watanabe}}, \bibinfo {author} {\bibfnamefont {K.}~\bibnamefont
  {Yoneda}}, \ and\ \bibinfo {author} {\bibfnamefont {A.}~\bibnamefont
  {Yoshida}}} (\bibinfo {year} {1999}),\ \href {\doibase
  10.1016/S0370-2693(99)00015-5} {\bibfield  {journal} {\bibinfo  {journal}
  {Physics Letters B}\ }\textbf {\bibinfo {volume} {448}}~(\bibinfo {number}
  {3}),\ \bibinfo {pages} {180 }}\BibitemShut {NoStop}%
\bibitem [{\citenamefont {Sammarruca}(2010)}]{sammarruca2010}%
  \BibitemOpen
  \bibfield  {author} {\bibinfo {author} {\bibnamefont {Sammarruca},
  \bibfnamefont {F.}}} (\bibinfo {year} {2010}),\ \href {\doibase
  10.1142/S0218301310015874} {\bibfield  {journal} {\bibinfo  {journal}
  {International Journal of Modern Physics E}\ }\textbf {\bibinfo {volume}
  {19}}~(\bibinfo {number} {07}),\ \bibinfo {pages} {1259}}\BibitemShut
  {NoStop}%
\bibitem [{\citenamefont {Sch\"onhammer}\ and\ \citenamefont
  {Gunnarsson}(1978)}]{SchGun78}%
  \BibitemOpen
  \bibfield  {author} {\bibinfo {author} {\bibnamefont {Sch\"onhammer},
  \bibfnamefont {K.}}, \ and\ \bibinfo {author} {\bibfnamefont
  {O.}~\bibnamefont {Gunnarsson}}} (\bibinfo {year} {1978}),\ \href {\doibase
  10.1103/PhysRevB.18.6606} {\bibfield  {journal} {\bibinfo  {journal} {Phys.
  Rev. B}\ }\textbf {\bibinfo {volume} {18}},\ \bibinfo {pages}
  {6606}}\BibitemShut {NoStop}%
\bibitem [{\citenamefont {Shavitt}\ and\ \citenamefont
  {Bartlett}(2009)}]{shavittbartlett2009}%
  \BibitemOpen
  \bibfield  {author} {\bibinfo {author} {\bibnamefont {Shavitt}, \bibfnamefont
  {I.}}, \ and\ \bibinfo {author} {\bibfnamefont {R.~J.}\ \bibnamefont
  {Bartlett}}} (\bibinfo {year} {2009}),\ \href@noop {} {\emph {\bibinfo
  {title} {Many-body Methods in Chemistry and Physics}}}\ (\bibinfo
  {publisher} {Cambridge University Press})\BibitemShut {NoStop}%
\bibitem [{\citenamefont {Shlomo}\ and\ \citenamefont
  {Bertsch}(1975)}]{shlomo1975}%
  \BibitemOpen
  \bibfield  {author} {\bibinfo {author} {\bibnamefont {Shlomo}, \bibfnamefont
  {S.}}, \ and\ \bibinfo {author} {\bibfnamefont {G.}~\bibnamefont {Bertsch}}}
  (\bibinfo {year} {1975}),\ \href {\doibase 10.1016/0375-9474(75)90292-4}
  {\bibfield  {journal} {\bibinfo  {journal} {Nuclear Physics A}\ }\textbf
  {\bibinfo {volume} {243}}~(\bibinfo {number} {3}),\ \bibinfo {pages} {507
  }}\BibitemShut {NoStop}%
\bibitem [{\citenamefont {Sieja}\ and\ \citenamefont
  {Nowacki}(2012)}]{sieja2012}%
  \BibitemOpen
  \bibfield  {author} {\bibinfo {author} {\bibnamefont {Sieja}, \bibfnamefont
  {K.}}, \ and\ \bibinfo {author} {\bibfnamefont {F.}~\bibnamefont {Nowacki}}}
  (\bibinfo {year} {2012}),\ \href {\doibase 10.1103/PhysRevC.85.051301}
  {\bibfield  {journal} {\bibinfo  {journal} {Phys. Rev. C}\ }\textbf {\bibinfo
  {volume} {85}},\ \bibinfo {pages} {051301}}\BibitemShut {NoStop}%
\bibitem [{\citenamefont {Som\`a}\ \emph {et~al.}(2013)\citenamefont {Som\`a},
  \citenamefont {Barbieri},\ and\ \citenamefont {Duguet}}]{soma2013}%
  \BibitemOpen
  \bibfield  {author} {\bibinfo {author} {\bibnamefont {Som\`a}, \bibfnamefont
  {V.}}, \bibinfo {author} {\bibfnamefont {C.}~\bibnamefont {Barbieri}}, \ and\
  \bibinfo {author} {\bibfnamefont {T.}~\bibnamefont {Duguet}}} (\bibinfo
  {year} {2013}),\ \href {\doibase 10.1103/PhysRevC.87.011303} {\bibfield
  {journal} {\bibinfo  {journal} {Phys. Rev. C}\ }\textbf {\bibinfo {volume}
  {87}},\ \bibinfo {pages} {011303}}\BibitemShut {NoStop}%
\bibitem [{\citenamefont {{Som{\`a}}}\ \emph {et~al.}(2013)\citenamefont
  {{Som{\`a}}}, \citenamefont {{Cipollone}}, \citenamefont {{Barbieri}},
  \citenamefont {{Navr{\'a}til}},\ and\ \citenamefont {{Duguet}}}]{soma2013b}%
  \BibitemOpen
  \bibfield  {author} {\bibinfo {author} {\bibnamefont {{Som{\`a}}},
  \bibfnamefont {V.}}, \bibinfo {author} {\bibfnamefont {A.}~\bibnamefont
  {{Cipollone}}}, \bibinfo {author} {\bibfnamefont {C.}~\bibnamefont
  {{Barbieri}}}, \bibinfo {author} {\bibfnamefont {P.}~\bibnamefont
  {{Navr{\'a}til}}}, \ and\ \bibinfo {author} {\bibfnamefont {T.}~\bibnamefont
  {{Duguet}}}} (\bibinfo {year} {2013}),\ \href@noop {} {\bibfield  {journal}
  {\bibinfo  {journal} {ArXiv e-prints}\ }}\Eprint
  {http://arxiv.org/abs/1312.2068} {arXiv:1312.2068 [nucl-th]} \BibitemShut
  {NoStop}%
\bibitem [{\citenamefont {Stanton}\ and\ \citenamefont
  {Bartlett}(1993)}]{stanton1993}%
  \BibitemOpen
  \bibfield  {author} {\bibinfo {author} {\bibnamefont {Stanton}, \bibfnamefont
  {J.~F.}}, \ and\ \bibinfo {author} {\bibfnamefont {R.~J.}\ \bibnamefont
  {Bartlett}}} (\bibinfo {year} {1993}),\ \href {\doibase 10.1063/1.464746}
  {\bibfield  {journal} {\bibinfo  {journal} {The Journal of Chemical Physics}\
  }\textbf {\bibinfo {volume} {98}}~(\bibinfo {number} {9}),\ \bibinfo {pages}
  {7029}}\BibitemShut {NoStop}%
\bibitem [{\citenamefont {Steppenbeck}\ \emph
  {et~al.}(2013{\natexlab{a}})\citenamefont {Steppenbeck}, \citenamefont
  {Takeuchi}, \citenamefont {Aoi}, \citenamefont {Doornenbal}, \citenamefont
  {Lee}, \citenamefont {Matsushita}, \citenamefont {Wang}, \citenamefont
  {Baba}, \citenamefont {Fukuda}, \citenamefont {Go}, \citenamefont {Honma},
  \citenamefont {Matsui}, \citenamefont {Michimasa}, \citenamefont
  {Motobayashi}, \citenamefont {Nishimura}, \citenamefont {Otsuka},
  \citenamefont {Sakurai}, \citenamefont {Shiga}, \citenamefont
  {S{\"o}derstr{\"o}m}, \citenamefont {Sumikama}, \citenamefont {Suzuki},
  \citenamefont {Taniuchi}, \citenamefont {Utsuno}, \citenamefont
  {Valiente-Dob{\'o}n},\ and\ \citenamefont {Yoneda}}]{steppenbeck2013}%
  \BibitemOpen
  \bibfield  {author} {\bibinfo {author} {\bibnamefont {Steppenbeck},
  \bibfnamefont {D.}}, \bibinfo {author} {\bibfnamefont {S.}~\bibnamefont
  {Takeuchi}}, \bibinfo {author} {\bibfnamefont {N.}~\bibnamefont {Aoi}},
  \bibinfo {author} {\bibfnamefont {P.}~\bibnamefont {Doornenbal}}, \bibinfo
  {author} {\bibfnamefont {J.}~\bibnamefont {Lee}}, \bibinfo {author}
  {\bibfnamefont {M.}~\bibnamefont {Matsushita}}, \bibinfo {author}
  {\bibfnamefont {H.}~\bibnamefont {Wang}}, \bibinfo {author} {\bibfnamefont
  {H.}~\bibnamefont {Baba}}, \bibinfo {author} {\bibfnamefont {N.}~\bibnamefont
  {Fukuda}}, \bibinfo {author} {\bibfnamefont {S.}~\bibnamefont {Go}}, \bibinfo
  {author} {\bibfnamefont {M.}~\bibnamefont {Honma}}, \bibinfo {author}
  {\bibfnamefont {K.}~\bibnamefont {Matsui}}, \bibinfo {author} {\bibfnamefont
  {S.}~\bibnamefont {Michimasa}}, \bibinfo {author} {\bibfnamefont
  {T.}~\bibnamefont {Motobayashi}}, \bibinfo {author} {\bibfnamefont
  {D.}~\bibnamefont {Nishimura}}, \bibinfo {author} {\bibfnamefont
  {T.}~\bibnamefont {Otsuka}}, \bibinfo {author} {\bibfnamefont
  {H.}~\bibnamefont {Sakurai}}, \bibinfo {author} {\bibfnamefont
  {Y.}~\bibnamefont {Shiga}}, \bibinfo {author} {\bibfnamefont {P.-A.}\
  \bibnamefont {S{\"o}derstr{\"o}m}}, \bibinfo {author} {\bibfnamefont
  {T.}~\bibnamefont {Sumikama}}, \bibinfo {author} {\bibfnamefont
  {H.}~\bibnamefont {Suzuki}}, \bibinfo {author} {\bibfnamefont
  {R.}~\bibnamefont {Taniuchi}}, \bibinfo {author} {\bibfnamefont
  {Y.}~\bibnamefont {Utsuno}}, \bibinfo {author} {\bibfnamefont {J.~J.}\
  \bibnamefont {Valiente-Dob{\'o}n}}, \ and\ \bibinfo {author} {\bibfnamefont
  {K.}~\bibnamefont {Yoneda}}} (\bibinfo {year} {2013}{\natexlab{a}}),\ \href
  {http://stacks.iop.org/1742-6596/445/i=1/a=012012} {\bibfield  {journal}
  {\bibinfo  {journal} {Journal of Physics: Conference Series}\ }\textbf
  {\bibinfo {volume} {445}}~(\bibinfo {number} {1}),\ \bibinfo {pages}
  {012012}}\BibitemShut {NoStop}%
\bibitem [{\citenamefont {Steppenbeck}\ \emph
  {et~al.}(2013{\natexlab{b}})\citenamefont {Steppenbeck}, \citenamefont
  {Takeuchi}, \citenamefont {Aoi}, \citenamefont {Doornenbal}, \citenamefont
  {Matsushita}, \citenamefont {Wang}, \citenamefont {Baba}, \citenamefont
  {Fukuda}, \citenamefont {Go}, \citenamefont {Honma}, \citenamefont {Lee},
  \citenamefont {Matsui}, \citenamefont {Michimasa}, \citenamefont
  {Motobayashi}, \citenamefont {Nishimura}, \citenamefont {Otsuka},
  \citenamefont {Sakurai}, \citenamefont {Shiga}, \citenamefont {Soderstrom},
  \citenamefont {Sumikama}, \citenamefont {Suzuki}, \citenamefont {Taniuchi},
  \citenamefont {Utsuno}, \citenamefont {Valiente-Dobon},\ and\ \citenamefont
  {Yoneda}}]{steppenbeck2013b}%
  \BibitemOpen
  \bibfield  {author} {\bibinfo {author} {\bibnamefont {Steppenbeck},
  \bibfnamefont {D.}}, \bibinfo {author} {\bibfnamefont {S.}~\bibnamefont
  {Takeuchi}}, \bibinfo {author} {\bibfnamefont {N.}~\bibnamefont {Aoi}},
  \bibinfo {author} {\bibfnamefont {P.}~\bibnamefont {Doornenbal}}, \bibinfo
  {author} {\bibfnamefont {M.}~\bibnamefont {Matsushita}}, \bibinfo {author}
  {\bibfnamefont {H.}~\bibnamefont {Wang}}, \bibinfo {author} {\bibfnamefont
  {H.}~\bibnamefont {Baba}}, \bibinfo {author} {\bibfnamefont {N.}~\bibnamefont
  {Fukuda}}, \bibinfo {author} {\bibfnamefont {S.}~\bibnamefont {Go}}, \bibinfo
  {author} {\bibfnamefont {M.}~\bibnamefont {Honma}}, \bibinfo {author}
  {\bibfnamefont {J.}~\bibnamefont {Lee}}, \bibinfo {author} {\bibfnamefont
  {K.}~\bibnamefont {Matsui}}, \bibinfo {author} {\bibfnamefont
  {S.}~\bibnamefont {Michimasa}}, \bibinfo {author} {\bibfnamefont
  {T.}~\bibnamefont {Motobayashi}}, \bibinfo {author} {\bibfnamefont
  {D.}~\bibnamefont {Nishimura}}, \bibinfo {author} {\bibfnamefont
  {T.}~\bibnamefont {Otsuka}}, \bibinfo {author} {\bibfnamefont
  {H.}~\bibnamefont {Sakurai}}, \bibinfo {author} {\bibfnamefont
  {Y.}~\bibnamefont {Shiga}}, \bibinfo {author} {\bibfnamefont {P.-A.}\
  \bibnamefont {Soderstrom}}, \bibinfo {author} {\bibfnamefont
  {T.}~\bibnamefont {Sumikama}}, \bibinfo {author} {\bibfnamefont
  {H.}~\bibnamefont {Suzuki}}, \bibinfo {author} {\bibfnamefont
  {R.}~\bibnamefont {Taniuchi}}, \bibinfo {author} {\bibfnamefont
  {Y.}~\bibnamefont {Utsuno}}, \bibinfo {author} {\bibfnamefont {J.~J.}\
  \bibnamefont {Valiente-Dobon}}, \ and\ \bibinfo {author} {\bibfnamefont
  {K.}~\bibnamefont {Yoneda}}} (\bibinfo {year} {2013}{\natexlab{b}}),\ \href
  {http://dx.doi.org/10.1038/nature12522} {\bibfield  {journal} {\bibinfo
  {journal} {Nature}\ }\textbf {\bibinfo {volume} {502}}~(\bibinfo {number}
  {7470}),\ \bibinfo {pages} {207}}\BibitemShut {NoStop}%
\bibitem [{\citenamefont {Stetcu}\ \emph {et~al.}(2007)\citenamefont {Stetcu},
  \citenamefont {Barrett},\ and\ \citenamefont {van Kolck}}]{stetcu2007}%
  \BibitemOpen
  \bibfield  {author} {\bibinfo {author} {\bibnamefont {Stetcu}, \bibfnamefont
  {I.}}, \bibinfo {author} {\bibfnamefont {B.}~\bibnamefont {Barrett}}, \ and\
  \bibinfo {author} {\bibfnamefont {U.}~\bibnamefont {van Kolck}}} (\bibinfo
  {year} {2007}),\ \href {\doibase 10.1016/j.physletb.2007.07.065} {\bibfield
  {journal} {\bibinfo  {journal} {Physics Letters B}\ }\textbf {\bibinfo
  {volume} {653}}~(\bibinfo {number} {2–4}),\ \bibinfo {pages} {358
  }}\BibitemShut {NoStop}%
\bibitem [{\citenamefont {Suzuki}(1992)}]{suzuki1992}%
  \BibitemOpen
  \bibfield  {author} {\bibinfo {author} {\bibnamefont {Suzuki}, \bibfnamefont
  {K.}}} (\bibinfo {year} {1992}),\ \href {\doibase 10.1143/PTP.87.937}
  {\bibfield  {journal} {\bibinfo  {journal} {Progress of Theoretical Physics}\
  }\textbf {\bibinfo {volume} {87}}~(\bibinfo {number} {4}),\ \bibinfo {pages}
  {937}}\BibitemShut {NoStop}%
\bibitem [{\citenamefont {Suzuki}\ \emph {et~al.}(2000)\citenamefont {Suzuki},
  \citenamefont {Okamoto}, \citenamefont {Kohno},\ and\ \citenamefont
  {Nagata}}]{suzuki2000}%
  \BibitemOpen
  \bibfield  {author} {\bibinfo {author} {\bibnamefont {Suzuki}, \bibfnamefont
  {K.}}, \bibinfo {author} {\bibfnamefont {R.}~\bibnamefont {Okamoto}},
  \bibinfo {author} {\bibfnamefont {M.}~\bibnamefont {Kohno}}, \ and\ \bibinfo
  {author} {\bibfnamefont {S.}~\bibnamefont {Nagata}}} (\bibinfo {year}
  {2000}),\ \href {\doibase 10.1016/S0375-9474(99)00399-1} {\bibfield
  {journal} {\bibinfo  {journal} {Nuclear Physics A}\ }\textbf {\bibinfo
  {volume} {665}}~(\bibinfo {number} {1–2}),\ \bibinfo {pages} {92
  }}\BibitemShut {NoStop}%
\bibitem [{\citenamefont {Suzuki}\ \emph {et~al.}(1994)\citenamefont {Suzuki},
  \citenamefont {Okamoto},\ and\ \citenamefont {Kumagai}}]{suzuki1994}%
  \BibitemOpen
  \bibfield  {author} {\bibinfo {author} {\bibnamefont {Suzuki}, \bibfnamefont
  {K.}}, \bibinfo {author} {\bibfnamefont {R.}~\bibnamefont {Okamoto}}, \ and\
  \bibinfo {author} {\bibfnamefont {H.}~\bibnamefont {Kumagai}}} (\bibinfo
  {year} {1994}),\ \href {\doibase 10.1016/0375-9474(94)90770-6} {\bibfield
  {journal} {\bibinfo  {journal} {Nuclear Physics A}\ }\textbf {\bibinfo
  {volume} {580}}~(\bibinfo {number} {2}),\ \bibinfo {pages} {213
  }}\BibitemShut {NoStop}%
\bibitem [{\citenamefont {Szalay}\ \emph {et~al.}(1995)\citenamefont {Szalay},
  \citenamefont {Nooijen},\ and\ \citenamefont {Bartlett}}]{szalay1995}%
  \BibitemOpen
  \bibfield  {author} {\bibinfo {author} {\bibnamefont {Szalay}, \bibfnamefont
  {P.~G.}}, \bibinfo {author} {\bibfnamefont {M.}~\bibnamefont {Nooijen}}, \
  and\ \bibinfo {author} {\bibfnamefont {R.~J.}\ \bibnamefont {Bartlett}}}
  (\bibinfo {year} {1995}),\ \href {\doibase 10.1063/1.469641} {\bibfield
  {journal} {\bibinfo  {journal} {The Journal of Chemical Physics}\ }\textbf
  {\bibinfo {volume} {103}}~(\bibinfo {number} {1}),\ \bibinfo {pages}
  {281}}\BibitemShut {NoStop}%
\bibitem [{\citenamefont {Takahashi}\ and\ \citenamefont
  {Paldus}(1986)}]{TakPal86}%
  \BibitemOpen
  \bibfield  {author} {\bibinfo {author} {\bibnamefont {Takahashi},
  \bibfnamefont {M.}}, \ and\ \bibinfo {author} {\bibfnamefont
  {J.}~\bibnamefont {Paldus}}} (\bibinfo {year} {1986}),\ \href {\doibase
  10.1063/1.451241} {\bibfield  {journal} {\bibinfo  {journal} {The Journal of
  Chemical Physics}\ }\textbf {\bibinfo {volume} {85}}~(\bibinfo {number}
  {3}),\ \bibinfo {pages} {1486}}\BibitemShut {NoStop}%
\bibitem [{\citenamefont {Tanihata}\ \emph {et~al.}(2013)\citenamefont
  {Tanihata}, \citenamefont {Savajols},\ and\ \citenamefont
  {Kanungo}}]{tanihata2013}%
  \BibitemOpen
  \bibfield  {author} {\bibinfo {author} {\bibnamefont {Tanihata},
  \bibfnamefont {I.}}, \bibinfo {author} {\bibfnamefont {H.}~\bibnamefont
  {Savajols}}, \ and\ \bibinfo {author} {\bibfnamefont {R.}~\bibnamefont
  {Kanungo}}} (\bibinfo {year} {2013}),\ \href {\doibase
  10.1016/j.ppnp.2012.07.001} {\bibfield  {journal} {\bibinfo  {journal}
  {Progress in Particle and Nuclear Physics}\ }\textbf {\bibinfo {volume}
  {68}}~(\bibinfo {number} {0}),\ \bibinfo {pages} {215 }}\BibitemShut
  {NoStop}%
\bibitem [{\citenamefont {Tarasov}\ \emph {et~al.}(2009)\citenamefont
  {Tarasov}, \citenamefont {Morrissey}, \citenamefont {Amthor}, \citenamefont
  {Baumann}, \citenamefont {Bazin}, \citenamefont {Gade}, \citenamefont
  {Ginter}, \citenamefont {Hausmann}, \citenamefont {Inabe}, \citenamefont
  {Kubo}, \citenamefont {Nettleton}, \citenamefont {Pereira}, \citenamefont
  {Portillo}, \citenamefont {Sherrill}, \citenamefont {Stolz},\ and\
  \citenamefont {Thoennessen}}]{tarasov2009}%
  \BibitemOpen
  \bibfield  {author} {\bibinfo {author} {\bibnamefont {Tarasov}, \bibfnamefont
  {O.~B.}}, \bibinfo {author} {\bibfnamefont {D.~J.}\ \bibnamefont
  {Morrissey}}, \bibinfo {author} {\bibfnamefont {A.~M.}\ \bibnamefont
  {Amthor}}, \bibinfo {author} {\bibfnamefont {T.}~\bibnamefont {Baumann}},
  \bibinfo {author} {\bibfnamefont {D.}~\bibnamefont {Bazin}}, \bibinfo
  {author} {\bibfnamefont {A.}~\bibnamefont {Gade}}, \bibinfo {author}
  {\bibfnamefont {T.~N.}\ \bibnamefont {Ginter}}, \bibinfo {author}
  {\bibfnamefont {M.}~\bibnamefont {Hausmann}}, \bibinfo {author}
  {\bibfnamefont {N.}~\bibnamefont {Inabe}}, \bibinfo {author} {\bibfnamefont
  {T.}~\bibnamefont {Kubo}}, \bibinfo {author} {\bibfnamefont {A.}~\bibnamefont
  {Nettleton}}, \bibinfo {author} {\bibfnamefont {J.}~\bibnamefont {Pereira}},
  \bibinfo {author} {\bibfnamefont {M.}~\bibnamefont {Portillo}}, \bibinfo
  {author} {\bibfnamefont {B.~M.}\ \bibnamefont {Sherrill}}, \bibinfo {author}
  {\bibfnamefont {A.}~\bibnamefont {Stolz}}, \ and\ \bibinfo {author}
  {\bibfnamefont {M.}~\bibnamefont {Thoennessen}}} (\bibinfo {year} {2009}),\
  \href {\doibase 10.1103/PhysRevLett.102.142501} {\bibfield  {journal}
  {\bibinfo  {journal} {Phys. Rev. Lett.}\ }\textbf {\bibinfo {volume} {102}},\
  \bibinfo {pages} {142501}}\BibitemShut {NoStop}%
\bibitem [{\citenamefont {Taube}\ and\ \citenamefont
  {Bartlett}(2008)}]{taube2008}%
  \BibitemOpen
  \bibfield  {author} {\bibinfo {author} {\bibnamefont {Taube}, \bibfnamefont
  {A.~G.}}, \ and\ \bibinfo {author} {\bibfnamefont {R.~J.}\ \bibnamefont
  {Bartlett}}} (\bibinfo {year} {2008}),\ \href {\doibase 10.1063/1.2830236}
  {\bibfield  {journal} {\bibinfo  {journal} {The Journal of Chemical Physics}\
  }\textbf {\bibinfo {volume} {128}}~(\bibinfo {number} {4}),\ \bibinfo {eid}
  {044110}}\BibitemShut {NoStop}%
\bibitem [{\citenamefont {Thirolf}\ \emph {et~al.}(2000)\citenamefont
  {Thirolf}, \citenamefont {Pritychenko}, \citenamefont {Brown}, \citenamefont
  {Cottle}, \citenamefont {Chromik}, \citenamefont {Glasmacher}, \citenamefont
  {Hackman}, \citenamefont {Ibbotson}, \citenamefont {Kemper}, \citenamefont
  {Otsuka}, \citenamefont {Riley},\ and\ \citenamefont {Scheit}}]{thirolf2000}%
  \BibitemOpen
  \bibfield  {author} {\bibinfo {author} {\bibnamefont {Thirolf}, \bibfnamefont
  {P.~G.}}, \bibinfo {author} {\bibfnamefont {B.~V.}\ \bibnamefont
  {Pritychenko}}, \bibinfo {author} {\bibfnamefont {B.~A.}\ \bibnamefont
  {Brown}}, \bibinfo {author} {\bibfnamefont {P.~D.}\ \bibnamefont {Cottle}},
  \bibinfo {author} {\bibfnamefont {M.}~\bibnamefont {Chromik}}, \bibinfo
  {author} {\bibfnamefont {T.}~\bibnamefont {Glasmacher}}, \bibinfo {author}
  {\bibfnamefont {G.}~\bibnamefont {Hackman}}, \bibinfo {author} {\bibfnamefont
  {R.~W.}\ \bibnamefont {Ibbotson}}, \bibinfo {author} {\bibfnamefont {K.~W.}\
  \bibnamefont {Kemper}}, \bibinfo {author} {\bibfnamefont {T.}~\bibnamefont
  {Otsuka}}, \bibinfo {author} {\bibfnamefont {L.~A.}\ \bibnamefont {Riley}}, \
  and\ \bibinfo {author} {\bibfnamefont {H.}~\bibnamefont {Scheit}}} (\bibinfo
  {year} {2000}),\ \href {\doibase 10.1016/S0370-2693(00)00720-6} {\bibfield
  {journal} {\bibinfo  {journal} {Physics Letters B}\ }\textbf {\bibinfo
  {volume} {485}}~(\bibinfo {number} {1}),\ \bibinfo {pages} {16 }}\BibitemShut
  {NoStop}%
\bibitem [{\citenamefont {Tsang}\ \emph {et~al.}(2012)\citenamefont {Tsang},
  \citenamefont {Stone}, \citenamefont {Camera}, \citenamefont {Danielewicz},
  \citenamefont {Gandolfi}, \citenamefont {Hebeler}, \citenamefont {Horowitz},
  \citenamefont {Lee}, \citenamefont {Lynch}, \citenamefont {Kohley},
  \citenamefont {Lemmon}, \citenamefont {M\"oller}, \citenamefont {Murakami},
  \citenamefont {Riordan}, \citenamefont {Roca-Maza}, \citenamefont
  {Sammarruca}, \citenamefont {Steiner}, \citenamefont {Vida\~na},\ and\
  \citenamefont {Yennello}}]{tsang2012}%
  \BibitemOpen
  \bibfield  {author} {\bibinfo {author} {\bibnamefont {Tsang}, \bibfnamefont
  {M.~B.}}, \bibinfo {author} {\bibfnamefont {J.~R.}\ \bibnamefont {Stone}},
  \bibinfo {author} {\bibfnamefont {F.}~\bibnamefont {Camera}}, \bibinfo
  {author} {\bibfnamefont {P.}~\bibnamefont {Danielewicz}}, \bibinfo {author}
  {\bibfnamefont {S.}~\bibnamefont {Gandolfi}}, \bibinfo {author}
  {\bibfnamefont {K.}~\bibnamefont {Hebeler}}, \bibinfo {author} {\bibfnamefont
  {C.~J.}\ \bibnamefont {Horowitz}}, \bibinfo {author} {\bibfnamefont
  {J.}~\bibnamefont {Lee}}, \bibinfo {author} {\bibfnamefont {W.~G.}\
  \bibnamefont {Lynch}}, \bibinfo {author} {\bibfnamefont {Z.}~\bibnamefont
  {Kohley}}, \bibinfo {author} {\bibfnamefont {R.}~\bibnamefont {Lemmon}},
  \bibinfo {author} {\bibfnamefont {P.}~\bibnamefont {M\"oller}}, \bibinfo
  {author} {\bibfnamefont {T.}~\bibnamefont {Murakami}}, \bibinfo {author}
  {\bibfnamefont {S.}~\bibnamefont {Riordan}}, \bibinfo {author} {\bibfnamefont
  {X.}~\bibnamefont {Roca-Maza}}, \bibinfo {author} {\bibfnamefont
  {F.}~\bibnamefont {Sammarruca}}, \bibinfo {author} {\bibfnamefont {A.~W.}\
  \bibnamefont {Steiner}}, \bibinfo {author} {\bibfnamefont {I.}~\bibnamefont
  {Vida\~na}}, \ and\ \bibinfo {author} {\bibfnamefont {S.~J.}\ \bibnamefont
  {Yennello}}} (\bibinfo {year} {2012}),\ \href {\doibase
  10.1103/PhysRevC.86.015803} {\bibfield  {journal} {\bibinfo  {journal} {Phys.
  Rev. C}\ }\textbf {\bibinfo {volume} {86}},\ \bibinfo {pages}
  {015803}}\BibitemShut {NoStop}%
\bibitem [{\citenamefont {Tshoo}\ \emph {et~al.}(2012)\citenamefont {Tshoo},
  \citenamefont {Satou}, \citenamefont {Bhang}, \citenamefont {Choi},
  \citenamefont {Nakamura}, \citenamefont {Kondo}, \citenamefont {Deguchi},
  \citenamefont {Kawada}, \citenamefont {Kobayashi}, \citenamefont {Nakayama},
  \citenamefont {Tanaka}, \citenamefont {Tanaka}, \citenamefont {Aoi},
  \citenamefont {Ishihara}, \citenamefont {Motobayashi}, \citenamefont {Otsu},
  \citenamefont {Sakurai}, \citenamefont {Takeuchi}, \citenamefont {Togano},
  \citenamefont {Yoneda}, \citenamefont {Li}, \citenamefont {Delaunay},
  \citenamefont {Gibelin}, \citenamefont {Marqu\'es}, \citenamefont {Orr},
  \citenamefont {Honda}, \citenamefont {Matsushita}, \citenamefont {Kobayashi},
  \citenamefont {Miyashita}, \citenamefont {Sumikama}, \citenamefont
  {Yoshinaga}, \citenamefont {Shimoura}, \citenamefont {Sohler}, \citenamefont
  {Zheng},\ and\ \citenamefont {Cao}}]{tshoo2012}%
  \BibitemOpen
  \bibfield  {author} {\bibinfo {author} {\bibnamefont {Tshoo}, \bibfnamefont
  {K.}}, \bibinfo {author} {\bibfnamefont {Y.}~\bibnamefont {Satou}}, \bibinfo
  {author} {\bibfnamefont {H.}~\bibnamefont {Bhang}}, \bibinfo {author}
  {\bibfnamefont {S.}~\bibnamefont {Choi}}, \bibinfo {author} {\bibfnamefont
  {T.}~\bibnamefont {Nakamura}}, \bibinfo {author} {\bibfnamefont
  {Y.}~\bibnamefont {Kondo}}, \bibinfo {author} {\bibfnamefont
  {S.}~\bibnamefont {Deguchi}}, \bibinfo {author} {\bibfnamefont
  {Y.}~\bibnamefont {Kawada}}, \bibinfo {author} {\bibfnamefont
  {N.}~\bibnamefont {Kobayashi}}, \bibinfo {author} {\bibfnamefont
  {Y.}~\bibnamefont {Nakayama}}, \bibinfo {author} {\bibfnamefont {K.~N.}\
  \bibnamefont {Tanaka}}, \bibinfo {author} {\bibfnamefont {N.}~\bibnamefont
  {Tanaka}}, \bibinfo {author} {\bibfnamefont {N.}~\bibnamefont {Aoi}},
  \bibinfo {author} {\bibfnamefont {M.}~\bibnamefont {Ishihara}}, \bibinfo
  {author} {\bibfnamefont {T.}~\bibnamefont {Motobayashi}}, \bibinfo {author}
  {\bibfnamefont {H.}~\bibnamefont {Otsu}}, \bibinfo {author} {\bibfnamefont
  {H.}~\bibnamefont {Sakurai}}, \bibinfo {author} {\bibfnamefont
  {S.}~\bibnamefont {Takeuchi}}, \bibinfo {author} {\bibfnamefont
  {Y.}~\bibnamefont {Togano}}, \bibinfo {author} {\bibfnamefont
  {K.}~\bibnamefont {Yoneda}}, \bibinfo {author} {\bibfnamefont {Z.~H.}\
  \bibnamefont {Li}}, \bibinfo {author} {\bibfnamefont {F.}~\bibnamefont
  {Delaunay}}, \bibinfo {author} {\bibfnamefont {J.}~\bibnamefont {Gibelin}},
  \bibinfo {author} {\bibfnamefont {F.~M.}\ \bibnamefont {Marqu\'es}}, \bibinfo
  {author} {\bibfnamefont {N.~A.}\ \bibnamefont {Orr}}, \bibinfo {author}
  {\bibfnamefont {T.}~\bibnamefont {Honda}}, \bibinfo {author} {\bibfnamefont
  {M.}~\bibnamefont {Matsushita}}, \bibinfo {author} {\bibfnamefont
  {T.}~\bibnamefont {Kobayashi}}, \bibinfo {author} {\bibfnamefont
  {Y.}~\bibnamefont {Miyashita}}, \bibinfo {author} {\bibfnamefont
  {T.}~\bibnamefont {Sumikama}}, \bibinfo {author} {\bibfnamefont
  {K.}~\bibnamefont {Yoshinaga}}, \bibinfo {author} {\bibfnamefont
  {S.}~\bibnamefont {Shimoura}}, \bibinfo {author} {\bibfnamefont
  {D.}~\bibnamefont {Sohler}}, \bibinfo {author} {\bibfnamefont
  {T.}~\bibnamefont {Zheng}}, \ and\ \bibinfo {author} {\bibfnamefont {Z.~X.}\
  \bibnamefont {Cao}}} (\bibinfo {year} {2012}),\ \href {\doibase
  10.1103/PhysRevLett.109.022501} {\bibfield  {journal} {\bibinfo  {journal}
  {Phys. Rev. Lett.}\ }\textbf {\bibinfo {volume} {109}},\ \bibinfo {pages}
  {022501}}\BibitemShut {NoStop}%
\bibitem [{\citenamefont {Tsukiyama}\ \emph {et~al.}(2011)\citenamefont
  {Tsukiyama}, \citenamefont {Bogner},\ and\ \citenamefont
  {Schwenk}}]{tsukiyama2011}%
  \BibitemOpen
  \bibfield  {author} {\bibinfo {author} {\bibnamefont {Tsukiyama},
  \bibfnamefont {K.}}, \bibinfo {author} {\bibfnamefont {S.~K.}\ \bibnamefont
  {Bogner}}, \ and\ \bibinfo {author} {\bibfnamefont {A.}~\bibnamefont
  {Schwenk}}} (\bibinfo {year} {2011}),\ \href {\doibase
  10.1103/PhysRevLett.106.222502} {\bibfield  {journal} {\bibinfo  {journal}
  {Phys. Rev. Lett.}\ }\textbf {\bibinfo {volume} {106}},\ \bibinfo {pages}
  {222502}}\BibitemShut {NoStop}%
\bibitem [{\citenamefont {Tsukiyama}\ \emph {et~al.}(2012)\citenamefont
  {Tsukiyama}, \citenamefont {Bogner},\ and\ \citenamefont
  {Schwenk}}]{tsukiyama2012}%
  \BibitemOpen
  \bibfield  {author} {\bibinfo {author} {\bibnamefont {Tsukiyama},
  \bibfnamefont {K.}}, \bibinfo {author} {\bibfnamefont {S.~K.}\ \bibnamefont
  {Bogner}}, \ and\ \bibinfo {author} {\bibfnamefont {A.}~\bibnamefont
  {Schwenk}}} (\bibinfo {year} {2012}),\ \href {\doibase
  10.1103/PhysRevC.85.061304} {\bibfield  {journal} {\bibinfo  {journal} {Phys.
  Rev. C}\ }\textbf {\bibinfo {volume} {85}},\ \bibinfo {pages}
  {061304}}\BibitemShut {NoStop}%
\bibitem [{\citenamefont {Utsuno}\ \emph {et~al.}(2004)\citenamefont {Utsuno},
  \citenamefont {Otsuka}, \citenamefont {Glasmacher}, \citenamefont
  {Mizusaki},\ and\ \citenamefont {Honma}}]{utsuno2004}%
  \BibitemOpen
  \bibfield  {author} {\bibinfo {author} {\bibnamefont {Utsuno}, \bibfnamefont
  {Y.}}, \bibinfo {author} {\bibfnamefont {T.}~\bibnamefont {Otsuka}}, \bibinfo
  {author} {\bibfnamefont {T.}~\bibnamefont {Glasmacher}}, \bibinfo {author}
  {\bibfnamefont {T.}~\bibnamefont {Mizusaki}}, \ and\ \bibinfo {author}
  {\bibfnamefont {M.}~\bibnamefont {Honma}}} (\bibinfo {year} {2004}),\ \href
  {\doibase 10.1103/PhysRevC.70.044307} {\bibfield  {journal} {\bibinfo
  {journal} {Phys. Rev. C}\ }\textbf {\bibinfo {volume} {70}}~(\bibinfo
  {number} {4}),\ \bibinfo {pages} {044307}}\BibitemShut {NoStop}%
\bibitem [{\citenamefont {Vary}\ \emph {et~al.}(2009)\citenamefont {Vary},
  \citenamefont {Maris}, \citenamefont {Ng}, \citenamefont {Yang},\ and\
  \citenamefont {Sosonkina}}]{vary2009}%
  \BibitemOpen
  \bibfield  {author} {\bibinfo {author} {\bibnamefont {Vary}, \bibfnamefont
  {J.~P.}}, \bibinfo {author} {\bibfnamefont {P.}~\bibnamefont {Maris}},
  \bibinfo {author} {\bibfnamefont {E.}~\bibnamefont {Ng}}, \bibinfo {author}
  {\bibfnamefont {C.}~\bibnamefont {Yang}}, \ and\ \bibinfo {author}
  {\bibfnamefont {M.}~\bibnamefont {Sosonkina}}} (\bibinfo {year} {2009}),\
  \href {http://stacks.iop.org/1742-6596/180/i=1/a=012083} {\bibfield
  {journal} {\bibinfo  {journal} {Journal of Physics: Conference Series}\
  }\textbf {\bibinfo {volume} {180}}~(\bibinfo {number} {1}),\ \bibinfo {pages}
  {012083}}\BibitemShut {NoStop}%
\bibitem [{\citenamefont {Volya}\ and\ \citenamefont
  {Zelevinsky}(2005)}]{volya2005}%
  \BibitemOpen
  \bibfield  {author} {\bibinfo {author} {\bibnamefont {Volya}, \bibfnamefont
  {A.}}, \ and\ \bibinfo {author} {\bibfnamefont {V.}~\bibnamefont
  {Zelevinsky}}} (\bibinfo {year} {2005}),\ \href {\doibase
  10.1103/PhysRevLett.94.052501} {\bibfield  {journal} {\bibinfo  {journal}
  {Phys. Rev. Lett.}\ }\textbf {\bibinfo {volume} {94}},\ \bibinfo {pages}
  {052501}}\BibitemShut {NoStop}%
\bibitem [{\citenamefont {Voss}\ \emph {et~al.}(2012)\citenamefont {Voss},
  \citenamefont {Baugher}, \citenamefont {Bazin}, \citenamefont {Clark},
  \citenamefont {Crawford}, \citenamefont {Dewald}, \citenamefont {Fallon},
  \citenamefont {Gade}, \citenamefont {Grinyer}, \citenamefont {Iwasaki},
  \citenamefont {Macchiavelli}, \citenamefont {McDaniel}, \citenamefont
  {Miller}, \citenamefont {Petri}, \citenamefont {Ratkiewicz}, \citenamefont
  {Rother}, \citenamefont {Starosta}, \citenamefont {Walsh}, \citenamefont
  {Weisshaar}, \citenamefont {Forss\'en}, \citenamefont {Roth},\ and\
  \citenamefont {Navr\'atil}}]{voss2012}%
  \BibitemOpen
  \bibfield  {author} {\bibinfo {author} {\bibnamefont {Voss}, \bibfnamefont
  {P.}}, \bibinfo {author} {\bibfnamefont {T.}~\bibnamefont {Baugher}},
  \bibinfo {author} {\bibfnamefont {D.}~\bibnamefont {Bazin}}, \bibinfo
  {author} {\bibfnamefont {R.~M.}\ \bibnamefont {Clark}}, \bibinfo {author}
  {\bibfnamefont {H.~L.}\ \bibnamefont {Crawford}}, \bibinfo {author}
  {\bibfnamefont {A.}~\bibnamefont {Dewald}}, \bibinfo {author} {\bibfnamefont
  {P.}~\bibnamefont {Fallon}}, \bibinfo {author} {\bibfnamefont
  {A.}~\bibnamefont {Gade}}, \bibinfo {author} {\bibfnamefont {G.~F.}\
  \bibnamefont {Grinyer}}, \bibinfo {author} {\bibfnamefont {H.}~\bibnamefont
  {Iwasaki}}, \bibinfo {author} {\bibfnamefont {A.~O.}\ \bibnamefont
  {Macchiavelli}}, \bibinfo {author} {\bibfnamefont {S.}~\bibnamefont
  {McDaniel}}, \bibinfo {author} {\bibfnamefont {D.}~\bibnamefont {Miller}},
  \bibinfo {author} {\bibfnamefont {M.}~\bibnamefont {Petri}}, \bibinfo
  {author} {\bibfnamefont {A.}~\bibnamefont {Ratkiewicz}}, \bibinfo {author}
  {\bibfnamefont {W.}~\bibnamefont {Rother}}, \bibinfo {author} {\bibfnamefont
  {K.}~\bibnamefont {Starosta}}, \bibinfo {author} {\bibfnamefont {K.~A.}\
  \bibnamefont {Walsh}}, \bibinfo {author} {\bibfnamefont {D.}~\bibnamefont
  {Weisshaar}}, \bibinfo {author} {\bibfnamefont {C.}~\bibnamefont
  {Forss\'en}}, \bibinfo {author} {\bibfnamefont {R.}~\bibnamefont {Roth}}, \
  and\ \bibinfo {author} {\bibfnamefont {P.}~\bibnamefont {Navr\'atil}}}
  (\bibinfo {year} {2012}),\ \href {\doibase 10.1103/PhysRevC.86.011303}
  {\bibfield  {journal} {\bibinfo  {journal} {Phys. Rev. C}\ }\textbf {\bibinfo
  {volume} {86}},\ \bibinfo {pages} {011303}}\BibitemShut {NoStop}%
\bibitem [{\citenamefont {Waltersson}\ \emph {et~al.}(2013)\citenamefont
  {Waltersson}, \citenamefont {Wessl\'en},\ and\ \citenamefont
  {Lindroth}}]{waltersson2013}%
  \BibitemOpen
  \bibfield  {author} {\bibinfo {author} {\bibnamefont {Waltersson},
  \bibfnamefont {E.}}, \bibinfo {author} {\bibfnamefont {C.~J.}\ \bibnamefont
  {Wessl\'en}}, \ and\ \bibinfo {author} {\bibfnamefont {E.}~\bibnamefont
  {Lindroth}}} (\bibinfo {year} {2013}),\ \href {\doibase
  10.1103/PhysRevB.87.035112} {\bibfield  {journal} {\bibinfo  {journal} {Phys.
  Rev. B}\ }\textbf {\bibinfo {volume} {87}},\ \bibinfo {pages}
  {035112}}\BibitemShut {NoStop}%
\bibitem [{\citenamefont {Wegner}(1994)}]{wegner1994}%
  \BibitemOpen
  \bibfield  {author} {\bibinfo {author} {\bibnamefont {Wegner}, \bibfnamefont
  {F.}}} (\bibinfo {year} {1994}),\ \href {\doibase 10.1002/andp.19945060203}
  {\bibfield  {journal} {\bibinfo  {journal} {Annalen der Physik}\ }\textbf
  {\bibinfo {volume} {506}}~(\bibinfo {number} {2}),\ \bibinfo {pages}
  {77}}\BibitemShut {NoStop}%
\bibitem [{\citenamefont {Weinberg}(1990)}]{weinberg1990}%
  \BibitemOpen
  \bibfield  {author} {\bibinfo {author} {\bibnamefont {Weinberg},
  \bibfnamefont {S.}}} (\bibinfo {year} {1990}),\ \href {\doibase
  10.1016/0370-2693(90)90938-3} {\bibfield  {journal} {\bibinfo  {journal}
  {Physics Letters B}\ }\textbf {\bibinfo {volume} {251}}~(\bibinfo {number}
  {2}),\ \bibinfo {pages} {288 }}\BibitemShut {NoStop}%
\bibitem [{\citenamefont {Weinberg}(1991)}]{weinberg1991}%
  \BibitemOpen
  \bibfield  {author} {\bibinfo {author} {\bibnamefont {Weinberg},
  \bibfnamefont {S.}}} (\bibinfo {year} {1991}),\ \href {\doibase
  10.1016/0550-3213(91)90231-L} {\bibfield  {journal} {\bibinfo  {journal}
  {Nuclear Physics B}\ }\textbf {\bibinfo {volume} {363}}~(\bibinfo {number}
  {1}),\ \bibinfo {pages} {3 }}\BibitemShut {NoStop}%
\bibitem [{\citenamefont {Wienholtz}\ \emph {et~al.}(2013)\citenamefont
  {Wienholtz}, \citenamefont {Beck}, \citenamefont {Blaum}, \citenamefont
  {Borgmann}, \citenamefont {Breitenfeldt}, \citenamefont {Cakirli},
  \citenamefont {George}, \citenamefont {Herfurth}, \citenamefont {Holt},
  \citenamefont {Kowalska}, \citenamefont {Kreim}, \citenamefont {Lunney},
  \citenamefont {Manea}, \citenamefont {Menendez}, \citenamefont {Neidherr},
  \citenamefont {Rosenbusch}, \citenamefont {Schweikhard}, \citenamefont
  {Schwenk}, \citenamefont {Simonis}, \citenamefont {Stanja}, \citenamefont
  {Wolf},\ and\ \citenamefont {Zuber}}]{wienholtz2013}%
  \BibitemOpen
  \bibfield  {author} {\bibinfo {author} {\bibnamefont {Wienholtz},
  \bibfnamefont {F.}}, \bibinfo {author} {\bibfnamefont {D.}~\bibnamefont
  {Beck}}, \bibinfo {author} {\bibfnamefont {K.}~\bibnamefont {Blaum}},
  \bibinfo {author} {\bibfnamefont {C.}~\bibnamefont {Borgmann}}, \bibinfo
  {author} {\bibfnamefont {M.}~\bibnamefont {Breitenfeldt}}, \bibinfo {author}
  {\bibfnamefont {R.~B.}\ \bibnamefont {Cakirli}}, \bibinfo {author}
  {\bibfnamefont {S.}~\bibnamefont {George}}, \bibinfo {author} {\bibfnamefont
  {F.}~\bibnamefont {Herfurth}}, \bibinfo {author} {\bibfnamefont {J.~D.}\
  \bibnamefont {Holt}}, \bibinfo {author} {\bibfnamefont {M.}~\bibnamefont
  {Kowalska}}, \bibinfo {author} {\bibfnamefont {S.}~\bibnamefont {Kreim}},
  \bibinfo {author} {\bibfnamefont {D.}~\bibnamefont {Lunney}}, \bibinfo
  {author} {\bibfnamefont {V.}~\bibnamefont {Manea}}, \bibinfo {author}
  {\bibfnamefont {J.}~\bibnamefont {Menendez}}, \bibinfo {author}
  {\bibfnamefont {D.}~\bibnamefont {Neidherr}}, \bibinfo {author}
  {\bibfnamefont {M.}~\bibnamefont {Rosenbusch}}, \bibinfo {author}
  {\bibfnamefont {L.}~\bibnamefont {Schweikhard}}, \bibinfo {author}
  {\bibfnamefont {A.}~\bibnamefont {Schwenk}}, \bibinfo {author} {\bibfnamefont
  {J.}~\bibnamefont {Simonis}}, \bibinfo {author} {\bibfnamefont
  {J.}~\bibnamefont {Stanja}}, \bibinfo {author} {\bibfnamefont {R.~N.}\
  \bibnamefont {Wolf}}, \ and\ \bibinfo {author} {\bibfnamefont
  {K.}~\bibnamefont {Zuber}}} (\bibinfo {year} {2013}),\ \href
  {http://dx.doi.org/10.1038/nature12226} {\bibfield  {journal} {\bibinfo
  {journal} {Nature}\ }\textbf {\bibinfo {volume} {498}}~(\bibinfo {number}
  {7454}),\ \bibinfo {pages} {346}}\BibitemShut {NoStop}%
\bibitem [{\citenamefont {Wiringa}\ \emph {et~al.}(1995)\citenamefont
  {Wiringa}, \citenamefont {Stoks},\ and\ \citenamefont
  {Schiavilla}}]{wiringa1995}%
  \BibitemOpen
  \bibfield  {author} {\bibinfo {author} {\bibnamefont {Wiringa}, \bibfnamefont
  {R.~B.}}, \bibinfo {author} {\bibfnamefont {V.~G.~J.}\ \bibnamefont {Stoks}},
  \ and\ \bibinfo {author} {\bibfnamefont {R.}~\bibnamefont {Schiavilla}}}
  (\bibinfo {year} {1995}),\ \href {\doibase 10.1103/PhysRevC.51.38} {\bibfield
   {journal} {\bibinfo  {journal} {Phys. Rev. C}\ }\textbf {\bibinfo {volume}
  {51}},\ \bibinfo {pages} {38}}\BibitemShut {NoStop}%
\bibitem [{\citenamefont {W\l{}och}\ \emph {et~al.}(2005)\citenamefont
  {W\l{}och}, \citenamefont {Dean}, \citenamefont {Gour}, \citenamefont
  {Hjorth-Jensen}, \citenamefont {Kowalski}, \citenamefont {Papenbrock},\ and\
  \citenamefont {Piecuch}}]{wloch2005}%
  \BibitemOpen
  \bibfield  {author} {\bibinfo {author} {\bibnamefont {W\l{}och},
  \bibfnamefont {M.}}, \bibinfo {author} {\bibfnamefont {D.~J.}\ \bibnamefont
  {Dean}}, \bibinfo {author} {\bibfnamefont {J.~R.}\ \bibnamefont {Gour}},
  \bibinfo {author} {\bibfnamefont {M.}~\bibnamefont {Hjorth-Jensen}}, \bibinfo
  {author} {\bibfnamefont {K.}~\bibnamefont {Kowalski}}, \bibinfo {author}
  {\bibfnamefont {T.}~\bibnamefont {Papenbrock}}, \ and\ \bibinfo {author}
  {\bibfnamefont {P.}~\bibnamefont {Piecuch}}} (\bibinfo {year} {2005}),\ \href
  {\doibase 10.1103/PhysRevLett.94.212501} {\bibfield  {journal} {\bibinfo
  {journal} {Phys. Rev. Lett.}\ }\textbf {\bibinfo {volume} {94}},\ \bibinfo
  {pages} {212501}}\BibitemShut {NoStop}%
\bibitem [{\citenamefont {Wuosmaa}\ \emph {et~al.}(2005)\citenamefont
  {Wuosmaa}, \citenamefont {Rehm}, \citenamefont {Greene}, \citenamefont
  {Henderson}, \citenamefont {Janssens}, \citenamefont {Jiang}, \citenamefont
  {Jisonna}, \citenamefont {Moore}, \citenamefont {Pardo}, \citenamefont
  {Paul}, \citenamefont {Peterson}, \citenamefont {Pieper}, \citenamefont
  {Savard}, \citenamefont {Schiffer}, \citenamefont {Segel}, \citenamefont
  {Sinha}, \citenamefont {Tang},\ and\ \citenamefont {Wiringa}}]{wuosma2005}%
  \BibitemOpen
  \bibfield  {author} {\bibinfo {author} {\bibnamefont {Wuosmaa}, \bibfnamefont
  {A.~H.}}, \bibinfo {author} {\bibfnamefont {K.~E.}\ \bibnamefont {Rehm}},
  \bibinfo {author} {\bibfnamefont {J.~P.}\ \bibnamefont {Greene}}, \bibinfo
  {author} {\bibfnamefont {D.~J.}\ \bibnamefont {Henderson}}, \bibinfo {author}
  {\bibfnamefont {R.~V.~F.}\ \bibnamefont {Janssens}}, \bibinfo {author}
  {\bibfnamefont {C.~L.}\ \bibnamefont {Jiang}}, \bibinfo {author}
  {\bibfnamefont {L.}~\bibnamefont {Jisonna}}, \bibinfo {author} {\bibfnamefont
  {E.~F.}\ \bibnamefont {Moore}}, \bibinfo {author} {\bibfnamefont {R.~C.}\
  \bibnamefont {Pardo}}, \bibinfo {author} {\bibfnamefont {M.}~\bibnamefont
  {Paul}}, \bibinfo {author} {\bibfnamefont {D.}~\bibnamefont {Peterson}},
  \bibinfo {author} {\bibfnamefont {S.~C.}\ \bibnamefont {Pieper}}, \bibinfo
  {author} {\bibfnamefont {G.}~\bibnamefont {Savard}}, \bibinfo {author}
  {\bibfnamefont {J.~P.}\ \bibnamefont {Schiffer}}, \bibinfo {author}
  {\bibfnamefont {R.~E.}\ \bibnamefont {Segel}}, \bibinfo {author}
  {\bibfnamefont {S.}~\bibnamefont {Sinha}}, \bibinfo {author} {\bibfnamefont
  {X.}~\bibnamefont {Tang}}, \ and\ \bibinfo {author} {\bibfnamefont {R.~B.}\
  \bibnamefont {Wiringa}}} (\bibinfo {year} {2005}),\ \href {\doibase
  10.1103/PhysRevLett.94.082502} {\bibfield  {journal} {\bibinfo  {journal}
  {Phys. Rev. Lett.}\ }\textbf {\bibinfo {volume} {94}},\ \bibinfo {pages}
  {082502}}\BibitemShut {NoStop}%
\bibitem [{\citenamefont {Zabolitzky}(1974)}]{zabolitzky1974}%
  \BibitemOpen
  \bibfield  {author} {\bibinfo {author} {\bibnamefont {Zabolitzky},
  \bibfnamefont {J.~G.}}} (\bibinfo {year} {1974}),\ \href {\doibase
  10.1016/0375-9474(74)90432-1} {\bibfield  {journal} {\bibinfo  {journal}
  {Nuclear Physics A}\ }\textbf {\bibinfo {volume} {228}}~(\bibinfo {number}
  {2}),\ \bibinfo {pages} {272 }}\BibitemShut {NoStop}%
\bibitem [{\citenamefont {Zuker}(2003)}]{zuker2003}%
  \BibitemOpen
  \bibfield  {author} {\bibinfo {author} {\bibnamefont {Zuker}, \bibfnamefont
  {A.~P.}}} (\bibinfo {year} {2003}),\ \href {\doibase
  10.1103/PhysRevLett.90.042502} {\bibfield  {journal} {\bibinfo  {journal}
  {Phys. Rev. Lett.}\ }\textbf {\bibinfo {volume} {90}},\ \bibinfo {pages}
  {042502}}\BibitemShut {NoStop}%
\end{thebibliography}%
